\documentclass[english,3p,floatfix]{elsarticle}
\usepackage{float}
\usepackage{amsmath}
\usepackage{bm}
\usepackage{graphicx}
\usepackage{amssymb}
\usepackage{esint}
\usepackage{array}

\DeclareMathOperator{\Tr}{Tr}
\DeclareMathOperator{\Real}{Re}
\DeclareMathOperator{\sgn}{sgn}
\DeclareMathOperator{\Imag}{Im}
\makeatletter

\providecommand{\tabularnewline}{\\}
\providecommand{\eprint}[2][]{\url{#2}}

\usepackage{mathrsfs}\usepackage{amsfonts}\@ifundefined{definecolor}
 {\@ifundefined{definecolor}
 {\@ifundefined{definecolor}
 {\@ifundefined{definecolor}
 {\usepackage{color}}{}
}{} }{} }{}
\usepackage[colorlinks,linkcolor=blue]{hyperref}

\makeatother

\usepackage{babel}

\makeatother

\usepackage{babel}

\makeatother

\usepackage{babel}

\makeatother

\usepackage{babel}

\begin{document}

\title{Electronic properties of graphene-based bilayer systems}

\author[RIKEN,ITAE,MIPT]{A.V. Rozhkov}

\author[RIKEN,ITAE]{A.O. Sboychakov}

\author[RIKEN,ITAE,MIPT,VNIIA]{A.L. Rakhmanov}

\author[RIKEN,UM]{Franco Nori}

\ead{fnori@riken.jp }

\address[RIKEN]{CEMS, RIKEN, Saitama 351-0198, Japan}

\address[ITAE]{Institute for Theoretical and Applied Electrodynamics,
Russian Academy of Sciences, 125412 Moscow, Russia}

\address[MIPT]{Moscow Institute of Physics and Technology, Dolgoprudny,
Moscow Region, 141700 Russia}

\address[VNIIA]{All-Russia Research Institute of Automatics, Moscow, 127055
Russia}

\address[UM]{Physics Department, The University of Michigan, Ann Arbor,
Michigan 48109-1040, USA}
\begin{abstract}
This article reviews the theoretical and experimental work related to the
electronic properties of bilayer graphene systems. Three types of bilayer
stackings are discussed: the AA, AB, and twisted bilayer graphene. This
review covers single-electron properties, effects of static electric and
magnetic fields, bilayer-based mesoscopic systems, spin-orbit coupling,
dc transport and optical response, as well as spontaneous symmetry
violation and other interaction effects. The selection of the material aims
to introduce the reader to the most commonly studied topics of theoretical
and experimental research in bilayer graphene.
\end{abstract}
\maketitle
\tableofcontents{}

~
\section*{Notation used in the text}
%
%
%
%


\textbf{Section 1}

$c_0$: interlayer distance

\vspace*{0.1in}

\textbf{Section 2}

$\sigma, \sigma_{\alpha\beta}$: conductivity

$n$: electron density

\vspace*{0.1in}

\textbf{Section 3}

$\mathbf{a}_{1,2}$: elementary lattice vectors

$a_0$: distance between nearest carbon atoms in the layer

$t$: hoping amplitude between in-plane nearest-neighbor carbon atoms

$t_0$: hoping amplitude between inter-plane nearest-neighbor carbon atoms

$t'$: hoping amplitude between in-plane next-nearest-neighbor carbon atoms

$t_g$: hoping amplitude between inter-plane next-nearest-neighbor carbon atoms

$t_3$: hoping amplitude between nearest-neighbor non-dimer carbon atoms

$t_4$: hoping amplitude between nearest-neighbor non-dimer to dimer carbon
atoms

$\sigma$: spin projection

${\cal N}$: number of unit cells

$f(\mathbf{k}) = \sum_{j=1}^3\exp(i\mathbf{k}\mathbf{\delta}_j)$

$F(\mathbf{k})=2\cos{(\sqrt{3}k_ya_0)}+ 4\cos{(\sqrt{3}k_ya_0/2)}\cos{(3k_xa_0/2)}$

$\mathbf{b}_{1,2}$: elementary reciprocal-lattice vectors

$\bm{\delta}_j$: three vectors ($j=1,2,3$) connecting nearest-neighbor
sites in honeycomb lattice

$\mathbf{K},\mathbf{K}'$: locations of the Dirac points in the reciprocal
space

$\varepsilon^{(s)}_\mathbf{k}$:  electronic bands

$\hat{\bm{\sigma}}=(\sigma_x,\sigma_y,\sigma_z)$: Pauli matrices

$\rho(\varepsilon)$: density of states

$\mu$: chemical potential

$V_{BZ}$: volume of Brillouin zone

$V$: bias voltage

$n_{\mathbf{j}\alpha\sigma}$: the electron number operator 

$\xi=\pm 1$: valley index

$m_0$: free electron mass

$m=t_0/2v_F^2$: effective charge carrier mass

$p_L$: shift of the cones from Brillouin zone corner due to trigonal warping

$\varepsilon_L$: energy of saddle points due to trigonal warping

\vspace*{0.1in}

\textbf{Section 4}

$\hat{\chi}_{1,2}$: chirality operators

$\eta_{1,2}$: band indices

$\gamma^5$: 5th Dirac matrix

$c=\eta_1\eta_2$: cone index

$\chi=\eta_1$: chirality index

$\hat{\pi}=\xi\partial/\partial x+i\partial/\partial y$

$J$: winding number

\vspace*{0.1in}

\textbf{Section 5}

$\Phi_0 = 2\pi \hbar c / e$: magnetic flux quantum

$\omega_c$: cyclotron frequency

$l_b$: magnetic length

$E_{LL}(n)$: $n$th Landau level energy

$N_f$: degeneracy factor

$N$: electron number

$\nu$: filling factor

$\Theta_F$: Faraday rotation angle

\vspace*{0.1in}

\textbf{Section 6}

$t_R$: Rashba spin-orbit strength

$\mathbf{d_{ij}}$: lattice vector from site $\mathbf{i}$ to site
$\mathbf{j}$

$M$: magnetization

\vspace*{0.1in}

\textbf{Section 8}

$\varepsilon_0=4\varepsilon_L$: energy scale characterizing the trigonal
warping

$\Sigma(\varepsilon)$: self-energy

$n_i$: impurity concentration

$u$: energy characterizing the impurity strength

$x$: doping

$\tilde{x}$: local doping

$x^2_{\textrm{rms}}=\langle\tilde{x}^2\rangle-x^2$

$\rho_\infty$: density of states asymptotic at high energy

$W=n_iu^2\rho_\infty/2\varepsilon_0$: dimensionless disorder strength

$\Gamma=\pi W\varepsilon_0/2$: scattering rate due to disorder

$\mu_c$: mobility

\vspace*{0.1in}

\textbf{Section 9}

$\hat{G}(z)$: Green's function

$\omega_n$: Matsubara frequency

$f(\varepsilon)$: Fermi function

$\mu$: chemical potential

$A_{ij}$: amplitudes of the spectral representation of Green's function

$\Gamma$: phenomenological broadening constant

\vspace*{0.1in}

\textbf{Section 10}

$U_0$: on-site Coulomb repulsion

$U_{01}$: nearest-neighbor in-plane Coulomb repulsion

$U_{02}$: next-nearest-neighbor in-plane Coulomb repulsion

$U_{11}$: nearest-neighbor inter-plane Coulomb repulsion

$\Delta_{k\sigma}^{xy}$: order parameters

$\Omega$: grand potential

$\mu'=\mu-U_0x/2$: shifted chemical potential

$\mathbf{q}$: vector characterizing incommensurability of the
antiferromagnetic state

$\mathbf{S}_{\mathbf{n}\alpha a}$: average electron spin at site $\mathbf{n}$

$V(\mathbf{q})$: Coulomb potential in the momentum space

\vspace*{0.1in}

\textbf{Section 11}

$\epsilon(\mathbf{k},\omega)$: dielectric function

$\Pi(\mathbf{k},\omega)$: irreducible polarization

\vspace*{0.1in}

\textbf{Section 13}

$\bar{A}$: in this section means the complex conjugate $A$

$\theta$: twist angle between two graphene layers

$L$: Moir\`{e} period

$m_0,r$: mutually prime positive integers defining a particular
``commensurate'' twist angle

$N(m_0,r)$ or $N$: number of sites in the elementary unit cell of the
superlattice

$\mathbf{R}_{1,2}$: superlattice vectors

$L_{SC}=|\mathbf{R}_{1,2}|$: supercell linear size

$\mathbf{G}_{1,2}$: basis vectors of the reciprocal superlattice

$\mathbf{K}_{\theta},\mathbf{K}^{\prime}_{\theta}$: Dirac points of the top (rotated) layer

$\Delta \mathbf{K}=\mathbf{K}_{\theta}-\mathbf{K}$

$\mathbf{K}_{1,2}$: non-equivalent Dirac points of the reciprocal superlattice

$t_\bot(\mathbf{r})$ and $\tilde{t}_\bot^{\alpha\beta}(\mathbf{G})$: interlayer hoping amplitude and its Fourier transform

$T^{\alpha\beta}_\bot$: interlayer hoping Hamiltonian

$\bm{\varrho}$: relative shift between the layers

$v_F^*$: renormalized Fermi velocity

$\theta_c$: twist angle at which $v_F^*(\theta_c)=0$

$\Delta E_{vHs}$: energy difference between two van Hove singularities

$V_{pp\pi}(r)$ and $V_{pp\sigma}(r)$: Slater-Koster parameters

$\Delta_s$: band splitting

$\omega_c^*$: twist-dependent cyclotron frequency

\vspace*{0.1in}


\section{Introduction}

Systems with only carbon atoms show a number of different structures with a
variety of physical properties because of the flexibility of its chemical
bonding. Among these systems there are allotropes of different
dimensionality from three-dimensional (3D) graphite and diamond to
low-dimensional structures, such as fullerene and carbon nanotubes, which
could be thought as zero-dimensional and one-dimensional objects
respectively. A two-dimensional (2D) carbon allotrope graphene was the
first produced stable 2D
crystal~\cite{NG1}.
Graphene is composed of carbon atoms forming a honeycomb lattice. This 2D
system has a number of intriguing physical properties that are of interest
for fundamental physics and important applications. It is also important
and interesting to increase the family of 2D solids. A possible approach in
this direction is to produce 2D structures of different chemical
compositions such as silicene, nitrogen boride, etc. A different method is
to manufacture few-layer carbon systems, of which the simplest and the most
obvious is the graphene bilayer. The electronic properties of the graphene
bilayer is the subject of the present review.

The graphene bilayer can exist in three modifications: AA, AB (or Bernal
phase), and twisted bilayer. The simplest form is the AA
bilayer~\cite{liu_aa_exp2009},
in which each carbon atom of the second layer is placed exactly above the
corresponding atom of the first carbon sheet. However, this structure is
likely to be metastable, and only few authors reported manufacturing AA
samples. In the AB bilayer, or Bernal phase, half of the carbon atoms of
the top layer are above the carbon atoms of the lower layer, while other
atoms located above the centers of the lower-layer hexagons (in
naturally-occurring graphite the carbon atoms are ordered in similar
manner). The Bernal (or AB) bilayer graphene is the most stable, and its
high-quality samples are produced and studied in many experiments. In the
third type of bilayer graphene structure, the top carbon layer is rotated
with respect to the lower layer by some angle
$\theta$~\cite{dSPRL}.
Such a structure is also stable, and the twisted samples are produced using
some special technological processes. The electronic properties of the
bilayer structures listed above are rather different. Thus, in the chapters
of this review we usually consider each of them separately.

The theories of electronic properties of the bilayer systems are built
upon the knowledge of the electronic structure of single-layer
graphene~\cite{CastrNrev}.
It is established that, for single-layer graphene, the
$sp^2$
hybridization between the carbon $s$ orbital and two $p$ orbitals enables
the formation of a $\sigma$ bond between neighboring atoms. The $\sigma$
bond is responsible for the robustness of the lattice structure in all
carbon
allotropes. Because of the Pauli principle, these bands have a filled
shell. As a result, a valence band appears. The
$p_z$
orbital, which is perpendicular to the graphene plane, binds covalently
with carbon atoms nearby. This leads to the formation of a $\pi$ band.
Since every $p$ orbital hosts one extra electron, the $\pi$ band is
half-filled.

In the equilibrium configuration, the interlayer spacing in the AB bilayer
is
estimated~\cite{mccann_kosh_rev2013}
as
$c_0=3.35$~{\AA}.
In the AA and twisted bilayer samples, the interlayer distance could be
slightly different. Furthermore, in twisted bilayer graphene the interlayer
distance is spatially
modulated~\cite{STM1}.
The nature of the interaction between graphene sheets in bilayer graphene
was analyzed by many authors (see, e.g.,
Refs.~\cite{bi1,bi2,bi3,bi4,bi5}).
The hybridization between the $s$ orbital and the
$p_z$
orbitals in the bilayers is more cumbersome than in the single-layer
graphene. However, as in the case of the single layer, the $\pi$ band is
formed in bilayer graphene and the $\pi$ band is half-filled if the sample
is not doped.

The charge carriers in single-layer graphene are massless chiral
quasiparticles with a linear dispersion, as described by a Dirac-like
effective
Hamiltonian~\cite{CastrNrev}.
This was confirmed in a number of experimental observations, particularly,
the Landau quantization in the magnetic field and the integer quantum Hall
effect. The low-energy Hamiltonian of bilayer graphene may be viewed
as a generalization of the Dirac-like Hamiltonian of monolayer graphene,
resulting in a more complicated picture of electron and hole dispersion.
For example, AB bilayer quasiparticles demonstrate parabolic dispersion at
low energies and linear dispersion at higher energies. As we will see
below, the electronic spectrum of bilayer graphene may be gapped or
gapless. There are several mechanisms, which could induce a single-electron
gap in a bilayer sample. Such a mechanism may, or may not,
rely on the electron-electron interaction. Depending on a particular
situation, the gap in the spectrum can be controlled by doping, gate
voltage, or some other parameter. The possibility of having a
graphene-based system with the gap in the spectrum is of interest for
applications in electronic devices.

Many macroscopic properties of bilayer graphene are similar to that of
the monolayer samples. In particular, the well-prepared samples have high
electric and thermal conductivity, high mechanical stiffness, high
transparency with respect to white light, impermeability to gases, and the
ability to be chemically functionalized (see, e.g., the review in
Ref.~\cite{mccann_kosh_rev2013}
and the papers cited therein). For some applications, the graphene bilayers
may have definite advantages over the monolayer due to larger possibilities
for tuning their physical properties. All this makes the study of bilayer
graphene a timely and important endeavor.

Currently, in graphene-related reviews, it is an accepted marketing device
to scare the readers with huge numbers of published papers dedicated to the
graphene research. Following this trend, let us remind that in years
2014-2015 at least 1000 publications mentioned bilayer graphene. Clearly,
for navigation in such a sea of information a role of review papers must
not be underestimated. While this review aims to offer a more general and
synergetic coverage of the field, a reader interested in more 
specialized areas may consider more topical works. For example, paper of
Das~Sarma 
et al.~\cite{das_sarma_rev2011}
is focused primarily on transport properties of both single-layer and AB
bilayer graphene. Bilayer photonics applications are discussed in
Ref.~\cite{yanRev}. 
Publications regarding plasmons in graphene-based systems are reviewed by
T.~Stauber in
Ref.~\cite{plasmonicsRev}.
Nonlinear optical phenomena of single-layer and AB bilayer were subject of
Ref.~\cite{glazovRev}.
Mini-review of the works on twisted bilayer is
Ref.~\cite{MeleReview}.
Finally, let us cite several reviews,
Refs.~\cite{Abergel,meso_review,Avouris2007},
which might be useful for understanding the subject.

Regarding the organization of the presented material, we would like to
comment that the AB and AA bilayers are reviewed in parallel. Due to
peculiar geometry of the twisted bilayer lattice, this type of the graphene
bilayer is discussed separately, in
Section~\ref{spectraTw}.

\section{AA and AB graphene bilayers: available samples}
\label{samples}
%
%

\subsection{AA-stacked bilayer graphene}
\label{subsec::sample::AA}

We would like to start the review with a brief discussion of the available
samples and their quality. To this date, both theoretical and experimental
reasearch of AA-stacked bilayer lags behind the studies of AB bilayer by a
significant margin. While several papers described preparation of AA
bilayer
samples~\cite{liu_aa_exp2009,
lee_aa_exp2008,
lauffer_aa_exp2008,
borysiuk2011_aa},
most of the available data are gathered either by transmission electron
microscopy
(TEM)~\cite{liu_aa_exp2009,
lee_aa_exp2008,
borysiuk2011_aa},
or by scanning tunneling microscopy
(STM)~\cite{lauffer_aa_exp2008}.
These are imaging methods. As a result, the lattice structure of the
samples is visualized with high precision. Unfortunately, these techniques
give virtually no information about the electronic properties. At present,
we cannot say how good or bad the available samples are in terms of
mobility or other electronic characteristics. However, witnessing the
growing theoretical interest in AA-stacked systems, we think that the
situation with experiment will improve.

\subsection{AB-stacked bilayer graphene}
\label{subsec::sample::AB}

The study of AB bilayer graphene is a very active and diverse area of
research. As a result, several publications reported the preparation of AB
samples of excellent quality. In modern graphene literature it is common to
characterize the quality of samples by their electron mobility (although,
there is an exception to this
rule~\cite{Martin2010}).
Interpreting the relevant data one has to remember two points. First, it is
well-known that single-layer graphene has large mobility: even in the
earliest 
samples~\cite{Novoselov2005}
the mobility was as high as
$10^4$\,cm$^2$/Vs.
This is about $10$ times larger than that of silicon. For a suspended
single-layer sample it is even higher (e.g.,
K.~Bolotin et~al.~\cite{Bolotin2008}
reported the value of
$2\times 10^5$\,cm$^2$/Vs).
Therefore, one should not be surprised by the extremely high mobility
values of graphene-based systems.
Second, since this property is not directly accessible from experiments, it
must be extracted from measurements of other quantities. For example,
A.\,S.~Mayorov et~al.~\cite{Mayorov2011}
estimated the mobility as 
$c/B_0$,
where $c$ is the velocity of light,
$B_0$
is the threshold magnetic field for the onset of quantum oscillations. Such
quantity is called ``quantum mobility". In other papers, e.g.,
Refs.~\cite{Mayorov2011,Feldman2009},
the mobility was equated with the expression
$(1/e) d \sigma/dn$,
where $e$ is the electron charge, $\sigma$ is conductivity, and $n$ is
electron density. The discrepancy between two definitions can be as small
as
30\%~\cite{Mayorov2011},
or as large as 200\% (see Supporting Information for
Ref.~\cite{Bao2012}).
Thus, comparing data from different experimental groups one must
keep in mind the possible ambiguity in the experimental definition of this
quantity.

The samples should be divided into two classes: suspended samples and
samples on a substrate [e.g.,
SiO$_2$,
boron nitride, polymethylmethacrylate (PMMA)]. The former usually have
higher mobility because they are free from disorder introduced by the
substrate. The experimental data is summarized in
Table~\ref{table::ab_experiment}.

\begin{table}[t]
\begin{center}
\begin{tabular*}{\textwidth}{>{}m{2cm}>{\centering}m{5.5cm}>{\centering}m{5.5cm}>{\centering}m{2cm}}
\hline
\hline
{\vspace{1.5mm}Substrate\vspace{1mm}} &
{\vspace{1.5mm}Mobility, $\times10^4$\,cm$^2$/Vs\vspace{1mm}} &
{\vspace{1.5mm}Temperature\vspace{1mm}} &
{\vspace{1.5mm}References\vspace{1mm}}
\tabularnewline
\hline
\hline
{\vspace{1.5mm}
SiO$_2$~and~PMMA
\vspace{1.5mm}} &
{\vspace{1.5mm}
0.7
\vspace{1.5mm}}  &
{\vspace{1.5mm}
from liquid He to room $T$
\vspace{1.5mm}} &
{\vspace{1.5mm}
\cite{Morozov2008}
\vspace{1.5mm}}
\tabularnewline
\hline
{\vspace{1.5mm}
SiO$_2$
\vspace{1.5mm}} &
{\vspace{1.5mm}
0.3
\vspace{1.5mm}}  &
{\vspace{1.5mm}
unspecified
\vspace{1.5mm}} &
{\vspace{1.5mm}
\cite{Oostinga2008}
\vspace{1.5mm}}
\tabularnewline
\hline
{\vspace{1.5mm}
h-BN
\vspace{1.5mm}} &
{\vspace{1.5mm}
4
\vspace{1.5mm}}  &
{\vspace{1.5mm}
room $T$
\vspace{1.5mm}} &
{\vspace{1.5mm}
\cite{dean2010boron}
\vspace{1.5mm}}
\tabularnewline
\hline
{\vspace{1.5mm}
substrate material was not specified
\vspace{1.5mm}} &
{\vspace{1.5mm}
0.05\,--\,0.2
\vspace{1.5mm}}  &
{\vspace{1.5mm}
1.5\,K
\vspace{1.5mm}} &
{\vspace{1.5mm}
\cite{Bao2012}
\vspace{1.5mm}}
\tabularnewline
\hline
{\vspace{1.5mm}
Suspended
\vspace{1.5mm}} &
{\vspace{1.5mm}
0.6\,--\,35
\vspace{1.5mm}}  &
{\vspace{1.5mm}
most data were taken at 1.5\,K, highest mobility data were taken at
0.3\,K
\vspace{1.5mm}} &
{\vspace{1.5mm}
\cite{Bao2012}
\vspace{1.5mm}}
\tabularnewline
\hline
{\vspace{1.5mm}
Suspended
\vspace{1.5mm}} &
{\vspace{1.5mm}
1.0\,--\,1.5
\vspace{1.5mm}}  &
{\vspace{1.5mm}
some data were taken at 450\,mK
\vspace{1.5mm}} &
{\vspace{1.5mm}
\cite{Feldman2009}
\vspace{1.5mm}}
\tabularnewline
\hline
{\vspace{1.5mm}
Suspended
\vspace{1.5mm}} &
{\vspace{1.5mm}
50--150
\vspace{1.5mm}}  &
{\vspace{1.5mm}
unspecified
\vspace{1.5mm}} &
{\vspace{1.5mm}
\cite{Mayorov2011}
\vspace{1.5mm}}
\tabularnewline
\hline
{\vspace{1.5mm}
Suspended
\vspace{1.5mm}} &
{\vspace{1.5mm}
8--10
\vspace{1.5mm}}  &
{\vspace{1.5mm}
unspecified
\vspace{1.5mm}} &
{\vspace{1.5mm}
\cite{Velasco2012}
\vspace{1.5mm}}
\tabularnewline
\hline
{\vspace{1.5mm}
Suspended
\vspace{1.5mm}} &
{\vspace{1.5mm}
$>2$
\vspace{1.5mm}}  &
{\vspace{1.5mm}
4.2\,K
\vspace{1.5mm}} &
{\vspace{1.5mm}
\cite{Elferen2012}
\vspace{1.5mm}}
\tabularnewline
\hline
\hline
\end{tabular*}
\end{center}
\caption{
The quality of a bilayer sample is characterized by the mobility of its
charge carriers. In this table the summary of reported mobility data for
AB-bilayer graphene samples is presented.
\label{table::ab_experiment}
}
\end{table}


%
%
\section{Electron spectra: free electron approximation}
\label{spectra}
%

In this section we analyze the single-electron spectra of bilayer graphene.
Two alternative approaches are usually applied to derive the electronic
spectra of graphene systems, as well as many other materials. The first is
the ``ab-initio'' density functional theory (DFT) calculations. The second
is a tight-binding approximation, using as input parameters appropriate
hopping amplitudes. In many cases these amplitudes are calculated by means
of the DFT method. The results obtained in the framework of these two
approaches are almost similar for the graphene-like systems. The
tight-binding approach is more simple and physically transparent. Thus, in
this review we will follow the tight-binding calculations, mentionning the
DFT approach where it is possible.

Electronic spectra of different graphene systems can be classified as the
so-called Dirac-type spectra. For example, in single-layer graphene the
dispersion of the low-energy electronic states is linear, similar to the
dispersion of relativistic massless Dirac electrons. Accordingly,
relativistic effect, such as the Klein paradox, is observed in graphene.
Recently, a number of other systems of this type became a focus of intense
investigation, including topological insulators, and Weil semimetals.
Note, however, that the physical reasons giving rise to the Dirac-like
behavior of the electronic spectra may be quite different in different
systems. For example, in topological insulators it arises due to a strong
spin-orbital interaction, while in graphene it occurs due to the specific
symmetry of the crystalline lattice.

A graphene bilayer consists of two connected single layers of graphene.
Thus, the tight-binding Hamiltonian of the bilayer graphene can be
written as a sum of the Hamiltonians of two single-layer graphene sheets
and a term describing the electron hopping between these sheets. For
readers' convenience, we start with a brief review of the electronic
spectrum of single-layer graphene. Then, we consider bilayer graphene with
AA stacking,
(Section~\ref{spectraAA})
and with AB stacking
(Section~\ref{spectraAB}).
The electronic spectrum of twisted bilayer graphene is analyzed separately,
in
Section~\ref{spectraTw}.
In this chapter we disregard the effects of
electron-electron interaction, which, however, is not small in graphene
systems. The applicability condition of such an approach as well as the
effects of electron-electron coupling will be discussed in detail in
Section~\ref{interaction}.

\subsection{Single-layer graphene}\label{spectraSLG}

\subsubsection{Single-electron tight-binding description}
\label{SlgEB}

\begin{figure}
\includegraphics[width=0.45\columnwidth]{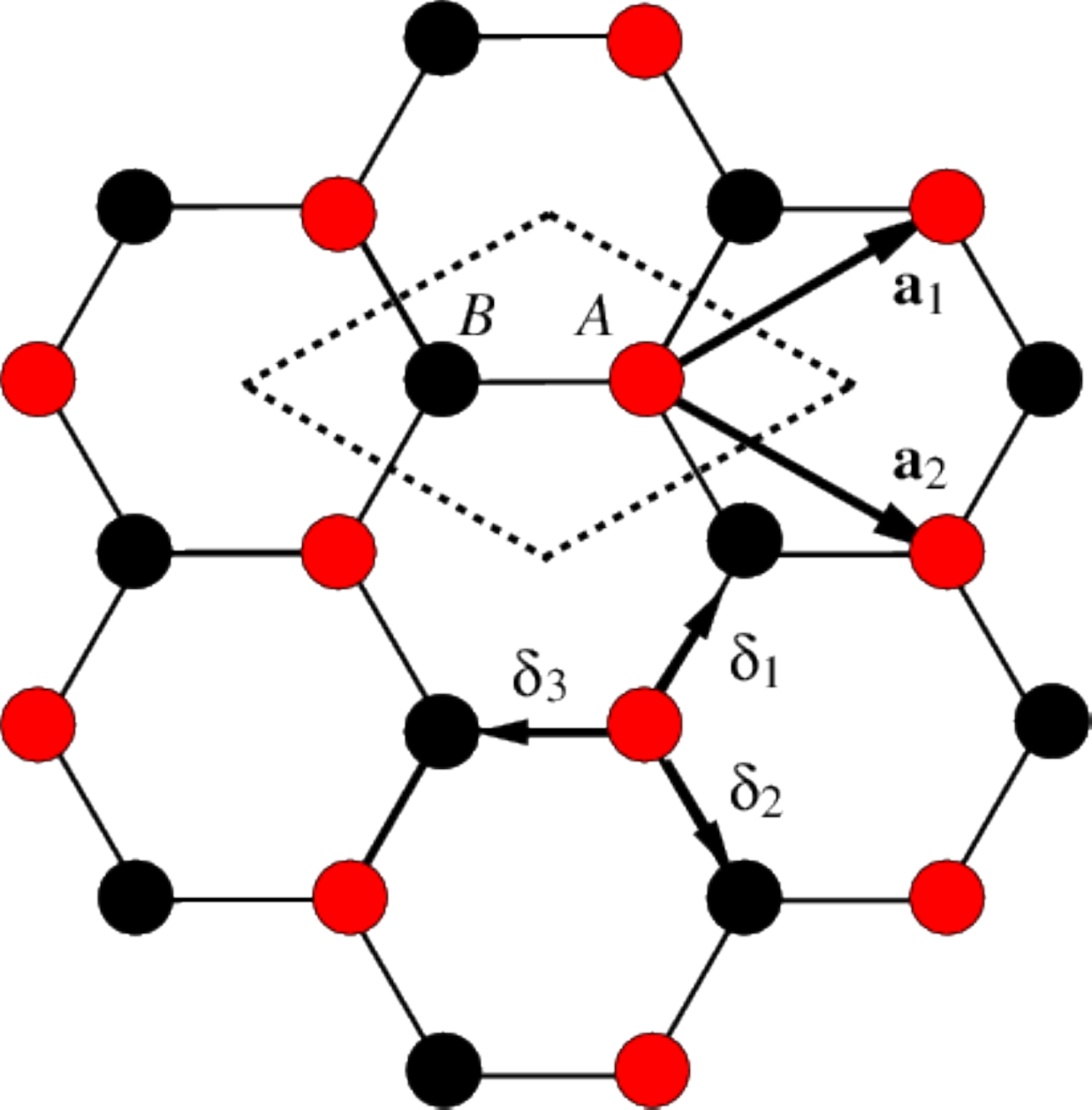}
\qquad
\includegraphics[width=0.35\columnwidth]{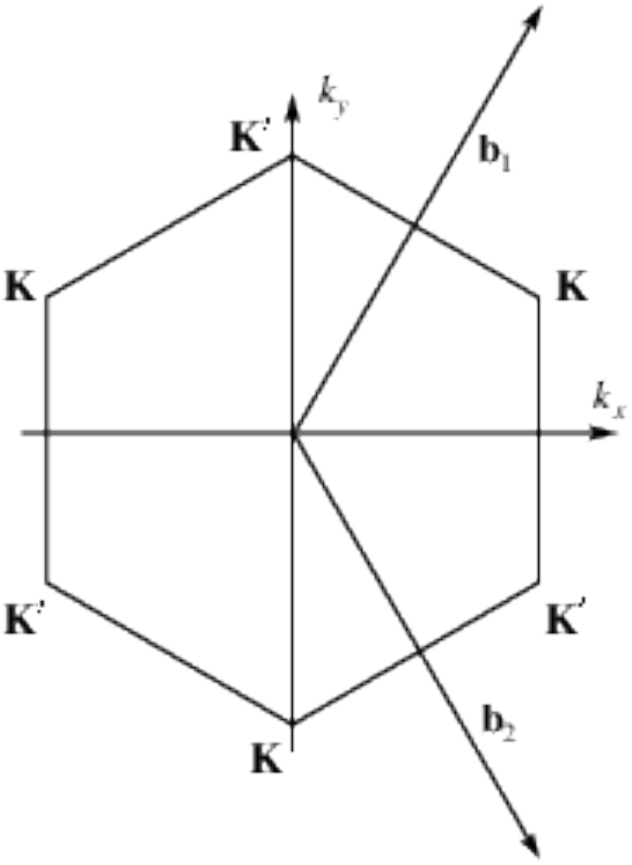}
\centering
\caption{(Color online) Honeycomb lattice of graphene and its Brillouin
zone. Left panel: lattice structure of single-layer graphene, made out of
two triangular sublattices, $A$ (red circles) and $B$ (black circles);
$\mathbf{a}_{1,2}$
are the lattice unit vectors and
$\boldsymbol{\delta}_i$
are the nearest-neighbor vectors. Right panel: the Brillouin zone of
graphene is a hexagon in momentum space. The reciprocal lattice
vectors are
${\bf b}_{1,2}$.
The Brillouin zone corners are denoted by
$\mathbf{K}$
and
$\mathbf{K}'$.
\label{GraphLatBrulfig}
}
\end{figure}

The electronic spectrum of single-layer graphene was investigated in
details in many papers (see, e.g., the
review~\cite{CastrNrev}).
The honeycomb lattice of graphene consists of two sublattices, $A$ and $B$
(see
Fig.~\ref{GraphLatBrulfig},
left panel). The lattice vectors are
\begin{equation}\label{a12}
\mathbf{a}_1=\frac{a_0}{2}(3,\sqrt{3}),\qquad \mathbf{a}_2=\frac{a_0}{2}(3,-\sqrt{3}),
\end{equation}
where
$a_0=1.42$~{\AA}
is the distance between the nearest carbon atoms. The simplest
tight-binding Hamiltonian of single-layer graphene can be written in the
form
\begin{equation}
\label{Hslg}
\hat{H}
=
-t\sum_{\langle \mathbf{ij}\rangle\sigma}
	a^\dag_{\mathbf{i}\sigma}
	b_{\mathbf{j}\sigma}^{\vphantom{\dagger}} +\text{h.c.}\,,
\end{equation}
where $t$ is the nearest-neighbor hopping amplitude,
$a^\dag_{\mathbf{i}\sigma}$
($b^\dag_{\mathbf{i}\sigma}$)
is the creation operator of the electron on site
$\mathbf{i}$
of the sublattice $A$ ($B$). Index $\sigma$ labels spin projection, the
brackets
$\langle...\rangle$
stand for nearest-neighbor hopping, and h.c. means the Hermitian
conjugate term. Since elementary crystalline unit of graphene includes two
carbon atoms $A$ and $B$, the single-electronic spectrum of
the Hamiltonian
Eq.~(\ref{Hslg})
consists of two bands, both of which are doubly degenerate with respect to
the spin projection $\sigma$.

To derive the spectrum we will follow the standard procedure. We rewrite
Eq.~\eqref{Hslg}
in
$\mathbf{k}$-space
with the help of the Fourier transform of the electron operators,
\begin{equation}
\label{fourier}
a_{\mathbf{k}\sigma}
=
\frac{1}{\sqrt{\cal N}}\sum_{\mathbf{n}}
e^{i\mathbf{k}\mathbf{r}^A_{\mathbf{n}}}a_{\mathbf{n}\sigma},
\qquad
b_{\mathbf{k}\sigma}
=
\frac{1}{\sqrt{\cal N}}\sum_{\mathbf{n}}
e^{i\mathbf{k}\mathbf{r}^B_{\mathbf{n}}}b_{\mathbf{n}\sigma},
\end{equation}
where the index
${\bf n}$
runs over all the unit cells of the honeycomb lattice,
${\cal N}$
is the number of unit cells in the lattice, and
$\mathbf{r}^\alpha_{\mathbf{n}}$
is the position of a carbon atom in the
$\mathbf{n}$-th
unit cell for sublattice $\alpha$ (sublattice index
$\alpha = A,B$).
The quasimomentum
${\bf k}$
belongs to the Brillouin zone. For the honeycomb lattice, the Brillouin
zone is a hexagon (see the right panel of
Fig.~\ref{GraphLatBrulfig}).
The corresponding reciprocal-lattice vectors are given by
\begin{equation}
\label{slgrecipV}
\mathbf{b}_1
=
\frac{2\pi}{3a_0}(1,\sqrt{3}),
\quad
\mathbf{b}_2=\frac{2\pi}{3a_0}(1,-\sqrt{3}).
\end{equation}

\begin{figure}[t]
\includegraphics[width=0.6\columnwidth]{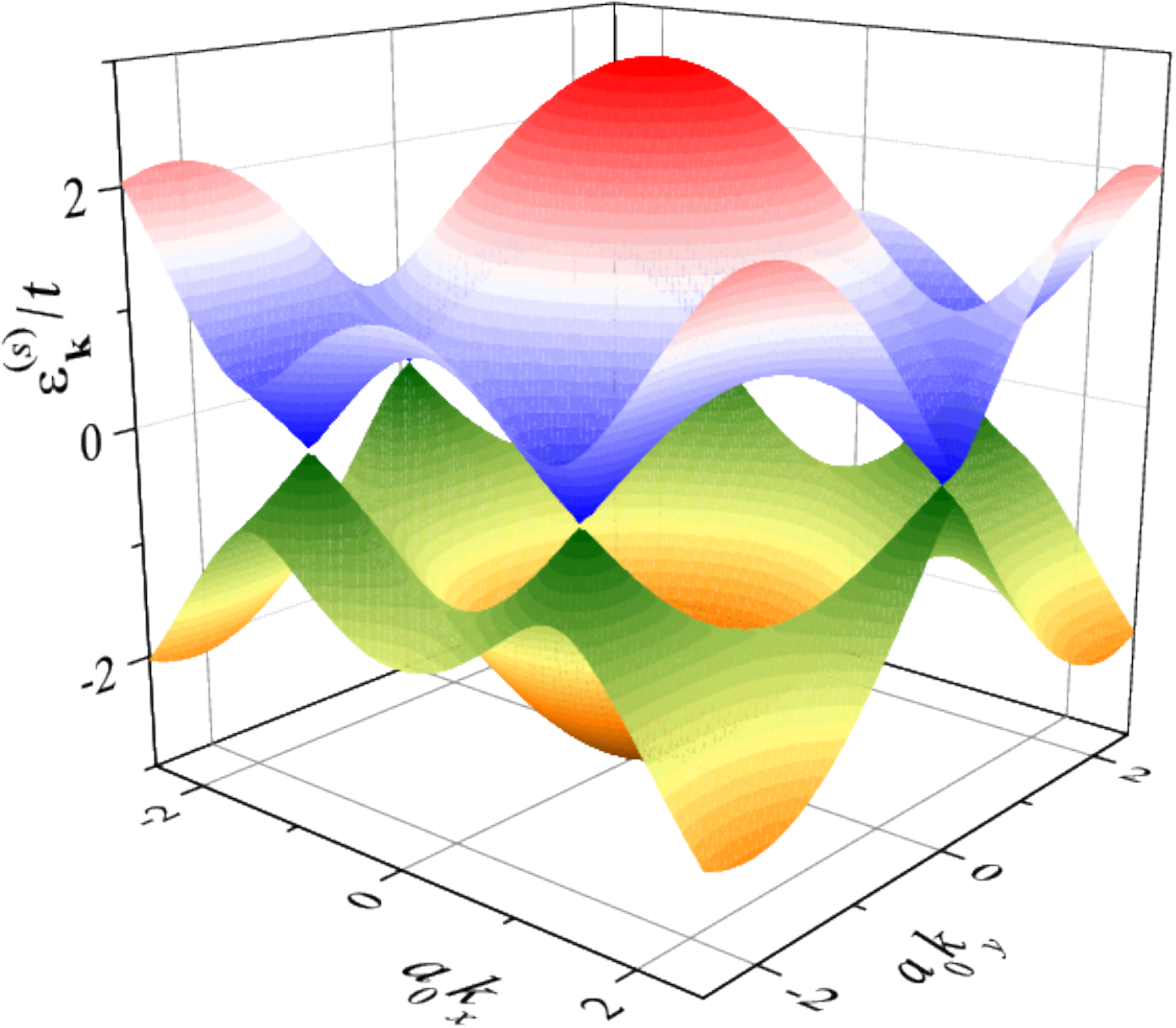}
\centering
\caption{(Color online) {\bf Electronic dispersion of single-layer graphene.}
The single-layer graphene conductance band (top red-and-blue surface)
touches the valence band (bottom yellow-and-green surface) at
${\bf K, K}'$
points in the reciprocal space.
\label{GraphSLGBandsfig}
}
\end{figure}

It is convenient to group two operators, introduced by
Eq.~(\ref{fourier}),
into a single spinor
$\Psi_{\mathbf{k}\sigma }=(a_{\mathbf{k}\sigma },b_{\mathbf{k}\sigma })^T$,
where superscript $T$ means transposed matrix. In this representation the
Hamiltonian
Eq.~(\ref{Hslg})
becomes
\begin{equation}
\label{HslgM}
H = \sum_{{\bf k} \sigma}
	\Psi^\dag_{{\bf k} \sigma}
	\hat{H}_{\mathbf{k}}^{\vphantom{\dagger}}
	\Psi_{\mathbf{k}}^{\vphantom{\dagger}},
\qquad
\text{where}
\qquad
H_{\bf k}
=
-t\left(
   \begin{array}{cc}
     0 & f(\mathbf{k}) \\
     f^*(\mathbf{k}) & 0 \\
   \end{array}
 \right).
\end{equation}
The function
$f(\mathbf{k})$
is very common in the theoretical description of the single-particle
graphene spectrum. It is defined as follows
$f(\mathbf{k})=\sum_{j=1}^3\exp(i\mathbf{k}\boldsymbol{\delta}_j)$,
with
$\boldsymbol{\delta}_j$
being the nearest-neighbor vectors (see
Fig.~\ref{GraphLatBrulfig}).
These vectors are equal to
\begin{equation}\label{deltaj}
\boldsymbol{\delta}_1
=
\frac{a_0}{2} (1,\sqrt{3}),
\qquad
\boldsymbol{\delta}_2
=
\frac{a_0}{2} (1,-\sqrt{3}),
\qquad
\boldsymbol{\delta}_3
=-a_0(1,0).
\end{equation}
Thus,
\begin{equation}\label{fk}
f(\mathbf{k})
=
\exp({-ia_0k_x})
\left[
	1
	+
	2\exp{\left(\frac{3ia_0k_x}{2}\right)\cos{\frac{\sqrt{3}a_0k_y}{2}}}
\right].
\end{equation}
The function
$f^*$
in
Eq.~(\ref{fk})
is the complex conjugate of $f$. The electron spectrum is obtained by
diagonalizing the matrix
$H_{\bf k}$.
It is given by
\begin{equation}
\label{slgBands}
\varepsilon^{(1,2)}_{\mathbf{k}}=\pm t|f(\mathbf{k})|.
\end{equation}
The functions $\varepsilon^{(1,2)}_{\mathbf{k}}$ are shown in
Fig.~\ref{GraphSLGBandsfig}.

As it follows from
Eqs.~\eqref{fk}
and~\eqref{slgBands},
the energies
$\varepsilon^{(1,2)}_{\mathbf{k}}$
become zero at the corners of the Brillouin zone. It is important to
remember that of six corners of the BZ only two non-equivalent points may
be chosen. These points are conventionally denoted as
${\bf K}$
and
${\bf K}'$
(see
Fig.~\ref{GraphLatBrulfig}).
A possible choice of two non-equivalent points is
\begin{equation}
\mathbf{K}=\frac{2\pi}{3a_0}\left(1,\frac{1}{\sqrt{3}}\right),
\qquad
\qquad
\mathbf{K}'=\frac{2\pi}{3a_0}\left(1,-\frac{1}{\sqrt{3}}\right).
\end{equation}
All other Brillouin zone corners are connected to either
${\bf K}$,
or
${\bf K}'$
by a reciprocal lattice vector.
Near the corners, the electronic dispersion is linear with zero effective
mass, similar to relativistic massless Dirac particles
\begin{equation}
\label{slgDirac}
\varepsilon_{\mathbf{K+q}}^{(1,2)}=\pm \hbar v_F|{\bf q}|,
\qquad
\qquad
v_{\rm F} = \frac{3a_0t}{2\hbar},
\end{equation}
where
$v_{\rm F}$
is the Fermi velocity,
$\mathbf{q}=\mathbf{k}-\mathbf{K}$
(or
$\mathbf{q}=\mathbf{k}-\mathbf{K'}$),
and
$|{\bf q}| \ll |{\bf K}|$.

Thus, at low energies the single-particle spectrum of graphene consists
of two Dirac cones located at the
${\bf K}$
and
${\bf K}'$
points of the BZ. This feature of the electronic spectrum has numerous
experimental
confirmations~\cite{CastrNrev},
and the measured value of the Fermi velocity is
$v_F\simeq 10^6$~m/s.

The Hamiltonian Eq.~\eqref{Hslg} is symmetric with respect to electron-hole
exchange. Thus, the Fermi level is zero if the sample is undoped, that
is, there is one electron per carbon atom. In this regime, the single layer
graphene is semimetal, because the energy gap in its electronic spectrum is
zero, but the density of states on the Fermi level is also zero. Graphene
becomes a metal upon electron or hole doping. The electron-hole symmetry is
an accidental symmetry, it breaks when the next-to-nearest neighbor hopping
amplitude $t'$ is taken into account. However, near the Dirac points,
a non-zero
$t'$
only renormalizes the Fermi level. The linear dispersion and the expression
for the Fermi velocity,
Eq.~(\ref{slgDirac}),
remain the same. The value of
$t'$
is estimated
as~\cite{CastrNrev}
\begin{eqnarray}
0.02 t \lesssim t' \lesssim 0.2t
\end{eqnarray}
(see also the discussion in
subsection~\ref{AAEB}).

\subsubsection{Effective Dirac equation}
\label{SlgChiral}

At low energy, the electron states of the single-layer graphene obey
massless Dirac equation. To obtain the Dirac equation one can proceed as
follows. Using the transformation of the electronic operators
Eq.~\eqref{fourier},
the electronic fields can be expressed as a sum of two terms, each
containing single-electron states with momenta close to a particular Dirac
point, either
${\bf K}$
or
${\bf K}'$
\begin{equation}
\label{{SlgTwoF}}
a_{\mathbf{n}} =
e^{-i\mathbf{K}R_{\mathbf{n}}}a_{1{\mathbf{n}}}
+
e^{-i\mathbf{K}'R_{\mathbf{n}}}a_{2{\mathbf{n}}},
\qquad
b_{\mathbf{n}} =
e^{-i\mathbf{K}R_{\mathbf{n}}}b_{1{\mathbf{n}}}
+
e^{-i\mathbf{K}'R_{\mathbf{n}}}b_{2{\mathbf{n}}},
\end{equation}
where the index
$i=1$
($i=2$)
refers to the
${\bf K}$
(${\bf K}'$)
point, and the spin label $\sigma$ is suppressed. These new fields are
assumed to vary slowly over the unit cell. To derive the desired low-energy
theory we substitute this representation in the tight-binding Hamiltonian and
expand the operators up to linear order in the
$\boldsymbol{\delta}$'s.
The resultant effective Hamiltonian splits into two copies of the massless
Dirac-like Hamiltonian, one applicable near
${\bf K}$,
and the other around
${\bf K}'$~\cite{Semenoff}.
In terms of first quantization, we have two 2D Dirac equations
for massless fermions valid close to the Dirac points
\begin{eqnarray}\label{SlgDiracEqs}
\nonumber
  -i\hbar v_F\hat{{\bm\sigma}}{\bm\nabla}\psi_1(\mathbf{r})
&=&\varepsilon\psi_1(\mathbf{r})\quad \textrm{(near ${\bf K}$)}, \\
  -i\hbar v_F\hat{{\bm\sigma}}^*{\bm\nabla}\psi_2(\mathbf{r})
&=&\varepsilon\psi_2(\mathbf{r})\quad \textrm{(near ${\bf K}'$)},
\end{eqnarray}
where
$\hat{\bm\sigma}=(\hat{\sigma}_x,\hat{\sigma}_y,\hat{\sigma}_z)$
are the Pauli matrices and
$\psi_i$
are the two-component electron wave functions.
In momentum space we have
\begin{equation}\label{SlgPsi}
\!\!\!\!\!\psi_1^{\pm}(\mathbf{q})\!=\!e^{i\mathbf{q}\mathbf{r}}\!\!\left(
                           \begin{array}{c}
                             \!e^{-i\theta_{\mathbf{q}}/2}\! \\
                             \!\pm e^{i\theta_{\mathbf{q}}/2}\! \\
                           \end{array}
                         \right),\,\,
\qquad
\qquad
\psi_2^{\pm}(\mathbf{q})\!=\!e^{i\mathbf{q}\mathbf{r}}\!\!\left(
                           \begin{array}{c}
                            \!e^{i\theta_{\mathbf{q}}/2}\! \\
                             \!\pm e^{-i\theta_{\mathbf{q}}/2}\! \\
                           \end{array}
                         \right),
\end{equation}
where the $\pm$ signs correspond to two bands
$\varepsilon_q^{(1)}$
and
$\varepsilon_q^{(2)}$
[see
Eq.~(\ref{slgDirac})].
The phase is defined according to
$\theta_{\mathbf{q}}=\arctan(q_y/q_x)$,
and the normalization factor is omitted.

\subsubsection{Density of states}\label{SlgDos}

\begin{figure}
\includegraphics[width=0.95\columnwidth]{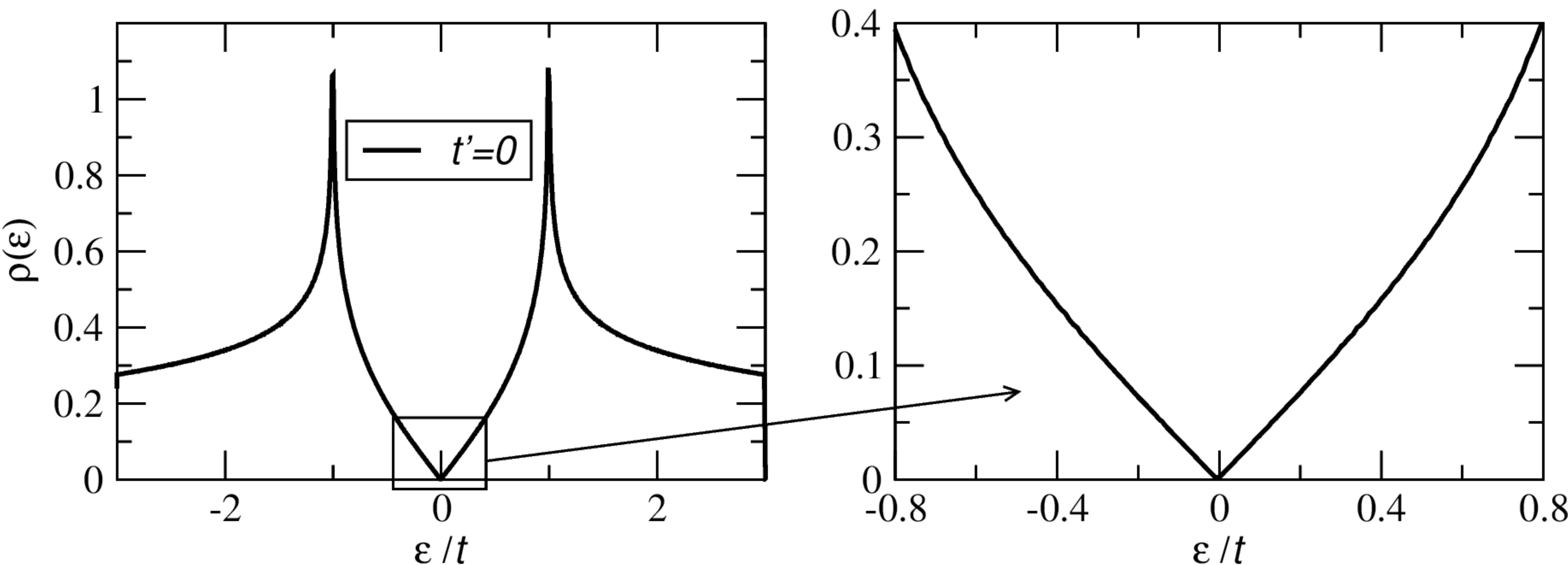}
\centering
\caption{{\bf Density of states for single-layer graphene.}
Left panel: the density of states per unit cell as a function of energy in
units of $t$. To calculate this density, the single-electron dispersion,
Eq.~(\ref{slgBands}),
was used. For one electron per carbon atom (undoped graphene) the Fermi
level corresponds to
$\varepsilon = 0$.
The right panel shows a zoom-in of the density of states close to the Dirac
point.
Reprinted figure with permission from
A.H.~Castro Neto et al.,
Rev. Mod. Phys. {\bf 81} 109 (2009).
Copyright 2009 by the American Physical Society.
\url{http://dx.doi.org/10.1103/RevModPhys.81.109}
\label{SlgDosfig}
}
\end{figure}

The density of states (DOS) per unit cell $\rho(\varepsilon)$, derived for
the Hamiltonian
Eq.~\eqref{Hslg},
is plotted in
Fig.~\ref{SlgDosfig}.
It shows semimetallic
behavior~\cite{CastrNrev}.
For the case considered, when we take into account only nearest-neighbors
hopping, it is possible to derive the analytical expression for the DOS per
unit cell, which has the
form~\cite{Hobson}
\begin{equation}
\label{SlgDOS}
\rho(\varepsilon)
=
\frac{4|\varepsilon|}{\pi^2t^2\sqrt{Z_0 (\varepsilon)}}\,
\mathbf{F}\!\left(
		\frac{\pi}{2},
		\sqrt{\frac{Z_1(\varepsilon)}{Z_0(\varepsilon)}}
             \right),
\end{equation}
where
$\mathbf{F}(\pi/2,x)$
is the complete elliptic integral of the first kind and
\begin{eqnarray}
  Z_0 = \left\{
            \begin{array}{ll}
              	\left(1+\frac{|\varepsilon|}{t}\right)^2
		-
		\frac{[(\varepsilon/t)^2-1]^2}{4},
		&
		 \hbox{$|\varepsilon|\leq t$,}
		\\
              	\frac{4|\varepsilon|}{t}\,,
		&
		\hbox{$t\leq|\varepsilon|\leq 3t$,}
            \end{array}
          \right.
   \\
\nonumber
    Z_1 = \left\{
            \begin{array}{ll}
              	\frac{4|\varepsilon|}{t}\,,
		&
		\hbox{$|\varepsilon|\leq t$,}
		\\
              	\left(1+\frac{|\varepsilon|}{t}\right)^2
		-
		\frac{[(\varepsilon/t)^2-1]^2}{4},
		&
		\hbox{$t \leq|\varepsilon|\leq 3t$,}
             \end{array}
          \right.
\end{eqnarray}
Close to the Dirac point
\begin{eqnarray}
\rho(\varepsilon) \approx
\frac{3\sqrt{3} a_0^2}{\pi v_{\rm F}^2} |\varepsilon|,
\end{eqnarray}
that is, the DOS vanishes linearly when the energy approaches the Fermi
level.

\subsection{AA bilayer graphene}\label{spectraAA}

\subsubsection{Single-electron tight-binding description}
\label{AAEB}

\begin{figure}
\includegraphics[width=0.55\columnwidth]{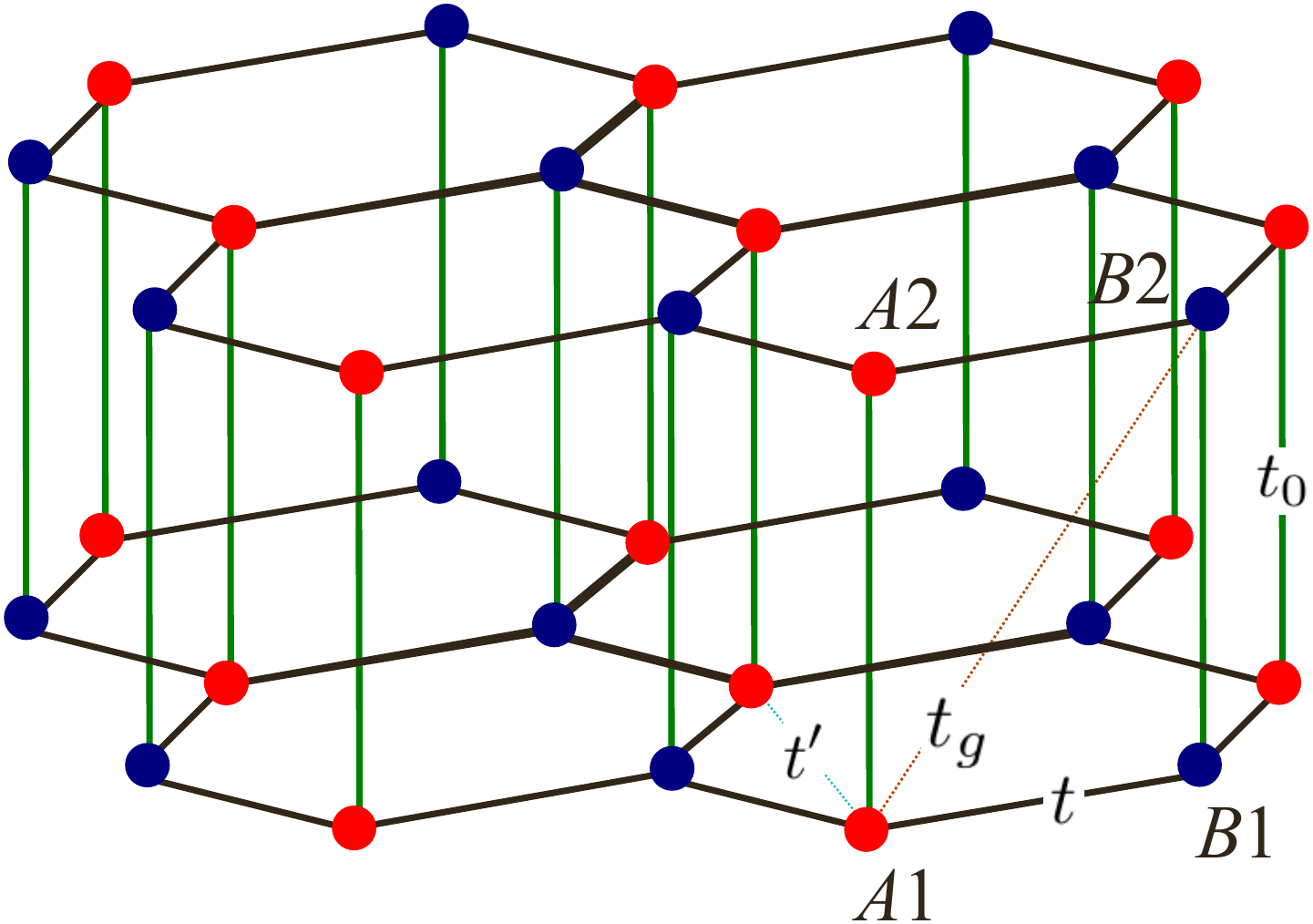}
\centering
\caption{(Color online) Crystal structure of the AA-stacked bilayer
graphene. The circles denote carbon atoms in the A (red) and B (blue)
sublattices in the bottom (1) and top (2) layers. The unit cell of the
AA-stacked bilayer graphene consists of four atoms $A_1$, $A_2$, $B_1$, and
$B_2$. Hopping integrals $t$ and $t_0$ correspond to the in-plane and
interplane nearest-neighbor hopping, $t'$ and $t_g$ correspond to the
in-plane and interplane next-nearest-neighbor hopping.
\label{AALatfig}
}
\end{figure}

First, we derive the electronic spectrum of the AA bilayer graphene. The
crystal structure of the AA bilayer graphene is shown in
Fig.~\ref{AALatfig}.
The AA-stacked bilayer consists of two graphene layers, 1 and 2. Each
carbon atom of the upper layer is located above the corresponding atom of
the lower layer. Since layer consists of two triangular sublattices, $A$
and $B$, the elementary unit cell of the AA-stacked bilayer graphene
contains four carbon atoms, $A_1$, $A_2$, $B_1$, and $B_2$. The system is
modeled by the tight-binding Hamiltonian for
$p_z$
electrons of carbon atoms, which includes in-plane and interplane
tunneling~\cite{PrlOur}
\begin{eqnarray}\label{Haa}
\nonumber
\hat{H}
&=&
 -t\!\!\sum_{\langle \mathbf{ij}\rangle\alpha\sigma}
\!\!\!
a^\dag_{\mathbf{i}\alpha\sigma}
b_{\mathbf{j}\alpha\sigma}^{\vphantom{\dagger}}
+
t_0\!\sum_{ \mathbf{i}\sigma}
\left(
	\!a^\dag_{\mathbf{i}1\sigma}a_{\mathbf{i}2\sigma}^{\vphantom{\dagger}}
	+
	\!b^\dag_{\mathbf{i}1\sigma}b_{\mathbf{i}2\sigma}^{\vphantom{\dagger}}
\right)
-t'\!\!\!\!\sum_{\langle\langle \mathbf{ij}\rangle\rangle\alpha\sigma}\!\!\!
\left(\!a^\dag_{\mathbf{i}\alpha\sigma}
a_{\mathbf{j}\alpha\sigma}^{\vphantom{\dagger}}
+b^\dag_{\mathbf{i}\alpha\sigma}
b_{\mathbf{j}\alpha\sigma}^{\vphantom{\dagger}}\right)
\\
&&
+t_g\!
\sum_{\langle \mathbf{ij}\rangle\sigma}
\left(
	\!a^\dag_{\mathbf{i}1\sigma}b_{\mathbf{j}2\sigma}^{\vphantom{\dagger}}
	+
	a^\dag_{\mathbf{i}2\sigma}b_{\mathbf{j}1\sigma}^{\vphantom{\dagger}}
\right)+\textrm{h.c.}
\end{eqnarray}
The first sum in this expression is the in-plane nearest-neighbor hopping,
as in a graphene sheet [see
Eq.~\eqref{Hslg}],
the second sum describes the nearest-neighbor inter-plane hopping with
amplitude
$t_0$,
the third and the fourth sums correspond to the in-plane and interplane
next-nearest-neighbor hopping with amplitudes
$t'$
and
$t_g$,
respectively. The symbol
$\langle\langle \ldots \rangle\rangle$
denotes summation over the next-nearest-neighbor sites.

The values of the hopping integrals can be obtained from the \textit{ab
initio} calculations and tight-binding fit to experiments. The value of $t$
is well-established and lies within
$t=2.5$--$3$~eV~\cite{CastrNrev,tunnel,tunnel1,tunnel2}.
The interplane nearest-neighbor hopping integral is much weaker,
$t_0=0.3$--0.4~eV~\cite{CastrNrev,tunnel,tunnel1,tunnel2}.
The integral
$t'$
is not well known. The {\it ab initio}
calculations performed by
S.~Reich et~al.~\cite{tunnel3}
indicate that
$0.02t<t'<0.2t$
depending on the tight-binding parametrization. A fit to cyclotron
resonance experiments in the single-layer
graphene done by R.\,S.~Deacon et~al.~\cite{tunnel4}
finds
$t'\approx 0.1$\,eV.
The interplane next-nearest-neighbor integral $t_g$ is significantly
smaller than the others.
\textit{Ab initio}
calculations performed in Ref.~\cite{tunnel1}
estimated that
$t_g\approx -0.03$\,eV.

\begin{figure}
\includegraphics[width=0.43\columnwidth]{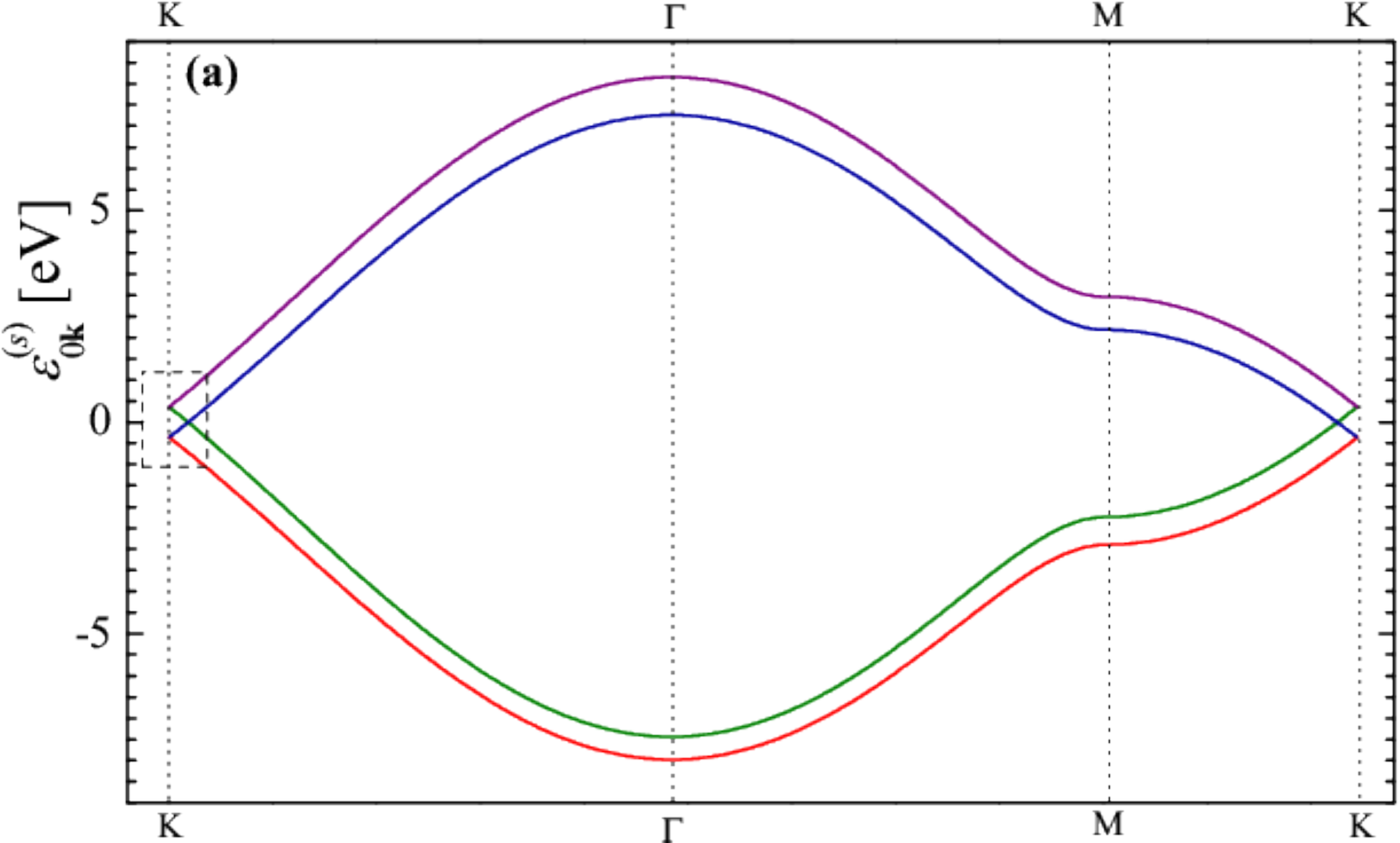}
\includegraphics[width=0.28\columnwidth]{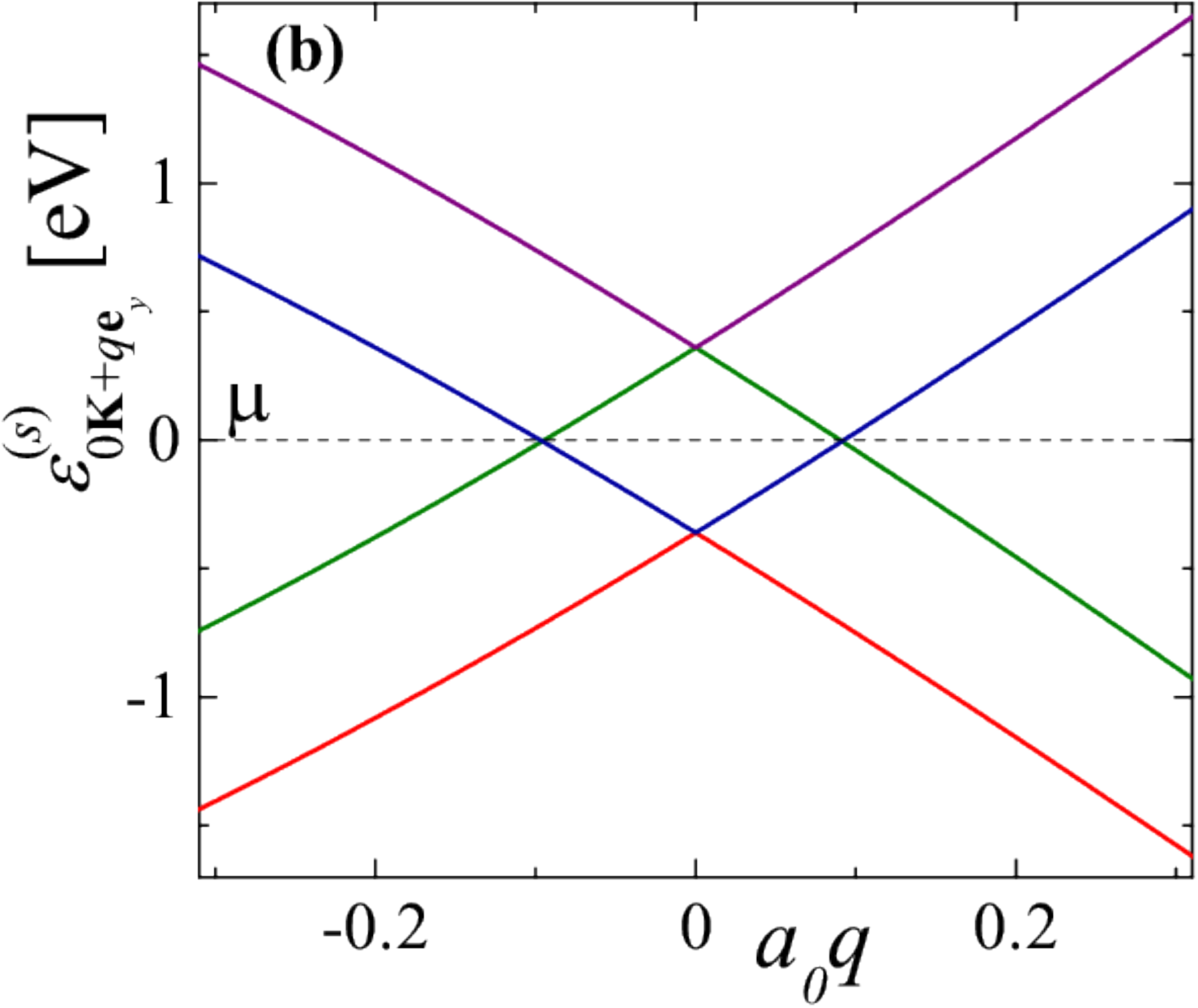}
\includegraphics[width=0.27\columnwidth]{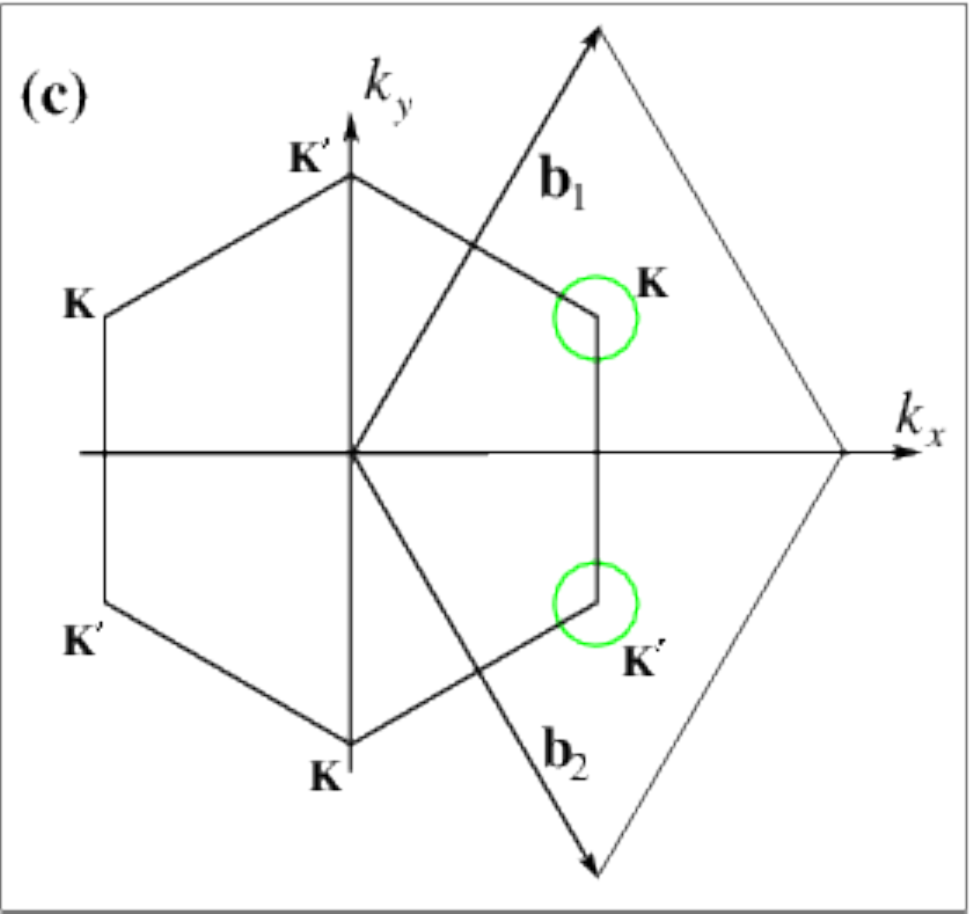}
\centering
\caption{(Color online) (a) Single-particle band structure of the
AA-stacked bilayer
graphene.
It consists of two single-layered graphene spectra shifted relative to each
other by the energy
$2t_0$.
(b) The $k$ dependence of the spectra
$\varepsilon^{(s)}_{0 \mathbf{k}}$
near the Dirac point $\mathbf{K}$.
Here,
$\mathbf{k} = \mathbf{K} + q {\hat{\bf e}_y}$.
The bands
$s = 2$ and 3
cross each other exactly at zero energy, which corresponds to the Fermi
level of the undoped system. (c) The first Brillouin zone (hexagon) and the
reciprocal-lattice unit cell (rhombus) of the AA-stacked graphene bilayer.
The (green) circles around
$\mathbf{K}$
and 
$\mathbf{K}'$ 
points represent (degenerate) Fermi surfaces (Fermi arcs).
\label{AABandBrfig}
}
\end{figure}

At first, we disregard next-nearest-neighbor hoppings (which will be
considered at the end of this Section) keeping only the first two terms in
the Hamiltonian
Eq.~\eqref{Haa}.
The unit cell of the AA bilayer contains four atoms, and electronic
spectrum consists of four bands
$\varepsilon^{s}_{0\mathbf{k}}$, where $s=1,2,3,4$.
The eigenenergies
$\varepsilon^{s}_{0\mathbf{k}}$
can be obtained by the same method we used for the single-layer graphene.
We transform the Hamiltonian
Eq.~\eqref{Haa}
into the
$\mathbf{k}$-space
representation, and introduce the four-component spinor
\begin{equation}
\label{4spinor}
\Psi_\mathbf{k \sigma}
=
(a_{\mathbf{k}1\sigma},a_{\mathbf{k}2\sigma},
b_{\mathbf{k}1\sigma},b_{\mathbf{k}2\sigma})^T.
\end{equation}
Without the next-nearest-neighbor hopping, the Hamiltonian
Eq.~\eqref{Haa}
can be written as
$H = \sum_{{\bf k}\sigma}
	\Psi_{{\bf k} \sigma}^\dag
	{\hat H}_{\bf k}
	\Psi_{{\bf k} \sigma}^{\vphantom{\dagger}}$,
where the 4$\times$4 matrix
${\hat H}_{\bf k}$
has the form
\begin{equation}
\label{HaaK}
\hat{H}_\mathbf{k}=\left(
                     \begin{array}{cccc}
                       0 & t_0 & -tf(\mathbf{k}) & 0 \\
                       t_0 & 0 & 0 & -tf(\mathbf{k}) \\
                       -tf^*(\mathbf{k}) & 0 & 0 & t_0 \\
                       0 & -tf^*(\mathbf{k}) & t_0 & 0 \\
                     \end{array}
                   \right).
\end{equation}
The spectrum of this Hamiltonian is comprised of four energy bands
\begin{eqnarray}\label{aaBands}
\varepsilon^{(1)}_{0\mathbf{k}}
=
-t_0-t|f(\mathbf{k})|,
\quad
\varepsilon^{(2)}_{0\mathbf{k}}
=
+t_0-t|f(\mathbf{k})|,
\quad
\varepsilon^{(3)}_{0\mathbf{k}}
=
-t_0+t|f(\mathbf{k})|,
\quad
\varepsilon^{(4)}_{0\mathbf{k}}
=
+t_0+t|f(\mathbf{k})|,
\end{eqnarray}
which are plotted in
Fig.~\ref{AABandBrfig}.
The spectrum consists of two copies of the single-layer spectrum. One copy
(bands 2 and 4) is shifted to higher energies by the amount
$t_0$.
The other copy (bands 1 and 3) is shifted down by the
amount
$-t_0$.
In this approximation the spectrum possesses electron-hole symmetry. Near
the Dirac points, the bands 1 and 2 are hole-like, whiel the bands 3 and 4
are electron-like. Under the action of the electron-hole transformation,
bands 1 and 4 are transformed into each other. The same is true for bands 2
and 3. The first Brillouin zone, shown in panel~(c) of
Fig.~\ref{AABandBrfig},
is the same as that for single-layer graphene (see
Fig.~\ref{GraphLatBrulfig}).
If the bilayer is undoped, bands
$s = 2$
and 3 cross the Fermi level near the Dirac points
${\bf K}$
and
${\bf K}'$.
In such a situation, the Fermi surfaces are given by the equation
$|f(\mathbf{k})| = t_0/t$.
Since
$t_0/t\ll 1$,
we can expand the function
$|f(\mathbf{k})|$
near the Dirac points and find that the Fermi surface consists of two circles
with radius
$k_r = 2t_0/(3ta_0)$.
Thus, the AA-stacked graphene bilayer is a metal even in the case of zero
doping.


A striking feature of this tight-binding band structure is that for zero
doping the Fermi surfaces of both bands coincide. That is, the electron
and hole components of the Fermi surface are perfectly nested. This
single-electron property is quite stable against changes in the
tight-binding Hamiltonian. It survives even if longer-range hoppings are
taken into account, or a system with two nonequivalent layers is considered
(e.g., similar to the single-side hydrogenated
graphene~\cite{graphane1,graphane2,graphane3}). However, an arbitrarily small
electron interaction can destabilize such a degenerate spectrum, generating
a gap (see
subsection~\ref{subsect::broken_aa}).
On the other hand, upon doping the nesting is violated, and
the Fermi surfaces transform into four circles.

As in the case of the single-layer graphene, the spectrum of the AA-stacked
graphene bilayer is linear near the Dirac points. Expanding
Eq.~\eqref{aaBands}
near the Dirac points in powers of
$|\mathbf{q}|=|\mathbf{K}-\mathbf{k}|,
|\mathbf{K}'-\mathbf{k}|\ll |\mathbf{K}|$,
we derive in the lowest-order approximation
\begin{equation}\label{aaBandDir}
\varepsilon^{(1,2,3,4)}_{0\mathbf{K}+\mathbf{q}}
=
\mp t_0\mp \hbar v_F|\mathbf{q}|.
\end{equation}
The Fermi velocity for the AA-stacked graphene bilayer is the same as for
the single-layer graphene.

The electron-hole symmetry breaks, if we take into account the
next-nearest-neighbors hopping. In this case, the Hamiltonian in
$\mathbf{k}$-representation
becomes
\begin{equation}\label{HaaK1}
\hat{H}_\mathbf{k}=\left(
                     \begin{array}{cccc}
                       -t'F(\mathbf{k}) & t_0 & -tf(\mathbf{k}) & t_gf(\mathbf{k}) \\
                       t_0 & -t'F(\mathbf{k}) & t_gf(\mathbf{k}) & -tf(\mathbf{k}) \\
                       -tf^*(\mathbf{k}) & t_gf^*(\mathbf{k}) & -t'F(\mathbf{k}) & t_0 \\
                       t_gf^*(\mathbf{k}) & -tf^*(\mathbf{k}) & t_0 & -t'F(\mathbf{k}) \\
                     \end{array}
                   \right),
\end{equation}
where
\begin{equation}
\label{Fk}
F(\mathbf{k})\,
=
\,
|f(\mathbf{k})|^2 - 3
\,
=
\,
2\cos(\!{\sqrt{3}k_ya_0\!})\! +\! 4\!\cos{\!\left(\!\frac{\sqrt{3}k_ya_0}{2}\!\right)}\!\cos{\!\left(\frac{3k_xa_0}{2}\!\right)}.
\end{equation}
The electron bands satisfy the equations
\begin{eqnarray}
\label{aaBands1}
\nonumber
\varepsilon^{(1)}_{\mathbf{k}} = -t'F(\mathbf{k})-t_0-(t+t_g)|f(\mathbf{k})|,
\qquad
\nonumber
\varepsilon^{(2)}_{\mathbf{k}} = -t'F(\mathbf{k})+t_0-(t-t_g)|f(\mathbf{k})|,\\
\varepsilon^{(3)}_{\mathbf{k}} = -t'F(\mathbf{k})-t_0+(t+t_g)|f(\mathbf{k})|,
\qquad
\varepsilon^{(4)}_{\mathbf{k}} = -t'F(\mathbf{k})+t_0+(t-t_g)|f(\mathbf{k})|.
\end{eqnarray}
The expansion of the spectrum around the Dirac point, up to second order in
$|{\bf q}|/|{\bf K}|$,
is given by
\begin{eqnarray}
\label{aaBands2}
\varepsilon^{(1,2)}_{\mathbf{q}}
\!\!\!\! &=&\!\!\! 3t'\!\mp\! t_0\!-\!\hbar
v_F\!\!\left(\!1\!\pm\! \frac{t_g}{t}\!\right)\!\!|\mathbf{q}|\!
-
\!\frac{3}{4}\!\!
\left(
	\!3t'\!-\!\frac{t\!\pm\! t_g}{2}\sin{\!(3\theta_\mathbf{q})}\!\!
\right)\!\!a_0^2|\mathbf{q}|^2,\\
\nonumber
\varepsilon^{(3,4)}_{\mathbf{q}}\!\!\!\!
&=&
\!\!\! 3t'\!\mp\! t_0\!
+\!\hbar v_F\!\!\left(\!1\!\pm\! \frac{t_g}{t}\!\right)\!\!|\mathbf{q}|\!
-
\!\frac{3}{4}\!\!
\left(
	\!3t'\!+\!\frac{t\!\pm\! t_g}{2}\sin{\!(3\theta_\mathbf{q})}\!\!
\right)\!\!a_0^2 |\mathbf{q}|^2.
\end{eqnarray}
These equations demonstrate that, to lowest order in
$|{\bf q}|$,
the next-nearest-neighbor hopping
$t'$
shifts the position of the Fermi level: the first term
$3t'$
is the same for all four bands, and can be absorbed into the chemical
potential $\mu$.
The interlayer next-nearest-neighbor hopping
$t_g$
breaks the electron-hole symmetry more dramatically: if
$t_g \ne 0$,
the renormalized electron and hole Fermi velocities become unequal to each
other. Corrections to the linear dispersion are also electron-hole
asymmetric. Note that up to order
${\bf q}^2/{\bf K}^2$ the dispersion depends on the
direction in momentum space and has a threefold symmetry (so-called
trigonal warping of the electronic
spectrum~\cite{CastrNrev}).

Despite electron-hole asymmetry and the trigonal warping, the nesting of
the Fermi surface persists. To prove this claim one must solve the
equations for the electron Fermi surface
($\varepsilon^{(3)}_{\bf k} = \mu$),
and the hole Fermi surface
($\varepsilon^{(2)}_{\bf k} = \mu$),
and demonstrate that the resultant surfaces (lines) coincide. The required
equations are:
\begin{eqnarray}
\left[
	-t'F(\mathbf{k}) + t_g|f(\mathbf{k})|
\right]
+t_0-t|f(\mathbf{k})| = \mu,
\\
\left[
	-t'F(\mathbf{k})+t_g|f(\mathbf{k})|
\right]
-t_0+t|f(\mathbf{k})| = \mu.
\end{eqnarray}
We can demonstrate that the solution of these equations correspond to the
perfect nesting: both Fermi surfaces are defined by the relation
\begin{eqnarray} 
|f({\bf k})| = \frac{t_0}{t},
\nonumber 
\end{eqnarray}
with the chemical potential
\begin{eqnarray}
\mu = -t'(t_0^2/t^2 - 3) + (t_0 t_g/t).
\nonumber
\end{eqnarray} 
These statements are easy to check, if one remembers that the function
$F(\mathbf{k})$
can be expressed in terms of the function
$f(\mathbf{k})$
according to
Eq.~\eqref{Fk}.
The stable nesting has important implications for the robustness of the
broken symmetry phases, which will be discussed below.

\subsubsection{Density of states}\label{AaDos}

\begin{figure}
\centering
\includegraphics[width=0.65\columnwidth]{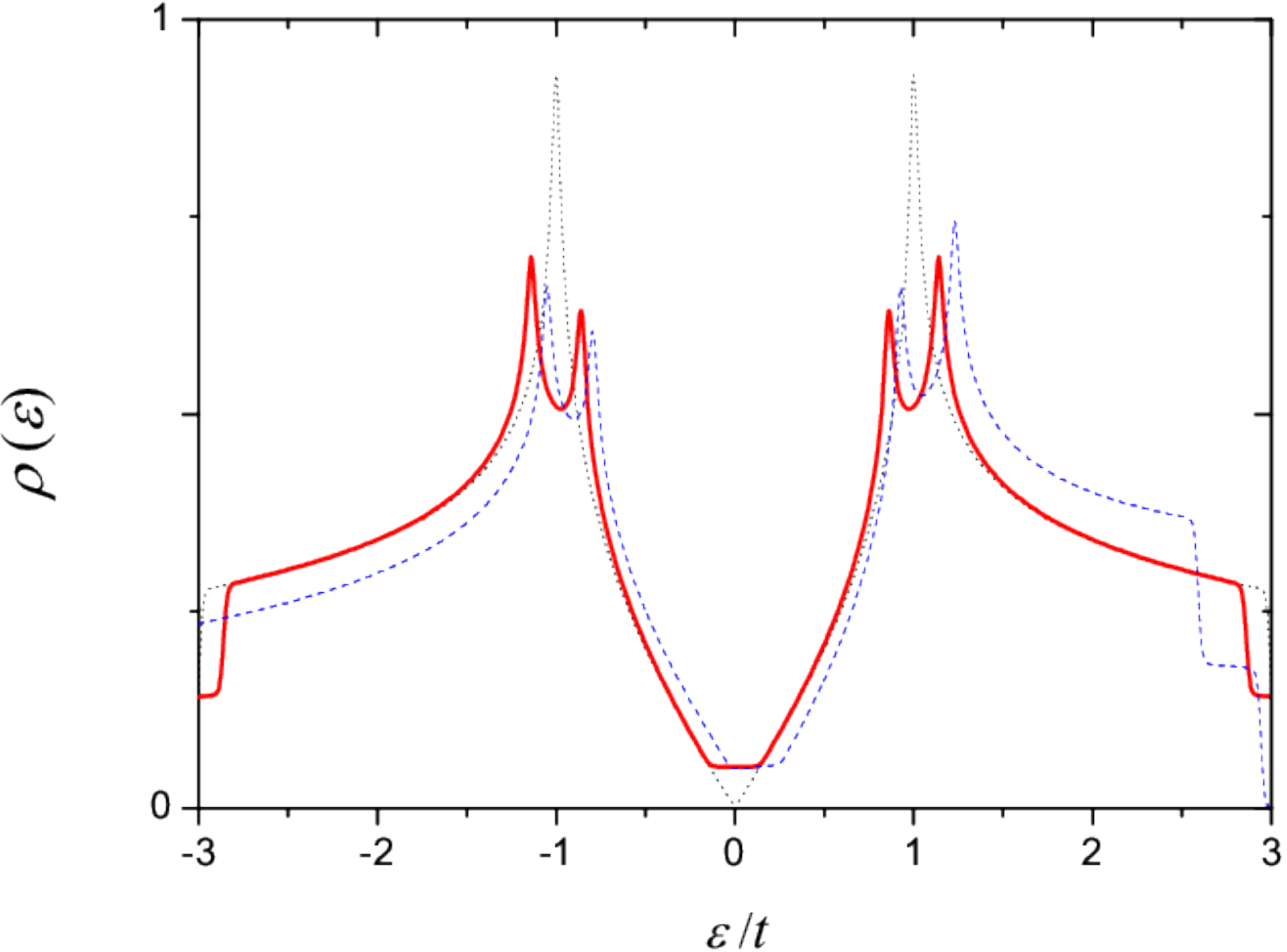}
\caption{(Color online)  Density of states of the AA-stacked bilayer
graphene. Red solid curve is calculated for
$t_g=0$
and
$t'=0$,
while blue dashed curve corresponds to
$t_g=-0.03$~eV
and
$t'=0.1$~eV.
For both cases
$t=2.57$~eV,
$t_0=0.36$~eV.
Density of states of the
single-layer graphene ($t'=0$) is shown by black dotted curve for
comparison.
\label{AaDOSfig}
}
\end{figure}

The density of states of the AA-stacked bilayer graphene 
can be calculated using the expression
\begin{equation}\label{DOSFormula}
\rho(\varepsilon)=\sum_s\!\int\limits_{V_{BZ}}\!\!
\frac{d^2\mathbf{k}}{V_{BZ}}\,
\delta\!\left(\varepsilon-\varepsilon^{(s)}_{\mathbf{k}}\right)\,,
\end{equation}
where the integration is taken over the volume of the Brillouin zone
$V_{BZ}=8\pi^2/3\sqrt{3}a_0^2$,
and
$\delta(x)$
is the Dirac delta-function. The results are shown in
Fig.~\ref{AaDOSfig}.
As it is seen from this figure, the DOS is non-zero near the Fermi-level
even in the case of the undoped system. The function
$\rho(\varepsilon)$
is flat near
$\varepsilon=0$,
where energy bands
$s = 2$ and 3
intersect: the DOS is almost constant in the energy range
$|\varepsilon| < t_0$,
and
\begin{eqnarray}
\rho(0)= \frac{4\pi t_0}{\hbar^2 v_F^2 V_{\rm BZ}}
=
\frac{2 t_0}{\pi \sqrt{3} t^2}.
\nonumber
\end{eqnarray} 
Each band of the bilayer has one van Hove peak, inherited from a
single-layer graphene spectrum. As a result, the bilayer DOS has four van
Hove singularities, clearly seen in the graph. If we take into account
next-nearest-neighbor hopping, the electron-hole symmetry of the DOS is
violated. This effect, however, is not strong, especially in the energy
range smaller than van Hove singularities.

\subsubsection{Effect of the bias voltage}\label{aaVolt}

The electronic properties of graphene systems can be tuned by the
application of electric field transverse to the
layers~\cite{PrbVOur}.
For the bilayer systems, the corresponding term in the Hamiltonian
reads
\begin{equation}\label{HV0}
\hat{H}_V
=
\frac{eV}{2}
\sum_{\mathbf{j}\sigma}
	\left(
		n_{\mathbf{j}1\sigma}-n_{\mathbf{j}2\sigma}
	\right),
\qquad
\text{where\ \ }
n_{\mathbf{j}\alpha\sigma}
=
a^\dag_{\mathbf{j}\alpha\sigma}
a_{\mathbf{j}\alpha\sigma}^{\vphantom{\dagger}}
+
b^\dag_{\mathbf{j}\alpha\sigma}
b_{\mathbf{j}\alpha\sigma}^{\vphantom{\dagger}},
\end{equation}
and $V$ is the bias voltage produced by the applied electric field in the
bilayer. Operator
$n_{\mathbf{j}\alpha\sigma}$
is the electron occupation number corresponding to
spin $\sigma$ and lattice site 
${\bf R}_{\bf j}$
in the layer
$\alpha=1,2$.
In the bi-spinor representation, the corresponding
4$\times$4
matrix is
\begin{equation}\label{HV0M}
\hat{H}_V=\frac{e}{2}\left(
                          \begin{array}{cccc}
                            V & 0 & 0 & 0 \\
                            0 & -V & 0 & 0 \\
                            0 & 0 & V & 0 \\
                            0 & 0 & 0 & -V \\
                          \end{array}
                        \right).
\end{equation}
The quantity $V$ is the effective, not ``bare", voltage. It differs from
the bare voltage
$V_0$
due to the partial screening of the external electric field by the
redistribution of the charge between the layers. This screening may
significantly modify the resultant spectrum, as seen experimentally in,
e.g.,
Ref.~\cite{Henriksen2010}.
Combining
$\hat{H}_V$
with the tight-binding
Hamiltonian~\eqref{HaaK},
we derive the spectrum of biased graphene, which coincides with
Eq.~\eqref{aaBands},
if we substitute the interlayer hopping integral
$t_0$
by a renormalized value
\begin{equation}
\label{t0aaV0}
t_0\rightarrow \sqrt{t_0^2+\left(\frac{eV}{2}\right)^2}.
\end{equation}
Thus, as far as the band energies
$\varepsilon_{\bf k}^{(s)}$
are concerned, the application of the bias voltage to the bilayer
effectively increases the interlayer hopping, but the qualitative structure
of the electronic spectrum remains the same. However, the quantitative
details of the band structure may be affected by the transverse electric
field of realistic strength: if the applied electric field
$E_0$
is
$10^7$~V/cm,
then, the value
$eV_0=eE_0c_0\approx 0.33$~eV
is comparable to
$t_0$.
Nonetheless, one has to remember that the substitution rule
Eq.~(\ref{t0aaV0})
may be inapplicable for evaluation of other physical properties of the
biased AA bilayer graphene. For example, the wave functions should not be
calculated with the help of
Eq.~(\ref{t0aaV0}).

%
%
\subsection{AB bilayer graphene}
\label{spectraAB}
\subsubsection{Single-electron tight-binding description}
\label{ab::energy_bands}
\begin{figure}[t]
\begin{centering}
\includegraphics[width=8cm]{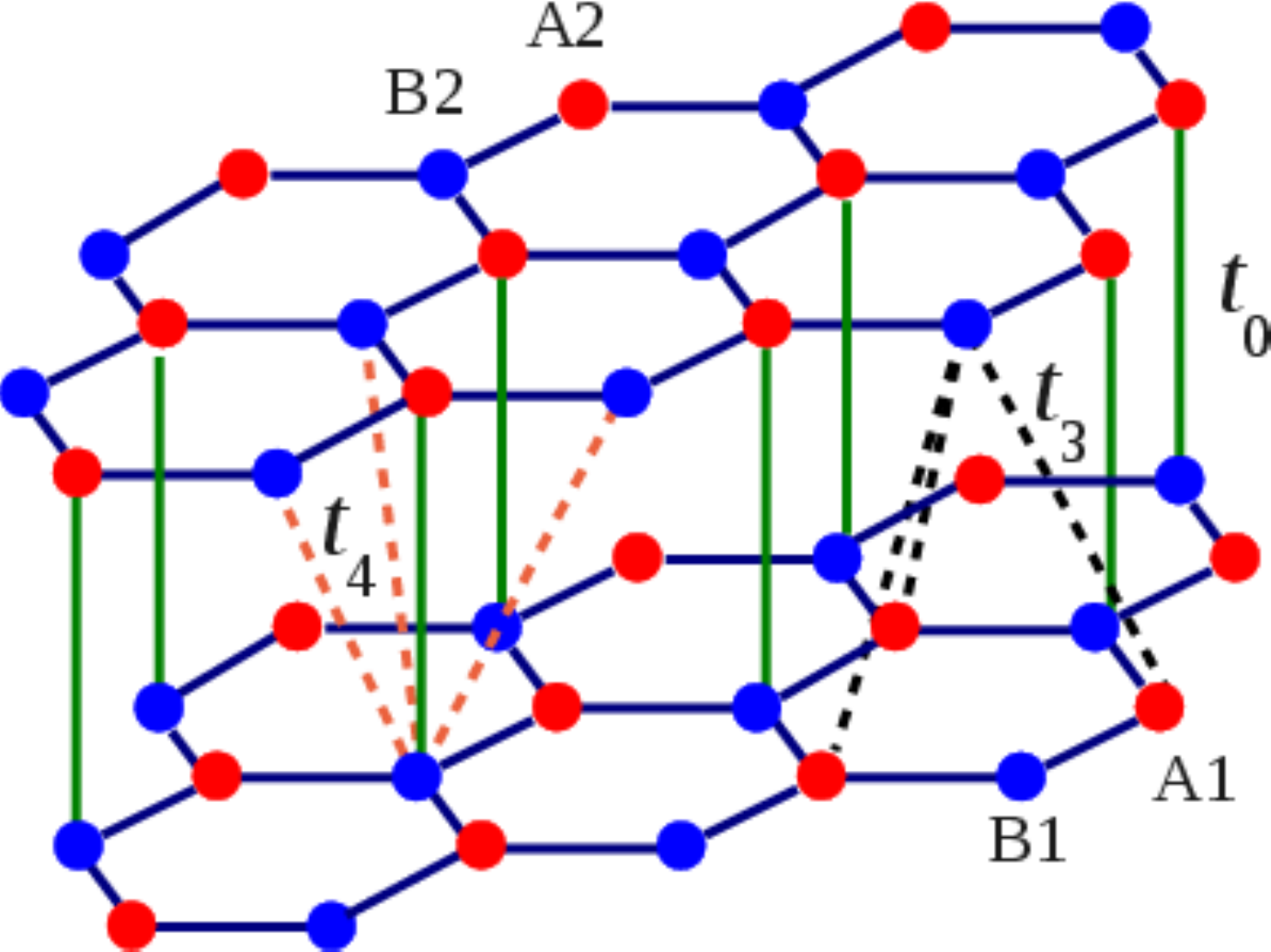}
\par\end{centering}
\caption{{\bf Bernal stacking.} If we start from AA stacking (see
Fig.~\ref{AALatfig})
and then shift the top layer by the vector
$- \boldsymbol{\delta}_3$,
we arrive at the AB, or Bernal, stacking of the bilayer. In this
arrangement $A_2$ sublattice lies opposite to $B_1$ sublattice. ``The dimer
sites" from these sublattices are connected by green lines. The
``non-dimer" sites lie against the centers of the hexagons in the opposite
layer. Besides the hopping between the dimer sites
$t_0$,
one can introduce the hopping
$t_3$
from a non-dimer site to nearest non-dimer sites in the opposite layer
(shown by black dashed lines), and hopping
$t_4$
from a dimer site to nearest non-dimer sites of the opposite layer (shown
by dashed red lines). The amplitude
$t_3$
introduces so-called ``trigonal warping" to the single-electron
dispersion, while
$t_4$
breaks the electron-hole symmetry of the single-electron states.
\label{fig::ab_lattice}
}
\end{figure}

Starting from the AA-stacked bilayer it is possible to construct a bilayer
with AB, or Bernal, stacking. To obtain the AB stacking (see 
Fig.~\ref{fig::ab_lattice})
we must shift one of two layers by the vector
$-\boldsymbol{\delta}_3$,
shown in
Fig.~\ref{GraphLatBrulfig}.
Under such a displacement sublattice {\it A} of the shifted layer ends up
over the sublattice {\it B} of the immobile layer. A pair of opposite sites
from these sublattices is called ``dimer sites". Two other sublattices
({\it B} of the shifted layer and {\it A} of the immobile layer) will be
over the centers of the hexagons of the opposite layer. The sites from
these sublattices are ``non-dimer".

\begin{figure}[t!]
\begin{centering}
\includegraphics[width=8cm]{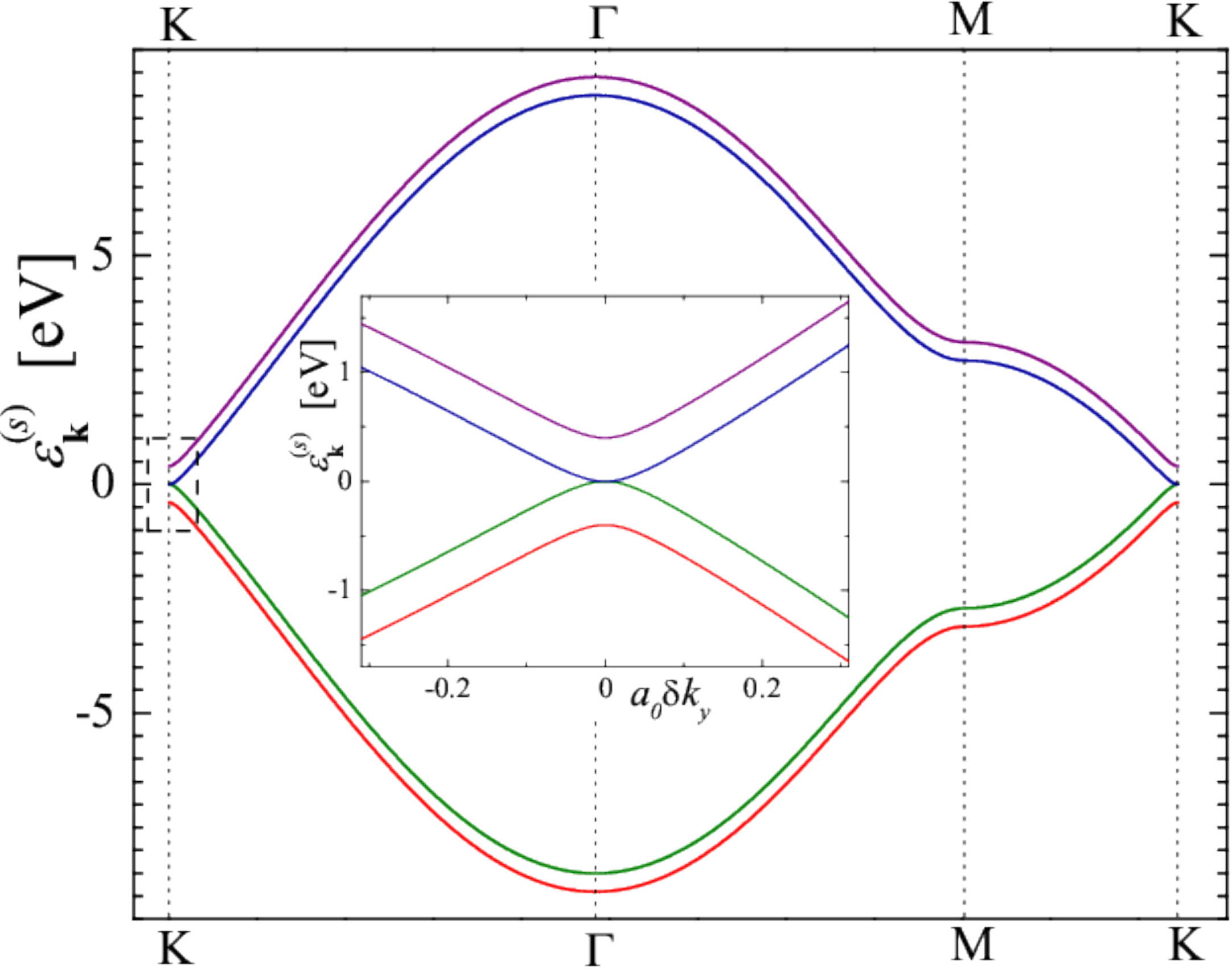}
\par\end{centering}
\caption{
(Color online) Single-particle band structure of AB-stacked bilayer
graphene. At large energy the spectra of AB and AA bilayers resemble each
other. However, near the Dirac points thier structures are quite
dissimilar. Instead of degenerate Fermi surface of AA-stacked graphene (see
Fig.~\ref{AABandBrfig}),
the AB bilayer has two Fermi points at which two parabolic bands touch each
other (see inset, which zooms on the area bounded by the dashed line).
\label{ab::fig::bands}
}
\end{figure}

From the general geometry of carbon $p$-orbitals one expects that the
overlap between the orbitals of the dimer sites is the largest.
Consequently, the hopping
$t_0$
between the dimer sites is the strongest of all possible interlayer
tunneling processes. Thus, the simplest possible description of the
AB bilayer is to include only these tunneling amplitudes. In such an
approximation, using the notation of
Eq.~(\ref{HaaK}),
one can write for the Hamiltonian in
${\bf k}$-space
\begin{eqnarray}
\label{ab::H}
H^{AB}_{\bf k}
=
\left(
	\begin{matrix}
		0& 0& -tf({\bf k})& 0\cr
		0& 0& t_0 &-tf({\bf k})\cr
		-tf^*({\bf k})& t_0& 0& 0\cr
		0& -tf^*({\bf k})& 0& 0\cr
	\end{matrix}
\right).
\end{eqnarray}
For the AB bilayer, the Hamiltonian parameters are estimated as
follows~\cite{Malard2007,
Zhang2008,
Li2009}
\begin{eqnarray}
2.9\,{\rm eV} \leq t \leq 3.16\,{\rm eV},
\qquad
0.3\,{\rm eV} \leq t_0 \leq 0.4\,{\rm eV}.
\end{eqnarray}
The matrix from
Eq.~(\ref{ab::H})
is easy to diagonalize. The corresponding eigenenergies satisfy the
equation
\begin{eqnarray}
\varepsilon^4
-
2 \varepsilon^2 \left( t^2 |f({\bf k})|^2 + \frac{1}{2} t_0^2 \right)
+
t^4 |f({\bf k})|^4
= 0\,,
\end{eqnarray}
which can be solved to find the following four bands
\begin{eqnarray}
\label{ab::energy1234}
\left(
	\varepsilon^{(1,4)}_{\bf k}
\right)^2
=
t^2 |f({\bf k})|^2
+
\frac{1}{2} t_0^2
+
\sqrt{t_0^2 t^2 |f({\bf k})|^2 + \frac{1}{4} t_0^4}\,,
\\
\left(
	\varepsilon^{(2,3)}_{\bf k}
\right)^2
=
t^2 |f({\bf k})|^2
+
\frac{1}{2} t_0^2
-
\sqrt{t_0^2 t^2 |f({\bf k})|^2 + \frac{1}{4} t_0^4}\,.
\end{eqnarray}
These bands are plotted in
Fig.~\ref{ab::fig::bands}.

In the vicinity of the Dirac points
${\bf K}$
(or
${\bf K}'$),
where
\begin{eqnarray}
\label{ab::near_dirac}
|f({\bf k})| \ll \frac{t_0}{2t},
\end{eqnarray}
equation~(\ref{ab::energy1234})
can be expanded, and the approximate formulas for the energy bands are
derived
\begin{eqnarray}
\label{ab::bands_approx}
\varepsilon^{(1,4)}_{\bf K+q}
\approx
\pm \left(
	t_0 + \frac{t^2}{t_0} |f({\bf k})|^2
    \right)
\approx
\pm \left(
	t_0 + \frac{\hbar^2 v_{\rm F}^2}{t_0} |{\bf q}|^2
    \right),
\\
\label{ab::bands_approx2}
\varepsilon^{(2,3)}_{\bf K+q}
\approx
\pm \frac{t^2}{t_0} |f({\bf k})|^2
\approx
\pm \frac{\hbar^2 v_{\rm F}^2}{t_0} |{\bf q}|^2,
\end{eqnarray}
where we used
Eq.~(\ref{slgDirac})
for
$v_{\rm F}$,
valid near the corners of the Brillouin zone.
Unlike both the monolayer and AA bilayer graphene, AB bilayer bands have
parabolic dispersion near
${\bf K}$
and
${\bf K}'$
points. The bands
$s=2,3$
touch each other at the Dirac points. If the AB bilayer is not doped, the
bands touch exactly at the Fermi energy. Two other bands
($s=1,4$)
do not reach the Fermi energy.

If one is interested in the effective description of the bilayer at
$|\varepsilon| \ll t_0$,
the bands
$s=1,4$
may be neglected, since
$|\varepsilon_{\bf k}^{(1,4)}| \geq t_0$.
Further, one can easily check that, if
${\bf k} \approx {\bf K}$
(${\bf k} \approx {\bf K}'$),
the wave functions for
$s=1,4$
bands are non-zero only at the dimer sites. Conversely, the low-energy
states of
$s=2,3$
bands are localized at the non-dimer sites. Therefore, only the electrons
at the non-dimer atoms participate in the low-energy processes.  The
two-band (thus, superscript `2b') effective Hamiltonian is equal
to~\cite{mccann_kosh_rev2013,mccann_falko2006}
\begin{eqnarray}
\label{ab::2b}
H^{\rm 2b}_{\xi \bf q}
\approx
-\frac{\hbar^2}{2m}
\left(
	\begin{matrix}
		0 & (i q_x + \xi q_y)^2 \cr
		(i q_x - \xi q_y)^2& 0 \cr
	\end{matrix}
\right),
\qquad
m = \frac{t_0}{2 v_{\rm F}^2},
\end{eqnarray}
where the integer-valued index $\xi$ is equal to +1 (-1) at the
${\bf K}$
(${\bf K}'$)
Dirac point. The effective mass $m$ can be expressed in terms of the
free-electron mass
$m_0$
as
$m = 0.033 m_0$.

The described band structure, however, is easy to spoil. For example, the
gap opens in the spectrum if we apply an electric field perpendicular to
the bilayer plane. We will discuss this type of perturbation below, in
subsection~\ref{subs::bernal_bias_gap}.
In addition to the transverse field, the dispersion given by
Eq.~(\ref{ab::bands_approx})
may be destroyed by hopping from non-dimer sites in one layer to non-dimer
sites in another layer. In
Fig.~\ref{fig::ab_lattice}
the tunneling processes of this type are shown by dashed lines labeled by
$t_3$.
The corresponding Hamiltonian matrix reads
\begin{eqnarray}
\label{ab::H_trig}
H^{AB}_{\bf k}
=
-\left(
	\begin{matrix}
		0& 0& -tf({\bf k})& t_3 f^* ({\bf k}) \cr
		0& 0& t_0 &-tf({\bf k})\cr
		-tf^*({\bf k})& t_0& 0& 0\cr
		t_3 f ({\bf k}) & -tf^*({\bf k})& 0& 0& \cr
	\end{matrix}
\right).
\end{eqnarray}
\begin{figure}[t!]
\begin{centering}
\includegraphics[width=0.45\textwidth]{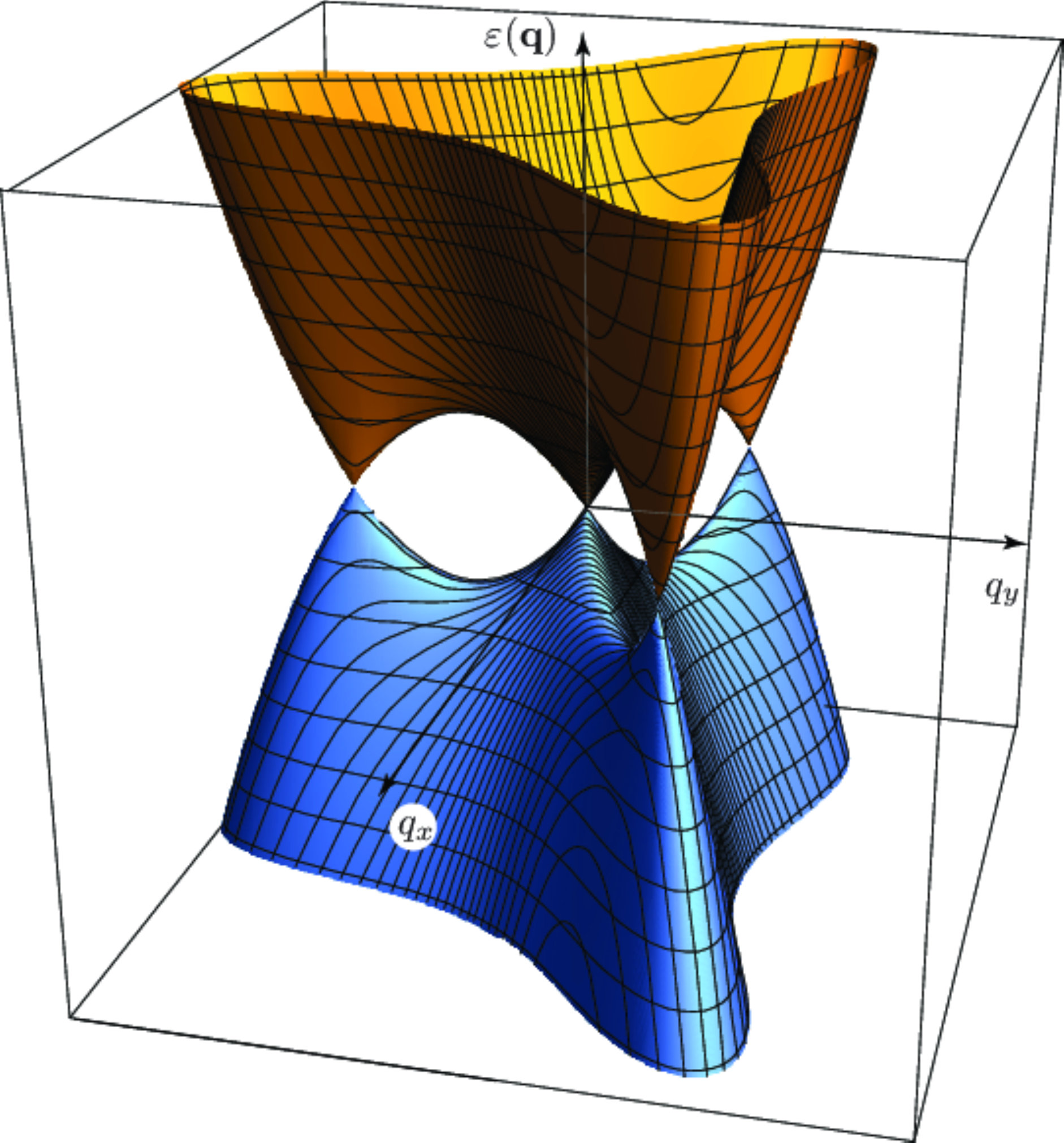}
\includegraphics[width=0.45\textwidth]{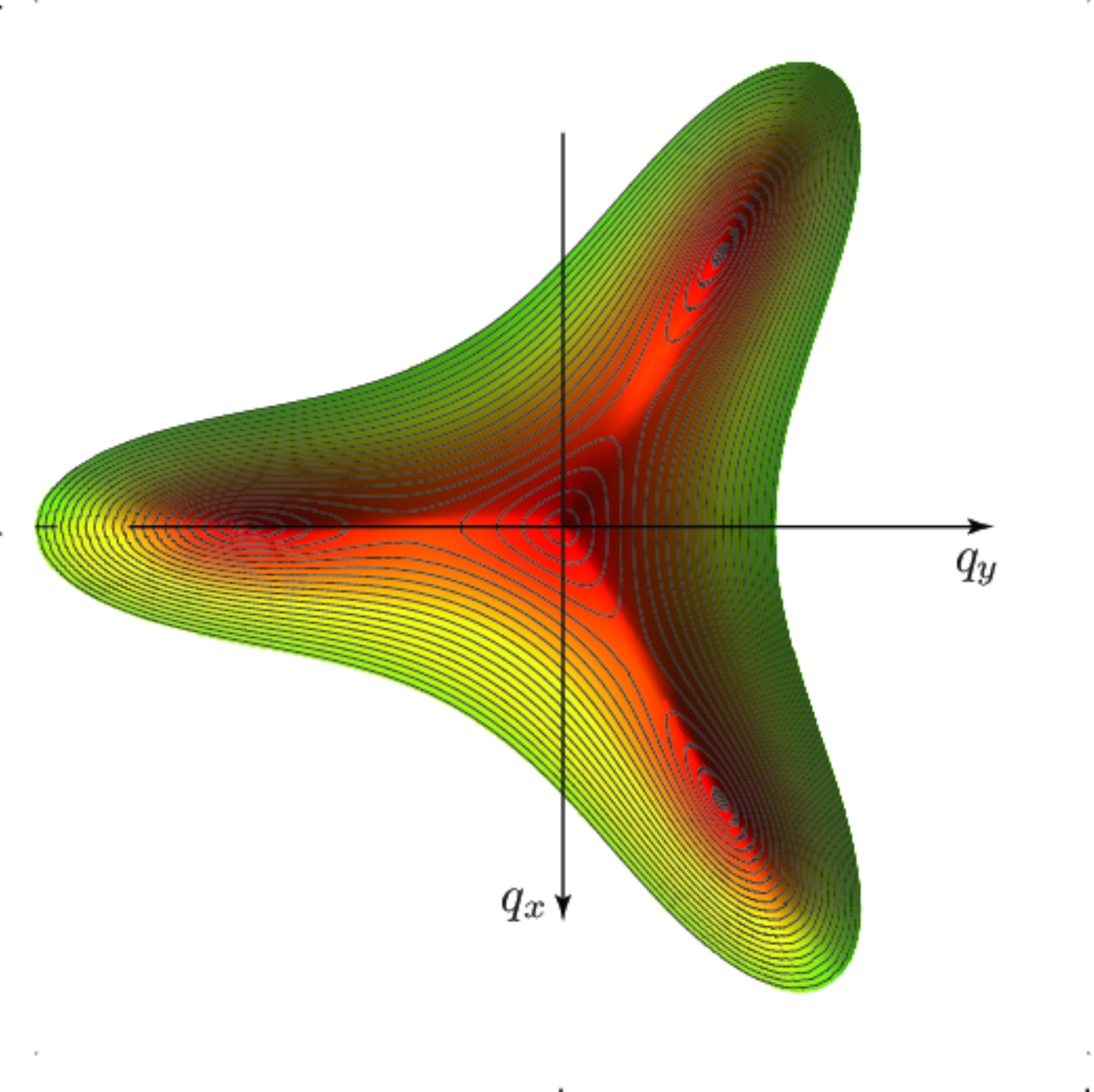}
\par\end{centering}
\caption{Trigonal warping. When the interlayer tunneling between
non-dimer sites is taken into account, the parabolic dispersion at the Dirac
point is split into four Dirac cones with linear dispersion. This distorted
dispersion surface is shown on the left panel, while the right panel 
shows top view of the same surface, with equal-energy contours visible. The
point (0,0) corresponds to a corner of the Brillouin zone, and
$q_{x,y}$
are momenta measured from this corner. Of the four emerged Dirac cones, the
central is located at the corner of the Brillouin zone, while three others
are shifted from the corner by the amount $p_L$. The resultant structure
complies with the underlying geometrical symmetry of the bilayer. This
distortion of the original parabolic dispersion is called ``trigonal
warping". Note, however, that for
$t_3 = 0.3$\,eV
(the value of
$t_3$
has been estimated in
Refs.~\cite{Kuzmenko2009,dft_graphIte1991})
the warping effect is very weak. The energy
$\varepsilon_L$,
at which the saddle points are located, is
estimated~\cite{mccann_kosh_rev2013}
to be of the order of 1\,meV. The quantity
$p_L$
is
$\sim 0.007a_0^{-1}$.
%
\label{fig::trig_warp}
}
\end{figure}
The spectrum of this matrix can be found
exactly~\cite{mccann_falko2006}.
However, since the analytical formulas are quite complicated, we will not
discuss them. Instead, let us direct our attention to
Fig.~\ref{fig::trig_warp}.
In this figure the two bands reaching the Fermi level are plotted in the
vicinity of a Brillouin zone corner. The bands are calculated for non-zero
$t_3$,
whose value was extracted from
experiment~\cite{Kuzmenko2009}
and DFT
calculations~\cite{dft_graphIte1991}:
\begin{eqnarray}
t_3 = 0.3\,\text{eV}.
\end{eqnarray}
As we can see, the interlayer hopping between the non-dimer sites
introduces a qualitative distortion of the single-electron bands: the
parabolic dispersion near the
${\bf K}$
and
${\bf K}'$
points is replaced at low energy by a structure consisting of four Dirac
cones. The apexes of these cones lie at the Fermi energy. The apex of the
central cone coincides with a corner of the Brillouin zone
(${\bf K}$,
or
${\bf K}'$).
Three others are shifted from the corner by a small quantity
$p_L$.
This low-energy fine structure is called ``trigonal warping"
\cite{mccann_falko2006}.
To account the warping within the two-band approximation one may work with
the following effective Hamiltonian
\cite{mccann_kosh_rev2013}:
\begin{eqnarray}
\label{ab::trig_warp}
\tilde H^{\rm 2b}_{\xi {\bf q}}
=
H^{\rm 2b}_{\xi {\bf q}}
+
H^{\rm 2b}_{\xi {\bf q}{\rm w}},
\qquad
\text{where}
\qquad
H^{\rm 2b}_{\xi {\bf q}{\rm w}}
=
\hbar v_3
\left(
	\begin{matrix}
		0 & \xi q_y-iq_x \cr
		\xi q_y+iq_x& 0 \cr
	\end{matrix}
\right),
\quad
v_3
=
\frac{3 a_0 t_3}{2\hbar}.
\end{eqnarray}
It is necessary to remember, however, that this warping is a very weak
effect, which become noticeable only very close to the
${\bf K, K}'$
points at small energies. Specifically, the six saddle points (three for
$\varepsilon>0$
and another three for
$\varepsilon<0$),
clearly visible in
Fig.~\ref{fig::trig_warp},
are at the
energies
$\varepsilon = \pm \varepsilon_L$,
where~\cite{mccann_kosh_rev2013,
mccann_falko2006}
\begin{eqnarray}
\label{ab::lifshits}
\varepsilon_L = \frac{t_0}{4} \left( \frac{v_3}{v_{\rm F}} \right)^2
\approx 1\,\text{meV},
\quad
\text{and}
\quad
p_L=\frac{t_0v_3}{\hbar v_{\rm F}^2}\sim 0.007 a_0^{-1}
\end{eqnarray}
Both
$\varepsilon_L$
and
$p_L$
are quite small. As a result, in experiment, disorder may easily mask the
warping.

Finally, let us mention that to fit experimental data to the
theoretically-calculated dispersion additional tunneling amplitudes and
on-site energies are often
included~\cite{Malard2007,
Zhang2008,
Li2009,
Kuzmenko2009,
Mafra2009,
Kuzmenko2009b,
Zou2011}.
As an example, in
Fig.~\ref{fig::ab_lattice}
the tunneling
$t_4$
between dimer and non-dimer sites is shown. This tunneling process
introduces small, however, experimentally measurable asymmetry between hole
and electron states of the AB bilayer.

\subsubsection{Effect of the bias voltage}
\label{subs::bernal_bias_gap}

The asymmetry between the layers due to, for example, the application of
the transverse voltage, opens a gap in the single-electron
spectrum~\cite{mccann_falko2006,McCann2006}.
To account for the applied voltage we can generalize
Eq.~(\ref{ab::H})
\begin{eqnarray}
\label{ab::H_U}
H^{AB}_{\bf k}(V)
=
\left(
	\begin{matrix}
		eV/2 & 0& -tf({\bf k})& 0\cr
		0& -eV/2 & t_0 &-tf({\bf k})\cr
		-tf^*({\bf k})& t_0& eV/2 & 0\cr
		0& -tf^*({\bf k})& 0& -eV/2 \cr
	\end{matrix}
\right).
\end{eqnarray}
%
The eigenenergies of
$H_{\bf k}^{AB} (V)$
satisfy the equation:
\begin{eqnarray}
\left(\varepsilon^2 - \frac{e^2 V^2}{4} \right)^2
-
\left( 2 \varepsilon^2 - \frac{e^2 V^2}{2} \right)
\left( t^2 |f({\bf k})|^2 + \frac{1}{2} t_0^2 \right)
+
t^4 |f({\bf k})|^4
-e^2 V^2 t^2 |f({\bf k})|^2
= 0,
\end{eqnarray}
which has four solutions:
\begin{eqnarray}
\label{ab::band_U}
\varepsilon^{(1,2,3,4)}_{\bf k}
=
\pm
\sqrt{
	\frac{e^2 V^2}{4}
	+
	\frac{1}{2} t_0^2
	+
	t^2 |f({\bf k})|^2
	\pm
	\sqrt{
		t^2|f({\bf k})|^2(e^2 V^2 + t_0^2) + \frac{1}{4} t_0^4
	}
}.
\end{eqnarray}
The bands
$s=2,3$,
which touch at the Fermi level when
$V=0$,
do not reach the Fermi energy for
$V > 0$.
To prove this, it is convenient to expand the expression
Eq.~(\ref{ab::band_U}),
assuming that the condition
Eq.~(\ref{ab::near_dirac})
is valid. Thus, near the Dirac point we derive
\begin{eqnarray}
\label{ab::bands_U_approx}
\varepsilon^{(2,3)}_{\bf q}
\approx
\pm\sqrt{
	\frac{e^2 V^2}{4}
	-
	\frac{e^2 \hbar^2 v_{\rm F}^2 V^2}{4 t^2_0}|{\bf q}|^2
	+
	\frac{\hbar^4 v_{\rm F}^4}{t_0^2}
	\left(
		1 + \frac{e^2 V^2}{4 t_0^2}
	\right)
	|{\bf q}|^4
}.
\end{eqnarray}
It is easy to check that the expression under the root sign is always
positive when
$V \ne 0$.
If we put
$V=0$
in this formula,
Eq.~(\ref{ab::bands_approx2})
is recovered.

For the low-energy description, we can use the generalized
Eq.~(\ref{ab::2b})
\begin{eqnarray}
\label{ab::2b_U}
H^{\rm 2b}_{\xi \bf k}(V)
\approx
\left(
	\begin{matrix}
		eV/2 & -\frac{\hbar^2 v_{\rm F}^2}{t_0} (\xi q_y+iq_x)^2 \cr
		-\frac{\hbar^2 v_{\rm F}^2}{t_0} (\xi q_y-iq_x)^2 & -eV/2 \cr
	\end{matrix}
\right).
\end{eqnarray}
The dispersion relation for this Hamiltonian is as follows
\begin{eqnarray}
\label{ab::band_U_crude}
\varepsilon^{2b\pm}_{\xi \bf k}
=
\pm \sqrt{
		\frac{e^2 V^2}{4}
		+
		\frac{\hbar^4 v_{\rm F}^4}{t_0^2}|{\bf q}|^4
         }.
\end{eqnarray}
Apparently, this expression does not coincide with
Eq.~(\ref{ab::bands_U_approx}).
The effective Hamiltonian given by
Eq.~(\ref{ab::2b_U})
is a fairly coarse approximation to
Eq.~(\ref{ab::H_U}).
Despite that it is being used in the theoretical literature: the
Hamiltonian
$H^{\rm 2b}_{\xi \bf k}(V)$,
Eq.~(\ref{ab::2b_U}),
is able to reproduce the gapped dispersion and significantly simplify the
formalism.

As we have commented earlier,
Eqs.~(\ref{ab::bands_U_approx})
and~(\ref{ab::band_U_crude})
show that, for finite transverse electric field, the bilayer becomes an
insulator with gap $\Delta$ of the order of $|eV|$. Experiments confirmed
the existence of this
gap~\cite{Henriksen2010,
Kuzmenko2009,
Kuzmenko2009b,
Ohta2006,
Castro2007,
Jing2010,
Hao2016}.
For example, A.\,B.~Kuzmenko
et~al.~\cite{Kuzmenko2009,Kuzmenko2009b}
observed the band gap as large as
$0.1$\,eV.
As an additional support, the emergence of the gap was also reported by
workers employing density functional methods (e.g.,
Ref.~\cite{Nanda2009}
discussed free-standing samples,
Refs.~\cite{Slawinska2010,Ramasubramaniam2011}
investigated the samples on a substrate). The ability to engineer the gap
attracted a lot of attention, both from theorists and experimentalists. The
possibility of turning a graphene sample into a semiconductor with
controlled gap appears to be a promising notion with a variety of
applications, some of which will be discussed below.

\subsubsection{Density of states}

\begin{figure}[t!]
\begin{centering}
\includegraphics[width=16cm]{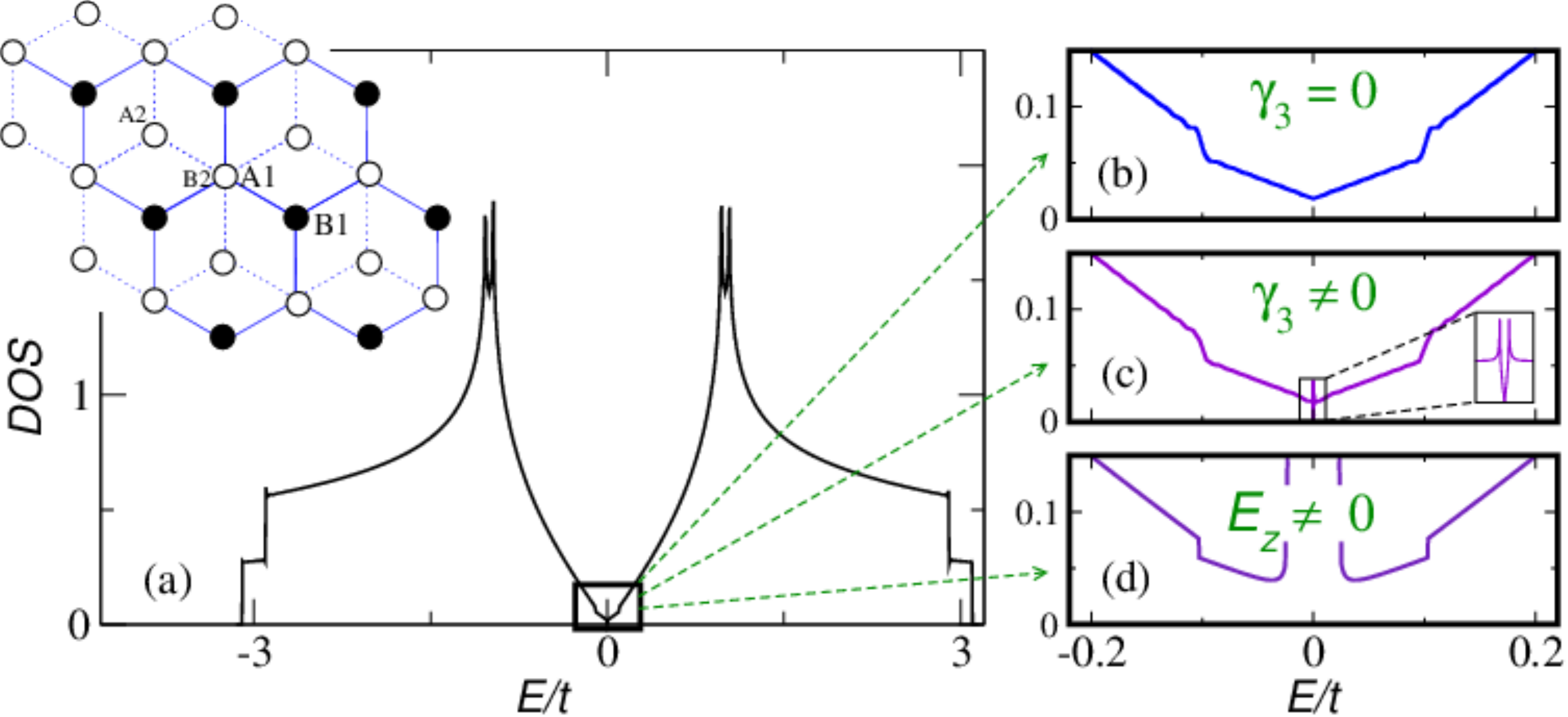}
\par\end{centering}
\caption{{\bf Density of states for graphene bilayer with AB stacking.} In
panel~(a) the general shape of the DOS is demonstrated. Unlike 
single-layer graphene, the AB bilayer has a finite density of states at the
Fermi energy. This is a consequence of the parabolic dispersion, which in
two-dimensional systems leads to finite density of states. In panel~(b) the
DOS for
$t_3=0$
(no trigonal warping) is presented ($\gamma_3$ in the Figure
is the same as our amplitude
$t_3$).
In panel~(c) the trigonal warping is accounted for. In the presence of the
warping, the density of states vanishes linearly near the zero energy.
However, this effect becomes significant only at very small temperatures.
In panel~(d) the spectrum with a gap due to a transverse voltage is
plotted.
Reprinted from
E.V.~Castro et al.,
New J. Phys. {\bf 11}, 095017 (2009).
\label{fig::ab_dos}
}
\end{figure}

The density of states for AB graphene calculated by E.\,V.~Castro et~al. in Ref.~\cite{dos2009} is shown in
Fig.~\ref{fig::ab_dos}.
In panel~(a) the general shape of the density of states is demonstrated.
Unlike the single-layer graphene, the AB bilayer has finite density of
states at the Fermi energy. This is a consequence of the parabolic
dispersion in two-dimensional systems:
\begin{eqnarray}
\int \ldots
	\frac{dq_xdq_y}{(2\pi)^2}
\ \  
=
\ \ 
\int \ldots
	\frac{d(q^2)}{4\pi}
\ \  
=
\ \  
\int \ldots \rho(0) \, d \varepsilon.
\end{eqnarray}
This point is illustrated more explicitly in
Fig.~\ref{fig::ab_dos}(b),
where the density of states for
$t_3=0$
(no trigonal warping) is presented (in the notation of E.\,V.~Castro et~al.~\cite{dos2009}
the quantity
$\gamma_3$
is the same as our amplitude
$t_3$).

In
Fig.~\ref{fig::ab_dos}(c),
the trigonal warping is accounted for. In the presence of the warping, the
density of states vanishes linearly near the zero energy. However, this
effect becomes significant only at very small energy/temperature.

In panel~(d), the spectrum with a gap due to the transverse voltage is
plotted. When the field is applied the gap opens, see
Eq.~(\ref{ab::bands_U_approx}).
The electron states pushed from the sub-gap region pile up into two peaks
at the edges of the gap.

\subsubsection{Immediate consequences of the single-electron band theory}
\label{subs::bernal_single-electron_props}

As we have mentioned above, significant amount of experimental data
concerning the electronic properties of the AB bilayer may be explained
with the help of single-electron concepts. Thus, it is often assumed that,
at not too low temperatures, single-particle models are sufficient to
describe how the external perturbations (strain, disorder, radiation,
electric or magnetic field, etc.) affect the electronic properties of a
graphene bilayer.

For example, the AB bilayer in a magnetic field was discussed in
Refs.~\cite{Castro2007,
Castro2007a,
Castro2010,
Nemec2007}.
For low magnetic field, the semi-classical cyclotron mass for a biased AB
bilayer was calculated by E.\,V.~Castro et~al. in
Ref.~\cite{Castro2007a}
for different carrier densities and bias electric fields. The effects of
disorder on the bias-induced gap were investigated. The
authors~\cite{Castro2007a}
concluded that the gap induced by the transverse electric field is quite
insensitive to the diagonal disorder. The paper also discusses the
formation of Landau levels in bilayer nanoribbons, using the tight-binding
approximation.

Reference~\cite{Castro2007}
compared the theoretical results for the cyclotron mass with experiments.
The theory was consistent with the data for samples with electron doping.
Tangible disagreement between the theory and the experiment for the hole
doping was attributed to poor performance of the Hartree approximation used
for calculating the electrostatic screening of the bias field. In addition,
Ref.~\cite{Castro2007}
consider quantum Hall conductivity of a bilayer sample (see also
Section~\ref{MagnF}).
The gap was estimated from the data. It was
concluded~\cite{Castro2007}
that the gap could be as large as
$\sim 0.1$\,eV.
The results of
Refs.~\cite{Castro2007,Castro2007a}
were summarized and generalized in
Ref.~\cite{Castro2010}.

In
Ref.~\cite{Nemec2007},
N.~Nemec and G.~Cuniberti calculated the magnetic-field-induced Hofstadter
butterfly patterns for
samples of different stackings (AA, AB, and intermediate) at zero bias.

The effects of strain on the single-electron dispersion were discussed in
Refs.~\cite{Nanda2009,
Choi2010,
Mucha-Kruczynski2011}.
In particular, the uniaxial strain perpendicular to the layers was studied by B.\,R.\,K.~Nanda and S.~Satpathy~\cite{Nanda2009}
using DFT. Such a perturbation changes the interlayer distance. The
authors~\cite{Nanda2009}
concluded that the strain cannot open the gap for AB and for AA bilayers.

The situation in which the strain of the top layer is unequal to the strain
of the bottom layer was investigated by S.-M.~Choi et~al. in
Ref.~\cite{Choi2010}
with the help of DFT. The authors noticed that the asymmetric strain
induces and electric field perpendicular to the layers. The electric field
appears because the deformations significantly alter the layers work
function.  According
to~\cite{Choi2010},
this field generates gap in the single-electron spectrum. The authors
proposed that the studied system may be used as an electromechanical
switch.

The modifications to the single-electron dispersion and the Landau levels
in a strained AB bilayer were discussed by M.~Mucha-Kruczy{\'{n}}ski et~al.
in
Ref.~\cite{Mucha-Kruczynski2011}.
The strain splits the parabolic bands
$\varepsilon^{(2,3)}_{\bf k}$,
Eq.~(\ref{ab::bands_approx}),
into several Dirac cones (this splitting resembles the trigonal warping
phenomena). The number of these cones, as well as other details of
the dispersion surface, depends on the strain. The energy scale for these
effects was estimated to be of order of a few meV. The changes in the
dispersion affect the behavior of the Landau levels energies.

%
%
\section{Dirac electrons: chirality and Klein paradox}\label{ChiralKlein}
%

\subsection{Chiral electrons}\label{Chiral}

As we have already seen, the wave functions of single-layer and bilayer graphene are spinors. These spinors are solutions of the Dirac equation for massless fermions, which implies specific symmetries and related topological properties of the charge carriers in these systems. These effects have been intensively studied in single-layer graphene. Let us briefly review them following Ref.~\cite{CastrNrev}.

Consider once more the wave functions $\psi^\pm_{1,2} ({\bf q})$ of single-layer graphene near two Dirac points $\mathbf{K}$ and $\mathbf{K}'$, Eqs.~\eqref{SlgPsi}. According to these equations, if the phase of $\theta_{\bf q}$ is rotated by $2\pi$, the wave function changes sign, since it acquires a phase of $\pi$ (this is commonly called Berry's phase). This change of phase under rotation is characteristic of spinors of the particle with spin $1/2$. However, this property is unrelated to a real spin of the electron, which enters the problem in a trivial manner. It is associated with a pseudospin variable, related to the two components of the wave function, and usually referred to as helicity, or chirality, which is defined as the projection of the momentum on the pseudospin direction. To work with the chirality it is necessary to introduce the corresponding operator. The states near $\mathbf{K}$, and the states near $\mathbf{K}'$ have different chirality operators:
\begin{equation}\label{Chir}
\text{near\ \ }
{\bf K}:
\text{\ \ }
\hat{\chi}_{1}=\frac{\hat{\bm\sigma}\mathbf{q}}{2 |{\bf q|}};
\qquad\quad
\text{near\ \ }
{\bf K}':
\text{\ \ }
\hat{\chi}_{2}=\frac{\hat{\bm\sigma}^{*}\mathbf{q}}{2 |{\bf q|}}.
\end{equation}
These operators commute with the respective Dirac Hamiltonians Eqs.~\eqref{SlgDiracEqs}. It follows from the above definition that the states $\psi_{1,2}^{\pm}$ are also eigenstates of $\hat{\chi}_{1,2}$,
\begin{equation}\label{pmChi}
\hat{\chi}_1\psi_1^{\pm}=\pm\frac{1}{2}\psi_1^{\pm}\,,
\quad
\hat{\chi}_2\psi_2^{\pm}=\pm\frac{1}{2}\psi_2^{\pm}\,.
\end{equation}
Therefore, electrons (holes) have a positive (negative) chirality, or helicity. Thus, the pseudospin can be either parallel, or antiparallel to the momentum $\mathbf{q}$. The chirality values are good quantum numbers \textit{only} near the Dirac points, as long as Eqs.~\eqref{SlgDiracEqs} are valid, and the spectrum is linear. Thus, it is only an asymptotic property, and at larger energies the chirality stops being a good quantum number.

The states near a particular Dirac point, either $\mathbf{K}$ or $\mathbf{K}'$, are commonly referred to as valleys~\cite{CastrNrev}. Each valley can be characterized by a valley index $\xi=\pm 1$, where $\xi=+1$ corresponds to $\mathbf{K}$, and $\xi=-1$ to $\mathbf{K}'$, respectively. Using this index, we can rewrite the Dirac equations Eq.~\eqref{SlgDiracEqs} for electrons in single-layer graphene in the form
\begin{equation}\label{SlgDiracEqs_Valley}
\hat{H}\psi_\xi(\mathbf{r})
=
\varepsilon\psi_\xi(\mathbf{r}),
\qquad
\hat{H}=-i\hbar v_{\rm F}\left(
\begin{array}{cc}
0 & \frac{\partial}{\partial x}-i\xi\frac{\partial}{\partial y} \\
\frac{\partial}{\partial x}+i\xi\frac{\partial}{\partial y} & 0 \\
\end{array}
\right)
\end{equation}

Thus, if we consider only low-energy processes and disregard hopping, which
can break electron-hole symmetry, then, the intervalley scattering is
absent, and the valley index is conserved. However, the interactions, which
break both the inversion symmetry of the lattice and time-reversal, also
break the symmetry between the two
valleys~\cite{CastrNrev}.
Therefore, the electron scattering becomes valley dependent, and, in
general, electrons from different valleys propagate along different paths.
The chiral Dirac nature of fermions in graphene might be of use for
applications where one can control the valley flavor of the electrons
besides its charge, the so-called
valleytronics~\cite{CastrNrev,Rycerz,Shimazaki2015}.

\subsubsection{AA bilayer graphene}\label{AAChiral}

Here we consider the chiral properties of AA bilayer graphene. Note, however, that up to now only a few works treated this subject.

As it is readily seen from the previous consideration, the low-lying energy spectrum ($\varepsilon\ll t_0$) of the AA graphene is made up of two Dirac cones, one shifted to higher energy, and other one shifted to lower energy, see panel~(b) of Fig.~\ref{AABandBrfig}. These cones intersect each other at $\varepsilon=0$, forming a circular Fermi surface, instead of the Fermi point for single-layer graphene. Similar to single-layer graphene, the low-lying electronic excitations in AA-stacked bilayer graphene can be treated asymptotically as massless Dirac fermions. Expanding the function $f(\mathbf{k})$ [see Hamiltonian of Eq.~\eqref{HaaK}] to lowest order in either $\mathbf{q}=\mathbf{K}-\mathbf{k}$, or $\mathbf{q}=\mathbf{K}'-\mathbf{k}$, one obtains $f(\mathbf{k}) \approx \hbar v_{\rm F} |{\bf q}| e^{\mp i\theta_\mathbf{q}}$, where the `minus' (`plus') sign corresponds to the $\mathbf{K}$ ($\mathbf{K}'$) valley. It is convenient to express the four-component spinor $\Psi_{\mathbf{q}} = (a_{\mathbf{q}1}, b_{\mathbf{q}1}, a_{\mathbf{q}2}, b_{\mathbf{q}2})^T$ in the form $\Psi_{\mathbf{q}}=(\psi_{1\mathbf{q}},\psi_{2\mathbf{q}})^T$, where $\psi_{i\mathbf{q}}=(a_{\mathbf{q}i},b_{\mathbf{q}i})^T$ are the two-component spinors constructed from the electron operators in layer $i=1,2$. We also introduce two sets of Pauli matrices ${\hat{\bm\sigma}}$ and ${\hat{\bm\tau}}$, where the ${\hat{\bm\sigma}}$ matrices act on the sublattice index, while the ${\hat{\bm\tau}}$ matrices act on the layer index. Then, in
the real-space representation, the corresponding Dirac
equation can be written as
\begin{equation}\label{AaDirEq}
\hat{H}\Psi_\mathbf{q}(\mathbf{r})=\varepsilon\Psi_\mathbf{q}(\mathbf{r})\,,
\end{equation}
with
\begin{equation}\label{AaHDir}
\hat{H}=\hat{\tau}_x\otimes\left(t_0\hat{I}-i\hbar
v_{\rm F} \hat{\bm\sigma}{\bm\nabla}\right)\,\,\,
(\textrm{near}\,\, \mathbf{K}),
\qquad \hat{H}=\hat{\tau}_x\otimes\left(t_0\hat{I}-i\hbar
v_{\rm F} \hat{\bm\sigma}^*{\bm\nabla}\right)\,\,\,
(\textrm{near}\,\, \mathbf{K'}),
\end{equation}
where the symbol $A \otimes B$ denotes the direct product of the matrices $A$ and $B$, and symbol $\hat{I}$ is the $2\times2$ unit matrix.

Following notations by M.~Sanderson et~al. in Ref.~\cite{ChirAaKlein}, we introduce the band indexes $\eta_1=\pm 1$ and $\eta_2=\pm 1$. In this notation the spectrum Eq.~\eqref{aaBandDir} can be rewritten as
\begin{equation}\label{epsilonqnn}
\varepsilon_{\mathbf{q}\eta_1\eta_2}=\eta_1\left(\eta_2t_0+\hbar v_{\rm F} |{\bf q}|\right),
\end{equation}
with eigenvectors
\begin{equation}\label{aaEigenV}
\psi_{\mathbf{q}\eta_1\eta_2}=\frac{e^{i\mathbf{q}\mathbf{r}}}{2}\left(
                      \begin{array}{c}
                        \eta_1e^{-i\theta_\mathbf{q}} \\
                        \eta_1\eta_2 \\
                        \eta_2e^{-i\theta_\mathbf{q}} \\
                        1 \\
                      \end{array}
                    \right).
\end{equation}
For the Hamiltonians of Eq.~\eqref{AaHDir} we can define two operators for two conserved quantities
\begin{equation}\label{AaOperTop}
\hat{C} = \gamma^5 = \hat{\tau}_x\otimes\hat{I}, \qquad \hat{\chi}_1 = \hat{\tau}_x\otimes\frac{\hat{\bm\sigma}\mathbf{q}}{q}\,\,
(\textrm{near}\,\,\mathbf{K}), \qquad \textrm{or} \qquad
\hat{\chi}_2=\hat{\tau}_x\otimes\frac{\hat{\bm\sigma}^*\mathbf{q}}{q}\,\,
(\textrm{near}\,\,\mathbf{K}').
\end{equation}
where $\gamma^5$ is the fifth Dirac gamma matrix. The operators $\hat{\chi}_{1,2}$ can be considered as chirality operators (the analog of the chirality operators for single-layer graphene). Regarding $\hat{C}$, it can be referred to as the cone operator~\cite{ChirAaKlein}. Both $\hat{C}$, and $\hat{\chi}_{1,2}$ are the direct products of $\tau_x$, with either $\hat{I}$ or chirality operators for single-layer graphene, Eq.~\eqref{pmChi}. The authors of Ref.~\cite{ChirAaKlein} introduced the cone $c\equiv\eta_1\eta_2$, and chirality $\chi\equiv\eta_1$ indexes. The energy bands can now be labeled by these indexes as
\begin{equation}\label{eqcx}
\varepsilon_{\mathbf{q}c\chi}=ct_0+\chi\hbar v_{\rm F} |{\bf q}|,
\end{equation}
with the eigenvectors
\begin{equation}\label{aaEigenVchi}
\psi_{\mathbf{q}c\chi}=e^{i\mathbf{q}\mathbf{r}}\left(
                      \begin{array}{c}
                        \chi e^{-i\theta_\mathbf{q}} \\
                        c \\
                        c\chi e^{-i\theta_\mathbf{q}} \\
                        1 \\
                      \end{array}
                    \right).
\end{equation}
Thus, each band is characterized by two numbers ($c,\chi$): the first band has ($c=\chi=-1$), the second ($c=+1,\chi=-1$), the third ($c=-1,\chi=+1$), and the fourth ($c=\chi=+1$).

It can be readily shown that
$\hat{C}\psi_{\mathbf{q}c\chi}=c\psi_{\mathbf{q}c\chi}$,
and
$\hat{\chi}_{1,2}\psi_{\mathbf{q}c\chi}=\chi\psi_{\mathbf{q}c\chi}$.
A quasiparticle state belongs to the upper (lower) cone, if $c = +1$ ($c =
-1$). The physical significance of the $\chi$ eigenvalue becomes
immediately obvious if we consider the group
velocity
$\mathbf{v}_{\mathbf{q}c\chi} \equiv\partial\varepsilon_{\mathbf{q}c\chi}/\partial\mathbf{q}=\chi v_{\rm F} \mathbf{q}/|{\bf q}|$. A quasiparticle state is electron-like (i.e., $\mathbf{v}_{\mathbf{q}c\chi}$ is parallel to $\mathbf{q}$) if $\chi=+1$, and hole-like (i.e., $\mathbf{v}_{\mathbf{q}c\chi}$ is anti-parallel to $\mathbf{q}$) if $\chi= -1$. Finally, let us note that these discrete quantum numbers can also be defined for the lattice model~\cite{PrlOur,PrbOur}.

In addition, as in the case of single-layer graphene, we could introduce also the valley index $\xi$, and rewrite the Hamiltonian Eq.~\eqref{AaHDir} as
\begin{equation}\label{AaHDir_Valley}
\hat{H}=\hat{\tau}_x\otimes\left[t_0\hat{I}-i\hbar v_{\rm F}\left(
\begin{array}{cc}
0 & \frac{\partial}{\partial x}-i\xi\frac{\partial}{\partial y} \\
\frac{\partial}{\partial x}+i\xi\frac{\partial}{\partial y} & 0 \\
\end{array}
\right)
\right].
\end{equation}
Here $\xi=+1$ corresponds to the point $\mathbf{K}$ and $\xi=-1$ corresponds to the $\mathbf{K'}$ point.

\subsubsection{AB bilayer graphene}\label{ABChiral}

If we are interested only in electrons whose energy is low ($\varepsilon\ll t_0$), but not too low [$\varepsilon \gg \varepsilon_L$, see Eq.~(\ref{ab::lifshits})], we can neglect the trigonal warping and use the effective Hamiltonian of the AB-stacked graphene bilayer in the form of Eq.~\eqref{ab::2b}. In this effective Hamiltonian, the valley index $\xi$ is already included. Hamiltonian~(\ref{ab::2b}), together with the Hamiltonian of single-layer graphene, may be viewed as the first two members of the infinite sequence of chiral Hamiltonians $H_J$ ($J=1,2,\dots$), corresponding to Berry's phase $J\pi$~\cite{mccann_kosh_rev2013} 
\begin{equation}\label{BerryJHam}
 H_J\propto \left(
              \begin{array}{cc}
                0 & (\hat{\pi}^\dag)^J \\
                (\hat{\pi})^J & 0 \\
              \end{array}
            \right)\,,
\end{equation}
where
\begin{equation}\label{oneline}
\hat{\pi}=\xi\partial/\partial x+i\partial/\partial y \qquad \textrm{and}\qquad \hat{\pi}^\dag=\xi\partial/\partial x-i\partial/\partial y.
\end{equation}
It is convenient to generalize this series for negative $J$ as well
\begin{equation}\label{neg_BerryJHam}
 H_J\propto \left(
              \begin{array}{cc}
                0 & (\hat{\pi})^{|J|} \\
                (\hat{\pi}^\dag)^{|J|} & 0 \\
              \end{array}
            \right)\,,
\text{ for\ } J < 0.
\end{equation}
The chiral Hamiltonian~\eqref{ab::2b} has a single Dirac point at $\mathbf{q} = 0$, the spectrum is rotationally invariant, and the phase difference between sublattices changes by $2\times2\pi$ when the quasiparticle-momentum circles the Dirac point. This property is referred to as $J = 2$ chirality, or winding number $+2$. When trigonal warping is included, the conduction and valence bands touch each other not only at $\mathbf{q} = 0$, but also at three additional $\mathbf{q} \neq 0$ points (see Fig.~\ref{fig::trig_warp}). The central Dirac point at $\mathbf{q} = 0$ has winding number $J=-1$, whereas the three surrounding points have $J = 1$.

\subsection{Klein tunneling and chiral scattering}\label{Klein}

\begin{figure}[t]
\includegraphics[width=0.8\columnwidth]{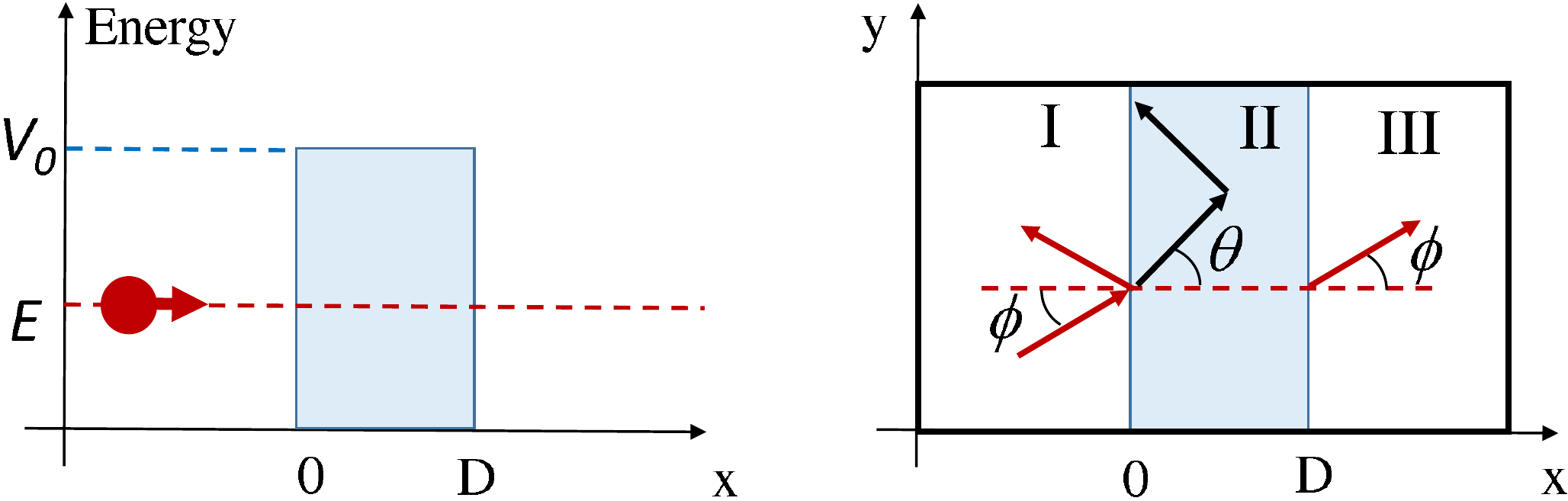}
\centering
\caption{(Color online) \textbf{Klein tunneling in single-layer graphene.} Left panel: schematic of the scattering of Dirac electrons by a rectangular potential. Right panel shows the definition of the angles $\theta$ and $\phi$ used in the scattering formalism in regions I, II, and III.
\label{SLGBarKleinfig}}
\end{figure}

A striking peculiarity of electron scattering in graphene systems is the so-called Klein tunneling, which is a characteristic of electrons with linear dispersion. We start by calculating the transparency of the rectangular barrier in single-layer graphene for electrons near the Dirac point~\cite{CastrNrev}. In Fig.~\ref{SLGBarKleinfig} we depict the scattering process due to the square barrier of width $D$ and height $V_0$. We further assume that the scattering does not mix the momenta around the $\mathbf{K}$ and $\mathbf{K}'$ points.

\begin{figure}[t]
\includegraphics[width=0.8\columnwidth]{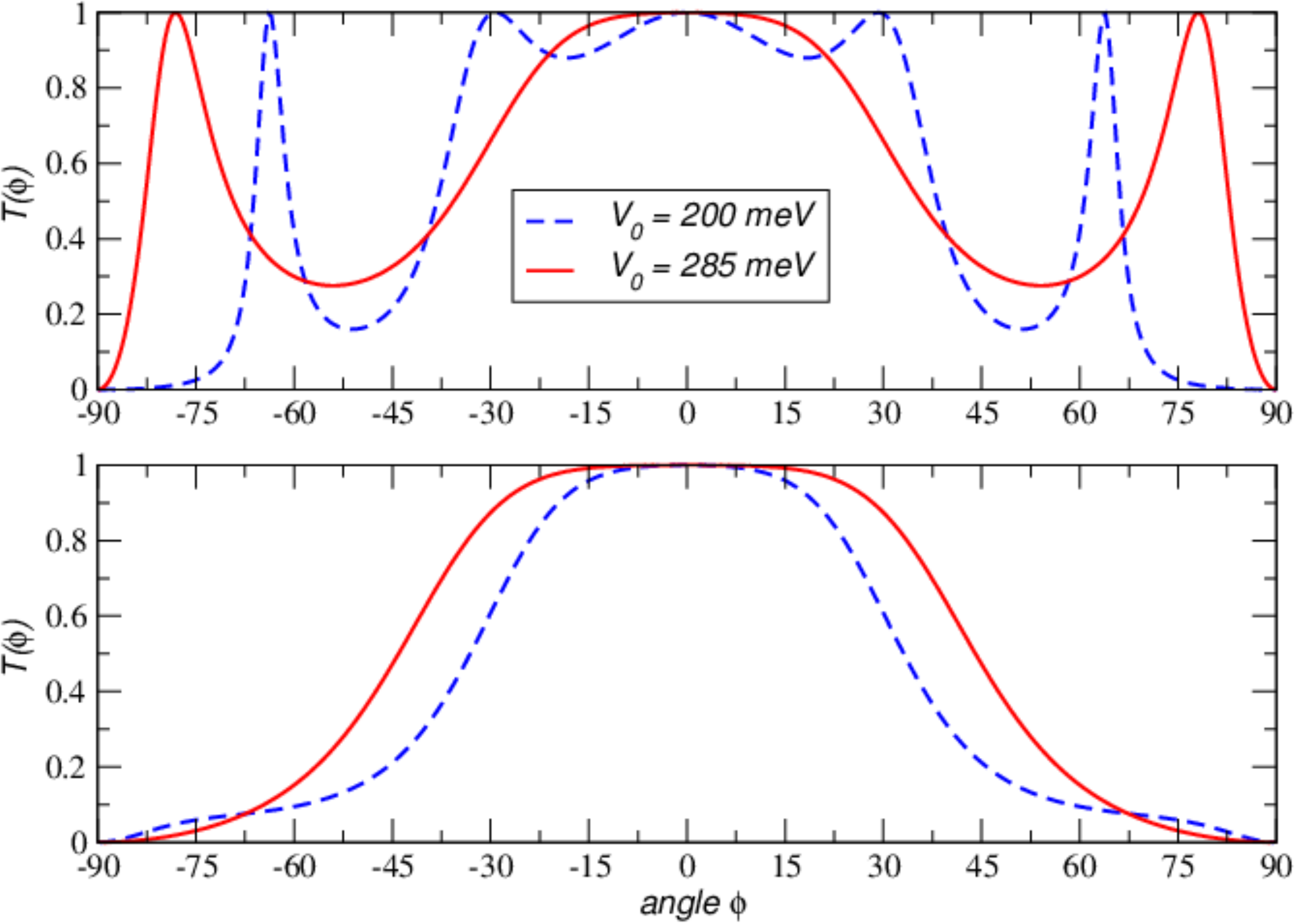}
\centering
\caption{(Color online) Angular behavior of the transmission coefficient
$T(\phi)$ for single-layer graphene calculated for two different values of
the barrier height $V_0$: $V_0=200$\,meV, dashed (blue) curve;
$V_0=285$\,meV, solid (red) curve. The barrier width is $D=110$\,nm for the
top panel and $D =50$\,nm for the bottom panel. For all curves the energy
of the transmitted electrons is $E=80$\,meV, and the Fermi momentum is
$k_F=2\pi/50$\,\,nm$^{-1}$
Reprinted figure with permission from
A.H.~Castro Neto et al.,
Rev. Mod. Phys. {\bf 81} 109 (2009).
Copyright 2009 by the American Physical Society.
\url{http://dx.doi.org/10.1103/RevModPhys.81.109}
\label{Klein_SLG_AAfig}}
\end{figure}

Using the spinor representation Eq.~\eqref{SlgPsi}, we can rewrite the wave function of the electron with momentum $\mathbf{k}$ (near the $\mathbf{K}$ point) and energy $E$ in the different regions $\psi_{\textrm{I,II,III}}$ in terms of incident and reflected waves
\begin{eqnarray}\label{ParadKlein}
\nonumber
\psi_{\textrm{I}} &=& \psi(\chi,\phi)e^{ik_xx}+r\psi(\chi,\pi-\phi)e^{-ik_xx}, \\
\nonumber
\psi_{\textrm{II}} &=& a\psi(\chi',\theta)e^{iq_xx}+b\psi(\chi',\pi-\theta)e^{-iq_xx}, \\
\psi_{\textrm{III}} &=& \tau\psi(\chi,\phi)e^{ik_xx},
\end{eqnarray}
where
\begin{equation}\label{SLGKleinSpinor}
\psi(\chi,\phi) =\left(\!
                               \begin{array}{c}
                                 1 \\
                                 \chi e^{i\phi}  \\
                               \end{array}
                             \right)e^{ik_yy},
   \end{equation}
and $\phi=\arctan(k_y/k_x)$, $k_x=k_F\cos{\phi}$, $k_y=k_F\sin{\phi}$, $k_F$ is the Fermi momentum, $\theta=\arctan(k_y/q_x)$, $q_x=\sqrt{(V_0-E)^2/v_{\rm F}^2-k_y^2}$, with chiralities $\chi=\textrm{sign}(E)$ and $\chi'=\textrm{sign}(E-V_0)$. Unlike the non-relativistic Schr\"{o}dinger equation, we only need to match the wave function, but not its derivative. As a result, we find the transmission coefficient of the barrier $T(\phi)=|\tau|^2$ in the form~\cite{CastrNrev}:
\begin{equation}\label{SLGKleinTrans}
T(\phi)
=
\frac{\cos^2{\phi}\cos^2{\theta}}
{[\cos{\!(Dq_x)}\cos{\phi}\cos{\theta}]^2
+
\sin^2{\!(Dq_x)}(1\!-\!\chi\chi'\sin{\phi}\sin{\theta})^2}.
\end{equation}
The transmission is ideal [$T(\phi)=1$] for normal incident electrons ($\phi=0$, and $\theta = 0$) even for very high ($V_0\gg E$) and long ($D q_x \gg 1$) barriers. This property is referred to as the Klein paradox, and it does not occur for non-relativistic electrons. Note, that chirality is not conserved in the process of scattering. The dependence of the transmission coefficient $T$ on the incident angle $\phi$ is shown in Fig.~\ref{Klein_SLG_AAfig}.

As it follows from Eq.~\eqref{ParadKlein}, the electrons moving through a barrier separating the regions of n- and p-doped graphene (regions I and II in Fig.~\ref{SLGBarKleinfig}) are transmitted as holes. For a hole, the momentum and velocity are opposite vectors. This behavior is the same as that of photons moving in a medium with negative refraction index~\cite{CastrNrev}.

Klein tunneling and related phenomena in graphene were discussed in
numerous
papers~\cite{klein_tunn2006,graphene_veselago2007,bliokhFrei,yam_klein1,
yam_klein2}
and
reviews~\cite{CastrNrev,chiral_tunn_tudorovskiy2012,meso_review} 
Note, however, that typically the authors assume a ballistic regime; that
is, they suppose that the mean free path of the charge carriers exceeds
$D$. Otherwise, some delicate effects might
disappear~\cite{fogler_pn_disorder}.

\subsubsection{Klein scattering in AA-stacked bilayer graphene}\label{aaKlein}

The scattering of the charge carriers by barriers in AA bilayer graphene was studied theoretically by several authors. The scattering of electrons by a rectangular potential barrier in AA-stacked bilayer graphene was analyzed in detail by M.~Sanderson et~al. in Ref.~\cite{ChirAaKlein}. A standard barrier geometry shown in Fig.~\ref{SLGBarKleinfig} was considered. The wave function of the electron with momentum $\mathbf{k}$ near the ${\bf K}$ point has the same structure as the wave function in Eq.~\eqref{ParadKlein}. This time, however, instead of the two-component spinor $\psi(\chi,\phi)$, Eq.~\eqref{SLGKleinSpinor}, we have the four-component spinor Eq.~\eqref{aaEigenVchi}, which is convenient to write in the form \begin{equation}\label{aaKleinSpinor}
\psi(c,\chi,\phi)=\left(
                      \begin{array}{c}
                        \chi e^{i\phi} \\
                        c \\
                        c\chi e^{i\phi} \\
                        1 \\
                      \end{array}
                    \right)e^{ik_yy}.
\end{equation}
As compared with the single-layer graphene, the electron in AA bilayer graphene is characterized by the second topological number $c$ (cone index).

The calculation described in the previous subsection can be adopted for the derivation of the transmission coefficient for AA bilayer graphene. As a result, we obtain that Eq.~\eqref{SLGKleinTrans} is valid for the AA-stacked bilayer graphene, if we consider electrons with the same cone index~\cite{ChirAaKlein}. While the electron can change its chirality in the scattering process, the cone index is conserved. This can be understood by considering the step potential as a sudden perturbation~\cite{ChirAaKlein}. The probability for an initial wave function $\psi(c,\chi,\phi)$ to undergo a sudden transition into a new state of $\psi(c',\chi',\phi')$ is given by
\begin{equation}\label{aaConsCon}
P= 
\langle\psi^\dag\!(c',\chi',\phi')
\psi(c,\chi,\phi)\rangle
\equiv(1\!+\!cc')(1\!+\!\chi\chi'e^{i(\phi'-\phi)}).
\end{equation}
This means that the transition probability $c\rightarrow -c$ is strictly zero.

Due to the orthogonality of states with different cone indexes, electron transport across a potential barrier must conserve the cone index, and this leads to protected cone transport, which is unique in AA bilayer graphene. Together with the negative refraction of electrons, electrons residing in different cones can be spatially separated according to their cone index when transmitted across an n-p junction. This suggests the possibility of `cone-tronic' devices based on AA-stacked bilayer~\cite{ChirAaKlein}. A detailed study of conductance of the junctions is also performed in Ref.~\cite{ChirAaKlein}

The Klein tunneling in AA bilayer graphene in the presence of a magnetic
field and a bias voltage was studied by D.~Wang and G.~Jin in
Ref.~\cite{DWang}.

\subsubsection{Klein scattering in AB-stacked bilayer graphene}\label{abKlein}

A comprehensive study of chiral Klein tunneling in AB-stacked bilayer graphene is presented in the review paper by T.~Tudorovskiy et~al., Ref.~\cite{chiral_tunn_tudorovskiy2012}, where different barrier configurations were analyzed. Here we only present some general ideas and apply these to calculate the transmission coefficient for the rectangular barrier in Fig.~\ref{SLGBarKleinfig}.

As usual for the problem of Klein tunneling of chiral electrons in graphene, we consider low-energy excitations, $|E|,\, |E - V_0|\ll t_0$. In this case, the effective Hamiltonian is given by Eq.~\eqref{ab::2b}. Further, for definiteness, we consider the valley with $\xi=1$. For not-too-small wave vectors $\mathbf{k}$, the effects of trigonal warping may be neglected~\cite{chiral_tunn_tudorovskiy2012}. Indeed, if $k\gg p_L$, the warping is not important (see
subsection~\ref{spectraAB}).

Let us now consider an electron incident on the potential step barrier $u(x)$ shown in Fig.~\ref{SLGBarKleinfig} at an angle $\phi$. Since the potential is constant in the $y$-direction, we can write the solution as $\psi(x,y)=\psi(x)\exp{(ik_yy)}$. Inserting this into the Schr\"{o}dinger equation with the Hamiltonian Eq.~\eqref{ab::2b}, we obtain
\begin{equation}\label{SchrABKlein}
\left(\frac{d^2}{dx^2}-k_y^2\right)^2
\psi
=
\left[\frac{m\left(E-u(x)\right)}{\hbar^2}\right]^2
\psi,
\end{equation}
where $m$ is the effective mass of the quasiparticle in the AB bilayer [see Eq.~(\ref{ab::2b})]. Equation~(\ref{SchrABKlein}) has four solutions: two propagating waves $\exp{\!(\pm ik_xx)}$, one exponentially growing, and one exponentially decaying modes $\exp{\!(\pm \kappa_x x)}$. The presence of evanescent modes is markedly different from both the Schr\"{o}dinger case, and the Dirac case~\cite{chiral_tunn_tudorovskiy2012}. The parameters $k_{x,y}$ and $\kappa_x$ satisfy the relations
\begin{equation}\label{WVectabKlein}
k_x^2+k_y^2=\frac{m\left|E-u(x)\right|}{\hbar^2},
\qquad
\kappa_x^2-k_y^2=\frac{m\left|E-u(x)\right|}{\hbar^2}.
\end{equation}
In what follows, we assume that the barrier is high, $V_0>E+\hbar^2k_y^2/2m$.

Now let $k = \sqrt{2mE}/\hbar$ be the wave vector for the propagating modes in the region I and III (see Fig.~\ref{SLGBarKleinfig}), while $q = \sqrt{2m(E-V_0)}/\hbar$ is the wave vector in the region II. Then, the solution in regions I, II, and III is given by
\begin{eqnarray}\label{abWaveKlein}
\nonumber
  \psi_\textrm{I} &=& \left(
               \begin{array}{c}
                 1 \\
                 \chi e^{2i\phi} \\
               \end{array}
             \right)e^{ik_xx}+r\left(
               \begin{array}{c}
                 1 \\
                 \chi e^{-2i\phi} \\
               \end{array}
             \right)e^{-ik_xx}+c_1\left(
               \begin{array}{c}
                 1 \\
                 -\chi h_1(\phi) \\
               \end{array}
             \right)e^{\kappa_xx},
   \\
  \psi_{\textrm{II}} &=& a_2\left(
                              \begin{array}{c}
                                1 \\
                                \chi'e^{2i\theta} \\
                              \end{array}
                            \right)e^{iq_xx}+b_2\left(
                              \begin{array}{c}
                                1 \\
                                \chi'e^{-2i\theta} \\
                              \end{array}
                            \right)e^{-iq_xx}
	\\
	\nonumber
			&&+ c_2\left(
                              \begin{array}{c}
                                1 \\
                                -\chi'h_1(\theta) \\
                              \end{array}
                            \right)e^{i\lambda_xx}+d_2\left(
                              \begin{array}{c}
                                1 \\
                                -\chi'/h_1(\theta) \\
                              \end{array}
                            \right)e^{-i\lambda_xx}
    , \\
\nonumber
  \psi_{\textrm{III}} &=& \tau\left(
                               \begin{array}{c}
                                 1 \\
                                 \chi e^{2i\phi}\\
                               \end{array}
                             \right)e^{ik_xx}+c_3\left(
                                                   \begin{array}{c}
                                                     1 \\
                                                     -\chi /h_1(\phi) \\
                                                   \end{array}
                                                 \right)e^{-\kappa_xx},
\end{eqnarray}
where $k_y=k\sin{\phi}$, $k_x=k\cos{\phi}$, $\kappa_x=k\sqrt{1+\sin^2{\phi}}$, $h_1(\phi)=\left(\sqrt{1+\sin^2{\phi}}-\sin{\phi}\right)^2$, $q_x=q\sin{\theta}$, and $\lambda_x=q\sqrt{1+\sin^2{\theta}}$. To obtain constants in Eqs.~\eqref{abWaveKlein}, we have to match the spinor $\psi$ and its derivative $d\psi/dx$, as in the case of the usual Schr\"{o}dinger equation.

For the case of normal incidence $\phi = \theta = 0$, we can solve the problem analytically. The transmission coefficient is given by~\cite{chiral_tunn_tudorovskiy2012}
\begin{equation}\label{ABTr_Klein}
T(0)=\frac{8k^2q^2}{(k^2+q^2)^2\cosh{(2qD)}+6k^2q^2-k^4-q^4}.
\end{equation}
When $2qD\gg 1$, the transmission probability becomes exponentially small, even at normal incidence.

In the general case, the transmission $T(\phi)=|\tau|^2$, or reflection $R(\phi)=|r|^2$
coefficients can be found only numerically~\cite{chiral_tunn_tudorovskiy2012}. Similar to the case of single-layer and AA bilayer graphene, there are angles, at which the transmission is perfect. The existence of such angles in bilayer graphene means that we cannot lock a conventional transistor made of AB-stacked bilayer graphene.

The results obtained in considering the Klein tunneling can be applied for analyzing pn-junctions~\cite{chiral_tunn_tudorovskiy2012} (see also Section~\ref{subsect::meso::pn}).

\section{Bilayer graphene in a magnetic field: Landau levels and
quantum Hall effect}\label{MagnF}
%

One of the most common approaches for investigating the band structure of metals and semiconductors is to study Landau levels using magneto-optics or magneto-transport measurements. In this section we consider the properties of bilayer graphene in a magnetic field. In the tight-binding approximation, the hopping integrals are replaced by a Peierls substitution
\begin{equation}\label{Magnticghtbi}
t_{\mathbf{RR}'}\Rightarrow t_{\mathbf{RR}'}\exp{\!\!\left(\!\frac{ie}{\hbar c}\!\int_\mathbf{R}^{\mathbf{R}'}{\!\!\!\!\mathbf{A}\cdot d\mathbf{r}}\right)}= t_{\mathbf{RR}'}\exp{\!\!\left(\!\!\frac{2\pi i}{\Phi_0}\!\int_\mathbf{R}^{\mathbf{R}'}{\!\!\!\!\mathbf{A}\cdot d\mathbf{r}}\right)},
\end{equation}
where $t_{\mathbf{RR}'}$ is the hopping integral between the sites located at $\bf R$ and $\bf R'$ (calculated at zero magnetic field), the quantity $\mathbf{A}$ is the vector potential of the magnetic field, and $2\pi\hbar c/e=\Phi_0$ is the flux quantum. Correspondingly, for the effective Hamiltonian, near the Dirac points, we should replace
\begin{equation}\label{MagnticDirEq}
-i\hbar{\bm\nabla}\,\,\Rightarrow\,\, -i\hbar{\bm\nabla}+e\mathbf{A}/c\,.
\end{equation}

When a uniform magnetic field $\mathbf{B}$ is applied perpendicular to the graphene plane, we can use the Landau gauge: $\mathbf{A}=B(-y,0)$. For the case of single-layer graphene, such a problem was analyzed in details in many papers (see the review Ref.~\cite{CastrNrev}). Here we are interested in the low-energy electron states with momenta near the Dirac points. For massless charge carriers with linear dispersion, the cyclotron frequency is~\cite{CastrNrev}
\begin{equation}\label{CyclFre}
\omega_c=\sqrt{2}\frac{v_F}{l_B},
\end{equation}
where
\begin{equation}\label{MagnLength}
l_B=\sqrt{\frac{c\hbar}{eB}}
\end{equation}
is the magnetic length. The energies of the Landau levels are given by
\begin{equation}\label{LanLev}
E_{LL}^\pm(n) = \pm\hbar\omega_c\sqrt{n},
\end{equation}
where $n=0,1,2, ...$. The Landau levels for both Dirac points, ${\bf K}$ and ${\bf K}'$, have exactly the same spectrum. Hence each Landau level is doubly-degenerate due to the valley factor, and also doubly degenerate due to the spin factor. Thus, the total degeneracy of each level in single-layered graphene is four. The Landau level energies for graphene scale as $\sqrt{Bn}$ (while for the electrons with quadratic dispersion: $E_{LL} \propto Bn$). Such a behavior for single-layer graphene was confirmed in a number of experiments. The energy scale associated with the magnetic field for Dirac fermions is rather different from that found in the ordinary 2D gas of massive non-relativistic electrons. For instance, for fields of the order of $B=10$\,T, the cyclotron energy $\hbar\omega_c$ in the 2D electron gas is of the order of $10$\,K, while in graphene it is about $400$\,K. This implies that the quantum Hall effect can be observed at
room temperature~\cite{NovoQHE}.

The Landau level spectrum of a conventional two dimensional semiconductor is
\begin{equation}\label{LL2Dg}
E_{LL} = \hbar\,\frac{eB}{mc}\left(n + \frac{1}{2}\right), \qquad n\geq 0,
\end{equation}
where $eB/mc$ is the cyclotron frequency for massive charge carriers with
quadratic dispersion~\cite{MacDonald}. As the electron density is changed,
there is a step in the Hall conductivity whenever a Landau level crosses
the Fermi level, and the separation of steps on the density axis is equal
to the maximum carrier density per Landau level, $N_fB/\Phi_0$, where $N_f$
is a degeneracy factor. Each plateau of the Hall conductivity  occurs at a
quantized value of $nN_fe^2/2\pi\hbar$, where $n$ is an integer labeling
the plateau. Steps between adjacent plateaus have height
$N_fe^2/2\pi\hbar$. In single-layer graphene there exist a zero-level state
and all of the levels are fourfold degenerate. The corresponding steps of
the Hall conductivity have the height $2e^2/\pi\hbar$ between each plateau,
but the plateaus occur at half-integer values of $2e^2/\pi\hbar$ instead of
integer ones, which was observed experimentally~\cite{Novoselov2005,Dubonos2}. This is due to the existence of the level at zero energy, which contributes to a step of height $2e^2/\pi\hbar$ at zero density.

\subsection{Landau levels and integer quantum Hall effect in AA-stacked
bilayer graphene}\label{LandauLAA}

\subsubsection{Landau levels}

The effects on the AA-stacked bilayer graphene due to a constant uniform magnetic field were discussed by several authors (see, e.g., Refs.~\cite{QE_AA,Opt_LLAA,LL_QE_AA,LL_Gate_AA,LL_NonUnF_AA}). In all of these studies the electron-electron interaction was neglected and the electronic spectrum considered was gapless (see Fig.~\ref{AABandBrfig}). One must remember, however, that the interaction could open a gap in the bilayer single-electron spectrum (Section~\ref{interaction}). The presence of such a gap may significantly affect the system behavior in a magnetic field.

Following Refs.~\cite{QE_AA,LL_QE_AA} we consider electrons near the Dirac cone in AA-stacked bilayer graphene placed in a constant uniform transverse magnetic field $B$ with a bias voltage $V$ across the layers. Then, in the Landau gauge we have to replace in the Hamiltonian Eq.~\eqref{AaHDir} the term
\begin{equation}\label{REPLACE}
-i\hbar\,\partial/\partial x\,\Rightarrow \, -[i\hbar\,\partial/\partial x+(e/c)yB]
\end{equation}
and renormalize the interlayer hopping integral $t_0$ according to Eq.~\eqref{t0aaV0}. The solution $\Psi(x,y)$ of the Dirac equation $\hat{H}\Psi=E\Psi$ with renormalized Hamiltonian Eq.~\eqref{AaHDir} is sought in the form $e^{ikx}\Psi(y)$. Solving the eigenvalue problem, we obtain the Landau levels for AA bilayer graphene~\cite{QE_AA,LL_QE_AA}
\begin{equation}\label{LL_AA}
E_{LL}^{\eta\lambda}(n)=-\eta\hbar\omega_c\sqrt{n}-\lambda\sqrt{t_0^2+\left(\frac{eV}{2}\right)^2}.
\end{equation}
The corresponding eigenfunctions are
\begin{eqnarray}\label{LLEigF_AA}
\Psi_{k0}^{1\lambda}(y)\!\!\!
&=&\!\!\!
\frac{1}{\sqrt{2}}\left[0,Q_0 (\zeta),0,\lambda Q_0 (\zeta)\right]^T,\qquad
\textrm{if}\;n=0\,,\nonumber\\
\Psi_{kn}^{\eta\lambda}(x)\!\!\!
&=&\!\!\!
\frac{1}{2}\!\left[Q_{n-1} (\zeta),\eta Q_n (\zeta),
\lambda Q_{n-1} (\zeta),\eta\lambda Q_n (\zeta) \right]^T\!\qquad
\textrm{if}\;n\geq 0\,.
\end{eqnarray}
In these equations, $\eta =\pm1$, $\lambda =\pm1$, and
\begin{equation}\label{Hermite}
 Q_n (\zeta) =2^{-n/2}(n!)^{-1/2}\exp{\left(-\zeta^2/2\right)}H_n(\zeta),
\end{equation}
where $H_n(\zeta)$ are the Hermite polynomials and dimensionless variable $\zeta=y/l_B+kl_B$. Clearly, the Landau levels of the AA bilayer graphene are nothing but two copies of the Landau levels of single-layer graphene shifted up and down by $\sqrt{t_0^2+e^2V^2_0/4}$. The Landau levels in the AA bilayer are four-fold degenerate, as in the case of the single-layer graphene.

\begin{figure}[t]
\includegraphics[width=0.6\columnwidth]{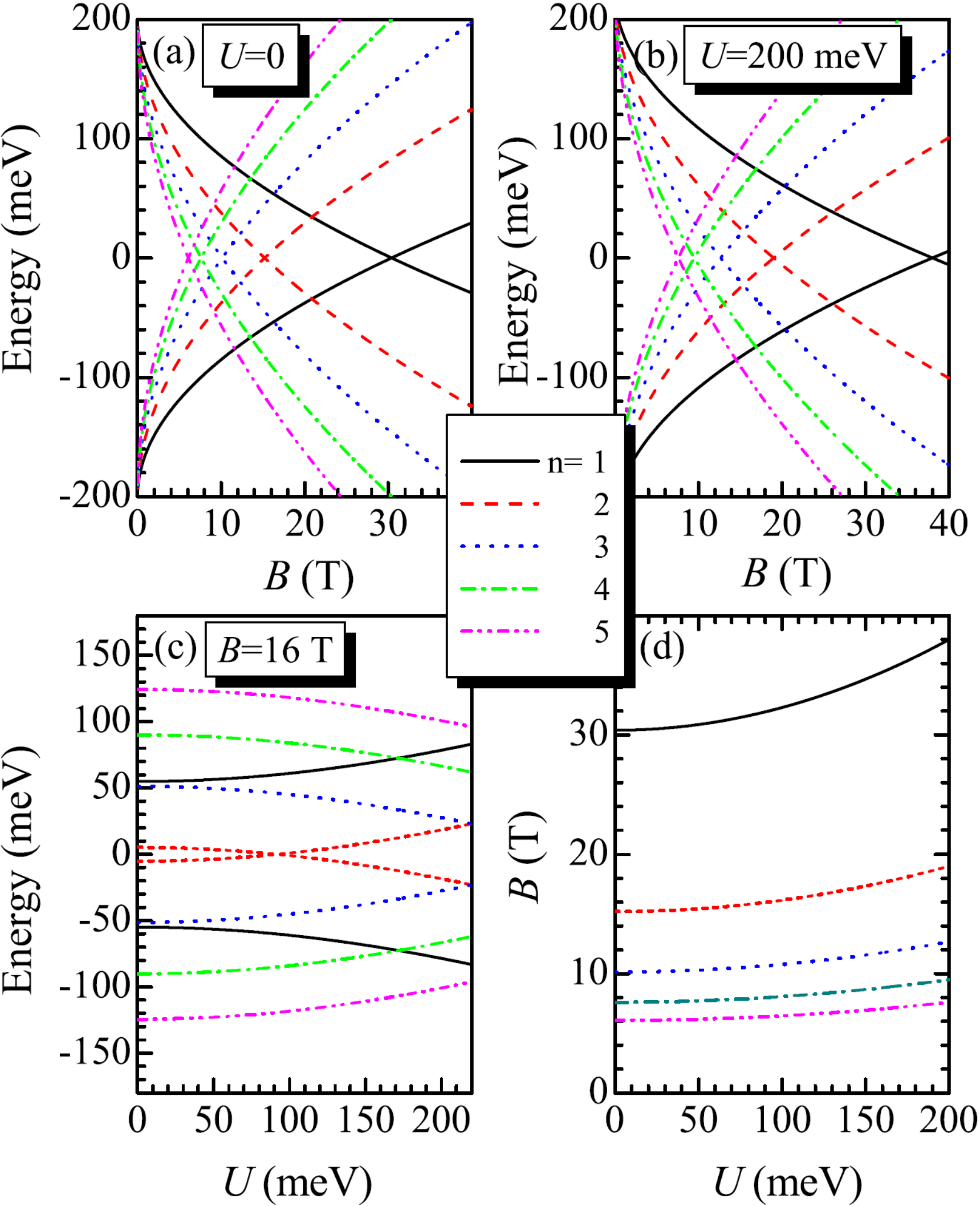}
\centering
\caption{(Color online) Five ($n=1,...5$) Landau levels in the AA-stacked bilayer graphene (from Ref.~\cite{LL_QE_AA}) with $\eta\lambda=-1$ as functions of the magnetic field $B$ for two interlayer potential differences $U=eV$: (a) $U =0$ and (b) $U = 200$~meV. (c) Landau levels in AA-stacked bilayer graphene as a function of $U$ for $B = 16$~T. (d) It is possible to adjust the bias voltage and the transverse magnetic field in such a manner that the energy of a particular Landau level becomes equal to the Fermi energy. For this to happen $V$ and $B$ must satisfy Eq.~(\ref{LL_AA_0}). The latter condition defines a function $B(n, V)$, which is plotted in this panel for several values of the index $n$. All curves in this figure are calculated for $v_F=10^6$~m/s.
Reprinted from Physics Letters A, {\bf 375}, 
D.~Wang, ``Electric- and magnetic-field-tuned Landau levels and Hall
conductivity in AA-stacked bilayer graphene",
page~4070, Copyright 2011, with permission from Elsevier.
\url{http://dx.doi.org/10.1016/j.physleta.2011.09.025}
\label{LL_AAfig}}
\end{figure}

Five Landau energy levels ($n=1,...5$) with $\eta\lambda=-1$ are shown in Fig.~\ref{LL_AAfig}~(a) and~(b) as functions of the magnetic field (figures are taken from the D.~Wang, Ref.~\cite{LL_QE_AA}).The levels with $\eta\lambda=-1$ cross each other. The Landau levels of the AA-stacked bilayer graphene can be effectively modified by the bias potential between the layers [see Fig.~\ref{LL_AAfig} (c)]. In general, one could adjust $B$ and $V$, and fine tune the energy of a particular Landau level to Fermi energy. For this to happen, $V$ and $B$ must satisfy the equation~\cite{LL_QE_AA}
\begin{equation}\label{LL_AA_0}
B=\frac{c}{2n\hbar v_F^2e}\sqrt{t_0^2+\frac{e^2V^2}{4}}.
\end{equation}
The obtained dependence $B(eV=U)$ is shown in Fig.~\ref{LL_AAfig}~(d).

The effect of the strong spatial inhomogeneity of the magnetic field on the Landau levels in the AA-stacked bilayer graphene was studied by D.~Wang and G.~Jin in Ref.~\cite{LL_NonUnF_AA}.

The magneto-optical absorption spectra of AA-stacked bilayer graphene was analyzed theoretically by Y.-H.~Ho et~al. in Ref.~\cite{Opt_LLAA}. They showed that these spectra exhibit two kinds of absorption peaks resulting from two groups of Landau levels classified from the selection rule based on electronic band symmetry. Only intragroup excitations that satisfy the selection rule take place. The magneto-optical absorption spectra of AA-stacked bilayer graphene are different from the spectra of AB-stacked bilayer graphene. This result could offer a way to distinguish AA-stacked bilayer graphene from AB stacking and monolayer graphene. However, it should be taken into account that the results for the absorption spectra of AA-stacked bilayer graphene were obtained in Ref.~\cite{Opt_LLAA} disregarding the possible existence of the gap in the electronic spectrum.

\subsubsection{Conductivity in the magnetic field and integer quantum Hall effect}\label{aa:iqhe}

\begin{figure}[t]
\includegraphics[width=1\columnwidth]{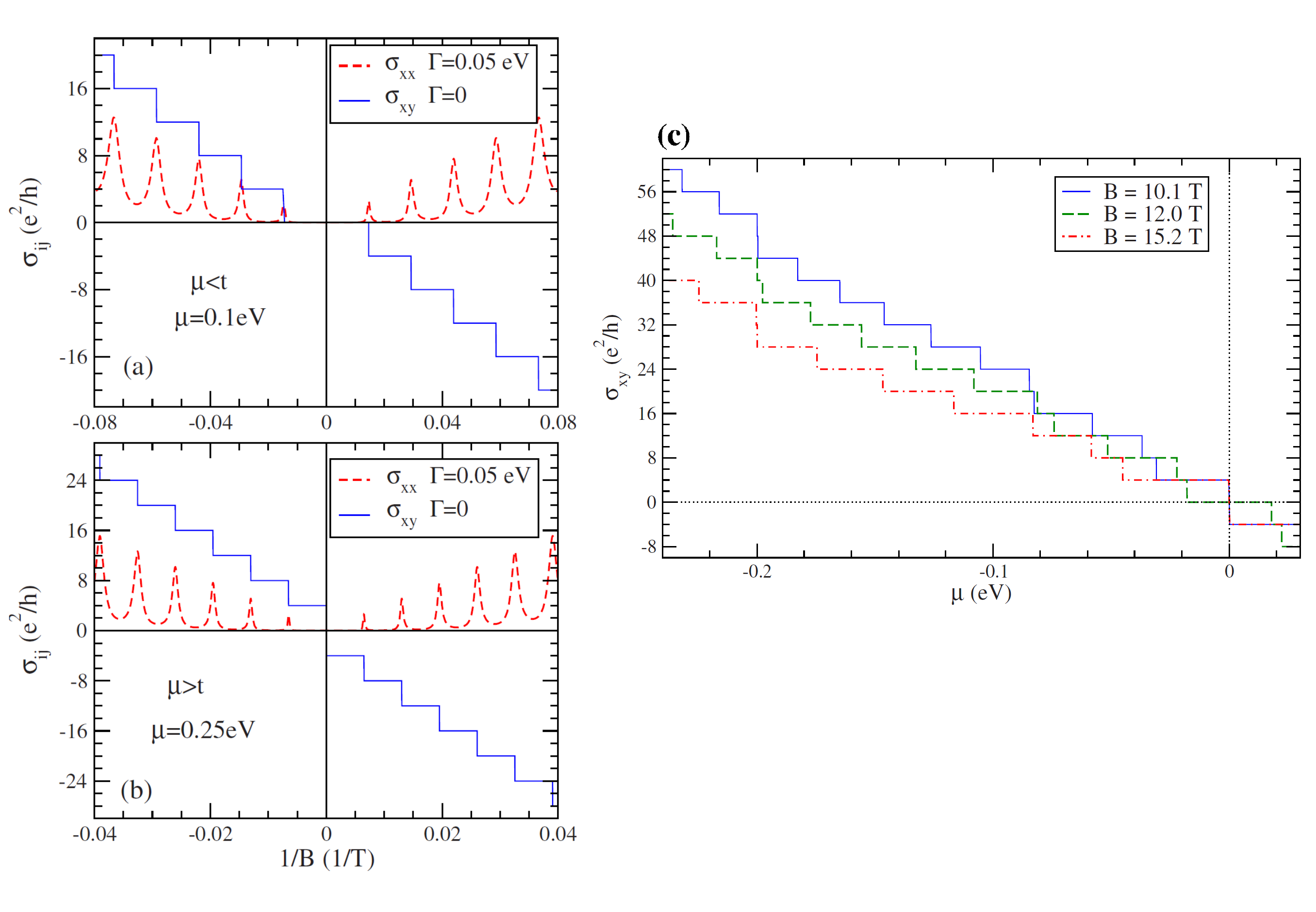}
\centering
\caption{(Color online) Quantized transverse conductivity $\sigma_{xy}$ for
the AA-stacked bilayer graphene (from Ref.~\cite{QE_AA}): (a) as a function
of $1/B$ for $\mu<t_0$, (b) as a function of $1/B$ for $\mu>t_0$, and (c)
as a function of $\mu$ for different $B$. The longitudinal conductivity
$\sigma_{xx}$ as a function of $1/B$ is shown by the dashed (red) curve in
figures (a) and (b) when taking into account the Landau level broadening
$\Gamma$. The data were calculated for $T=0$, $t_0=0.2$\,eV, and
$v_F=10^6$\,m/s.
Reprinted figure with permission from
Ya-Fen Hsu and Guang-Yu Guo,
Phys. Rev. B {\bf 82}, 165404 (2010).
Copyright 2010 by the American Physical Society.
\label{QEHall_AA}}
\end{figure}

The conductivity of AA-stacked bilayer graphene in a magnetic field can be easily derived in the clean limit using the Kubo formalism described in Section~\ref{DynCond}. The corresponding calculations were done in Refs.~\cite{QE_AA,LL_QE_AA} in the case of zero temperature (that is, $T\ll \hbar\omega_c/k_B$). The resultant expression for the Hall conductivity is

\begin{equation}\label{CondHall_AA}
\sigma_{xy}=
-\frac{e^2}{\pi \hbar}\textrm{sgn}(eB)\textrm{sgn}(\mu)
	\left\{\theta(g_+)\theta(g_-)+\left[\frac{g_+^2}{2}\right]
+\left[
			\theta(g_++\sqrt{2})
			+\theta(g_--\sqrt{2})
			-1\right]
		\left[\frac{g_-^2}{2}\right]\right\},
\end{equation}
where ${e^2}/{\pi \hbar}$ is the conductance quantum, $[x]$ denotes the
integer part of $x$, $\theta(x)$ is the Heaviside step-function, and the functions $g_{\pm}$ are defined as
\begin{equation}\label{gpmAA}
g_{\pm}=\frac{l_B}{\hbar v_F}\left[\mu\pm\sqrt{t_0^2+\frac{e^2V_0^2}{4}}\right].
\end{equation}
Thus, the Hall conductance in AA-stacked bilayer graphene is quantized according to $\pm({e^2}/{\pi \hbar})n$ ($n=0,1,2,...$), and the positions of the steps in the Hall conductivity depend on both the chemical potential $\mu$ and the magnetic field $B$. A step occurs when a certain Landau level crosses the chemical potential. The dependence of the quantized Hall conductivity versus magnetic field and chemical potential (that is, doping) is presented in Fig.~\ref{QEHall_AA}. Naturally, the height of the steps in the Hall conductivity corresponds to the four-fold degeneracy of the Landau levels in AA-bilayer graphene.

The magnetic field also affects the longitudinal conductivity $\sigma_{xx}$ of AA-stacked bilayer graphene. The calculation of $\sigma_{xx}$ was performed by Y.-F.~Hsu and G.-Y.~Guo in Ref.~\cite{QE_AA} using the same Kubo approach. At zero temperature, this conductivity is non-zero only if one of the Landau levels coincides with the chemical potential $\mu$. In particular, this means that the application of an infinitesimally small magnetic field can give rise to a metal-insulator transition in the system. To overcome this unphysical result, one should take into account the Landau level broadening due to the presence of disorder~\cite{QE_AA}. This disorder can be characterized by a parameter $\Gamma$, which is defined~\cite{Mahan} as a correction to the Matsubara frequency
\begin{equation}\label{Matsubara}
i\tilde{\omega}_n = i\omega_n+\mu/\hbar+i\,\textrm{sgn}(\omega_n)\Gamma/\hbar.
\end{equation}
The dependence of $\sigma_{xx}$ on $1/B$, calculated in Ref.~\cite{QE_AA}, is shown in Fig.~\ref{QEHall_AA} (a) and (b) by the dashed (red) curves.

\subsection{Landau levels and integer quantum Hall effect in AB-stacked bilayer graphene}

\subsubsection{Landau levels}

\label{LandauLAB}

The study of Landau quantization in AB-stacked bilayer graphene began soon after the fabrication of graphene samples (see, e.g., Refs.~\cite{mccann_falko2006,McCann2006,Pereira2007,Lai}), since it was then recognized that the bilayer system could display even more intriguing properties than the single-layer one. As we already saw in the previous section, the bias voltage affects significantly the electronic properties of the AA bilayer graphene placed in a magnetic field. In the case of the AB bilayer, this effect is even more pronounced because the small bias voltage opens a gap in the electronic spectrum of this system. Below we consider an AB-stacked bilayer graphene in a transverse dc magnetic field $B$ with bias voltage $V$. As in the case of the AA bilayer, we consider only low-energy states. This is a common approximation used in the original papers cited in this subsection.

The exact solution for the band structure problem in magnetic fields for
AB-bilayer graphene cannot has not been found in the general case, even in
the tight-binding approach. The problem can be solved
numerically~\cite{Partoens,Wirtz}, by semiclassical
methods~\cite{Arovas,Ozerin}, or by using simplified
Hamiltonians~\cite{mccann_falko2006,McCann2006,Pereira2007,Falkovsky,andrei,
Chuang,Zhang2008,Aoki}.

Here, following Refs.~\cite{Pereira2007,Falkovsky},  we will use a simplified analysis disregarding trigonal warping (that is, neglecting the $t_3$ hopping amplitude) and electron-hole asymmetry (that is, neglecting $t_4$). In this approach, we rewrite the Hamiltonian of the AB bilayer Eq.~\eqref{ab::H_U} in the form~\cite{Pereira2007,Falkovsky}
\begin{equation}\label{AB_V_B}
H^{AB}= - \left(
         \begin{array}{cccc}
           eV/2 & 0 & \hat{\Pi} & 0 \\
           0 & -eV/2 & t_0 & \hat{\Pi} \\
           \hat{\Pi}^* & t_0 & eV/2 & 0 \\
           0 & \hat{\Pi}^* & 0 & -eV/2 \\
         \end{array}
       \right),
\end{equation}
where in the gauge chosen above the operator $\hat{\Pi}$ is defined as
\begin{equation}\label{Pihat}
\hat{\Pi}=-v_F\left\{i\hbar\left(\partial/\partial x+\partial/\partial y\right)+eBy/c\right\}.
\end{equation}
As in the case of the AA bilayer, we seek a solution of the Schr\"odinger equation corresponding to this Hamiltonian in the form $\Psi(x,y)=e^{ikx}\Psi(y)$. The appropriate eigenvalue problem can be solved explicitly~\cite{Pereira2007,Falkovsky}. In the notations of the present review, the energies $E^n_{LL}$ of the Landau levels are determined by the equation
\begin{equation}\label{AB_LL_eq}
\left[\left(\epsilon_n+u_0\right)^2-n)\right]\left[\left(\epsilon_n-u_0\right)^2-(n-1)\right]
=\frac{\left(\epsilon_n^2-u_0^2\right)t_0^2}{\hbar^2\omega_c^2},
\end{equation}
where
\begin{equation}\label{nnn}
\epsilon_n=E_{LL} (n) /\hbar\omega_c \quad \textrm{and} \quad u_0=eV/2\hbar\omega_c.
\end{equation}

The eigenfunctions for $n\geq2$ can be expressed as ~\cite{Pereira2007}
\begin{equation}\label{LL_EigF_AB}
\Psi_{kn}(y)=d_n\left[Q_{n-1}(\zeta),
	\frac{\sqrt{n-1}}{\epsilon_n-u_0}Q_{n-2}(\zeta),
	f_nQ_{n-1}(\zeta),
	-f_n\frac{\sqrt{n}}{\epsilon_n+u_0}Q_{n}(\zeta)
\right]^T,
\end{equation}
where the functions $Q_n$ and variable $\zeta$ are defined in the text after Eq.~\eqref{LLEigF_AA},
\begin{equation}\label{dnfn}
f_n=\frac{\hbar\omega_c}{t_0}
\frac{\left(\epsilon_n-u_0\right)^2-(n-1)}{\epsilon_n-u_0},
\quad
d_n
=
\left\{f_n^2\left[1+\frac{n}{\left(\epsilon_n+u_0\right)^2}\right]
	+1+\frac{n-1}{\left(\epsilon_n-u_0\right)^2}
\right\}^{-1/2}.
\end{equation}
These results for the Landau levels become invalid for large $n$, when $E_{LL}(n)\gtrsim t_0$, and the Hamiltonian in Eq.~\eqref{AB_V_B} is not applicable. As it follows from Eqs.~\eqref{AB_LL_eq} and~\eqref{LL_EigF_AB}, for each integer $n$ there exist four states with unequal energies. Within such an approach, the trigonal warping can be taken into account using perturbation theory with respect to the small parameter $(t_3/t)^2$~\cite{Falkovsky,Falkovsky1}.

In the case of an unbiased bilayer, $u_0=0$,
Eq.~\eqref{AB_LL_eq}
simplifies considerably, and the eigenenergies can be expressed as
\begin{equation}\label{AB_LL_eq_U0}
\epsilon_n^{(1,2,3,4)}=\pm\left\{\frac{(2n-1)+
(t_0/\hbar\omega_c)^2}{2}\pm\sqrt{\frac{\left[(2n-1)+
(t_0/\hbar\omega_c)^2\right]^2}{4}-n(n-1)}\right\}^{1/2}\!.
\end{equation}
If $n=0$ or $n=1$, we have
\begin{equation}\label{AB_LL_eq_U00}
\epsilon_{0,1}^{(1,2)}=0,\qquad \epsilon_{0,1}^{(3,4)}=\pm\sqrt{\frac{t_0^2}{(\hbar\omega_c)^2}+1}\,.
\end{equation}

The Landau levels have four-fold degeneracy due to spin and valley degrees of freedom. In the unbiased case ($V=0$), zero level is eight-fold degenerate, which corresponds to experimental observations (see below). However, if we take into account the trigonal warping, the total degeneracy of the zero-energy Landau level is 16, due to four Dirac minicones near the zero energy level~\cite{mccann_falko2006}. As argued by S.~Berciaud et~al. in
Ref.~\cite{Berciaud}, the electron-electron coupling destroys the ideal picture of the trigonal warping, e.g., reducing the number of minicones near each ${\bf K}$ point from four to two, restoring eight-fold degeneracy.

In the biased AB bilayer, the valley degeneracy is lifted~\cite{mccann_falko2006}. In particular, the zero-energy level splits in two levels independent on the magnetic field $\epsilon_{0,1}^{(1)}=-\xi u_0$, where $\xi=\pm 1$ is the valley index.

\begin{figure}[t]
\includegraphics[width=0.5\columnwidth]{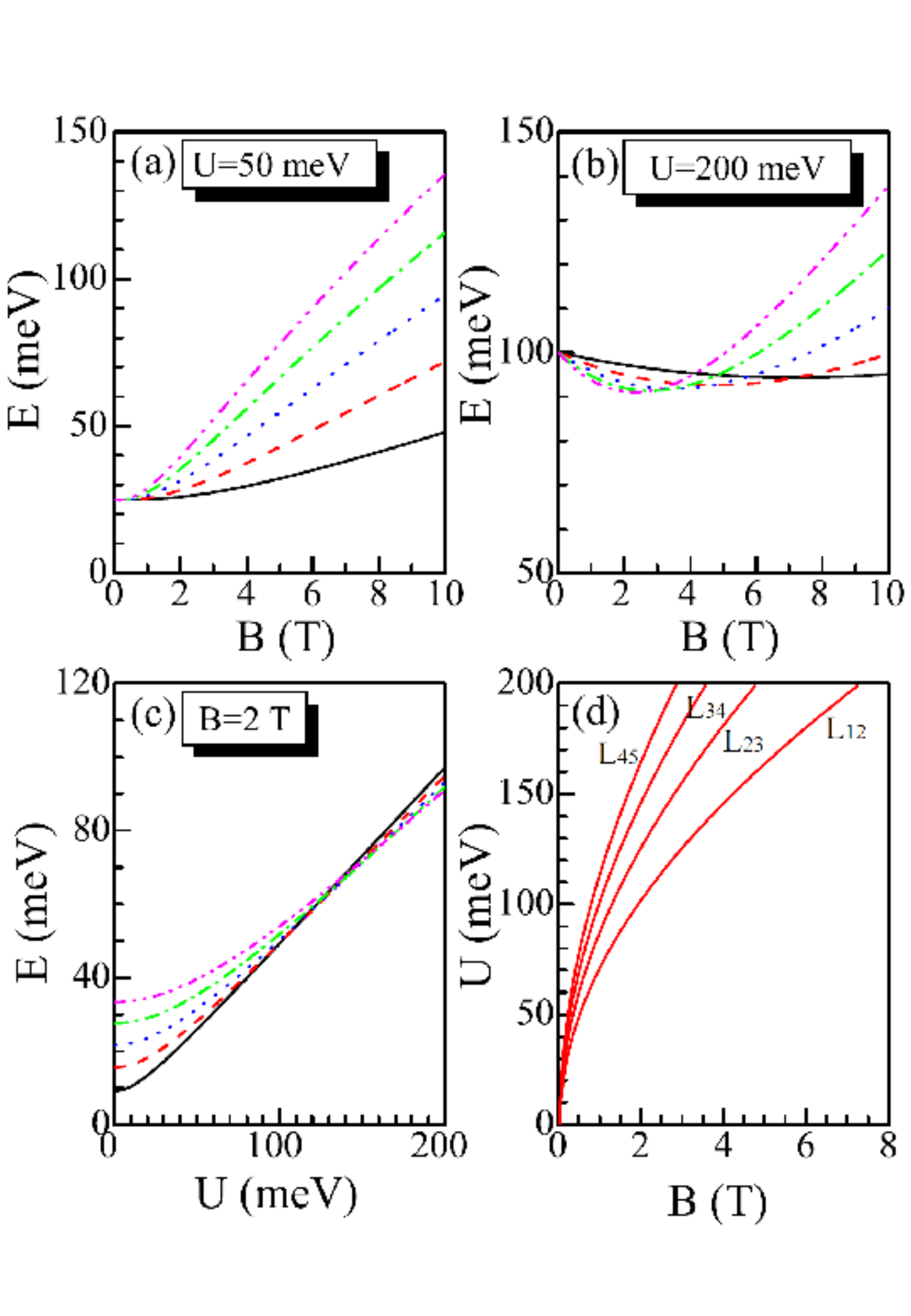}
\centering
  \caption{(Color online) First five electron-like Landau levels
($n=1,2,3,4,5$) in biased ($U=eV$) AB graphene bilayers from
Ref.~\cite{Wang_Jin}. Panels (a, b): Landau levels as function of magnetic
field. The level corresponding to $n=1$ is shown by the solid black curve,
$n=2$ -- (red) dashed curve, $n =3$ -- (blue) dotted curve, $n = 4$ --
(green) dot-dashed curve, and $n =5$ -- (purple) double-dot-dashed curve.
Two different bias potentials are used: (a) $U =50$~meV, and (b) $U
=200$~meV. (c) First five electron-like Landau levels in the graphene
bilayer as a function of the interlayer potential difference $U=eV$ at $B =
2$~T. (d) Dependence of the crossings of the adjacent Landau levels on the
combination of applied magnetic field and interlayer potential difference.
The symbol $L_{n_1n_2}$ denotes the crossing of the Landau levels with
indices $n_1$ and $n_2$.
Figure from EPL {\bf 92}, 57008 (2010), 
by Dali Wang and Guojun Jin.
Reprinted with permission
from IOP.
\label{LL_ABfig}
}
\end{figure}

The dependence of the Landau levels eigenenergies on the combination of the magnetic field and the interlayer voltage bias is illustrated in Fig.~\ref{LL_ABfig}. This dependence is different from that of the AA bilayer, Fig.~\ref{LL_AAfig}. This discrepancy is due to the different electron energy spectra of these two systems. In particular, the applied interlayer voltage opens a gap in the spectrum of the AB graphene bilayer, in contrast to the AA system.

The Landau quantization in AB bilayer graphene was observed and
experimentally studied in many experiments, not only devoted to the
analysis of the Landau levels
themselves~\cite{Mayorov2011,Berciaud,Henriksen,Henriksen2010, Dillen,Kratz}, but also in numerous papers studying the quantum Hall effect, which will be discussed in the next subsection. The experimental results are in a satisfactory agreement with the theoretical analysis presented above. The discrepancies between the theory and experiment can be attributed to experimental errors, sample quality and, which is more important, to the approximations made when deriving Eq.~\eqref{AB_LL_eq}, especially, neglecting the $t_3$ and $t_4$ hopping amplitudes responsible for the trigonal warping and electron-hole symmetry breaking, disregarding electron-electron interaction, and continuous medium approach. For example, E.\,A.~Henriksen and J.\,P.~ Eisenstein in Ref.~\cite{Henriksen} measured the Landau levels in AB bilayer using infrared spectroscopy in magnetic fields up to $18$\,T. They observed four distinct intraband transitions in both the conduction and valence bands. The transition energies are roughly linear in $B$ between the lowest Landau levels, whereas they follow $\sqrt{B}$ for the higher transitions. This behavior represents a change from a parabolic to a linear energy dispersion. These observations, as well as the density of states derived from the data, generally agrees with the existing lowest order tight-binding calculation for AB bilayer graphene discussed above. However, in comparing data to theory, a single set of fitting parameters fails to describe the experimental results quantitatively.

\begin{figure}
\includegraphics[width=1.0\columnwidth]{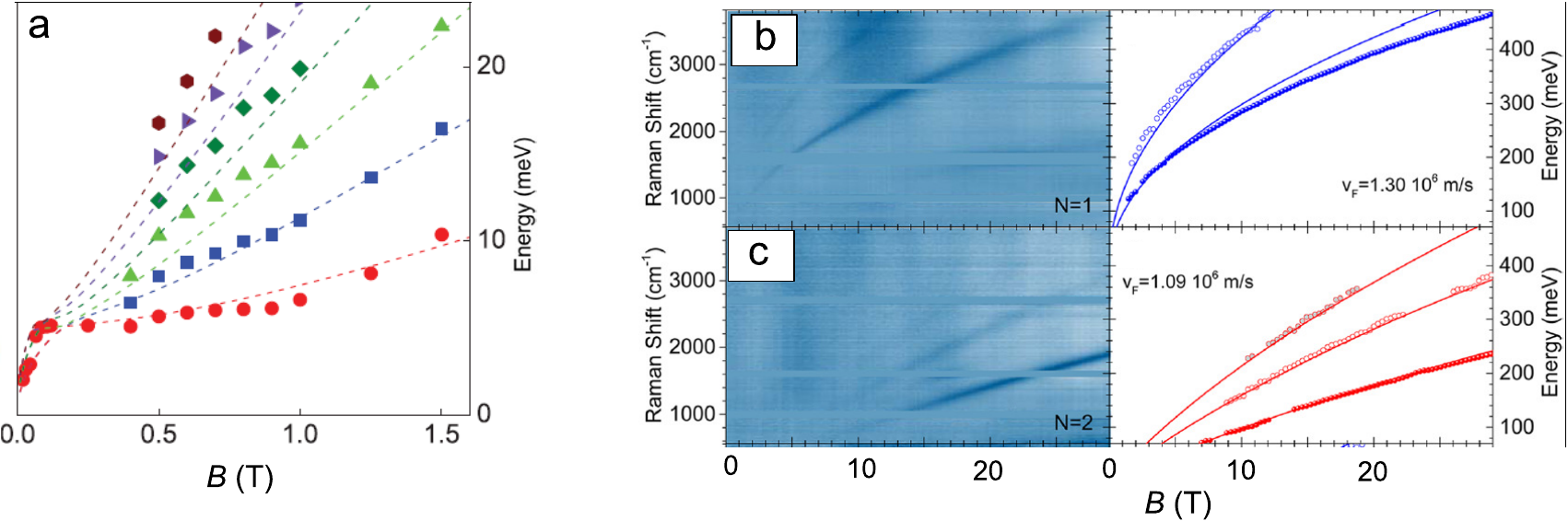}
\centering
\caption{(Color online) (a) Position of the Landau levels in an AB bilayer
as a function of the magnetic field extracted from the transport
measurements~\cite{Mayorov2011}.
Symbols represent experimental data points, and curves correspond to
theoretical calculations.
From 
A.S. Mayorov et al., Science, {\bf 333}, 860 (2011).
Reprinted with permission from AAAS.
Panels (b) and (c): The Landau levels in a single layer (b) and AB bilayer
(c) graphene from
Ref.~\cite{Berciaud}.
Left panels show false-color maps of the micro magneto-Raman scattering
spectra as a function of the magnetic field. The corresponding peak
frequencies of the electronic Raman features extracted from Lorentzian fits
are shown in the right panels. The theoretical results are shown by solid
lines. The interlayer hopping was
$t_0 = 0.4$\,eV,
while the Fermi velocity
$v_{\rm F}$
that best fits the results is indicated in each panel.
Reprinted with permission from 
S.~Berciaud et al.,
Nano Letters, {\bf 14}, 4548 (2014).
Copyright 2014 American Chemical Society.
\label{LLAB_expfig}
}
\end{figure}

In Ref.~\cite{Mayorov2011}, A.\,S.~Mayorov et~al. extracted the position of
the Landau levels from experiments on electronic transport in free-standing
AB bilayer graphene in magnetic fields, see Fig.~\ref{LLAB_expfig}(a). The
calculated values are shown in this figure by dashed lines. In contrast to
results by E.\,A.~Henriksen and J.\,P.~ Eisenstein~\cite{Henriksen}, the
range of parameters that fit the theoretical and experimental results was
very narrow, which indicates that the proposed tight-binding Hamiltonian is
appropriate for the analysis of the electronic properties of the AB
bilayer. From Fig.~\ref{LLAB_expfig} (a) it is seen that at energies below
$5$\,meV, the almost parabolic electronic dispersion at higher energies is
replaced with a linear spectrum at lower energies. Such a behavior could be
attributed to the effect of the trigonal warping, which becomes significant
at low energies. However, the authors of Ref.~\cite{Mayorov2011} show that
a more detailed analysis of the data reveals a significant deviation from
the ideal picture of trigonal warping, which could be attributed, e.g., to
the effects of electron-electron coupling. In Ref.~\cite{Berciaud} by
S.~Berciaud et~al., the electronic excitations between Landau levels in
free-standing N-layer graphene were measured using micro magneto-Raman
scattering spectroscopy. A characteristic evolution of electronic bands in
up to five Bernal-stacked graphene layers was observed. Their behavior is
described remarkably well by the simple tight-binding theoretical approach.
The results for single-layer and AB bilayer samples are shown in
Figs.~\ref{LLAB_expfig} (b) and (c). The calculations done with the help of
Eq.~\eqref{AB_LL_eq} are shown by solid curves. As it is seen, the
theoretical and experimental results are in a good agreement. Note also
that the measurements of the quantum Hall conductivity confirmed that the
zero Landau level is eight-fold degenerate, while other levels are
four-fold degenerate (see
subsection~\ref{subsect::iqhe}).

Thus, we can conclude, the commonly-used simple tight-biding approach presented above is valid for the description of the main electronic properties of the AB bilayer graphene.

Several theoretical papers generalized the models discussed above. The Landau level spectrum for AB bilayer graphene in a tilted magnetic field was studied by Y.-H.~Hyun et~al.~in Ref.~\cite{Hyun}. In the low-energy approximation of the tight-binding model, the authors found analytically the Landau level spectrum in terms of spheroidal functions and their respective eigenvalues. In the limit of large in-plane field, this spectrum becomes two-fold degenerate, which is a consequence of the Dirac point splitting, induced by the in-plane field. The effect of disorder was analyzed in Refs.~\cite{Castro2007a,Kawarabayashi}. It was shown that the Landau spectrum, especially the zero-energy level, is rather stable against disorder.

\subsubsection{Conductivity in a magnetic field and magneto-optic phenomena}

\begin{figure}[t]
\includegraphics[width=0.5\columnwidth]{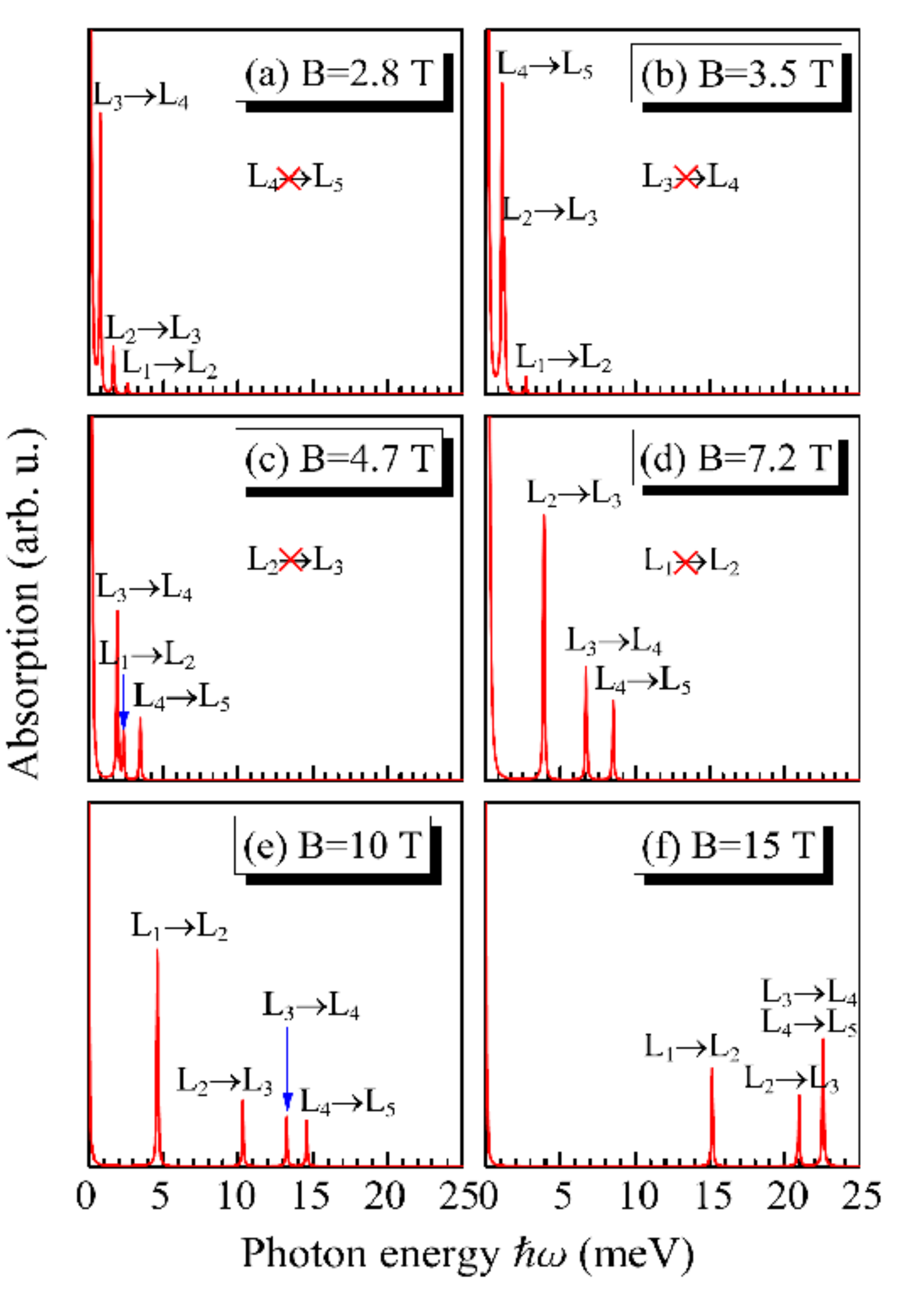}
\centering
\caption{(Color online) Calculated intraband magneto-optical absorption for
AB bilayer graphene as a function of photon energy at a fixed bias
potential (from Ref.~\cite{Wang_Jin} by D.~Wang and G.~Jin). The absorption
peaks correspond to transitions between adjacent Landau levels. The curves
are plotted for transverse bias voltage $eV = 200$\,meV, and the several
different values of the magnetic field $B$; the notation
$L_i\rightarrow L_j$
denotes transitions between levels $i$ and $j$.
Figure from EPL {\bf 92}, 57008 (2010), 
by Dali Wang and Guojun Jin.
Reprinted with permission
from IOP.
\label{ABMO_absfig}}
\end{figure}

The electronic structure can be probed by spectroscopic measurements in magnetic field. Naturally, this technique was applied to study AB graphene bilayer. In particular, in Refs.~\cite{Mayorov2011,Henriksen,Kratz}, the Landau levels energies were extracted from the magneto-optic transport measurements. On the theory side, the magneto-optical spectrum of the biased and unbiased AB graphene was calculated in the tight-binding approximation in several papers~\cite{Pereira2007,Falkovsky,Wang_Jin,Koshino,ChangLin}. Generalizing the approach used in Section~\ref{DynCond} [see Eq.~\eqref{condAA}] to the case of finite dc magnetic field, it is possible to compute the dynamical conductivity and absorption. The effect of disorder is included with the phenomenological constant $\Gamma$ as it was done in Section~\ref{DynCond}. In a magnetic field, an optical excitation by normally-incident light is only allowed between the Landau levels with index values $n$ and $n \pm 1$, since the matrix element of the velocity operator $\hat{v}_x$ in Eq.~\eqref{condAA} vanishes otherwise~\cite{Koshino}. Figure~\ref{ABMO_absfig} shows an example of such calculations of the magneto-optical absorption spectrum performed by D.~Wang and G.~Jin in Ref.~\cite{Wang_Jin}.

The Hall conductivity $\sigma_{xy}$ can be calculated the same way as $\sigma_{xx}$, Eq.~\eqref{condAA} with $(\alpha,\beta)=(x,y)$. The optical Hall conductivity is directly related to the Faraday rotation angle $\Theta_F$ of the incident light, $\Theta_F=2\pi\textrm{Re}(\sigma_{xy})/c$. The Faraday rotation in AB bilayer graphene was computed in Refs.~\cite{Falkovsky,Aoki}.

\subsubsection{Integer quantum Hall effect}\label{subsect::iqhe}

The integer quantum Hall effect in the AB bilayer has some specific peculiarities at low carrier density $N$ due to the eight-fold degeneracy of the zero-energy Landau level at $V=0$. It has been studied theoretically in several papers~\cite{mccann_falko2006,McCann2006,Falkovsky,Falkovsky1,Novoselov2006,Nakamura2008}. As it was mentioned above, the calculation of the Hall conductivity is a straightforward procedure based on the Kubo approach, Eq.~\eqref{condAA}. However, even in the tight-binding approximation and neglecting the trigonal warping the result can be obtained only numerically.

\begin{figure}[t]
\includegraphics[width=0.55\columnwidth]{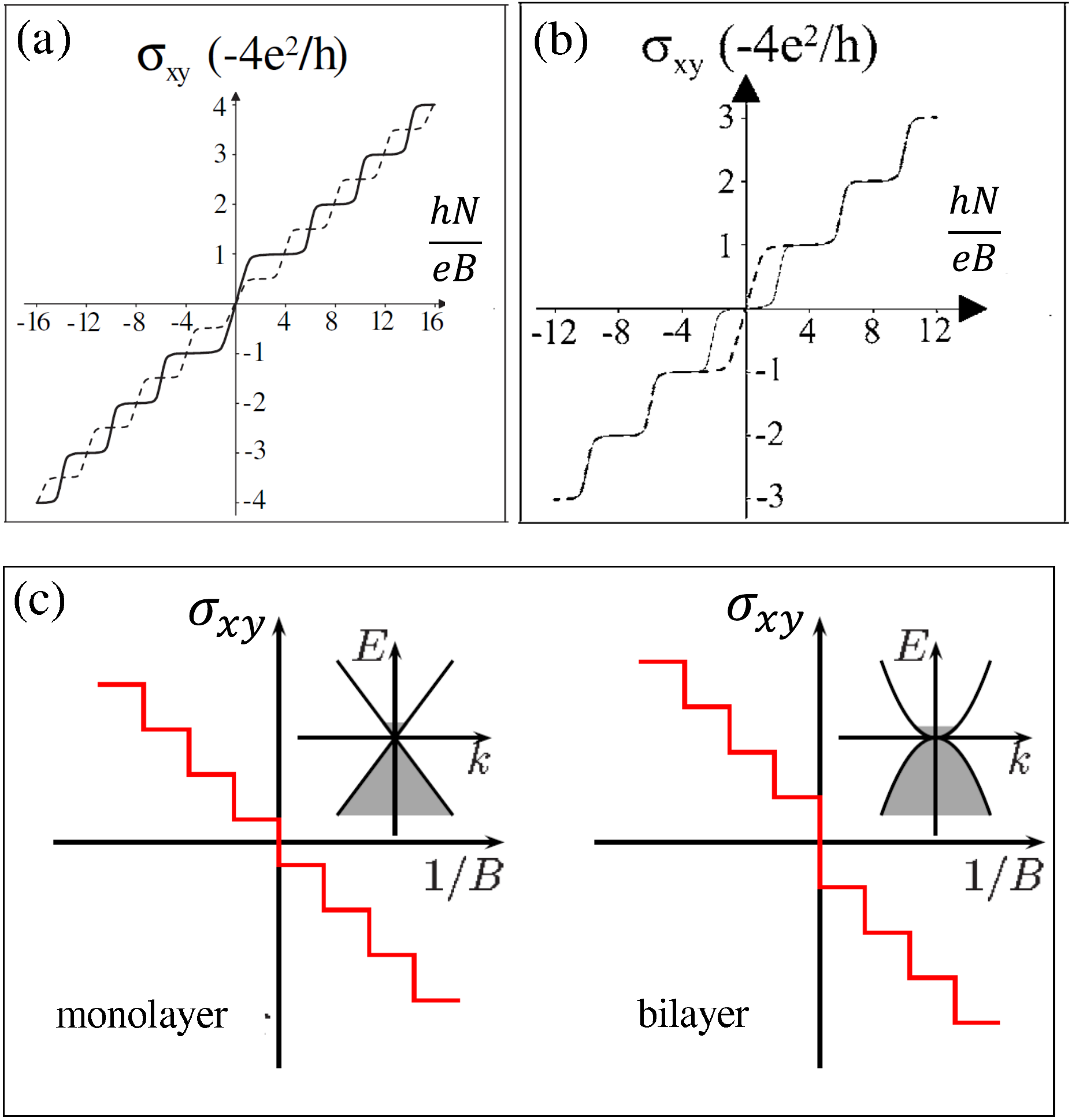}
\centering
\caption{(Color online) Integer quantum Hall effect in AB bilayer graphene.
(a) The calculated Hall conductivity $\sigma_{xy}$ as a function of carrier
density $N$ for unbiased AB bilayer (solid line) is compared to that of a
monolayer (dashed line), from Ref.~\cite{mccann_falko2006}. 
(b) The calculated Hall conductivity as a function of carrier density for
unbiased (dashed curve) is compared to that for large $V$ (solid curve),
from Ref.~\cite{McCann2006}.
(c) Dependence of $\sigma_{xy}$ on $1/B$ for monolayer (left
picture) and unbiased AB bilayer (right picture) graphene, from
Ref.~\cite{Nakamura2008}. The insets show the dispersion relations of both
systems. The twice larger step at $N=0$ and $1/B=0$ in the unbiased AB
bilayer is due to the additional two-fold degeneracy of the Landau level at
zero energy.  Disorder is phenomenologically taken into
account in (a) and (b), and disregarded in (c).
Credits.
Panel~(a):
Reprinted figure with permission from
E.~McCann, V.I.~Fal'ko, 
Phys. Rev. Lett., {\bf 96}, 086805 (2006).
Copyright 2006 by the American Physical Society.
\url{http://dx.doi.org/10.1103/PhysRevLett.96.086805}
Panel~(b):
Reprinted figure with permission from
E.~McCann,
Phys. Rev. B, {\bf 74}, 161403 (2006).
Copyright 2006 by the American Physical Society.
\url{http://dx.doi.org/10.1103/PhysRevB.74.161403}
Panel~(c):
Reprinted figure with permission from
M.~Nakamura et al.,
Phys. Rev. B, {\bf 78}, 033403 (2008).
Copyright 2008 by the American Physical Society.
\url{http://dx.doi.org/10.1103/PhysRevB.78.033403}
\label{AB_QHE_Theorfig}
}
\end{figure}

\begin{figure}[t]
\includegraphics[width=0.8\columnwidth]{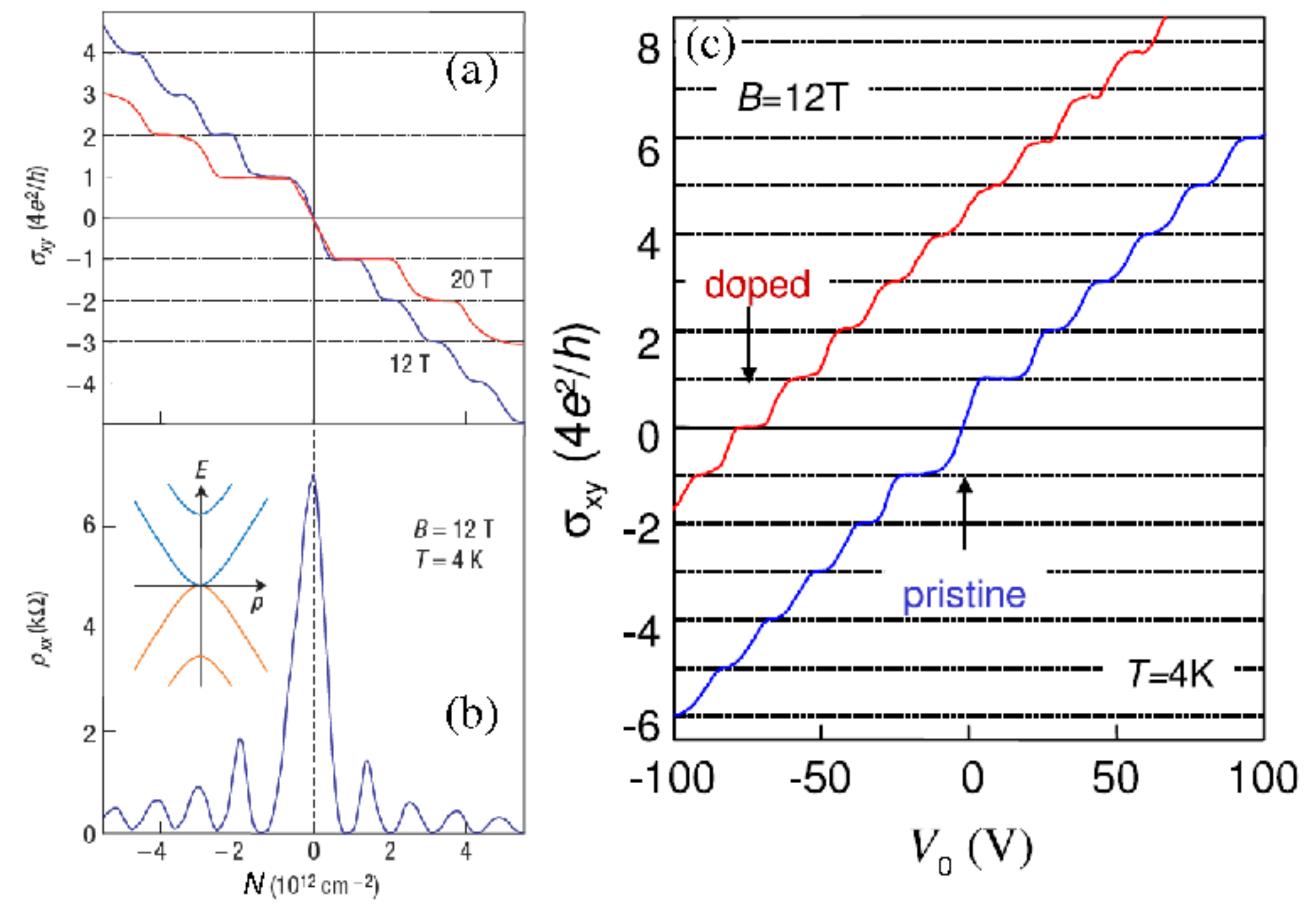}
\centering
\caption{(Color online) Experimental observation of the integer
quantum Hall effect in AB bilayer graphene.
The Hall
conductivity $\sigma_{xy}$~(a) and $\rho_{xx}$~(b) are plotted as functions
of the electron density $N$ at a fixed $B$ and temperature $T=4$~K.
Positive and negative $N$ correspond to field-induced electrons and holes,
respectively. The Hall conductivity crosses zero without any sign of the
zero-level plateau that would be expected for a conventional 2D system. The
inset shows the energy spectrum for AB bilayer graphene.
Reprinted by permission from Macmillan Publishers Ltd: Nature Physics,
{\bf 2}, 177 (2006), copyright 2006.
\url{http://www.nature.com/nphys/index.html}
Panel~(c): measured Hall conductivity as function of the applied voltage
$V_0$
of pristine (undoped) and chemically doped AB bilayer graphene
($N_0\approx5.4\times 10^{12}$~cm$^2$).
Reprinted figure with permission from
E.V.~Castro et al.,
Phys. Rev. Lett., 99, 216802 (2007).
Copyright 2007 by the American Physical Society.
\url{http://dx.doi.org/10.1103/PhysRevB.78.033403}
\label{AB_QHE_Expfig}
}
\end{figure}

Theoretical predictions regarding the integer quantum Hall effect in the AB
bilayer graphene (E.~McCann and V.\,I.~Fal'ko~\cite{mccann_falko2006},
E.~McCann~\cite{McCann2006}, M.~Nakamura et~al.~\cite{Nakamura2008}) are
illustrated in Fig.~\ref{AB_QHE_Theorfig}. Experimental results
(E.\,V.~Castro et~al.~\cite{Castro2007}, K.\,S.~Novoselov
et~al.~\cite{Novoselov2006}) are shown in Fig.~\ref{AB_QHE_Expfig}. In
Figs.~\ref{AB_QHE_Theorfig} (a), (b), and Fig.~\ref{AB_QHE_Expfig} (a) the
Hall conductivity is shown as function of the carrier density $N$, while in
Fig.~\ref{AB_QHE_Theorfig} (c) it is shown as function of $1/B$. The
measured dependencies of $\sigma_{xy}$ on the applied voltage $V_0$ are
shown in Fig.~\ref{AB_QHE_Expfig} (c) for undoped and chemically doped AB
bilayer samples. Note that physically the applied voltage changes the
carrier density, and the dependencies of $\sigma_{xy}$ on both $N$ and
$V_0$ are in some sense similar. The measured dependence of the
longitudinal resistivity $\rho_{xx}$ on $N$ is shown in
Fig.~\ref{AB_QHE_Expfig}~(b). As it follows from the theory,
Eq.~\eqref{condAA}, the steps in the quantum Hall conductivity corresponds
to peaks in the longitudinal resistance (see also Fig.~\ref{QEHall_AA}).
For comparison, the dependencies $\sigma_{xy}$ on $N$ and $1/B$ for
single-layer graphene are shown schematically in
Figs.~\ref{AB_QHE_Theorfig} (a) and (c).

The data shown in Figs.~\ref{AB_QHE_Theorfig} and~\ref{AB_QHE_Expfig} demonstrate that, on both theoretical and experimental pictures, the plateaus in the Hall conductivity $\sigma_{xy}$ occur at integer multiples of $4e^2/h$, and each plateau corresponds to the quantized change in the charge density value of $4B/\Phi_0$. This behavior is due to the four-fold degeneracy of the AB bilayer arising from the spin and valley degrees of freedom. Deviation from the conventional integer quantum Hall effect occurs near $\sigma_{xy}=0$. In the unbiased AB bilayer, there is a step in $\sigma_{xy}$ of height $8e^2/h$ due to additional degeneracy of the zero-energy Landau level. This big step is accompanied by a plateau separation of $4B/\Phi_0$ in density, arising from the eight-fold degeneracy of the zero-energy Landau levels. These features are clearly seen on the experimental curves, see Fig.~\ref{AB_QHE_Expfig}~(a) and blue curve (undoped sample) in Fig.~\ref{AB_QHE_Expfig}~(c). In many papers studying Landau levels and quantum Hall effect, the value of the charge density $N$ is expressed as a dimensionless filling factor $\nu=N\Phi_0/B$. In terms of $\nu$, the steps in  the Hall conductivity in the AB bilayer are observed when $\nu$ changes by the value $\pm 4$, except $\nu=0$, when the step is 8.

Thus, the manifestation of the integer quantum Hall effect in AB bilayers differs significantly from that in AA bilayers and single layer graphene. First, there is no zero-energy Landau level in AA bilayer and single layer graphene, therefore, there is no step across zero density [the dashed line in Fig.~\ref{AB_QHE_Theorfig}~(a)]. All Landau levels in AA bilayer and single-layer graphene have four-folded degeneracy, therefore, there is no ``double'' step in the Hall conductivity $\sigma_{xy}$.

The application of the transverse voltage splits the valley degeneracy as shown theoretically by E.~McCann and V.\,I.~Fal'ko in Ref.~\cite{mccann_falko2006}. In particular, as it was mentioned in Section~\ref{LandauLAB}, zero-energy level splits as $E_{0,1}=\pm eV/2$. The splittings of the non-zero levels are weaker. When the energy scale $eV$ is large enough, then the splitting $eV$ of the zero energy levels from each valley results in a sequence of quantum Hall plateaus at all integer values of $4e^2/h$, including a plateau at zero density, Fig.~\ref{AB_QHE_Theorfig} (a) (dotted line), as observed experimentally, Fig.~\ref{AB_QHE_Expfig} (c), (red curve, for a doped sample).

Recall that in this section we have assumed that the degeneracy of the Landau levels is preserved, i.e., any splitting of the levels is negligible as compared to the temperature and level broadening in experiments.

\subsubsection{Fractional quantum Hall effect
}\label{FrHallEff}
%

Fractional quantum Hall effect is observed in two-dimensional systems with electron correlations~\cite{MacDonald}. In such systems the quasiparticles can have fractional charge. As a result, the steps in the Hall conductivity could arise at fractional filling factors. Usually such steps can be distinctly extracted from measurements only at very low temperatures. The fractional quantum Hall effect has been observed in monolayer graphene~\cite{Du,Bolotin,Dean}. The observation of this effect in a high-quality AB bilayer graphene samples was reported in Refs.~\cite{Jing2010,Bao2010}. More detailed experimental studies of the factional quantum Hall effect in the AB bilayer were performed~\cite{Feldman2,Ki}.

\begin{figure}[t]
\includegraphics[width=1\columnwidth]{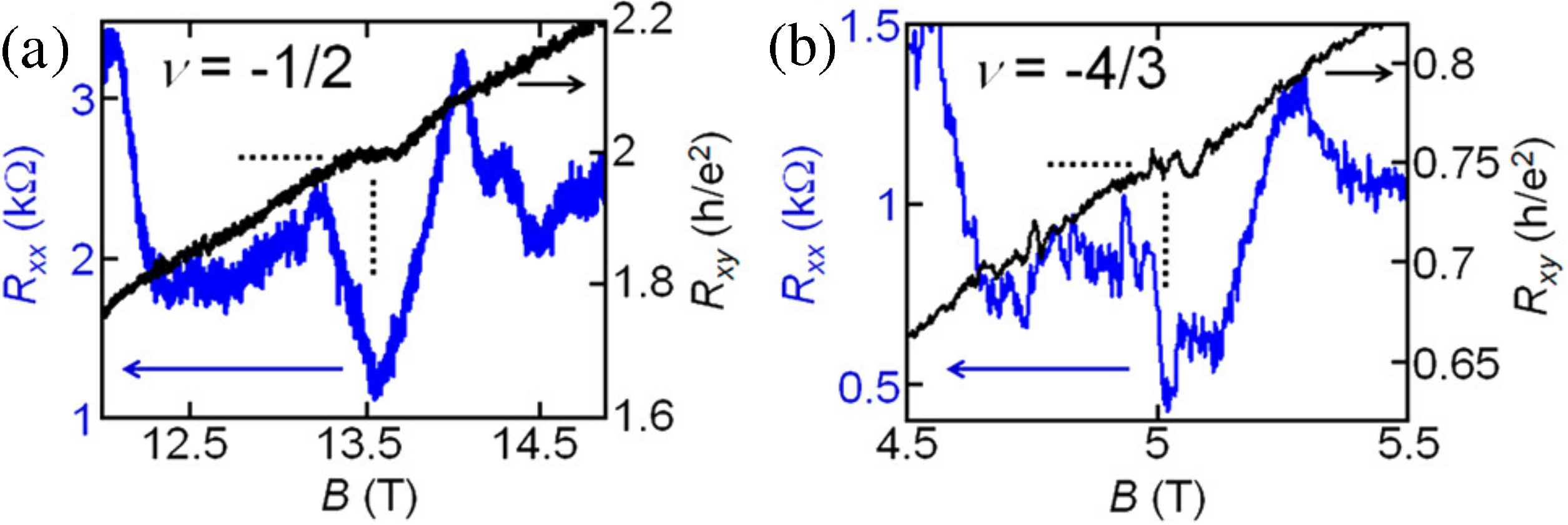}
\centering
\caption{(Color online)  Observation of fractional quantum Hall effect at
$T = 0.25$~K.
Magneto-resistance
$R_{xx}$ (blue curve) and $R_{xy}$ (black curve) at $V_0 = -27$~V. Panels
(a) and (b)
illustrate the features associated with the $\nu=-1/2$ and $\nu=-4/3$
fractional quantum Hall states, respectively.
Reprinted with permission from
D.-K. Ki et al.,
Nano Letters, {\bf 14}, 2135 (2014).
Copyright 2014 American Chemical Society.
\label{FrQHE_ABfig}}
\end{figure}

In Ref.~\cite{Bao2010}, W.~Bao et~al. performed magneto-transport measurements of a suspended bilayer graphene device with charge carrier mobilities up to 2.7$\times 10^5$\,cm$^2/$Vs in magnetic field up to $31$\,T. A small conductance plateau was observed at $\nu=1/3$. This plateau scales appropriately with $N$ and $B$ and disappears at temperature above $T\approx$ 2 -- 2.5\,K. In Ref.~\cite{Feldman2}, A.~Kou et~al. presented local electronic compressibility measurements of the fractional quantum Hall effect in the lowest Landau level of a bilayer graphene device fabricated on hexagonal boron nitride (h-BN). Peaks were observed in the compressibility, corresponding to the fractional quantum Hall effect states at filling factors $\nu = 2p + 2/3$, with hints of additional states appearing at $\nu = 2p + 3/5$, where $p = -2\,,-1,\,0,$, and $1$. According to Ref.~\cite{Feldman2}, this sequence breaks particle-hole symmetry and obeys a $\nu \rightarrow \nu + 2$ symmetry, which highlights the importance of the orbital degeneracy for many-body states in bilayer graphene. Robust fractional quantum Hall states at $\nu = -1/2$ and $-4/3$ in suspended AB samples were revealed in low-temperature magneto-transport measurements~\cite{Ki}. The experimental dependence of the longitudinal, $R_{xx}$, and Hall, $R_{xy}$, resistances on the magnetic field from Ref.~\cite{Ki} are shown in Fig.~\ref{FrQHE_ABfig}; panels (a) and (b) illustrate the fractional quantum Hall effect with $\nu=-1/2$ and $\nu=-4/3$, respectively. Note that the observation of the quantum Hall effect requires high-quality AB bilayer graphene samples, low temperatures (below 2 -- 2.5\,K), and strong magnetic fields (about 5 -- 10\,T and higher).

The nature of fractional quantum Hall states is determined by the interplay between the Coulomb interaction and the symmetries of the system. The distinct combination of spin, valley, and orbital degeneracies in AB bilayer graphene is predicted theoretically to produce an unusual and tunable sequence of fractional quantum Hall states in the lowest Landau level~\cite{Shibata,Abanin1,Apalkov,Abanin2,Barlas,Apalkov1,Snizhko}.

Strongly correlated states at fractional filling factors and the prospect of tuning their properties have been the focus of recent theoretical attention~\cite{Shibata,Apalkov,Apalkov1,Barlas,Snizhko,maher2014tunable,Geraedts}. Z.~Papi{\'{c}} et~al.~\cite{Abanin2} proposed a method to continuously tune the effective electron interactions in graphene and its bilayer by the dielectric environment of the sample. Using this method, the charge gaps of prominent fractional quantum Hall states, including $\nu = 1/3$ or $\nu = 5/2$ states, can be increased several times, or reduced to zero. The tunability of the interactions can be used to realize and stabilize various strongly-correlated phases and explore the transitions between them. The fact that different sample preparations result in different fractional quantum Hall states could be a sign of the theoretically predicted tunability of the fractional quantum Hall effect in bilayer graphene~\cite{Apalkov,Barlas,Snizhko}. Applying a perpendicular electric and/or a parallel magnetic field to the sample may provide insight into the conditions under which different fractional quantum Hall states are favored.

The effects of the electron-electron interaction on the properties of AB bilayer graphene in a strong magnetic field were the focus of a many theoretical studies. In particular, D.\,S.\,L.~Abergel and T.~Chakraborty predicted that the interactions can cause the mixing of Landau levels in moderate magnetic fields~\cite{Abergeland}. Broken-symmetry states induced by electron-electron coupling and phase diagram of the lowest Landau level in the Bernal or AB bilayer graphene in a magnetic field were investigated in Refs.~\cite{Kharitonov,haritonov_afm2012,Gorbar,Gorbar1,Toke2011,Barlas}.

\section{Spin-orbit coupling}
%

Spin-orbit coupling is a relativistic effect, which can be derived from Dirac's model of the electron. Spin-orbit coupling describes a process in which an electron changes simultaneously its spin and angular momentum or, in general, moves from one orbital wave function to another. The spin-orbital mixing is large in heavy ions, where the average velocity of the electrons is higher.

Single-layer graphene has been proposed to be a two-dimensional quantum spin Hall system when the intrinsic spin-orbit interaction is taken into account~\cite{Haldane,KaneMele1,KaneMele2}. Spin-orbit coupling (SOC) in graphene opens a gap at the Dirac points and the system becomes a topological insulator. However, carbon is a light atom, and the spin-orbit interaction is expected to be weak. Tight-binding and band-structure calculations provide estimates for the intrinsic and extrinsic spin-orbit interactions in the range 0.01 -- 0.2~K, and hence much smaller than the other energy scales in the problem~\cite{CastrNrev}. The artificial enhancement of the spin-orbit coupling strength in graphene can open up new possibilities in graphene-based spintronics. Several theoretical and experimental works have addressed the effects of enhanced spin-orbit coupling in graphene by doping with heavy adatoms such as indium or thallium~\cite{Weeks}, doping with $3d/5d$ transition metal atoms~\cite{Lazo1,Lazo2,Lazo3}, and interfacing with metal substrates, e.g., Ni(111).~\cite{Varykhalov1,Varykhalov2,Varykhalov3,Varykhalov4}

In bilayer graphene the effect of spin-orbit coupling was investigated in
several theoretical
works~\cite{Guinea2010,Gelderen,Prada,Qiao,Mireles,Konschuh,Qiao1,Zhai,dyrdal2014spin}.
The Hamiltonian of the problem is usually written in the form
\begin{equation}\label{HSO}
\hat{H}=\hat{H}_0+\hat{H}_R+\hat{H}_M.
\end{equation}
Here $\hat{H}_0$ is the Hamiltonian of bilayer graphene, either AB of AA, which includes the terms describing both in-plane and inter-plane hopping, the gate voltage contribution, and disregards all interactions (see Section~\ref{spectra}). The term $\hat{H}_R$ describes Rashba spin-orbit interaction. If one takes into account only nearest neighbor interaction this term can be expressed as~\cite{Qiao,Qiao1}
\begin{equation}\label{RSOHam}
\hat{H}_R=it_R
\sum_{\langle \mathbf{ij}\rangle a \alpha \beta}
	\mathbf{z}\cdot
	(\bm{\sigma}_{\alpha\beta}
	\times
	\mathbf{d}_{\mathbf{ij}})
	a^\dag_{\mathbf{i}a\alpha}b_{\mathbf{j}a\beta}^{\vphantom{\dagger}}
+\text{h.c.}\,,
\end{equation}
where $t_R$ is the Rashba spin-orbit coupling strength, $\mathbf{z}$ is the unit vector perpendicular to the layers, $\mathbf{d}_{\mathbf{ij}}$ is a vector pointing from sites $\mathbf{j}$ to $\mathbf{i}$, $a=1,2$ is the layer index, and $\bm{\sigma}_{\alpha\beta}$ are spin Pauli matrices with $\alpha$ and $\beta$ denoting up spin or down spin. The term $\hat{H}_M$ is the exchange field contribution with magnetization $M$
\begin{equation}\label{MSOHam}
\hat{H}_M = M\sum_{ \mathbf{i} a \alpha \beta}
\left(
	a^\dag_{\mathbf{i}a\alpha}
	{\sigma}^z_{\alpha\beta}
	a_{\mathbf{i}a\beta}^{\vphantom{\dagger}}
	+
	b^\dag_{\mathbf{i}a\alpha}
	{\sigma}^z_{\alpha\beta}
	b_{\mathbf{i}a\beta}^{\vphantom{\dagger}}
\right).
\end{equation}
Such an exchange field effect can arise due to the proximity coupling between graphene and magnetic adatoms or ferromagnetic substrate materials (see Ref.~\cite{Qiao1}). This term is of importance for several reasons, which will be discussed below.

According to Refs.~\cite{Prada,Qiao1}, the effect of spin-orbit coupling in AA-stacked bilayer graphene is insignificant, unless $t_R$ becomes of the order or larger than interlayer hopping amplitude $t_0$, which seems unrealistic. Thus, the main research efforts for this problem were devoted to the AB bilayer graphene.

The electronic band structure of the AB bilayer in the presence of spin-orbit coupling and a transverse electric field was calculated from ``first principles'' using the linearized augmented plane-wave method by S.~Konschuh et~al. in Ref.~\cite{Konschuh}. The main conclusion of the latter paper is that the spin-orbit effects in AB-stacked bilayer graphene differ essentially from the single-layer spin-orbit coupling. The intrinsic spin-orbit splitting (anticrossing) around the ${\bf K}$ points is about $24$\,$\mu$eV for the low-energy valence and conduction bands, which are closest to the Fermi level, similarly as in single-layer graphene. An applied transverse electric field breaks space-inversion symmetry, and leads to a Rashba spin-orbit splitting. This splitting is usually linearly proportional to the electric field. The peculiarity of AB-stacked bilayer graphene is that the low-energy bands remain split by $24$\,$\mu$eV, independently of the applied external field. The electric field, instead, opens a semiconducting band gap separating these low-energy bands. In the remaining two high-energy bands the spin degeneracy is lifted. The strength of this splitting is proportional to the electric field; the proportionality coefficient is given by the second intrinsic spin-orbit coupling, whose value is 20~$\mu$eV. All the band-structure effects and their spin splittings can be explained by a tight-binding model, in which the spin-orbit Hamiltonian is derived from symmetry considerations. The magnitudes of intra- and interlayer couplings (their values are similar to the single-layer graphene ones) are determined by fitting to first-principles results.

A systematic study on the influence of Rashba spin-orbit coupling, interlayer potential difference, and exchange field on the topological properties of bilayer graphene was performed by Z.~Qiao et~al. in Refs.~\cite{Qiao,Qiao1}. Due to the band gap opening from broken out-of-plane inversion symmetry, gated AB bilayer graphene is a quantum valley Hall insulator characterized by a quantized-valley Chern number. The presence of Rashba spin-orbit coupling turns the gated bilayer graphene system from a quantum valley Hall insulator into a topological insulator. The phase boundary is given by~\cite{Qiao}
\begin{equation}\label{trt0V}
9t_R^2=t_0^2+\left(eV/2\right)^2.
\end{equation}
In the presence of different Rashba spin-orbit coupling strengths on the top and bottom layers, $t_R^{(1)}\neq t_R^{(2)}$, the topological insulator phase remains robust as long as~\cite{Qiao1}
\begin{equation}\label{trt0}
9t_R^{(1)}t_R^{(2)}>t_0.
\end{equation}
When the time-reversal symmetry is broken by an exchange field $M$, Eq.~\eqref{MSOHam}, the AB-stacked bilayer graphene  hosts different topological phases characterized by different Chern numbers. The phase boundaries associated with the topological phase transitions are given by $eV=\pm M$ and 
\begin{equation}\label{tRt0VM}
9t_R^2=t_0^2+\left(eV/2\right)^2-M^2.
\end{equation}

The effects of next-nearest-neighbors spin-orbit coupling interaction and stability limits of the topological phases against trigonal warping and staggered sublattice potential were investigated by E.~Prada et~al. in Ref.~\cite{Prada}.

A theoretical study of the band structure and Landau levels in AB bilayer
graphene at low energies in the presence of a transverse magnetic field and
Rashba spin-orbit interaction was done by F.~Mireles and G.~Schliemann in
Ref.~\cite{Mireles}. These authors predicted an unexpected asymmetric spin
splitting and crossings of the Landau levels due to the interplay between
the Rashba interaction and the external bias voltage. The edge modes in
AB-stacked bilayer graphene arising due to spin-orbit interactions were
analyzed in Refs.~\cite{Guinea2010,Prada}.

Let us conclude this section with a reminder that all spin-orbit effects discussed above have not been observed experimentally. Thus, realistic values of the gaps, associated with the spin-orbit interaction, are not known.

\section{Bilayer-based mesoscopic systems} \label{meso}
%

The list of bilayer-based mesoscopic objects studied by theorists is quite similar to the list of single-layer graphene mesoscopic systems (for a summary of single-layer mesoscopic research, see, e.g., the review by A.~Rozhkov et~al.~\cite{meso_review}). The items on such a list are: quantum dots, nanoribbons (narrow stripes of graphene), pn-junctions, edge states, and conducting channels. However, since the electronic spectra of the bilayers are different from the single layer, the electronic properties of the bilayer mesoscopic objects may be quite dissimilar from their single-layer counterparts.

Most of this section is dedicated to mesoscopic structures made of AB bilayer graphene. The number of studies of AA-based system is fairly small and it is reviewed in subsection~\ref{meso::subsec::AA_nanoribbon} below. In addition, Klein tunneling for AA bilayer graphene was discussed in subsection~\ref{AAChiral}. Klein tunneling in twisted bilayer graphene is discussed in subsection~\ref{subsect::twisted::klein}. Finally, there is a small number of theoretical~\cite{PankratovFlakes,MorellNanoribons2014} and experimental~\cite{nanoribbon_exper_joule} studies where mesoscopic systems based on twisted bilayer graphene have been studied.

\subsection{AB bilayer graphene quantum dots}\label{meso::ab::qd}

Quantum dots are simplest mesoscopic objects. In the theoretical literature, both single-layer and AB bilayer graphene quantum dots are discussed. Studies of quantum dots made of AA bilayer are virtually absent.

A simple quantum dot can be a small piece of material. As a first example of a quantum dot let, us consider a nanoribbon of finite length. Sahu et~al.~\cite{Sahu2010} investigated a finite-length AB bilayer nanoribbon using DFT. According to this paper, the system ground state is magnetic: it demonstrates interlayer antiferromagnetic arrangement of magnetic moments. The confinement-induced gap is enhanced by the internal magnetism. The application of an external electric field weakens this gap. A related study of an infinite length nanoribbon, Ref.~\cite{Sahu2008}, will be reviewed below.

For single-layer and AB bilayer graphene, tearing a macroscopic sample into small pieces is not the only method to define a quantum dot. Alternatively, a dot may be created within a larger sample with the help of a position-dependent external potential, either chemical or electric. For AB bilayer samples, appropriate ideas were explored theoretically in Refs.~\cite{Matulis2008,Pereira2007a}. If the symmetry between the layers is broken (e.g., by a transverse electric field, see subsection~\ref{subs::bernal_bias_gap}), a gap opens up in the bulk of the bilayer. In such an insulating environment, one can induce a well-defined quantum dot. This can be done by introducing dopants to a small area of the sample, as discussed theoretically by J.~Pereira et~al. in Ref.~\cite{Pereira2007a}. If the donor or acceptor levels lie inside the gap, the wave functions associated with these states are localized near the region where the dopants are placed. As a result, several discrete subgap states emerge (in connection with single-layer graphene similar proposals were considered by G.~Giavaras and F.~Nori in Ref.~\cite{giavaras_nori2011}).

Experimentally, this type of dot was realized by M.\,T.~Allen et~al.~\cite{Allen2012} with the help of a complicated gating system in suspended AB bilayer graphene samples. The tunable bulk gap in this experiment was up to $250$\,meV. Inside this insulating bulk an electric-field-defined quantum dot was created. Dots with diameters ranging from $150$ to $450$\,nm were reported. Measurements performed on a particular dot at $0.1$\,K demonstrated well-defined Coulomb blockade with charging energy $E_{\rm C} \approx 0.4$\,meV. The same principle was used by A.M.~Goossens et al.~\cite{bilayer_dot_exper_bor_nitr_gate2012}, to define a quantum dot within a bilayer sample on a hexagonal boron nitride substrate. A related experiment was described in Ref.~\cite{gate_def_dot_exp}.

If the electric potential varies in space, but it is identical for both
layers, the gap does not emerge. In such a situation it is impossible to
create true bound states: the electron wave function would ``leak" away
from the dot into a gapless environment. In this situation, a set of
quasi-bound states (resonances) can be induced. This can occur both in
single-layer
graphene~\cite{Matulis2008,martino_qdot2007,giavaras2009,
elect_magn_qdot_single2012},
and in gapless bilayer graphene~\cite{Matulis2008}. However, the
application of the external magnetic field may confine these states.

This method of creating quantum dots is very appealing, since it allows one, at least in principle, to control the properties of the dots by varying external fields. Using an appropriately-designed gating system it is possible to engineer, for example, quantum dots of non-trivial shapes. Specifically, a ring-shaped quantum dot was studied theoretically by M.~Zarenia et~al. in Ref.~\cite{zarenia_blg_ring2009}. To generate such a dot we need a gating system creating a bias potential, which vanishes within a finite-width ring: $U(r) = eV(r) =0$ when $R_{\rm min} < r < R_{\rm max}$, where $r$ is a polar radius, $R_{\rm min}$ and $R_{\rm max}$ are the internal and external radii of the ring. The required potential is non-zero and has the same polarity for both $r < R_{\rm min}$, and $r>R_{\rm max}$. As a result, several discrete subgap states may localize at the ring. The behavior of these states in an external magnetic field and other properties are discussed in~\cite{zarenia_blg_ring2009}.

A ring with a different confining mechanism was proposed by L.\,J.\,P.~Xavier et~al. in Ref.~\cite{Xavier2010}. This ``topological" confinement is based on the idea of I.~Martin et~al.~\cite{Martin2008} (it will be discussed in more detail in subsection~\ref{meso::ab::topological}). According to Ref~\cite{Martin2008}, when the transverse electric potential applied to an AB bilayer sample changes sign as a function of coordinate [for example, $U(x>0)>0$, but $U(x<0)<0$], the electric potential kink at $x=0$ becomes the source of topological states. The wave functions of such subgap states are extended along the $y$-axis and localized in the $x$-direction. If the gate system creates the transverse potential $U(r)$, which changes sign at some finite $r=R>0$, then the topological states are localized at the ring of radius $R$. Due to the finite circumference of this ring, instead of a one-dimensional band, a set of quantized levels appears.

\subsection{Bilayer graphene nanoribbons}

\subsubsection{AA bilayer nanoribbons}\label{meso::subsec::AA_nanoribbon}

Nanoribbons (narrow stripes of the material) prepared from bilayer graphene are another popular object of research. Similar to single-layer graphene nanoribbons, AA bilayer nanoribbons are also classified according to their edge type: there are zigzag and armchair nanoribbons.

The Landauer-B\"uttiker transport through AA bilayer nanoribbon was studied
in
Refs.~\cite{sun_transport_aa_ab2008,xu_transport_aa2012}.
In
Ref.~\cite{xu_transport_aa2012},
N.~Xu et~al. studied a transport through the AA nanoribbons with both
zigzag and armchair edges. The effects of interlayer electron hopping and
ripples were discussed. For the armchair nanoribbon the current decreases
at fixed voltage when the interlayer hopping grows. This effect is
virtually absent for the zigzag nanoribbons. Ripples act to suppress the
current, especially for zigzag systems.
S.-J.~Sun and
C.~Chang~\cite{sun_transport_aa_ab2008}
compared AA-stacked and AB-stacked nanoribbons. The influence of both bias
voltage and perpendicular magnetic field were studied. The bias voltage
affects the AA and AB nanoribbons differently: the bias electric field in
the interval
$E<4$\,V/nm
suppress the current through the AB nanoribbon, but does not significantly
change the current through the AA nanoribbon. Current oscillations, which
are predicted to occur at higher $E$, are likely to be difficult to observe
due to electrical breakdown. As for the magnetic field, it appears that the
magnetoresistance is extremely weak
($<10^{-5}$
at $10$\,T) in both cases.

Transport between two overlapping single-layer nanoribbons was studied by
J.\,W.~Gonzalez et~al.~in
Ref.~\cite{Gonzalez2010}
within the framework of the Landauer-B\"uttiker formalism. Both AA and AB
types of stacking in the overlap region were discussed.
Reference~\cite{Gonzalez2010}
found that for both stackings the Landauer conductance oscillates as a
function of the overlap region width. The period of oscillations is about
$20$ unit cells. Since the conductance is very sensitive to the geometry of
the system, the authors proposed to use such a setup as an
electromechanical switch. Experimentally, electromechanical properties of
both AB bilayer and monolayer graphene were studied in
Ref.~\cite{Benameur2015}.

In Ref.~\cite{Habib2011}, K.\,M.\,M.~Habib et~al. investigated the same system of two overlapping single-layer nanoribbons in a non-equilibrium regime. Similar to~\cite{Gonzalez2010},  Ref.~\cite{Habib2011} assumed that the overlap area is either of AA, or AB type. Using DFT and non-equilibrium Green's function approach Ref.~\cite{Habib2011} proved that such a system would demonstrate a region of negative differential resistance at sufficiently large applied voltage. Details of the $IV$ curve, however, depend on the type of stacking.

\subsubsection{AB bilayer nanoribbons}\label{meso::ab::nanoribbon}

\begin{figure}[t]
\begin{centering}
\includegraphics[width=8cm]{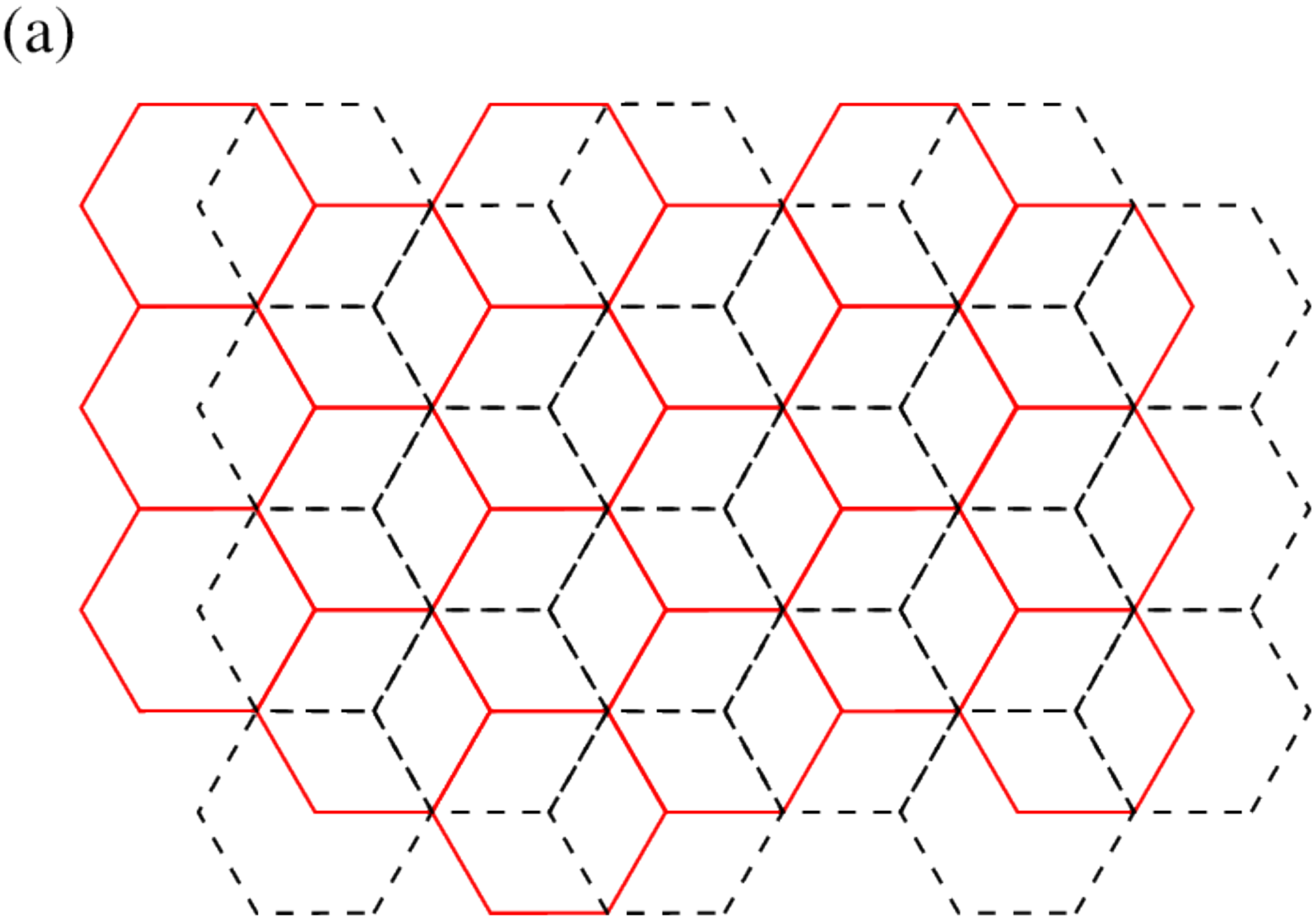}
\includegraphics[width=8cm]{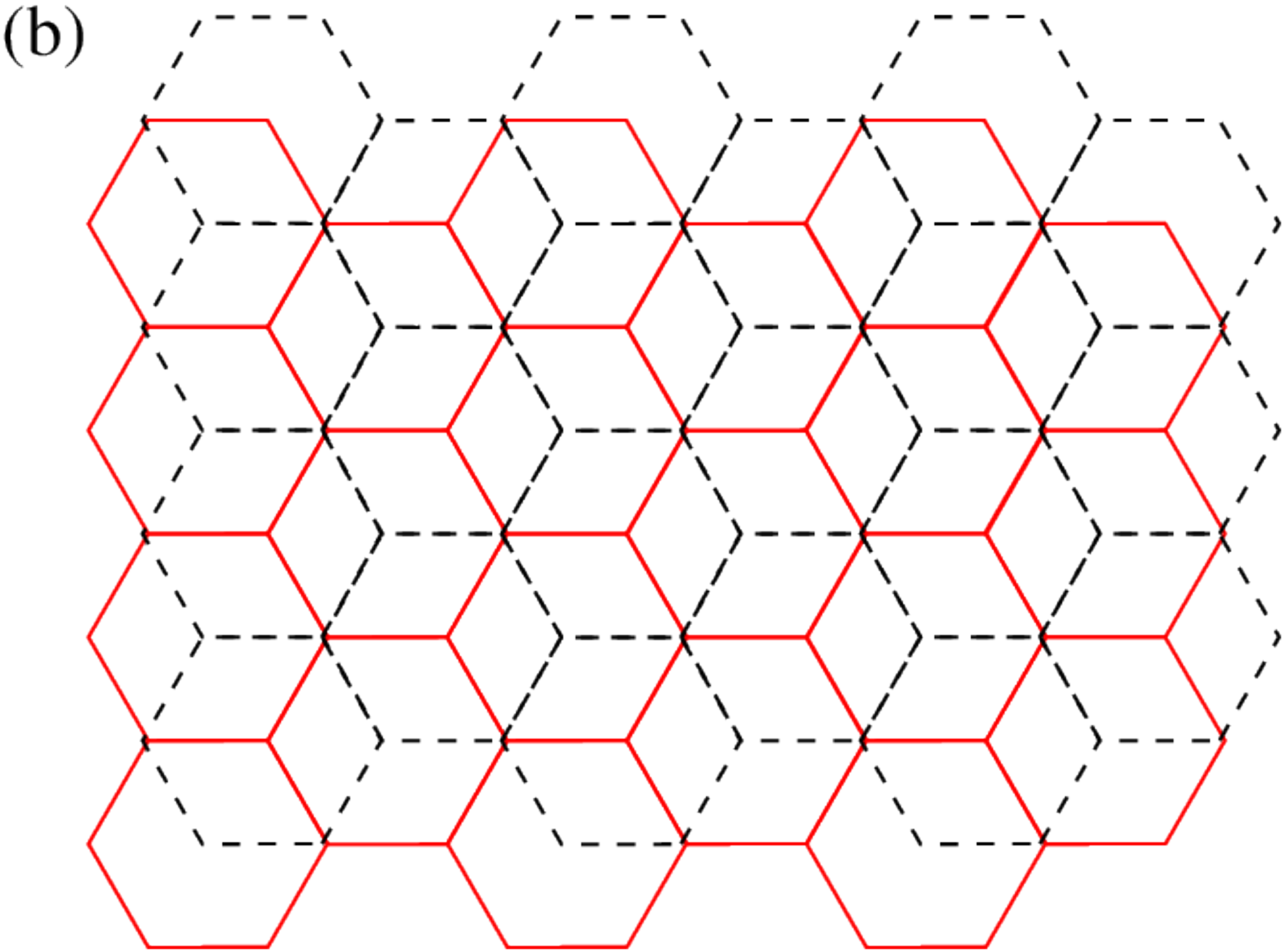}
\par\end{centering}
\caption{\textbf{AB bilayer edges.} Edge classification developed for
single-layer graphene is incomplete for AB-stacked bilayer. Both armchair
edge (horizontal edges in both panels) and zigzag edge (vertical edges) can
appear in any of the two varieties, $\alpha$-alignment [panel (a)] and 
$\beta$-alignment [panel (b)]. In both panels one layer is shown by (red)
solid hexagons, another is represented by (black) dashed hexagons.
\label{meso::bilayer_edge_types}
}
\end{figure}

For AB bilayer nanoribbons, the usual classification of the edge types (armchair versus zigzag edges) is not entirely complete: both types must be further split into $\alpha$-alignment and $\beta$-alignment (see Fig.~\ref{meso::bilayer_edge_types}).

It is known that single-layer graphene allows for a flat band of edge
states near the zigzag termination (localized states near the armchair
termination are also possible under suitable
circumstances~\cite{Pavel_armchair}). Similar localized states were
described theoretically by E.\,V.~Castro
et~al.~\cite{bilayer_zigzag_loc2008} for $\beta$-aligned AB bilayer zigzag
semiplane and AB bilayer zigzag nanoribbons. There are two edge bands per
edge; thus, a nanoribbon has four such bands. The wave functions
corresponding to different edge bands are quite dissimilar. Namely, in the
nearest-neighbor-hopping model of Ref.~\cite{bilayer_zigzag_loc2008} there
are two bands whose wave functions are localized exclusively on a single
layer of the bilayer. The wave functions for the two other bands arespread
over both layers. This disparity explains the different behavior of the
edge states under the action of the transverse bias voltage $V$. When
$V=0$, all four bands are dispersionless, with zero eigenenergy
$\varepsilon_k = 0$. For $V \neq0$, the single-layer localized bands shift
rigidly from $\varepsilon_k = 0$ to $\varepsilon_k = \pm eV/2$. The
localized states whose wave functions are spread over both layers remain in
the bulk gap, and acquire a pronounced dispersion.

This model was generalized by A.~Cortijo et~al.~in Ref.~\cite{bilayer_zigzag_loc_gen2010}. They included longer-range hopping and interaction into their Hamiltonian. The longer-range hopping lifted the degeneracy of the perfectly flat edge bands, introducing the dependence of the eigenenergies on momentum $k$ along the nanoribbon. The combined effect of the longer-range hopping and non-zero transverse bias $V$ creates a disparity between the nanoribbon edges. The authors identify ``fast" and ``slow" edge: the velocity of the band localized at the ``fast edge" of the nanoribbon is higher than that at the ``slow edges". The electron-electron interaction induces single-electron gaps at both edges. According~\cite{bilayer_zigzag_loc_gen2010}, ``the slow edge" must have a higher gap value. Measuring the conductance of an individual edge, an experimentalist could extract each gap value independently.

It is theoretically predicted for single-layer graphene that the edge-state band is unstable with respect to ferromagnetic order when considering the repulsive interaction between electrons (see, e.g., the review Ref.~\cite{meso_review}). Likewise, AB bilayer zigzag nanoribbons were actively researched in connection with their magnetic properties~\cite{Sahu2010,Sahu2008,Zhang_first_principle_2010a,Lima2009}. At the bilayer edge one might expect that both layers would be magnetized. Indeed, this type of magnetic structure has been reported in several theoretical papers where different versions of density functional theory (DFT) calculations were employed. However, one must remember that the relative stability of the magnetic and non-magnetic states turns out to be very sensitive to numerous details: the type of density-functional approximation, edge structure ($\alpha$-alignment versus $\beta$-alignment), relaxation of the atom positions at the edge. For example, Sahu et~al.~\cite{Sahu2008} concluded that for $\alpha$-aligned zigzag graphene nanoribbons, the DFT calculations cannot reliably determine if the stable state is magnetic or not. Also,~\cite{Sahu2008} claimed that $\beta$-aligned zigzag nanoribbons are magnetic: every edge of every layer carries finite magnetization. However, the method was unable to assess if the magnetization vectors at the same edge of the nanoribbon have the same (ferromagnetic) or opposite (antiferromagnetic) orientations.

The results of a similar DFT study were described by M.\,P.~Lima et~al. in Ref.~\cite{Lima2009}. Unlike Ref.~\cite{Sahu2008}, where the graphene strips comprising the nanoribbon were assumed to be perfectly flat, Ref.~\cite{Lima2009} allowed for bending of the nanoribbon layers. Therefore, they found that in the stable configuration the ``thickness" of the nanoribbon is not constant, but rather changes from point to point as one moves from one edge to another. This thickness variation is the result of the interlayer interaction variation: a local value of the interlayer interaction in the proximity of an edge becomes sensitive to the distance from the edge. For an $\alpha$-aligned nanoribbon, this additional interaction can be thought of as an extra attraction between the layers, while for a $\beta$-aligned nanoribbon acts as repulsion. Thus, the $\alpha$-aligned nanoribbon was determined to be more stable than its $\beta$-aligned counterpart. Regarding magnetism, $\alpha$-aligned nanoribbons were found to be non-magnetic. For $\beta$-aligned nanoribbon~\cite{Lima2009} confirmed the conclusion of \cite{Sahu2008} about the stability of a magnetic configuration. However, the energy differences between competing states, possessing dissimilar magnetic structures, were smaller than the temperature. As a result, which of theses states was the ``true" ground state, remained unknown.

The papers~\cite{Sahu2008,Lima2009} discussed the situation of a nanoribbon with no contact with the substrate. Thus, they are most relevant for suspended ribbons. The effect of silicone substrate on nanoribbon electron properties was investigated by Z.~Zhang et~al. in Ref.~\cite{Zhang_first_principle_2010a}. According to this DFT study, the carbon atoms at the edges of the nanoribbon bottom layer form covalent Si-C bonds to the silicone atoms of the substrate. Thus, the top layer, which interacts with the bottom layer only through weak van~der Waals forces, behaves as an almost isolated single-layer zigzag nanoribbon. The latter object is known to have a magnetic ground state. Consequently, a bilayer nanoribbon on a silicone substrate is magnetic: each edge carries a finite ferromagnetic magnetization, while the coupling between the edges is antiferromagnetic. Further, the authors proposed to control the magnetization of the nanoribbon by an external electric field.

Thus, it appears that the zigzag bilayer nanoribbons have several competing low-energy states, some of which are magnetic, some of which are not. Which state wins in a laboratory experiment may depend on a variety of conditions: type of nanoribbon edge, substrate, sample history, etc. Unfortunately, relevant experimental data are not available at the time of writing.

Optical properties of the zigzag and armchair nanoribbons were studied by A.\,R.~Wright et~al. in Ref.~\cite{Wright2009} using the tight-binding approximation. According to those calculations, AB bilayer armchair nanoribbons should demonstrate strong optical response in the far infrared and terahertz regimes. The frequency-dependent dimensionless optical conductivity $\sigma$ was predicted to be as large as $150 (e^2/4 \hbar)$. Such a response strength greatly exceeds the values typical for graphene. The authors claimed that the required nanoribbons were already created in laboratory experiment, Ref.~\cite{nanoribbon_production2008}.

Numerical calculations of the Landauer conductance of AB bilayer nanoribbons were performed by H.~Xu et~al. in Ref.~\cite{landauer_ab_ideal_disord_zozoul2009}. This paper discusses both zigzag and armchair nanoribbons with ideal and disordered edges, with and without a magnetic field. The electron-electron and electron-phonon interactions were disregarded. The authors reported that the dimensionless conductance $\pi \hbar \sigma /e^2$ for the bilayer nanoribbon quantizes as $n$ for the armchair edge and $(2n+1)$ for the zigzag edge. The edge disorder quickly destroys this quantization, and opens a so-called transport gap. This phenomenon is particularly strong for armchair nanoribbons. The transport through the zigzag nanoribbons is more robust, possibly due to edge states. However, that paper investigated a very idealized object. Since the model disregarded the effects of electron-electron and electron-phonon interaction, several mechanisms, which may affect the transport of single-layer ribbon (see, e.g., Refs.~\cite{lattice_distortion, sandler_coulomb_gap, our_nanoribbon_paper_2009}), were neglected. The authors themselves assessed their results as ``a necessary first step" toward developing a reliable theoretical description of nanoribbon transport. A related theoretical study of a bilayer nanoribbon in a spatially inhomogeneous magnetic field was reported in Ref.~\cite{nribb_theor_magn_landauer2015}.

As mentioned in the previous paragraph, Ref.~\cite{landauer_ab_ideal_disord_zozoul2009} reaffirmed the expected conclusion that edge disorder is very harmful for the ballistic propagation of the carriers. However, the fabrication of nanoribbons with smooth edges is a very difficult task, and experimental attempts in this direction are ongoing (see, for example, Ref.~\cite{nanoribbon_exper_joule}).

\subsection{Topologically-protected conducting channels in AB bilayer
samples}\label{meso::ab::topological}

As we already mentioned, the bias voltage applied to the AB bilayer opens a gap in the electron spectrum, turning the sample into a semiconductor. I.~Martin et~al.~\cite{Martin2008} noticed that, if the bias potential $U({\bf r}) = e V({\bf r})$ passes through zero and changes its sign upon crossing some line ${\bf r} = {\bf r}_\ell$, then a set of conducting single-electron subgap states binds to this line.

To demonstrate the existence of these modes, the authors of Ref.~\cite{Martin2008} modified the effective two-band Hamiltonian, Eq.~\eqref{ab::2b_U}. They assumed that the transverse bias potential is position-dependent $V = V({\bf r})$, and satisfies the following relations
\begin{eqnarray}
V({\bf r}) = V(x),
\qquad
V(x) = - V(-x).
\end{eqnarray}
That is, the bias voltage changes sign on the straight line $x=0$. The corresponding Schr\"odinger equation may be solved~\cite{Martin2008} rather straightforwardly, if we choose \begin{eqnarray}\label{meso::U_sgn}
U(x)=U_0 \,{\rm sgn}\,(x),
\text{\ \ where  }
U(x) = e V(x).
\end{eqnarray}
Since the system remains invariant under translations along the $y$-axis, one can replace $-i\partial/\partial y \rightarrow q_y$. Thus, for positive $x$ the following system of ordinary differential equations needs to be investigated:
\begin{eqnarray}\label{ab::top_channel_1a}
\begin{cases}
\frac{\hbar^2 v_{\rm F}^2}{t_0}
		(\xi q_y+\partial_x)^2 \varphi_b(x,q_y)
=
(\frac{U_0}{2} - \varepsilon) \varphi_a (x,q_y),
\\
\label{ab::top_channel_1b}
\frac{\hbar^2v_{\rm F}^2}{t_0}
(\xi q_y-\partial_x)^2 \varphi_a (x, q_y)
=
-(\frac{U_0}{2} + \varepsilon) \varphi_b (x, q_y),
\end{cases}
\end{eqnarray}
where $\xi = \pm 1$ is the valley index, defined in subsection~\ref{Chiral}. The system has four solutions. However, of these four only two are normalizable at $x \rightarrow +\infty$. The divergent solutions must be discarded. Likewise, for $x<0$ the following system of equations must be solved:
\begin{eqnarray}\label{ab::top_channel_2a}
\begin{cases}
\frac{\hbar^2 v_{\rm F}^2}{t_0}
		(\xi q_y+\partial_x)^2 \varphi_b(x,q_y)
=
-(\frac{U_0}{2} + \varepsilon) \varphi_a (x,q_y),
\\
\label{ab::top_channel_2b}
\frac{\hbar^2v_{\rm F}^2}{t_0}
(\xi q_y-\partial_x)^2 \varphi_a (x, q_y)
=
(\frac{U_0}{2} - \varepsilon) \varphi_b (x, q_y).
\end{cases}
\end{eqnarray}
Only two solutions are normalizable, and can be kept.

Once the solutions in each half-plane are found, they must be matched, together with their derivative, at $x=0$. The matching condition cans be cast in the form of a homogeneous system of four linear equations for four unknown coefficients. To have a non-zero solution the system's determinant has to vanish. This implicitly defines two dispersion relations $\varepsilon_{1,2}^\xi = \varepsilon_{1,2}^\xi (q_y)$ for two bands localized at the line $x=0$. In $k$-space these bands are located near the Dirac point specified by the value of $\xi$. Reference~\cite{Martin2008} explained that the presence of exactly two bands per Dirac point is related to the topological properties of the insulating bilayer state. This set of four bands may be visualized as a one-dimensional conducting channel inside a two-dimensional insulating host system.

\begin{figure}[t]
\begin{centering}
\includegraphics[width=10cm]{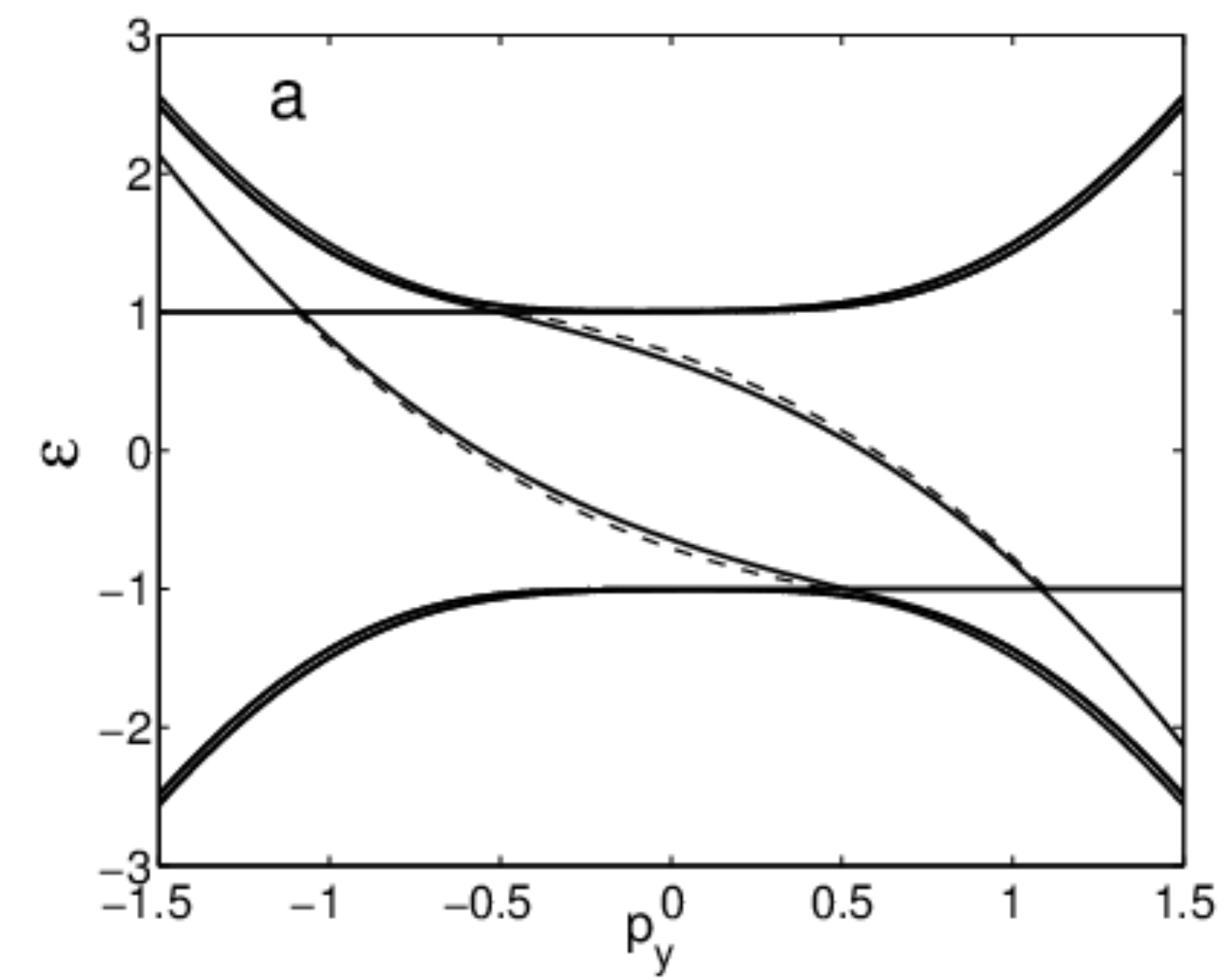}
\par\end{centering}
\caption{Dispersion relation for localized bands near a particular Dirac
point. The two solid curves crossing the $\varepsilon = 0$ level represent
the dependence of the energy [measured in units of $(\hbar v_{\rm
F})^2/(3a_0^2 t_0)$] versus dimensionless momentum $p_y = \sqrt{3} a_0 q_y$
(note that $\sqrt{3} a_0$ is the graphene lattice constant). These curves
were estimated numerically for the dimensionless bias potential $(3 a_0^2
t_0 U(x))/(\hbar^2 v_{\rm F}^2)$ equal to $2\tanh (x/w)$, where the length
scale $w$ was chosen to be $w = \sqrt{3} a_0/2$. Both bands demonstrate
negative group velocity. Their time-reversal partners are located at the
other Dirac point, both having positive group velocities. If $\tanh(x/w)$
is approximated by ${\rm sgn}\,(x)$, the dispersion curves for this case
are shown by dashed curves. One can see that for such choices of parameters
the latter approximation works extremely well.
%
Reprinted figure with permission from
I.~Martin et al.,
Phys. Rev. Lett., {\bf 100}, 036804 (2008).
Copyright 2008 by the American Physical Society.
\url{http://dx.doi.org/10.1103/PhysRevLett.100.036804}
\label{meso::topol_conf_bands}
}
\end{figure}

While the approximation Eq.~(\ref{meso::U_sgn}) accounts for many qualitative features of the topological bands, it is still quite crude. Unfortunately, this is the only known limit which admits an analytic solution. For more realistic choices one has to resort to numerical calculations. Let us examine Fig.~\ref{meso::topol_conf_bands}, where the dispersion of the localized bands are shown. These dispersion relations were obtained numerically in~\cite{Martin2008} for $U(x) \propto \tanh(x/w)$, where $w$ has the dimension of length. The figure shows two subgap modes (solid curves crossing the $\varepsilon = 0$ level) near a particular Dirac point. Both curves have negative group velocity. Their time-reversal partners are located at the other Dirac point, and they both have positive group velocities. The dashed curves represent the dispersions corresponding to the approximation Eq.~(\ref{meso::U_sgn}). One can see that for such a value of $w$ and $U$, the approximate dispersion gives a very accurate estimate for the numerical result. Similar results were obtained by M.~Zarenia et~al. in Ref.~\cite{Zarenia2011a}.

As can be seen, the localized modes have non-trivial valley structure: the
direction of propagation for a particular band depends entirely on its
valley index $\xi$.
Reference~\cite{Martin2008}
proposed to use this property to create ``valley filters" and ``valley
valves".
Reference~\cite{schroer_cooper_pair_split2015}
discussed the use of the topological bands to split Cooper pairs.

Electronic states localized near the bias voltage kinks were studied in
some detail in
Ref.~\cite{Zarenia2011a}.
The authors showed that, in addition to the topological bands crossing the
zero energy, for a sufficiently smooth function
$U(x)$,
non-topological subgap bands also appear. For zero doping, the
non-topological bands do not reach the Fermi energy, and remain separated
from the Fermi level by a gap
$\sim U_0$.
They also considered the more general case when
$U(x)$
changes sign at some finite
$x=d>0$,
and then again changes sign at
$x=-d<0$.
If $d$ is not too large, the topological bands at
$x=d$
hybridize with the bands at
$x=-d$.
Therefore, the bands acquire small, but finite, gaps, which may be useful
for the control of transport properties. The presence of the magnetic field
does not influence strongly the topological bands.

The transport properties of these bands were studied by Z.~Qiao et~al. in
Ref.~\cite{Qiao2011}. Unlike the papers mentioned above, they used a
tight-binding Hamiltonian, thus, avoiding the perils of the low-energy
effective models, such as the non-physical absence of intervalley
scattering. Of course, the study relies exclusively on numerical
calculations. Using the Landauer-B\"uttiker approach and non-equilibrium
Green's function technique, it was shown~\cite{Qiao2011} that the
topological bands have excellent transport properties. They propagate well
even when the channel bends: the authors report ``zero bend resistance"
even for sharp turns of the propagation paths. Disorder-induced
backscattering is also predicted to be very weak both for long-range and
short-range disorder. These properties may be useful for applications.

Topological bands could be used as a Tomonaga-Luttinger-like liquid with
tunable parameters. This proposal was formulated by M.~Killi et~al.~in
Ref.~\cite{Killi2010}. Since these systems have two bands propagating along
the $y$-axis and two bands propagating in the opposite direction, the
resultant one-dimensional liquid does not coincide with the standard
Tomonaga-Luttinger concept, which has only one propagating mode per
direction of propagation (so-called ``left-moving" and ``right-moving
electrons"). Nonetheless, this one-dimensional liquid is an interesting
object in its own right. Its effective low-energy description was outlined
in
Ref.~\cite{Killi2010}.
In particular, the authors noted that the effective parameters of such a
generalized Tomonaga-Luttinger liquid may be varied by the bias voltage.

Finally, we would like to mention two variations of the ideas reviewed in this subsection. (I)~It was predicted~\cite{zhang_defect_topolog_channel2013, vaezi_defect_topolog_channel2013} that certain stacking-fault domain walls may host similar modes in the presence of coordinate-independent bias $U({\bf r}) = U_0 \ne 0$. Experimentally, such domain walls were studied in Refs.~\cite{Alden2013,topological_experiment}. (II)~In the context of edge states, related concepts were discussed in Ref.~\cite{Li2011}.

\subsection{pn-junctions and similar structures}\label{subsect::meso::pn}

A junction of pn type is both a cornerstone of modern electronics and an interesting object for fundamental research. This type of mesoscopic structure has been studied for single-layer graphene (see Ref.~\cite{meso_review} for a review). It is no wonder that bilayer-based junctions are also discussed in the literature.

Similar to single-layer graphene, the physics of Klein tunneling, reviewed in subsection~\ref{Klein}, also affects the properties of bilayer junctions as well. However, the setups discussed in subsection~\ref{Klein} are too idealized to be directly relevant for  experiments. To develop a realistic description of the transport through the junctions other important factors, such as disorder and interaction, must be accounted for.

\begin{figure}[t]
\begin{centering}
\includegraphics[width=10cm]{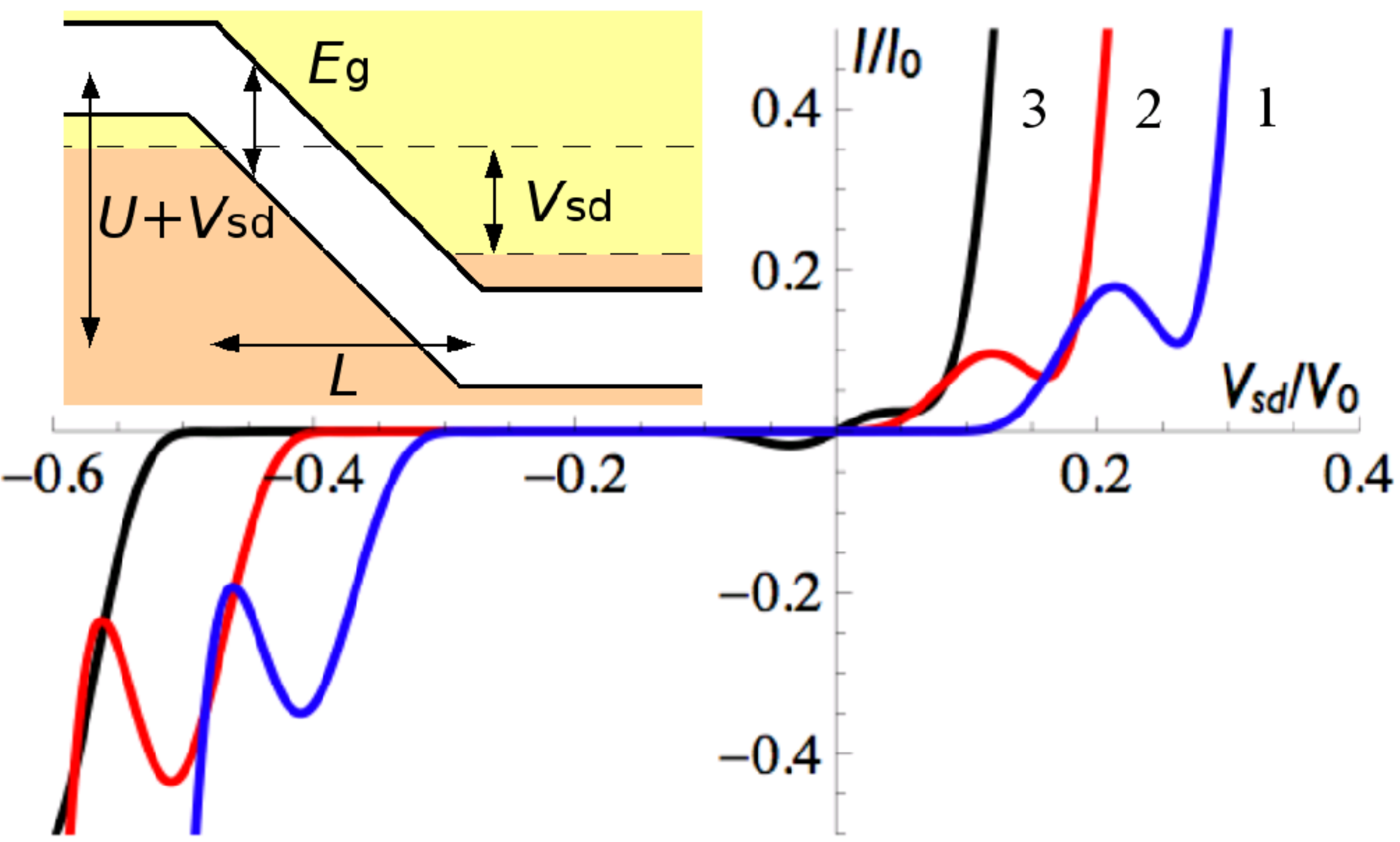}
\par\end{centering}
\caption{Calculated $IV$ characteristics for a pn-junction made from an AB graphene bilayer, Ref.~\cite{Nandkishore2011}. Three curves represent three values of the gate-induced potential difference across the junction $U$ (curve~1 corresponds to the smallest $U$, curve~3 corresponds to the largest $U$). The curves are asymmetric with respect to polarity change of the source-drain voltage $V_{\rm sd}$. Negative differential conductivity regions are clearly visible. Inset: Structure of the junction under finite $V_{\rm sd}$.
Reprinted from: R.~Nandkishore and L.~Levitov, PNAS {\bf 108}, 14021 (2011).
\label{meso::pn_nand_levit}}
\end{figure}

The AB bilayer graphene has two obvious differences when compared to a single-layer graphene: its dispersion is quadratic, and it can become a gapped insulator (semiconductor) when the transverse bias is present. Thus, the properties of a bilayer pn-junction differ from the properties of a single-layer junction. A semiclassical theoretical investigation of transport through a ballistic junction was presented by R.~Nandkishore and L.~Levitov in Ref.~\cite{Nandkishore2011}. The geometry of the system under study is shown in the inset for Fig.~\ref{meso::pn_nand_levit}.

As one can expect, the junction's $IV$ curve demonstrates asymmetry with respect to the source-drain voltage polarity change (see Fig.~\ref{meso::pn_nand_levit}). In addition, for a certain range of source-drain voltages the differential conductivity of the junction is negative. In Ref.~\cite{Nandkishore2011} it was explained that the negative differential conductivity is a consequence of quantum-mechanical interference between two semiclassical trajectories in a classically-forbidden region, separating $p$ and $n$ electrodes of the junction. Such under-the-barrier interference is uncommon for usual semiconducting junctions.

The interference is very sensitive to the disorder inside the sample. However, estimates suggest that the currently available experimental purity is sufficient to guarantee the observation of the theoretically-predicted behavior.

A model of a pn-junction with a step-like barrier was discussed in
Refs.~\cite{gradinar_pn_bilayer2012, four_band_tunn_duppen2013}. In
particular, the authors of Ref.~\cite{four_band_tunn_duppen2013} studied a
pn-junction and the tunneling across a potential barrier (the latter system
may be viewed as a pnp-junction). A disorder-free configuration was
assumed. The paper investigated the accuracy of different effective models
of the bilayer graphene. Predictions of the four-band and two-band models
were compared, and the importance of the longer-range single-electron
hopping processes was assessed. It was found that the longer-range hopping
amplitudes are unimportant for energies in excess of 4\,meV. On the other
hand, the two-band effective theory prediction may deviate significantly
from the four-band theory at higher energies.

The effects of strain on the properties of ballistic pn-, pnp-, and other
types of junctions were discussed in
Ref.~\cite{gradinar_pn_bilayer2012}.
Uniaxial strain, when applied to an AB bilayer sample, qualitatively
modifies the low-energy dispersion of electrons: the parabolic dispersion
surface is replaced by two Dirac cones separated by a saddle point. The
resultant dispersion surface is somewhat similar to the trigonal warping,
Fig.~\ref{fig::trig_warp},
but with two emergent cones instead of four. The saddle point at
$\pm 5$\,meV
may be induced by strain of about 1\%.
Reference~\cite{gradinar_pn_bilayer2012}
concluded that in the presence of strain the ballistic conductance of both
pn- and pnp-junctions may demonstrate non-monotonous dependence on doping
and temperature.

Above we have already mentioned
Refs.~\cite{gradinar_pn_bilayer2012,four_band_tunn_duppen2013}
investigating pnp-junctions. Another paper dedicated to this topic is
Ref.~\cite{chiral_tunn_tudorovskiy2012}.
It developed a semiclassical formalism for the theoretical study of a
ballistic AB bilayer pnp-junction. The bilayer was assumed to be in the
gapless regime (no transverse bias voltage). They showed that there is a
set of ``magic" incident angles for which an electron passes the junction
without reflection (for the single-layer graphene and rectangular barrier
these angles can be seen on the transmission plot of
Fig.~\ref{Klein_SLG_AAfig}, for more general case see, e.g.,
Ref.~\cite{bliokhFrei}).
This reflectionless transmission is a manifestation of the Klein tunneling.
The authors concluded that it would be difficult to lock a transistor based
on a gapless AB bilayer because of the perfect transmission at the ``magic
angles". Of course, one might try using a gapped biased bilayer to
circumvent this difficulty, but this route is not without a flaw either
(see
subsection~\ref{ab::transport_bias},
where the experiment on biased AB bilayer graphene is discussed). Yet
another opportunity is a relatively new theoretical
proposal,
Ref.~\cite{kleptsyn_no_magic2015}.
The latter paper demonstrated that the tunneling accross the barrier for
any angle can be small provided that the barrier shape is suitably chosen.
Recently, the tunneling through a smooth barrier was examined in
Ref.~\cite{smooth_barrier2015}.
The latter paper also discussed the effects of the trigonal warping on the
tunneling.

\begin{figure}[t]
\begin{centering}
\includegraphics[width=10cm]{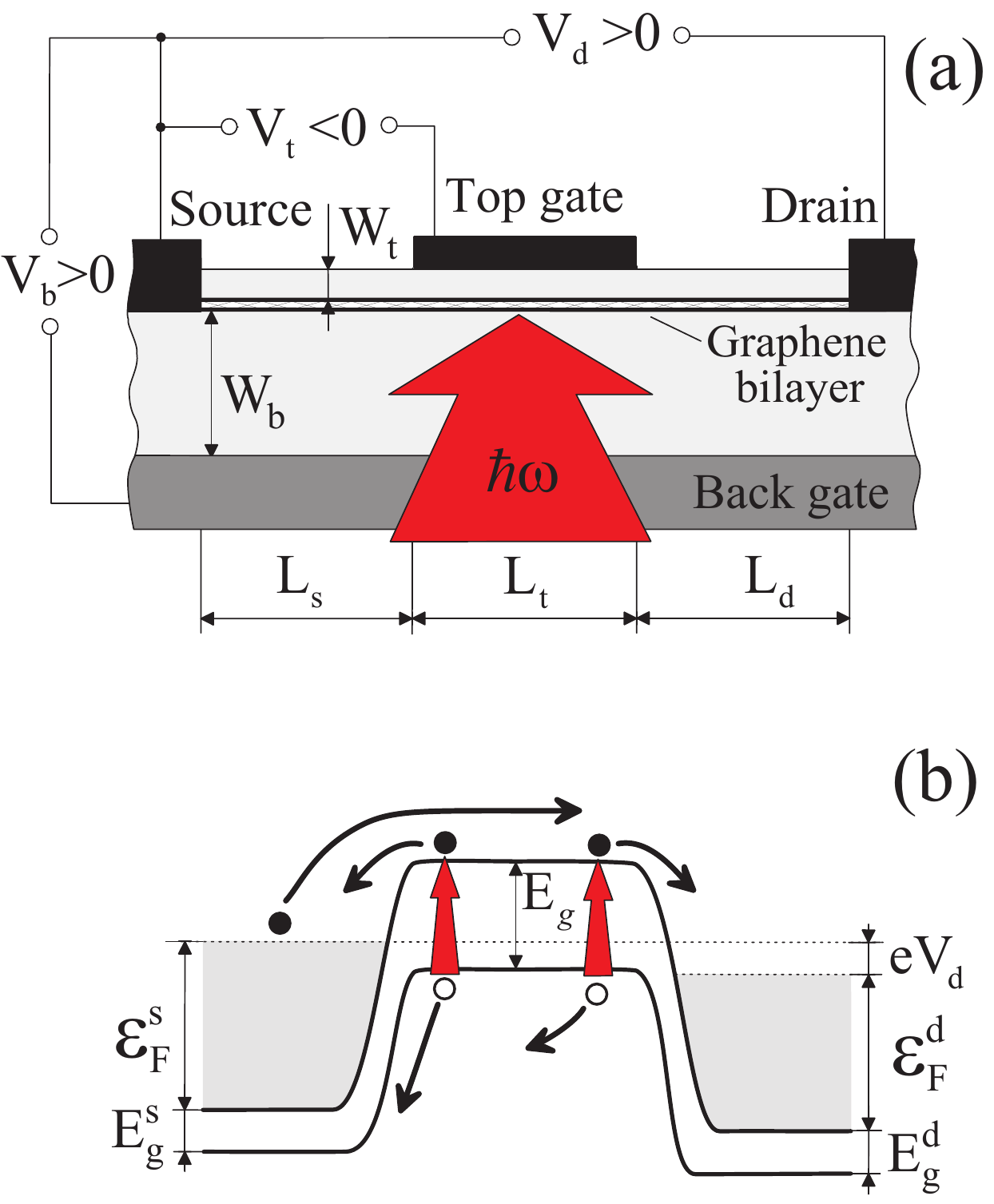}
\par\end{centering}
\caption{Photodetector based on gapped AB bilayer graphene. In panel (a)
$V_d$ is the source-drain voltage, $V_t$ ($V_b$) is the top-gate
(back-gate) voltage. The huge red arrow denotes the incoming radiation.
Panel (b) describes the operation of the detector. The incoming radiation
(red arrows) excites electron-hole pairs. Under the action of the
source-drain field the photoholes move mostly to the source, and the
photoelectrons move mostly to the drain. The motion in the opposite
direction is also possible due to thermal fluctuations. The latter are
responsible for ``the dark current".
Reprinted figure with permission from
V.~Ryzhii, M.~Ryzhii,
Phys. Rev. B, {\bf 79}, 245311 (2009).
Copyright 2009 by the American Physical Society.
\url{http://dx.doi.org/10.1103/PhysRevB.79.245311}
\label{meso::fet_radiation}}
\end{figure}

A theoretical proposal to use an AB bilayer heterojunction as a detector of
radiation in terahertz and infrared range was presented by V.~Ryzhii and
M.~Ryzhii in Ref.~\cite{Ryzhii2009}.
The setup is depicted in
Fig.~\ref{meso::fet_radiation}
and that heterojunction resembles a npn-junction. However, unlike the
latter, the electrochemical potential in the central area lies inside the
gap. Therefore, it is not a $p$-type doped semiconductor, but rather an
insulator with a small gap. The ``dark current" (the source-drain current
in the absence of any radiation) of the detector is small, since the charge
carriers from the source must overcome the gap in this insulating area to
reach the drain. The radiation excites electron-hole pairs across the gap,
and these pairs contribute to the source-drain current. Observation of the
photocurrent in excess of a dark current signifies the finite radiation
flux. The authors calculated different characteristics (spectral
characteristics, dark current, responsivity, dark current limited
detectivity) of their detector. They concluded that at optimal parameters
it can demonstrate better performance than the detectors of other types.
The experimental realization of a photodetector working in the terahertz
range was reported by D.~Spirito et
al.~\cite{terahertz_exp2014}.

Experimental realizations of pn-junctions and related heterostructures have
been described in many papers. For example, the properties of pn-junction
were studied experimentally by B.~Chakraborty et~al. in
Ref.~\cite{Chakraborty2009}.
The authors also offered a simple theory explaining the observed $IV$
curve. The formulated theory fits well the data. However, according to the
paper, the described device becomes pn-junction only under substantial
source-drain voltage: it was assumed that the areal charge density near the
drain electrode changes its polarity due to spillage of the charge from the
electrode itself. Clearly, such a situation does not correspond to any of
the discussed theoretical models.

Many workers are attempting to create a field-effect transistor, which may
serve as a cornerstone of future potential graphene-based electronics (for
a general discussion on this issue one may consult
Ref.~\cite{Avouris2007}).
For example, a series of
papers~\cite{Szafranek2010,Szafranek2011,Szafranek2012}
studied this type of devices. A theoretical proposal for a tunnel
field-effect transistor was put forward in
Ref.~\cite{Alymov2016}.

In Ref.~\cite{pn_miyazaki2012}
H.~Miyazaki et~al. investigated a setup consisting of several junctions
connected in series. The purpose of that study was to check if multiple
junctions can suppress leakage current in the ``off'' state of the
transistor. The junctions were created by a comb-like top gate. Depending
on the voltage on the top and back gates, the system may be in one of four
regimes: pn-, np-, nn'-, and pp'-junction. The ambipolar regimes (pn and
np) are indeed characterized by a lower source-drain current.

Unfortunately, these setups do not yet demonstrate the characteristics
required for successful commercial exploitation. For example, the on/off
ratio (the ratio of currents in ``on'' and ``off'' states) for the
transistors devised in
Ref.~\cite{Szafranek2011}
is less than 50, while for logic
applications~\cite{Avouris2007}
this ratio must be higher than
10$^4$.
Of course, at lower temperature this ratio grows:
Ref.~\cite{bilayer_gap_exp2015}
described a device with the on/off ratio of several thousands at
20\,K.
However, such a restriction on the operation temperature makes the device
useless for mass market applications.

\section{Electronic transport through an AB bilayer sample}
\label{sec::transport_ab}
%
%

Both theoretical and experimental research of the electric transport
through AA bilayer graphene is non-existent. Current experimental and
theoretical efforts are focused on the electrical conductivity of
AB bilayers. Consequently, in this section we will only review the transport
through AB bilayer samples.

The relevant papers may be roughly divided into three groups. The papers in
the first group study the conductivity of doped AB bilayer graphene.
A doped sample has a well-defined Fermi surface. Despite some
peculiarities, such a system may be viewed as a metal, and its transport
properties can be investigated accordingly. We review this topic in
subsection~\ref{transport::subsec::fin_dop}.

The second group involves work with the so-called minimum
conductivity, that is, the conductivity of an undoped sample. While the
undoped bilayer with no disorder has no charge carriers, they can be
induced (by random variations of an external potential, or by impurities)
and contribute to the transport. This and other mechanisms for the minimum
conductivity are discussed in
subsection~\ref{ab::transport::minimum_cond}.

The third group of papers is dedicated to transport in samples to where
the transverse bias is applied. While the reasoning outlined in
subsection~\ref{subs::bernal_bias_gap}
implies that a system of this kind should demonstrate insulating behavior,
transport measurements paint a more complicated picture. Theoretical
proposals explaining this discrepancy are reviewed below, in
subsection~\ref{ab::transport_bias}.

\subsection{Transport through doped bilayer}
\label{transport::subsec::fin_dop}
One of the earliest theoretical papers studying the conductivity through
AB-stacked bilayer graphene was
Ref.~\cite{Koshino2006} by M.~Koshino and T.~Ando.
There it was assumed that the band structure of the bilayer is described by
only two bands closest to the Dirac point energy. In addition to the
parabolic dispersion terms, the trigonal warping was also included into the
model. Speaking technically, we can say that such a model corresponds to
the low-energy sector of the Hamiltonian,
Eq.~(\ref{ab::H_trig}).

Within formalism, the doping level $x$ was controlled by a finite Fermi
energy
$\varepsilon_{\rm F}$,
measured from the Dirac point energy. The authors estimated that for a
typical experimentally-available sample, the doping is
$\sim 10^{12}$\,cm$^{-2}$,
or higher. Such electron concentration corresponds to the Fermi energy
$\varepsilon_{\rm F} \sim 10 \varepsilon_{0}$,
where the energy scale
$\varepsilon_{0} \approx 3.9$\,meV
used in
Ref.~\cite{Koshino2006}
characterizes the trigonal warping. It can be related to the quantity
$\varepsilon_{\rm L}$,
defined by
Eq.~(\ref{ab::lifshits}),
as the energy of the saddle points of the dispersion surface:
$\varepsilon_0 = 4 \varepsilon_{\rm L}$.
According to the estimates presented in the paper, when the Fermi energy
$\varepsilon_{\rm F} \sim 10 \varepsilon_{0}$
the trigonal warping exerts significant modifications to the
single-electron dispersion.

The only mechanism of current relaxation discussed in
Ref.~\cite{Koshino2006}
is disorder scattering. Two types of disorder were studied
separately: long-range and short-range potential disorder. The
difference between the two is the presence (absence) of intervalley
scattering for the short-range (long-range) scatterers.
The disorder potential was assumed to be weaker than the amplitude
$t_0$. The violation of this condition implies the necessity to use the full four-band
model, instead of the two-band description employed in the paper.
To calculate the electric conductivity the authors resorted to the
self-consistent Born approximation. Within this framework the
single-electron self-energy
$\Sigma (\varepsilon_{\rm F})$
can be found as a solution of the non-linear integral equation.

If we want to evaluate a transport coefficient, in a typical situation the
single-electron self-energy
is not sufficient: one needs to know vertex corrections as well. However,
for the case of short-range scatterers, the vertex corrections vanish, and
the knowledge of
$\Sigma (\varepsilon_{\rm F})$
is sufficient to derive $\sigma$. The conductivity calculated in such a
manner depends on the Fermi energy
$\varepsilon_{\rm F}$,
which, in turn, is a function of doping $x$.

For general value of disorder, the self-energy
may be found only numerically. Yet, for the case of weak short-range
disorder an analytic solution is available. Specifically, in the
$\Sigma(\varepsilon_{\rm F}) \rightarrow 0$
limit, the following expression for the self-energy was established in
Ref.~\cite{Koshino2006}:
\begin{eqnarray}
\Sigma (\varepsilon_{\rm F} + i0)
=
-\frac{i\pi}{2}
W \varepsilon_0
\frac{\rho_0 (\varepsilon_{\rm F})}{\rho_\infty},
\quad\text{where\quad}
W = \frac{n_i u^2}{2} \frac{\rho_\infty}{\varepsilon_0}.
\end{eqnarray}
Here
$\rho_0 (\varepsilon_{\rm F})$
is the density of states at the Fermi energy $\varepsilon_{\rm F}$ for a pristine
bilayer, the quantity
$\rho_\infty$
is the ``high-energy" asymptotic value of the density of states:
$\rho_0 (\varepsilon_{\rm F}) \approx \rho_\infty$,
if
$t_0 \gg \varepsilon_{\rm F} \gg \varepsilon_0$.
The disorder is characterized by the product
$n_i u^2$,
where
$n_i$
gives the concentration of scatterers, while $u$ stands for typical
strength of an individual scatterer. The parameter $W$ quantifies the strength
of the disorder relative to the strength of the trigonal warping: for
$W > 1$
the disorder scattering smears the fine effects due to the warping.

Using the above expression for the self-energy, Ref.~\cite{Koshino2006} derived for the
conductivity:
\begin{eqnarray}
\sigma
=
\frac{e^2}{\pi^2\hbar}
\times
\begin{cases}
\frac{1}{W}
\left(
	\frac{|\varepsilon_{\rm F}|}{\varepsilon_0}
	+
	1
\right),
&
\text{if}
\quad
|\varepsilon_{\rm F}| > \varepsilon_0,
\cr
\frac{3}{4W},
&
\text{if}
\quad
|\varepsilon_{\rm F}| \ll \varepsilon_0.
\cr
\end{cases}
\end{eqnarray}
These results demonstrate that for weak short-range disorder the
conductivity is non-universal, and depends on the disorder strength $W$.

The dependence on $W$ disappears for strong disorder: in the regime
$W > 1$ we have
\begin{eqnarray}
\Sigma (\varepsilon_{\rm F} + i0)
\approx
-i \Gamma,
\quad
\text{where}
\quad
\Gamma = \frac{\pi}{2} W \varepsilon_0.
\end{eqnarray}
The conductivity is equal to
\begin{eqnarray}
\sigma
=
\frac{e^2}{\pi^2 \hbar}
\left[
\left(
	\frac{\varepsilon_{\rm F}}{\Gamma}
	+
	\frac{\Gamma}{\varepsilon_{\rm F}}
\right)
\arctan\left(
	\frac{\varepsilon_{\rm F}}{\Gamma}
\right)
+
\frac{1}{W}
+
1
\right].
\end{eqnarray}
If
$\varepsilon_{\rm F} < \Gamma$
the conductivity becomes
\begin{eqnarray}
\sigma
=
\frac{e^2}{\pi^2 \hbar}
\left(
	2
	+
	\frac{1}{W}
\right).
\end{eqnarray}
As disorder grows, this expression approaches the finite universal value:
$\sigma \approx 2e^2/(\pi^2 \hbar)$
per spin projection.

In addition to these analytic estimates, the results for
$\sigma(\varepsilon_{\rm F})$
with numerically-calculated self-energy were also presented. For weak
disorder, the numerical data revealed that at very small
$\varepsilon_{\rm F}$
the conductivity drops from large non-universal value to the universal
value of
$6 e^2/(\pi^2 \hbar)$
per spin projection.

For long-range disorder the vertex corrections are finite, and must be
taken into account. However, the dependence of the conductivity on the disorder
strength and the Fermi energy are similar to the case of short-range
disorder. It was demonstrated~\cite{Koshino2006} that at weak disorder the conductivity is
inversely proportional to $W$. When the disorder is strong
($W > 1$),
the conductivity approaches the universal value $2e^2/(\pi^2 \hbar)$.

The theory predictions~\cite{Koshino2006} were compared against transport data in
Ref.~\cite{Novoselov2006}.
The authors of Ref.~\cite{Koshino2006} concluded that in the doping range
$<2\times 10^{12}$\,cm$^{-2}$,
their theory was consistent with the experiment, provided that
$1 \lesssim W \lesssim 2$. For larger doping, the experimental resistivity decreased faster
than anticipated by the theory. The authors of Ref.~\cite{Koshino2006} discussed briefly possible
reasons of such deviations.

The study of
Ref.~\cite{Koshino2006}
was later generalized in
Refs.~\cite{Koshino2009,ando_transport2011}.
The formalism of the latter papers included all four bands of the bilayer.
Reference~\cite{Koshino2009}
concluded that the two higher-energy bands contribute to the conductivity at
large Fermi energy. At the same time, these bands do not significantly
affect transport at the charge-neutrality point, even if the disorder is
large. In
Ref.~\cite{ando_transport2011}
the effects of the trigonal warping were neglected. However, Ref.~\cite{ando_transport2011} discussed more general
disorder models (impurities with Gaussian potential and impurities with
screened Coulomb potential). For the Gaussian-shape impurity
potential
$$
v({\bf r})=\frac{v_0}{\pi d^2}\exp(-r^2/d^2)\,,
$$
it was found that $\sigma$ is sensitive both to the strength $v_0$ of the
potential,
and to its range $d$. At zero Fermi energy, the conductivity becomes
non-universal at larger $d$. The screened charged impurities also destroy the
universality of the conductivity.

The conductivity of a sample with short-range scatterers was numerically studied by M.~Trushin et~al.~\cite{Trushin2010}
using the Kubo formula and kinetic equation formalism.
This paper~\cite{Trushin2010} generalized ideas, previously formulated for transport
in single-layer graphene~\cite{trushin_single_layer2007,auslander_single_layer2007}.

J.~Nilsson et~al.~\cite{Nilsson2006_transp} used the Kubo formula and the coherent-potential approximation
to discuss the effects of short-range impurity disorder on the
single-electron and transport properties of the AB bilayer and multilayers.
The coherent-potential approximation is considered to be superior to the
self-consistent Born approximation of
Ref.~\cite{Koshino2006}.
The paper~\cite{Nilsson2006_transp} presented the dependence of the dc conductivity on doping for
different impurity concentrations. The zero-doping value of the conductivity
was found to be universal.

\begin{figure}[t]
\begin{centering}
\includegraphics[width=0.6\textwidth]{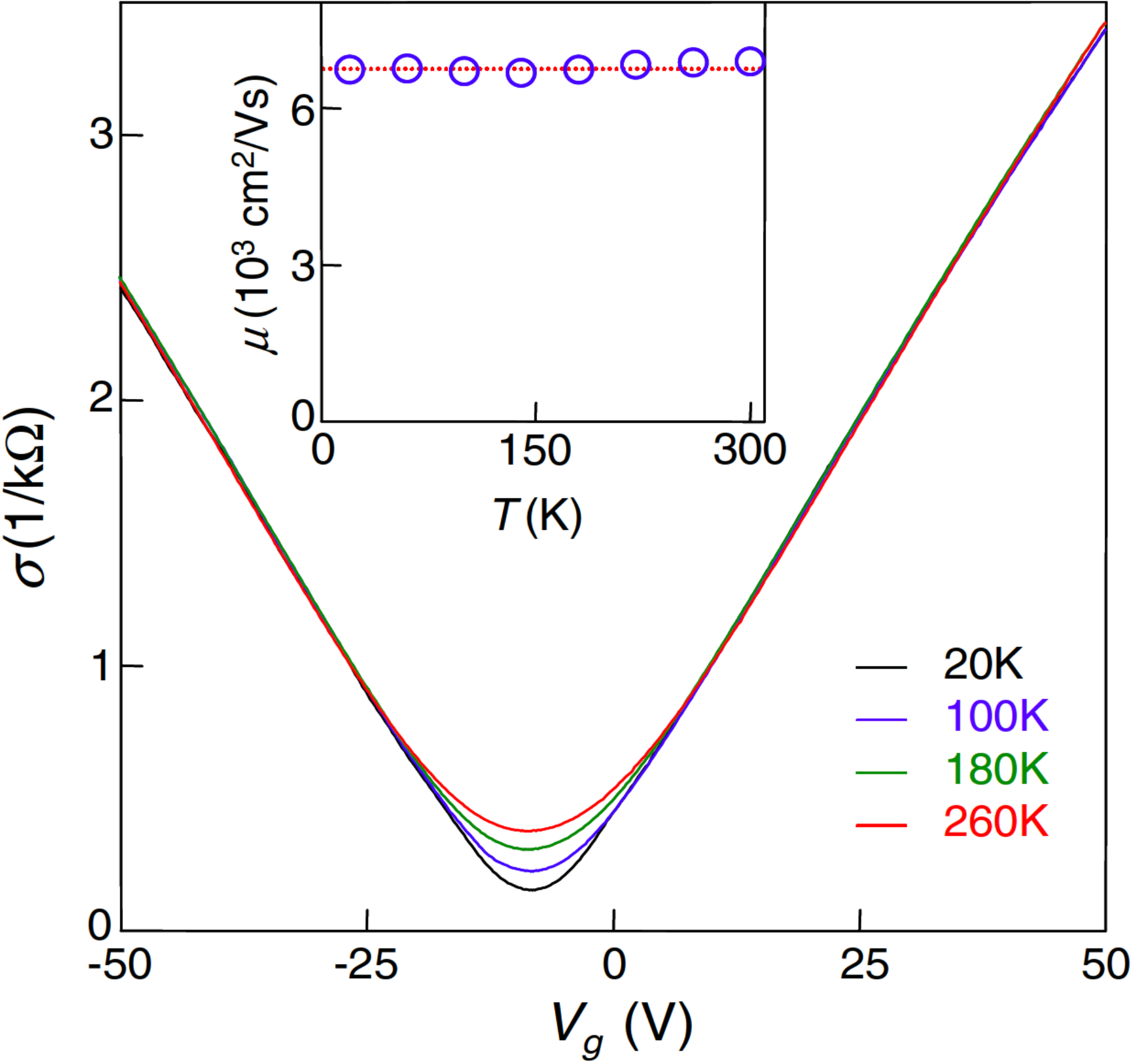}
\par\end{centering}
\caption{Measured AB bilayer graphene conductivity as a function of
gate voltage $v_g$ for different
temperatures~\cite{Morozov2008}.
The conductivity passes through a minimum at non-zero value of the gate
voltage. This minimum is believed to occur when the average doping level is
zero. Away from the minimum, the conductivity demonstrates almost ideal
linear behavior as a function of the gate voltage. The temperature
dependence of the conductivity is particularly pronounced at the minimum,
see
Fig.~\ref{transp::exp_min_cond_vs_temp}.
Reprinted figure with permission from
S.V.~Morozov et al.,
``Giant intrinsic carrier mobilities in graphene and its bilayer",
Phys. Rev. Lett., {\bf 100}, 016602 (2008).
Copyright 2008 by the American Physica Society.
\url{http://dx.doi.org/10.1103/PhysRevLett.100.016602}
\label{transp::exp_conduct_vs_doping}
}
\end{figure}

The electric transport through a bilayer in the presence of screened
charged impurities was studied theoretically by S.~Adam and S.~Das~Sarma in
Ref.~\cite{Adam2008}.
In the limit of higher carrier density the authors argued that the impurity
potential is strongly screened. Such an impurity can be treated as a
short-range scatterer with a doping-independent effective potential. The
influence of these impurities on the charge transport was accounted with
the help of the Boltzmann equation. For not-too-small doping $x$, the
calculated conductivity demonstrates a linear dependence on $x$. As the doping
$x$ grows, weak deviation from this linear behavior becomes observable.

If the doping $x$ drops below a certain disorder-dependent value, a crossover
to a different regime occurs: at low $x$ disorder-induced electron
and hole puddles appear. Such `charge puddles' are thought to be unipolarly
charged areas, which may appear within a sample, whose average doping $x$
is small or zero. In this limit the conductivity saturates near a finite
non-universal value and shows only very weak dependence on $x$. The
transport in the presence of puddles will be discussed in more detail in
subsection~\ref{ab::transport::minimum_cond}.

It was concluded in
Ref.~\cite{Adam2008}
that the developed formalism is consistent with the experimental
data (see
Fig.~\ref{transp::exp_conduct_vs_doping}).
Namely, the authors stressed that they explained both the saturation of the
conductivity at low carrier density, and obtained a
$\sigma(x) \propto x$
behavior at higher doping levels.

The analysis of Ref.~\cite{Adam2008} was re-examined in Ref.~\cite{DasSarma2010}. It was claimed that Ref.~\cite{Adam2008},
``considering only screened Coulomb disorder, included drastic approximations (e.g. complete screening) which are both unreliable and
uncontrolled". The authors of Ref.~\cite{DasSarma2010}
argued that not only the screened Coulomb disorder, but also short-range disorder must be accounted to describe
experiments~\cite{Morozov2008,Xiao2010}.
It was determined that in a wide range of dopings the conductivity
dependence on $x$ is approximately linear, and the temperature dependence
of $\sigma$ is very weak for higher $x$. Both features are consistent with
the experimental data. The conductivity at
$x=0$
was also analyzed.

Another theoretical study of the Coulomb disorder was presented by M.~Lv and S.~Wan in
Ref.~\cite{Lv2010}.
This paper investigated the screening of a charged impurity in a doped
bilayer sample. The obtained results were used to calculate the
conductivity as a function of temperature.

Consequences of a different type of disorder were explored in
Ref.~\cite{Ferreira2011}.
The authors noted that, for single-layer graphene, experimental
studies~\cite{Ni2010,Katoch2010,Monteverde2010}
suggested that radicals adsorbed on a sample become an important source of
scattering. According to
Ref.~\cite{Ferreira2011},
the same scattering mechanism is relevant in bilayer samples at large
doping levels. The paper developed a theoretical framework to describe such
a disorder, and discussed in detail several subtleties of the formalism. The
authors concluded that the proposed mechanism alone is sufficient to
explain the linear relation between the conductivity of a bilayer and the
sample's doping level observed in
experiments~\cite{Morozov2008}
for large doping.

The scattering by the adsorbed atoms, together with other types of disorder
(short-range and long-range impurities, vacancies, and others), were
considered in
Ref.~\cite{Yuan2010}.
The authors employed a numerical technique called time-evolution
method~\cite{time-evol}.
They claimed that the latter method allows ``to carry out calculations for
very large systems, up to hundreds of millions of sites, with a
computational effort that increases only linearly with the system size".
The paper numerically calculated the conductivity $\sigma$ as a
function of the doping level $x$. Depending on the type and strength of
the disorder, the function
$\sigma (x)$
demonstrated linear or sublinear behavior.

In
Ref.~\cite{Zhu2009}
the carrier mobility was measured as a function of both temperature and
carrier concentration. The experimental data for the mobility was analyzed
with the help of the scattering time approximation. The authors concluded
that for a bilayer on a substrate ``the mobility is dominated by Coulomb
scattering".

F. Kisslinger et~al.~\cite{Kissl} observed a linear behavior in the magnetoresistance of epitaxial AB bilayer graphene in a magnetic field up to $62$\,T at and above room temperature. The authors argued that such a behavior can be attributed to dislocations in epitaxial AB bilayer, ``a system that is frequently assumed to be dislocation-free''.

\subsection{Minimum conductivity of a bilayer}
\label{ab::transport::minimum_cond}

\begin{figure}[t]
\begin{centering}
\includegraphics[width=10cm]{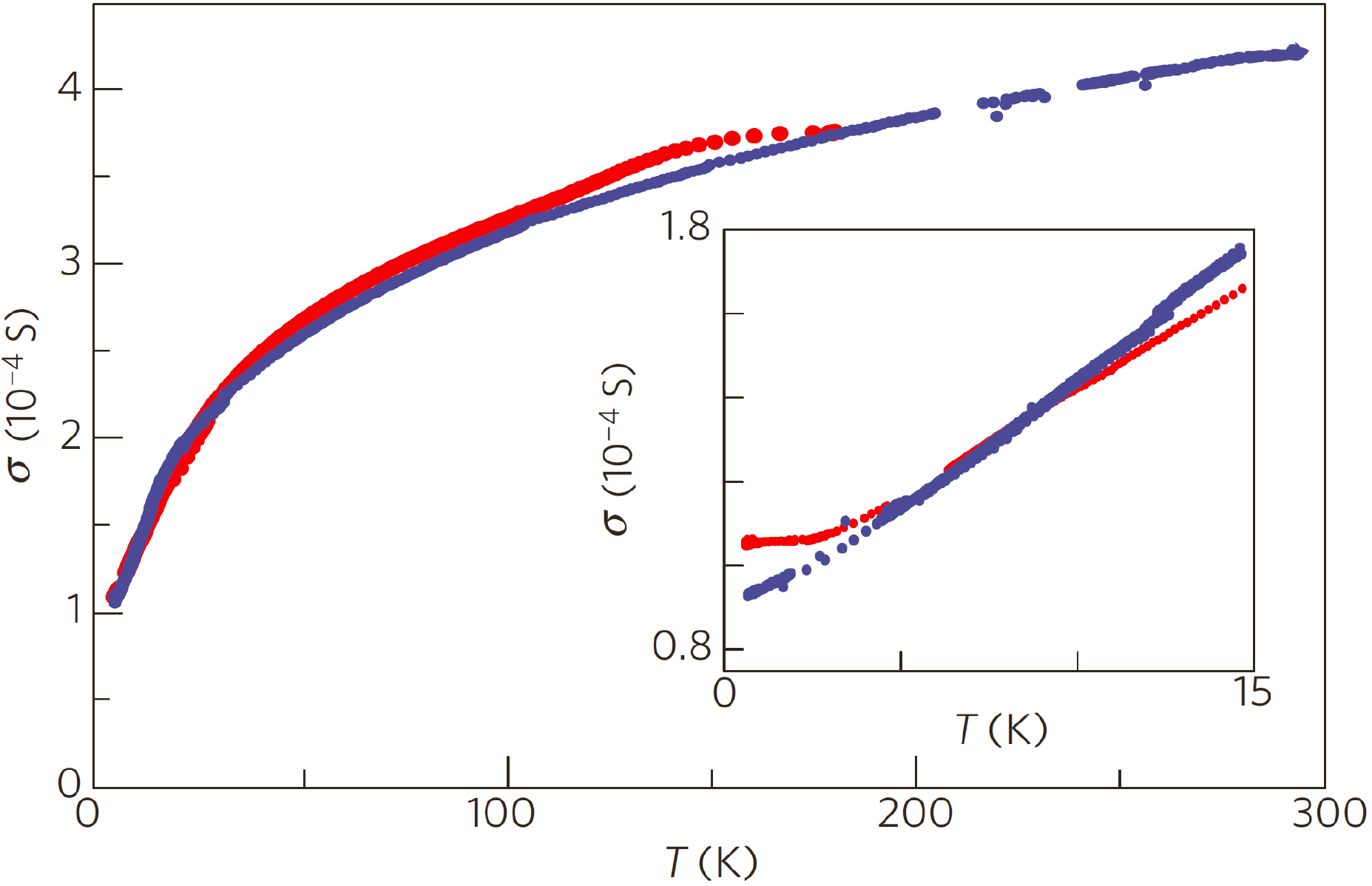}
\par\end{centering}
\caption{Experimental data for the minimum conductivity: The blue and red
curves demonstrate the measured minimum conductivity of two suspended AB
bilayer samples as a function of temperature. The inset shows fine details
of the low-temperature data. The conductivity of the bilayer remains finite
even for very low temperature.
%
Reprinted by permission from Macmillan Publishers Ltd: Nature Physics,
{\bf 5}, 889 (2009),
copyright 2009.
\url{http://www.nature.com/nphys/index.html}
\label{transp::exp_min_cond_vs_temp}
}
\end{figure}

As can be seen from
Fig.~\ref{transp::exp_conduct_vs_doping}
and
Fig.~\ref{transp::exp_min_cond_vs_temp},
even at zero doping, the experimentally-measured conductivity of an AB bilayer
sample is small, but finite even at very low temperatures. Apparently, in
such a regime, the sample is able to conduct current, although, formally
speaking, at zero doping and temperature the system has no charge carriers
(single-layer graphene also demonstrates a finite minimum conductivity~\cite{Novoselov2005}). This property was investigated theoretically using several approaches,
defined by nonequivalent sets of assumptions and technical tools.
Some of these methods have already been mentioned in
Section~\ref{transport::subsec::fin_dop}. Others will be reviewed below.
However, choosing a most suitable technique for a particular experimental
situation, one has to be aware of certain theoretical subtleties, important
for both monolayer, and AB bilayer graphene at low doping. Specifically,
Refs.~\cite{trushin_single_layer2007,auslander_single_layer2007}
developed a kinetic-equation approach to disordered single-layer
graphene. It was demonstrated that the use of a simple Boltzmann equation
might be unjustified. Instead, a more general kinetic equation accounting for
the ``chiral" nature of the single-electron states near the Dirac points
must be employed. The ideas put forward in these papers were extended for
a bilayer by M.~Trushin et~al. in
Ref.~\cite{Trushin2010}.
The function $\sigma(x)$ approaches finite minimum of the order of $e^2/(\pi \hbar)$, when $x \rightarrow 0$.~\cite{Trushin2010}

Further, if a sample with low or zero average doping is placed on a
substrate, charged impurities in the substrate affect the system in a highly
non-perturbative manner. The smooth random variation of the electrostatic
potential due to these impurities could induce so-called ``charge puddles"
in the sample. The puddles are defined in
subsection~\ref{transport::subsec::fin_dop}
as areas with finite unipolar charge density, generated by Coulomb
disorder in the substrate. Puddles of both polarities coexist within a
single sample. For monolayer graphene, the puddles were discussed
theoretically~\cite{Hwang_puddles2007}
and observed
experimentally~\cite{single_layer_puddles2008}.
It is only natural to also apply this concept to the bilayer. For
example, above we already discussed
Ref.~\cite{Adam2008} by S.~Adam and S.~Das~Sarma,
which invoked this notion in connection with the transport through the
bilayer. Using these ideas, W.~Zhu et~al.~\cite{Zhu2009}
offered a semi-phenomenological fitting expression to describe the behavior
of the minimum conductivity. A more rigorous approach was
adopted in
Ref.~\cite{Adam2010} by S.~Adam and M.\,D.~Stiles,
where an effective medium theory was used to investigate the conductivity
of a sample with a disordered ensemble of puddles (this approximation was
also employed in Refs.~\cite{DasSarma2010,Hwang2010}).

The effective medium theory is a fairly old method designed to study
macroscopic properties of a composite material. According to a version of
the effective medium theory used in
Ref.~\cite{Adam2010},
the effective conductivity of a macroscopic sample
$\sigma_{\rm EMT}$
must satisfy the following equation
\begin{eqnarray}
\label{ab::transport::emt}
\Big<
	\frac{\sigma (\tilde x) - \sigma_{\rm EMT}}
		{\sigma (\tilde x) + \sigma_{\rm EMT}}
\Big>
=
0.
\end{eqnarray}
Here
$\sigma (\tilde x)$
is the local conductivity of a small patch of the sample with local doping $\tilde x$.
Patches with different $\tilde x$
are randomly distributed over the sample. To account for such a disorder we
should view the local doping as a random variable characterized by some
distribution function
$P[\tilde x]$.
The triangular brackets in
Eq.~(\ref{ab::transport::emt})
denote averaging over
$P[\tilde x]$.
Overall,
Eq.~(\ref{ab::transport::emt})
defines
$\sigma_{\rm EMT}$
as an implicit functional of $P$ and the bilayer graphene dispersion.
Extracting
$\sigma_{\rm EMT}$
from
Eq.~(\ref{ab::transport::emt}), S.~Adam and M.\,D.~Stiles~\cite{Adam2010}
found that the conductivity
$\sigma_{\rm EMT}$
possesses interesting universal properties, which we now briefly review.

To discuss universality we need to introduce some notation. Obviously,
$\langle \tilde x \rangle = x$,
which may be treated as a mathematical definition of the averaged doping level $x$.
In addition, it is convenient to define the root-mean-square deviation
$x_{\rm rms}$
\begin{eqnarray}
x_{\rm rms}^2
=
\langle \tilde x^2 \rangle
-
x^2.
\end{eqnarray}
In the ensemble of puddles, the quantity
$x_{\rm rms}$
characterizes fluctuations of the local doping
$\tilde x$
around its average value $x$.
Disorder introduces additional characteristic scales into our system.
We can define the dimensionless doping
$\bar{x}$,
temperature
$\bar{T}$,
and conductivity
$\tilde \sigma_{\rm EMT}$
according to the equations
\begin{eqnarray}
\bar{x} = \frac{x}{x_{\rm rms}},
\\
\bar{T} = \frac{T}{\varepsilon_{\rm F} (x=x_{\rm rms})},
\\
\bar{\sigma}_{\rm EMT} = \frac{\sigma_{\rm EMT}}{e x_{\rm rms} \mu_c}.
\end{eqnarray}
In these formulas
$\varepsilon_{\rm F} (x=x_{\rm rms})$
is the doping-dependent Fermi energy of a clean sample with
$x=x_{\rm rms}$,
and the charge mobility is
$\mu_c$.
If
$x_{\rm rms}$
is sufficiently large, we need an additional dimensionless parameter
to characterize the deviation of the single-electron dispersion from a simple
parabola. Assuming that the distribution function
$P[\tilde x]$
is completely characterized by $x$ and
$x_{\rm rms}$
(for example, in
Ref.~\cite{Adam2010} the
distribution $P$ was taken to be a Gaussian), the authors demonstrated that
$\bar{\sigma}_{\rm EMT}$
is a universal function of
$\bar{x}$
and
$\bar{T}$.

\begin{figure}[t]
\begin{centering}
\includegraphics[width=10cm]{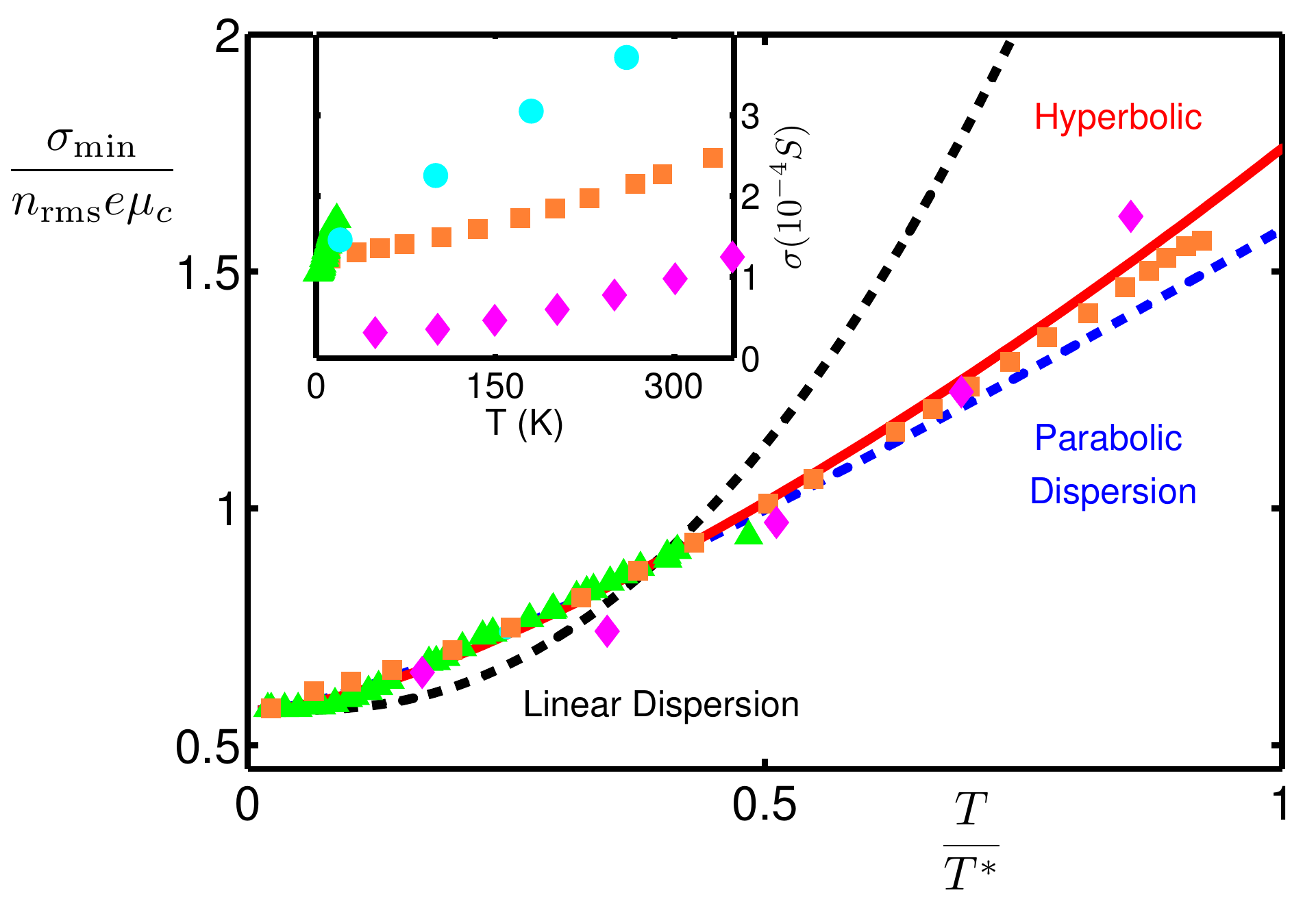}
\par\end{centering}
\caption{
Scaling property of the minimum conductivity of the AB bilayer graphene, as
described by the effective medium theory by S.~Adam and M.\,D.~Stiles~\cite{Adam2010}.
Points are experimental data, lines are calculations for different
band dispersions specified in the picture [linear dispersion $\varepsilon(k)\propto k$, parabolic
$\varepsilon(k)\propto k^2$, and hyperbolic $\varepsilon(k)\propto 1/(k+k_0)$].
The raw minimum conductivity data from different
papers~\cite{Feldman2009,Morozov2008,Zhu2009}
are shown in the inset. After the scaling procedure developed in
Ref.~\cite{Adam2010}
the experimental points collapse on a single theoretical (red solid) curve.
%
Reprinted figure with permission from
S.~Adam, M.D.~Stiles, 
``Temperature dependence of the diffusive conductivity of bilayer
graphene",
Phys. Rev. B, {\bf 82}, 075423 (2010).
Copyright 2010 by the American Physical Society.
\url{http://dx.doi.org/10.1103/PhysRevB.82.075423}
\label{figure::collapse}
}
\end{figure}

Once this universal function is determined, numerically or analytically,
the theoretical predictions can be compared against experimental data.
As a crude approximation, one can ignore the dependence of the mobility on
temperature. Under such a premise one can use experimental low-temperature
values for
$\mu_c$.
In principle, the quantity
$x_{\rm rms}$
could be measured experimentally. However, due to lack of the required data,
$x_{\rm rms}$
was treated in
Ref.~\cite{Adam2010}
as a fitting parameter. With these approximations, the experimental
conductivity reported in
Refs.~\cite{Feldman2009,Morozov2008,Zhu2009},
when plotted on a
($\bar{T}$, $\bar{\sigma}_{\rm EMT}$)
plane, collapsed on a single universal curve, see
Fig.~\ref{figure::collapse}.
The authors of
Ref.~\cite{Adam2010}
commented that ``in the appropriate limits" their results agree with the
findings of
Refs.~\cite{Nilsson2006_transp,Zhu2009}.

A different mechanism for the minimum conductivity was discussed in
Refs.~\cite{Katsnelson2006,Cserti2007,Snyman2007,Cserti_trig2007}.
These papers evaluated the conductivity of bilayer samples in the ballistic regime.
Assuming that the sample's length is smaller than its width, the
conductivity of
$e^2/(4\pi \hbar)$
per valley per spin projection was derived by M.\,I.~Katsnelson~\cite{Katsnelson2006}.
In such a regime, the charge is transported by evanescent modes to which the
Landauer-B\"uttiker formalism was applied.

Similar Landauer-B\"uttiker study was reported by I.~Snyman and C.\,W.\,J.~Beenakker in
Ref.~\cite{Snyman2007},
where the conductivity of a disorder-free bilayer sample was evaluated as a
function of the Fermi energy
$\varepsilon_{\rm F}$.
The conductivity demonstrated a non-monotonous dependence on the Fermi
energy. At the charge neutrality point,
$\varepsilon_{\rm F}=0$,
the conductivity value
$\sigma = 4e^2/(\pi^2 \hbar)$
of
Ref.~\cite{Snyman2007}
deviated from the minimum conductivity
$e^2/(\pi \hbar)$
of
Ref.~\cite{Katsnelson2006}.
The analysis of
Ref.~\cite{Snyman2007}
suggested that the discrepancy was due to nonidentical assumptions about
the metallic electrodes, which supply current to the sample. The authors of
Ref.~\cite{Snyman2007}
commented that their value for the
$\varepsilon_{\rm F}=0$
conductivity agreed with the result of
Ref.~\cite{Cserti2007},
where the Kubo formula was used.

Unlike
Refs.~\cite{Katsnelson2006,Cserti2007,Snyman2007},
which all neglected the trigonal warping, J.~Cserti
et~al.~\cite{Cserti_trig2007}
accounted for such a complication. The trigonal warping drastically changes
the minimum conductivity:
$\sigma = 12 e^2/(\pi^2 \hbar)$.

Recently, P. San-Jose
et~al.~\cite{san2014stacking}
presented a comprehensive experimental study of the minimal conductivity in
AB bilayers. Authors argued that the non-zero minimal conductivity in AB
bilayers can be explained by scattering on boundaries between domains with
different stacking order (AB and BA). They also presented experimental
evidence, reinforcing their interpretation, of reversible switching between
a metallic and an insulating regime in suspended bilayers when subjected to
thermal cycling or high current annealing.

\subsection{Transport through an AB bilayer in a transverse electric field}
\label{ab::transport_bias}

As we explained in
subsection~\ref{ab::energy_bands},
in a transverse electric field, the bilayer becomes an insulator with a
field-dependent gap in the single-electron spectrum. Clearly, if the
temperature is small, the presence of this gap suppresses charge transport.
In an ideal undoped sample with gap $\Delta$, one expects that when
$T<\Delta$
the conductivity would demonstrate the activation behavior
\begin{eqnarray}
\label{transp::activat_cond}
- \ln \sigma (T) \propto \Delta/T\,.
\end{eqnarray}
However, experimental data does not support such a simple picture. For
example, Ref.~\cite{Oostinga2008} reported that at zero doping the measured resistivity
$R(T)$
as a function of temperature is best fitted by the law
\begin{eqnarray}
\label{transport::variable_range}
\ln R (T) \sim (T_0/T)^{1/3},
\end{eqnarray}
where the characteristic temperature scale $T_0 \sim 0.5$--$0.8$\,K.
This behavior is observed for
$T < 5$\,K
and strong transverse fields. The exponent of $1/3$ in
Eq.~(\ref{transport::variable_range})
is consistent with variable-range hopping transport via in-gap impurity
states in two dimensions. The emergence of such states in a bilayer sample
with the gap was discussed theoretically in
Ref.~\cite{Nilsson2007}.

The authors of
Ref.~\cite{Zou2010}
used a more complicated fitting expression
\begin{eqnarray}
\label{transport::three_exp}
\frac{1}{R(T)}
=
\frac{1}{R_1\exp(E_1/T)}
+
\frac{1}{R_2\exp(E_2/T)}
+
\frac{1}{R_3\exp[(T_3/T)^{1/3}]}
\end{eqnarray}
to describe their experimental data. Here
$R_{1,2,3}$
are the fitting parameters with dimension of resistivity, while
$E_{1,2}$ and $T_3$ have the dimension of energy. This formula aims to describe three
transport channels: variable-range hopping and two activation channels.
One of the activation channels corresponds to the transport via the
carriers excited thermally above the transport gap
$E_1$.
Reference~\cite{Zou2010}
reported that
$E_1 > \Delta/2$
for small
$\Delta$.
When the gap $\Delta$ grows, however,
$E_1 \sim \Delta/2$.
The second activation channel is ``associated with the hopping conduction
between nearest neighboring impurity states". It is characterized by
$E_2$
and
$R_2$,
and dominates the transport in the interval between $50$\,K and $5$\,K. Below
$5$\,K, the variable-range hopping (characterized by $T_3$ and $R_3$)
is the strongest contribution to the conductivity.

Another experimental study,
Ref.~\cite{Taychatanapat2010},
investigated transport in the regime of large electric fields. The
strong electric field induces a large single-electron gap; thus, the
activation behavior becomes more pronounced. The authors estimated that the
field in their setup was three times stronger than the field reported in
Ref.~\cite{Oostinga2008}.
The resistivity data obtained in
Ref.~\cite{Taychatanapat2010}
was fitted by a formula similar to
Eq.~(\ref{transport::three_exp}).
However, the activation energy extracted from the transport data was two
orders of magnitude smaller than the single-electron gap, whose value was
obtained by infrared spectroscopy. The authors used an impurity band model
to qualitatively explain the data.

To address these obvious deviations from the simple
Eq.~(\ref{transp::activat_cond}), various theorists discussed several mechanisms. In Ref.~\cite{rossi2011}
the interplay between the smoothly-varying disorder potential and the gap
was studied. If the disorder is strong then charge puddles may appear. The
authors~\cite{rossi2011} identified several transport regimes for the proposed model. Among
them, they particularly focused on a case of strong disorder in which
the variation of the disorder potential is larger than the local value of
the gap. In such a situation, most of the sample surface is covered by
puddles, with narrow charge-free insulating strips separating puddles
of opposite polarities. Neither electron nor hole puddles form percolating
clusters. Thus, the gapped bilayer ``is expected to behave like a bad metal
in which transport is dominated by hopping processes between electron and
hole puddles that cover most of the sample". It was hypothesized that this
theoretical picture is relevant for the experimental situation of
Refs.~\cite{Oostinga2008,Zou2010,Taychatanapat2010}.

Reference~\cite{trushin2012}
investigated theoretically the conductivity of the gapped bilayer in the
presence of short-range disorder. The paper concluded that, due to a
certain interband coherence effect, specific to the AB bilayer, the
short-range disorder generates a finite conductivity even at
$T=0$.
As a function of temperature, the conductivity may demonstrate a
non-monotonous behavior. Related ideas were reviewed in
Section~\ref{ab::transport::minimum_cond}, in our discussion of Ref.~\cite{Trushin2010}.

It was demonstrated
experimentally~\cite{Alden2013}
that the AB bilayer sample may possesses certain topological stacking defects.
These defects were already mentioned in
Section~\ref{meso::ab::topological}:
they may host subgap topological
modes~\cite{vaezi_defect_topolog_channel2013}.
These modes give additional contributions to the conductivity at low
temperatures. Similar modes bound to the sample
edges~\cite{Li2011}
also affect the low-$T$ conductivity.

The phase diagram summarizing the stacking-dependent gap openings of biased bilayer graphene
was obtained by C. Park et~al.~\cite{Park} using density functional theory and perturbation analysis. The authors also identified high-density midgap states, localized on grain boundaries, even under a strong field, which can considerably reduce the overall transport gap.

Thus, it appears that both theory and experiment agree that for a
non-ideal bilayer sample there are several mechanisms contributing to
the low-temperature transport violating
Eq.~(\ref{transp::activat_cond}).
This indicates that creating field-effect transistors based on AB
bilayers may be a challenging task: in a gaped ``off'' state of the
transistor, unless we have a very clean sample, there are various transport
pathways, which allow the flow of the unwanted leak currents.

In recent paper~\cite{YZouIn}, transport measurements were reported
in low-doped biased AB bilayer
graphene in a perpendicular magnetic field and at low temperatures. Near charge neutrality point
the resistivity was found to increase exponentially with decreasing temperature from $10$\,K to $1.4$\,K
at $B = 14$\,T, ``indicating an emergence of transport energy gap induced by magnetic field''.

\section{Dynamic conductivity and optical spectra}\label{DynCond}
%

The dynamic (or optical) conductivity of single-layer graphene has been
extensively studied
theoretically~\cite{DyCondSLGT1,DyCondSLGT2,DyCondSLGT3,DyCondSLGT4,
DyCondSLGT5},
and experiments have largely verified the expected
behavior~\cite{DyCondSLGExp1,DyCondSLGExp2,DyCondSLGExp3,DyCondSLGExp4}.
The dynamic conductivity of graphene could elucidate features that would
demonstrate unique properties of this system and allow for the
identification of characteristic energy scales associated with the band
structure. Moreover, as the optical properties of graphene are of
considerable importance for technological applications, all variants of
graphene are also of potential interest and should be examined.

\subsection{Dynamic conductivity in AA-stacked bilayer graphene}\label{DynCondAA}

Let us analyze theoretically the dynamic conductivity of the AA and AB
bilayers within the framework of the Kubo formalism for the current-current
response function and the Green's function
approach~\cite{Mahan}.
To derive an expression for the dynamic conductivity it is useful to determine the
Green's function $\hat{G}(z)$ from $\hat{G}^{-1}(z) = z\hat{I} - \hat{H}$,
where $\hat{I}$ is the unit matrix and $\hat{H}$ is the system's
Hamiltonian. Following Ref.~\cite{NicCarb2} by C.\,J.~Tabert and E.\,J.~Nicol, we choose our Hamiltonian
in its simplest form, given by Eq.~\eqref{HaaK}. This Hamiltonian accounts for nearest-neighbor hopping only, and
neglects the possibility of symmetry-breaking and energy gap generation
(symmetry breaking and related issues in AA-stacked system are
discussed below, in
subsection~\ref{subsect::broken_aa}).
Keeping these assumptions in mind, one can
write~\cite{NicCarb2}
\begin{equation}\label{GreenAA}
\hat{G}^{-1}(z, {\bf k}) =\left(
             \begin{array}{cccc}
               z & -t_0 & tf(\mathbf{k}) & 0 \\
               -t_0 & z & 0 & tf(\mathbf{k})  \\
               tf^*(\mathbf{k}) & 0 & z & -t_0 \\
               0 & tf^*(\mathbf{k})  & -t_0 & z \\
             \end{array}
           \right),
\end{equation}
where $z=i\omega_n$,
with
$\omega_n=\pi T(2n+1)$
being the $n$th ($n=0,\pm 1,\pm 2, ...$)
fermionic Matsubara frequency. The finite frequency conductivity is
calculated through the standard procedure based on the Kubo
formula~\cite{Mahan}.
Within this formalism, the conductivity is expressed in terms of the retarded
current-current correlation function. The real part of the conductivity can
be written
as~\cite{NicCarb2,NicCarb1}
\begin{eqnarray}\label{condAA}
\sigma_{\alpha\beta}(\omega)
= \frac{N_fe^2}{2\omega}
\int\limits_{-\infty}^{+\infty}
{\frac{d\varepsilon}{2\pi}}
\left[f(\varepsilon-\mu)-f(\varepsilon+\hbar\omega-\mu)\right]
\int\limits_{V_{BZ}}
{\frac{d^2\mathbf{k}}{(2\pi)^2}
Tr\left[\hat{v}_\alpha\hat{A}(\varepsilon+\omega,
   \mathbf{k})\hat{v}_\beta\hat{A}(\varepsilon,\mathbf{k})\right]},
\end{eqnarray}
where, $N_f=4$ is a degeneracy factor, $\alpha$ and $\beta$ represent the
spatial coordinates $x,y,z$
(the axis $Oz$ is perpendicular to the graphene plane), $f(\varepsilon)=1/[\exp{(\varepsilon/T)}+1]$
is the Fermi function, $\mu$ is the chemical potential, and the velocity operator
is defined as
$\hbar {\hat v}_\alpha=\partial\hat{H}_\mathbf{k}/\partial k_\alpha$.
The spectral function
${\hat A} ( \varepsilon, {\bf k} )$
is related to the Green's function according to the usual spectral
representation
\begin{equation}\label{Griin_spectr}
{\hat G} ( z, {\bf k})
=
\int\limits_{-\infty}^{+\infty}
\frac{d\varepsilon}{2\pi}
\frac{{\hat A}(\varepsilon, {\bf k})}{z-\varepsilon}.
\end{equation}

If $T=0$, straightforward calculations in the continuum approximation
around the
${\bf K}$
point of the graphene Brillouin zone
($|\mathbf{v}(\mathbf{k})|^2=v_F^2$)
give the longitudinal
$\sigma_{xx}$
and transverse $\sigma_{zz}$
conductivities~\cite{NicCarb1}
\begin{eqnarray}\label{condAAxxzz}
\sigma_{xx}(\omega)& = &\frac{8e^2v_F^2}{\omega}
\int\limits_{\mu-\hbar\omega}^{\mu}{\frac{d\varepsilon}{2\pi}}
\int\limits_{V_{BZ}}{\frac{d^2\mathbf{k}}{(2\pi)^2}}
\Big[A_{11}(\varepsilon+\hbar\omega)A_{11}(\varepsilon)+A_{12}(\varepsilon+\hbar\omega)A_{12}(\varepsilon)\Big]\,,\\
\nonumber
\sigma_{zz}(\omega)& =& \frac{8e^2v_F^2}{\omega}
\int\limits_{\mu-\hbar\omega}^{\mu}{\frac{d\varepsilon}{2\pi}}
\int\limits_{V_{BZ}}{\frac{d^2\mathbf{k}}{(2\pi)^2}}
\Big[A_{11}(\varepsilon+\hbar\omega)A_{11}(\varepsilon)+A^*_{13}(\varepsilon+\hbar\omega)A_{13}(\varepsilon)\\
&&-A_{12}(\varepsilon+\hbar\omega)A_{12}(\varepsilon)-A^*_{14}(\varepsilon+\hbar\omega)A_{14}(\varepsilon)\Big]\,.
\end{eqnarray}
The components of the spectral function are equal to
\begin{eqnarray}\label{condAA_Aij}
\nonumber
\! A_{11} \!\!\!&=&\!\!\!\frac{\pi}{2}
\left[\delta(\varepsilon\!+\!\varepsilon^{(3)}_{0\mathbf{k}})\!
+\!
\delta(\varepsilon\!-\!\varepsilon^{(3)}_{0\mathbf{k}})
\!+\!
\delta(\varepsilon\!+\!\varepsilon^{(4)}_{0\mathbf{k}})
\!+\!
\delta(\varepsilon\!-\!\varepsilon^{(4)}_{0\mathbf{k}})
\right], \\
\nonumber
\! A_{12} \!\!\!&=&\!\!\!
\frac{\pi}{2}
\left[
	\delta(\varepsilon\!+\!\varepsilon^{(3)}_{0\mathbf{k}})
	\!-\!
	\delta(\varepsilon\!-\!\varepsilon^{(3)}_{0\mathbf{k}})
	\!-\!
	\delta(\varepsilon\!+\!\varepsilon^{(4)}_{0\mathbf{k}})
	\!+\!
	\delta(\varepsilon\!-\!\varepsilon^{(4)}_{0\mathbf{k}})
\right],
\\
\nonumber
\! A_{13} \!\!\!&=&\!\!\! \frac{\pi f(\mathbf{k})}{2|f(\mathbf{k})|}
\left[
	\delta(\varepsilon\!-\!\varepsilon^{(3)}_{0\mathbf{k}})
	\!-\!
	\delta(\varepsilon\!+\!\varepsilon^{(3)}_{0\mathbf{k}})
	\!-\!
	\delta(\varepsilon\!+\!\varepsilon^{(4)}_{0\mathbf{k}})
	\!+\!
	\delta(\varepsilon\!-\!\varepsilon^{(4)}_{0\mathbf{k}})
\right],
\\
\! A_{14} \!\!\!&=&\!\!\! \frac{\pi f(\mathbf{k})}{2|f(\mathbf{k})|}
\left[
	\delta(\varepsilon\!+\!\varepsilon^{(4)}_{0\mathbf{k}})
	\!+\!
	\delta(\varepsilon\!-\!\varepsilon^{(4)}_{0\mathbf{k}})
	\!-\!
	\delta(\varepsilon\!+\!\varepsilon^{(3)}_{0\mathbf{k}})
	\!-\!
	\delta(\varepsilon\!-\!\varepsilon^{(3)}_{0\mathbf{k}})
\right],
\end{eqnarray}
where the energy bands
$\varepsilon^{(3,4)}_{0\mathbf{k}}$
are defined by
Eqs.~\eqref{aaBands}.

Following Ref.~\cite{NicCarb1} we can perform integration in
Eqs.~\eqref{condAAxxzz} and obtain analytical formulas for the real parts
of the dynamic conductivities
\begin{eqnarray}\label{condAAxxzz_anal}
\nonumber
\sigma_{xx}&=&\frac{e^2}{4\hbar}\big[8\delta(\hbar\omega)\textrm{max}(\mu,t_0)+
\theta(\hbar\omega-2|\mu-t_0|)+\theta(\hbar\omega-2|\mu+t_0|)\big]\,,\\ \sigma_{zz}&=&\frac{e^2t_0c_0^2}{\hbar^3v_F^2}\,\delta(\hbar\omega-2t_0)\left[(t_0-\mu)^2\theta(t_0-\mu)+2t_0\mu\right]\,,
\end{eqnarray}
where $\theta(x)$ is the Heaviside step-function. To obtain the
conductivity of single layer graphene we shall put
$t_0=0$
and take a half of the obtained value of $\sigma_{xx}$. The curve
$\sigma_{xx}(\omega)$
for the AA-stacked bilayer graphene has a zero frequency Drude peak for any
doping and two steps at
$\omega=|\mu\pm t_0|/\hbar$,
due to the existence of two conductivity bands. In contrast, the conductivity
$\sigma_{xx}(\omega)$
for the single-layer graphene is a smooth function of $\omega$ in the case of
zero doping,
$\mu=0$.
The Drude peak and a step at
$\omega=|\mu|/\hbar$ in single-layer graphene
arise only if
$\mu\neq0$
and the DOS at the Fermi level becomes non-zero. The Kramers-Kronig
relations allows one to calculate the imaginary parts of the
conductivities~\cite{NicCarb1}.
The bias voltage $V$ acts to renormalize the interlayer hopping parameter $t_0$:
for finite voltage it becomes
$$
t'_0=\sqrt{t_0^2+(e V)^2/4}\,.
$$
Thus, the bias voltage does not introduce new features
into the
conductivity~\cite{NicCarb1}.
Note that the results reviewed here need further generalization, because they
are valid only if
$T=0$,
and the possible existence of a gap in a single-particle spectrum due to
spontaneous symmetry breaking
(see
subsection~\ref{subsect::broken_aa})
is disregarded.

\subsection{Dynamic conductivity in AB-stacked bilayer graphene}
\label{DynCondAB}

We start our discussion of the dynamic (or optical) conductivity of AB
bilayer graphene using a theoretical approach similar to the one employed
in the previous subsection for AA system. As the simplest approximation, we
neglect trigonal warping (hopping amplitude
$t_3$)
and electron-hole asymmetry (hopping amplitude
$t_4$)
as it was done in
Refs.~\cite{NicCarb2,McCann2007,Nilsson2008,Abergel2007,Benfatto}.
This model successfully explains the major features of frequency-dependence
of
the conductivity as well as its dependence on the gate voltage $V$. In
this approximation, the conduction and valence bands are symmetric. In
the absence of an electrostatic potential difference between the layers,
the two conduction (valence) bands have the same shape and are shifted by
$t_0$. Except for the range of very small momenta $k$, where the trigonal
warping is of importance, their shape remains nearly identical even in the
presence of a finite $V$. As a result, there is a high optical density of
states for transitions between the two pairs of bands at frequency
$\omega_m=t_0/\hbar$, which gives rise to a sharp peak in the real part of
the conductivity at
$\omega = \omega_m$.
Taking for estimate $t_0=0.40$\,eV, we obtain
that $\omega_m=3200$\,cm$^{-1}$ or $96$\,THz. Other transitions give more
gradually-varying contributions to
${\rm Re}\,\sigma$,
eventually leading to asymptotic ``universal'' value at high frequency equal to $e^2/(2\hbar)$,
which is twice the value for the monolayer graphene. 
Finally, in real graphene systems the conductivity features are never sharp
because of finite lifetime due to, e.g., disorder scattering. This
broadens the peaks and can also merge together several features that are
close in energy.

We use the tight-binding Hamiltonian given by the sum of Hamiltonians in
Eqs.~(\ref{HV0})
and~\eqref{ab::H}
to account for the effect of the gate voltage, which opens the gap in the
spectrum of the AB bilayer. The equation for the corresponding Green's
function can be expressed in the following
form~\cite{NicCarb2}
\begin{equation}\label{GreenAB}
\hat{G}^{-1}(z) =\left(
             \begin{array}{cccc}
               z+eV/2 & 0 & tf(\mathbf{k}) & 0 \\
               0 & z-eV/2 & -t_0 & tf(\mathbf{k})  \\
               tf^*(\mathbf{k}) & -t_0 & z+eV/2 & 0 \\
               0 & tf^*(\mathbf{k})  & 0 & z-eV/2 \\
             \end{array}
           \right).
\end{equation}
Applying Eqs.~\eqref{condAA} and \eqref{Griin_spectr} in the continuum
approximation around the Dirac
${\bf K}$
point, we derive the formula for the longitudinal
conductivity~\cite{NicCarb2}
\begin{eqnarray}
\label{condABxx}
\sigma_{xx}&=&\frac{2e^2v_F^2}{\omega}\!\!
\int\limits_{-\infty}^{\infty}
{\frac{d\varepsilon}{2\pi}}\!\!
\int\limits_{V_{BZ}}\!\!{\frac{d^2\mathbf{k}}{(2\pi)^2}}
\Big\{A_{11}(\varepsilon,V)A_{33}(\varepsilon+\hbar\omega,V)
+
A_{33}(\varepsilon,V)A_{11}(\varepsilon+\hbar\omega,V)
\nonumber\\
&&
+
A_{11}(\varepsilon,-V)A_{33}(\varepsilon+\hbar\omega,-V)
+
A_{33}(\varepsilon,-V)A_{11}(\varepsilon+\hbar\omega,-V)
\nonumber\\
&&
+2\left[
	A_{12}(\varepsilon,V)A^*_{12}(\varepsilon+\hbar\omega,-V)
	+
	A^*_{12}(\varepsilon,-V)A_{12}(\varepsilon+\hbar\omega,V)
\right]\Big\}\,.
\end{eqnarray}
If we take the limit of this expression for
$t_0=V=\mu=0$,
we find a constant equal to
$2e^2/(4\hbar)$,
which is twice the value of the single-layer graphene conductivity. The
spectral functions
$A_{ij}$
can be written
explicitly~\cite{NicCarb2}.
However, these formulas are too cumbersome, and we do not present them here.

\begin{figure}[t]
\centering
  \includegraphics[width=0.9\columnwidth]{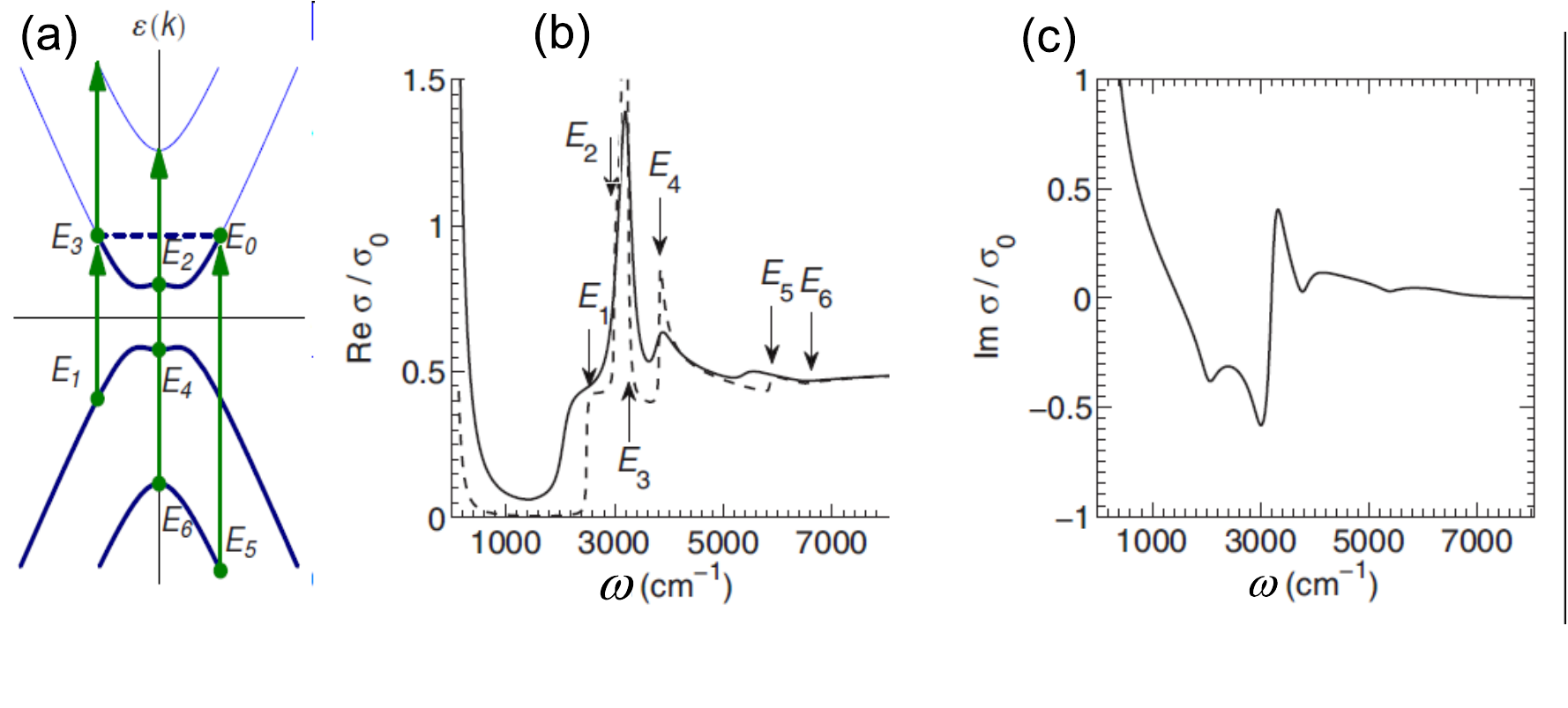}
  \caption{(Color online) The calculated dynamic conductivity of AB
bilayer graphene as a function of the
frequency~\cite{Zhang2008}.
(a) Examples of the allowed optical transitions for the chemical potential
indicated by the dashed line. Occupied states are shown by the thicker
lines. The dots and the arrows mark the initial and the final states,
respectively, of the transitions that produce features at frequencies
$E_j$,
$j=1,2,...,6$
in panel (b).
(b) Real and (c) imaginary parts of the conductivity in units of
$\sigma_0=e^2/\hbar$
for the gate voltage
$V=-100$\,V.
The solid curves are for broadening
$\Gamma=0.02t_0$.
The dashed curve is for
$\Gamma=0.002t_0$.
Reprinted figure with permission from
L.M.~Zhang et al.,
Phys. Rev. B, {\bf 78}, 235408 (2008).
Copyright 2008 by the American Physical Society.
\url{http://dx.doi.org/10.1103/PhysRevB.78.235408}
\label{DynCond_ABfig}
}
\end{figure}

\begin{figure}[t]
  \includegraphics[width=0.99\columnwidth]{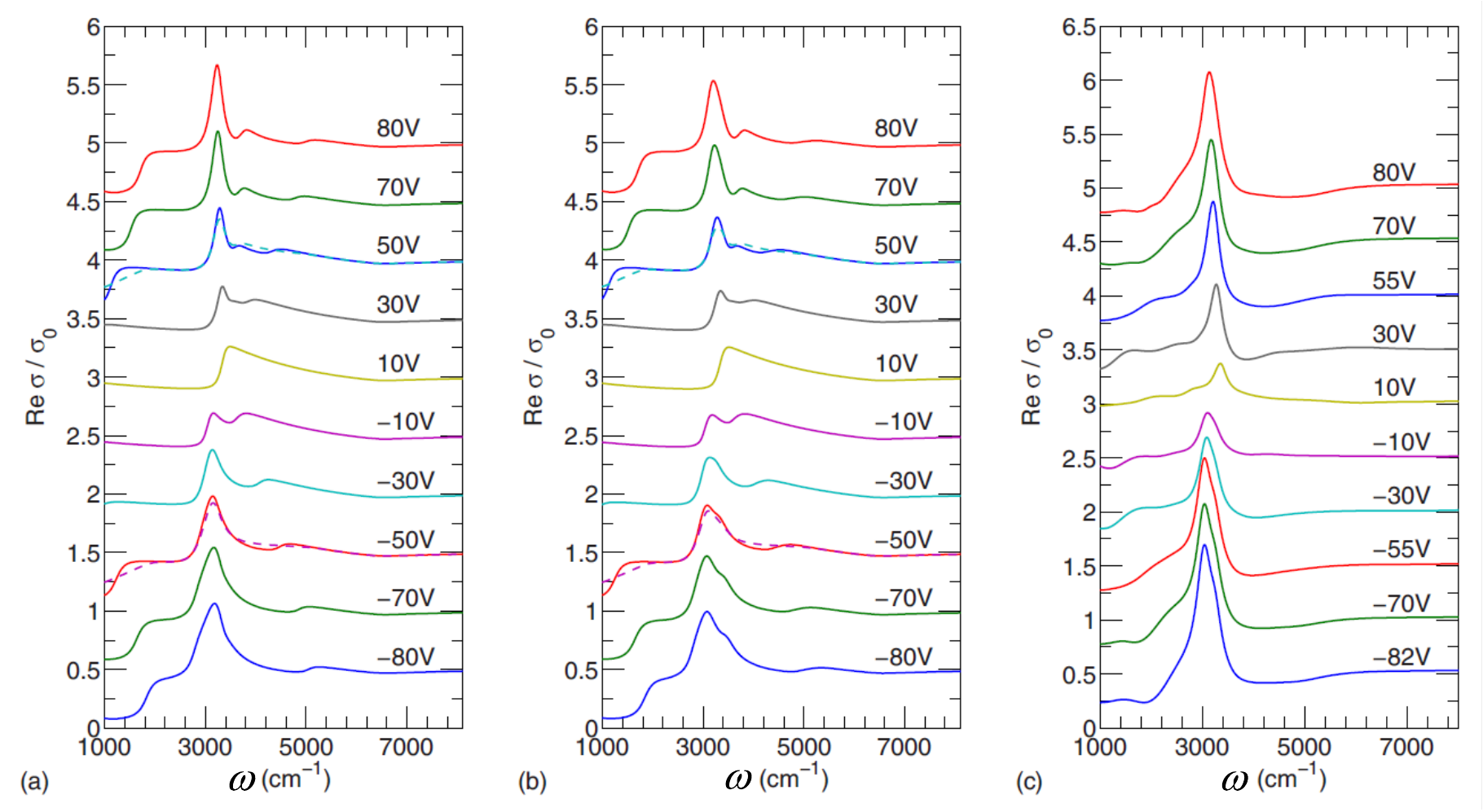}
  \centering
  \caption{(Color online) The dynamic conductivity of AB bilayer
graphene as a function of the
frequency~\cite{Zhang2008}.
Panels (a) and (b) present theoretical, and (c) experimental results for
the conductivity
${\rm Re}\,\sigma(\omega)$.
The deviation $V$ of the gate voltage from the
charge-neutrality point is indicated next to each curve. For clarity, the
curves are offset vertically by $0.5\sigma_0$ from one another, where $\sigma_0=e^2/\hbar$. For the
plots in panel~(b) the parameters are chosen as follows
$t=3$\,eV, $t_0=0.4$\,eV, $t_3=0.3$\,eV, and $t_4=0.15$\,eV.
In panel~(a) the parameters are the same, except that
$t_3=0$.
The dashed curves superimposed on the
$V= +50$ Volts  ($-50$ Volts)
traces in (a) and (b) are the arithmetic means of all the positive
(negative) $V$ curves.
Reprinted figure with permission from
L.M.~Zhang et al.,
Phys. Rev. B, {\bf 78}, 235408 (2008).
Copyright 2008 by the American Physical Society.
\url{http://dx.doi.org/10.1103/PhysRevB.78.235408}
\label{DynCondExpfig}
}
\end{figure}

The analytical expression for
$\sigma(\omega)$
can be derived if the temperature is zero and the sample is unbiased
($T=0$, $V=0$)~\cite{Abergel,NicCarb2}
\begin{eqnarray}
\sigma_{xx}&=&\frac{e^2}{8\hbar}\left\{
\left[\frac{\hbar\omega+2t_0}{\hbar\omega+t_0}+\frac{\hbar\omega-2t_0}{\hbar\omega-t_0}\theta(\hbar\omega-2t_0)\right]
\theta(\hbar\omega-2\mu)\right.\\\nonumber
&&\left.
+\frac{t_0^2}{\hbar^2\omega^2}\big[\theta(\hbar\omega-2\mu-t_0)+\theta(\hbar\omega-2\mu+t_0)\big]\theta(\hbar\omega-t_0)
+a(\mu)\delta(\hbar\omega)+b(\mu)\delta(\hbar\omega-t_0)\right\}\,,
\end{eqnarray}
where the functions $a(\mu)$ and $b(\mu)$ are defined according to the equations
\begin{eqnarray}
\nonumber
  a(\mu) \!\! &=&\!\!\frac{4\mu(\mu+t_0)}{2\mu+t_0}+\frac{4\mu(\mu-t_0)}{2\mu-t_0}\theta(\mu-t_0)\,,
  \qquad  b(\mu)= \frac{t_0}{2}\left[\ln{\frac{2\mu+t_0}{t_0}}-\ln{\frac{2\mu-t_0}{t_0}}\right]\theta(\mu-t_0)\,.
\end{eqnarray}
In the general case, the function $\sigma(\omega)$ can be calculated
numerically. Note also that we derive here only the real part of the
conductivity; the imaginary part can be calculated using a similar
approach~\cite{Zhang2008,Falkovsky1}.

A calculated dependence of the real and imaginary parts of the conductivity
of AB bilayer
graphene~\cite{Zhang2008}
is shown in
Fig.~\ref{DynCond_ABfig}.
The calculations were performed with a phenomenological broadening constant
$\Gamma$ (i.e., to mimic the effects of the impurity scattering, the
delta-function
$\delta(x)$
is replaced by
$(\Gamma/\pi)/(x^2+\Gamma^2)$~\cite{Zhang2008,NicCarb2}).
The real part of the conductivity
${\rm Re}\, \sigma(\omega)$
has a Drude peak at
$\omega=0$,
a characteristic high peak at
$\omega=\omega_m$,
and several steps and peaks correspond to the interband optical
transitions shown in
Fig.~\ref{DynCond_ABfig}(a).

Recently, a theoretical study of infrared absorption of graphene samples,
both single-layer and bilayer, on hexagonal boron nitride substrate was
published by D.S.L.~Abergel and
M.~Mucha-Kruczy{\'{n}}ski~\cite{optics2015}.

An experimental study of the infrared conductivity, transmission, and
reflection of AB gated bilayer graphene has been performed by L.\,M.~Zhang
et~al. in
Ref.~\cite{Zhang2008}.
In addition, this paper presented theoretical calculations of the real and
imaginary parts of the dynamic conductivity. The authors took into account
both
$t_3$,
and
$t_4$
hopping amplitudes. Such an approach allowed not only a more complete
interpretation of the experimental observations, but also to derive from
the experimental data the values of
$t_0=0.4$\,eV
and
$t_4=0.15$\,eV,
and obtain the estimates $t\sim3$\,eV and $t_3\sim0.3$\,eV.
For illustration, the results of calculations and measurements of the real
part of the dynamic conductivity taken from
Ref.~\cite{Zhang2008}
are shown in
Fig.~\ref{DynCondExpfig}.
The correspondence of the measurements and calculations is satisfactory. As
it is seen from the comparison of the data in
Figs.~\ref{DynCondExpfig} (a)
and (b), the effect of the trigonal warping (that is, $t_3$) on the
conductivity is practically negligible.

\section{Broken symmetry phases}\label{interaction}

The electron-electron coupling in carbon systems can be rather large.
However, the single-electron approximation is a good theoretical approach for
solving many problems in the case of single-layer graphene.  Indeed,
the most striking effect of the strong Coulomb repulsion between charge
carriers is the opening of the insulating gap in the electronic spectrum.
The DOS at the Fermi level in undoped single-layer graphene is zero
(see Fig.~\ref{SlgDosfig}). Thus, the insulating gap should arise in such a
system only if the electron-electron coupling is larger than some threshold
value~\cite{single_layer_afm2009,polikarpov2013}.
It is commonly accepted that for single-layer graphene this coupling is
close to, but lower than the critical value. In undoped bilayer graphene,
the DOS is non-zero at the Fermi level. Thus, the correlation effects are
much more pronounced. Moreover, the hole and electron Fermi surfaces (Fermi
arcs) of undoped AA bilayer graphene are perfectly nested (see
Fig.~\ref{AaDOSfig}).
The latter feature means that at zero temperature a symmetry breaks and the
gap opens at an arbitrary weak electron-electron repulsion.

For graphene systems, a specific difficulty arises when we
wish to take into account the electron-electron coupling. These materials
are two-dimensional and the electrons interact through
three-dimensional electric fields. As a result, the long range part of the
Coulomb potential is not strongly screened in graphene, as it usually
occurs in 3D materials. The second problem is the strong effect of the
substrate on the electron coupling, especially on the long-range part of
this coupling.

To avoid serious mathematical difficulties associated with the description
of the long-range Coulomb interaction, the majority of researchers consider
either only on-site Coulomb repulsion
$U_0$,
or work in the continuum-media (or long-wave) limit. The latter approach is
the most convenient when the effect of the substrate is of importance.
However, the nearest-neighbor in-plane repulsion
$U_{01}$,
or inter-plane
$U_{11}$
terms are smaller, but comparable to
$U_0$.
Thus, to obtain realistic results one has to take these terms in
consideration either directly, or by an appropriate renormalization of
$U_0$.
Note also that the continuum-media approach is valid only to study 
the low-lying electronic states, and could be inappropriate in the case of
high doping or for the calculation of thermodynamic properties.

The values of the Coulomb potentials are usually calculated~\cite{Wehling} using some
{\it ab initio}
methods or estimated from the comparison of the theoretical results with
the experimental findings. In
table~\ref{CoulPot}
we present data on the Coulomb potentials in graphene taken from
Ref.~\cite{Wehling}.
In this table
$U_{02}$
is the in-plane next-nearest-neighbors repulsion potential, other $U$'s are
defined in the text above.

\begin{table}[t]
\begin{center}
\begin{tabular*}{\textwidth}{>{}m{3cm}>{\centering}m{3cm}>
{\centering}m{3cm}>{\centering}m{3cm}>{\centering}m{3cm}}
\hline
\hline
{\vspace{1.5mm}Reference\vspace{1mm}} &
{\vspace{1.5mm}$U_0$ [eV] \vspace{1mm}} &
{\vspace{1.5mm}$U_{01}$ [eV]\vspace{1mm}} &
{\vspace{1.5mm}$U_{02}$ [eV]\vspace{1mm}} &
{\vspace{1.5mm}Method\vspace{1mm}}
\tabularnewline
\hline
{\vspace{1.5mm}\cite{Wehling}\vspace{1mm}} &
{\vspace{1.5mm} 9.3 \vspace{1mm}} &
{\vspace{1.5mm} 5.5 \vspace{1mm}} &
{\vspace{1.5mm} 4.1 \vspace{1mm}} &
{\vspace{1.5mm} cRPA \vspace{1mm}}
\tabularnewline
\hline
\hline
\end{tabular*}
\end{center}
\caption{Coulomb potentials in graphene. Parameters $U_0$, $U_{01}$, and $U_{02}$ are the on-site, nearest-neighbor, and next-nearest-neighbor in-plane Coulomb repulsions, respectively. Here cRPA stands for the constrained random phase approximation.}\label{CoulPot}
\end{table}

\subsection{Low-temperature broken symmetry phases of AA bilayer graphene}
\label{subsect::broken_aa}

In the AA-stacked bilayer graphene the effect of the electron-electron
correlations is the most pronounced. The presence of two bands with
identical Fermi surfaces makes the AA-stacked bilayer graphene unstable
with respect to spontaneous symmetry breaking. This instability can open a
gap in the electronic spectrum to decrease the free energy of the system
for arbitrary small interaction. Since there is no experimental data on the
electronic properties of the AA graphene bilayer, we below review
only theoretical results.

\subsubsection{Electron correlations: symmetry analysis}
\label{spectraAA_symmetry}

The specific symmetries of the Hamiltonian can be used to narrow the
choices of the possible symmetry breakings and order parameters. Such an
analysis has been performed at the mean-field level by A.\,L.~Rakhmanov et~al. in
Ref.~\cite{PrlOur}.
In the mean-field approximation, the interaction operator
$$
\hat{H}_{\text{int}}\propto\sum\psi^\dag_\alpha\psi^{\phantom{\dag}}_\beta\psi^\dag_\gamma\psi^{\phantom{\dag}}_\delta
$$
is replaced by a single-particle operator
$$
\hat{H}^{\text{MF}}_{\text{int}}\propto\sum\langle\psi^\dag_\alpha\psi^{\phantom{\dag}}_\beta\rangle\psi^\dag_\gamma\psi^{\phantom{\dag}}_\delta\,,
$$
where the average
$\langle\psi^\dag_\alpha\psi^{\phantom{\dag}}_\beta\rangle$ represents
different types of possible order parameters. To be at least metastable,
the order parameter must open a gap at the Fermi level. The authors of
Ref.~\cite{PrlOur} started from a tight-binding Hamiltonian similar to
Eq.~\eqref{HaaK1} (with $t'=0$), but written in the basis
$$
\Psi_{\mathbf{k}\sigma}=(a_{\mathbf{k}1\sigma},a_{\mathbf{k}2\sigma},
e^{-i\varphi_{\mathbf{k}}}b_{\mathbf{k}1\sigma},e^{-i\varphi_{\mathbf{k}}}b_{\mathbf{k}2\sigma})^T\,,
$$
where $\varphi_{\mathbf{k}}=\arg f(\mathbf{k})$. In this basis, the matrix Hamiltonian~\eqref{HaaK1} becomes real and equals
\begin{equation}\label{HaaK2}
\hat{H}_{\mathbf{k}}=\left(
                     \begin{array}{cccc}
                       0 & t_0 & -t\zeta_{\mathbf{k}} & t_g\zeta_{\mathbf{k}} \\
                       t_0 & 0 & t_g\zeta_{\mathbf{k}} & -t\zeta_{\mathbf{k}} \\
                       -t\zeta_{\mathbf{k}} & t_g\zeta_{\mathbf{k}} & 0 & t_0 \\
                       t_g\zeta_{\mathbf{k}} & -t\zeta_{\mathbf{k}} & t_0 & t0 \\
                     \end{array}
                   \right),
\end{equation}
where $\zeta_{\mathbf{k}}=|f(\mathbf{k})|$. The Hamiltonian is invariant
under the transpositions of both the layers and the sublattices. The
authors showed that, to open a gap at the Fermi level for arbitrary small
interaction, it is necessary to break both of these symmetries. The general
expression for the mean-field interaction Hamiltonian
$\hat{H}^{\text{MF}}_{\text{int}}$,
which breakes both symmetries and induces a gap at the Fermi level,
is
\begin{equation}\label{genGap}
\hat{H}^{\text{MF}}_{\text{int}}= \sum_{\mathbf{k}\sigma}\psi^\dag_{\mathbf{k}\sigma}\delta\hat{H}_{\mathbf{k}\sigma}\psi^{\phantom{\dag}}_{\mathbf{k}\sigma},\qquad
\delta\hat{H}_{\mathbf{k}\sigma} =\left(
                                    \begin{array}{cccc}
                                      \Delta^{zz}_{\mathbf{k}\sigma} & -i\Delta^{yz}_{\mathbf{k}\sigma} & -i\Delta^{zy}_{\mathbf{k}\sigma} & -\Delta^{yy}_{\mathbf{k}\sigma} \\
                                      i\Delta^{yz}_{\mathbf{k}\sigma} & -\Delta^{zz}_{\mathbf{k}\sigma} & \Delta^{yy}_{\mathbf{k}\sigma} & i\Delta^{zy}_{\mathbf{k}\sigma} \\
                                      i\Delta^{zy}_{\mathbf{k}\sigma} & \Delta^{yy}_{\mathbf{k}\sigma} & -\Delta^{zz}_{\mathbf{k}\sigma} & i\Delta^{yz}_{\mathbf{k}\sigma} \\
                                      -\Delta^{yy}_{\mathbf{k}\sigma} & -i\Delta^{zy}_{\mathbf{k}\sigma} & -i\Delta^{yz}_{\mathbf{k}\sigma} & \Delta^{zz}_{\mathbf{k}\sigma} \\
                                    \end{array}
                                  \right),
\end{equation}
where the real-valued functions $\Delta^{yy}_{\mathbf{k}\sigma}$, $\Delta^{zz}_{\mathbf{k}\sigma}$, $\Delta^{yz}_{\mathbf{k}\sigma}$, and $\Delta^{zy}_{\mathbf{k}\sigma}$, are different components of the multi-component order parameter.

To calculate the renormalized spectra near the Fermi level, one should diagonalize the matrix $\hat{H}_{\mathbf{k}\sigma}=\hat{H}_\mathbf{k}+\delta
\hat{H}_{\mathbf{k}\sigma}$. This gives
\begin{equation}\label{aaRenormEn}
E_{\mathbf{k}\sigma}^{(2,3)}=\frac{\varepsilon_{\mathbf{k}}^{(2)}+\varepsilon_{\mathbf{k}}^{(3)}}{2}\mp \sqrt{\!\left(\frac{\varepsilon_{\mathbf{k}}^{(2)}-\varepsilon_{\mathbf{k}}^{(3)}}{2}\right)^2  \!+\left|\Delta_{\mathbf{k}\sigma}\right|^2},
\end{equation}
where $\varepsilon_{\mathbf{k}}^{(2,3)}$ are given by Eq.~\eqref{aaBands1} (these two bands cross the Fermi level when interactions are neglected), and
\begin{equation}\label{ElemaaMatrix}
\Delta_{\mathbf{k}\sigma}=\Delta^{zz}_{\mathbf{k}\sigma}+\Delta^{yy}_{\mathbf{k}\sigma}+
i\left(\Delta^{zy}_{\mathbf{k}\sigma}+\Delta^{yz}_{\mathbf{k}\sigma}\right).
\end{equation}
The gap between renormalized bands is equal to 
$$
\Delta_0=2 \textrm{min}_{\mathbf{k}}\left|\Delta_{\mathbf{k}\sigma}\right|\,.
$$

The order parameters in Eq.~\eqref{ElemaaMatrix} are found by minimization
of the grand potential. Which type of order corresponds to the ground state
depends on the features of the interaction Hamiltonian. The parameter
$\Delta^{zz}_{\mathbf{k}\sigma}$ can be related to the so called $G$-type
antiferromagnetic (AFM) order (all nearest neighboring spins in the lattice
are antiparallel to each other). It is controlled by the on-site Coulomb
repulsion energy. The order parameter $\Delta^{yy}_{\mathbf{k}\sigma}$ can
be attributed to the instability toward the homogeneous shift of one
graphene layer with respect to another one. To consider this type of order
one has to take into account the electron-phonon interaction. The
order parameters $\Delta^{zy}_{\mathbf{k}\sigma}$ and
$\Delta^{yz}_{\mathbf{k}\sigma}$ can be viewed as excitons, which produce
circular currents flowing inside and between the layers, respectively.
These parameters are controlled by the in-plane and out-of-plane nearest
neighbor Coulomb repulsion energies.
All these types of order compete against each other. The ground state of
the system corresponds to antiferromagnetism, when the on-site Coulomb
repulsion energy $U_0$ is the strongest interaction constant.

\subsubsection{Electron correlations and antiferromagnetic order}
\label{spectraAA_AFM}

The AFM order in AA-stacked bilayer graphene was considered first in
Ref.~\cite{PrlOur},
and then studied in detail in
Refs.~\cite{PrbOur,PrbROur}.
The authors used a Hubbard-like model, where the single-particle part of
the Hamiltonian is given by
Eq.~\eqref{HaaK},
while the Hubbard interaction term has a form
\begin{equation}\label{HaaintU0}
\hat{H}_{\text{int}}
=
\frac{U_0}{2}
\sum_{\mathbf{i}\alpha a\sigma}
	\left(n_{\mathbf{i}\alpha a \sigma}-\frac{1}{2}\right)\!\!
	\left(n_{\mathbf{i}\alpha a \bar{\sigma}}-\frac{1}{2}\right).
\end{equation}
Longer range Coulomb interactions were neglected. It is commonly
accepted that in this limit one should use the estimate for $U_0$ smaller
than that predicted by the
{\it ab initio}
(see Table~\ref{CoulPot})
calculations~\cite{EffCoulomb}. The authors used the characteristic values
$U_0=5$--$7$\,eV.

\begin{figure}[t]
\includegraphics[width=0.6\columnwidth]{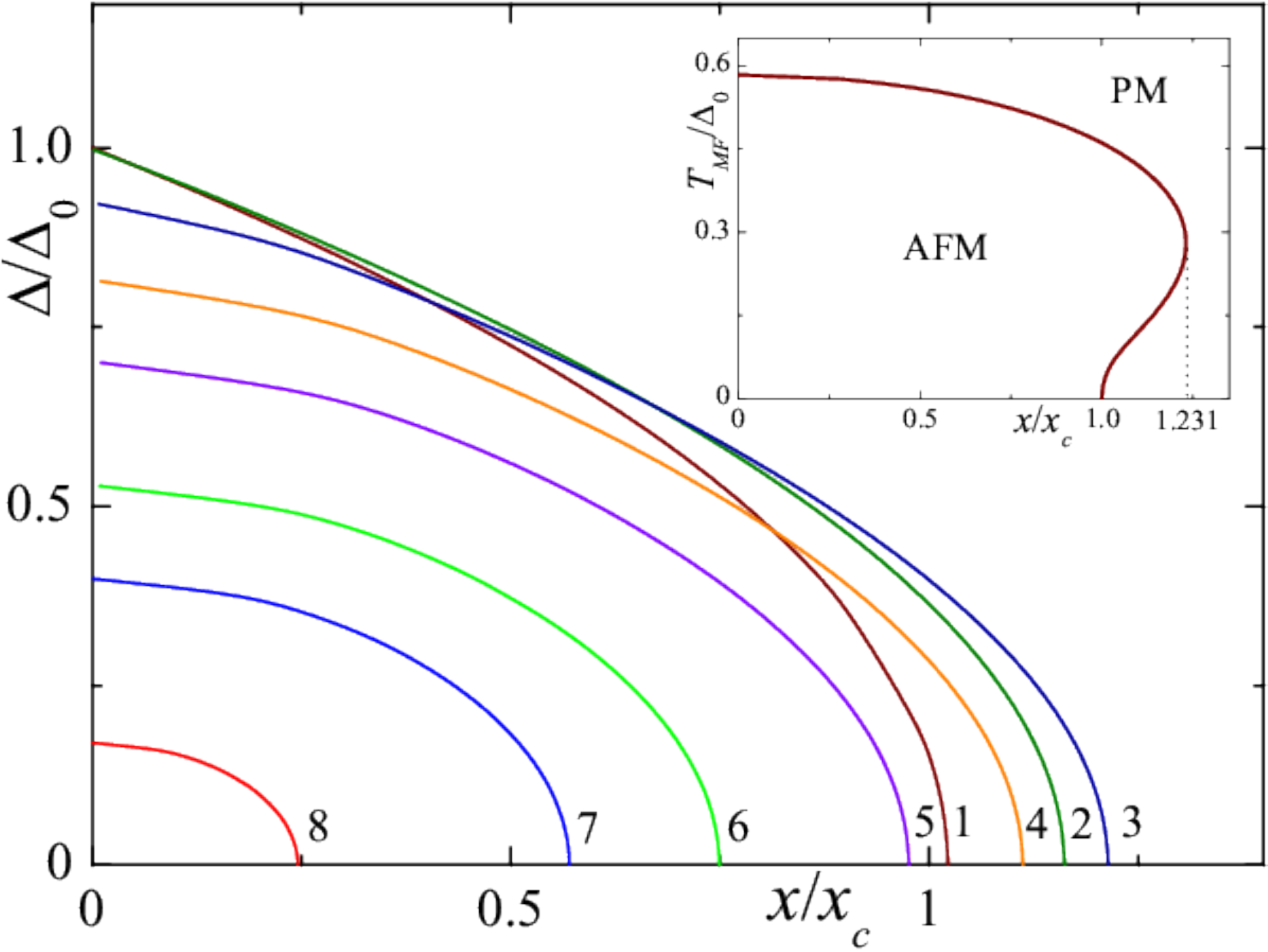}
\centering
\caption{(Color online) Dependence of the AFM gap $\Delta(x,T)$ on doping $x$, calculated for $U_0 = 5.5$\,eV and different values of the normalized temperature $T/\Delta_0$, from Ref.~\cite{PrbOur}: (1) $T/\Delta_0 = 0.06$, (2) $T/\Delta_0 = 0.17$, (3) $T/\Delta_0 = 0.33$, (4) $T/\Delta_0 = 0.41$, (5) $T/\Delta_0 = 0.47$, (6) $T/\Delta_0 = 0.52$, (7) $T/\Delta_0 = 0.55$, and (8) $T/\Delta_0 = 0.58$; note that $\Delta_0=\Delta(0,0)$ is the value of the AFM gap at zero temperature and doping. Inset: The dependence of the mean-field transition temperature $T_{\text{MF}}$ on doping $x$. The reentrance from the PM to AFM state exists in the doping range $x_c < x < 1.231x_c$; $x_c = 0.128$ and $\Delta_0 = 0.124$\,eV. PM means paramagnetic phase, and AFM means antiferromagnetic phase.\label{AAGap_dopfig}}
\end{figure}

In Refs.~\cite{PrlOur,PrbOur,PrbROur}
it was shown that the on-site Coulomb repulsion stabilizes so-called G-type
AFM order. In such a state the localized spin at any given site is
antiparallel to spins at nearest-neighbor sites. Other types of AFM order
are either unstable or metastable. Thus, the order parameter can be written
as
\begin{equation}
\label{aaOderPar}
\Delta_{\alpha A}\equiv U_0\left\langle a^{\dag}_{\mathbf{n}\alpha\uparrow}a^{\phantom{\dag}}_{\mathbf{n}\alpha\downarrow}\right\rangle,\qquad
\Delta_{\alpha B}\equiv U_0\left\langle b^{\dag}_{\mathbf{n}\alpha\uparrow}b^{\phantom{\dag}}_{\mathbf{n}\alpha\downarrow}\right\rangle,\qquad
\Delta_{1 A} = \Delta_{2 B}=-\Delta_{2 A}=-\Delta_{1 B}\equiv\Delta\,,
\end{equation}
and $\Delta$ is real. In the mean-field approximation, the interaction Hamiltonian has the form~\cite{PrbOur}
\begin{equation}\label{aaMFH}
\hat{H}_{\text{int}} = {\cal
N}\left[\frac{4\Delta^2}{U_0}-U_0(n^2-1)\right]+\frac{U_0x}{2}\sum_{\mathbf{n}\alpha
a\sigma}n_{\mathbf{n}\alpha a\sigma} -
\sum_{\mathbf{n}\alpha}\left(\Delta_{\alpha A}a^{\dag}_{\mathbf{n}\alpha
\uparrow}a^{\phantom{\dag}}_{\mathbf{n}\alpha\downarrow}+\Delta_{\alpha
B}b^{\dag}_{\mathbf{n}\alpha\uparrow}b^{\phantom{\dag}}_{\mathbf{n}\alpha\downarrow}+\text{h.c.}\right),
\end{equation}
where $x = n - 1$ is the doping, $n$ is the number of electrons per site,
and ${\cal N}$ is the number of unit cells. Mean-field spectra are obtained
by diagonalizing of $\hat{H}_0+\hat{H}_{\text{int}}$.
The spectra are
\begin{equation}
\label{aaMFBands}
\varepsilon_{\mathbf{k}}^{(1,4)}=\mp\sqrt{\Delta^2+\left(t|f(\mathbf{k})|+t_0\right)^2},\;\;\;
\varepsilon_{\mathbf{k}}^{(2,3)}=\mp\sqrt{\Delta^2+\left(t|f(\mathbf{k})|-t_0\right)^2}.
\end{equation}
To determine the AFM gap $\Delta$ one has to minimize the grand potential per unit cell
\begin{equation}
\label{aaGP}
\Omega=\frac{4\Delta^2}{U_0}-U_0(n^2-1)-2T\sum_{s=1}^4\int{\frac{d\mathbf{k}}{V_{BZ}}\ln{\left[1+e^{\left(\mu'-\varepsilon_{\mathbf{k}}^{(s)}\right)/T}\right]}}\,,
\end{equation}
where $\mu'$ is the renormalized chemical potential,
$\mu'=\mu-U_0x/2$.
The equation
\begin{equation}\label{aaGap}
\frac{\partial\Omega}{\partial\Delta}=0
\end{equation}
specifies the gap as a function of the chemical potential and temperature.
To find $\Delta$ as a function of doping $x$, one needs to relate the
doping and the chemical
potential~\cite{PrbOur}
\begin{equation}\label{aaMuvsx}
n=1+x=-\frac{1}{4}\frac{\partial\left(\Omega-E_0\right)}{\partial\mu'}.
\end{equation}
Equations~\eqref{aaGap}
and~\eqref{aaMuvsx}
implicitly define the AFM gap
$\Delta(x,T)$
and the chemical potential
$\mu(x,T)$.
If only the nearest-neighbor hopping is taken into account, the gap obeys
the particle-hole symmetry relation,
$$
\Delta(-x,T)=\Delta(x,T)\,.
$$
The next-nearest-neighbor hopping breaks this symmetry; however, this effect is small. The gap as a function of doping $x$($>0$) is shown in Fig.~\ref{AAGap_dopfig}. The gap decreases
monotonously when the doping level $x$ grows. The gap becomes zero at some $x=x_c(T)$.

In two-dimensional systems no long-range order is possible if $T > 0$. In
such a situation, the mean-field solutions characterize the short-range
order, which survives for sufficiently low $T$. The effects beyond the
mean-field approximation are discussed in Ref.~\cite{PrbOur}. The gap
$\Delta(T)$ decreases when temperature increases, and the crossover from
(short-range) AFM insulating state to the paramagnetic (PM) metallic state
(see inset of
Fig.~\ref{AAGap_dopfig})
occurs at temperatures about mean-field transition temperature
$T_{\text{MF}}(x)$.
It is interesting to observe that the transition temperature is not a
single-valued function of doping. Instead, it exhibits a pronounced
reentrant behavior. This feature will be discussed below.

The maximum value of the crossover temperature depends on the on-site
Coulomb repulsion
$U_0$.
For example, AFM ordering can exist up to temperatures of about 50~K for
$U_0 = 5.5$\,eV
and to temperatures much higher than room temperature for
$U_0\gtrsim 6.5$\,eV.
The critical doping value
$x_c$,
at which the AFM state is replaced by the PM one, also strongly depends on
$U_0$,
changing from about
$0.1\%$ for $U_0 = 5.5$\,eV
to about
$10\%$
for
$U_0 \simeq 8$\,eV.~\cite{PrbOur}

\subsubsection{Incommensurate antiferromagnetic order}\label{AA_AFM_Incom}

\begin{figure}[t]
\includegraphics[width=0.6\columnwidth]{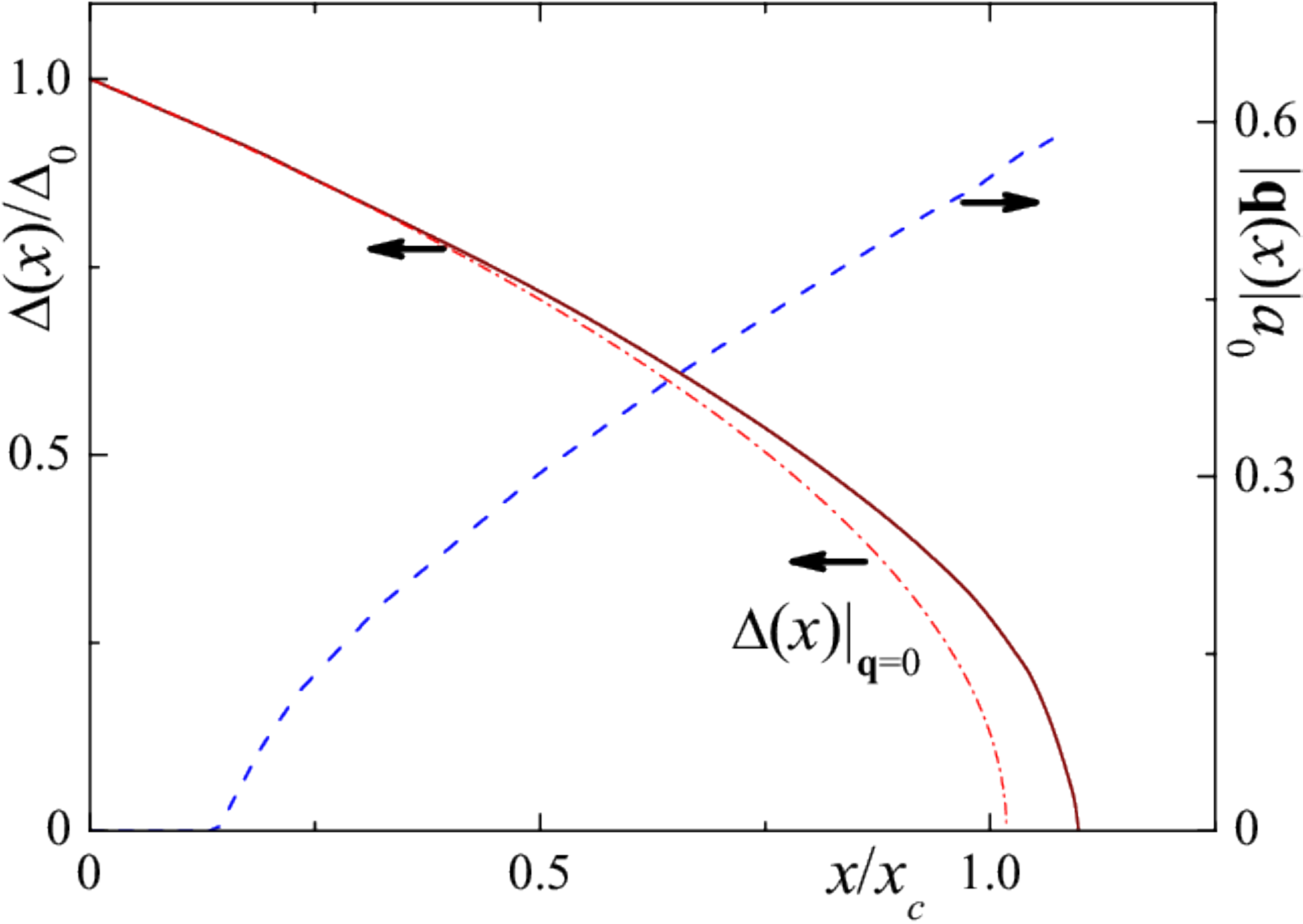}
\centering
\caption{(Color online) The dependence of the AFM gap $\Delta$ (red solid curve) and
$|\mathbf{q}|$ (blue dashed curve) on the doping $x$, calculated for $T/\Delta_0
= 0.06$ and $U_0 = 8$\,eV, from Ref.~\cite{PrbOur}. The dot-dashed curve is
the gap calculated for $\mathbf{q}= 0$. The doping $x$ is normalized by the
critical doping $x_c$, calculated for the commensurate AFM state. The
incommensurate AFM exists in a slightly larger doping range than the
commensurate AFM.
\label{AAIncomGQfig}}
\end{figure}

\begin{figure}
\centering
\includegraphics[width=0.49\textwidth]{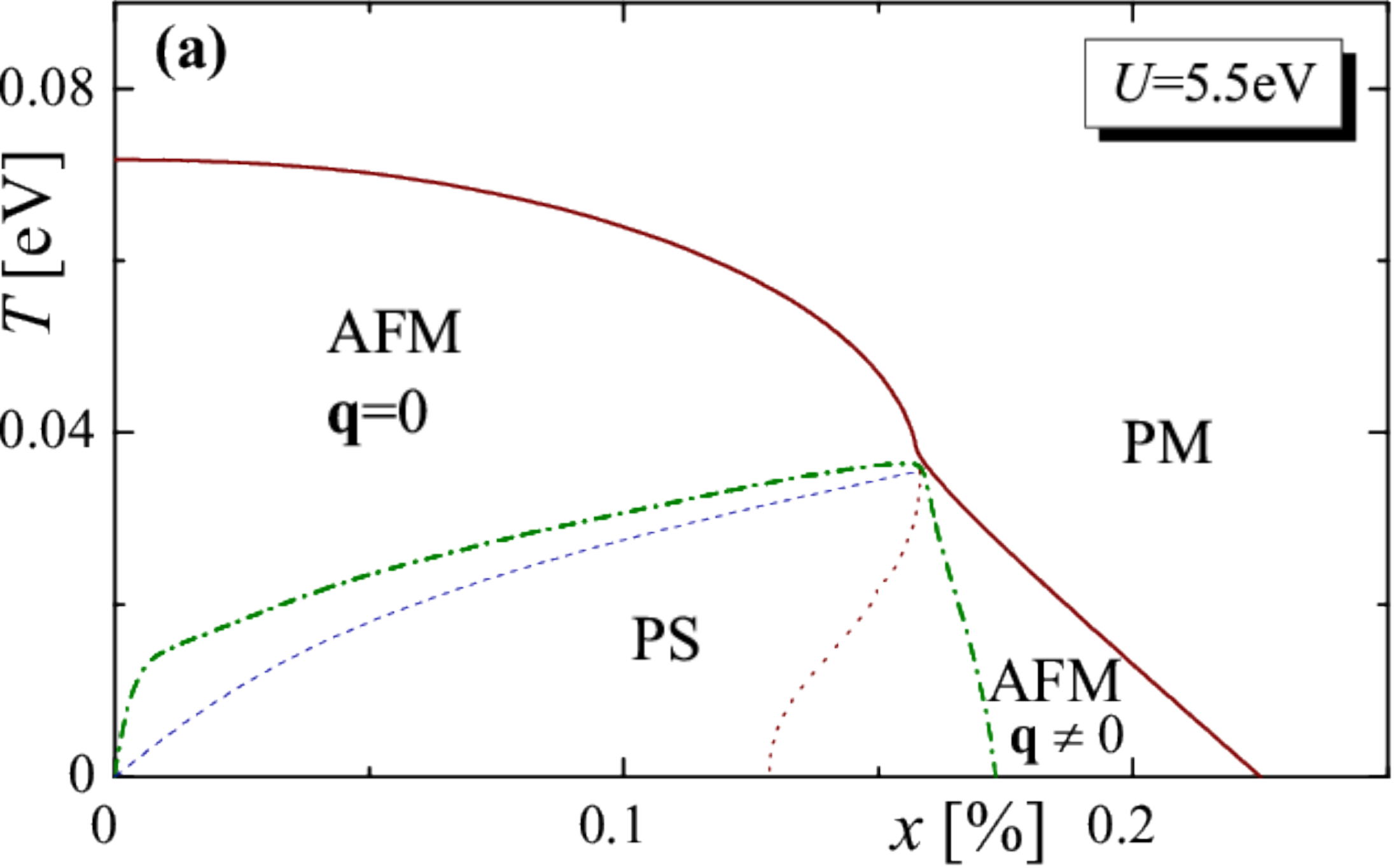}
\includegraphics[width=0.49\textwidth]{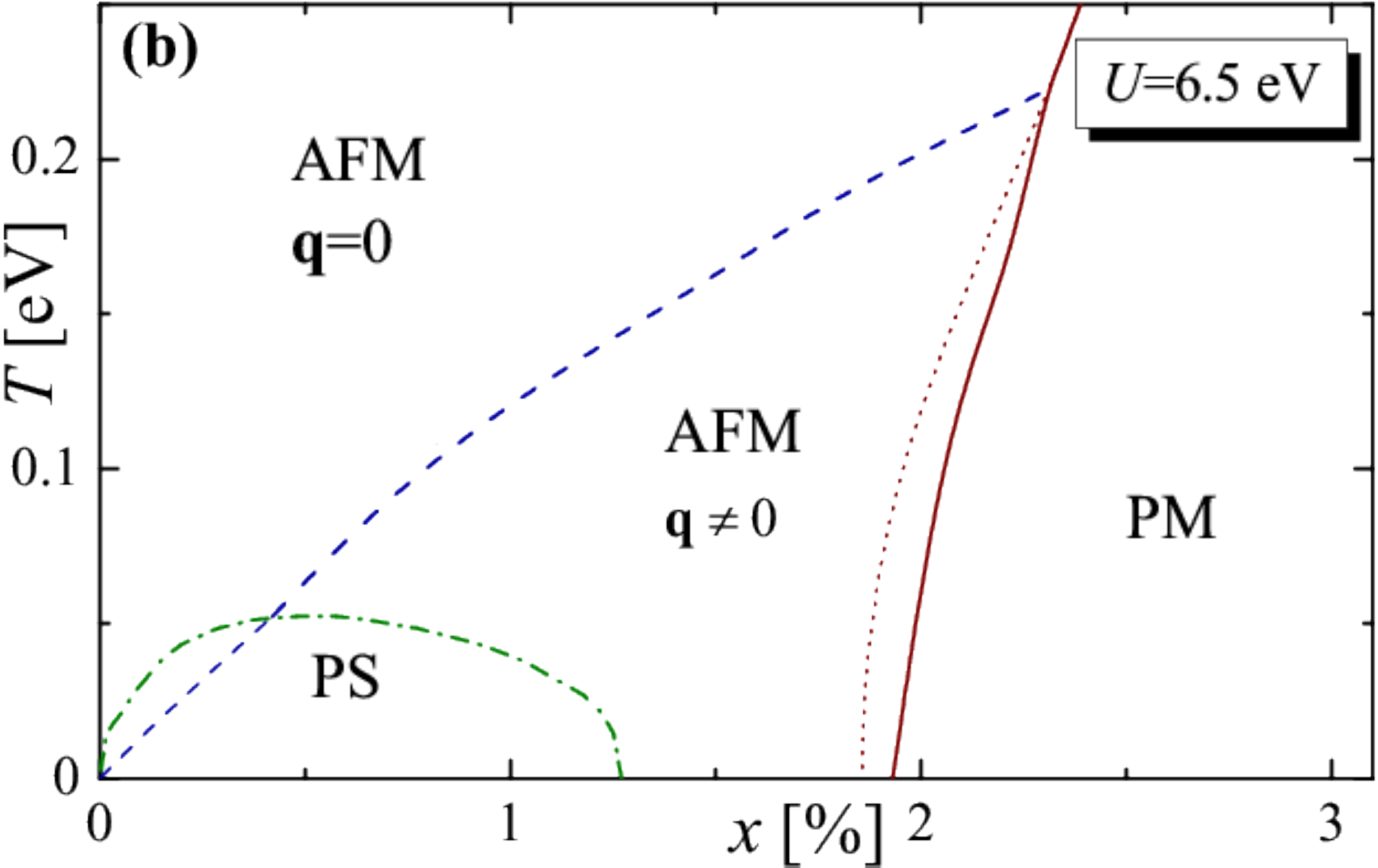}
\caption{(Color online) The phase diagram of the AA-stacked bilayer
graphene~\cite{PrbOur},
calculated for
$U_0 = 5.5$\,eV
(a) and
$U_0 = 6.5$\,eV
(b). Solid (red) curves are mean-field AFM to PM transition temperature
$T_{\text{MF}}(x)$;
(blue) dashed curves are
$T^q(x)$,
at which the commensurate-incommensurate transition occurs. The dotted
(red) curves are
$T_{\text{MF}}(x)$,
calculated without taking into account the incommensurate AFM state. The
dot-dashed (green) curves show the region of phase separation (see
subsection~\ref{AA_AFM_PS}).
\label{AAPhaseDiagfig}
}
\end{figure}

The G-type AFM state has the smallest value of the thermodynamic
potential $\Omega$ among states with commensurate magnetic order. Further
optimization of $\Omega$ can be achieved if the local direction of the
antiferromagnetic vector is allowed to rotate slightly from site to
site~\cite{Rice}.
Such a state is referred to as incommensurate (or helical) AFM. The complex
order parameter for this state has the form
\begin{equation}\label{aaIncomOdPar}
\Delta_{\mathbf{n}\alpha A}=U_0\langle a^{\dag}_{\mathbf{n}\alpha\uparrow}a^{\phantom{\dag}}_{\mathbf{n}\alpha \downarrow}\rangle=e^{i\mathbf{q}\mathbf{n}}\Delta_{\alpha A},\;\;\;
\Delta_{\mathbf{n}\alpha B}=U_0\langle b^{\dag}_{\mathbf{n}\alpha\uparrow}b^{\phantom{\dag}}_{\mathbf{n}\alpha \downarrow}\rangle=e^{i\mathbf{q}\mathbf{n}}\Delta_{\alpha B},
\end{equation}
where
$\mathbf{q}$
describes the spatial variation of the direction of the AFM vector. The
averaged electron spin
$\mathbf{S}_{\mathbf{n}\alpha a}$
in the unit cell
$\mathbf{n}$,
layer $\alpha$, sublattice $a$ lies in the
$x$--$y$
plane and is related to the order
parameter by
\begin{equation}
\mathbf{S}_{\mathbf{n}\alpha a}=\frac{\Delta_{\alpha a}}{U_0}\left[\cos{(\mathbf{q}\mathbf{n})},\,\sin{(\mathbf{q}\mathbf{n})}\right]\,.
\end{equation}
Therefore, the grand potential becomes a function of $\mathbf{q}$. Thus,
one should add the minimization condition
$\partial\Omega/\partial \mathbf{q}=0$
to derive equilibrium value of
$\mathbf{q}$.
The corresponding calculations were done in
Ref.~\cite{PrbOur}.
The plots of the AFM gap $\Delta(x)$ and $|\mathbf{q}(x)|$ are shown in Fig.~\ref{AAIncomGQfig}.
For comparison, the curve $\Delta(x)$
calculated for the commensurate AFM ($\mathbf{q}=0$) is also presented. We see that the
incommensurate AFM state exists in a slightly wider doping range than the
commensurate one. The phase diagram of the model in the $x$--$T$ plane is
shown in
Fig.~\ref{AAPhaseDiagfig}
for two different values of
$U_0$.
The diagrams for
$U_0\lesssim6$\,eV
and
$U_0\gtrsim6$\,eV
demonstrate a qualitative difference. Namely, for smaller
$U_0$
the reentrance, seen in the inset of
Fig.~\ref{AAGap_dopfig},
disappears. For larger
$U_0$
it survives. On the other hand, for large
$U_0$
the mean-field approximation is not an accurate method. Thus, it is not
clear, whether the reentrance is a genuine feature of the system, or it is
an artifact of the mean-field approximation.

Note, the incommensurate AFM phase is mathematically equivalent to the
Fulde-Ferrel-Larkin-Ovchinni\-kov state in
superconductors~\cite{FFLO1,FFLO2,FFLO3,FFLO4,FFLO5}.
Such incommensurate order is known to be sensitive to disorder. However,
the effect of disorder on the incommensurate AFM state was not analyzed in
Ref.~\cite{PrbOur}.

\subsubsection{Phase separation}\label{AA_AFM_PS}

\begin{figure}[btp]
\includegraphics[width=0.6\columnwidth]{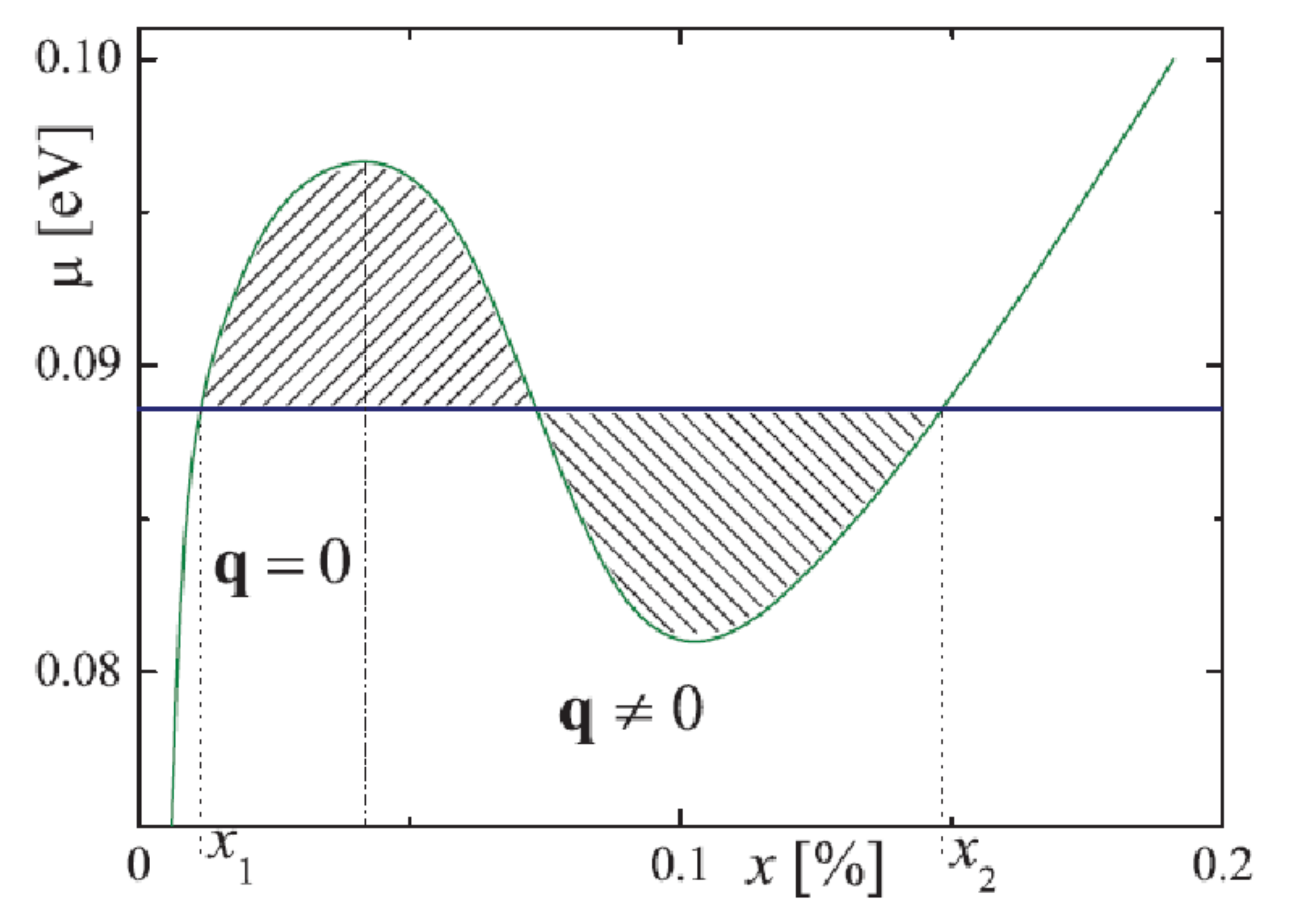}
\centering
\caption{(Color online)  Chemical potential $\mu$ versus doping $x$ from
Ref.~\cite{PrbOur};
$U_0 = 5.5$\,eV and $T = 0.014$\,eV. The vertical
dot-dashed line separates the AFM states with $\mathbf{q} = 0$ and
$\mathbf{q} \neq 0$. In the doping range $x_1 < x < x_2$, phase separation
occurs. The values $x_{1,2}$ are determined by the Maxwell construction:
the horizontal (black) line is drawn in such a manner that the areas of the
shaded regions are equal to each other.  \label{AAMuXfig}
}
\end{figure}

The formation of the inhomogeneous or phase separated states is an inherent
property of multiband systems with significant electron-electron
coupling. In Refs.~\cite{PrbOur} and \cite{PrbROur} it was predicted that
at non-zero doping, the AA bilayer graphene can separate in two phases
with unequal electron densities $n_{1,2} = 1 + x_{1,2}$. A typical
dependence of the chemical potential on doping, $\mu(x)$, for nonzero temperature is shown in
Fig.~\ref{AAMuXfig}.
The derivative
$\partial\mu(x)/\partial x$
is negative in some range of doping. This indicates that the system is
unstable, and can experience phase separation into commensurate
($\mathbf{q} = 0$, $x_1 < x$)
and incommensurate
$(\mathbf{q} \neq 0, x_2 > x)$
AFM phases. The doping levels
$x_{1,2}$
are found using the Maxwell
construction~\cite{Thermo}:
the (black) horizontal line is drawn in such a manner that the areas of the
shaded regions in
Fig.~\ref{AAMuXfig}
are equal to each other.
The region of phase separation in the
$(x$--$T)$-phase
diagram is bounded, in
Fig.~\ref{AAPhaseDiagfig},
by (green) dot-dashed curves. If the possibility of an incommensurate AFM
is ignored, then phase separation occurs between the AFM insulator
($x_1 = 0$)
and the PM
($U_0 \gtrsim 6$\,eV),
or the AFM
($U_0 \lesssim 6$\,eV)
metal
($x_2 > 0$).

The separated phases have different electron concentrations, and the phase
separation will be frustrated by long-range Coulomb
repulsion~\cite{PSsructure1,PSsructure2,PSsructure3,PSsructure4}.
Unless the Coulomb interaction completely arrests the separation, the
formation of  nanoscale inhomogeneities is likely. If the electron-rich
phase (incommensurate AFM) is a metal, and the electron-poor phase is an
insulator or ``bad'' metal, then, the percolative insulator-metal
transition could occur when the doping $x$ exceeds some threshold value,
which is about
$0.5(x_1+ x_2)$
in 2D systems. Phase separation can exist in the doping range
$x_1 < x < x_2$.
Depending on $U_0$, phase separation could be observed in a temperature
range from
$30$--$40$\,K
to room or even higher temperatures (see
Fig.~\ref{AAPhaseDiagfig}).

\subsubsection{Nearest-neighbor Coulomb repulsion, bias voltage, and exciton order parameter}\label{AA_AFM_Exiton}

The nearest neighbor terms in the interaction part of the Hamiltonian
can lead to the appearance of different ordered states. However, according to Ref.~\cite{PrlOur} all these order parameters
compete against antiferromagnetism, and only the AFM
order parameter survives~\cite{PrlOur,PrbOur,PrbROur}, because $U_0$ is the
strongest interaction constant (see Table~\ref{CoulPot}). A nonzero bias
voltage $V$ breaks the symmetry between two graphene layers. It was shown by R.\,S.~Akzyanov et~al.~\cite{PrbVOur}, that in this case,
there exists an order parameter driven by the inter-layer interaction
coexisting with antiferromagnetism. The analysis based on symmetry
considerations (see
subsection~\ref{spectraAA_symmetry})
shows that this order parameter should have the
form~\cite{PrbVOur}
\begin{equation}\label{aaExcGap}
\Delta^{A}_{exc} = U_{11}\left\langle a^{\dag}_{\mathbf{n}1\uparrow}a^{\phantom{\dag}}_{\mathbf{n}2\downarrow}\right\rangle,\;\;
\Delta^{B}_{\text{exc}} = U_{11}\left\langle b^{\dag}_{\mathbf{n}1\uparrow}b^{\phantom{\dag}}_{\mathbf{n}2\downarrow}\right\rangle,\;\;
\Delta^A_{\text{exc}} = -\Delta^B_{\text{exc}}\equiv\Delta_{\text{exc}},
\end{equation}
and the $\Delta_{\text{exc}}$ is real. This order parameter corresponds to a bound state of an electron and a hole in different layers, and can be referred to as the exciton order parameter.

The calculation of
$\Delta_{\text{exc}}$
was performed in
Ref.~\cite{PrbVOur}
for an undoped sample at
$T=0$
using the mean-field approach similar to that described in
Section~\ref{spectraAA_AFM}.
The exciton order parameter was found to increase when the bias voltage
$V$ grows. However, the full gap in the spectrum decreases when $V$
increases, since the AFM gap is much larger than the exciton one,
$|\Delta|\gg|\Delta_{\text{exc}}|$.
At non-zero temperature, the applied voltage can fully suppress the gap
giving rise to a voltage-driven metal-insulator transition. For the typical
values of parameters
$U_0=2.2t$,
$U_{11}/U_0 = 1/4$,
and
$eV/t_0 = 0.1$,
the values of the order parameters are estimated as
$\Delta\approx 0.17$\,eV
and
$\Delta_{\text{exc}}\approx8$\,meV.~\cite{PrbVOur}

The expressions for the order parameters can be rewritten in terms of
magnetization: $S=\Delta/U_0$ is the averaged spin on the site $A1$, and
$\phi=\Delta_{\text{exc}}/U_{11}$ can be viewed as the spin located on the
link connecting the nearest sites $A1$ and $A2$ in different layers. The
spin on the link connecting carbon atoms $B1$ and $B2$ has opposite sign.
The dimensionless exciton magnetization $\phi(V)$ (and
the order parameter as well) increases almost linearly with $V$, while $S$
decreases with increasing bias voltage. Nevertheless, $\phi\ll S$
(and, consequently, $\Delta_{\text{exc}}\ll\Delta$), even for relatively large
$V$. The exciton condensation could be observed by measuring the Coulomb
drag~\cite{ExcDrag1,ExcDrag2,ExcDrag3}.
The experimental observation of the Coulomb drag in an artificial structure of
two graphene sheets with a dielectric barrier between them has been
reported by R.~Gorbachev et~al.~in Ref.~\cite{DragGr}.

The crossover temperature $T^*$, at which the transition from ordered phase to PM state occurs, is of the order of $\Delta$. Consequently, $\Delta_{\text{exc}} \ll T^*$. However, the exciton order parameter is correlated with the AFM one, and it could be expected that they both have the same crossover temperature. According to results presented in the previous sections, doping suppresses the AFM ordering. It is natural to expect that it will also suppress the exciton gap~\cite{PrbVOur}.

\subsubsection{Long-range Coulomb interaction}\label{AA_AFM_LjngR}

Since in 2D systems the screening of the long-range Coulomb repulsion is
not so effective as in 3D systems, taking it into account is of special
interest. Proper treatment of the long-range effects improves the accuracy
of the calculations. Further, this long-range interaction also stabilizes
additional order parameters, which can be important for the analysis of the
phase separated state. However, for AA bilayer graphene this problem was
considered only in a few theoretical papers.

In Ref.~\cite{Honerkamp} D.\,S.~de~la Pe{\~n}a et~al. analyzed ordering in undoped AA-stacked bilayer graphene taking into
consideration the on-site Coulomb repulsion, both in-plane and out-of-plane nearest-neighbor Coulomb repulsion terms, and in-plane next-nearest-neighbor Coulomb repulsion. They studied the possible ground states of AA-stacked bilayer graphene using
a functional renormalization group approach. The authors analyzed a set of
possible electronic instabilities: AFM ordering of the type considered
above, charge density wave (CDW), quantum spin Hall instability (QSH), etc.
They concluded that, in the range of parameters characteristic of
AA-stacked bilayer graphene, the only stable phase is the AFM order, while
additional orders could exist in the AA honeycomb bilayer at a smaller value
of the on-site interaction. However, a possible existence of significant
QSH fluctuations in the AA-stacked bilayer graphene was predicted.

The long-range Coulomb repulsion was taken into account by L.~Brey and H.\,A.~Fertig in
Ref.~\cite{BreyFertig}.
The authors consider a single-valley model Hamiltonian for electrons in
undoped AA bilayer near the Dirac point using the Hartree-Fock approximation.
They observe the instability of a single-electron spectrum with the
formation of the energy gap and analyzed several possible order parameters
arising due to the long-range Coulomb repulsion. However, they neglect the spin
of the electrons, which automatically excludes the AFM ordering from their
consideration. As it was stated above, the AFM instability competes with
other possible instabilities and could suppress them. Thus, the question
about possible additional order parameters in the AA-stacked bilayer
graphene and their coexistence with the AFM ordering is still open.

A Monte Carlo study of the AFM ordering of AA bilayer graphene was reported
by A.~Nikolaev and N.~Ulybyshev
in Ref.~\cite{ulybyshev,ulybyshev_conf}.
The authors used long-range interaction Hamiltonian. They concluded that
long-range terms act to destroy the AFM order on finite clusters. It is not
clear at the moment how to apply this result to an infinite sample.
However, it seems to imply that a longer-range interaction weakens the AFM
order.

\subsection{Low-temperature broken symmetry phases of AB bilayer graphene}

Since the AB bilayer has a finite density of states at the Fermi energy, it
is natural to inquire about the stabilization of a broken symmetry phase in
such a system at sufficiently low temperature. Indeed, there are several
experimental
observations~\cite{Martin2010,
Mayorov2011,
Feldman2009,
Bao2012,
Velasco2012,
Weitz2010,
Freitag2012,
Freitag20122053,
veligura2012,
freitag2013},
which support this notion. However, the exact nature of the ordering
transition is not known. For example, some experiments found that the
low-temperature state has a single-electron gap, while others reported
gapless behavior. This may indicate an experimental artifact, or could be a
manifestation of a competition between several non-equivalent ordered states.
Below we will discuss some candidate broken-symmetry states, which are
considered by theorists.

\subsubsection{Ferromagnetic and antiferromagnetic states}

One of the earliest papers on the subject is
Ref.~\cite{Nilsson2006} by J.~Nilsson et~al.
The authors investigated the effects of the long-range and short-range repulsion on
the low-temperature symmetry breaking for the unbiased AB bilayer. To study
the long-range Coulomb interaction the authors used the two-band
Hamiltonian (at both Dirac points only two bands touching the Fermi energy
are kept) augmented by the following interaction term
\begin{eqnarray}
\label{eq::H_long}
H_{\rm int}^{\rm long}
=
\frac{1}{2S}
\sum_{\alpha=\pm, {\bf q}}
	V_\alpha ({\bf q}) \rho_{\alpha {\bf q}}\rho_{\alpha -{\bf q}}\,,
\end{eqnarray}
where
\begin{eqnarray}
V_\pm ({\bf q}) =
\frac{ 1 \pm e^{-qc_0} }{2}
V ({\bf q}) =
\frac{  \pi e^2}{q} ( 1 \pm e^{-qc_0} ),
\text{ and }
\rho_\pm = \rho_1 \pm \rho_2\,.
\end{eqnarray}
In these equations, the symbol $S$ denotes the sample area, and
$c_0$
is the distance between the layers, the bare Coulomb potential in the
momentum space is
\begin{eqnarray}
\label{eq::bare_coulomb}
V ({\bf q})
=
\frac{2\pi e^2}{|{\bf q}|}.
\end{eqnarray}
The layer density operators
$\rho_{1,2}$
correspond to smooth components of the total electron density:
$\rho_{1,2}$
are sums of the terms
$\psi_{\bf K}^\dag \psi_{\bf K}^{\vphantom{\dagger}}$
and
$\psi_{\bf K'}^\dag \psi_{\bf K'}^{\vphantom{\dagger}}$,
which conserves the valley index. The oscillating components of the density,
$\psi_{\bf K}^\dag \psi_{\bf K'}^{\vphantom{\dagger}}
e^{-i({\bf K - K'}) {\bf r}}$
and
$\psi_{\bf K'}^\dag \psi_{\bf K}^{\vphantom{\dagger}}
e^{i({\bf K - K'}) {\bf r}}$,
were initially omitted.

To investigate the stability of a ferromagnetic state the authors~\cite{ferro_bloch} applied
the variational approach due to Bloch,
adopting it to the AB bilayer. The corresponding wave function was
constructed in such a way that at each Dirac point the density of spin-up
electrons was unequal to the spin-down density, while the total density was
kept constant. The authors shown that the state with finite magnetization
in each valley is more stable than the paramagnetic state. The phase
diagram at finite doping was also discussed.

Unlike the case of a single Fermi surface, for the bilayer one has to worry
about
{\it two}
${\bf K}$-points.
Thus, finite magnetizations at individual valleys do not immediately imply
that the sample has finite magnetic moment, since the magnetizations of two
valleys may be either add up, or cancel each other out, resulting in either a
finite, or zero total magnetic moment of the whole system. Within the
variational calculations outlined above, these states are degenerate. To
lift this degeneracy, the authors~\cite{ferro_bloch} added the exchange term describing the
interaction between the oscillating components of the electron density. For
such a model, the most stable state is the one with a finite ferromagnetic
moment (related theoretical studies of ferromagnetism in biased AB
bilayers were reported in
Refs.~\cite{Castro2008,bilayer_ferro2007}).

Thus, the results of J.~Nilsson et~al.~\cite{Nilsson2006}
suggest that the long-range interaction favors a ferromagnetic ground
state. To study the effects of the short-range interaction, the authors of
Ref.~\cite{Nilsson2006}
used the Hubbard model on the lattice of the AB bilayer. Applying the mean-field approximation to such a Hamiltonian it was shown that, at one electron
per carbon atom, the mean-field ground state is antiferromagnet. Since the
graphene bilayer lattice with nearest-neighbor hopping is bipartite, this
result is not that surprising.

The described investigation demonstrated that the competing long-range and 
short-range interactions may act to stabilize different types of ordered
states. This means that the phase diagram of a particular sample could
depend on delicate details of the system, which shift the balance between
the competing states.

However, one must not take the results of
Ref.~\cite{Nilsson2006}
too literally. It is necessary to remember that the justification of the
two-band model may be problematic when the effects of the interaction are
studied. Indeed, the single-site interaction constant for graphene is
estimated to be of the order of several eV, while the discarded
``high-energy" bands are separated from the Fermi level by a fraction of
eV. Further, the discussed variational ferromagnetic wave function is only
one possible type (among many) of ordered state promoted by the long-range
interaction. Consequently, other options must also be explored.

\subsubsection{``Pseudospin magnetic" states}
\label{sub::pseudomag}

Several ordered states are considered theoretically as alternative
to ferromagnetic and AFM ones. For
example, H.~Min et~al.
argued in Ref.~~\cite{min_pseudo_fm2008} that the long-range Coulomb interaction establishes a gapped broken
symmetry state, which they called a ``pseudospin magnet". The paper
suggested that, by neglecting the chiral structure of the single-electron
states, the authors of
Ref.~\cite{Nilsson2006}
ignored a very effective route to minimize the electron interaction energy.
The Hartree-Fock calculations of
Ref.~\cite{min_pseudo_fm2008}
demonstrated that, at each valley and for each spin projection, the bilayer
may experience the spontaneous generation of ``pseudospin polarization",
which may be viewed as a finite expectation value of some ``pseudospin"
operator. In the context of the latter study, the pseudospin is the layer index,
or, equivalently, the sublattice index. Therefore, the finite pseudospin
polarization (for a given spin projection $\sigma$ and a given valley $\xi$)
implies that the electron density
$$
\rho_{\sigma\xi}=\psi^\dag_{\sigma \xi}\psi^{\vphantom{\dagger}}_{\sigma\xi}
$$
associated with $\sigma$ and $\xi$ is spontaneously shifted toward a
particular layer. Summing the pseudospin polarizations over all values of
$\sigma$ and $\xi$, the total polarization for a given many-electron trial
state is determined. It could be either zero (``pseudospin
antiferromagnet"), or finite  (``pseudospin ferromagnet/ferrimagnet").
These order parameters are of excitonic origin in the sense that they
describe condensation of electron-hole pairs. A particular ordered phase
corresponds to electron-hole pairs with specific spin and orbital quantum
numbers.
At the charge neutrality point the most stable state is ``pseudospin
antiferromagnet". In Ref.~\cite{zhang_rg2010}
these conclusions were checked with the help of renormalization group
calculations. It was judged that the results of this renormalization group
study were consistent with the Hartree-Fock findings of H.~Min et~al.~\cite{min_pseudo_fm2008}.
A related renormalization group investigation was reported in Ref.~\cite{zhang_rg_order2012}.

R.~Nandkishore and L.~Levitov~\cite{Nandkishore2010}
also studied long-range interactions, but used a different
technique. They argued in favor of the stability of ``the ferroelectric"
state. The ferroelectric state is characterized by a finite value of the electric
charge polarization vector, directed normally to the bilayer plane. In
other words, some electrons from one layer are spontaneously shifted to the
other layer, creating a local violation of charge neutrality. The energy
associated with the ordering was estimated to
be~$\sim 4$\,meV,
that is, within the experimentally accessible range. In the language of
Ref.~\cite{min_pseudo_fm2008},
this ferroelectric state is a ``pseudospin ferromagnet". The results of
both papers are consistent with each other.

It is worth noting that an earlier
paper by E.~McCann et~al.~\cite{McCann2007}
reasoned against stability of such a ferroelectric state, pointing out that
the electric polarization increases the Hartree energy. In
Ref.~\cite{Nandkishore2010}
this argument was criticized: the exchange interaction, neglected in
Ref.~\cite{McCann2007},
overcomes the Hartree term. This makes the transition into the
ferroelectric state possible. Such a conclusion agrees with the
calculations of H.~Min et~al.~\cite{min_pseudo_fm2008},
which also identified the exchange contribution as the driving force behind
the symmetry breaking.

In Ref.~\cite{Nandkishore2010b} R.~Nandkishore and L.~Levitov
pointed out that if $V_- = 0$ in
Eq.~(\ref{eq::H_long}),
then the Hamiltonian symmetry is significantly enhanced. For a generic
value of $V_-$,
the Hamiltonian symmetry group is
SU(2)$\times$SU(2),
where one instance of SU(2) corresponds to the rotation of electron spin,
while the other describes the rotation in the valley space. When
$V_-$
vanishes, the symmetry group is augmented to SU(4).

Working with such a highly symmetric Hamiltonian, 
Ref.~\cite{Nandkishore2010b}
developed a classification of the broken symmetry states. It covers the
states discussed in
Refs.~\cite{min_pseudo_fm2008,Nandkishore2010},
as well as some others. The classification formalized and extended the
ideas in Ref.~\cite{min_pseudo_fm2008}
about ``pseudospin (anti)ferromagnet".
To explain the origin of this classification more rigorously, let us
examine the following mean-field
Hamiltonian~\cite{Nandkishore2010b}:
\begin{eqnarray}
H=-\frac{v_{\rm F}^2}{t_0}
\left(
        \begin{matrix}
                0 & (q_y+iq_x)^2\otimes{\hat 1}\otimes{\hat 1} \cr
                ( q_y-iq_x)^2\otimes{\hat 1}\otimes{\hat 1}& 0\cr
        \end{matrix}\right)
+\Delta\left(
        \begin{matrix}
                {\hat Q} & 0 \cr
			0 & - {\hat Q} \cr
        \end{matrix}\right).
\end{eqnarray}
In this expression, the first term corresponds to the kinetic energy of the
bilayer. The product
${\hat 1}\otimes{\hat 1}$
is an identity matrix in the spin-valley space. While the kinetic energy
differs from
Eq.~(\ref{ab::2b}),
the two representations are unitary
equivalent~\cite{Nandkishore2010b}.
The second term shows the mean-field contribution due to broken symmetry
caused by the Coulomb interaction. Specifically, the parameter $|\Delta|$
represents the absolute value of the order parameter. It is proportional to
the single-electron gap. The hermitian
$4\times4$
matrix
${\hat Q}$,
acting in the spin-valley space, and satisfying the condition
${\hat Q}^2 = 1$,
sets a ``direction" of the multicomponent order parameter. This
quantity is analogous to the vector
${\bf n}$, with
${\bf n}^2 = 1$,
which describes, for example, the direction of the magnetization for a
ferromagnet. Unlike the magnetization, however, the number of independent
components of
${\hat Q}$
is much higher.

The value of $|\Delta|$ is found by solving the mean-field
equation~\cite{Nandkishore2010}.
However, the sign of $\Delta$ and the
matrix elements of
${\hat Q}$,
beyond the constraint
${\hat Q}^2 = 1$,
are completely arbitrary within the framework of the mean-field theory. We
can say that all mean-field states form a degenerate manifold parametrized
by matrices
${\hat Q}$,
such that
${\hat Q}^2 = 1$,
and a specific matrix
${\hat Q}$
encodes a particular mean-field state from this manifold.

It was noticed in
Ref.~\cite{Nandkishore2010b}
that the latter manifold has some internal structure: it can be split into
three equivalence classes. To explain the division into the classes, let us
note that, because of the constraint, all eigenvalues of
${\hat Q}$
are equal to
$\pm 1$.
Thus, one can define three equivalence classes, which will be denoted below
as (4,\,0), (3,\,1), and (2,\,2).
The first class corresponds to matrices with four positive eigenvalues. The
second class, to matrices with three positive eigenvalues and one
negative. Finally, the matrices with two positive and two negative
eigenvalues constitute the third class. Within a particular class, any two
matrices are unitary equivalent, but a pair of matrices from different
classes cannot be connected by an SU(4) transformation. For example, if the
eigenvalues of some ${\hat Q}_1$ are all positive, while the eigenvalues of some other
${\hat Q}_2$ are both positive and negative, it is impossible to find a unitary matrix
such that ${\hat Q}_2=U {\hat Q}_1 U^\dag$.

Therefore, the SU(4)-symmetric Hamiltonian has only three types of
non-equivalent mean-field ground states, corresponding to the three
equivalence classes. Specifically, the first and the second classes
contain the so-called quantum anomalous Hall (QAH) states. These states
have finite non-diagonal (Hall) conductivity
$\sigma_{xy}$
(at zero doping). Depending of the type of QAH state, it equals to either
$4e^2/h$, or to $2e^2/h$. The finite value of the Hall conductivity is a manifestation of the
time-reversal symmetry breaking.

As for the third class, it is called the quantum flavor Hall state. For
this class $\sigma_{xy} = 0$,
thus, time-reversal symmetry is preserved. Instead, these states are
characterized by either finite non-diagonal spin conductivity, 
non-diagonal valley conductivity, or non-diagonal spin-valley conductivity.
All quantum flavor Hall states are equivalent in the sense that they can be
connected by a transformation from the Hamiltonian SU(4) symmetry group.
This equivalence is an artifact of the initial approximation, which mixes
spin and valley spaces, making them indistinguishable from each other. A
more detailed analysis reveals that different flavor Hall states are quite
dissimilar, and have non-identical physical properties: the valley Hall
state has finite interlayer electric polarization (it is the ferroelectric
state of Ref.~\cite{Nandkishore2010}), the spin-valley Hall state is layer
antiferromagnet~\cite{Jung2011} (total magnetization of the layer antiferromagnet is zero, but the
magnetizations of each layer is finite). Therefore, the less symmetric
Hamiltonian lifts the degeneracy between these states. The degeneracy
between the SU(4) multiplets may be lifted by fluctuations and a weak
external magnetic field~\cite{Nandkishore2010b}.

To overcome the shortcomings of the continuous effective models, the same set
of pseudospin-ordered phases was studied with the help of a lattice model by J.~Jung et~al.
in Ref.~\cite{Jung2011}.
It was found that the ordered states are stable even in the lattice model.
The quantum anomalous Hall states and flavor Hall states have different
energies; however, this difference is very small, of the order of
$10^{-9}$\,eV per carbon atom.
The paper~\cite{Jung2011} evaluated the energy gap as
$\sim 10$\,meV.
The latter value is roughly consistent with the estimate of
$\sim 4$\,meV
reported in
Ref.~\cite{Nandkishore2010}.
It was also claimed that the developed theoretical picture is in agreement
with experimental
results~\cite{Martin2010,Weitz2010}.

Experimental signatures of these broken symmetries were discussed by theorists in Refs.~\cite{Nandkishore_kerr2011a,zhang_quant_hall2012}.

\subsubsection{Antiferromagnetic state}
\label{sub::af}

As one can see from the previous discussion, the antiferromagnetic phase is
considered by many theorists to be a viable candidate for the
low-temperature broken-symmetry state of the bilayer. Thus, its properties
have been investigated in several publications. For example, a
first-principle calculations reported in
Ref.~\cite{af_1princ2013}
confirmed the stability of the antiferromagnetic ordering. The gap was
found to be
$\sim 1.8$\,meV,
and the surface magnetization was
$\sim 10^{-2}\mu_{\rm B}$\,nm$^{-2}$.

A very popular tool to study antiferromagnetism is the Hubbard model. It
was used, for example, in
Refs.~\cite{Nilsson2006,vafek_rg2010,lang_af_hubb2012,
af_ab_hubb2013,
sun_hub_mc_afm2014,
scherer_hub_frg2012}.
The latter references employed Monte Carlo
simulations~\cite{lang_af_hubb2012,sun_hub_mc_afm2014},
different versions of the renormalization group
approach~\cite{vafek_rg2010,lang_af_hubb2012,scherer_hub_frg2012},
and mean-field theories~\cite{Nilsson2006,lang_af_hubb2012,af_ab_hubb2013}.
In particular, the perturbative renormalization group study of
Ref.~\cite{vafek_rg2010}
concluded that the antiferromagnetic instability is the strongest
instability of the Hubbard model.

A continuous model was used in
Ref.~\cite{haritonov_afm2012},
which offered a mean-field treatment of the antiferromagnetic state in a
magnetic field. It was found that the insulating gap grows as a function of
the magnetic field. The author~\cite{haritonov_afm2012} reported that, after adjusting two
parameters of the proposed theory, the experimental data was described
quantitatively by the theory.

Experimentally, the low-temperature layer-antiferromagnetic order was
mentioned as a possibility in
Refs.~\cite{veligura2012,freitag2013}.

\subsubsection{Nematic phase}

A completely different way to eliminate the finite density of states at the
Fermi energy was proposed by O.~Vafek and K.~Yang in
Ref.~\cite{vafek_nemat_rg2010}.
This paper used a perturbative renormalization group approach, which was
applied to a two-band effective Hamiltonian with interactions. The authors
assumed that the electron interaction potential
$V({\bf r})$
vanishes, if
$|{\bf r}| > r_0$,
where
$r_0 \gg a$
is some finite length scale. According to this study, the strongest
instability corresponds to the order parameter, which breaks rotation
symmetry of the Hamiltonian. The mean-field Hamiltonian in the
broken-symmetry phase can be written as follows:
\begin{eqnarray}
\label{nem::ham}
H_{\rm nem}
=
-\frac{v_{\rm F}^2}{t_0}
\left(
        \begin{matrix}
                0 & (\xi q_y + iq_x)^2 \cr
                ( \xi q_y - iq_x)^2 & 0\cr
        \end{matrix}
\right)
+
\left(
        \begin{matrix}
                 0  & \Delta' - i \xi \Delta'' \cr
                 \Delta' + i \xi \Delta'' & 0 \cr
        \end{matrix}
\right),
\end{eqnarray}
where different values of
$\xi = \pm 1$
correspond to different valleys,
${\bf K}$
or
${\bf K}'$.
The complex number
$$
\Delta = \Delta' + i \Delta''\,,
$$
with
\begin{eqnarray}
\Delta'=\langle
	\Psi^\dag_{\bf K} \sigma_x \Psi^{\vphantom{\dagger}}_{\bf K}+
	\Psi^\dag_{\bf K'} \sigma_x \Psi^{\vphantom{\dagger}}_{\bf K'}\rangle\,,\qquad
\Delta''=\langle
	\Psi^\dag_{\bf K} \sigma_y \Psi^{\vphantom{\dagger}}_{\bf K}-
	\Psi^\dag_{\bf K'} \sigma_y \Psi^{\vphantom{\dagger}}_{\bf K'}
\rangle\,,
\end{eqnarray}
is the order parameter. This order parameter changes its sign, when the
real space is rotated by $\pi/2$.
In analogy with the nematic state of liquid crystals, which has the same
symmetry, this order parameter is called nematic. Note that this ordered
state is insensitive to the electron spin: the rotation of spins does not
change
$\Delta$.

The eigenenergies of the Hamiltonian~(\ref{nem::ham})
are given by the relation:
\begin{eqnarray}
\varepsilon = \pm |(\xi q_y - i q_x)^2 + \Delta|\,.
\end{eqnarray}
Upon the transition into the nematic phase, the Fermi point at
${\bf q}=0$
with parabolic dispersion splits into two Fermi points at
\begin{eqnarray}
\label{broken::ab:nematic_q}
q^*_x + i\xi q^*_y = \pm \sqrt{\Delta}\,,
\end{eqnarray}
both with linear dispersion. Unlike the states discussed in
subsections~\ref{sub::pseudomag},~\ref{sub::af},
the nematic state is gapless. However, below the energy scale set by
$|\Delta|$,
the electron density of states vanishes linearly near the Fermi energy.

Within the framework of the two-band model, the absolute value of the order
parameter $|\Delta|$ is fixed, but its complex phase is completely undetermined. Unlike the
superconducting order parameter phase, whose value is of no significance,
the phase of the nematic order parameter has a physical consequence: as one
can see from Eq.~(\ref{broken::ab:nematic_q}),
the locations of the emergent Dirac points in ${\bf k}$-space
depends on $\arg \Delta$. The presence of
the lattice, however, reduces the degeneracy relative to
$\arg \Delta$ down to a set of three discrete angles, defined with respect to the
crystallographic axis. Once the direction of the vector $(q^*_x, q^*_y)$
is chosen (randomly) from this discrete set, in the resultant nematic state
the effective hopping along the bonds parallel to a particular
$\boldsymbol{\delta}_i$, (for definition of $\boldsymbol{\delta}_i$
see Fig~\ref{GraphLatBrulfig}) is enhanced. In this regard Y.~Lemonik et~al. in
Ref.~\cite{lemonic_rg_nemat_long2012}
pointed out that the nematic order parameter is ``mimicking the effect of
anisotropic hopping along bonds with different directions on the honeycomb
lattice".

The nematic order, its stability, and competition against other ordered
states is a subject of several theoretical publications. In a comprehensive
renormalization group study of O.~Vafek, Ref.~\cite{vafek_rg2010},
the nematic, antiferromagnetic, and quantum anomalous Hall states were
listed as possible broken-symmetry states for a model with short-range
interaction. According to the latter paper, which state wins depends on
values of the ``microscopic" Hamiltonian coupling constants. For example, the
very short-range interaction of the Hubbard model leads to
antiferromagnetic order. These conclusions are consistent with the results
of Ref.~\cite{vafek_rg2012}, which argued that short-range interactions favor antiferromagnetic
order, while longer-range interactions stabilize the nematic state. The
antiferromagnetic phase is associated with short-range scattering, since
such a phase requires inter-valley scattering, which decreases as the
interaction range grows. A related perturbative renormalization group study
was reported by V.~Cvetkovic et~al. in
Ref.~\cite{cvetkovic_multi2012}.
It was shown that the two-band model with generic choice of interaction
constants has a very diverse phase diagram. However, ``nematic appears to
be the unique dominant instability when forward scattering dominates.
Similarly, the layer antiferromagnet appears upon inclusion of sufficiently
strong back and layer imbalance scattering".

Analogous conclusions were reached in
Refs.~\cite{lemonic_rg_nemat_long2012,Lemonik2010},
where a different version of the perturbative renormalization group was used
(among these two papers
Ref.~\cite{lemonic_rg_nemat_long2012}
presented the latest and most extensive account of the calculations, it
also corrected some misprints of an earlier publication,
Ref.~\cite{Lemonik2010}).
These papers derived the renormalization group flow in the presence of
screened Coulomb interaction. According to
Ref.~\cite{lemonic_rg_nemat_long2012},
the nematic, antiferromagnetic, and quantum spin Hall phases are the most
probable ground states of the bilayer.

A different approach to the question of the relative stability of the
nematic order and other ordered phases was adopted by E.V.~Gorbar et~al.~in
Ref.~\cite{gorbar_compet_nemta2012}.
That paper, using a two-band model with Coulomb long-range
interaction, obtained, in the random-phase approximation, the energies
of different ordered phases in the presence of a transverse bias voltage
and external strain. The authors showed that at zero strain the ground state
is gapped: one of the ``pseudospin magnetic" phases discussed in
subsection~\ref{sub::pseudomag}
wins over the nematic state. The transverse bias voltage can be used to
stabilize the otherwise unstable/metastable ferroelectric state (also referred
to as ``pseudospin ferromagnet", or quantum valley Hall state). Since the
polarization vector couples to the electric field, the energy of the
ferroelectric state decreases when the bias grows, and at sufficiently
strong bias the ferroelectric state becomes the ground state.
The strain, on the other hand, couples to the nematic order parameter.
Thus, if the bias is zero, a sufficiently strong external strain destroys the
pseudospin order, which is replaced by a gapless phase with a finite
expectation value corresponding to the nematic order parameter. This, of
course, is not spontaneous nematic order, since the related symmetry is
already broken in the Hamiltonian by the outside condition (strain). At
arbitrary strain and bias, the ground state is one of the ``pseudospin
magnetic" states with an admixture of the nematic order parameter. This phase diagram in the plane strain versus bias-voltage contains two
phases ``separated by a critical line of first- and second-order phase
transitions"~\cite{gorbar_compet_nemta2012}.

There are several experimental
papers~\cite{Martin2010,
Mayorov2011,Weitz2010},
whose findings were interpreted as consistent with the nematic
low-temperature state. However, there are other experiments (e.g.,
Refs.~\cite{Velasco2012,uultrafast}),
which are inconsistent with this hypothesis.
J. Zhang et~al.~\cite{zhangDisorder} suggested that
disorder might explain the discrepancy between experiments.
They demonstrated that, in mean-field approach, the ground state
of AB bilayer undergoes a transition between different ordered
states as a function of the disorder strength. In particular,
they predicted that small amounts of disorder can drive the gapped
ground state into a nematic state, and both gapped and nematic
orders vanish if the impurity density exceeds some critical value.

\subsubsection{Other types of order}

Other types of ordered states have been investigated in the theoretical
literature~\cite{cvetkovic_multi2012,Dillenschneider2008,Dahal_cdw2010,
zhu_loop_order2013}.
For example,
Refs.~\cite{Dahal_cdw2010}
discussed the charge density wave. The phase diagram of
Ref.~\cite{cvetkovic_multi2012}
contains numerous ordered states, some of which were mentioned above, some
others, like superconducting phases, were not.

The authors of
Ref.~\cite{zhu_loop_order2013}
argued that the bilayer lattice Hamiltonian may stabilize a particular
``current loop order", provided that the only interaction term included into
the Hamiltonian is the nearest-neighbor repulsion. While such a choice of
the interaction seems highly peculiar, some reasoning behind this
approximation was offered. The proposed symmetry-breaking state was called
``magnetoelectric". It has some similarities with the quantum anomalous
Hall state, but, unlike the latter, breaks layer-inversion symmetry.
The authors stated that the phenomenology of the magnetoelectric symmetry
violation offered a more comprehensive explanation to the experimental data
of Refs.~\cite{Bao2012,Velasco2012}
than the explanation in terms of the antiferromagnetic order, or quantum
anomalous Hall state.

Exciton order was investigated in
Refs.~\cite{Dillenschneider2008},
where the authors performed a mean-field analysis of a lattice model with
Hubbard-like interaction. To stabilize the exciton phase the repulsion
between the electrons located in different layers was taken into account.
Conversely, the discussed excitons are formed by electrons in one layer and
holes in the other. The authors concluded that, to stabilize the exciton
phase, the interaction constant must exceed a threshold value of the order
of $10t \sim 30$\,eV. Such high value of the coupling constant is unrealistic. The paper noticed
that this threshold may be reduced by a transverse electric field. However,
for realistic fields the effect is quite weak. Thus, it is likely that such
an order parameter cannot be stabilized in graphene bilayers.

As the above discussions demonstrate, the non-interacting ground state of
AB bilayer graphene is unstable with respect to a variety of symmetry
breakings. In such a situation, a theorist's ability to predict the ``true"
ground state is quite limited: the theory allows one to compile a list
enumerating various possible ordered states, which could be stabilized under
realistic conditions. Such a list may serve as a guide to an
experimentalist. However, which ordered state wins under a particular set
of external conditions can be determined only experimentally.

%

\section{Many-body and non-Fermi liquid effects in bilayer graphene}

Highly doped AB bilayer behaves almost as an ordinary metal
and is described by Fermi liquid theory. However,
if one is interested in the many-body properties of the undoped AB bilayer
graphene, it is necessary to remember that, unlike a metal with its
well-developed Fermi surface, it has no Fermi
surface, only Fermi points.
This circumstance has dramatic consequences for
the many-body physics of the electronic liquid of the bilayer. Even at weak
doping, when the Fermi points are replaced by a Fermi surface, many
electronic properties are dominated by the proximity of the
charge-neutrality regime.

Since the undoped AB bilayer does not have the Fermi surface, it lacks the
quasiparticles as defined by the Landau theory of the Fermi
liquid. This feature of the system was investigated in
Refs.~\cite{Barlas2009,Nandkishore2010a}.
Assuming that the RPA adequately describes the effective interaction between charge
carriers, Y.~Barlas and K.~Yang~\cite{Barlas2009}
calculated the self-energy to second-order in the screened interaction.
It was found that the quasiparticle spectral weight $Z$ vanishes as the
energy approaches zero
\begin{eqnarray}
\label{ab::Z_vanish}
Z \propto \frac{1}{(\ln \varepsilon )^2} \rightarrow 0\,,\;\text{ when }\,\varepsilon\rightarrow0\,.
\end{eqnarray}
The disappearance of the quasiparticle residue is inconsistent with the
quasiparticle notion of the Landau theory. This means that the undoped
bilayer is not a Fermi liquid. At the same time, while the residue $Z$
vanishes, the effective mass remains unchanged by the interaction. 

This perturbation-theory treatment was improved by R.~Nandkishore and L.~Levitov in
Ref.~\cite{Nandkishore2010a},
where the renormalization group approach was used. Those
renormalization group calculations confirmed the destruction of the
quasiparticles: as the energy approaches zero, the quasiparticle residue
vanishes as the inverse of the square of $\ln \varepsilon$,
in agreement with
Eq.~(\ref{ab::Z_vanish}).
The disappearance of the quasiparticles implies the suppression of the
tunneling density of states, which may be measured experimentally. A weaker
(logarithmic) renormalization of the mass was also discovered.  Since the
mass renormalization is a subleading effect, the perturbation theory of
Ref.~\cite{Barlas2009}
was unable to capture it. Interactions also introduce logarithmic
corrections to the compressibility. However, the authors of
Ref.~\cite{Nandkishore2010a}
concluded that these corrections are weak, and difficult to detect in
experiments.

For doped bilayer, the electronic liquid properties were discussed in
Refs.~\cite{Kusminskiy2008,Borghi2009a,Sensarma2011}.
Specifically, it was argued in
Ref.~\cite{Sensarma2011}
that, upon doping, the Fermi surface emerges, and the Fermi liquid behavior
is restored. In
Ref.~\cite{Kusminskiy2008}
the compressibility of the bilayer electronic liquid was calculated within
the framework of the Hartree-Fock approximation for different values of the
Fermi momentum
$k_{\rm F}$. In general, the compressibility was found to be a non-monotonous function
of $k_{\rm F}$. At low density it was negative. These results have to be contrasted with
those of
Ref.~\cite{Nandkishore2010a},
where only weak corrections to the compressibility were found.
This discrepancy may be related to the fact that
Ref.~\cite{Kusminskiy2008}
disregarded the screening of the Coulomb
interaction~\cite{Nandkishore2010a}.
The renormalization of mass as a function of the doping density was
discussed in
Ref.~\cite{Borghi2009a}.
The ground state wave function was constructed with the help of the
Hartree-Fock approximation. To account for the screening, the Coulomb
potential in the Hamiltonian was replaced by a static Thomas-Fermi potential.
The authors concluded that the mass is strongly suppressed by the
interaction but, nonetheless, it remains finite even at low doping levels.

Reference~\cite{Shizuya2010}
investigated many-body corrections to the cyclotron resonance in the
monolayer and AB bilayer graphene, and compared the theoretical conclusions
with the experimental
data~\cite{Henriksen}.
As the author admitted, the data ``defy good fit by theory but certainly
suggest nontrivial features of many-body corrections". The high resolution ARPES measurements of low doped AB bilayer samples done in Ref.~\cite{cheng2015}
confirmed directly non-Fermi liquid quasiparticle behavior in AB bilayer. 

The dielectric function
$\epsilon ({\bf k}, \omega)$
is among the physical quantities affected by the peculiar band structure of
AB bilayer graphene. Most generally, $\epsilon$ may be written as
follows
\begin{eqnarray}
\label{eq::diel_func}
\epsilon ({\bf k}, \omega) = 1 - V ({\bf k}) \Pi ({\bf k}, \omega)\,.
\end{eqnarray}
In this expression, $\Pi$ is the irreducible polarization, and
$V({\bf k})$
is the Fourier transform of the bare Coulomb potential,
Eq.~(\ref{eq::bare_coulomb}).
The dielectric function, due to its importance for both fundamental and
applied research, was investigated very
actively~\cite{DasSarma2010,Lv2010,
Hwang2008,
Sensarma2010,
Gamayun2011,
Tirola2012,Wang2007,Wang2010a}.
For example, the static Coulomb screening was discussed in
Refs.~\cite{DasSarma2010,Lv2010}
in the context of electric transport, see
subsection~\ref{transport::subsec::fin_dop}.

As another instance of this type of study, we would like to mention
Ref.~\cite{Wang2007} by X.-F.~Wang and T.~Chakraborty, where
$\epsilon ({\bf k}, \omega)$
was determined with the help of the extremely common random-phase
approximation (RPA). In the framework of RPA, the full polarization $\Pi$ in
Eq.~(\ref{eq::diel_func})
is replaced by the one-loop polarization
$\Pi^0$,
which may be viewed as the polarization of the system with no
electron-electron interaction. To find
$\Pi^0$, 
the authors of
Ref.~\cite{Wang2007}
used a two-band effective Hamiltonian with the trigonal warping term, see
Eq.~(\ref{ab::trig_warp}).
Because of the warping, the RPA static dielectric function
$\epsilon ({\bf k}, 0)$
demonstrated noticeable anisotropy for
$k \sim p_L$
[the momentum $p_L$ characterizes the scale below which the trigonal warping is important, see Eq.~(\ref{ab::lifshits})].
At $k=0$ the value of the static $\epsilon$ was as high as $10^2$ at zero temperature, and even bigger at finite temperature.

Studies of Coulomb screening were presented in
Refs.~\cite{Hwang2008,Sensarma2010,Gamayun2011,Tirola2012,
pisarra_screening_plasmon2016}.
In
Ref.~\cite{Hwang2008},
besides screening, the Kohn anomaly, RKKY interaction, and Friedel
oscillations were investigated for both doped and undoped bilayer, modeled
by the two-band Hamiltonian. Finite temperature RKKY interaction for both
bilayer and single-layer graphene was calculated in
Ref.~\cite{klier_rkky2015}.

The many-body properties of AA bilayer were recently studied by
Y. Mohammadi~\cite{mohammadi}. He derived an analytical formula for
the static polarization function for a biased AA bilayer using random phase
approximation and the simplest tight-binding
Hamiltonian taking into account only nearest neighbors hopping and
neglecting electron-electron interaction. The obtained results were
applied to calculate Coulomb screening and the electrical conductivity.
It was argued that the short-range scattering is not affected by the
perpendicular electric filed, while the conductivity limited
by the (long-range) Coulomb scattering is enhanced by the bias voltage.

\section{Dielectric function and plasmonics in AB bilayer}
\label{sec::plasmon}

Numerous investigations have covered photonics of graphene systems and AB
bilayer in particular, including plasmonics, the nonlinear optical, and
optoelectronic properties. Here we discuss these topics very briefly.
Interested reader may consult recent
reviews~\cite{yanRev,plasmonicsRev,glazovRev}.
We also restrict our focus to AB bilayer as being experimentally relevant.
Studies of plasmon physics in AA bilayer are limited to theoretical
publications~\cite{plasmon_theor_aa_ab,plasmon_theor_aa_mult2012,Roldan2013,
wang_plasmon_aa_ab2016}.

The knowledge of the dynamic dielectric function
$\epsilon({\bf k}, \omega)$
was used by several authors to determine the dispersion of the plasmons.
Briefly recall that plasmons are quanta of the longitudinal collective
charge density oscillations. Their dispersion relation
$\omega = \omega_{\bf k}$
is a solution to the equation
\begin{eqnarray}
\epsilon({\bf k}, \omega_{\bf k}) = 0\,.
\end{eqnarray}
Reference~\cite{Wang2007}
reported that the undoped AB bilayer at
$T=0$
has a single plasmon mode. When the temperature is finite, however, an
additional weakly-damped lower-frequency mode emerges. The effects of doping on
the plasmons were briefly mentioned as well.
Reference~\cite{Wang2010a}
of the same authors discussed plasmons in a situation when the
bilayer is subjected to a transverse bias voltage, which generates a
gap in the single-particle spectrum.

To analyze screening and the plasmon mode, the author of
Ref.~\cite{Gamayun2011}
used a four-band model at finite chemical potential, but neglected the
trigonal warping. This paper reported the technical details of the the
polarization operator calculations.
References~\cite{Sensarma2010,Tirola2012}
discussed different plasmon properties. Plasmons in the presence of a
magnetic field were considered in
Ref.~\cite{Tahir2008}.

In 
Ref.~\cite{transvers_plasmon_slg2007}
S.A.~Mikhailov and K.~Ziegler argued theoretically that, besides the
longitudinal plasmons, a single-layer graphene sample can support an
additional transverse electrodynamic mode. Their analysis was based on
previous theoretical study of electrodynamics of a thin metallic
film~\cite{cond_vs_speed_of_light}.
In the context of AB~bilayer such "transverse plasmons" were discussed in
Ref.~\cite{Jablan2011},
for the twisted bilayer in
Ref.~\cite{SigmaPlasmons}.

Interesting results were obtained in recent studies of plasmons in the AB
bilayer~\cite{LowP,FeiBasov,YanTr,pisarra_screening_plasmon2016,
wang_plasmon_aa_ab2016}.
T.~Low et~al.~\cite{LowP}
theoretically studied plasmonic response in biased Bernal-stacked bilayer
graphene. The authors found that the AB bilayer ``accommodates optically
active phonon modes and a resonant interband transition at infrared
frequencies". As a result, the plasmonic properties of bilayer graphene are
strongly modified ``leading to Fano-type resonances, giant plasmonic
enhancement of infrared phonon absorption, a narrow window of optical
transparency, and a new plasmonic mode at higher energy than that of the
classical plasmon.'' This suggests that bilayer graphene may be considered
as an interesting and important plasmonic material. These conclusions were
supported in the experiments by Z. Fei
et~al.~\cite{FeiBasov}.
Using infrared nanoimaging, the authors determined that ``bilayer graphene
supports plasmons with a higher degree of confinement" when compared to
single-layer graphene or an artificial double-layer (subnanometer separated)
graphene characterized by a random stacking order. The plasmons mode were
attributed to an interlayer tunneling process. The authors were able to tune
the plasmonic mode by a gate voltage and conclude that their observation
``uncovers essential plasmonic properties in bilayer graphene and suggests
a possibility to achieve novel plasmonic functionalities in graphene
few-layers''. It should be also mentioned here that an analog of the effect
of plasmon-induced transparency was observed by H. Yan
et~al.~\cite{YanTr}
in AB nanoribbons.

%
%
%
\section{Twisted bilayer graphene}
\label{spectraTw}

In addition to bilayer graphene with AB and AA stacking, there is a class of
bilayers characterized by a non-zero (twist) angle
$\theta$ between two
graphene layers. Such twisted bilayers can be synthesized  using
different preparation techniques. For example, bilayer samples with rotation
misorientation can be produced in the process of growth on the SiC
substrate (see, e.g., Refs.~\cite{STM1,STM_DFT,MengSTM,HicksARPES}).
They can be synthesized also by chemical vapor deposition
(CVD)~\cite{STM2,STM_VHS2,DF_TEM1,DF_TEM2,RighiRaman1,RighiRaman2,
RobinsonRaman,Liu2015}. Another approach is the folding of single graphene
sheet~\cite{PoncharalRaman,NiRamanVrenorm,CarozoRaman}.
Reference~\cite{SynthesysMoire}
reported the preparation of twisted bilayer graphene (tBLG) by
transferring the graphene monolayer synthesized by CVD onto another
monolayer epitaxially grown on SiC. Similar techniques were used also in Ref.~\cite{Othmen2015}. Note also that in graphite crystals
the top layer is often rotated with respect to the deeper
layers~\cite{MoirePattern}.

Twisted bilayer graphene has a more complex crystal structure than the AA and
AB bilayers. The review of the geometrical properties of the tBLG requires a
separate subsection. The  crystal structure of the twisted bilayer affects
significantly its electronic properties. Due to its complex geometry, even
the single-particle models for the tBLG are quite involved. Because of this
complexity, the majority of the published papers on the subject are devoted
to study electron spectra (both at zero or non-zero magnetic field) in
the single-electron approximation. The papers recently studying the effects of electron-electron  interactions in twisted
bilayer graphene are virtually absent.  We hope this gap will be addressed
in future studies. The peculiarities of the system and theoretical methods
used for its study require to review the twisted bilayer
graphene in a separate Section.

\subsection{Geometrical properties: Moir\'{e} pattern and superstructure.}
\label{SubSectGeom}

\begin{figure}
\centering
\includegraphics[width=0.5\textwidth]{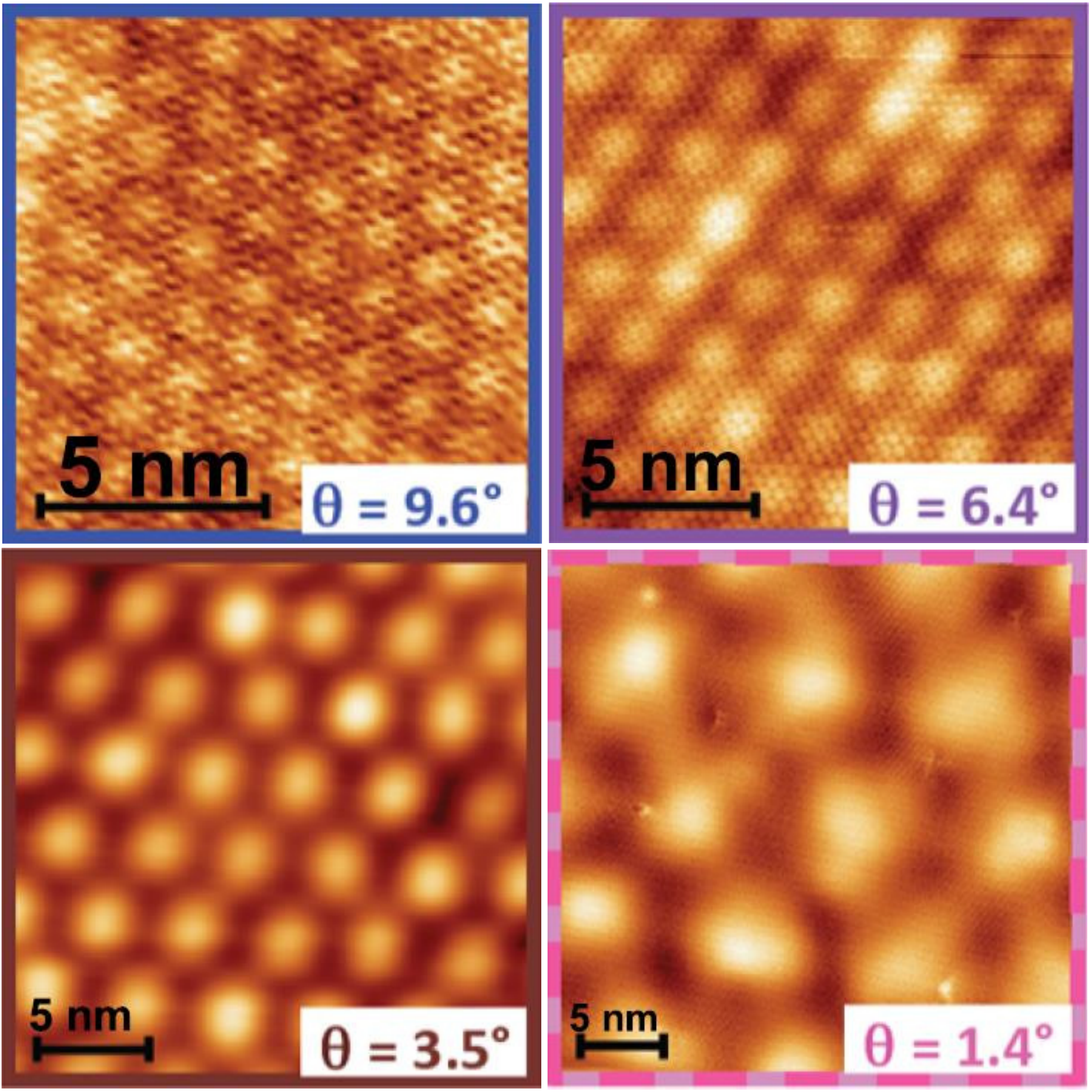}
\caption{(Color online) Scanning tunneling microscope images of bilayer graphene samples with
different twist angles $\theta$ revealing a Moir\'{e} pattern. The twist
angle is estimated by measuring the Moir\'{e} period.
Reprinted figure with permission from I. Brihuega et al., Phys. Rev.
Lett., {\bf 109}, 196802 (2012). Copyright 2012 by the American Physical
Society.
\url{http://dx.doi.org/10.1103/PhysRevLett.109.196802}.
\label{FigSTM}
}
\end{figure}

The twist of one graphene layer with respect to another one manifests
itself in the appearance of a Moir\'{e} pattern, which can be visualized
using STM techniques. Such structures have been observed in many experiments
using scanning
microscopy~\cite{STM1,STM_DFT,MengSTM,STM2,STM_VHS2,STM_VHS1,CisternasSTM}
and dark-field transmission electron
microscopy~\cite{DF_TEM1,DF_TEM2}
(DF-TEM). For different values of the twist angle
$\theta$, the STM images of the Moir\'{e} patterns are shown in Fig.~\ref{FigSTM}.
The value of the twist angle is determined from the measured Moir\'{e} period
$L$.
The latter length is defined as the distance between light (or dark)
regions in the STM images. The angle
$\theta$
and
$L$
are related according
to the following
formula~\cite{MoirePattern}:
\begin{equation}\label{MoireP}
L=\frac{a_0\sqrt{3}}{2\sin(\theta/2)}\,,\;\;\theta<30^{\circ}.
\end{equation}
where
$a_0=1.42$\,{\AA} is the in-plane C-C distance.
The Moir\'{e} pattern is a peculiar feature of twisted bilayer graphene,
which has no analog for the AA or AB bilayers. In general, for non-zero
$\theta$,
the lattice properties of the twisted bilayer are quite singular.
This subsection is dedicated to their description.

The details of the crystal structure of tBLG have been considered in many papers
(see, e.g., Refs.~\cite{dSPRL,MeleReview,dSPRB,Pankratov1,PankratovPRL}).
Here, we will follow, mainly, Ref.~\cite{dSPRB} by J.\,M.\,B.~Lopes~dos Santos et~al. Each graphene layer in tBLG consists of two sublattices ($A1$,
$B1$ in the layer $1$, and $A2$, $B2$ in the layer $2$). The positions of the carbon atoms in each sublattice in the bottom layer $1$ are
\begin{equation}
\mathbf{r}_{\mathbf{n}}^{1A}\equiv\mathbf{r}_{\mathbf{n}}=n\mathbf{a}_1+m\mathbf{a}_2\,,\;\;\;\;
\mathbf{r}_{\mathbf{n}}^{1B}=\mathbf{r}_{\mathbf{n}}+\bm{\delta}\,,
\end{equation}
where $\bm{\delta}=a_0(1,\,0)$. The quantity $\mathbf{n}=(n,m)$ is an integer-valued vector, and $\mathbf{a}_{1,2}$ are the basis vectors of
the graphene elementary unit cell,
Eq.~\eqref{a12}. The distance between graphene layers is $c_0=3.35$\,{\AA}. When the layers are not rotated ($\theta = 0$), we assume that the system is a perfect AB bilayer.

Let us, for definiteness, assume that layer $2$ is rotated around the axis connecting the atoms $A1$ and $B2$ with $\mathbf{n}=0$, while layer $1$ remains fixed. In this case, the atoms of the top layer (layer $2$) have the positions $\hat{\mathbf{e}}_zc_0+\mathbf{r}_{\mathbf{n}}^{2\alpha}$, where $\hat{\mathbf{e}}_z$ is the unit vector in the $z$ direction, and
\begin{equation}
\mathbf{r}_{\mathbf{n}}^{2B}\equiv\mathbf{r}'_{\mathbf{n}}=n\mathbf{a}'_1+m\mathbf{a}'_2\,,\;\;\;\;
\mathbf{r}_{\mathbf{n}}^{2A}=\mathbf{r}'_{\mathbf{n}}-\bm{\delta}'\,.
\end{equation}
In these equations,
\begin{equation}
\mathbf{a}'_{1,2}=\mathbf{a}_{1,2}\left(\cos\theta\mp\frac{\sin\theta}{\sqrt{3}}\,\right)\pm\mathbf{a}_{2,1}\frac{2\sin\theta}{\sqrt{3}}\,,
\end{equation}
and
$\bm{\delta}'=a_0(\cos\theta,\sin\theta)$.

While the Moir\'{e} exists for any
$\theta$,
the superstructure, that is,
strictly periodic repetition of some large multiatomic supercell, occurs
for the so-called ``commensurate"
$\theta$
only. The supercell may coincide
with the Moir\'{e} cell, but, generically, contains many Moir\'{e} cells.
For the superstructure to emerge, after the rotation, a certain atom
of the sublattice $B2$ from layer $2$
must end up exactly over an atom of the sublattice
$A1$ from layer $1$.
This happens when the twist
angle satisfies the following
relation~\cite{dSPRL,MeleReview,dSPRB,Pankratov1}
\begin{equation}
\label{theta}
\cos\theta=\frac{3m_0^2+3m_0r+r^2/2}{3m_0^2+3m_0r+r^2}\,,
\end{equation}
where $m_0$ and $r$ are mutually-prime positive integers. The number of sites in the elementary unit cell of the superlattice $N$ is
\begin{equation}\label{Nsc}
N(m_0,r)=\frac{4}{g}\left(3m_0^2+3m_0r+r^2\right),
\end{equation}
where $g=1$ if $r\neq3n$, and $g=3$ if $r=3n$ ($n$ is integer). In the notation used by E.\,J. Mele~\cite{MeleReview,MelePRB1,MelePRB2}, the structures with
$r\neq3n$ ($r=3n$) are called `odd' (`even'). The superlattice vectors can be expressed in terms of the single-layer graphene lattice vectors as:
\begin{eqnarray}
\mathbf{R}_1&=&m_0\mathbf{a}_1+(m_0+r)\mathbf{a}_2\,,\nonumber\\
\mathbf{R}_2&=&-(m_0+r)\mathbf{a}_1+(2m_0+r)\mathbf{a}_2\,,
\end{eqnarray}
if $r\neq3n$, and
 \begin{eqnarray}
\mathbf{R}_1&=&\left(m_0+\frac{r}{3}\right)\mathbf{a}_1+\frac{r}{3}\mathbf{a}_2\,,\nonumber\\
\mathbf{R}_2&=&-\frac{r}{3}\mathbf{a}_1+\left(m_0+\frac{2r}{3}\right)\mathbf{a}_2\,,
\end{eqnarray}
if $r=3n$. In both cases, the
size of the superlattice cell
$L_{SC}\equiv|\mathbf{R}_{1,2}|=a_0\sqrt{3N}/2$

\begin{figure}[t]
\centering
\includegraphics[width=0.5\columnwidth]{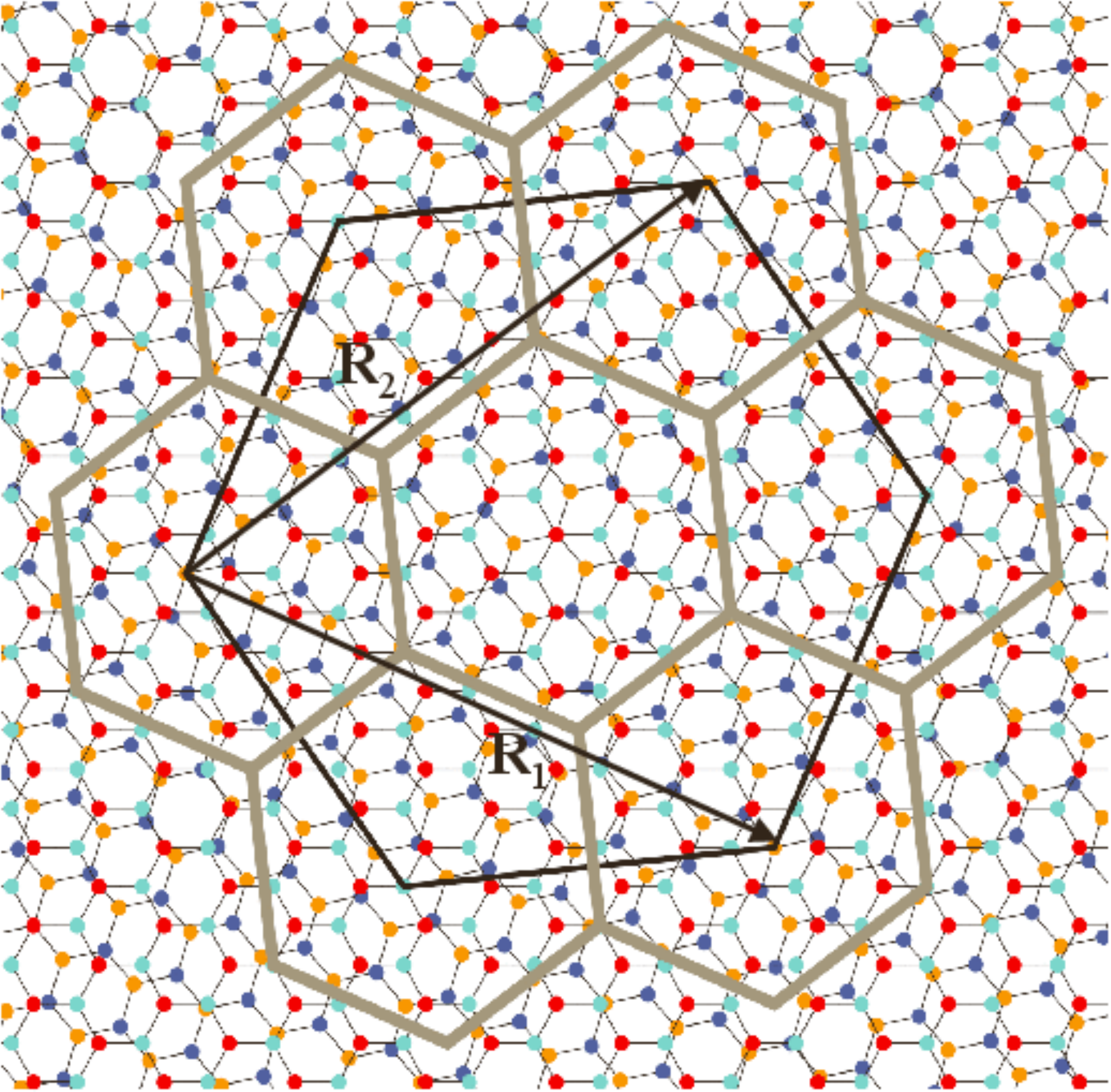}
\caption{(Color online) Structure of the twisted bilayer graphene with
$(m_0,r)=(7,3)$ ($\theta\cong12.95^{\circ}$). The large black hexagon is
the Wigner-Seitz cell of the superlattice, while $\mathbf{R}_{1,2}$ are the
superlattice vectors. This structure is an almost periodic repetition of
the structure with $(m_0,r)=(2,1)$ ($\theta\cong13.17^{\circ}$) having a
smaller size of the superlattice (gray hexagons). Reprinted figure with
permission from J.\,M.\,B. Lopes dos Santos et al.,
Phys. Rev. B, {\bf 86}, 155449 (2012).
Copyright 2012 by the American Physical Society.
\url{http://dx.doi.org/10.1103/PhysRevB.86.155449}.\label{FigTBLG}}
\end{figure}

Due to the symmetry of the single-layer graphene lattice, the rotation on
the angle
$\theta=2\pi/3$
transforms the AB bilayer to itself. In addition
to that, rotations on angles
$\theta$
and
$-\theta$
are equivalent to each
other. As a result, one can consider only twist angles in the range
$0<\theta<60^{\circ}$.
A rotation on the angle
$\theta=\pi/3$
transforms the
bilayer from AB to AA stacking. Moreover, if the angle
$\theta$
corresponds to
some commensurate structure
$(m_0,r)$,
the angle
$\theta'=\pi/3-\theta$
is
also commensurate with
$m_0'=r/g$
and
$r'=3m_0/g$,
where
$g=1$
if
$r\neq3n$,
and
$g=3$
otherwise. Thus, rotations for angles
$\theta>30^{\circ}$
can be considered as rotations for the angle
$\theta'=\pi/3-\theta<30^{\circ}$,
but starting from the AA-stacked bilayer
graphene as a reference structure. Note that
Eq.~\eqref{MoireP}
for the Moir\'{e} period is valid only for
$\theta<30^{\circ}$:
for larger
$\theta$
we have to replace
$\theta\to\theta'$.

In addition to the twist angle, the bilayer graphene can also be characterized
by a non-zero shift of one layer with respect to another one. In the
majority of the papers, however, this shift is not considered.

The authors of Ref.~\cite{PNAS} constructed an effective Hamiltonian in the continuum (low-energy) approximation for tBLG with a non-zero shift $\bm{\varrho}$ between the layers. They found, however, that this effective Hamiltonian coincides with that for $\bm{\varrho}=0$ upon a unitary transformation, so the electronic spectrum is independent on $\bm{\varrho}$. This is because, in the continuum approximation, the detailed information about the superlattice is averaged out, and the effect of non-zero $\bm{\varrho}$ is reduced only to shifting the Moir\'{e} pattern in space~\cite{PNAS}, which does not affect the electronic spectrum.

Among the superstructures
$(m_0,r)$
with
$\theta<30^{\circ}$,
there is a special subset corresponding to
$r=1$.
It can be shown from
Eq.~\eqref{theta} that in the vicinity of the angle
$\theta$,
corresponding
to some
$m_0$
and
$r=1$,
there is an infinite set of commensurate angles
with different
$m'_0$
and
$r'>1$.
These structures have {\it larger} size
supercells. However, as it was shown in
Ref.~\cite{dSPRB},
such structures are almost periodic repetitions of the structure with
$m_0$
and
$r=1$.
Figure~\ref{FigTBLG}
illustrates this fact. The structure
$(7,3)$
shown there is quite similar to the structure
$(2,1)$
with the smaller size of supercell.

For structures with
$r=1$,
the size of the superlattice
$L_{SC}$
coincides with the Moir\'{e} period
$L$,
while for all other structures
$L_{SC}>L$.
This can be easily checked using
Eq.~\eqref{theta}
for the twist angle and the
relation~\eqref{MoireP}
for the Moir\'{e}
period~\cite{dSPRB}.
The supercell of the structure
$(m_0,r)$
with
$r>1$
contains a
$r^2/g$
number of Moir\'{e} cells, where
$g=1$,
if
$r\neq3n$,
or
$g=3$
otherwise. The arrangements of carbon atoms inside these cells are
slightly different from each other and approximately correspond to the
superstructure
$([m_0/r],1)$,
where
$[a]$
means the integer part of
$a$.
For
angles
$\theta>30^{\circ}$,
the corresponding subset is
$(1,3m_0)$,
and
\begin{equation}
\theta(m_0,1)=\pi/3-\theta(1,3m_0)\,.
\end{equation}

The basis vectors of the reciprocal superlattice can be written as
\begin{eqnarray}
\mathbf{G}_1&=&\frac{4}{N}[(2m_0+r)\mathbf{b}_1+(m_0+r)\mathbf{b}_2]\,,\nonumber\\
\mathbf{G}_2&=&\frac{4}{N}[-(m_0+r)\mathbf{b}_1+m_0\mathbf{b}_2]\,,
\end{eqnarray}
if
$r\neq3n$,
and
\begin{eqnarray}
\mathbf{G}_1&=&\frac{4}{N}\left[\left(m_0+\frac{2r}{3}\right)\mathbf{b}_1+\frac{r}{3}\,\mathbf{b}_2\right]\,,\nonumber\\
\mathbf{G}_2&=&\frac{4}{N}\left[-\frac{r}{3}\,\mathbf{b}_1+\left(m_0+\frac{r}{3}\right)\mathbf{b}_2\right]\,,
\end{eqnarray}
if
$r=3n$.
In these equations,
$\mathbf{b}_{1,2}$
are the reciprocal lattice vectors of the bottom layer,
Eq.~\eqref{slgrecipV}.

\begin{figure}[t]
\centering
\includegraphics[width=0.45\columnwidth]{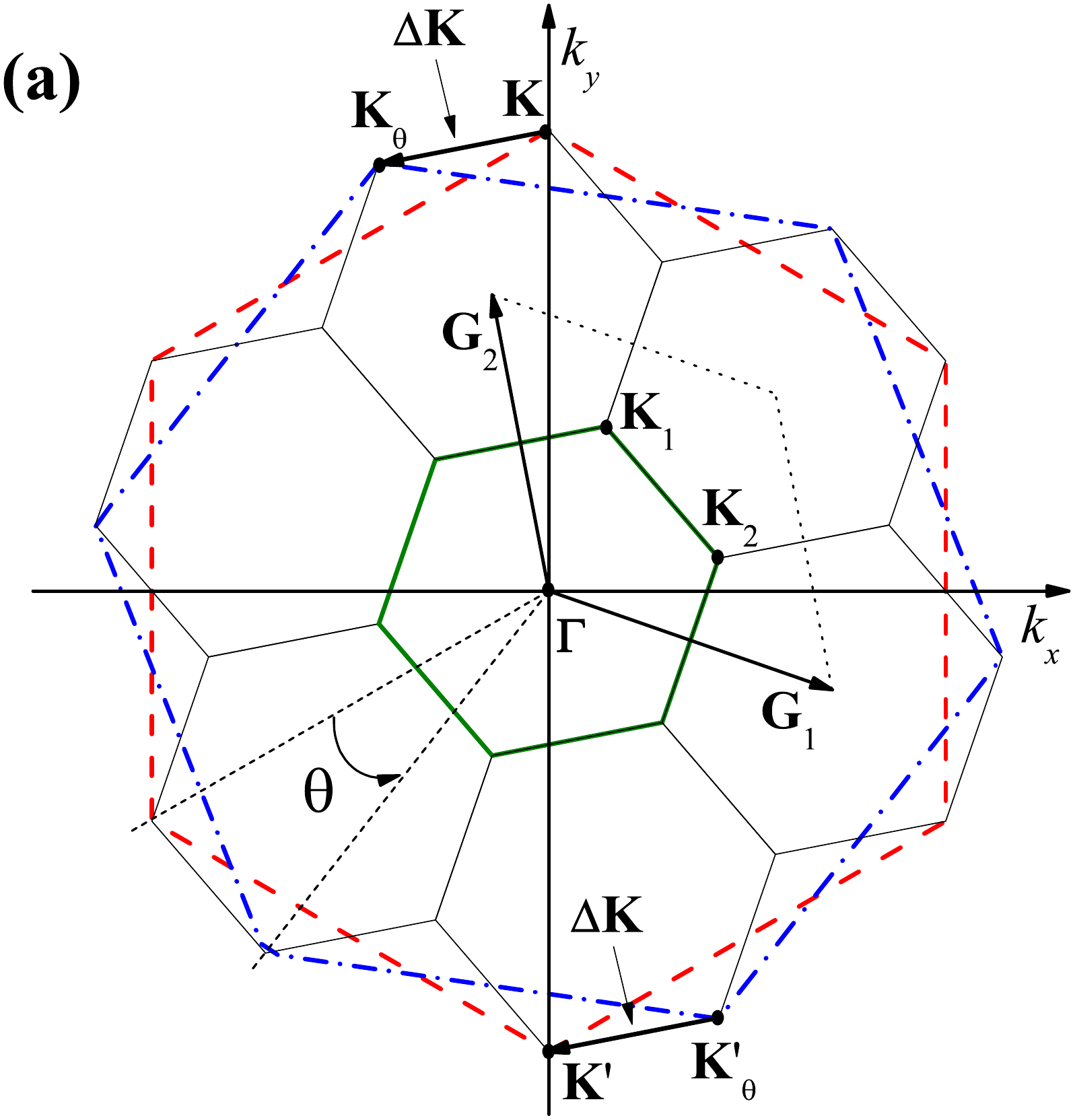}
\includegraphics[width=0.48\columnwidth]{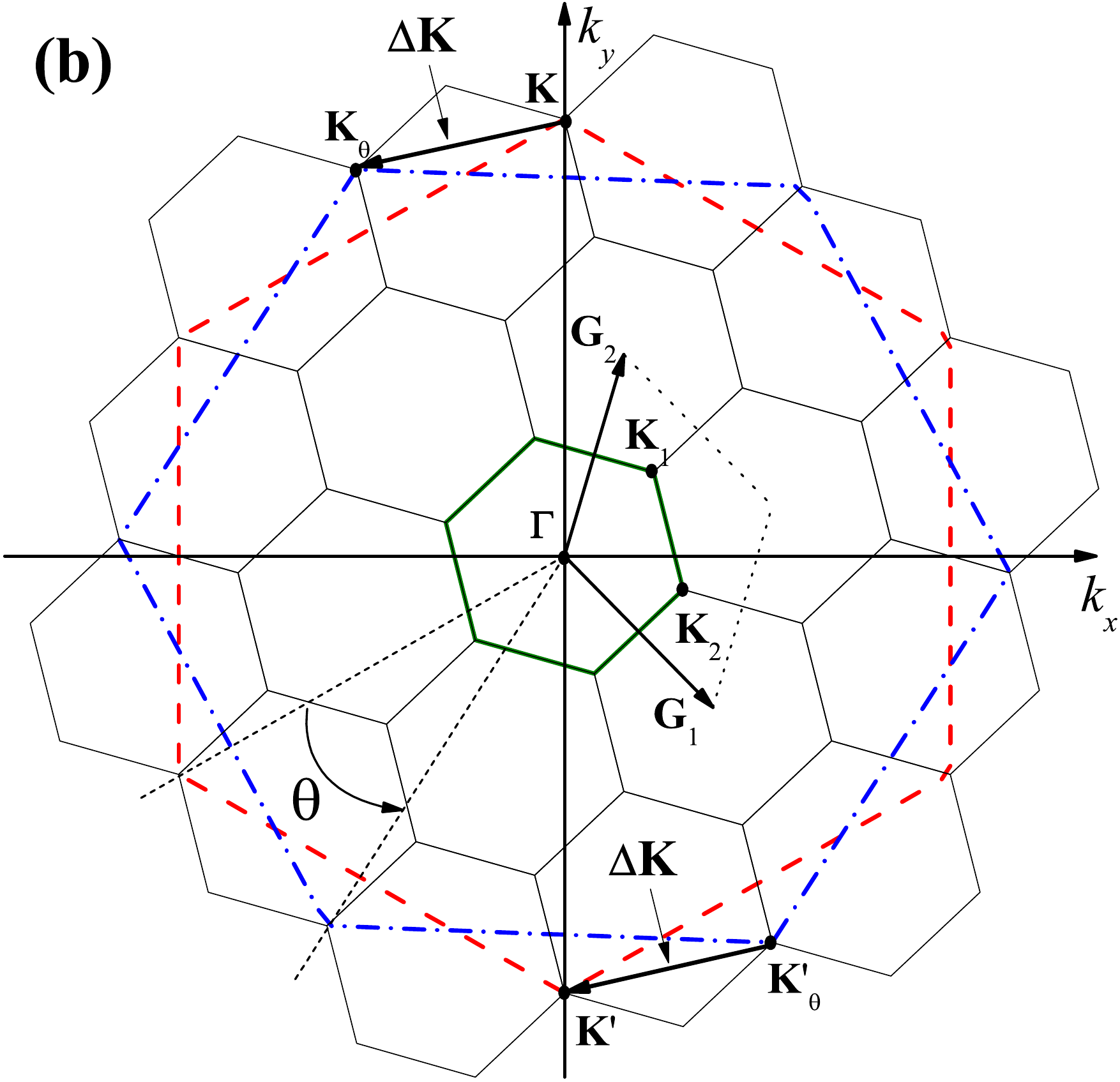}
\caption{(Color online) Geometry of the reciprocal lattice for the twisted
bilayer graphene. The green thick solid hexagon new the center shows the Brillouin zone of
the tBLG with the superstructures
$(1,1)$ [panel (a), $\theta\cong21.798^{\circ}$] and $(2,3)$ [panel (b), $\theta\cong27.796^{\circ}$].
The large hexagons show the Brillouin zones of individual layers: the red
dashed hexagon corresponds to the bottom layer, the blue dot-dashed hexagon
corresponds to the top layer. The next several Brillouin zones of the tBLG
are depicted by (black) thin solid hexagons. For the twisted bilayer, the
Dirac point
$\mathbf{K}'$
($\mathbf{K}'_{\theta}$)
is equivalent to the point
$\mathbf{K}_{\theta}$
($\mathbf{K}$)
if
$r\neq3n$.
When
$r=3n$,
$\mathbf{K}_{\theta}\sim\mathbf{K}$
and
$\mathbf{K}'_{\theta}\sim\mathbf{K}'$
(see the text), where the symbol `$\sim$' means equivalent. The twisted bilayer graphene Dirac
points $\mathbf{K}_{1,2}$
are doubly degenerate: each of them is equivalent to one of two Dirac
points of each graphene layer. For the particular case of the $(1,1)$
superstructure [panel (a)],
$\mathbf{K}_1\sim\mathbf{K}\sim\mathbf{K}'_{\theta}$
and
$\mathbf{K}_2\sim\mathbf{K}'\sim\mathbf{K}_{\theta}$,
while for the $(2,3)$
superstructure [panel (b)], we have
$\mathbf{K}_1\sim\mathbf{K}\sim\mathbf{K}_{\theta}$
and
$\mathbf{K}_2\sim\mathbf{K}'\sim\mathbf{K}'_{\theta}$.
The vector
$\Delta {\bf K} = {\bf K}_\theta - {\bf K}$
is shown for both structures. For the structure in panel~(b) it coincides
with the reciprocal vector of the superlattice, in panel~(a) vector
$\Delta {\bf K}$
does not equal to the reciprocal vector of the superlattice.
\label{FigTBLGBZ}
}
\end{figure}

The Brillouin zone of the superlattice has the shape of a hexagon with
sides
$Q_{BZ}=|\mathbf{G}_{2}-\mathbf{G}_{1}|/3$.
In the particular case
$r=1$,
this side is equal to
$\Delta K=|\mathbf{K}_{\theta}-\mathbf{K}|$,
where
$\mathbf{K}=4\pi(0,\,1)/(3\sqrt{3}a_0)$
and
$\mathbf{K}_{\theta}=4\pi(-\sin\theta,\,\cos\theta)/(3\sqrt{3}a_0)$
are the Dirac points of the bottom and top layers, respectively. For
$r>1$, $Q_{BZ}$
is smaller than
$\Delta K$
and is equal to
$Q_{BZ}=\sqrt{g}\Delta K/r$,
where
$g=1$
if
$r\neq3n$,
and
$g=3$
otherwise.
Figure~\ref{FigTBLGBZ} shows the Brillouin zones of the superstructures
$(1,1)$
and
$(2,3)$
having twist angles
$\theta\cong21.798^{\circ}$
and
$\theta\cong27.796^{\circ}$,
respectively. For comparison, the Brillouin zones
of the rotated and static layers are also shown in this figure. Recall that each graphene layer has two non-equivalent Dirac points,
$\mathbf{K}$
and
$\mathbf{K}'$($=-\mathbf{K}$)
for the bottom layer, and
$\mathbf{K}_{\theta}$
and
$\mathbf{K}_{\theta}'$($=-\mathbf{K}_{\theta}$)
for the top (rotated) layer. To find out where these Dirac points are
located in the Brillouin zone of the superstructure one can express their
co-ordinates in terms of reciprocal superlattice vectors. For
$r\neq3n$
we have
\begin{eqnarray}
\mathbf{K}&=&-\mathbf{K}'=m_0\mathbf{G}_2+\frac{r}{3}\left(\mathbf{G}_1+2\mathbf{G}_2\right),\nonumber\\
\mathbf{K}_{\theta}&=&-\mathbf{K}'_{\theta}=m_0\mathbf{G}_2+\frac{r}{3}\left(\mathbf{G}_2-\mathbf{G}_1\right)\,,\label{K1}
\end{eqnarray}
while for
$r=3n$
\begin{eqnarray}
\mathbf{K}&=&-\mathbf{K}'=\frac{r}{3}\mathbf{G}_2+\frac{m_0}{3}\left(\mathbf{G}_2-\mathbf{G}_1\right),\nonumber\\
\mathbf{K}_{\theta}&=&-\mathbf{K}'_{\theta}=-\frac{r}{3}\mathbf{G}_1+\frac{m_0}{3}\left(\mathbf{G}_2-\mathbf{G}_1\right)\,.\label{K3}
\end{eqnarray}
It follows from these formulas, that the points
$\mathbf{K}'$
and
$\mathbf{K}_{\theta}$
(as well as $\mathbf{K}$ and $\mathbf{K}'_{\theta}$) are equivalent to each other if $r\neq3n$
[see Fig.~\ref{FigTBLGBZ}(a)], that is, the difference
$\mathbf{K}'-\mathbf{K}_{\theta}$
is the reciprocal vector of the superlattice. When $r=3n$,
other pairs of Dirac points are equivalent:
$\mathbf{K}\sim\mathbf{K}_{\theta}$ and $\mathbf{K}'\sim\mathbf{K}'_{\theta}$
[see Fig.~\ref{FigTBLGBZ}(b)]. Thus, for any commensurate angle, there are two doubly-degenerate non-equivalent Dirac points of the twisted bilayer. Inside the reciprocal cell of the superlattice, these two non-equivalent tBLG Dirac points are located at:
\begin{equation}\label{K12}
\mathbf{K}_1=\frac{\mathbf{G}_1+2\mathbf{G}_2}{3}\,,\;\;\;\;\mathbf{K}_2=\frac{2\mathbf{G}_1+\mathbf{G}_2}{3}\,,
\end{equation}
for any superstructure (see
Fig.~\ref{FigTBLGBZ}).
This double degeneracy affects the electronic structure of the tBLG,
leading to band splitting and band-gap formation (see
subsection~\ref{lowenergysplitting}).

\subsection{Effective low-energy theories of twisted bilayer graphene
electronic spectrum}

\subsubsection{Low-energy Hamiltonians}

The specific electronic properties of tBLG are mostly determined by the
interlayer single-electron hopping and the twist angle. For
commensurate structures, numerical studies based on density functional
theory and tight-binding calculations were performed in numerous
papers~\cite{PankratovFlakes,STM_DFT,Pankratov1,PankratovPRL,PankratovDFT,
NanoLettTB,TramblyTB_Loc,Morell1,Morell2,PankratovPRB2013,
LatilDFT,DeltaUeffDFT,RamanTheory,LargeScaleDFT_PRB2014,LargeScaleDFT2015}.
Since the elementary unit cell of the tBLG superlattice contains a large
number of sites, especially at small twist angles, the {\it ab initio}
calculations incur a significant computational cost.

To avoid this difficulty, several semi-analytical theories have been
developed for describing the low-energy electronic properties of
tBLG~\cite{dSPRL,dSPRB,MelePRB1,MelePRB2,PNAS,alaMele,deGail,
LL_lowEnergy,NonAbelianGaugePot,ChuEffTheor,TBLGneutrino,
low_en_pankratov2016}.
These theories operate mainly on the electronic states near the Dirac
cones, which the tBLG inherits from its two constituent layers. In
tBLG, the Dirac cones having their origins at
$\mathbf{K}$ and $\mathbf{K}_{\theta}$ (as well as $\mathbf{K}'$ and $\mathbf{K}'_{\theta}$)
are located close to each other in momentum space (for
$\theta<30^{\circ}$).
This allows constructing an effective low-energy Hamiltonian for the Dirac
electrons moving in each graphene layer and hybridized by inter-layer
hopping.

The first such low-energy theory was proposed by J.\,M.\,B.~Lopes dos Santos et
al.~\cite{dSPRL}
and further developed in
Ref.~\cite{dSPRB}.
The authors consider the following effective Hamiltonian of tBLG:
\begin{eqnarray}
\hat{H}&=&\hbar v_F\sum_{\mathbf{k}\alpha\beta}
	d^{\dag}_{1\mathbf{k}\alpha}
	\bm{\sigma}_{\alpha\beta}
	\left(
		\mathbf{k}+\frac{\Delta\mathbf{K}}{2}
	\right)
	d^{\phantom{\dag}}_{1\mathbf{k}\beta}
	+
	\hbar v_F\sum_{\mathbf{k}\alpha\beta}
	d^{\dag}_{2\mathbf{k}\alpha}
	\bm{\sigma}^{\theta}_{\alpha\beta}
	\left(
		\mathbf{k}-\frac{\Delta\mathbf{K}}{2}
	\right)
	d^{\phantom{\dag}}_{2\mathbf{k}\beta}
+\nonumber\\
&&
\sum_{\mathbf{k}\alpha\beta} \sum_{\mathbf{G}}
	\left(
		\tilde{t}_{\bot}^{\beta\alpha}(\mathbf{G})
		d^{\dag}_{1\mathbf{k}+\mathbf{G}\alpha}
		d^{\phantom{\dag}}_{2\mathbf{k}\beta}
		+ \text{h.c.}
	\right)\,.
\label{HdS}
\end{eqnarray}
The first two terms here describe the Dirac quasiparticles moving in layers
$1$ and $2$ respectively, where
$d^{\dag}_{i\mathbf{k}\alpha}$
and
$d^{\phantom{\dag}}_{i\mathbf{k}\alpha}$
($i=1,2$) are the creation and
annihilation operators of the electron with the momentum
$\mathbf{k}$,
in layer $i$,
in the sublattice
$\alpha$.
The momentum
$\mathbf{k}$
is
measured from the point
$\mathbf{K}_0=(\mathbf{K}_{\theta}+\mathbf{K})/2$,
and additional terms
$\pm\Delta\mathbf{K}/2$
in brackets (where
$\Delta\mathbf{K}=\mathbf{K}_{\theta}-\mathbf{K}$),
appear due to the position of the Dirac points of the bottom and top layers
with respect to
$\mathbf{K}_0$.
The labels
$\alpha,\beta$
are the sublattice indices,
$\bm{\sigma}$
are Pauli matrices acting in the sublattice space, while
$\bm{\sigma}^{\theta}$
is defined as
$\bm{\sigma}^{\theta}=
e^{i\theta\sigma_z/2}\bm{\sigma}e^{-i\theta\sigma_z/2}$.
The use of
$\bm{\sigma}^{\theta}$
instead of
$\bm{\sigma}$
is associated with the rotation of the second layer.
Spin and valley degeneracy are assumed in this
model, and the corresponding indices in the electronic operators are omitted.

The last term in
Eq.~\eqref{HdS} describes inter-layer hopping. The second summation is performed over the vectors
$\mathbf{G}=n\mathbf{G}_1+m\mathbf{G}_2$
($n,m$
are integers) lying inside the single-layer graphene Brillouin zone, and
$\tilde{t}_{\bot}^{\beta\alpha}(\mathbf{G})$
are the Fourier transform of the interlayer hopping amplitudes, which can be written in the following form
\begin{equation}
\tilde{t}_{\bot}^{\beta\alpha}(\mathbf{G})=\frac{4}{N}\!\!\sum_{\mathbf{n}\in SC}t_{\bot}[\bm{\delta}^{\beta\alpha}(\mathbf{r}_{\mathbf{n}}^{1\alpha})]e^{-i\mathbf{G}'\mathbf{r}_{\mathbf{n}}^{1\alpha}}\,,
\end{equation}
where
$\mathbf{G}'=\Delta\mathbf{K}+\mathbf{G}$
and the summation is performed over the sites inside the superlattice cell. The function
$t_{\bot}[\bm{\delta}^{\beta\alpha}(\mathbf{r}_{\mathbf{n}}^{1\alpha})]$
is the amplitude of the electron hopping from the site in the layer
$1$
located at
$\mathbf{r}_{\mathbf{n}}^{1\alpha}$
to the nearest-neighbor site in the layer
$2$
located at
$c_0\hat{\mathbf{e}}_z+\mathbf{r}_{\mathbf{n}}^{1\alpha}+\bm{\delta}^{\beta\alpha}$.
Only nearest-neighbor hopping is considered, even though the authors mention that results are not changed qualitatively if the more distant hopping will be taken into account. These hopping amplitudes are calculated numerically using the procedure proposed in
Ref.~\cite{Tang} (see also Section~\ref{TB_tBLG} for more details).

The authors showed that the amplitudes
$\tilde{t}_{\bot}^{\beta\alpha}(\mathbf{G})$
decay rapidly with increasing $|\mathbf{G}|$.
As a result, in the first approximation, one can keep only three terms in
the sum over $\mathbf{G}$ in
Eq.~\eqref{HdS} with $\mathbf{G}=0$, $\mathbf{G}=\mathbf{G_1}$,
and $\mathbf{G}=\mathbf{G}_1+\mathbf{G}_2$. The corresponding values of $\mathbf{G}'=\Delta\mathbf{K}+\mathbf{G}$
have the same magnitude, $|\mathbf{G}|=|\mathbf{G}'|=\Delta K=2K\sin(\theta/2)$, where
$K=|\mathbf{K}|=4\pi/(3\sqrt{3}a_0)$. The tunneling amplitudes for these values of
$\mathbf{G}$ are not independent, and satisfy the following symmetry relations:
\begin{equation}
\tilde{t}_{\bot}^{\alpha\beta}(0)=\tilde{t}_{\bot}
\left(\begin{array}{cc}1&1\\1&1\end{array}\right)\,,\;\;
\tilde{t}_{\bot}^{\alpha\beta}(\mathbf{G_1})=
\tilde{t}_{\bot}
\left(
	\begin{array}{cc}z&1\\\bar{z}&z\end{array}
\right)\,,\;\;
\tilde{t}_{\bot}^{\alpha\beta}(\mathbf{G_1}+\mathbf{G_2})
=\tilde{t}_{\bot}
\left(
	\begin{array}{cc}\bar{z}&1\\z&\bar{z}\end{array}
\right)\,,
\label{Tsym}
\end{equation}
where
$z=e^{2\pi i/3}$
and the bar over
$z$
means complex conjugation. Thus, all interlayer
hopping amplitudes depend on the single parameter
$\tilde{t}_{\bot}$,
which the authors calculated numerically. Their result,
$\tilde{t}_{\bot}\cong0.4t_0$,
where
$t_0$
is the inter-plane nearest neighbor hopping amplitude for the AB bilayer,
shows very weak dependence on
$\theta$.

Due to the hybridization between electrons with momenta
$\mathbf{k}$
and
$\mathbf{k}+\mathbf{G}$
with different
$\mathbf{G}$,
the
Hamiltonian~\eqref{HdS} cannot be diagonalized analytically. The authors
diagonalize it numerically by truncating the matrix Hamiltonian (the
minimum rank of truncated matrix used is
$12$), but some results can be
obtained by perturbation theory on parameter
$\tilde{t}_{\bot}$.
An analysis of the spectrum for the commensurate structures with
$r=1$
shows that, as we explained above, all different structures are almost
periodic repetitions of the structures with
$r=1$~\cite{dSPRB}.
As a result, the low-energy Hamiltonian for structures with
$m_0'$
and
$r'\neq1$
can be effectively reduced to the
Hamiltonian~\eqref{HdS}
for the structure with
$m_0$
and
$r=1$
such that
$\theta(m_0',r')\approx\theta(m_0,1)$.

\begin{figure}[t]
\centering
\includegraphics[width=0.66\columnwidth]{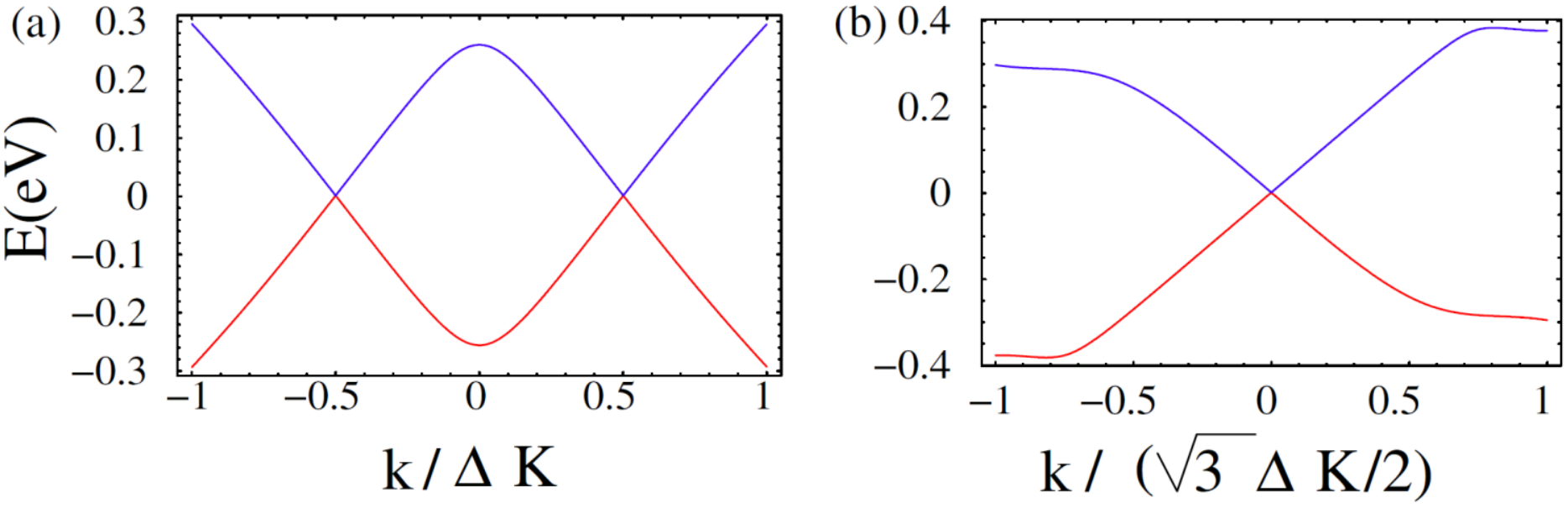}\hspace{5mm}
\includegraphics[width=0.25\columnwidth]{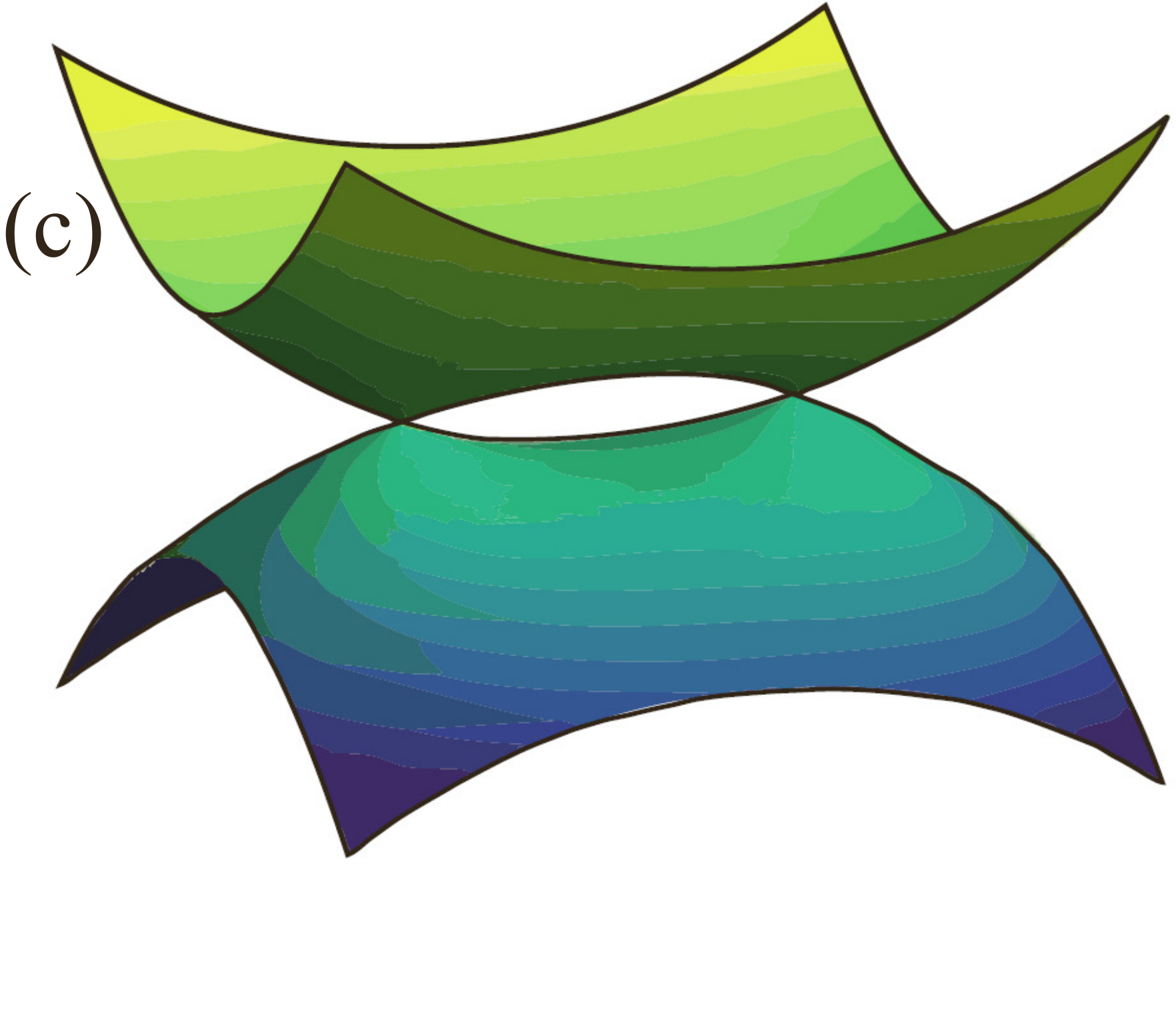}
\caption{(Color online) (a) Spectra of the low-energy tBLG bands calculated for $\theta=3.9^{\circ}$ ($m_0=8$, $r=1$) in Ref.~\cite{dSPRL}; the momentum $\mathbf{k}$ changes from $\mathbf{k}=\mathbf{K}_0-\Delta\mathbf{K}$ to $\mathbf{k}=\mathbf{K}_0+\Delta\mathbf{K}$ passing the Dirac points, $\mathbf{K}$
and $\mathbf{K}_{\theta}$ (points $k/\Delta K=\mp1/2$ in the plot). (b) The same as for (a), but for the momentum $\mathbf{k}$ varying in the perpendicular direction and passing the Dirac point $\mathbf{K}$ ($k=0$ in the plot).
Reprinted figures with permission from J.\,M.\,B. Lopes dos Santos,  et
al., Phys. Rev. B, {\bf 99}, 256802 (2007). Copyright 2007 by the American
Physical Society.  \url{http://dx.doi.org/10.1103/PhysRevLett.99.256802}.
(c) Schematic of the low-energy spectrum according to
Eq.~\eqref{VrenormGail}.
Reprinted figure with permission from M.-Y. Choi et al.,
Phys. Rev. B, {\bf 84}, 195437 (2011). Copyright 2011 by the American
Physical Society.  \url{http://dx.doi.org/10.1103/PhysRevB.84.195437}.
\label{FigSpecDS}}
\end{figure}

Similar approach based on continuum  approximation was proposed by R.~Bistritzer and A.H.~MacDonald~\cite{PNAS}.
In the latter reference the interlayer hopping Hamiltonian in real space
was expressed as
\begin{equation}\label{TPNAS}
T^{\alpha\beta}(\mathbf{r})
=
\sum_{j=1}^{3}e^{-i\mathbf{q}_{j}\mathbf{r}}T_j^{\alpha\beta}\,,
\end{equation}
where
$\mathbf{q}_1=\Delta\mathbf{K}$,
vectors
$\mathbf{q}_{2,3}$
are obtained by the rotation of the vector
$\mathbf{q}_1$
by angles
$\pm2\pi/3$,
respectively, and matrices
$T_j^{\alpha\beta}$
are given by
Eq.~\eqref{Tsym}. In the limit of
$\theta=0$,
corresponding to the AB bilayer, the matrix
$T^{\alpha\beta}$
is reduced to
$$
3\tilde{t}_{\bot}\!\left(\begin{array}{cc}0&1\\0&0\end{array}\right).
$$
This gives the value
$\tilde{t}_{\bot}=t_0/3$,
which deviates somewhat from the result
$\tilde{t}_{\bot}\cong0.4t_0$,
obtained by J.\,M.\,B.~Lopes dos Santos et al. The total Hamiltonian can be written as
\begin{equation}
\label{HPNAS}
\hat{H}
=
\left(
	\begin{array}{cc}
		-i\hbar v_F\bm{\sigma}^{\frac{-\theta}{2}}
		\bm{\nabla}&\hat{T}(\mathbf{r})\\
		\hat{T}^{\dag}(\mathbf{r})&
		-i\hbar v_F\bm{\sigma}^{\frac{\theta}{2}}\bm{\nabla}
	\end{array}
\right)\,.
\end{equation}
Hamiltonian~\eqref{HPNAS} is periodic with a Moir\'{e} period $L$.
The authors~\cite{PNAS} calculated the tBLG spectrum employing a technique similar to that used in Ref.~\cite{dSPRB}. This approach~\cite{PNAS} is valid for arbitrary twist angles,
not necessarily commensurate ones. If the twist angle corresponds to the
commensurate structure with
$r=1$, the Hamiltonian~\eqref{HPNAS}
becomes equivalent to the Hamiltonian~\eqref{HdS}
(when only three terms shown by
Eq.~(\ref{Tsym})
in the summation over
$\mathbf{G}$
are taken into account).

The Hamiltonian~(\ref{HPNAS}) can be generalized~\cite{PNAS} to include not only the relative rotation of the layers, but also the relative shift between the layers $\bm{\varrho}$. For non-zero shift, the matrices $T_j^{\alpha\beta}$ acquire addition phase factors
$$
e^{-i\bm{{\cal B}}_j\bm{\varrho}}\,,\;\;\text{where}\;\;\bm{{\cal B}}_1=0\,,\;\bm{{\cal B}}_2=\mathbf{b}_1\,,\;\text{and}\;\;
\bm{{\cal B}}_3=-\mathbf{b}_2\,.
$$
These factors, however, can be eliminated by a unitary transformation.
Therefore, the effective Hamiltonian $H$ for non-zero $\bm{\varrho}$ is equivalent to that for $\bm{\varrho}=0$,
Eq.~\eqref{HPNAS}.
Thus, in the continuum (low-energy) approximation, the tBLG spectrum is
independent of the shift
$\bm{\varrho}$.

Typical curves of the low-energy spectrum of tBLG for $r=1$,
calculated by diagonalizing the Hamiltonian~\eqref{HdS}, are
presented in Fig.~\ref{FigSpecDS}. The dispersion surface
$\varepsilon = \varepsilon ({\bf k})$
forms two Dirac cones near the Dirac points of two graphene layers,
$\mathbf{K}$  and $\mathbf{K}_{\theta}$.
The cones merge at higher energies. For tBLG, the Fermi velocity
$v_F^{*}$
near the Dirac points turns out to be smaller than the Fermi velocity
$v_F$
for the single-layer graphene. The authors of
Ref.~\cite{dSPRB}
calculated the dependence
$v_F^{*}=v_F^{*} (\theta)$,
both numerically and analytically, using perturbation theory on the parameter
$\tilde{t}_{\bot}$.
In the latter approach, they obtained
\begin{equation}\label{VrenormDS}
v_F^{*}=v_F\left[1-\left(\frac{3\tilde{t}_{\bot}}{\hbar v_F\Delta K}\right)^2\right]\,.
\end{equation}
This formula shows that the accuracy of the perturbation theory is controlled by
the dimensionless parameter
$\tilde{t}_{\bot}/\hbar v_F\Delta K$.
For
$\tilde{t}_{\bot}=0.11$\,eV,
the deviation of the numerical result from
Eq.~(\ref{VrenormDS})
is significant if
$\theta\lesssim5^{\circ}$,
and higher-order contributions should be taken into account.
The renormalized Fermi velocity goes to zero at some critical angle
$\theta_c$.
The value of
$\theta_c$
cannot be found by perturbation theory, and numerical calculations are
needed. R. Bistritzer and A.H. MacDonald~\cite{PNAS} obtained the relation
for the renormalized Fermi velocity
\begin{equation}\label{VrenormPNAS}
v_F^{*}=v_F\left[1-\frac{1-3\alpha^2}{1+6\alpha^2}\right]\,,\;\;\;\alpha=\frac{\tilde{t}_{\bot}}{\hbar v_F\Delta K}\,,
\end{equation}
which coincides with
Eq.~\eqref{VrenormDS} in the limit of small $\alpha$. It is interesting to note that such a spectrum of the tBLG is equivalent to the spectrum of the AB bilayer subject to the large in-plane magnetic field~\cite{PseudoMagneticField2014}.

Using further simplifications of the low-energy Hamiltonian, R. de Gail
with coauthors~\cite{deGail} constructed a simple effective
$2\times2$
Hamiltonian of tBLG allowing to obtain analytical expressions for the
low-energy spectrum. They started from a $4\times4$
Hamiltonian in the form
\begin{equation}\label{HdeGail0}
\hat{H}=\left(\begin{array}{cc}H_0(\mathbf{k}+\Delta\mathbf{K}/2)&\hat{H}_{\bot}^{\phantom{\dag}}\\
\hat{H}_{\bot}^{\dag}&\hat{H}_0(\mathbf{k}-\Delta\mathbf{K}/2)\end{array}\right)\,,
\end{equation}
where
$\hat{H}_0(\mathbf{k})=\hbar v_F\bm{\sigma}\mathbf{k}$,
and
$\hat{H}_{\bot}$
is the operator describing interlayer hopping. In the real-space representation,
$\hat{H}_{\bot}$
has a form of
Eq.~\eqref{TPNAS}. The authors consider the limit of small angles, when
$t_{\bot}\gg\hbar v_F\Delta K$.
As a first simplification, the operator
$\hat{H}_{\bot}$
is replaced to that corresponding to the AB bilayer ($\theta=0$),
\begin{equation}\label{TdeGail0}
\hat{H}_{\bot}\to t_0\!\left(\begin{array}{cc}0&1\\0&0\end{array}\right).
\end{equation}
The system of Schr\"odinger equations for the bi-spinor wave function
$\psi=(\psi_{A1},\,\psi_{B1},\,\psi_{A2},\,\psi_{B2})^{\text{T}}$
then becomes
\begin{eqnarray}
\hbar v_F\bar{k}_{+}\psi_{B1}+t_{\bot}\psi_{B2}&=&E\psi_{A1}\label{psiA1}\,,\\
\hbar v_Fk_{+}\psi_{A1}&=&E\psi_{B1}\label{psiB1}\,,\\
\hbar v_F\bar{k}_{-}\psi_{B2}&=&E\psi_{A2}\label{psiA2}\,,\\
\hbar v_Fk_{-}\psi_{A2}+t_{\bot}\psi_{A1}&=&E\psi_{B2}\,,\label{psiB2}
\end{eqnarray}
where
\begin{equation}\label{kpm}
k_{\pm}=(k_x\pm\Delta K_{x}/2)+i(k_y\pm\Delta K_{y}/2).
\end{equation}
For
$t_{\bot}\gg\hbar v_F\Delta K$
and small
$|E|$,
we determine from the above equations that
$\psi_{A1}\sim\psi_{B2}\ll\psi_{A2}\sim\psi_{B1}$.
Thus, the right-hand sides of
equations~\eqref{psiA1}
and~\eqref{psiB2}
can be set to zero. As a result, one obtains from these equations
$\psi_{A1}=-\hbar v_Fk_{-}\psi_{A2}/t_{\bot}$
and
$\psi_{B2}=-\hbar v_F\bar{k}_{+}\psi_{A2}/t_{\bot}$.
Substituting these relations to
Eqs.~\eqref{psiA2} and~\eqref{psiB1},
the following expression is derived
\begin{equation}
\label{EqdeGailEff}
\hat{H}^{\text{eff}}(\mathbf{k})
\left(\begin{array}{c}\psi_{A2}\\\psi_{B1}\end{array}\right)
=
E\left(\begin{array}{c}\psi_{A2}\\\psi_{B1}\end{array}\right)\,.
\end{equation}
Here, the effective
$2\times2$
Hamiltonian is
\begin{equation}\label{HTBLGEff}
\hat{H}^{\text{eff}}(\mathbf{k})=-\frac{\hbar v_F^{*}}{\Delta K}\left(\begin{array}{cc}0&\bar{k}_{+}\bar{k}_{-}\\
k_{+}k_{-}&0\end{array}\right),\;\;\;v_F^{*}=\frac{\hbar v_F^2\Delta K}{t_0}\,.
\end{equation}
The eigenvalues of this Hamiltonian are
\begin{equation}\label{VrenormGail}
\varepsilon(\mathbf{k})=\pm\frac{\hbar v_F^{*}}{\Delta K}\left|\mathbf{k}+\frac{\Delta\mathbf{K}}{2}\right|
\left|\mathbf{k}-\frac{\Delta\mathbf{K}}{2}\right|\,.
\end{equation}
The spectrum is schematically shown in
Fig.~\ref{FigSpecDS}(c).
It qualitatively captures the main features of the low-energy bands
calculated by the numerical diagonalization of the Hamiltonians~\eqref{HdS} or~\eqref{HPNAS}:
there are two Dirac cones located at
$\pm\Delta\mathbf{K}/2$ merging at higher energies, which are characterized by the renormalized
Fermi velocity. However, the dependence of
$v_F^{*}$
on
$\theta$
differs from the result of the numerical
calculations~\cite{dSPRL,PNAS}.
The estimates given by
Eq.~\eqref{VrenormDS}
or
Eq.~\eqref{VrenormPNAS}
are consistent with
Eq.~\eqref{VrenormGail}
only in the range
$\theta_c<\theta\lesssim5^{\circ}$.
Moreover, numerical calculations~\cite{dSPRB,PNAS}
show that the picture with Dirac cones becomes completely irrelevant for angles
below the critical
$\theta_c\approx1^{\circ}$,
while
Eq.~\eqref{VrenormGail}
predicts the existence of Dirac cones for arbitrary-small twist angles.

In Ref.~\cite{MelePRB2}, E.\,J.~Mele considered the effects of the lattice symmetry on the electronic properties of the tBLG in the framework of the continuum approximation. He derived a more general expression for the matrices describing the interlayer hopping [cf. with
Eq.~\eqref{Tsym}]
\begin{equation}
\tilde{t}_{\bot}^{\alpha\beta}(0)=\left(\begin{array}{cc}c_{aa}&c_{ab}\\c_{ba}&c_{bb}\end{array}\right)\,,\;\;
\tilde{t}_{\bot}^{\alpha\beta}(\mathbf{G_1})=\left(\begin{array}{cc}zc_{aa}&c_{ab}\\\bar{z}c_{ba}&zc_{bb}\end{array}\right)\,,\;\;
\tilde{t}_{\bot}^{\alpha\beta}(\mathbf{G_1}+\mathbf{G_2})=\left(\begin{array}{cc}\bar{z}c_{aa}&c_{ab}\\zc_{ba}&\bar{z}c_{bb}\end{array}\right)\,,\label{Tmele}
\end{equation}
where the parameters
$c_{aa}$,
$c_{ab}$,
$c_{ba}$,
and
$c_{bb}$
are expressed via the interlayer hopping amplitudes for the AB bilayer graphene,
$t_0$,
$t_3$,
and
$t_4$
(for definition, see
Fig.~\ref{fig::ab_lattice}), as
\begin{equation}
c_{aa}=c_{bb}=\frac{t_0-t_3}{3}+t_4\,,\;\;\;c_{ab}=c_{ba}=\frac{t_0+2t_3}{3}\,.
\end{equation}
The form of the matrices in
Eq.~\eqref{Tmele} coincides with the previously discussed
result~\eqref{Tsym} only for the `isotropic' case $t_3=t_4$. For $t_3=t_4=0$,
the matrices in
Eq.~\eqref{Tmele}
exactly coincide with that found by R.~Bistritzer and A.H.~MacDonald~\cite{PNAS}.
It is known, however, that the parameters $t_3$ and $t_4$
can substantially deviate from each other, despite the fact that the
distances separating the corresponding pairs of carbon atoms are the
same~\cite{tunnel2,AB::Mendez,AB::Mucha}.
For example, the estimate of these parameters for graphite~\cite{tunnel2,AB::Mendez}
produces a $t_3$ which is only slightly smaller than $t_0$
(the largest interlayer hopping amplitude), while $t_4$
is about 10 times smaller.
E.\,J.~Mele~\cite{MelePRB2} calculated the electron spectrum for
different values of two independent parameters
$c_{aa}$($=c_{bb}$)
and
$c_{ab}$($=c_{ba}$).
He reproduced the results of
Refs.~\cite{dSPRL,dSPRB,PNAS}
for the `isotropic' case. However, when
$c_{aa}\ll c_{ab}$,
or
$c_{aa}\gg c_{ab}$,
the spectrum structure differs from the picture with Dirac cones described
above, especially when $c_{aa}$ or $c_{ab}$ is comparable to
$\hbar v_F\Delta K$.
For
$c_{aa}>c_{ab}$,
that is, when
$t_4>t_3$,
the reduction of the Fermi velocity of the Dirac cones is strongly
suppressed.
E.\,J.~Mele~\cite{MelePRB2}
suggested that this can be a possible explanation of the lack of Fermi
velocity reduction experimentally observed for some tBLG
samples~\cite{HicksARPES,STM_VHS2,MillerLLexp,SprinkleARPES,
razado2016nanoarpes_tblg}.

\subsubsection{Fermi velocity reduction and low-energy van Hove singularities}\label{VHS_VfRenorm}

\begin{figure}[t]
\centering
\includegraphics[width=0.99\columnwidth]{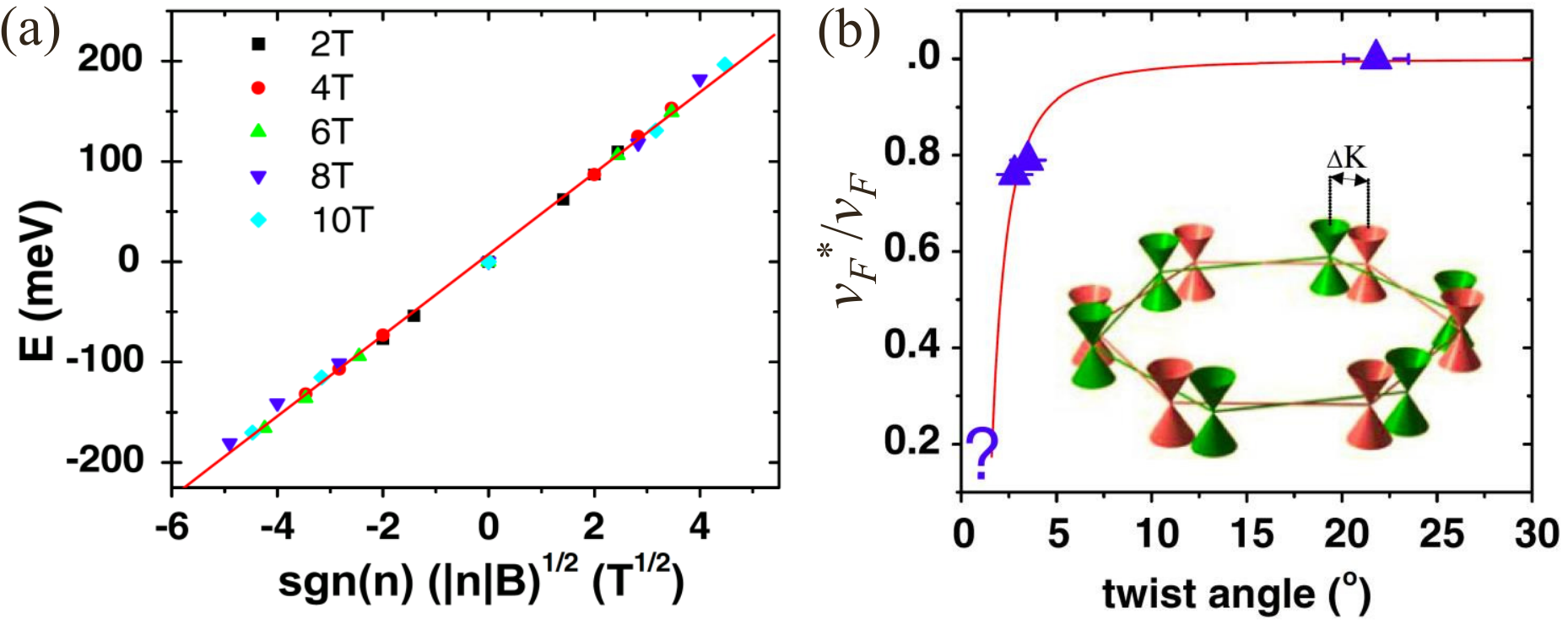}
\caption{(Color online) (a) Energies of the Landau levels $E_{n}$ measured
at different values of the magnetic field $B$ in Ref.~\cite{STM2}. The
twist angle is $\theta=21.8^{\circ}\pm1.7^{\circ}$. The solid line is a
linear fit following
Eq.~\eqref{LLspec}. (b) Dependence of the renormalized
Fermi velocity $v_F^{*}$ on the twist angle $\theta$. Triangles are the
experimental data in Ref.~\cite{STM2}. The solid (red) curve is the
theoretical prediction,
Eq.~\eqref{VrenormDS}. Reprinted figures with
permission from A.~Luican et al., Phys. Rev. Lett., {\bf 106}, 126802
(2011).
Copyright 2011 by the American Physical Society.
\url{http://dx.doi.org/10.1103/PhysRevLett.106.126802}.\label{FigVrenorm}}
\end{figure}

Many experimental studies, using
STM~\cite{STM1,STM_DFT,MengSTM,STM2,STM_VHS2,STM_VHS1,CisternasSTM,
MillerLLexp,NatureLL,tBLG_QHE,STM_VHS2014,STM_VHSdoping2015,HePRB2015}
(both at zero and non-zero magnetic field),
ARPES~\cite{HicksARPES,SprinkleARPES,OhtaARPES},
and
Raman~\cite{DF_TEM2,RighiRaman1,RighiRaman2,RobinsonRaman,
PoncharalRaman,NiRamanVrenorm,CarozoRaman,NiRaman,HeRaman,JorioRaman,RotDisorder2014,WuRaman2014}
spectroscopies revealed the persistence of the Dirac cones in tBLG. The
effect of the Fermi velocity reduction was also confirmed using a variety experimental techniques, e.g.,
ARPES~\cite{OhtaARPES},
Raman~\cite{NiRamanVrenorm},
and Landau level (LL)
spectroscopies~\cite{STM2,NatureLL,tBLG_QHE,HePRB2015}.
For example, the STM topography made by A.~Luican et~al. ~in Ref.~\cite{STM2}
for bilayer graphene samples prepared by chemical vapor deposition
shows large regions characterized by different twist angles, which are
estimated by measuring the Moir\'{e} period. In each such region, the LL
spectroscopy reveals the sequence of the Landau levels
$E_{n}$ characteristic for the massless Dirac fermions:
\begin{equation}\label{LLspec}
E_{n}={\sgn}(n)v_F^{*}\sqrt{2e\hbar |n|B}\,.
\end{equation}
The dependence of the
$E_{n}$ on the level's number $n$ for the region characterized by $\theta\cong21.8^{\circ}$ is shown in
Fig.~\ref{FigVrenorm}(a). From these data, one can extract the Fermi velocity
$v_F^{*}$ using
Eq.~\eqref{LLspec}. It turns out that
$v_F^{*}$
is different for the regions characterized by different
$\theta$,
and correlates well with the theoretical prediction
Eq.~\eqref{VrenormDS} [see
Fig.~\ref{FigVrenorm}(b)].

\begin{figure}[t]
\centering
\includegraphics[width=0.99\columnwidth]{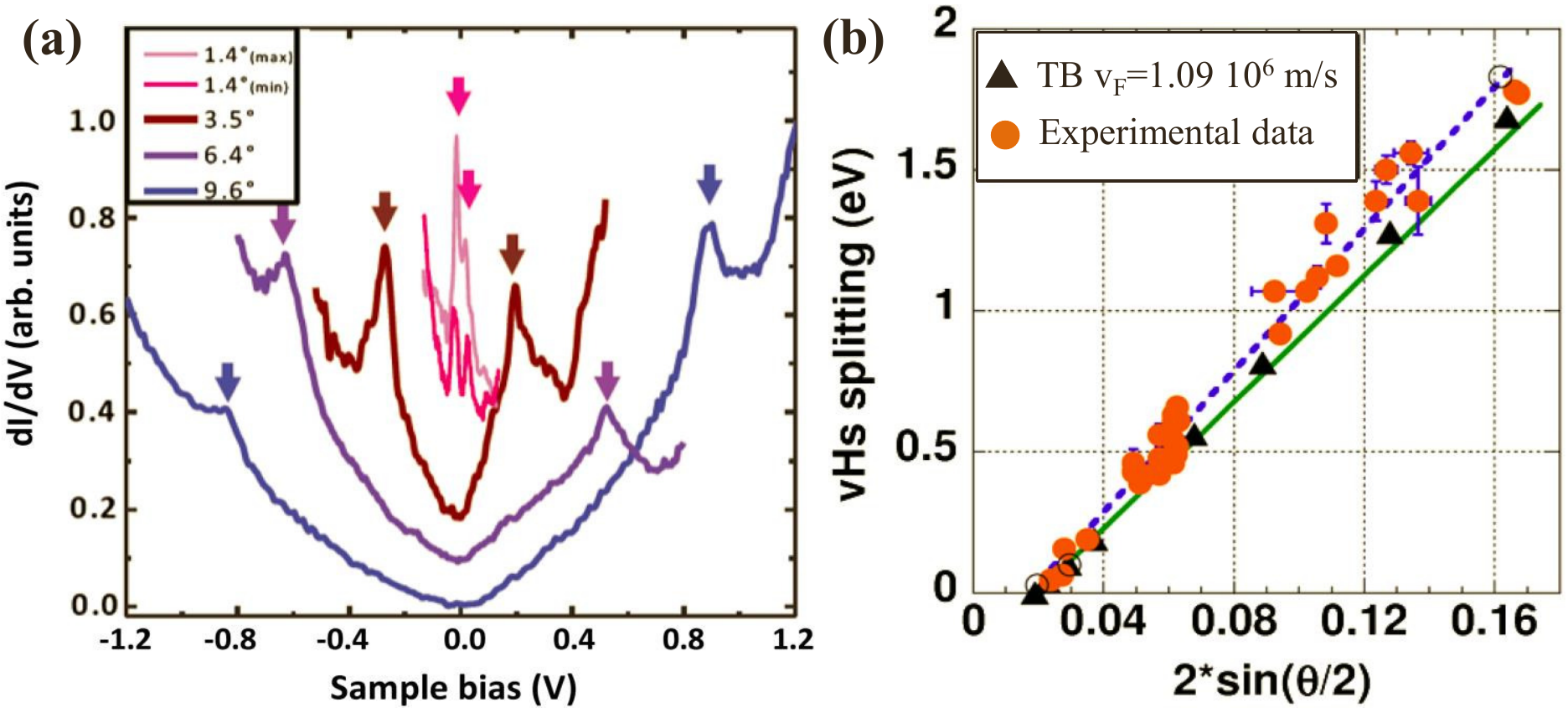}
\caption{(Color online) (a) Local DOS spectra for samples with different
twist angles measured by STM in
Ref.~\cite{STM1}.
(b) The dependence of the van Hove singularities splitting
$\Delta E_{\text{vHs}}$ on $2\sin\theta/2$.
Orange circles are the experimental data. The dashed (blue) line is the best
fit of the experimental data by
Eq.~\eqref{DeltaEvhs}
with $v_F^{*}=1.12\times10^6$\,m/s and $\tilde{t}_{\bot}=0.108$\,eV. The (black) triangles and solid (green) line are described in Ref.~\cite{STM1}.
Reprinted figures with permission from I.~Brihuega et al., Phys. Rev.
Lett., {\bf 109}, 196802 (2012). Copyright 2012 by the American Physical
Society.
\url{http://dx.doi.org/10.1103/PhysRevLett.109.196802}.\label{FigVHS}}
\end{figure}

Another important feature of the tBLG spectrum is the existence of van Hove singularities (VHS) at energies much lower than that for the single-layer graphene. The VHS of the twisted bilayer appear due to the merging of two Dirac cones leading to saddle points in the low-energy bands, both below and above the Fermi level [see Fig.~\ref{FigSpecDS}(a)]. Van Hove singularities manifest themselves as peaks in the density of states  in STM
measurements~\cite{STM1,STM_VHS2,STM_VHS1,STM_VHS2014}.
The energy difference between two VHS peaks, $\Delta E_{\text{vHs}}$, is smaller for smaller twist angles. In the framework of the continuum
approximation~\cite{dSPRL,dSPRB,PNAS}, this energy difference can be estimated as
\begin{equation}\label{DeltaEvhs}
\Delta E_{\text{vHs}}=2\hbar v_F^{*}K\sin(\theta/2)-2\tilde{t}_{\bot}\,.
\end{equation}
For a rough estimate of the $\Delta E_{\text{vHs}}$ at angles not close to the critical value $\theta_c$, one can neglect the dependence of $v_F^{*}$ on
$\theta$. The decrease of the $\Delta E_{\text{vHs}}$
with a twist angle was observed in different experiments~\cite{STM1,STM_VHS2,STM_VHS1,STM_VHS2014}. For example,
Fig.~\ref{FigVHS}(a) shows the DOS spectra for different tBLG, obtained by STM measurements in
Ref.~\cite{STM1}. The two peaks are closer to each other for samples with smaller twist angle. The VHS splitting,
$\Delta E_{\text{vHs}}$, as a function of $\theta$ is shown in
Fig.~\ref{FigVHS}(b). The fitting of the experimental data with formula~\eqref{DeltaEvhs} with a constant
$v_F^{*}$ is also shown in this figure.

Similar results were obtained in
Ref.~\cite{STM_VHS1}.
In the range of the twist angles
$2^{\circ}<\theta<5^{\circ}$,
the VHS splitting measured in the latter reference is approximately
described by
Eq.~\eqref{DeltaEvhs}
with constant
$v_F^{*}$.
At the same time, the best fitting of the experimental data reported in
Ref.~\cite{STM_VHS2} corresponds to
$\Delta E_{\text{vHs}}=2\hbar v_FK\sin(\theta/2)$,
with
$v_F$
equal to the Fermi velocity of single-layer graphene. The authors
concluded that the Fermi velocity remained non-renormalized for their tBLG
samples. From the viewpoint of low-energy
theories~\cite{dSPRL,dSPRB,PNAS},
this corresponds to the limit of uncoupled graphene layers,
$\tilde{t}_{\bot}\to0$.
No Fermi velocity reduction was observed also in ARPES measurements of
Refs.~\cite{HicksARPES,SprinkleARPES},
and by the Landau level spectroscopy in
Ref.~\cite{MillerLLexp}.
These results are in contradiction with those obtained by
Landau level~\cite{STM2}
and
Raman~\cite{NiRamanVrenorm}
spectroscopies (see the previous paragraph). It is likely that these
discrepancies can be explained by different sample preparation techniques,
influence of the substrate, disorder, etc.

Related results were obtained in Ref.~\cite{STM_VHS2014} by W.~Yan et~al.
Scanning tunneling microscopy data gathered in this paper demonstrated the
existence of low-energy VHS for bilayer samples with twist angles
$\theta\lesssim3.5^{\circ}$. For these samples, the energy difference
between VHS peaks, $\Delta E_{\text{vHs}}$, with a good accuracy follows
Eq.~\eqref{DeltaEvhs}. At the same time, no traces of the low-energy VHS were found for samples with $\theta\gtrsim5.5^{\circ}$: the measured $dI/dV$ spectra are ``almost identical to those of the graphene monolayer''. In the intermediate range, $3.5^{\circ}\lesssim\theta\lesssim5.5^{\circ}$, some samples show VHS peaks, but some others do not.

Measurements of both the Fermi velocity and the VHS splitting were
performed in
Ref.~\cite{HePRB2015}
for samples with different twist angles prepared on two different
substrates. The Fermi velocity was determined with the help of LL
spectroscopy and
Eq.~\eqref{LLspec}.
Using
Eq.~\eqref{DeltaEvhs}
and measured
$\Delta E_{\text{vHs}}$,
the authors extracted the inter-layer hopping parameter
$\tilde{t}_{\bot}$
for each sample. They found that the tBLG properties demonstrated
significant sample-to-sample variation, which cannot be explained by
the sample-to-sample variation of the twist angle. For example, in several
systems
$\tilde{t}_{\bot}\approx0$.
The authors speculated that this variation may be caused by the influence
of ``the stacking fault, tilt grain boundary, atomic defects, and roughness
of substrate'' on the inter-layer distance. However, in agreement with the
theory, strong correlation between Fermi velocity reduction and the
inter-layer coupling was found: the
authors~\cite{HePRB2015}
showed that the
$v_{\rm F}^{*}$
is well described by
Eq.~\eqref{VrenormDS}
with experimentally obtained
$\tilde{t}_{\bot}$
and $\theta$.

The effect of doping on the VHS has been studied using STM in
Ref.~\cite{STM_VHSdoping2015}.
For these bilayers a doping of the order of
$10^{12}$\,cm$^{-2}$
arose by charge transfer from the substrate. The doping leads to asymmetric
positions of the VHS peaks with respect to the Fermi level, while the
energy difference,
$\Delta E_{\text{vHs}}$,
is almost doping independent, as it is expected from theory. For the doping
$\sim5\times10^{12}$\,cm$^{-2}$,
the Fermi level exceeds the energy of the upper VHS for the tBLG with
$\theta\lesssim3^{\circ}$.
They~\cite{STM_VHSdoping2015}
also mentioned that doping leads to small
($<100$\,mV)
potential difference between layers.

In 
Ref.~\cite{wong_tblg2015}
both VHS as well as higher-energy spectroscopic features were detected
using STM technique. Consistency between experimental data and theoretical
predictions was found. 

\subsubsection{Low-energy spectrum at very small angles}

The band structure with Dirac cones described above is valid only for angles larger than the critical angle
$\theta_c$, below which the renormalized Fermi velocity $v_F^{*}$ vanishes. For angles close to $\theta_c$
the low-energy bands become almost flat, and for
$\theta<\theta_c$
the cone-like band structure becomes completely irrelevant. The value of
$\theta_c$,
which actually  depends on the interlayer hopping, was estimated by different groups, using both low-energy theories~\cite{dSPRB,PNAS} and tight-binding calculations~\cite{NanoLettTB,ourTBLG}. For
$\tilde{t}_{\bot}\approx0.1$\,eV, J.\,M.\,B.~Lopes dos Santos et al.~\cite{dSPRB}
concluded that
$\theta_c \approx1^{\circ}$.
This value correlates well with that found by R. Bistritzer and A.H.
MacDonald~\cite{PNAS}.
Tight-binding studies reported similar value (e.g., calculations done in
Ref.~\cite{ourTBLG} give
$\theta_c\cong1.89^{\circ}$).

\begin{figure}[t]
\centering
\includegraphics[width=0.48\textwidth]{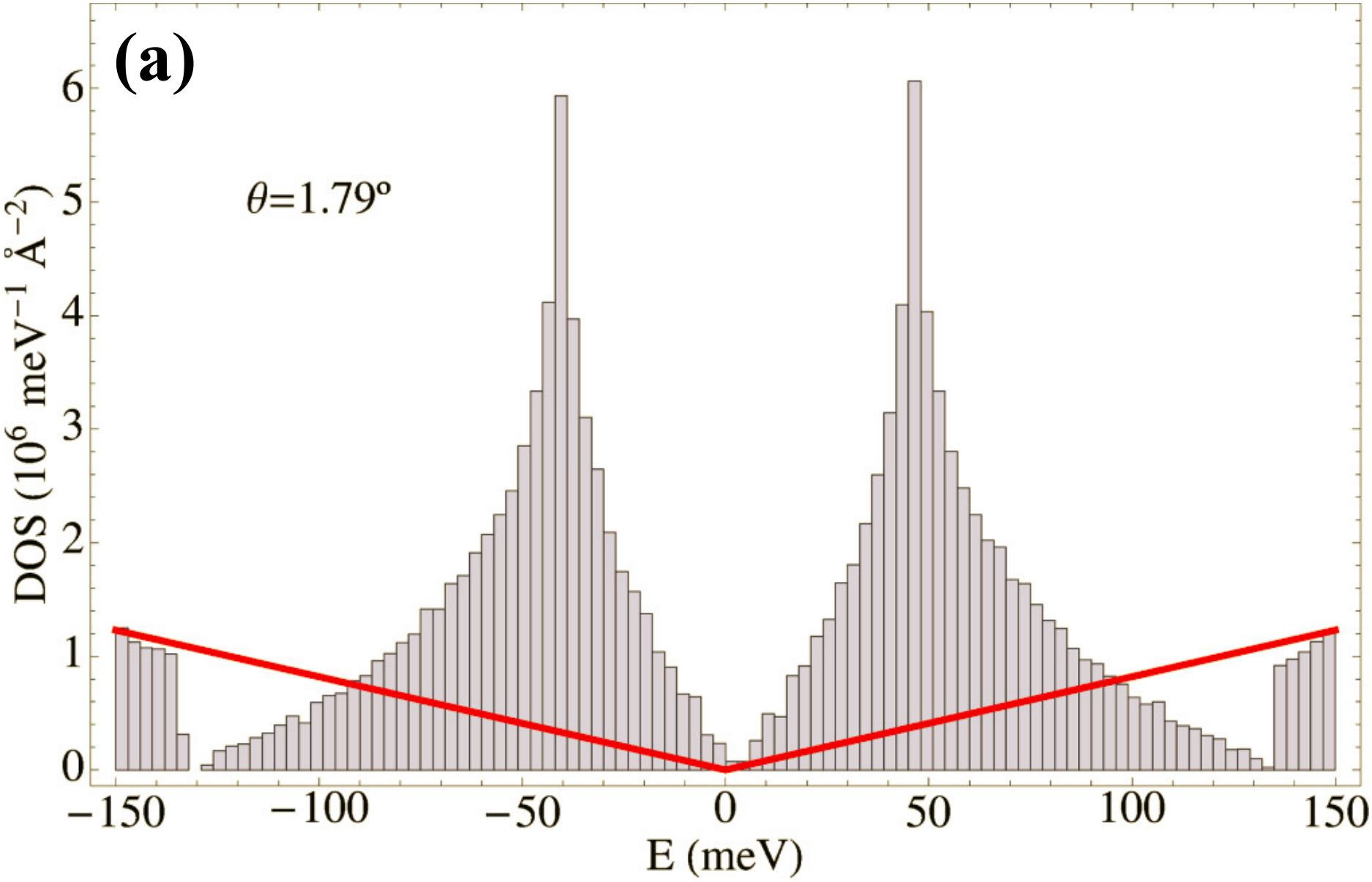}
\includegraphics[width=0.48\textwidth]{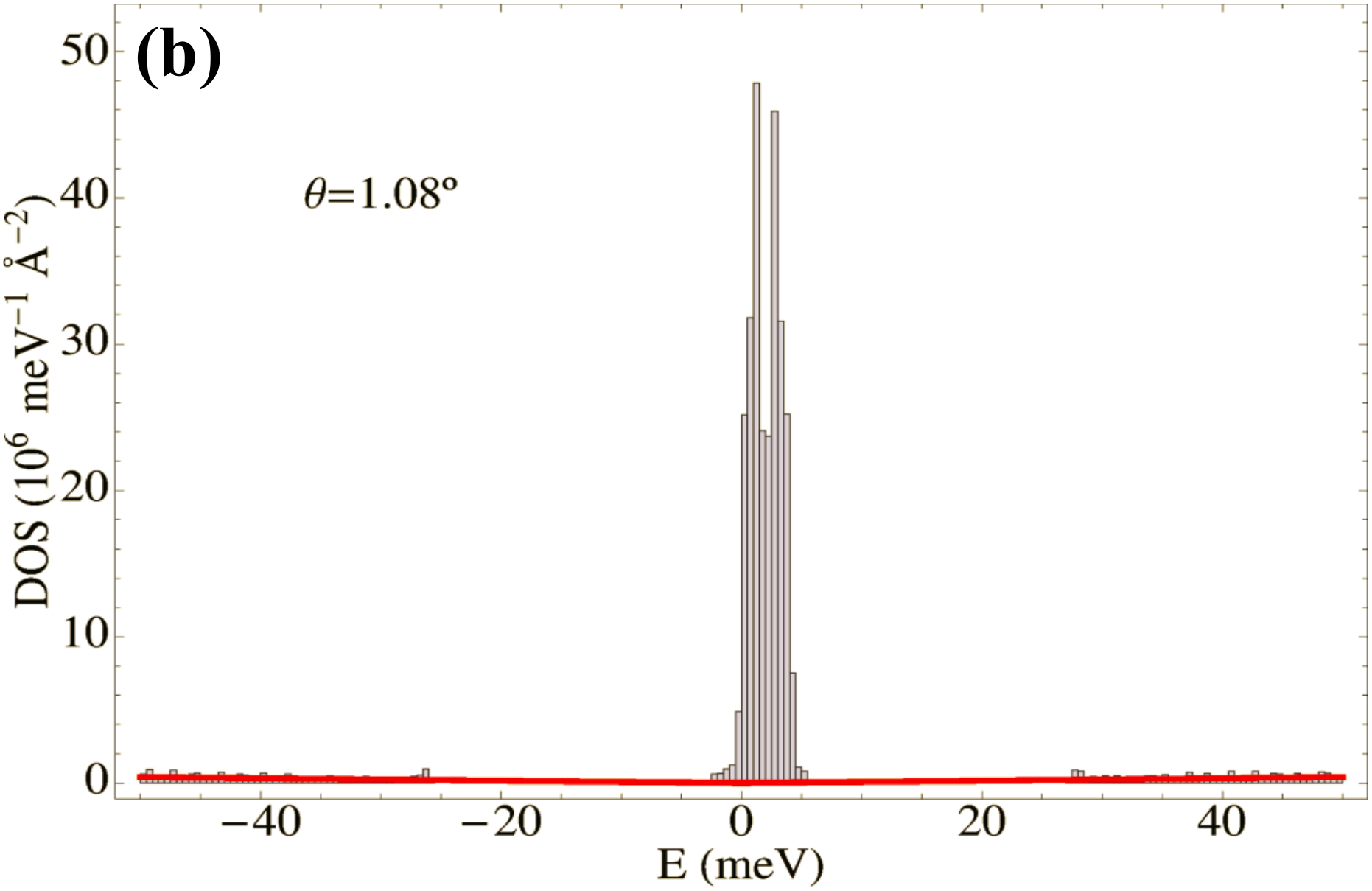}\\
\includegraphics[width=0.48\textwidth]{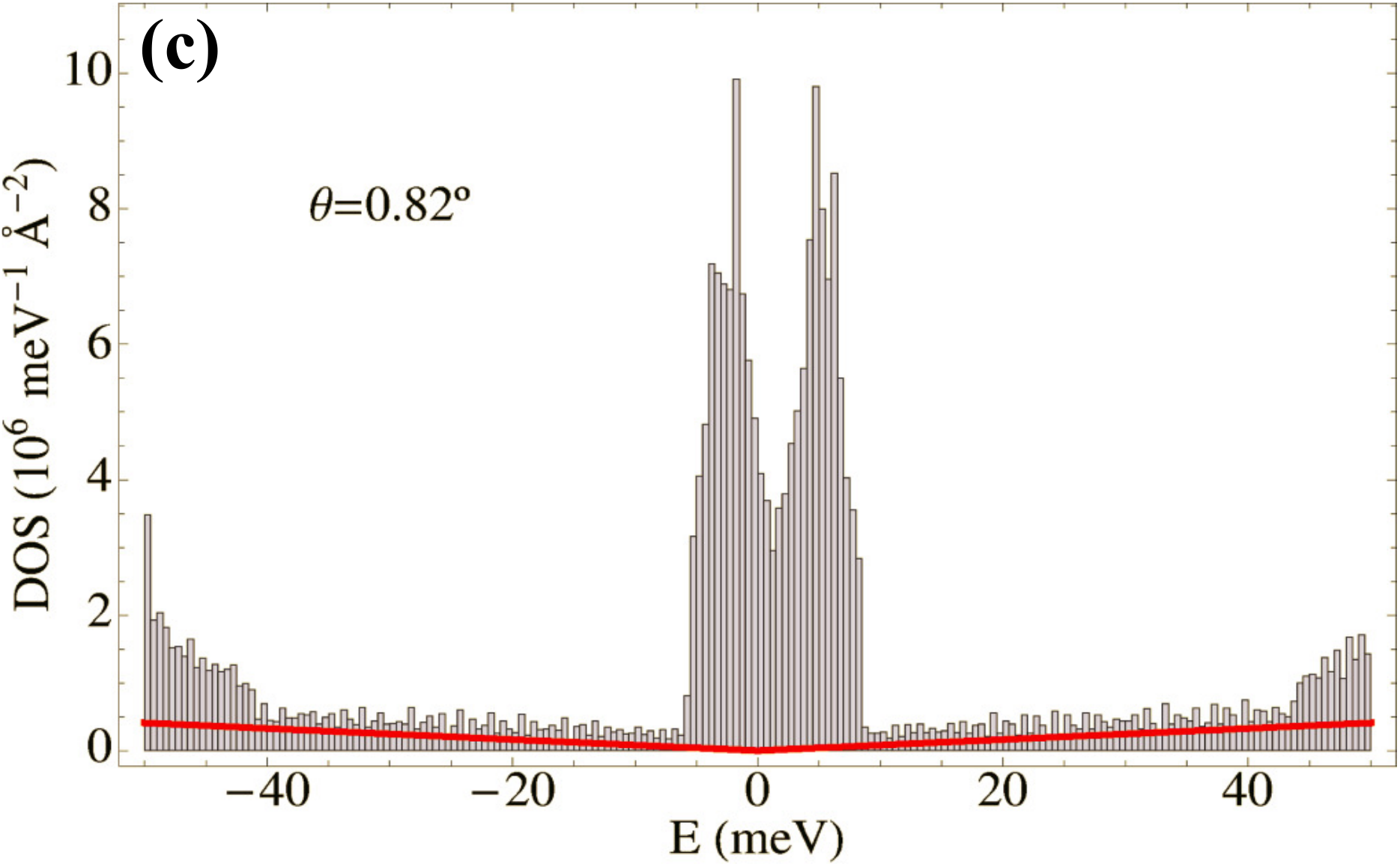}
\includegraphics[width=0.48\textwidth]{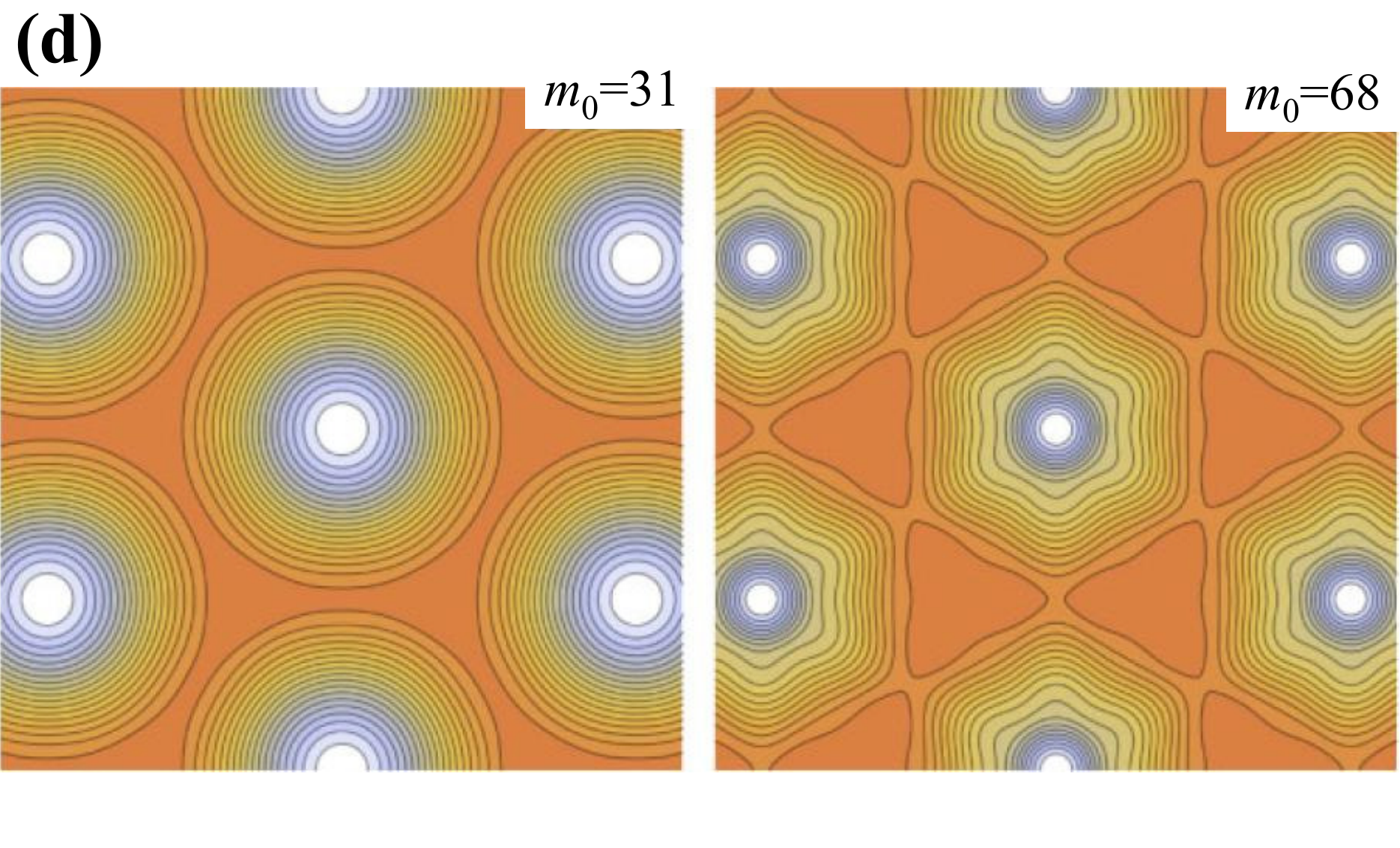}
\caption{(Color online) (a)\,--\,(c) The low-energy DOS, calculated for three
twist angles, is shown by gray rectangles. The solid (red) line is the DOS of
the single-layer graphene. (a):
$\theta>\theta_c$.
The DOS has an almost linear dependence on $E$ at small energies, consistent
with the existence of Dirac cones. It is zero at zero energy (Fermi level).
(b) and (c):
$\theta<\theta_c$.
The DOS has a double-peak structure at small energies, and it is non-zero
at zero energy.
Reprinted figures with permission from J.\,M.\,B. Lopes dos Santos,  et
al., Phys. Rev. B, {\bf 86}, 155449 (2012).
Copyright 2012 by the American Physical Society.
\url{http://dx.doi.org/10.1103/PhysRevB.86.155449}.
(d) Patterns of the local density of states at the Fermi
level calculated for two different structures
$(m_0,1)$
corresponding to the `magic' angles,
Eq.~\eqref{thetaN}. The twist angles are
$\theta\cong1.05^{\circ}$
($m_0=31$) and
$\theta\cong0.48^{\circ}$
($m_0=68$).
A logarithmic scale is used, the white color corresponds to the maxima of
the local DOS. These patterns show single-electron localization inside the
regions with approximately $AA$ stacking.
Reprinted figure with permission from P.~San-Jose et al., Phys. Rev. Lett,
{\bf 108}, 216802 (2012).
Copyright 2012 by the American Physical Society.
\url{http://dx.doi.org/10.1103/PhysRevLett.108.216802}.\label{FigTBLG_DOS}}
\end{figure}

The physical properties of tBLG differ qualitatively for $\theta$ above and
below
$\theta_c$.
Specifically, Ref.~\cite{dSPRB} calculated the density
of states for different twist angles. When
$\theta>\theta_c$,
the DOS is zero at the Fermi level and increases linearly at small energies,
consistent with the existence of Dirac cones. For
$\theta<\theta_c$, the density of states has a very pronounced double-peak structure at small
energies (see Fig.~\ref{FigTBLG_DOS}).
Moreover, the DOS $\rho_0$ at the Fermi level is non-zero. The width of this double-peak, as well as the value of
$\rho_0$,
oscillate as a function of $\theta$. For small twist angles, the tBLG can be
considered as composed of large regions with almost $AB$ and $AA$ stacking.
They~\cite{dSPRB} explain the peak in the density of states by the localization of
the electrons at the regions with $AA$ stacking (for details, see Section~V
in
Ref.~\cite{dSPRB}).
This interpretation was confirmed by the calculation of the local density
of states (LDOS) at the Fermi level, which demonstrated that almost all electrons
are located inside regions with $AA$ stacking. The oscillations of
$\rho_0$
with the twist angle are related to the quantization conditions for the
confined electrons:
$k_FL(\theta)\sim2\pi n$,
where
$n$
is integer,
$k_F=t_{\bot}/(\hbar v_F)$
is the Fermi momentum of electrons at zero energy localized inside the
$AA$
regions, and
$L(\theta)\approx a_0\sqrt{3}/\theta$
is the Moir\'{e} period. As a result, the maxima of
$\rho_0$
(and the minima of the peak's width) occur at angles
\begin{equation}\label{thetaN}
\theta_n\sim\frac{t_{\bot}a_0\sqrt{3}}{2\pi\hbar v_Fn}\,.
\end{equation}

Similar conclusions were reached in
Ref.~\cite{TramblyTB_Loc},
which relied on tight-binding calculations. They also showed
the electron localization inside the regions with almost $AA$ stacking (for details of the tight-binding calculations, see the next
Section). Further, the band flatness and the localization  of the
single-electron states inside the $AA$ regions was also confirmed by P.
San-Jose et
al.~\cite{NonAbelianGaugePot}
in the framework of the continuum approximation. The authors proposed an
approach treating the inter-layer hopping as a spatially-modulated
non-Abelian gauge potential. The calculations~\cite{NonAbelianGaugePot} showed the appearance of
flat bands for certain values of the twist angles. For these angles, the local
density of states has pronounced peaks in the regions
with $AA$ stacking [see
Fig.~\ref{FigTBLG_DOS}(d)].
The sequence of these `magic' angles satisfies approximately the law
$\theta_n\propto1/n$,
in agreement with
Eq.~\eqref{thetaN}.

The oscillation of the DOS at the Fermi level is also consistent with the
results of R. Bistritzer and
A.H.~MacDonald~\cite{PNAS}
showing oscillations of the Fermi velocity
$v_F^{*}$
calculated at one of the Dirac point with the twist angle. The Fermi
velocity vanishes at discrete angles
$\theta_n$,
whose values are approximately described by
Eq.~\eqref{thetaN},
indicating the electron localization. The value
$\theta_1$
corresponds to
$\theta_c$.

Such a behavior of the density of states at small twist angles has been
confirmed in STS and STM measurements in Ref.~\cite{SmallAngleSTM2015} by
L.-J.~Yin et~al. The authors observed a pronounced peak in the $dI/dV$
spectra at the charge-neutrality point for bilayer with
$\theta=1.11^{\circ}$. The authors associated this angle with the critical
angle $\theta_c$ described above. For bilayers with
$\theta\leqslant\theta_c$ (data for $\theta=0.88^{\circ}$ and
$\theta=1.11^{\circ}$ are presented), the peak is observed only in the
region with AA stacking, in agreement with theoretical predictions. For the
sample with $\theta=0.88^{\circ}$ the peak is broader than the peak for the
critical angle, which is also consistent with theoretical calculations. For
angles $\theta>\theta_c$ no peak at the charge neutrality point is
observed, and the $dI/dV$ spectra are almost position independent. For
these angles, the $dI/dV$ spectra have a minimum at the charge-neutrality
point, and two peaks below and above it, which is consistent with the Dirac
cones picture.

\subsubsection{Fine structure of the low-energy bands: band splitting and band gap}
\label{lowenergysplitting}

In the low-energy theories described above, the valley degeneracy is
assumed. For any tBLG superstructure, however, the two Dirac points of
one layer coincide with the Dirac points of another layer upon
translations of the reciprocal superlattice vectors.
Which pairs of Dirac points are
equivalent depends on the type of superstructure. It was shown in
Section~\ref{SubSectGeom}
that when
$r\neq3n$,
$\mathbf{K}'\sim\mathbf{K}_{\theta}$
and
$\mathbf{K}\sim\mathbf{K}'_{\theta}$.
When
$r=3n$,
other pairs of Dirac points are equivalent:
$\mathbf{K}\sim\mathbf{K}_{\theta}$
and
$\mathbf{K}'\sim\mathbf{K}'_{\theta}$.
In a generic situation, the single-electron states near equivalent cones are
hybridized by some finite matrix element. E.\,J.~Mele in
Ref.~\cite{MelePRB1}
took this into account by adding a mass term into the effective low-energy
Hamiltonian. He considered the Hamiltonian at the momenta close to one of
two tBLG Dirac points. The Hamiltonian has the form of a$4\times4$
matrix. It is different for `odd'
($r\neq3n$)
and `even'
($r=3n$)
superstructures. For
$r\neq3n$,
the effective Hamiltonian is
\begin{equation}\label{Hodd}
\hat{H}_{\text{odd}}=\left(\begin{array}{cc}-i\hbar v_F^{*}\bm{\sigma}{\bm\nabla}&\hat{H}_{\text{int}}^{-}\\
(\hat{H}_{\text{int}}^{-})^{\dag}&i\hbar v_F^{*}\bar{\bm{\sigma}}\bm{\nabla}
\end{array}\right)\,,
\end{equation}
where $\bar{\bm{\sigma}}$ are the complex conjugate Pauli matrices ${\bm{\sigma}}$. The mass term $\hat{H}_{\text{int}}^{-}$
is a $2\times2$ matrix which couples the electron states near the Dirac points $\mathbf{K}$
and $\mathbf{K}'_{\theta}$\,($\sim\mathbf{K}$). Since these Dirac points belong to different valleys, the diagonal
blocks in $\hat{H}_{\text{odd}}$
have a different structure.
The velocity $v_F^*$
is the {\it renormalized} Fermi velocity, not the bare
$v_F$.
By retaining
$v_F^*$,
as opposed to the bare
$v_{\rm F}$,
the proposed technique attempts to account ``phenomenologically" for more
robust effects discussed in the previous subsection. For `even' structures,
the mass term,
$\hat{H}_{\text{int}}^{+}$,
couples the electrons near the Dirac points in the same valley,
$\mathbf{K}$
and
$\mathbf{K}_{\theta}$\,($\sim\mathbf{K}$).
Therefore, the effective Hamiltonian has identical diagonal blocks:
\begin{equation}
\label{Heven}
\hat{H}_{\text{even}}
=
\left(
	\begin{array}{cc}
	-i\hbar v_F^{*}\bm{\sigma\nabla}&\hat{H}_{\text{int}}^{+}\\
	(\hat{H}_{\text{int}}^{+})^{\dag}&
	-i\hbar v_F^{*}\bm{\sigma}\bm{\nabla}
	\end{array}
\right)\,.
\end{equation}

\begin{figure}[t]
\centering
\includegraphics[width=0.4\columnwidth]{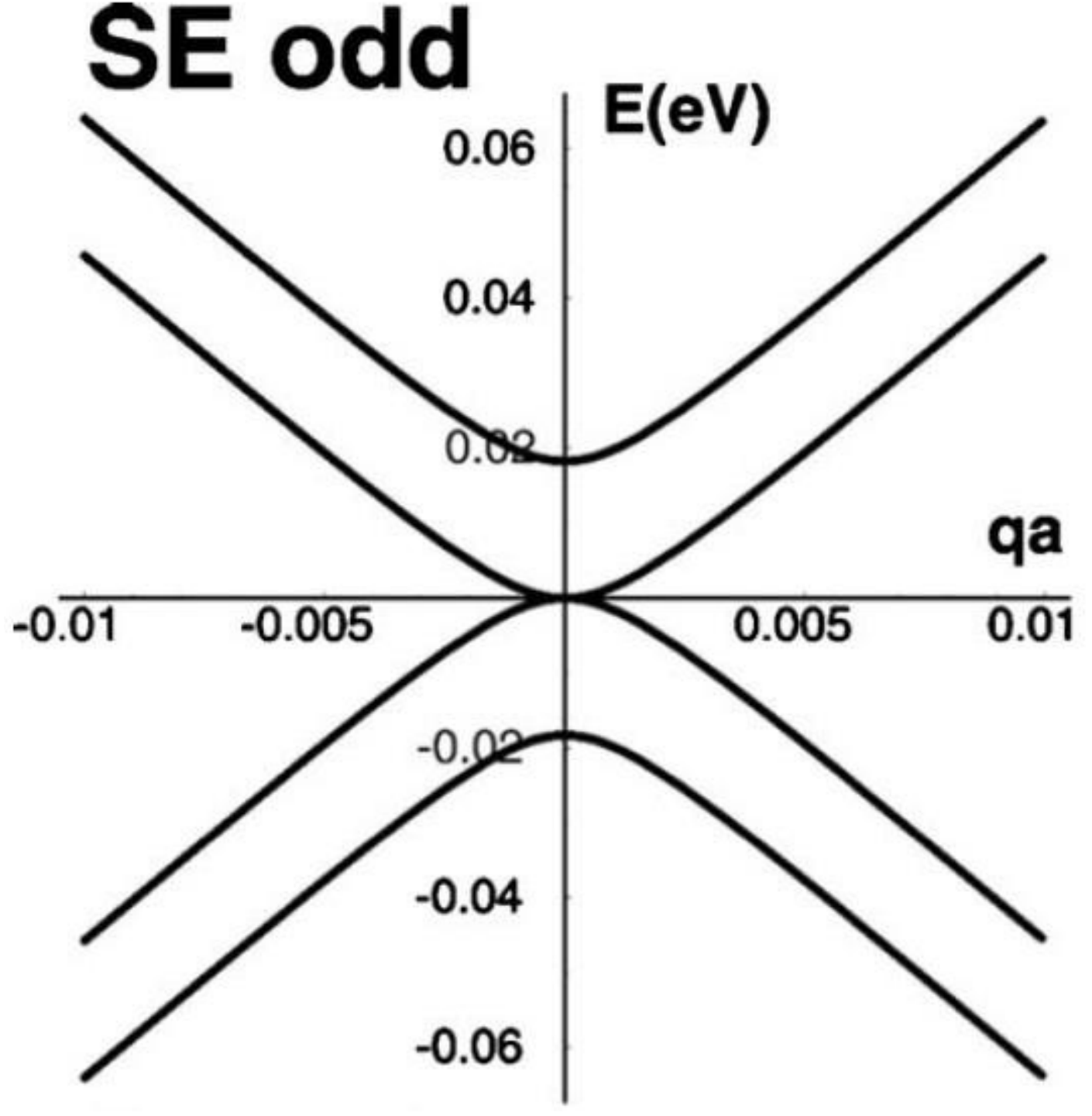}
\includegraphics[width=0.4\columnwidth]{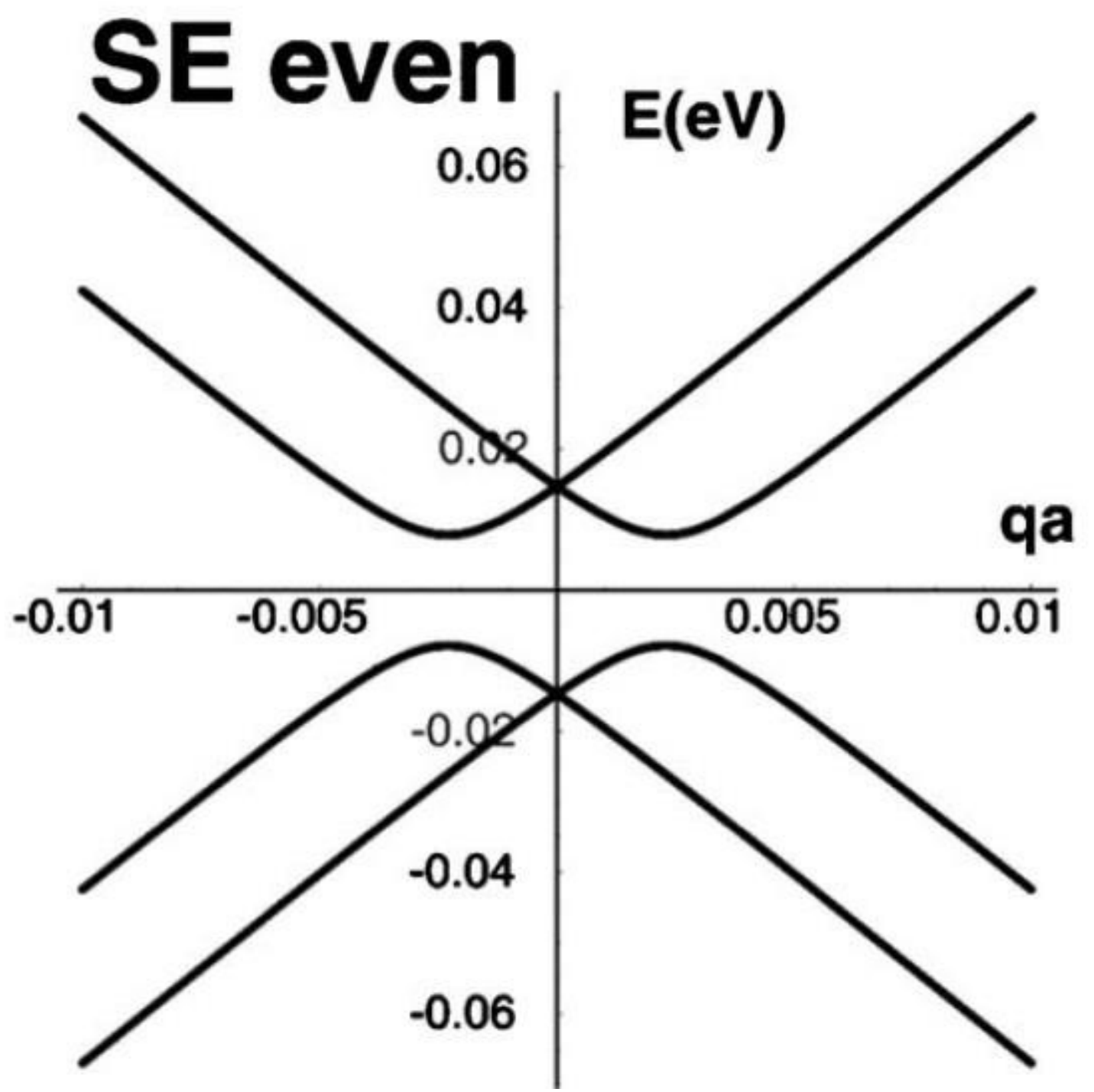}
\caption{(Color online) Low-energy spectra of tBLG for `odd' (left
panel) and `even' (right panel) superstructures.
Reprinted figures with permission from E.\,J.~Mele, Phys. Rev. B, {\bf 81},
161405 (2010). Copyright 2010 by the American Physical Society.
\url{http://dx.doi.org/10.1103/PhysRevB.81.161405}.\label{FigSpecOddEven}}
\end{figure}

Furthermore, the structure of the matrices
$\hat{H}_{\text{int}}^{+}$
and
$\hat{H}_{\text{int}}^{-}$
is different. Using symmetry considerations, E.\,J.~Mele~\cite{MelePRB1} proposed the
following general form of these matrices:
\begin{equation}\label{Hmass}
\hat{H}_{\text{int}}^{+}={\cal V}e^{i\vartheta}\!\!\left(\!\!\begin{array}{cc}e^{i\varphi/2}&0\\0&e^{-i\varphi/2}\!\end{array}\!\!\right),\;\;\;
\hat{H}_{\text{int}}^{-}={\cal V}e^{i\vartheta}\!\!\left(\!\!\begin{array}{cc}1&0\\0&0\end{array}\!\!\right).
\end{equation}
The parameter
${\cal V}$
describes the strength of the cone hybridization, while $\vartheta$ and
$\varphi$ are phase factors allowed by symmetry. E.\,J.~Mele estimated the parameter ${\cal V}$ as
${\cal V}\sim10$\,meV
for superstructures
$(m_0,r)=(1,1)$,
$\theta\cong21.787^{\circ}$
and
$(m_0,r)=(1,3)$,
$\theta\cong38.213^{\circ}$.

The low-energy spectrum consists of four bands. It is different for `odd'
and `even' structures. Typical curves are shown in
Fig.~\ref{FigSpecOddEven}. For
$r=3n$,
the spectrum is gapped with a gap proportional to
${\cal V}$.
For
$r\neq3n$,
all four bands have parabolic dispersion near the Dirac point, and the spectrum is gapless: two middle bands touch each other at the Dirac point. The band splitting was experimentally found by ARPES~\cite{OhtaARPES} and Raman spectroscopy~\cite{NiRaman}. The band gap in the tBLG was also observed in transport measurements in Ref.~\cite{TransportGap2015} (for more details about the latter reference, see next subsection).

\subsection{Tight-binding calculations of the twisted bilayer graphene spectrum}
\label{TB_tBLG}

The applicability of the approximations described above is restricted to
low energies and relatively small twist angles.
To lift up these restrictions, different approaches are required. One
possibility is {\it ab initio} calculations. Unfortunately, the
{\it ab initio} band structure calculations of the tBLG require huge
computation efforts because of the large number of atoms inside the
supercell $N$. Numerical calculations based on tight-binding
approximations are also quite time-consuming for the same reason,
especially for small twist angles. Indeed, to find the tBLG spectrum, one
needs to diagonalize the
$N\times N$
matrix of the tight-binding Hamiltonian on a 2D momentum grid, covering the
tBLG Brillouin zone. For example, for the structure
$(16,1)$
with the twist angle
$\theta\cong2^{\circ}$,
the rank of this matrix is
$N=3286$.
For these reasons, in many papers the authors combine DFT and tight-binding
calculations~\cite{PankratovFlakes,NanoLettTB,TramblyTB_Loc,Morell1,PankratovPRB2013}.
They perform DFT calculations only for a few structures with moderate $N$,
and then use these results to extract
the hopping amplitudes in the tight-binding Hamiltonian. In the general case,
the second-quantized tight-binding Hamiltonian of the tBLG can be written as
\begin{equation}\label{HTB}
H=\sum_{\mathbf{nm}i\atop\alpha\beta\sigma}
\left[t_{\|}^{(i)}\!(\mathbf{r}_{\mathbf{n}}^{i\alpha};\mathbf{r}_{\mathbf{m}}^{i\beta})d^{\dag}_{i\mathbf{n}\alpha\sigma}
d^{\phantom{\dag}}_{i\mathbf{m}\beta\sigma}+\text{h.c.}\right]+
\sum_{{\mathbf{n}\mathbf{m}\atop\alpha\beta\sigma}}\left[t_{\bot}(\mathbf{r}_{\mathbf{n}}^{1\alpha};\mathbf{r}_{\mathbf{m}}^{2\beta})
d^{\dag}_{1\mathbf{n}\alpha\sigma}d^{\phantom{\dag}}_{2\mathbf{m}\beta\sigma}+\text{h.c.}\right],
\end{equation}
where
$d^{\dag}_{i\mathbf{n}\alpha\sigma}$
and
$d^{\phantom{\dag}}_{i\mathbf{n}\alpha\sigma}$
are the creation and annihilation operators of an electron with spin projection
$\sigma$
in the site
$\mathbf{n}$
in the layer
$i\,(=1,2)$
on the sublattice
$\alpha\,(=A,B)$.
The first term describes the in-plane hopping, with
$t_{\|}^{(i)}\!(\mathbf{r};\mathbf{r}')$
being the in-plane hopping amplitude. The second term describes the
interlayer hopping, with
$t_{\bot}(\mathbf{r};\mathbf{r}')$
being the hoping amplitude between sites in layers
$1$
and
$2$,
located in positions
$\mathbf{r}$
and
$\hat{\mathbf{e}}_zd+\mathbf{r}'$,
respectively.

Different groups use different parametrizations for the hopping amplitudes in
Eq.~\eqref{HTB}. In the simplest case of two-center approximation, the functions
$t_{\|}^{(i)}\!(\mathbf{r};\mathbf{r}')$
and
$t_{\bot}(\mathbf{r};\mathbf{r}')$
depend entirely
on the relative positions of the orbitals participating in the hopping
event. The amplitudes can be expressed via Slater-Koster
parameters~\cite{SlaterKoster},
$V_{pp\sigma}$
and
$V_{pp\pi}$,
as follows:
\begin{eqnarray}
\label{SlaterKoster}
t_{\|}^{(i)}\!(\mathbf{r};\mathbf{r}')
&=&
V_{pp\pi}(|\mathbf{r}-\mathbf{r}'|)\,,\nonumber\\
t_{\bot}(\mathbf{r};\mathbf{r}')
&=&
\cos^2\!\gamma\,V_{pp\sigma}
\big(\sqrt{c_0^2+(\mathbf{r}-\mathbf{r}')^2}\big)
+
\sin^2\!\gamma\,V_{pp\pi}
\big(\sqrt{c_0^2+(\mathbf{r}-\mathbf{r}')^2}\big)\,,
\end{eqnarray}
where the functions
$V_{pp\sigma}(r)$
and
$V_{pp\pi}(r)$
depend only on the distance between two sites, and
$\gamma$
is the angle between the $z$ axis and the line connecting the sites in the positions
$\mathbf{r}$ and
$c_0\hat{\mathbf{e}}_z+\mathbf{r}'$:
\begin{equation}
\cos^2\!\gamma=\frac{c_0^2}{c_0^2+(\mathbf{r}-\mathbf{r}')^2}\,.
\end{equation}

\begin{figure}[t]
\centering
\includegraphics[width=0.9\columnwidth]{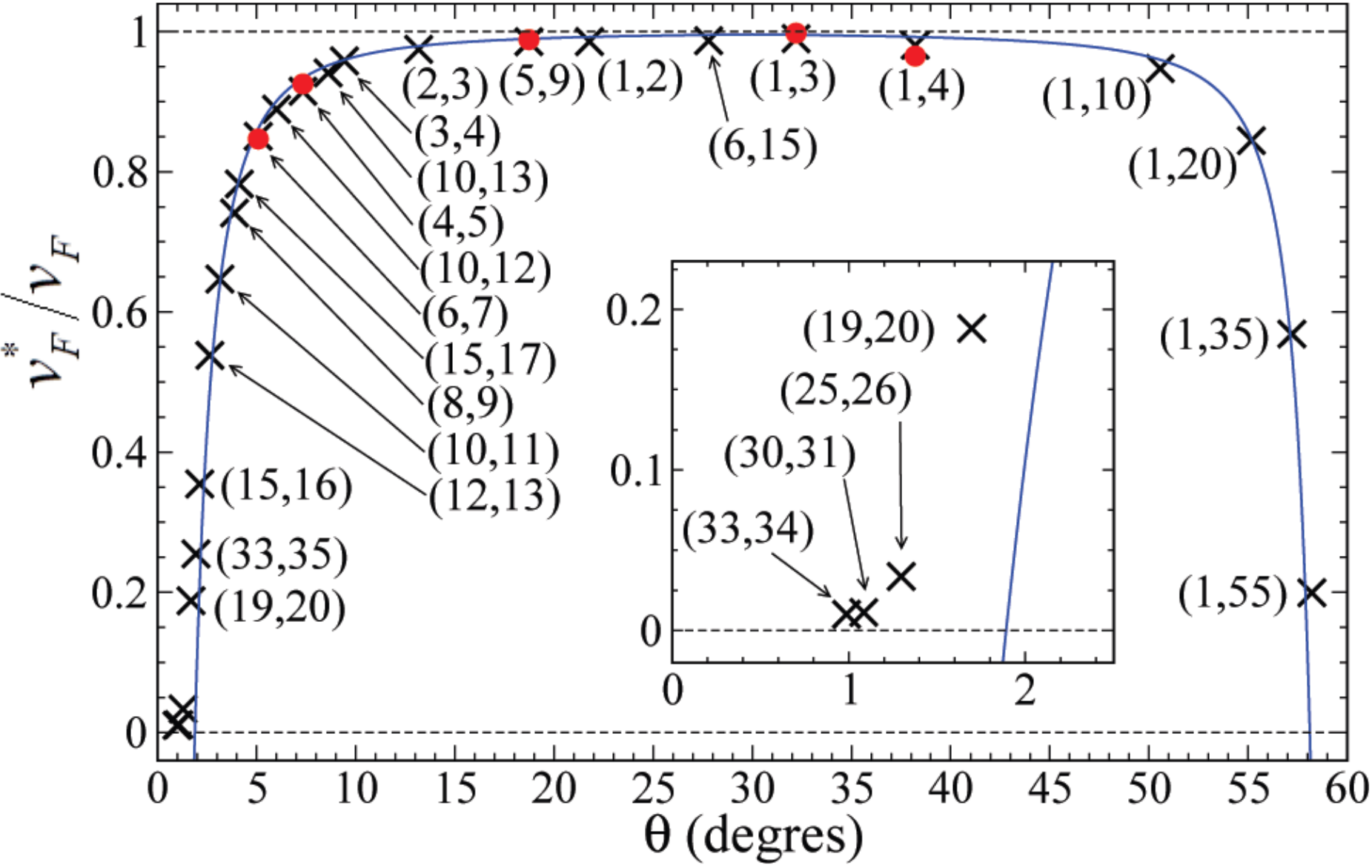}
\caption{(Color online) Fermi velocity versus twist angle $\theta$,
calculated for several superstructures in
Ref.~\cite{NanoLettTB}.
The pairs of integers denote superstructures, but the superstructure
indices
$(n,m)$
in the figure correspond to
$(n,m-n)$
in our notation. Crosses are the tight-binding calculations, (red) circles
correspond to {\it ab initio} calculations. The solid (blue) curve corresponds
to the approximate formula~\eqref{VrenormDS} by J.\,M.\,B.~Lopes dos Santos et
al.~\cite{dSPRL}. The inset shows the data for small angles.
Reprinted with permission from G.~Trambly de Laissardi\`{e}re et al., Nano
Letters, {\bf 10}, 804 (2010).
Copyright 2010 American Chemical Society.
\url{http://dx.doi.org/10.1021/nl902948m}.\label{FigV_TB}}
\end{figure}

In Refs.~\cite{NanoLettTB,TramblyTB_Loc}, Trambly de Laissardi\`ere et al. exploited the exponentially-decreasing function $V_{pp\pi}$ and $V_{pp\sigma}$
\begin{eqnarray}
\label{VExp1}
V_{pp\pi}(r)&=&-t\exp\left[q_{\pi}\left(1-r/a_{0}\right)\right]F_c(r)\,,\\
V_{pp\sigma}(r)&=&t_0\exp\left[q_{\sigma}\left(1-r/c_0\right)\right]F_c(r)\,,\;
\label{VExp2}
\end{eqnarray}
with equal spatial exponential-decreasing coefficients
\begin{eqnarray}
\label{decay_coeffs}
\frac{q_{\pi}}{a_{0}}=\frac{q_{\sigma}}{c_0}\,.
\end{eqnarray}
In these equations,
$F_c(r)$
is the cutoff function, reducing the long-range hopping amplitudes. In
Ref.~\cite{TramblyTB_Loc} it is taken to be
\begin{equation}\label{Fcutoff}
F_c(r)=\frac{1}{1+\exp[(r-r_c)/l_c]}\,,
\end{equation}
where the authors used
$r_c=2.5\sqrt{3}a_0=6.14$\,\AA,
and
$l_c=0.265$\,\AA.
In an earlier paper,
Ref.~\cite{NanoLettTB}, the cutoff function is not introduced, that is,
$F_c(r)=1$.
To fix the parameters
$t$
and
$q_{\pi}$
in
Eq.~\eqref{VExp1},
the authors take the in-plane nearest-neighbor and the next nearest-neighbor hopping amplitudes to be equal to their characteristic values in
the single-layer
graphene~\cite{CastrNrev}:
$t=2.7$\,eV,
$t'=0.1t$.
This yields for the ratio
$$
q_{\pi}/a_0=q_{\sigma}/c_0=\ln(t/t')/[a_0(\sqrt{3}-1)]=2.22\,\text{\AA}^{-1}.
$$
The largest inter-plane hopping amplitude is chosen to be
$t_0=0.48$\,eV. Thus, three hopping amplitudes,
$t$,
$t'$,
and
$t_0$,
together with
Eq.~\eqref{decay_coeffs},
allows to fix all the fitting parameters in functions
$t_{\|}^{(i)}\!(\mathbf{r};\mathbf{r}')$
and
$t_{\bot}(\mathbf{r};\mathbf{r}')$
describing all the hoppings in the twisted bilayer for any rotation angle.
Density functional calculations were also performed for several tBLG
superstructures, and a good agreement was found between tight-binding and
DFT results.

\begin{figure}[t]
\centering
\includegraphics[width=0.48\textwidth]{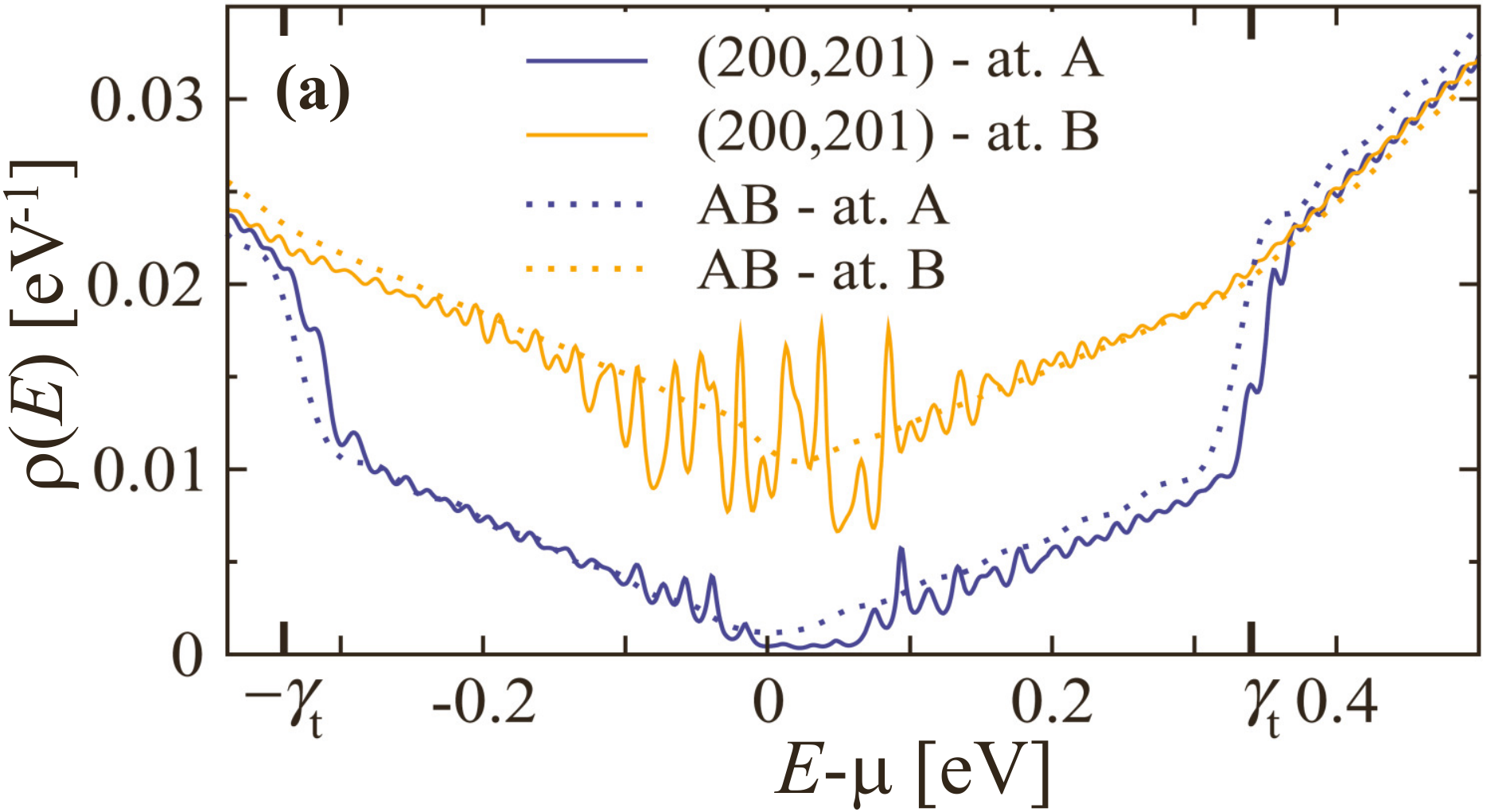}\vspace{2mm}
\includegraphics[width=0.48\textwidth]{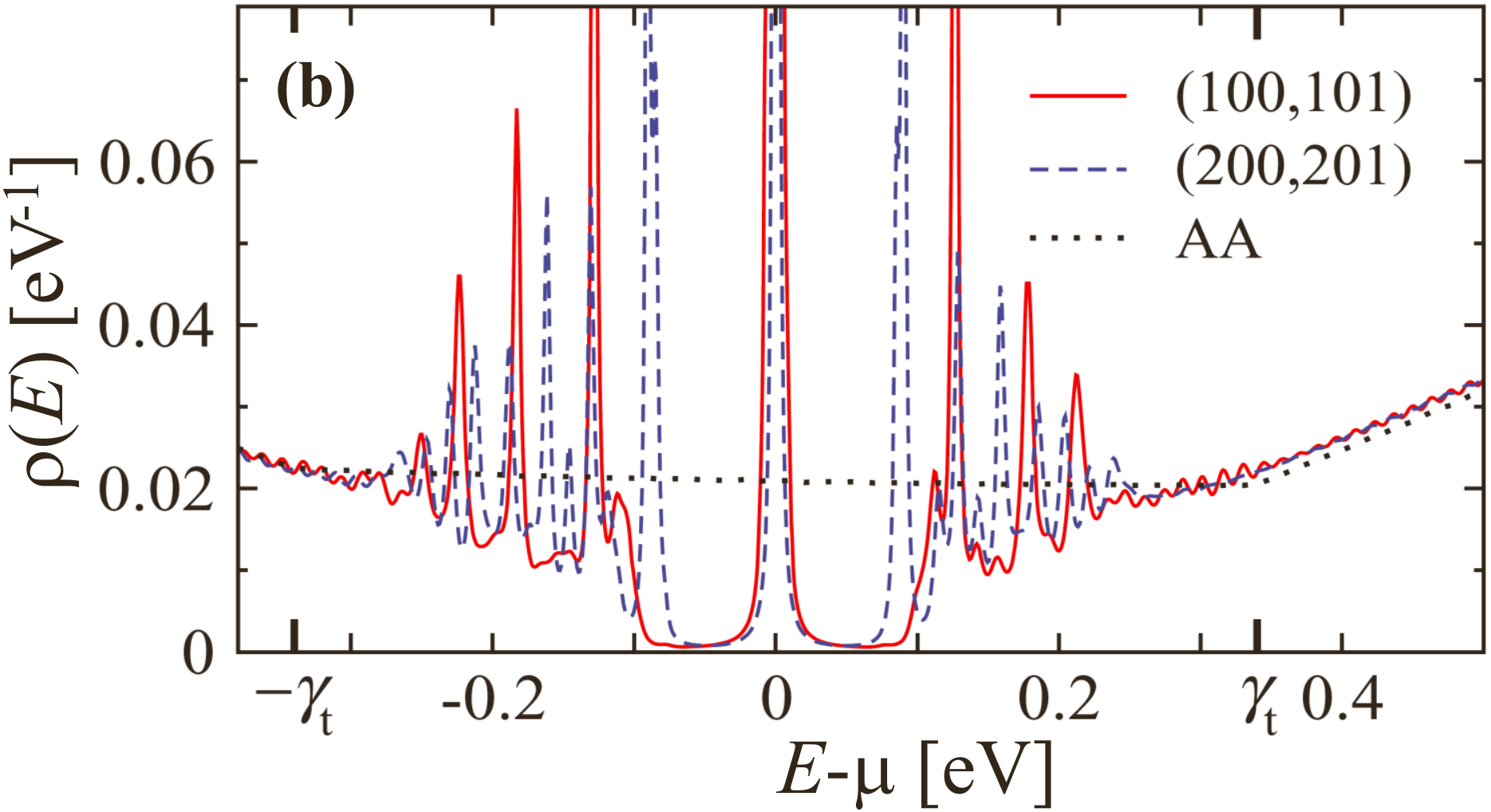}
\caption{(Color online) Local densities of states (LDOS) for two different
superstructures calculated for atoms located in the center of the nearly AB
(a) and nearly AA (b) regions. For the left panel: the curve ``at. A'' (``at. B'')
corresponds to the $A1$ or $B2$ ($B1$ or $A2$) atoms. Dotted curves are the LDOS
for pure AB (a) and AA (b) bilayers.
Note that G.~Trambly de Laissardi\`{e}re et al., Ref.~\cite{TramblyTB_Loc}, used different parametrization for the superstructures: $(n,m)$ in the figures corresponds to $(n,m-n)$ in our notation. The superstructures $(100,101)$ and $(200,201)$ have twist angles $\theta\cong0.33^{\circ}$ and $\theta\cong0.16^{\circ}$, respectively.
Reprinted figures with permission from G.~Trambly de Laissardi\`{e}re, et
al., Phys. Rev. B, {\bf 86}, 125413 (2012). Copyright 2012 by the American
Physical Society.
\url{http://dx.doi.org/10.1103/PhysRevB.86.125413}.
\label{FigDOS_TB}}
\end{figure}

The findings of
Refs.~\cite{NanoLettTB,TramblyTB_Loc},
in general, confirm the conclusions of the low-energy
theories~\cite{dSPRL,dSPRB,PNAS,NonAbelianGaugePot},
such as the Fermi velocity reduction at larger twist angles, the band
flattening, and wave function localization at smaller angles (the band
splitting and the deviation from the linear dispersion near the Dirac
points were not discussed in these papers). The calculated Fermi velocity
at the Dirac cones is found to be a smooth function of the twist angle. The
dependence of the Fermi velocity versus $\theta$ is symmetric around the
angle
$\theta=30^{\circ}$,
see
Fig.~\ref{FigV_TB}.
For not too small angles
$\theta\gtrsim3^{\circ}$
the Fermi velocity is very well described by the approximate
formula~\eqref{VrenormDS}.
In further work,
Ref.~\cite{TramblyTB_Loc},
the authors focused on the small angles. They demonstrated the appearance
of flat bands, and the non-zero density of states at the Fermi level
for $\theta<\theta_c\sim2^{\circ}$.
The authors calculated the local density of states of the tBLG in regions
with almost AB and AA stacking [see
Fig.~\ref{FigDOS_TB}(a,b)].
The local DOS at the AA region shows a pronounced peak at the Fermi level,
while this peak is absent for sites in the AB region. The oscillations of
the DOS at the Fermi level, as well as `magic' angles, were also found.

\begin{figure}[t]
\centering
\includegraphics[width=0.9\textwidth]{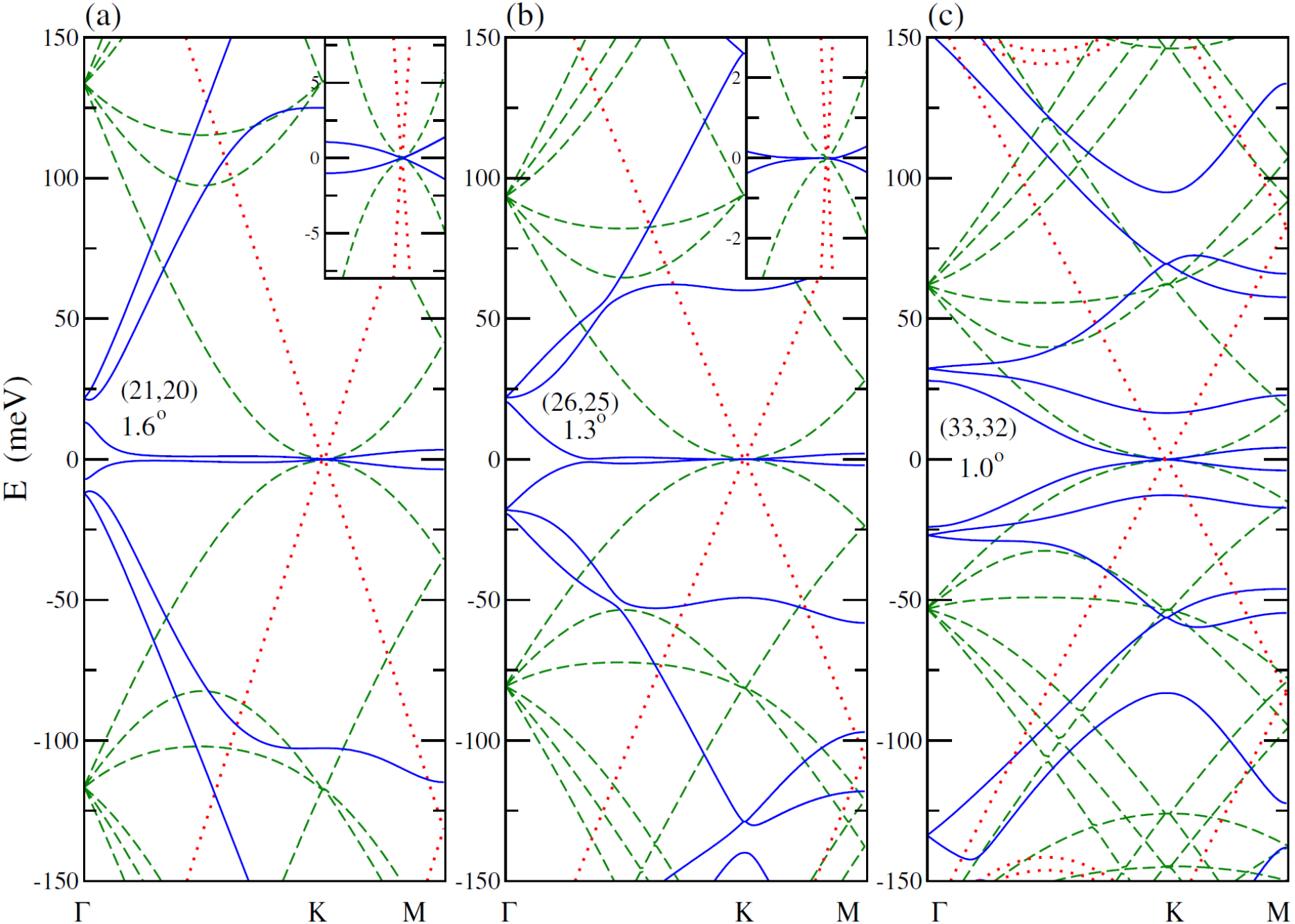}
\caption{(Color online) Twisted bilayer graphene spectra for three twist angles of about $1.5^{\circ}$ calculated in Ref.~\cite{Morell1} (blue solid curves).
For comparison, the spectra of the AB bilayer (green
dashed curves) and single-layer graphene (red dotted curves) are added.
In panels (a) and (b), there are two nearly-flat bands closest to zero
energy. For $\theta\cong1^{\circ}$ [panel(c)]
the bandwidth of these low-energy bands is relatively large. This is
consistent with the oscillatory behavior of the peak's width of the DOS at
small angles (see
Fig.~\ref{FigTBLG_DOS}).
Insets in panels (a) and (b) show the spectra close to the Dirac point. The
numbers
$(n,m)$
in the figures correspond to the structures
$(n,m-n)$
in our notation.
Reprinted figures with permission from S.~Morell et al., Phys.~Rev.~B,
{\bf 82}, 121407 (2010). Copyright 2010 by the American Physical Society.
\url{http://dx.doi.org/10.1103/PhysRevB.82.121407}.  \label{FigTBspec}
}
\end{figure}

The existence of flat bands at small angles is also shown by E.~Su\'arez Morell et
al.~\cite{Morell1}. They used a simpler dependence for the interlayer hopping:
\begin{equation}
t_{\bot}(\mathbf{r};\mathbf{r}')=t_0\exp\left\{\frac{1}{\beta}\left[c_0-\sqrt{c_0^2+(\mathbf{r}-\mathbf{r}')^2}\right]\right\}\,.
\end{equation}
To check the accuracy of the tight-binding calculations, they also compared the
tight-binding spectra with {\it ab initio} data for several twist angles
and found no difference at low energies. Typical spectra for small angles
are shown in
Fig.~\ref{FigTBspec}.
Two low-energy flat bands exist for
$\theta\lesssim1.5^{\circ}$.
No Dirac spectrum near
$\mathbf{K}$
point is observed for these
$\theta$.
The bandwidth
$\Delta E$
of the low-energy flat bands is determined as the difference in energies of the lower and the upper bands at the
$\bm{\Gamma}$
point (see
Fig.~\ref{FigTBspec}). The bandwidth
$\Delta E$
is a non-monotonous function of the twist angle, which is consistent with
the oscillations of the width of the DOS peak predicted in
Refs.~\cite{dSPRB,PNAS,TramblyTB_Loc}.

\begin{figure}[t]
\centering
\includegraphics[width=0.69\textwidth]{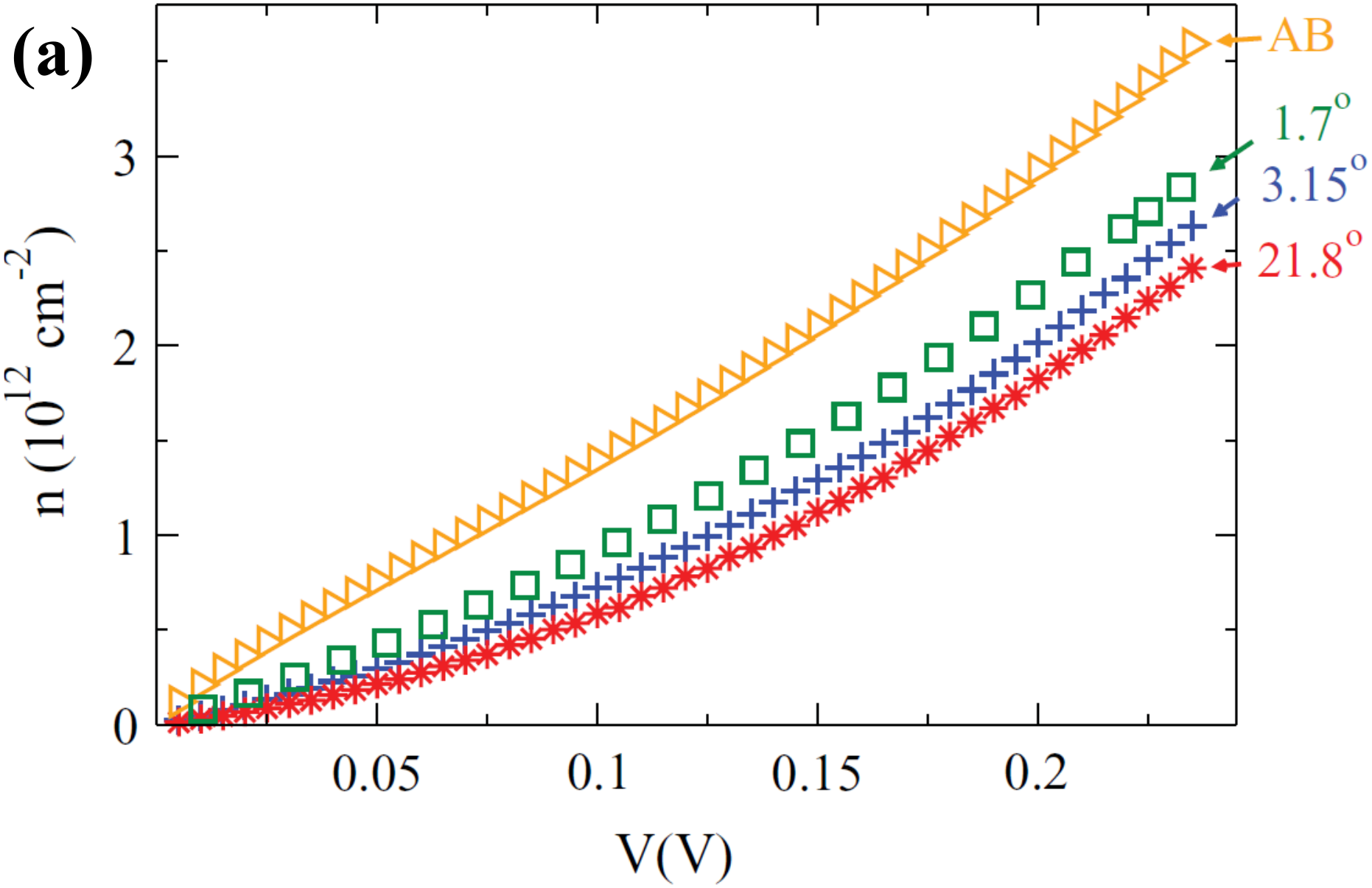}
\includegraphics[width=0.29\textwidth]{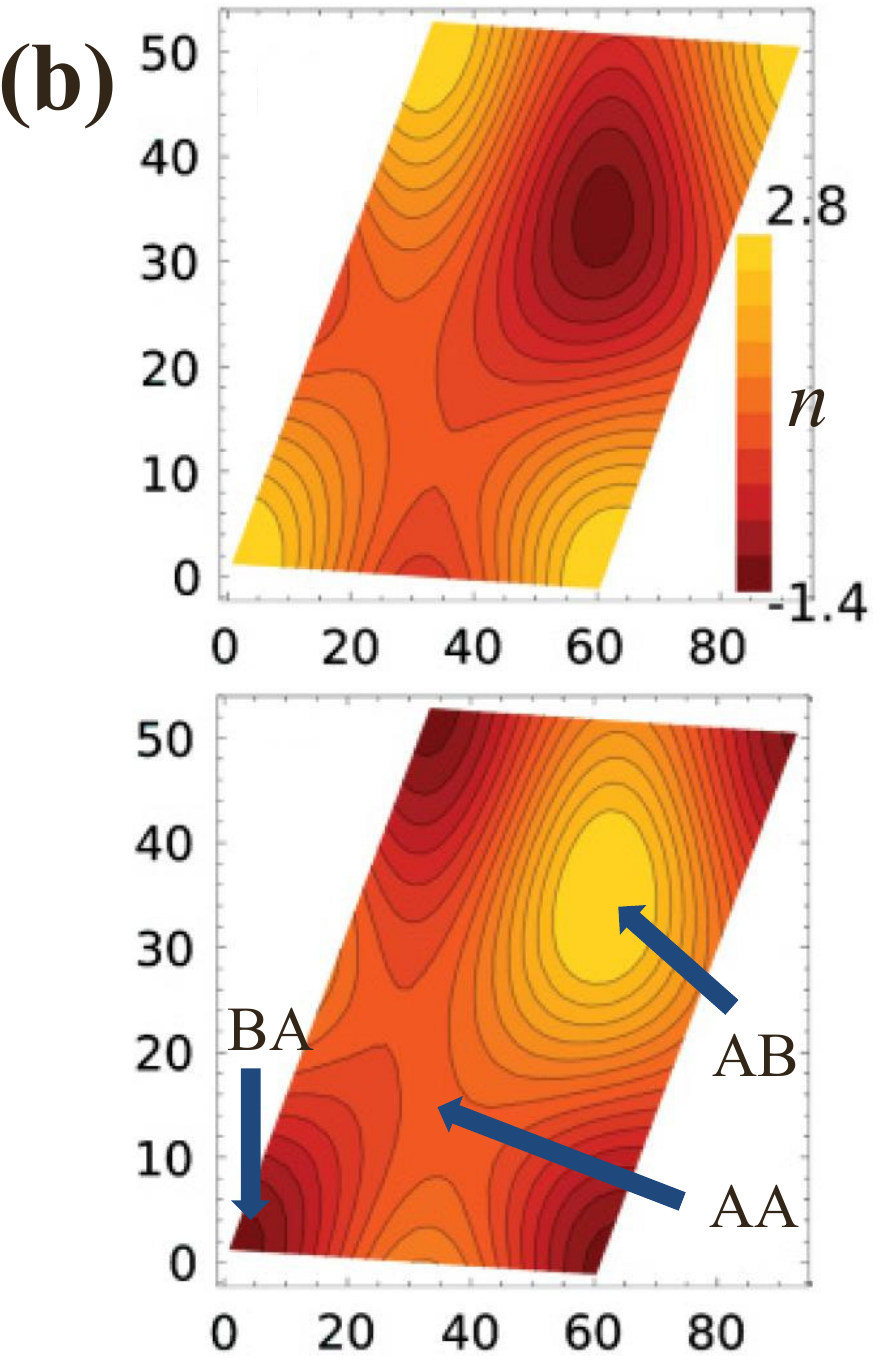}
\caption{(Color online) (a) The average density $n$ of extra electrons in
the top (negatively charged) tBLG layer as a function of the applied gate
voltage for different twist angles. The electron density for the AB bilayer
is also shown (orange triangles). (b) Spatial distribution of the electron
density in the A (upper panel) and B (lower panel) sublattices of the top
layer inside the tBLG supercell. The superstructure is
$(15,1)$,
$\theta\cong2.13^{\circ}$.
The gate voltage applied is
$V=0.08$\,V.
The electron density is in units
$10^{12}$\,cm$^{-2}$,
and the averaged value is
$n=0.7\times10^{12}$\,cm$^{-2}$.
The $x$ and $y$ axes are in {\AA}. Arrows show the regions with nearly
AB, BA, and AA stacking.
Reprinted figures with permission from S.~Morell et al., Phys. Rev. B,
{\bf 84}, 195421 (2011). Copyright 2011 by the American Physical Society.
\url{http://dx.doi.org/10.1103/PhysRevB.84.195421}.\label{FigNvsV}}
\end{figure}

The same parametrization of the inter-layer hopping is used in
Ref.~\cite{Morell2}.
The authors were interested in the effect of the bias voltage applied to
the twisted bilayer. The application of the gate voltage leads to the
redistribution of the charge carriers in the bilayer: one layer acquires
additional electrons with a density $n$, while another layer loses the same
amount of electrons (no doping is assumed).
They~\cite{Morell2}
did not analyze the modification of the electronic spectrum by the bias
voltage $V$, but instead discussed the dependence of the excess electron
density $n$ on the twist angle (the effect of the bias voltage on the tBLG
spectrum was studied in
Refs.~\cite{dSPRL,DeltaUeffDFT,DeltaUeff,MoonTBgate2014}).
Figure~\ref{FigNvsV}(a)
shows the dependence of $n$ on $V$ for bilayers with different $\theta$. At
a fixed bias voltage, the smaller $\theta$, the larger the charge
redistribution becomes. The authors interpreted this result as a
demonstration of the reduction of the effective interlayer coupling for
larger $\theta$.

The next interesting effect is the inhomogeneous spatial distribution of the
excess electron density.
Ref.~\cite{Morell2}
found that the spatial local distribution of electrons is substantially
different for the A and B sublattices in the layer. They plotted the
patterns of the electron density inside the supercell of the bilayer,
separately for A and B sites of the layer [see
Fig.~\ref{FigNvsV}(b)].
The authors distinguished three different regions inside the supercell.  In
the regions with nearly AA stacking, the electron densities at A and B
atoms are approximately equal to each other and correspond to the averaged
value $n$. In the regions with nearly AB stacking, the areas where the
local density near the A sites is larger than the local density at the B
sites is changed by areas where the local electron density at the B sites
exceeds the local density at the A sites (authors of
Ref.~\cite{Morell2}
referred the latter areas as BA stacking regions).  This observation
correlates very well with the calculations of the local density of states
of the twisted bilayer done in
Ref.~\cite{TramblyTB_Loc}.
Indeed, inside the nearly AB region, the LDOS at the B atoms is larger than
that for the A atoms [see Fig.~\ref{FigDOS_TB}(a)].

In
Ref.~\cite{MoonTBgate2014},
P.~Moon et~al. studied the effects of the bias voltage $V$ on the tBLG
spectrum and the dynamical conductivity using a tight-binding model with
hopping amplitudes described by
Eqs.~\eqref{SlaterKoster}\,--\,\eqref{VExp2}
[with
$F_c(r)=1$].
For the range of twist angles studied
($1^{\circ}\lesssim\theta\lesssim15^{\circ}$),
the spectrum is doubly degenerate near the tBLG Dirac points when
$V=0$.
The bias voltage lifts this degeneracy. For
$\theta>\theta_c$,
where Dirac cones exist, the resulting band structure is similar to that
for the AA-stacked bilayer graphene; near each Dirac point there are two
cones shifted upwards and downwards from the Fermi level (at zero doping)
by the energy value proportional to $V$. The coefficient of proportionality
depends on the twist angle. It decreases when $\theta$ decreases going to
zero when
$\theta\to\theta_c$.
Experimental investigation of the tBLG under the transverse bias was
reported in
Ref.~\cite{Liu2015}.

In all theoretical studies mentioned in this subsection, the authors relied
on the two-center approximation to describe the interlayer hopping, see
Eq.~(\ref{SlaterKoster}).
In addition, the Hamiltonians themselves were typically formulated with the
help of the second-quantization formalism.
The approach developed by O.~Pankratov et al. in
Refs.~\cite{PankratovFlakes,Pankratov1,PankratovPRL,PankratovPRB2013}
differs in several respects. First, the authors started from the
first-quantized single-electron Hamiltonian:
\begin{equation}\label{HTB1}
\hat{H}=\frac{\mathbf{p}^2}{2m}+V^{(1)}(\mathbf{r})+V^{(2)}(\mathbf{r})\,,
\end{equation}
where $m$ is the electron mass and
$V^{(i)}$
is the potential of the
$i$-th(=$1,2$) graphene layer. The potential
$V^{(1)}$ ($V^{(2)}$)
is invariant with respect to translations by the single-layer graphene lattice
vectors
$\mathbf{a}_{1,2}$
($\mathbf{a}'_{1,2}$),
while the total
Hamiltonian~\eqref{HTB1}
is invariant with respect to translations by the superlattice vectors
$\mathbf{R}_{1,2}$.
The authors used the basis constructed of the single-layer graphene
eigenstates of each layer,
$|\phi^{(i)}_{s_i\mathbf{k}_i}\rangle$
satisfying the equation
\begin{equation}
\hat{H}^{(i)}|\phi^{(i)}_{s_i\mathbf{k}_i}\rangle\equiv\left(\frac{\mathbf{p}^2}{2m}+V^{(i)}\right)|\phi^{(i)}_{s_i\mathbf{k}_i}\rangle=
\epsilon^{(i)}_{s_i\mathbf{k}_i}|\phi^{(i)}_{s_i\mathbf{k}_i}\rangle\,,
\end{equation}
where
$\epsilon^{(i)}_{s_i\mathbf{k}_i}$
is the eigenenergy, with
$s_i$
being the band index and
$\mathbf{k}_i$
is the momentum.
For every momentum
$\mathbf{k}$
belonging to the Brillouin zone of the twisted bilayer graphene, we can
define a complete basis
$|\Psi_{\mathbf{k}}\rangle
=
\{|\phi^{(1)}_{s_1\mathbf{k}_1}\rangle,
\,
|\phi^{(2)}_{s_2\mathbf{k}_2}\rangle\}$,
where the momenta
$\mathbf{k}_1$
belong to the Brillouin zone of the first layer, momenta
$\mathbf{k}_2$
belong to the Brillouin zone of the second layer, and the admissible
$\mathbf{k}_{1,2}$
satisfy the following relation
\begin{equation}
\label{kset}
\mathbf{k}_{1,2}=\mathbf{k}+n_{1,2}\mathbf{G}_1+m_{1,2}\mathbf{G}_2\,,
\end{equation}
where
$n_{1,2}$
and
$m_{1,2}$
are some properly chosen integers. It is easy to show that the number of eigenfunctions in the set
$|\Psi_{\mathbf{k}}\rangle=\{|\phi^{(1)}_{s_1\mathbf{k}_1}\rangle,\,|\phi^{(2)}_{s_2\mathbf{k}_2}\rangle\}$
is equal to the number of atoms in the supercell $N$. Thus,
$|\Psi_{\mathbf{k}}\rangle$
is the
$N$-component spinor (only
$p_z$
electrons are considered, the number of single-layer graphene bands is equal to
$2$; that is, $s_i=1,\,2$).

In the basis
$\{|\phi^{(1)}_{s_1\mathbf{k}_1}\rangle,\,|\phi^{(2)}_{s_2\mathbf{k}_2}\rangle\}$,
the Hamiltonian~\eqref{HTB1} is a
$2\times2$
block matrix
\begin{equation}\label{Hmatrix}
\hat{H}=\left(\begin{array}{cc}\hat{H}^{11}&\hat{H}^{12}\\(\hat{H}^{12})^{\dag}&\hat{H}^{22}\end{array}\right)\,,
\end{equation}
where the matrices
$\hat{H}^{ii}$
are diagonal,
$(\hat{H}^{ii})_{s_i\mathbf{k}_i,s'_i\mathbf{k}'_i}=\delta_{s_is'_i}\delta_{\mathbf{k}_i\mathbf{k}'_i}\epsilon^{(i)}_{s_i\mathbf{k}_i}$,
and the matrix
$\hat{H}^{12}$
describing the inter-layer hopping has the form:
\begin{equation}\label{H12}
(\hat{H}^{12})_{s_1\mathbf{k}_1,s_2\mathbf{k}_2}=\frac12\left\langle\phi^{(1)}_{s_1\mathbf{k}_1}\left|\left(V^{(1)}+V^{(2)}\right)\right|\phi^{(2)}_{s_2\mathbf{k}_2}\right\rangle\,.
\end{equation}
According to
Eq.~\eqref{kset}, the difference
$\mathbf{k}_2-\mathbf{k}_1$
is the reciprocal vector of the superlattice. Using geometrical considerations, the authors showed that the matrix elements in
Eq.~\eqref{H12} are negligible if
\begin{equation}
\mathbf{k}_2-\mathbf{k}_1\neq n_1\mathbf{G}_1^{(c)}+n_2\mathbf{G}_2^{(c)}\,,
\end{equation}
where
$n_{1,2}$
are integers and
\begin{equation}
\mathbf{G}_1^{(c)}=r\mathbf{G}_1\,,\;\;\mathbf{G}_2^{(c)}=r\mathbf{G}_2\,,
\end{equation}
if
$r\neq3k$,
and
\begin{equation}
\mathbf{G}_1^{(c)}=\frac{r}{3}\left(\mathbf{G}_1+2\mathbf{G}_2\right)\,,\;\;\mathbf{G}_2^{(c)}=\frac{r}{3}\left(2\mathbf{G}_1+\mathbf{G}_2\right)\,,
\end{equation}
otherwise\footnote{The authors of
Refs.~\cite{PankratovFlakes,Pankratov1,PankratovPRL,PankratovPRB2013}
characterized a superstructure by two co-prime integers $p$ and $q$, which
can be connected to our notation
$(m_0,r)$
as follows:
$p=r/2$,
$q=m_0+r/2$
for even
$r$
and
$p=r$,
$q=2m_0+r$
for odd
$r$.}. The authors of
Ref.~\cite{PankratovPRB2013} emphasized the appearance of a new momentum scale in the tight-binding model, which is described by
the $\mathbf{G}_{1,2}^{(c)}$
vectors. Their magnitude
$|\mathbf{G}_{1,2}^{(c)}|=8\pi\sin(\theta/2)/(a\sqrt{3})$
depends only on the twist angle, and the reciprocal vectors
$\mathbf{G}_{1,2}^{(c)}$
correspond to the real-space hexagonal superlattice with the size
$L$
equal to the Moir\'{e} period,
Eq.~\eqref{MoireP}.

For
$\mathbf{k}_2-\mathbf{k}_1=n_1\mathbf{G}_1^{(c)}+n_2\mathbf{G}_2^{(c)}$,
the matrix elements
$(\hat{H}^{12})_{s_1\mathbf{k}_1,s_2\mathbf{k}_2}$
decrease exponentially when
$|n_{1,2}|$
grows, and in numerical calculations one can keep only matrix elements with the
first several
$|n_{1,2}|$.
Since the
matrix~\eqref{Hmatrix}
is sparse, it is possible to apply the efficient Lanczos algorithm to find
the tBLG spectrum at low energies. The efficiency of the numerical scheme
can be increased substantially by using the truncated basis
$\{|\phi^{(1)}_{s_1\mathbf{k}_1}\rangle,\,|\phi^{(2)}_{s_2\mathbf{k}_2}\rangle\}$,
which includes only states with energies within certain window
$|\epsilon^{(n)}_{s_n\mathbf{k}_n}|\lesssim W$
\cite{Pankratov1,PankratovPRB2013}.
The authors claim, that at small twist angles, their algorithm works about
$10^{3}$ times faster than standard tight-binding methods~\cite{PankratovPRB2013}.

In general, the numerical calculations of the spectrum in
Refs.~\cite{Pankratov1,PankratovPRL,PankratovPRB2013} are in agreement with
the results of both tight-binding calculations and the low-energy
theories described above. The calculations show the existence of the Dirac
cones and the Fermi velocity reduction at sufficiently large angles, as well as the band
flatness and the charge localization at small angles. The authors~\cite{PankratovPRB2013} obtained also
several analytical results. Specifically, using perturbation theory
and their tight-binding model, they re-derived the
formula for
$v_F^* = v_F^* (\theta)$,
Eq.~\eqref{VrenormDS}, and demonstrated that it works well for $\theta\gtrsim5^{\circ}$.

\begin{figure}[t]
\centering
\includegraphics[width=0.7\textwidth]{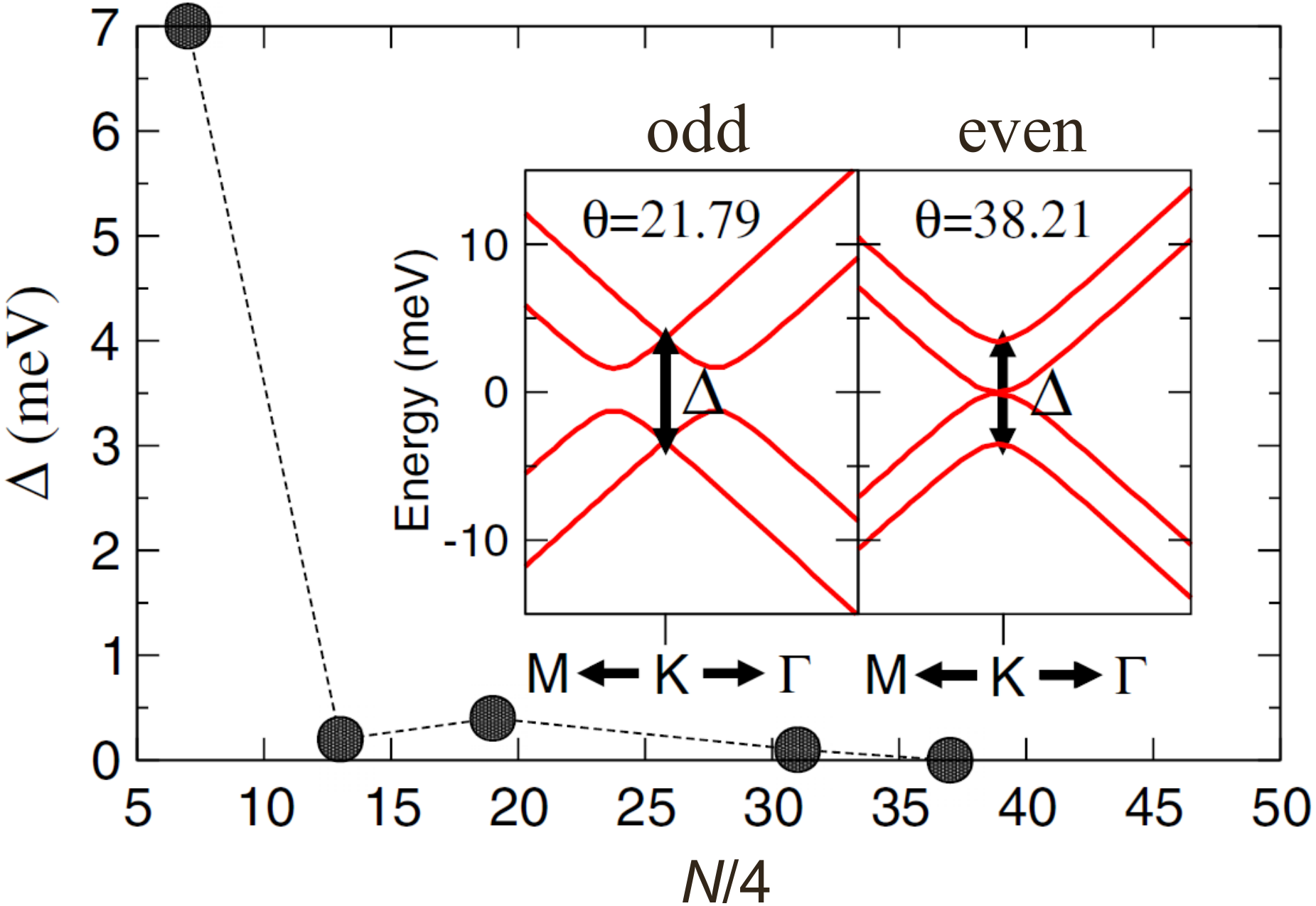}
\caption{(Color online) Dependence of the band splitting ($\Delta_s$ in the text) on the number
of atoms $N$ in the supercell calculated in
Ref.~\cite{PankratovPRL}. Inset shows the low-energy spectrum near the Dirac point for the
$(1,1)$
(`odd') and
$(1,3)$
(`even') superstructure.
Reprinted figure with permission from S.~Shallcross et al., Phys. Rev.
Lett., {\bf 101}, 056803 (2008).
Copyright 2008 by the American Physical Society.
\url{http://dx.doi.org/10.1103/PhysRevLett.101.056803}.\label{FigDeltaN}}
\end{figure}

The important analytical result confirmed by numerical calculations is
the twofold degeneracy of the Dirac cones when the number of atoms in the
supercell $N$ is large~\cite{PankratovPRL}. The analysis is based on the calculation of the
inter-layer matrix elements
$(\hat{H}^{12})_{s_1\mathbf{k}_1,s_2\mathbf{k}_2}$,
for
$\mathbf{k}_1=\mathbf{K}'$,
$\mathbf{k}_2=\mathbf{K}_{\theta}$,
if
$r\neq3n$,
or for
$\mathbf{k}_1=\mathbf{K}$,
$\mathbf{k}_2=\mathbf{K}_{\theta}$
otherwise. According to
Eqs.~\eqref{K1}
and~\eqref{K3},
the difference
$\mathbf{k}_2-\mathbf{k}_1$
is the reciprocal vector of the superlattice for both cases, and, strictly
speaking, the corresponding matrix elements, describing the band splitting,
are non-zero. However, the value
$(\hat{H}^{12})_{s_1\mathbf{k}_1,s_2\mathbf{k}_2}$
decays fast when $N$ grows, and the splitting of Dirac cones becomes
negligible for $N$ larger than some critical value. For smaller $N$, where
the band splitting is relevant, the low-energy spectrum deviates from the
linear dispersion near the Dirac cones, in agreement with the low-energy
theory by
E.\,J.~Mele~\cite{MelePRB1}.
Figure~\ref{FigDeltaN}
presents the dependence of the band splitting
$\Delta_s$
on the number of atoms in the supercell $N$ (for a definition of
$\Delta_s$,
see the inset in
Fig.~\ref{FigDeltaN}). The value of
$\Delta_s$
depends on $N$
non-monotonously, and it becomes negligible when
$N\gtrsim10^2$.
The maximum splitting,
$\Delta_s\approx7$\,meV,
corresponds to the superstructures
$(1,1)$
with
$\theta = 21.79^\circ$,
and
$(1,3)$
with
$\theta=38.21^{\circ}$,
both having
$N=28$
atoms per unit supercell. A similar estimate
$\Delta_s\sim10$\,meV
was obtained by
E.J.~Mele~\cite{MelePRB1}.
The character of the spectrum near the Dirac point, as it was reported in
Ref.~\cite{PankratovPRL},
is qualitatively different for the
$r\neq3n$
(`odd') and
$r=3n$
(`even') superstructures. In the former case, the spectrum is gapped, while
for
$r=3n$
it is gapless, with two parabolic bands touching each other at the Dirac
point (see the inset in
Fig.~\ref{FigDeltaN}).
Such a situation is opposite to that was found by
E.\,J.~Mele~\cite{MelePRB1}
(cf. inset in
Fig.~\ref{FigDeltaN}
with
Fig.~\ref{FigSpecOddEven}).

\begin{figure}[t]
\centering\includegraphics[width=0.7\columnwidth]{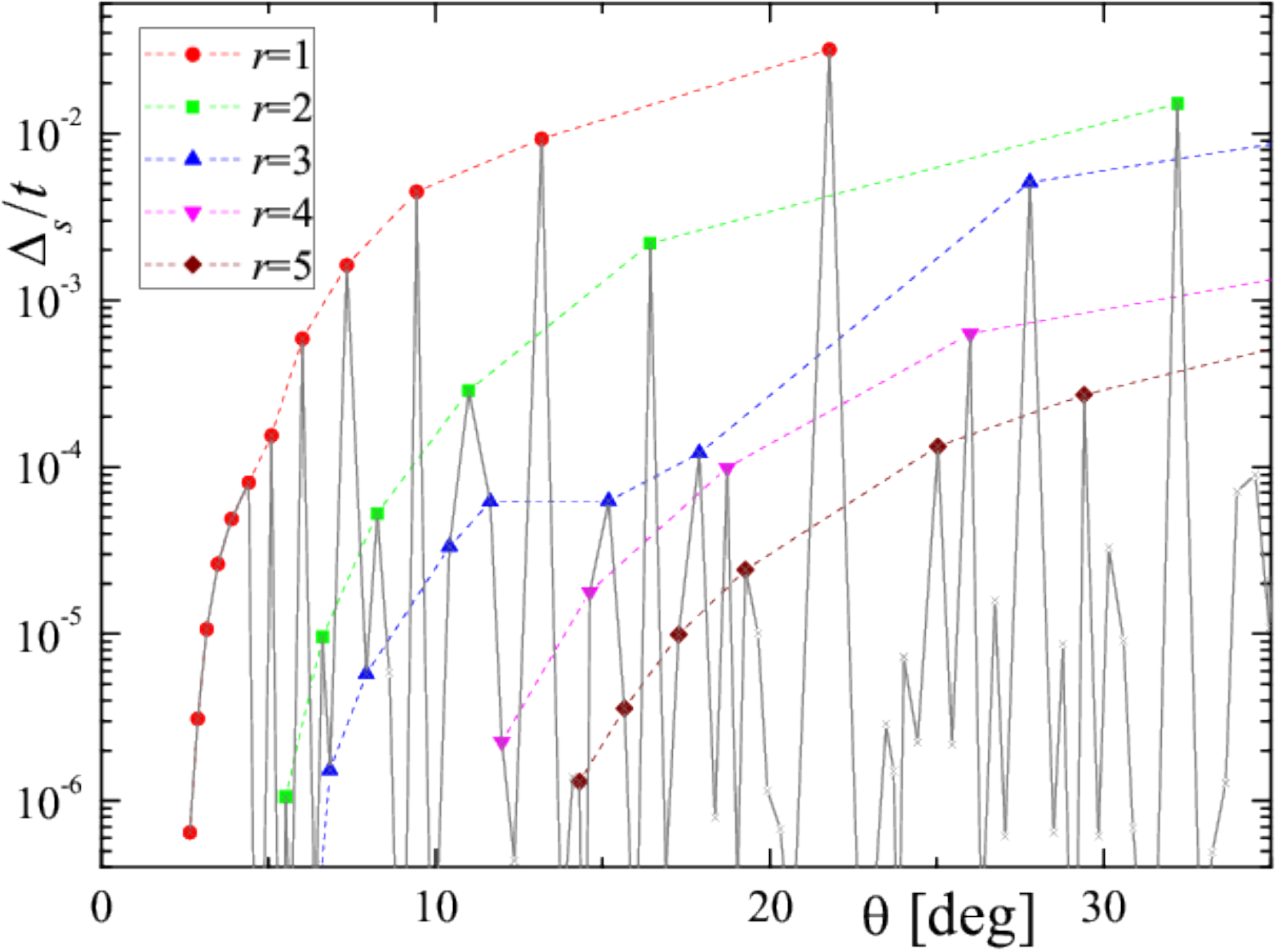}
\caption{(Color online) Dependence of the band splitting
$\Delta_s$
on the twist angle $\theta$, calculated for all superstructures with
$N<2000$
(gray solid line). Different dashed curves connect the points corresponding
to superstructures with a fixed $r$. Figure is from Ref.~\cite{ourTBLG}.\label{FigSplitting}}
\end{figure}

A different analysis of the band splitting has been done by A.\,O.~Sboychakov
et al.,
Ref.~\cite{ourTBLG}.
The authors studied the tight-binding
Hamiltonian~\eqref{HTB},
where the in-plane hopping amplitudes
$t_{\|}^{(i)}\!
(\mathbf{r}_{\mathbf{n}}^{i\alpha};\mathbf{r}_{\mathbf{m}}^{i\beta})$
are non-zero only for a nearest-neighbor hopping equal to
$t=2.57$\,eV.
To calculate the interlayer hopping amplitudes
$t_{\bot}\!(\mathbf{r}_{\mathbf{n}}^{1\alpha};\mathbf{r}_{\mathbf{m}}^{2\beta})$,
the authors use the approach proposed by M.\,S. Tang et al.~\cite{Tang},
which takes into account the environment dependence of the hopping. The
inter-layer hopping amplitude,
$t_{\bot}\!(\mathbf{r};\mathbf{r}')$,
is expressed via Slater-Koster parameters according to
Eq.~\eqref{SlaterKoster}.
However, in contrast to the widely-used two-center approximation, the functions
$V_{pp\sigma}(\mathbf{r};\mathbf{r}')$
and
$V_{pp\pi}(\mathbf{r};\mathbf{r}')$
depend also on the positions of other sites in the lattice via the factor
$[1-S(\mathbf{r};\mathbf{r}')]$,
where the screening function is
\begin{equation}\label{S}
S(\mathbf{r}_i;\mathbf{r}_j)
=
\tanh\left(\beta_1
	\sum_{l}
		\exp\left[
			-\beta_2\left(
					\frac{R_{il}+R_{lj}}{R_{ij}}
				\right)^{\beta_3}\right]\right)\,.
\end{equation}
In this equation,
$R_{ij}$
is the distance between sites
$i$
and
$j$,
the summation is performed over all sites in the lattice, and
$\beta_1,\beta_2$,
and
$\beta_3$
are fitting parameters. It follows from this equation, that the closer some of the neighboring atoms are to the line connecting the sites
$\mathbf{r}_i$
and
$\mathbf{r}_j$,
the larger is the screening. The inclusion of the screening helps to describe more correctly the longer-range inter-layer hopping amplitudes in bilayer graphene systems. Indeed, it is known from experiment and {\it ab initio} calculations~\cite{tunnel2,AB::Mendez,AB::Mucha}, that the next-nearest-neighbor interlayer hopping amplitudes for the AB graphene bilayer and graphite,
$t_3$
and
$t_4$
(for definitions, see
Fig.~\ref{fig::ab_lattice}), are different by about an order of magnitude, even though the distance between the corresponding pairs of sites are equal to each other. Taking the first three largest inter-layer hopping amplitudes of the AB bilayer,
$t_0=0.4$\,eV,
$t_3=0.254$\,eV, and
$t_4=0.051$\,eV, the authors of
Ref.~\cite{ourTBLG}, found the fitting parameters
$\beta_1$,
$\beta_2$,
and
$\beta_3$
of the screening function. The interlayer hopping amplitudes for the tBLG
are calculated then by using these $\beta$s. The same approach was used
also by O.A.~Pankratov et al. in
Refs.~\cite{PankratovFlakes,Pankratov1,PankratovPRB2013};
the fitting parameters are found in these papers by comparison of the
tight-binding spectra with the DFT results. The parametrization of M.S.
Tang et
al.~\cite{Tang}
for inter-layer hopping amplitudes was used also by J.\,M.\,B.~Lopes dos Santos et al.
in
Refs.~\cite{dSPRL,dSPRB}.

\begin{figure}[t]
\includegraphics[width=0.3\textwidth]{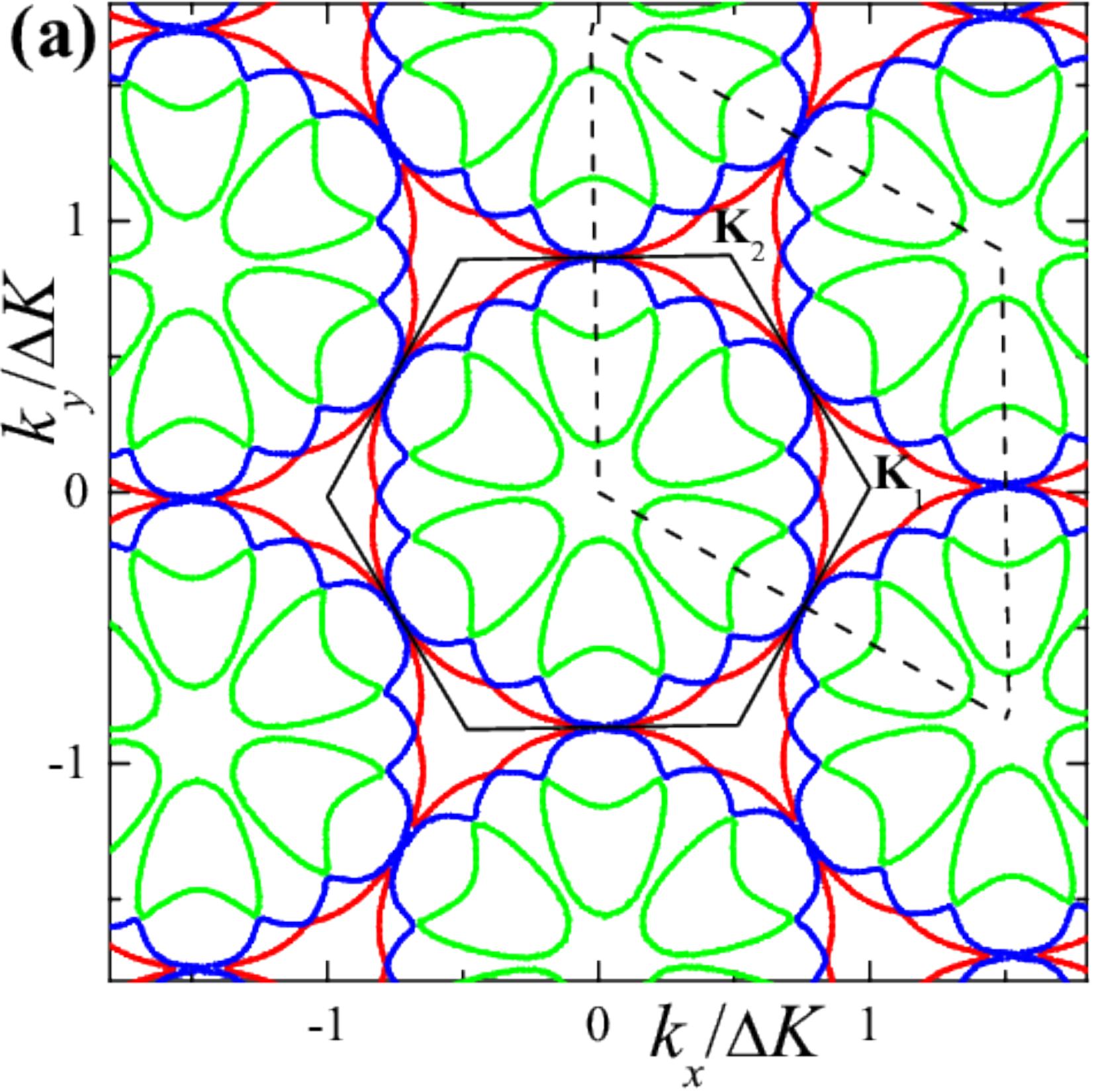}\hspace{2mm}
\includegraphics[width=0.3\textwidth]{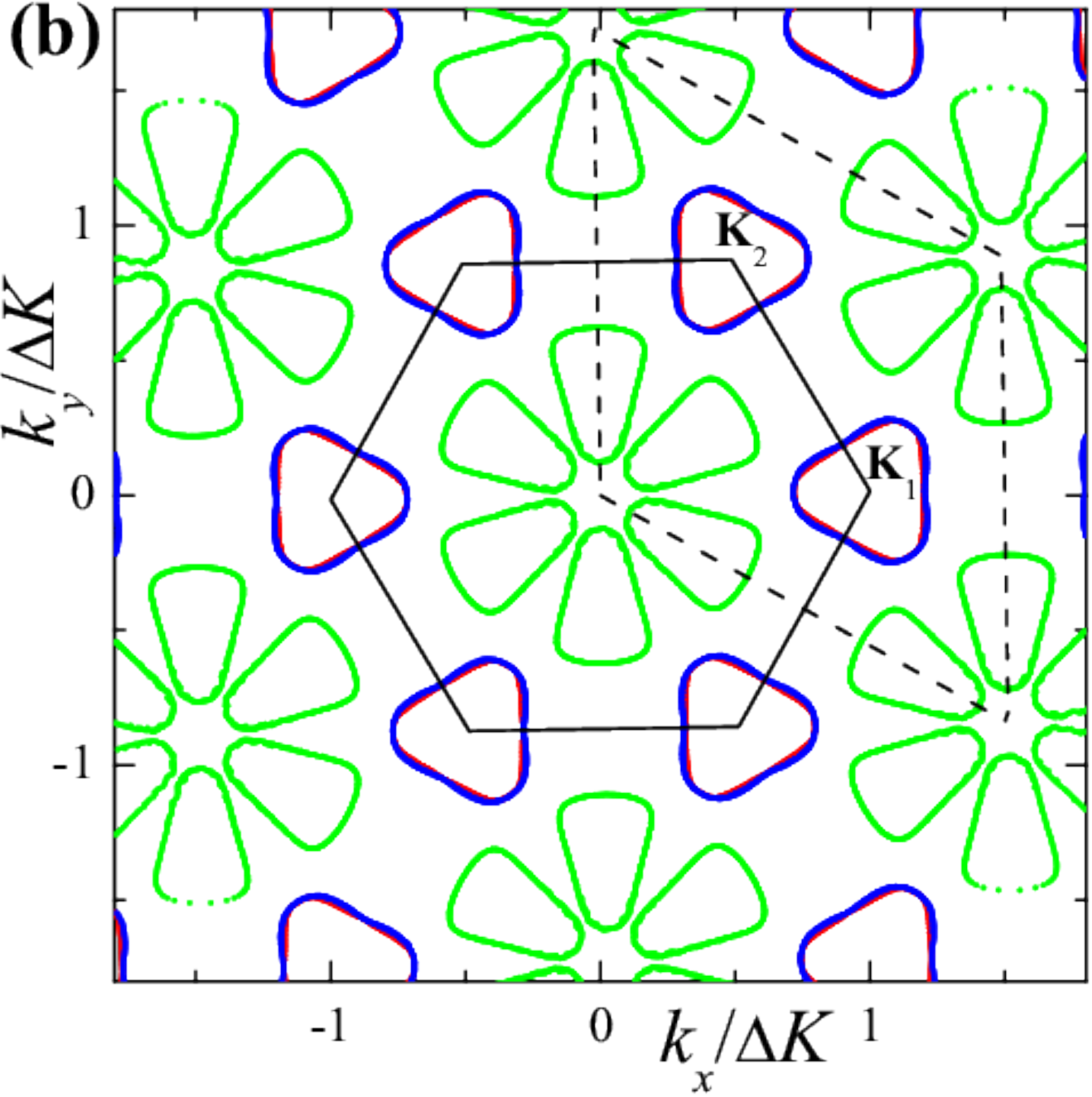}\hspace{2mm}
\includegraphics[width=0.34\textwidth]{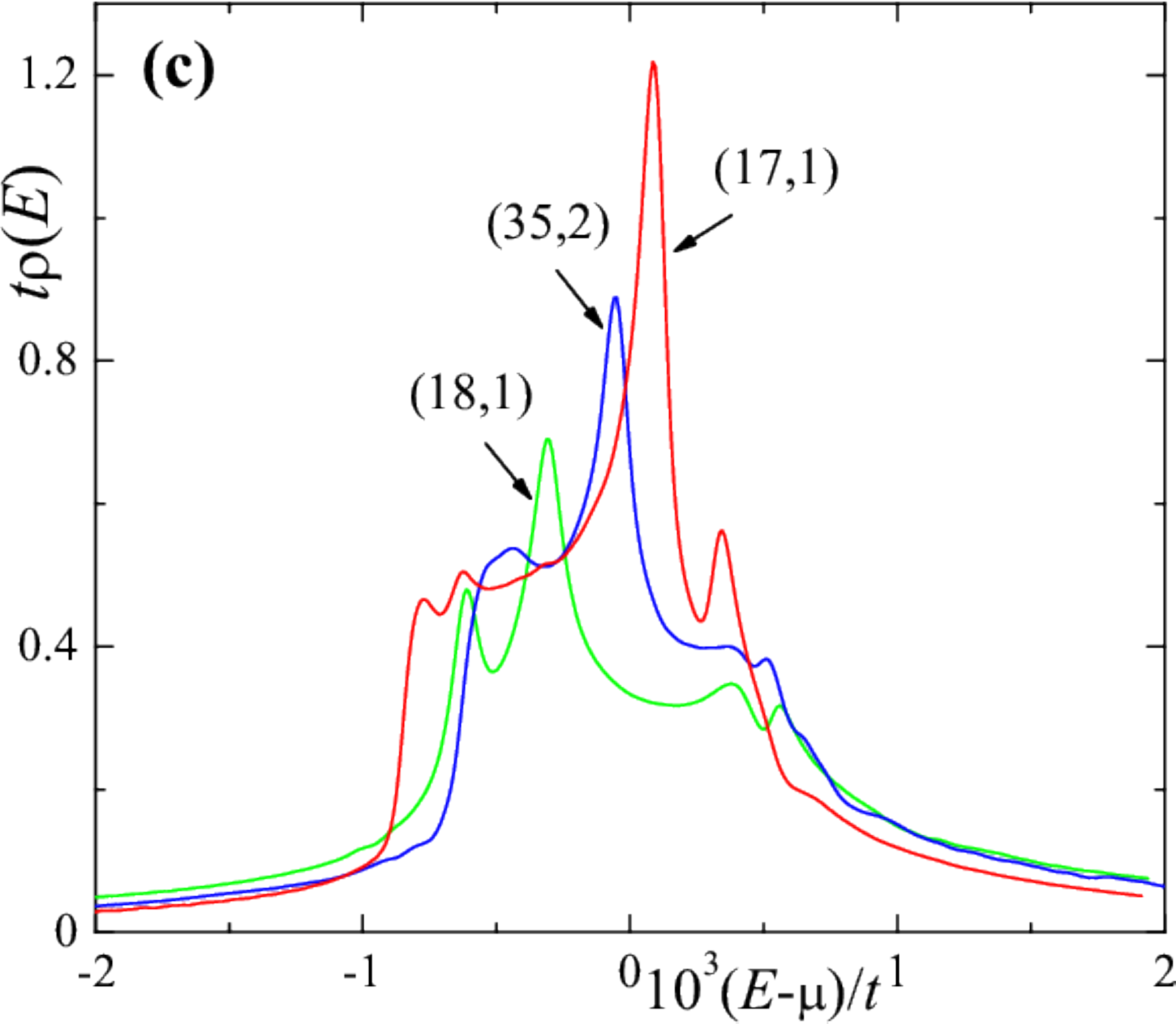}
\caption{(Color online) (a) and (b) Fermi surfaces of the superstructures
$(17,1)$
[$\theta\cong1.89^{\circ}$,
panel (a)] and
$(18,1)$
[$\theta\cong1.79^{\circ}$,
panel (b)] calculated at half-filling in
Ref.~\cite{ourTBLG}. Different colors correspond to different bands intersecting the Fermi level. Brillouin zone (hexagon), the reciprocal supercell (rhombus) and the Dirac points
$\mathbf{K}_{1,2}$
are also shown. (c) Low-energy density of states calculated for three superstructures with close twist angles
$\theta<\theta_c$.
The density of states is calculated at finite temperature
$T/t=10^{-5}$
by numerical integration over the momentum. Figures are from
Ref.~\cite{ourTBLG}.
\label{FigFStBLG}
}
\end{figure}

The authors of
Ref.~\cite{ourTBLG} showed that near the Dirac point
$\mathbf{K}$
the low-energy tight-binding spectrum can be approximated as
\begin{equation}\label{Efit}
E_{\mathbf{K}+\mathbf{k}}^{(\nu)}=\mu\pm\sqrt{\Delta^2+v_F^2\left(|\mathbf{k}|\pm k_0\right)^2}\,,\;\;\nu=1,\dots,4\,,
\end{equation}
if
$r\neq3n$,
or
\begin{eqnarray}\label{EfitReq3}
E_{\mathbf{K}+\mathbf{k}}^{(1,4)}=\mu\mp\sqrt{\Delta_s^2+v_F^2\mathbf{k}^2}\,,\;\;\;
E_{\mathbf{K}+\mathbf{k}}^{(2,3)}=\mu\mp\left(\sqrt{\Delta_s^2+v_F^2\mathbf{k}^2}-\Delta_s\right)\,.
\end{eqnarray}
otherwise. Qualitatively, these dependencies are similar to those found by O.\,A.~Pankratov et al.~\cite{Pankratov1,PankratovPRL,PankratovPRB2013} (see the inset in
Fig.~\ref{FigDeltaN}). The band splitting is defined as
$\Delta_s =E_{\mathbf{K}}^{(4)}-E_{\mathbf{K}}^{(1)}$,
and for
$r\neq3n$
structures, it is related to the band gap according to
$\Delta_s=\sqrt{\Delta^2+v_F^2k_0^2}$.
The dependence of
$\Delta_s$
on the twist angle is shown in
Fig.~\ref{FigSplitting}. It is not a continuous function of $\theta$: any
small deviation of the twist angle from a given value changes drastically
the band splitting.  However, the band splitting for superstructures with
a given $r$ monotonously decreases with $\theta$. Among all superstructures
in some range of the twist angle, the maximum band splitting corresponds to
$r=1$
superstructures. The maximum value
$\Delta_s\cong80$\,meV
corresponds to the
$(1,1)$
superstructure, and it is one order of magnitude larger than estimate of
O.\,A.~Pankratov et al.~\cite{Pankratov1,PankratovPRL}.
For
$m_0<7$
(and
$r=1$) the band splitting exceeds
$1$
Kelvin, which is an experimentally observable quantity. However, the
discontinuous behavior of
$\Delta_s$
on
$\theta$
can make the band splitting (and a band gap) non-observable, because
any infinitesimal deviation
$\delta\theta$
of the twist angle from its value
$\theta_0\equiv\theta(m_0,1)$
totally suppresses the band splitting. In real situation, however, the twist angle is known with a finite error.
Rotational disorder of the twisted bilayer graphene was studied experimentally in Ref.~\cite{RotDisorder2014}. They observed local variations of the twist angle $\Delta\theta\sim2^{\circ}$ in samples with $\theta\gtrsim15^{\circ}$.

The authors of Ref.~\cite{ourTBLG} pointed out that taking into
account the finite mean-free-path of electrons or the sample's size will
make the band splitting a continuous function of the twist angle. Indeed,
calculations of the band gap performed for a finite sample with
$\theta \cong 21.79^{\circ}$
showed~\cite{ourTBLG},
that the band gap changes continuously with $\theta$. The value of $\Delta$
has a peak at $\theta$ corresponding to the $(1,1)$
superstructure, the width of this peak is
$\Delta\theta\sim a_0/\tilde{L}$,
where
$\tilde{L}$
is the sample's size.

Reference~\cite{ourTBLG} also mentioned the presence of low-energy edge states with a characteristic localization length much smaller than the sample's size. In the band gap calculations these states were ignored. The existence of edge states was reported also by O.\,A.~Pankratov et al. in Ref.~\cite{PankratovFlakes}, where they studied the electronic properties of the tBLG flakes. Edge states in the twisted bilayer graphene nanoribbons were studied theoretically in more detail in Ref.~\cite{MorellNanoribons2014}. The authors found, in particular, that the edge states with energies close to the Fermi level (for undoped bilayer) are localized at the AB-type regions of the edge, while states localized at the AA-type regions of the edge become split; they have energies either above or below the Fermi level by a value of about $0.2$\,eV.

The band gap in twisted bilayer graphene has been observed experimentally in transport measurements by J.~Park et~al. in Ref.~\cite{TransportGap2015}. They estimated the twist angle of their sample as $\theta\sim30^{\circ}$. The band gap was extracted both from the $IV$ characteristics and from the temperature dependence of the conductivity and the carriers concentration obtained by Hall measurements. Thus, the authors~\cite{TransportGap2015} showed that both conductivity and the concentration of the charge carriers follow the thermal activation law, $\ln\sigma,\,\ln n\propto\Delta/T$, with gap $\Delta\sim80$\,meV (the estimate of the gap from the width of the flat region in the $IV$ curve gives a much larger value of the gap). The authors explain such a big value of the band gap by the local formation of the $sp^3$ interlayer $C$-$C$ bonding arising due to the peculiarities of the sample's preparation techniques they used. The theoretical background for such a possibility can be found in Refs.~\cite{Muniz1,Muniz2}. Note, however, that the estimate $\Delta\sim80$\,meV coincides with the gap found in Ref.~\cite{ourTBLG} using the tight-binding calculations described above.

Band gap engineering by the intercalation of twisted bilayer graphene by alkali metal ions has been proposed in Ref.~\cite{GapFabrication2015}. Using DFT calculations, the authors found the equilibrium positions for the intercalated ions inside the supercell of the bilayer. Intercalation modifies the electronic spectrum leading to the gap opening at the Dirac points. They mentioned that the gap can be ``several times larger than room temperature''. The modification of the tBLG electron spectrum by point defects has been studied, also using DFT, in Ref.~\cite{PointDefectsDFT2014}.

The authors of
Ref.~\cite{ourTBLG}
also found the estimate for the critical value of the twist angle
$\theta_c\cong1.89^{\circ}$.
If
$\theta\leqslant\theta_c$,
no Dirac spectrum exists, and the tBLG has a Fermi surface and a finite
density of states at the Fermi level.
Figure~\ref{FigFStBLG}
shows the Fermi surfaces calculated at half-filling for the superstructures
$(17,1)$
($\theta=\theta_c\cong1.89^{\circ}$) and
$(18,1)$
($\theta\cong1.79^{\circ}$).
The peak in the density of states existing near the Fermi level for
$\theta\leqslant\theta_c$
has a complex structure [see
Fig.~\ref{FigFStBLG}(c)].
Calculating the density of states allows a direct comparison of the
electronic properties of superstructures with different
$r$.~\footnote{Structures with different $r$ but similar $\theta$ have
Brillouin zones of greatly dissimilar sizes. Thus, to compare the
spectra and/or Fermi surfaces of these structures one has to work with
folded Brillouin zones.}
The authors found that, neglecting some fine structures, the electronic
properties of the tBLG change continuously with the twist angle when
$\theta<\theta_c$.
Another theoretical study of the Fermi surface structure and properties in
the regime of small $\theta$ was reported in
Ref.~\cite{low_en_pankratov2016}.

In all theoretical investigations described above it is assumed that the
graphene layers are flat, and the distance between layers is a constant.
Experiments shows, however, that the interlayer distance $c_0$ is spatially
modulated~\cite{STM1}.
Using STM,
Ref.~\cite{STM1}
showed that $c_0$ is larger at regions with almost AA stacking and smaller
in regions with almost AB stacking. The corrugation of graphene layers in
tBLG was found theoretically using DFT in
Refs.~\cite{LargeScaleDFT_PRB2014,LargeScaleDFT2015}.
The authors used their own efficient computation code allowing to perform
DFT calculations for systems consisting of
$10^4$-$10^5$
atoms. In particular, they showed that for twist
angles~\cite{LargeScaleDFT2015}
$\theta\lesssim20^{\circ}$
(or $\theta\gtrsim40^{\circ}$)
the layers are corrugated; the intralayer distance
$c_0$
is a Moir\'{e} periodic function of the real-space coordinate
$\mathbf{r}$,
having a maximum (minimum) for
$\mathbf{r}$
in the region with almost AA (AB) stacking, which correlates with the
experiment~\cite{STM1}.

\subsection{Twisted bilayer graphene in a magnetic field}

\subsubsection{Landau levels and quantum Hall effect}

Landau quantization and quantum Hall effect are two issues which are the
main focus for both theoretical and experimental studies of the tBLG in an
applied perpendicular magnetic field. As in the previous subsection, the
theoretical papers on this topic can be classified according to the level
of approximation of the tBLG Hamiltonian.
Here we follow the order of the previous section and consider first the
works using effective low-energy Hamiltonians. The analysis based on
tight-binding Hamiltonians will be considered afterwards.
We compare theoretical predictions with experimental results, both for
the Landau quantization and the quantum Hall effect.

In Refs.~\cite{deGail,LL_lowEnergy}, the structure of Landau levels is studied in the framework of a simplified
$2\times2$
effective Hamiltonian~\eqref{HTBLGEff}. In the presence of a magnetic field perpendicular to the bilayer, the Hamiltonian~\eqref{HTBLGEff} can be rewritten as
\begin{equation}\label{HTBLGeff_H}
\hat{H}=\hbar\omega_c^{*}\left(\begin{array}{cc}0&a^2-\bar{\beta}^2\\
a^{\dag2}-\beta^2&0\end{array}\right),
\end{equation}
where
\begin{equation}\label{OmegaCtblg}
\omega_c^{*}=\frac{2eBv_F^{*}}{\hbar\Delta K}
\end{equation}
is the cyclotron frequency, the complex number $\beta$ is defined as
\begin{equation}
\label{betaB}
\beta=\sqrt{\frac{\hbar}{8eB}}(\Delta K_x+i\Delta K_y)\,,
\end{equation}
and the operators
$a$
and
$a^{\dag}$,
such that
$[a,a^{\dag}]=1$,
are the usual lowering and raising operators of the harmonic
oscillator~\cite{LandauQM}.
Note that the cyclotron frequency
$\omega_c^{*}$,
given by
Eq.~\eqref{OmegaCtblg},
is different from
$\omega_c$
introduced  in Section~\ref{MagnF},
Eq.~\eqref{CyclFre}. The authors of
Ref.~\cite{deGail} solve the eigenvalue problem
$\hat{H}\psi=E\psi$ by
numerically expanding the spinor wave function
$\psi=(\psi_1,\,\psi_2)^T$
in the harmonic-oscillator basis $\{|m\rangle\}$:
\begin{equation}\label{psi1n}
\psi^{(n)}=\!\!\!\!\!\sum_{m=-[n/2]}^{\infty}\!\!\left(\begin{array}{c}\phi^1_{2m+n}\\\phi^2_{2m+n}\end{array}\right)|2m+n\rangle\,,
\end{equation}
where the index
$n=0,1,2,\dots$
enumerates the eigenvalues
and eigenfunctions,
and the ket-vectors
$|m\rangle$ satisfy the equations
\begin{equation}
a\,|m\rangle=\sqrt{m}|m-1\rangle\,,\;\;a^{\dag}|m\rangle=\sqrt{m+1}|m+1\rangle\,,\;\;m=0,\,1,\,2,\,\dots\,.
\end{equation}
For an odd (even) value of the index $n$, the
expansion~\eqref{psi1n}
contains only odd (even) basic states
$|m\rangle$.
The authors of Ref.~\cite{deGail} found that for any value of the magnetic field, the tBLG Hamiltonian has two zero-energy eigenstates with
$n=0$
and
$n=1$.
The corresponding eigenfunctions can be found analytically. They are~\cite{deGail}
\begin{equation}
\psi^{(0)}={\cal N}_0\cosh(\bar{\beta}a^{\dag})\left(\begin{array}{c}0\\\!\!\!|0\rangle\!\!\!\!\end{array}\right),\;\;\;
\psi^{(1)}={\cal N}_1\frac{\sinh(\bar{\beta}a^{\dag})}{\bar{\beta}}\left(\begin{array}{c}0\\\!\!\!|0\rangle\!\!\!\!\end{array}\right).
\end{equation}
Thus, together with the spin and valley degeneracy, the zero-energy eigenstate
is eight-fold degenerate.

The eigenvalue problem for the spinor wave function
$\psi=(\psi_1,\,\psi_2)^T$
can be reduced to a single equation of the form
\begin{equation}
(a^{\dag2}-\beta^2)(a^2-\bar{\beta}^2)\psi_1=\lambda\psi_1\,,\;\;\;
\lambda=\left(\frac{E}{\hbar\omega_c^{*}}\right)^2\,.
\end{equation}
The authors of
Ref.~\cite{LL_lowEnergy}
consider this equation in its holomorphic representation
\begin{equation}
a^{\dag}\mapsto\xi\,,\;\;a\mapsto\frac{d}{d\xi}\,.
\end{equation}
As a result, they derived a second-order differential equation, and solved
it numerically. In the general case, there is no analytical expressions for the
eigenenergies
$E_n$.
However, the authors of
Ref.~\cite{LL_lowEnergy}
were able to find asymptotic expansions for the Landau levels in two
limiting cases of small and large $|\beta|$. In the former case they
obtained
\begin{equation}
\label{ELLsmallb}
\left(\frac{E_n}{\hbar\omega_c^{*}}\right)^2
\cong
n(n-1)\left[
		1-\frac{2|\beta|^4}{(2n+1)(2n-3)}
		+
		\frac{4n(n-1)(4n(n-1)-39)+63}
			{(2n+3)(2n+1)^3(2n-3)^3(2n-5)}
			|\beta|^8
		+\dots
	\right]\,.
\end{equation}
When
$|\beta|\gg1$,
the result is
\begin{equation}
\label{ELLlargeb}
\left(\frac{E_n}{\hbar\omega_c^{*}}\right)^2
\cong
4\left[\frac{n}{2}\right]|\beta|^2
\left(
	1-\left[\frac{n}{2}\right]\!\frac{1}{2|\beta|^2}-
	\left[\frac{n}{2}\right]^2\!\!
	\frac{1}{4|\beta|^4}+\dots
\right)\,.
\end{equation}
For small $|\beta|$, the non-zero Landau levels are four-fold degenerate due
to the spin and valley degeneracy. The case
$|\beta|\ll1$
can be only realized either for very small twist angles or very high
magnetic fields. Indeed, since
$|\Delta\mathbf{K}|=8\pi\sin(\theta/2)/(3\sqrt{3}a_0)\cong4\pi\theta/(3\sqrt{3}a_0)$
and
$a_0=1.42\,{\AA}$,
one obtains from
Eq.~\eqref{betaB}:
\begin{equation}\label{betaBtheta}
|\beta|\cong2.7\frac{\theta[\text{deg}]}{\sqrt{B[\text{T}]}}\,.
\end{equation}
For the characteristic twist angle
$\theta=3^{\circ}$
for which the effective
$2\times2$
Hamiltonian~\eqref{HTBLGEff} is best applicable and the magnetic field
$B=10$\,T, one obtains
$|\beta|\approx2.57$,
which is closer to the limit of
$|\beta|\gg1$.
It follows from
Eq.~\eqref{ELLlargeb},
that a level with
$n=2\ell$
and another level with
$n=2\ell+1$,
where $\ell$ is an integer, have the same energy. Keeping only the leading term
in
Eq.~\eqref{ELLlargeb},
one obtains
\begin{equation}\label{ELLmassless}
E_{2\ell}
\cong
E_{2\ell+1}
\cong
\varepsilon_{LL}(\ell)\equiv\pm v_F^{*}\sqrt{2e\hbar |\ell|B}\,,
\end{equation}
which coincides with
Eq.~\eqref{LLspec}.
Thus, in the limit
$|\beta|\gg1$,
all Landau levels are eight-fold degenerate. Under more scrutiny, however,
one discovers that for large $|\beta|$, the difference
between
$E_{2\ell}$
and
$E_{2\ell+1}$
is finite, but exponentially small. The authors of
Ref.~\cite{LL_lowEnergy}
obtained an estimate
\begin{equation}\label{ELLdeg}
\frac{E_{2\ell+1}^2-E_{2\ell}^2}{\hbar^2\omega_c^{*2}}\approx\frac{2(2|\beta|)^{4\ell+2}}{(\ell-1)!\ell!}e^{-2|\beta|^2}\,.
\end{equation}
The parameter $|\beta|$ is smaller for smaller twist angles or for higher
magnetic fields. Thus, the degeneracy between the levels with
$n=2\ell$
and
$n=2\ell+1$
is lifted by stronger magnetic field. Due to the factor
$(2|\beta|)^{4\ell+2}$
in
Eq.~\eqref{ELLdeg},
the larger $\ell$ is, the smaller the magnetic field is required to lift the
degeneracy. The degeneracy of
the levels
$n=2\ell$
and
$n=2\ell+1$
is lifted when the right-hand-side of
Eq.~\eqref{ELLdeg} becomes of the order of unity. For
$\ell=1$
this happens when
$|\beta|\cong2.18$,
while for
$\ell=2$
one obtains
$|\beta|\cong2.99$.

\begin{figure}[t]
\centering
\includegraphics[width=0.48\textwidth]{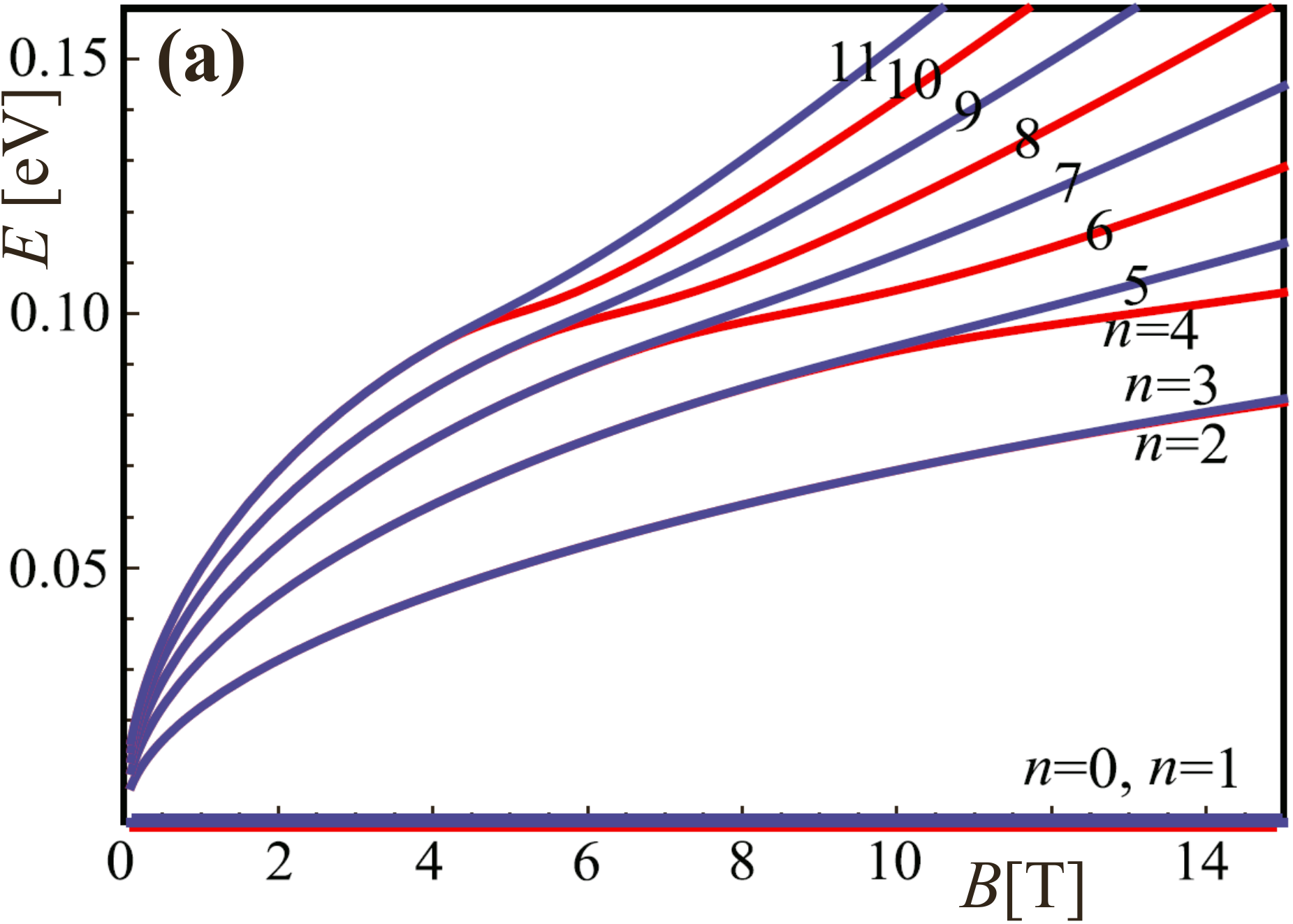}\hspace{2.5mm}
\includegraphics[width=0.49\textwidth]{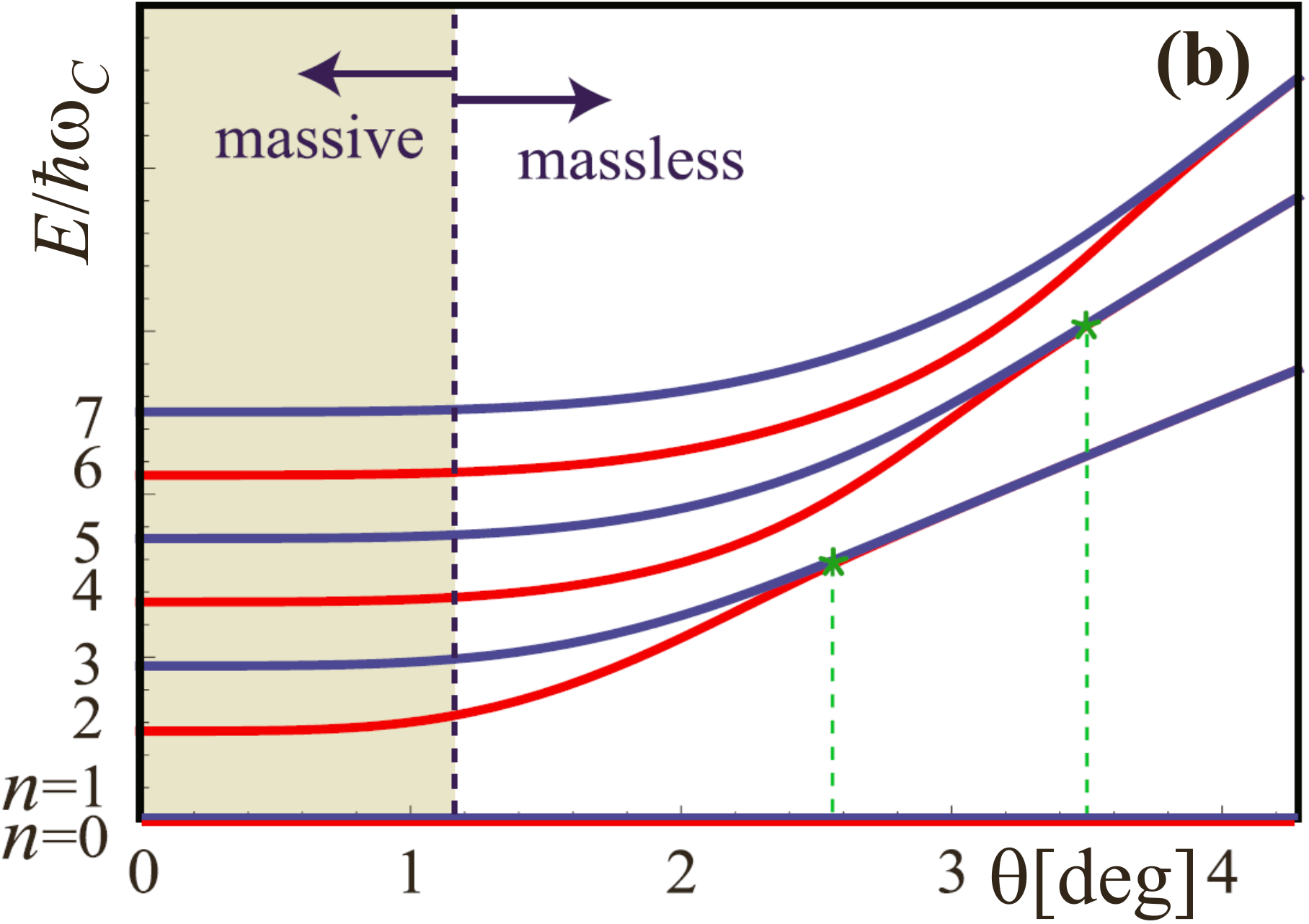}
\caption{(Color online) Energies of the first several Landau
levels  in tBLG versus applied perpendicular magnetic field (left panel, twist angle
$\theta=3.17^{\circ}$)
and on the twist angle at fixed magnetic field (right panel,
$B=10$\,T).
At small fields or large twist angles, the non-zero Landau levels are
eight-fold degenerate and form a sequence corresponding to massless
particles, as in
Eq.~\eqref{ELLmassless}.
In the opposite limit, Landau levels are approximately equidistant,
corresponding to massive particles. In this limit, the non-zero LLs are
four-fold degenerate. For any values of $B$ and $\theta$ there is one
eight-fold degenerate Landau level having zero energy.
Reprinted figures with permission from M.-Y.~Choi et al., Phys. Rev. B,
{\bf 84}, 195437 (2011). Copyright 2011 by the American Physical Society.
\url{http://dx.doi.org/10.1103/PhysRevB.84.195437}.\label{FigELLeff}}
\end{figure}

Figure~\ref{FigELLeff}
shows the dependencies of the Landau levels energies on the magnetic field
[panel (a)] and on the twist angle at fixed magnetic field [panel (b)]
calculated in
Ref.~\cite{LL_lowEnergy}.
The same results are obtained also in
Ref.~\cite{deGail}.
In these figures we see a crossover from the massless behavior at small
fields (or relatively large twist angles) to the equidistant massive behavior
at larger fields.

Measurements of the Landau levels in twisted graphene samples fabricated by chemical vapor deposition have been performed in
Ref.~\cite{STM2}.
The magnetic field was varied in the range
$2\,{\rm T} < B < 10$\,T.
The bilayer graphene sample studied had large regions within which the
Moir\'{e} pattern preserved its periodicity. That allowed the authors~\cite{STM2} to
extract the local values of åðó twist angles.
For twist angles larger than about
$2^{\circ}$,
the data are presented in
Fig.~\ref{FigVrenorm}(a).
The sequence of Landau levels is well described  by a square-root
dependence, given by
Eq.~\eqref{ELLmassless}.
This work~\cite{STM2} reported Fermi velocity renormalization for the samples [see
Fig.~\ref{FigVrenorm}(b)],
with the dependence of
$v_F^{*}$
on $\theta$ correlating well with the theoretical prediction
Eq.~\eqref{VrenormDS}.
A deviation from
Eq.~\eqref{ELLmassless}
was observed when
$\theta\lesssim1.17^{\circ}$.
For these angles, the form of the tunneling spectra depends substantially
on the position of the STM tip inside the Moir\'{e} cell, which is
consistent with the theoretical calculations of the local density of states
for small angles. A square root dependence of the Landau levels is reported
also in
Ref.~\cite{MillerLLexp}
for multilayer samples grown on SiC. No Fermi velocity reduction is found
in these samples (for a discussion on this issue, see page $3$ of
Ref.~\cite{STM2}).

\begin{figure}[t]
\centering
\includegraphics[width=0.7\textwidth]{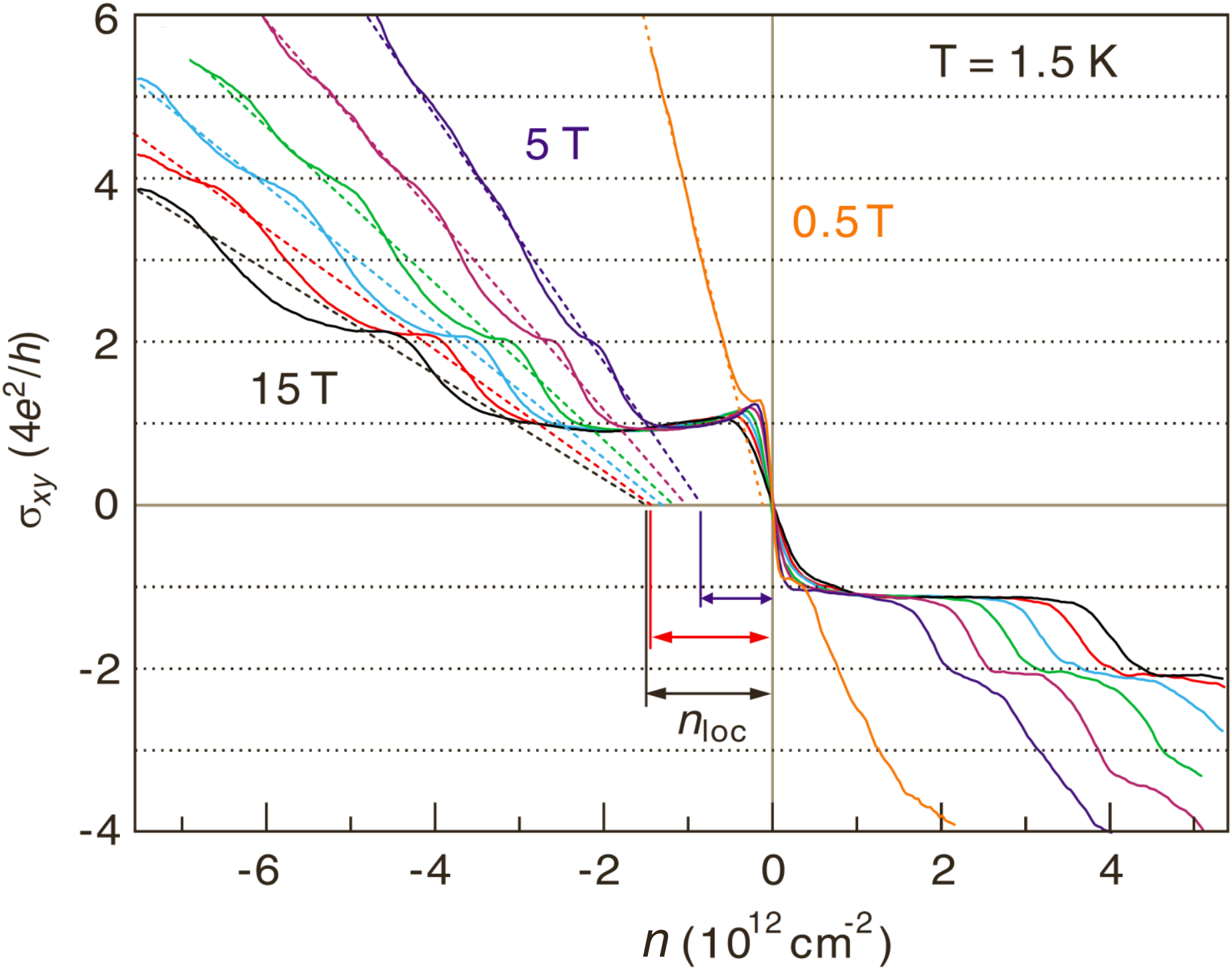}
\caption{(Color online) Hall conductivity of the tBLG
sample versus the electron density, measured for different values of the magnetic field.
Reprinted figure with permission from D.\,S.~Lee et al., Phys. Rev. Lett.,
{\bf 107}, 216602 (2011). Copyright 2011 by the American Physical Society.
\url{http://dx.doi.org/10.1103/PhysRevLett.107.216602}.
\label{FigQHE_TBLGprl2011}}
\end{figure}

Landau quantization can also be observed in measurements of both the longitudinal,
$\sigma_{xx}$,
and Hall,
$\sigma_{xy}$,
conductivities. At zero temperature, the Hall conductivity is a step function of the chemical potential (or the induced charge carrier density),
$\sigma_{xy}=(e^2/2\pi\hbar)\nu$,
where the filling factor
$\nu$
takes integer values which depend on the degeneracy of the Landau levels. When
$|\beta|\ll1$,
the zero-energy Landau level is eight-fold degenerate, while all other levels
are four-fold degenerate, and the filling factor is
\begin{equation}\label{QHE_ab}
\nu=\pm4,\,\pm8,\,\pm12,\dots.
\end{equation}
In the opposite case of
$|\beta|\gg1$,
when
Eq.~\eqref{ELLmassless}
is valid, all Landau levels are eight-fold degenerate, and
\begin{equation}\label{QHE_2slg}
\nu=\pm4,\,\pm12,\,\pm20,\dots.
\end{equation}

The Hall conductivity of the tBLG samples was measured in several works~\cite{tBLG_QHE,LeeQHE,FallahazadQHE}. Figure~\ref{FigQHE_TBLGprl2011} shows the dependence of the
$\sigma_{xy}$
on the electron density
$n$
at different values of the magnetic field (from D.\,S.~Lee et~al.~\cite{LeeQHE}).
The electron density is induced by the applying a gate voltage.
From Fig.~\ref{FigQHE_TBLGprl2011} we see that the Hall conductivity is quantized according
to
Eq.~\eqref{QHE_ab},
corresponding to the limit of
$|\beta|\ll1$.
For example, the step at filling factor
$\nu=\pm8$,
(which is absent in the opposite limit
$|\beta|\gg1$), is clearly seen in this figure for $B=5$\,T and larger. The twist angle for samples studied in
Ref.~\cite{LeeQHE} is estimated as
$\theta\cong2.2^{\circ}$,
and for
$B=5$\,T one obtains from
Eq.~\eqref{betaBtheta}
that
$|\beta|\cong2.66$.
This estimate is slightly below the value
$|\beta|\cong2.99$
Where the eight-fold degeneracy of the second Landau level is lifted.
The authors concluded that the regime of small $|\beta|$ was
realized in their experiments.

\begin{figure}[t]
\centering
\includegraphics[height=0.32\textwidth]{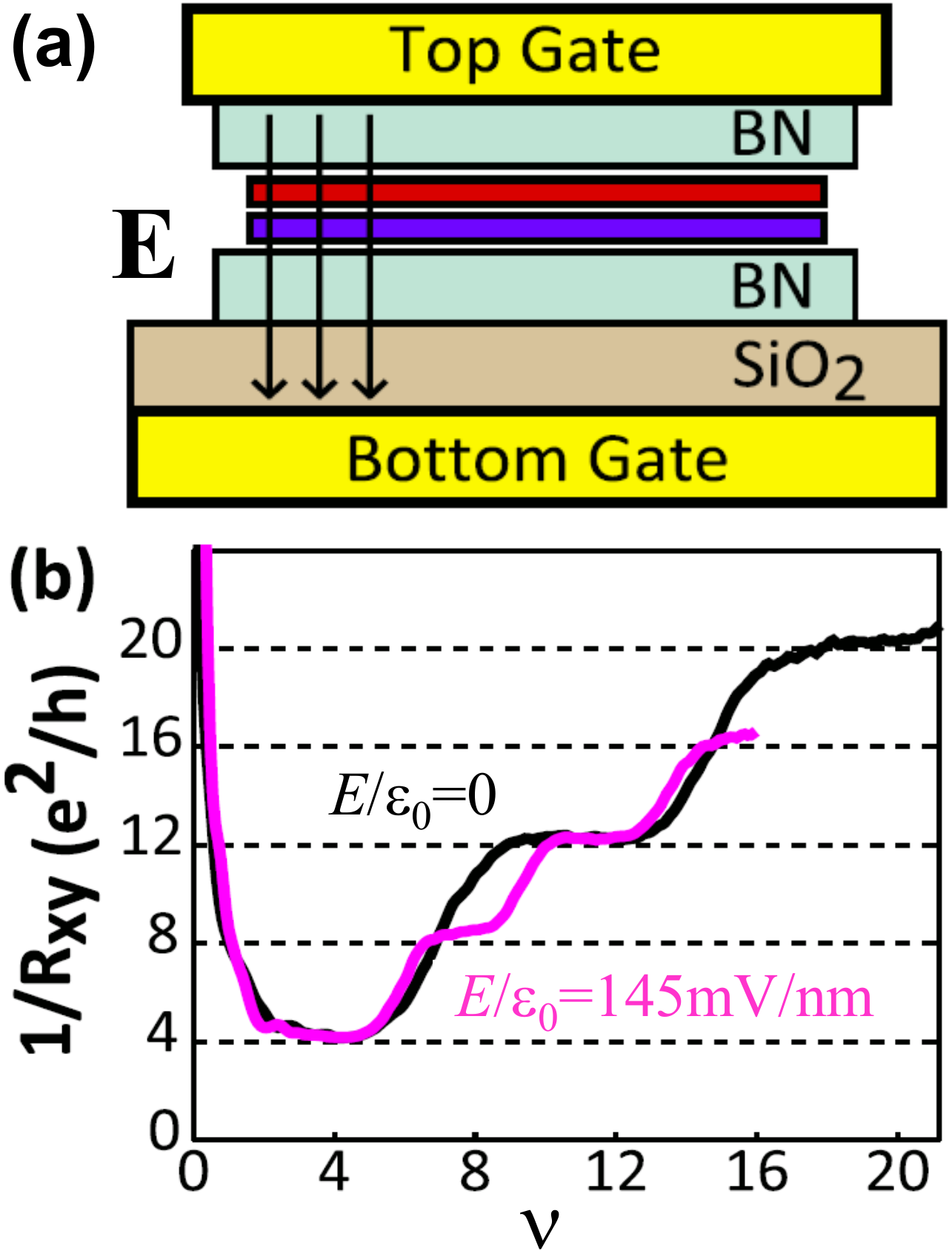}\hspace{3mm}
\includegraphics[height=0.32\textwidth]{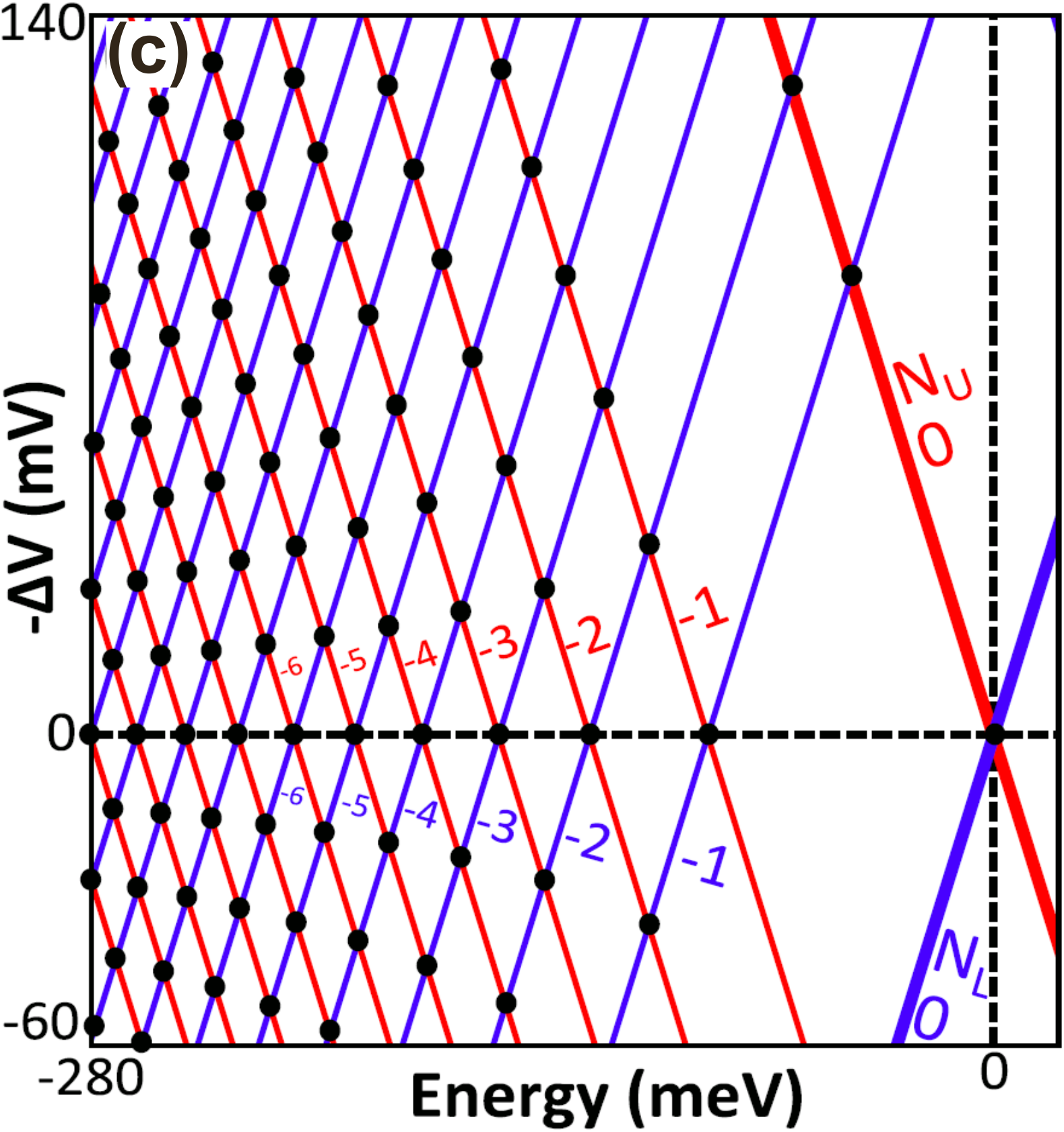}\hspace{3mm}
\includegraphics[height=0.32\textwidth]{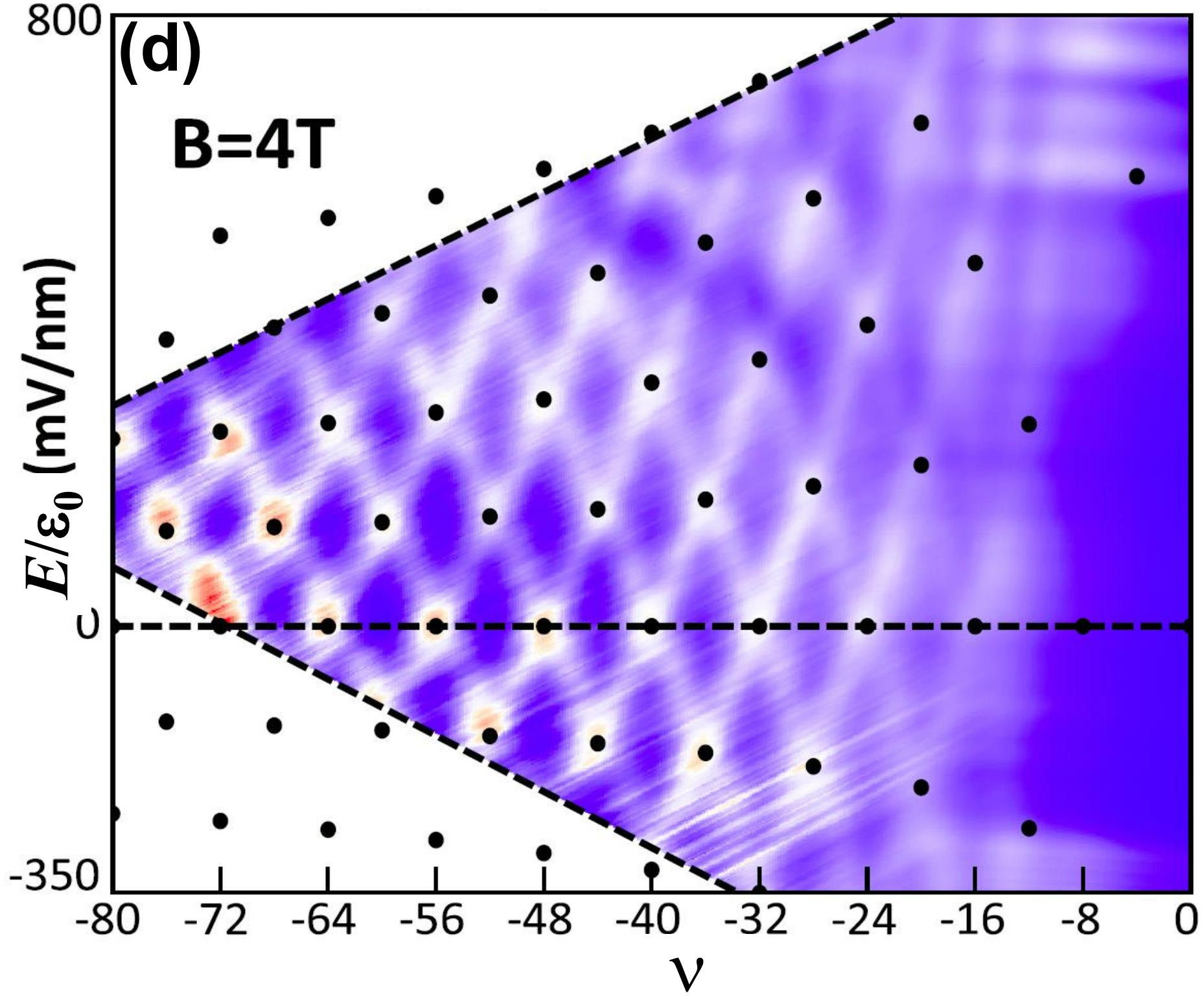}
\caption{(Color online) (a) Schematic of the dual-gated twisted bilayer
graphene device. Two central (red and blue) layers correspond to the tBLG.
BN -- hexagonal boron nitride layer. (b) Hall conductivity in units of
$e^2/2\pi\hbar$
measured at
$B=9$\,T.
For
$E=0$
the steps follow
Eq.~\eqref{QHE_2slg}.
If
$E\neq0$
addition steps exist, indicating the lifting of the eight-fold Landau levels
degeneracy. (c) The Landau level energy spectra as a function of the
interlayer potential difference
$\Delta V$
calculated in the uncoupled layer approximation;
$N_U$,
$N_L$
are integers corresponding to the upper and lower layers, respectively. The
Landau levels crossings are indicated by black dots. (d) Longitudinal
resistance with subtracted background value as a function of $E$ and
$\nu=nh/eB$
measured at
$B=4$\,T.
Peaks indicate the LL crossing. Black dots correspond to the theoretical
prediction.
Reprinted figures with permission from J.\,D.~Sanchez-Yamagishi et al.,
Phys. Rev. Lett., {\bf 108}, 076601 (2012). Copyright 2012 by the American
Physical Society.
\url{http://dx.doi.org/10.1103/PhysRevLett.108.076601}.
\label{FigQHE_TBLGprl2012}}
\end{figure}

J.\,D.~Sanchez-Yamagishi et~al.~\cite{tBLG_QHE}
reported results on transport measurements for tBLG, obtained
using the dual-gate device schematically shown in
Fig.~\ref{FigQHE_TBLGprl2012}(a).
The existence of the top and bottom gates allows one to control both the
total electron density $n$ in the bilayer and the potential difference $\Delta V$
between two layers
or the difference in the electron density,
$\Delta n$,
between two layers. The induced total electron density is related to the top
$V_{TG}$
and bottom
$V_{BG}$
gate voltages as
$$
n=(C_TV_{TG}+C_BV_{BG})/e\,,
$$
where
$C_T$, $C_B$ are the capacitances per unit area of the top and bottom gates, respectively. The applied electric field
$$
E=(C_TV_{TG}-C_BV_{BG})/2
$$
leads to the redistribution of the charge carriers between two graphene layers, which partially screens the electric field. As a result, the potential difference between two layers
$\Delta V$
can be written as~\cite{tBLG_QHE}
\begin{equation}
-\Delta V=\left(E-\frac{e\Delta n}{2}\right)\frac{1}{C}\,,
\end{equation}
where
$C=\epsilon/c_0$
is the interlayer capacitance per unit area and
$\epsilon$
is the interlayer dielectric constant.

In the case of completely uncoupled graphene layers, the potential difference gives rise to the shift of the Dirac cones of the top (bottom) layer by the value
$-e\Delta V/2$
($+e\Delta V/2$). The calculations done in
Ref.~\cite{dSPRL} in the framework of low-energy theory show that similar situation takes place in tBLG with finite interlayer hybridization. In magnetic field, the shift of the Dirac cones leads to the lifting of the eight-fold degeneracy of the tBLG Landau levels, existing in the regime $|\beta|\gg1$, and to the appearance of additional steps in the Hall conductivity. This was experimentally observed in
Ref.~\cite{tBLG_QHE}:
Figure~\ref{FigQHE_TBLGprl2012}(b)
demonstrates distinctive steps of
$2e^2/\pi\hbar$
in the Hall conductivity at non-zero $E$, while for unbiased bilayer
only the steps of
$4e^2/\pi\hbar$
exist. Further increasing
$\Delta V$
leads to Landau level crossing. The crossing occurs when the following
condition is satisfied:
\begin{equation}
-e\Delta V=\varepsilon_{LL}(\ell_1)-\varepsilon_{LL}(\ell_2)\,,
\end{equation}
where
$\ell_{1,2}$
are integers. For uncoupled layers,
$\ell_1$
and
$\ell_2$
are the numbers of Landau levels of the top and bottom layer,
respectively. In tBLG, the interlayer hybridization is non-zero, and
this works only when
$\varepsilon_{LL}(\ell_{1,2})$
is less than the energy of the van Hove singularity.
Figure~\ref{FigQHE_TBLGprl2012}(c)
presents the LL energy spectra as a function of the interlayer potential
difference, as they were calculated in
Ref.~\cite{tBLG_QHE}
in the approximation of uncoupled layers. When LL crossing appears, the
longitudinal conductivity has a peak, which was observed experimentally in
Ref.~\cite{tBLG_QHE}
[see
Fig.~\ref{FigQHE_TBLGprl2012}(d)].
The positions of these peaks are in a good agreement with theoretical
predictions.

\subsubsection{Fine structure of the Landau levels}

\begin{figure}[t]
\centering
\includegraphics[height=0.38\textheight]{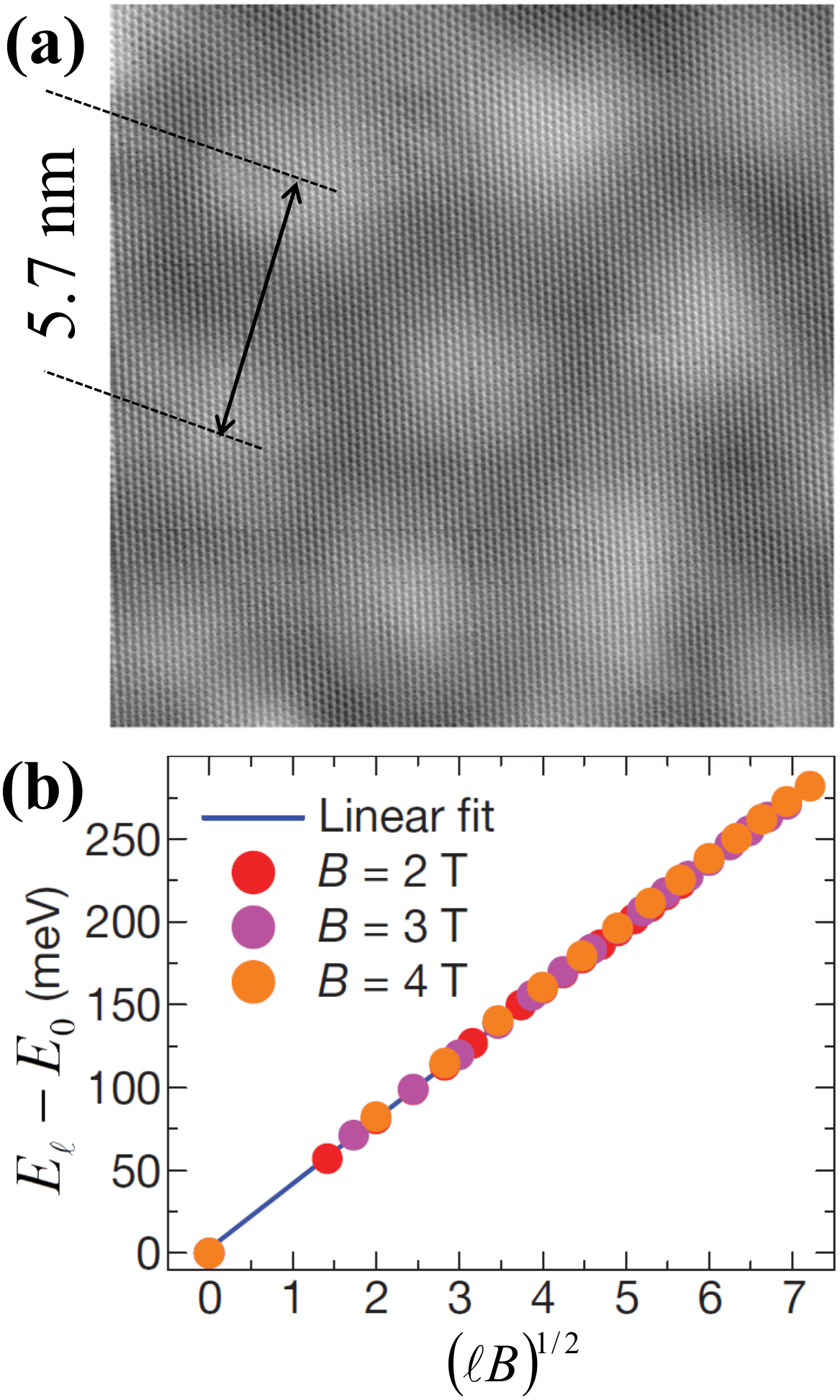}\hspace{5mm}
\includegraphics[height=0.38\textheight]{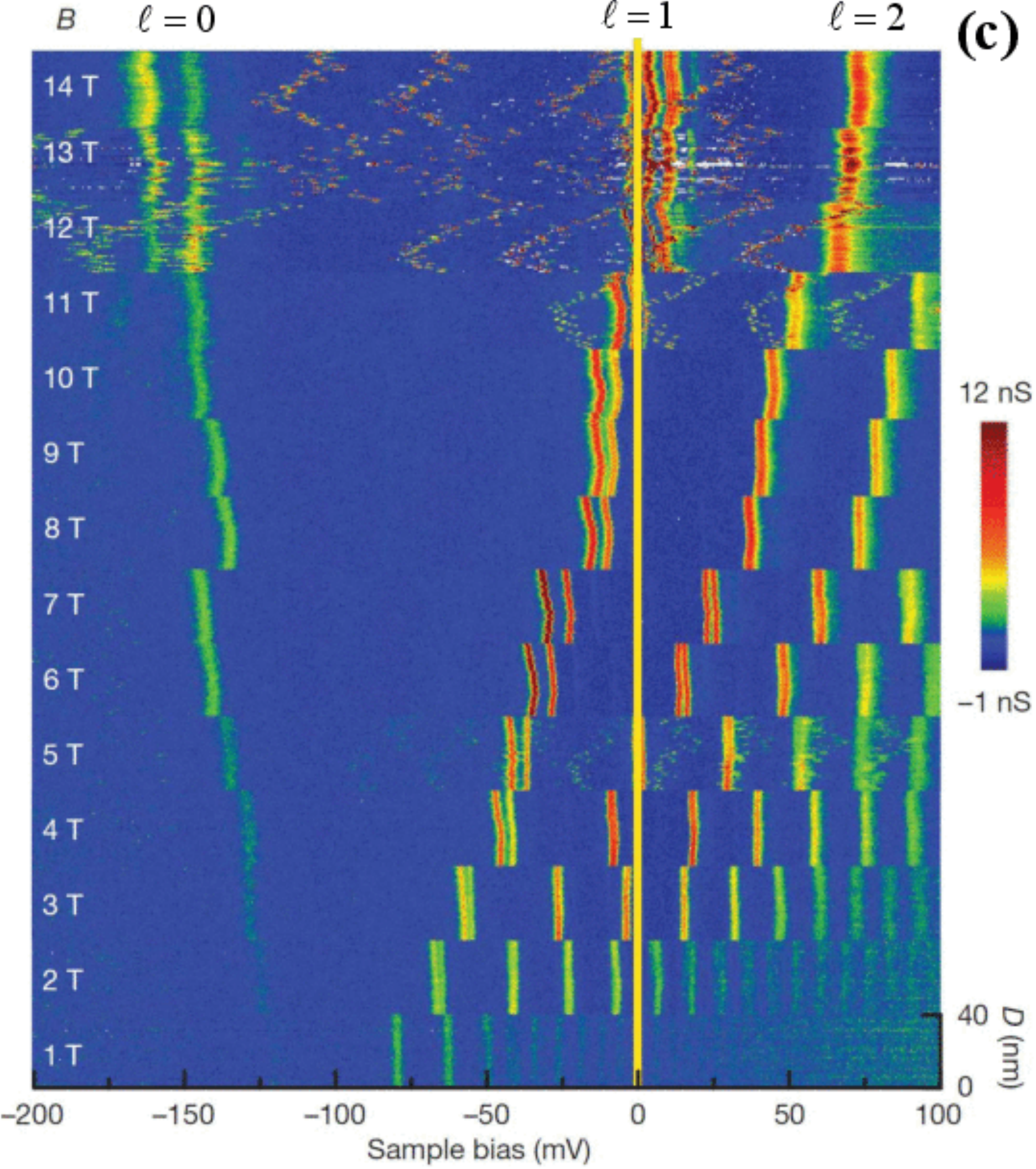}
\caption{(Color online) (a) STM image of epitaxial graphene grown on C-face of SiC showing a Moir\'{e} pattern with the period
$L=5.7$\,nm, corresponding to the twist angle between two top layers
$\theta=2.3^{\circ}$.
Total number of graphene layers is six. (b) Energies of Landau levels
$\ell$
versus the square root of
$\ell B$.
Solid (blue) line is the linear fit according to
Eq.~\eqref{ELLmassless} with the Fermi velocity
$v_F^{*}=(1.08\pm0.03)\times10^{6}$\,m\,s$^{-1}$.
(c) A series of tunneling spectra as a function of the magnetic field taken along the sample. Each panel shows the
$dI/dV$
intensity in a color scale. The vertical axis within each panel is distance
from $0$ to
$40$\,nm.
A splitting of the Landau levels in different field ranges is clearly seen.
Reprinted by permission from Macmillan Publishers Ltd: Y.\,J.~Song et al.,
Nature, {\bf 467}, 185 (2010), copyright (2010).
\url{http://dx.doi.org/10.1038/nature09330}.
\label{FigTBLG_LLsplit}}
\end{figure}

According to the simple model,
Eq.~\eqref{HTBLGeff_H},
the Landau levels in tBLG are eight-fold degenerate, when
$|\beta|\propto\theta/\sqrt{B}\gg1$.
This picture is confirmed experimentally by both Landau-level spectroscopy~\cite{STM2,MillerLLexp} and transport measurements~\cite{FallahazadQHE,tBLG_QHE}. However, a high-resolution LL spectroscopy measurements performed at low temperatures ($T\sim10$\,mK) in Ref.~\cite{NatureLL} by Y.\,J.~Song et~al.
revealed the fine structure of the tBLG Landau levels. The authors studied
multilayer graphene samples grown on the carbon face of SiC. The
estimated total number of layers in the sample was six. The STM
images show the Moir\'{e} pattern with period
$5.7$\,nm [see the Fig.~\ref{FigTBLG_LLsplit}(a)],
which corresponds to twist angle between the two top layers being
$\theta\cong2.3^{\circ}$.
Peaks in the tunneling spectra measured at low magnetic fields are well
described by the square root dependence of
$B$
and of the level's number, according to
Eq.~\eqref{ELLmassless} [see Fig.~\ref{FigTBLG_LLsplit}(b)].
However, already at
$B\gtrsim2$\,T, the fine structure of the Landau levels is clearly seen [see the
Fig.~\ref{FigTBLG_LLsplit}(c)]. Each Landau level with
$\ell\neq0$
splits into two pairs of levels. Thus, there are four sub-levels with energies
$E_{\ell L}\pm\Delta E_{\ell L}/2$
and
$E_{\ell R}\pm\Delta E_{\ell R}/2$.
The energy splitting
$\Delta E_{\ell}=E_{\ell L}-E_{\ell R}$
between two pairs of sub-levels is much higher than the splitting within
each pair,
$\Delta E_{\ell L}$
and
$\Delta E_{\ell R}$.
The splitting demonstrates an almost-linear dependence on the magnetic field,
$$
\Delta E_{\ell}\approx g_{\ell}\mu_{\text{B}}B\,,\;\;\;\Delta E_{\ell L,R}\approx g_{\ell L,R}\mu_{\text{B}}B\,,
$$
where the $g$-factors  are estimated as
$g_{\ell}\sim20$
and
$g_{\ell L,R}\sim2$.
The zeroth Landau level splits at higher magnetic fields,
$B\sim12$\,T.
The authors attributed the splitting
$\Delta E_{\ell}$
to the lifting of the valley degeneracy, while the splitting within
sub-levels can be explained by Zeeman coupling to electron spin. The
spatially-modulated splitting of the zeroth Landau level was also observed in
Ref.~\cite{MillerLL0split}.

The theoretical analysis of the fine structure of the Landau levels in tBLG has been done in
Refs.~\cite{KindermanLL,PalLL,ApalkovPRB2011,KindermanLL2}. In
Ref.~\cite{KindermanLL} M.~Kindermann and E.\,J.~Mele calculated the tBLG Landau levels in framework of the
low-energy theory. They took into account the hybridization of electron
states in different layers if these states are equivalent in the Brillouin
zone of the superlattice. The result depends on the type of
superstructure.
The effective Hamiltonian used is a
$4\times4$
block matrix having a form similar to that of
Eq.~\eqref{Hodd} for `odd' structures ($\hat{H}_{\text{odd}}$) or
Eq.~\eqref{Heven} for `even' structures ($\hat{H}_{\text{even}}$)
\begin{equation}\label{HoddevenB}
\hat{H}_{\text{odd}}=\left(\begin{array}{cc}\hat{H}_{K}(-\frac{\theta}{2})&\hat{H}_{\text{int}}^{-}\\
(\hat{H}_{\text{int}}^{-})^{\dag}&\hat{H}_{K'}(\frac{\theta}{2})\end{array}\right),\;\;\;\;
\hat{H}_{\text{even}}=\left(\begin{array}{cc}\hat{H}_{K}(-\frac{\theta}{2})&\hat{H}_{\text{int}}^{+}\\
(\hat{H}_{\text{int}}^{+})^{\dag}&\hat{H}_{K}(\frac{\theta}{2})\end{array}\right),
\end{equation}
where the non-diagonal
$2\times2$
`mass' matrices
$\hat{H}_{\text{int}}^{\mp}$
are given by
Eq.~\eqref{Hmass}, while
\begin{equation}\label{HKKB}
\hat{H}_{K}(\theta)=-i\hbar\omega_c\left(\begin{array}{cc}0&ae^{-i\theta}\\
-a^{\dag}e^{i\theta}&0\end{array}\right),\;\;\;
\hat{H}_{K'}(\theta)=\sigma_{y}\hat{H}_{K}(\theta)\sigma_{y}=-i\hbar\omega_c\left(\begin{array}{cc}0&a^{\dag}e^{i\theta}\\
-ae^{-i\theta}&0\end{array}\right),
\end{equation}
where
$a$
and
$a^{\dag}$
are the Landau level lowering and raising operators, and cyclotron frequency
$\omega_c$
is given by
Eq.~\eqref{CyclFre}
with the renormalized Fermi velocity.
The Hamiltonian~\eqref{HoddevenB}
can be diagonalized analytically both for `odd' and `even' structures. The
mass terms in
Eqs.~\eqref{HoddevenB}
lead to the splitting of the eight-fold degenerate Landau levels $\ell$
(with the single exception of
$\ell=0$
for the `odd' structures) into two four-fold degenerate levels, which now
can be described by the integers $\ell$ and $\kappa=\pm1$.
The energies of these levels are~\cite{KindermanLL}:
\begin{eqnarray}
E^{\text{odd}}_{\ell\kappa}&=&\pm\left[\left(\ell-\frac{\kappa}{2}\right)\hbar^2\omega_c^2+{\cal V}^2+
\frac{\kappa}{2}\sqrt{\hbar^4\omega_c^4+4\left(\ell-\frac{\kappa}{2}\right)\hbar^2\omega_c^2{\cal V}^2+{\cal V}^4}\right]^{1/2}\!\!\!,\label{ELLodd}\\
E^{\text{even}}_{\ell\kappa}&=&\pm\left[\ell\hbar^2\omega_c^2+{\cal V}^2+2\kappa\hbar\omega_c{\cal V}\cos\frac{\varphi}{2}\sqrt{\ell}\right]^{1/2}\!\!\!,\label{ELLeven}
\end{eqnarray}
where
${\cal V}$
describes the strength of the LL splitting, and
$\varphi$
is the additional parameter entering through
$\hat{H}_{\text{int}}^{-}$.
If the structure is `odd', the
level with $\ell=0$ remains degenerate:
$E^{\text{odd}}_{0\kappa}=0$
both for
$\kappa=+1$
and
$\kappa=-1$.
When
$\hbar\omega_c\gg{\cal V}$,
which corresponds to the experimental conditions
[see Fig.~\ref{FigTBLG_LLsplit}(b)], the splitting of the
$\ell$-th LL level is
$\Delta E^{\text{even}}_{\ell}\cong2{\cal V}\cos(\varphi/2)$,
while
$\Delta E^{\text{odd}}_{\ell}\cong{\cal V}^2\sqrt{\ell}/\omega_c$
if
$\ell\ll\omega_c/{\cal V}$
or
$\Delta E^{\text{odd}}_{\ell}\cong{\cal V}/2$
if
$\ell\gg\omega_c/{\cal V}$.
Thus, this theory predicts the LL splitting, but it can not describe the linear dependence of
$\Delta E_{\ell}$
on magnetic field observed experimentally.

Similar ideas have been discussed in
Ref.~\cite{PalLL} in a more realistic model.
The authors considered both the bilayer and trilayer graphene samples. In
addition, they took into account the non-zero
potential energy difference
$V_{ij}$
between graphene layers ($i,j=1,\,2$
or
$i,j=1,\,2,\,3$). In the limit
$V_{12}\gg\omega_c\gg{\cal V}$,
the result for the `odd' structures is
\begin{equation}\label{DeltaELL}
\Delta E_{0}\cong\frac{{\cal V}^2}{V_{12}}\,,\;\;\;\Delta E_{\ell}\cong\frac{2{\cal V}^2v^{*2}_{F}}{V_{12}^3}eB\,,\;\;\ell\neq0\,.
\end{equation}
Thus, indeed, the splitting of Landau levels  linearly increases with the magnetic field. However,
the effective $g$-factor estimated from
Eq.~\eqref{DeltaELL}
turns out to be much smaller than the experimental value. This is because
of the smallness of ${\cal V}$,
which was estimated by the authors for the twisted bilayer as
${\cal V}\sim\tilde{t}_{\bot}\theta^2$.
For graphene trilayer, however, the splitting is given by similar
relations, in which one has to replace
${\cal V}\to t_0$. Then, the discrepancy between theory and experiment
can be attributed to the effect of lower carbon
layers in the samples used in Ref.~\cite{NatureLL}.

The authors of Ref.~\cite{ApalkovPRB2011} calculated the optical absorption spectra in the presence of the magnetic field,  using an approach related to that of Refs.~\cite{KindermanLL,PalLL}. The LL splitting described above gives rise to the multipeak structure of the absorption spectra, which could be observed experimentally.

\subsubsection{The Hofstadter butterfly spectrum}
\label{butterfly}

The analysis of the Landau levels described above does not take into account
the superlattice periodicity of the tBLG. It is well known, however, that
a periodic potential changes the Landau level spectrum of the electrons
in a magnetic field, leading to a spectrum with a fractal structure. This
phenomenon, the so-called Hofstadter's
butterfly~\cite{Hofstadter},
becomes experimentally observable when the magnetic flux through the
elementary unit cell $\Phi$ becomes comparable to the magnetic flux
quantum $\Phi_0$, or, alternatively, when the magnetic length satisfies
$$
l_B=\sqrt{\hbar c/eB}\lesssim L\,,
$$
where $L$ is the lattice period. For common materials, with $L$
about several angstroms, this happens when
$B\sim10^4$--$10^5$\,T,
which significantly exceeds normally achievable magnetic fields.
However, for superlattices with large supercells, the
Hofstadter butterfly can be observed experimentally. Using
Eq.~\eqref{MoireP}
for the Moir\'{e} period $L$, one can calculate the magnetic flux through
the Moir\'{e} cell and obtain the estimate for the field
required for the observation of the fractal spectrum in the
tBLG~\cite{MoonButterfly}
\begin{equation}\label{BB}
B\gtrsim\frac{\hbar\theta^2}{3ea_0^2}\approx3.3(\theta^{\circ})^2\,[\text{T}]\,.
\end{equation}

Fractal spectra have been observed experimentally in single-layer graphene~\cite{Hunt2013,Ponomarenko2013} and also in Bernal-stacked bilayer graphene~\cite{Dean2013} placed on an hexagonal boron nitride (hBN) substrate. The mismatch between the graphene and the hBN lattice constants (along with a small rotational misorientation between the graphene layers and the substrate) creates a Moir\'{e} pattern, which modifies the single layer or AB-stacked bilayer graphene's electron spectrum. As a result, a Hofstadter's butterfly spectrum was observed in  a magnetic field of the order of several Tesla. Unfortunately, similar observations for twisted bilayer graphene are still absent. Note, however, that in Ref.~\cite{alaHofstadter2014}, the magnetotransport measurements were reported for different samples of the tBLG in a magnetic field up to $B=13$\,T. The authors observed a complex behavior of the resistivity as a function of the magnetic field and the gate voltage if $\theta<3^{\circ}$, but the interpretation of the obtained data is not yet clear.

Theoretical studies of the fractal spectrum of tBLG were performed both in the framework of the continuum approximation~\cite{PNASbutterfly} and tight-binding models~\cite{MoonButterfly,WangButterfly,HasegawaButterfly}. In
Ref.~\cite{PNASbutterfly}, R. Bistritzer and A.H. MacDonald used the effective low-energy Hamiltonian having a block matrix of the form of
Eq.~\eqref{HPNAS}, in which one should replace
$-i\hbar\bm{\nabla}\to-i\hbar\bm{\nabla}+e\mathbf{A}/c$,
that is,
\begin{equation}\label{HPNASbutterfly}
\hat{H}
=
\left(
	\begin{array}{cc}
		\hat{H}_0(-\theta/2)&\hat{T}(\mathbf{r})\\
		\hat{T}^{\dag}(\mathbf{r})&\hat{H}_0(\theta/2)
	\end{array}
\right),\;\;
\hat{H}_0(\theta)
=
v_F\bm{\sigma}^{\theta}
\left[
	-i\hbar\bm{\nabla}+e\mathbf{A}/c
\right]\,,\;\;
\hat{T}(\mathbf{r})
=
\sum_{s=1}^{3}e^{-i\mathbf{q}_{s}\mathbf{r}}\,\hat{T}_s\,,
\end{equation}
where
$$
\mathbf{q}_1=\Delta K(0,-1)\,,\;\;\mathbf{q}_2=\Delta K(\sqrt{3},1)/2\,,\;\;\mathbf{q}_3=\Delta K(-\sqrt{3},1)/2\,,
$$
and $2\times2$ matrices $\hat{T}_s$ are given by
Eq.~\eqref{Tsym}. The Hamiltonian~\eqref{HPNASbutterfly} has a Moir\'{e} periodicity. The authors of Ref.~\cite{PNASbutterfly}
used the Landau gauge
$\mathbf{A}=B(-y,0,0)$.
They expressed the
Hamiltonian~\eqref{HPNASbutterfly}
in terms of the basis states
$|i\alpha n\bar{y}\rangle\equiv|i\alpha\rangle\otimes|n\bar{y}\rangle$,
where
$i=1,2$
and
$\alpha=A,B$
are the layer and sublattice indices, respectively.
The ket-vector
$|i\alpha\rangle$
can be considered as a
$4$-component spinor
[$|1A\rangle=(1,0,0,0)^{\text{T}}$,
etc].
The second ket
$|n\bar{y}\rangle$
represents the $n$th Landau level wave function. The variable
$\bar{y}$
being the $y$-coordinate of the center of the electron's
orbit~\cite{LandauQM}
is related to the conserved momentum in the $x$-direction as
$\bar{y}=k_xl_B^2$.
In the real-space representation, the $|n\bar{y}\rangle$ can be written as~\cite{LandauQM}
\begin{equation}
|n\bar{y}\rangle=
\frac{e^{i\bar{y}x/l_B^2}\exp\!\!\left[-\frac{(y-\bar{y})^2}{4l_B^2}\right]}{[2^nn!\sqrt{\pi}l_BL_x]^{1/2}}H_n\left(\frac{y-\bar{y}}{l_B}\right)\,,
\end{equation}
where
$H_n(\xi)$
are the Hermite polynomials and
$L_x$
is the sample's length in the
$x$-direction. In terms of these wave functions, the intralayer part of the Hamiltonian~\eqref{HPNASbutterfly}
$\hat{H}_0(\theta)$
has the form
\begin{equation}\label{htheta}
\hat{H}_0(\theta)=-\omega_c\sum_{n\bar{y}}\sqrt{n}\left(\begin{array}{cc}0&e^{-i\theta}|n-1\bar{y}\rangle\langle n\bar{y}|\\e^{i\theta}|n\bar{y}\rangle\langle n-1\bar{y}|&0\end{array}\right).
\end{equation}
The spectrum is degenerate with respect
to $\bar{y}$ if we neglect the interlayer hopping.
Diagonalizing $\hat{H}_0(\pm\theta/2)$
one obtains the spectrum of the Landau levels corresponding to the
decoupled graphene layers
$E_n=\pm\omega_c\sqrt{n}$.
The interlayer hopping, however, hybridizes the states with
$\bar{y}$
and
$\bar{y}\pm{\delta y}$,
where
\begin{equation}
\delta y=\frac{\sqrt{3}\,\Delta K\,l_B^2}{2}\,,
\end{equation}
due to the factors
$\exp(-i\mathbf{qr})$
in
$\hat{T}(\mathbf{r})$.
The states with
$\bar{y}=\bar{y}_0+n{\delta y}$,
where
$n=0,\,\pm1,\,\pm2,\dots$,
form a `one-dimensional' chain of states. Such a structure allows a partial
diagonalization of the Hamiltonian. Each chain is characterized by the
momentum
$k_1=\bar{y}_0/l_B^2$,
where
$k_1$
belongs to the interval
$$
0<k_1<{\delta y}/l_B^2=\sqrt{3}\Delta K/2\,.
$$
Each chain is coupled to other chains.
For the commensurate flux
$$
\Phi_0/\Phi=\Delta K^2l_B^2\sqrt{3}/(8\pi)=p/q\,,
$$
where $p$ and $q$ are co-prime
integers, one can define a `supercell', introduce the second momentum
$k_2$, and define new basis vectors,
$|i\alpha nj\mathbf{k}\rangle\equiv|i\alpha\rangle\otimes|nj\mathbf{k}\rangle$,
according to the relation
\begin{equation}
|nj\mathbf{k}\rangle=\sqrt{\frac{q{\delta y} }{L_y}}\sum_{m}\exp[ik_2{\delta y}(mq+j)]|n\,k_1l_B^2+{\delta y}(mq+j)\rangle\,,
\end{equation}
where
$j=0,\,1,\dots,\,q-1$,
$L_y$
is the sample's size in the
$y$-direction, and the momentum
$\mathbf{k}=(k_1,k_2)$
lies inside the ``magnetic'' Brillouin zone
\begin{equation}
0<k_1<\frac{\Delta K\sqrt{3}}{2}\,,\;\;0<k_2<\frac{\Delta K}{2p}\,.
\end{equation}

\begin{figure}[t]
\centering
\includegraphics[width=0.7\textwidth]{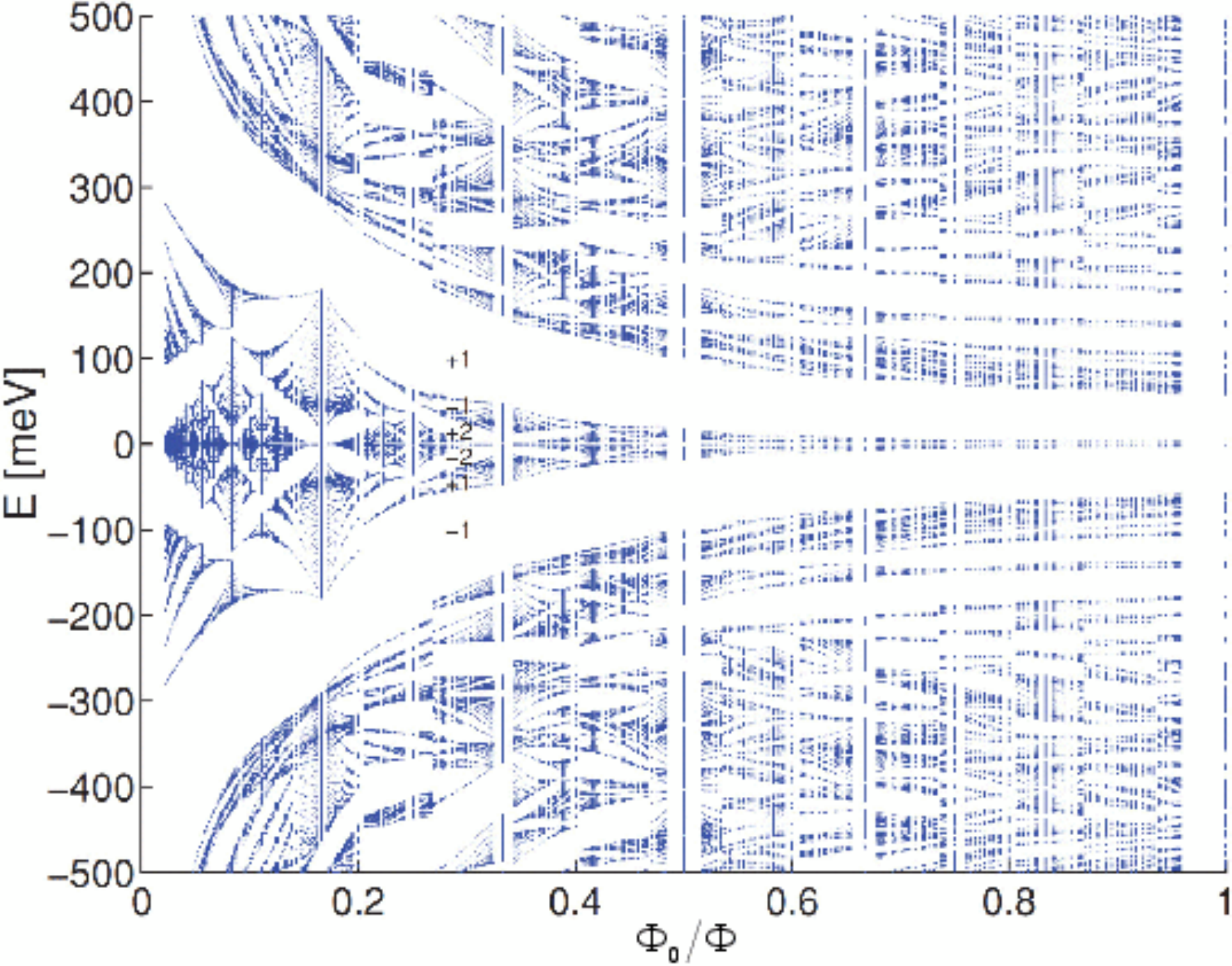}
\caption{(Color online) The low-energy spectrum of tBLG in magnetic field as a function of (rational) $\Phi_0/\Phi$ calculated for $\theta=2^{\circ}$ and $\tilde t_\perp = 110$\,meV. For any vertical cut of this image a tiny dot corresponds to a Landau level, white areas represent spectral gaps with no levels. The periodic inter-layer hopping leads to the Hofstadter-like structure of the spectrum.
Reprinted figure with permission from R.~Bistritzer and A.\,H.~MacDonald,
Phys. Rev. B, {\bf 84}, 035440 (2011). Copyright 2011 by the American
Physical Society.
\url{http://dx.doi.org/10.1103/PhysRevB.84.035440}.\label{FigButterfly}}
\end{figure}

In the latter basis, the intralayer term
$\hat{H}_0(\theta)$
is diagonal in
$\mathbf{k}$
and
$j$.
It is given by
Eq.~\eqref{htheta}
in which one substitute
$|n\bar{y}\rangle\to|nj\mathbf{k}\rangle$
and the summation over
$\bar{y}$
is replaced by the summation over
$\mathbf{k}$
and
$j$.
The interlayer term
$\hat{T}$
is diagonal in this basis in the wave vector
$\mathbf{k}$. It can be written as
\begin{eqnarray}
\hat{T}\!\!\!&=&\!\!\!\!\!\sum_{nm\mathbf{k}j}\sum_{\alpha\beta}\left[%
T_1^{\alpha\beta}F_{nm}\left(\frac{\mathbf{q}_1l_B}{\sqrt{2}}\right)e^{-i\Delta Kk_1l_B^2}e^{-4\pi i\frac{p}{q}j}|2\alpha nj\mathbf{k}\rangle\langle1\beta mj\mathbf{k}|\right.\nonumber\\
&&\phantom{\!\!\!\!\!\sum_{nm\mathbf{k}j}\sum_{\alpha\beta}\!\!\!}%
+T_2^{\alpha\beta}F_{nm}\left(\frac{\mathbf{q}_2l_B}{\sqrt{2}}\right)e^{i\Delta Kk_1l_B^2/2}e^{ik_2{\delta y}}e^{i\pi\frac{p}{q}(2j-1)}|2\alpha nj+1\mathbf{k}\rangle\langle1\beta mj\mathbf{k}|\label{TPNASLL}\\
&&\phantom{\!\!\!\!\!\sum_{nm\mathbf{k}j}\sum_{\alpha\beta}\!\!\!}\left.+%
T_3^{\alpha\beta}F_{nm}\left(\frac{\mathbf{q}_3l_B}{\sqrt{2}}\right)e^{i\Delta Kk_1l_B^2/2}e^{-ik_2{\delta y}}e^{i\pi\frac{p}{q}(2j+1)}|2\alpha nj-1\mathbf{k}\rangle\langle1\beta mj\mathbf{k}|%
\right]\nonumber\,,
\end{eqnarray}
where
$j$
is defined modulo
$q$,
and
\begin{equation}
F_{nm}(\mathbf{z})=\sqrt{\frac{m!}{n!}}\left(-z_x+iz_y\right)^{n-m}\exp\left(-\frac{\mathbf{z}^2}{2}\right){\cal L}_m^{n-m}\left(\mathbf{z}^2\right)\,,\;\;n\geqslant m\,,
\end{equation}
with
${\cal L}_m^{a}$
being the associated Laguerre polynomials. For
$n<m$,
the function
$F_{nm}$
is defined by
$F_{nm}(\mathbf{z})=[F_{mn}(-\mathbf{z})]^{*}$.

R. Bistritzer and A.H. MacDonald~\cite{PNASbutterfly} calculated the spectrum numerically by only considering first
\begin{eqnarray}
N_0\sim2[\max(v_F\Delta K,\,\tilde{t}_{\bot})/\omega_c]^2
\end{eqnarray} 
Landau levels. Thus, for any
$\mathbf{k}$,
the rank of the matrix, which has to be diagonalized, is equal to
$4qN_0$.
As it follows from the structure of
Eq.~\eqref{TPNASLL},
the interlayer hopping splits each Landau level into $q$ finite-width
sub-bands, and couples them to each other. Both of these effects change the
spectrum. However, the coupling of Landau levels is more important for
smaller magnetic fields.
The smaller the twist angle is, the smaller the magnetic field is required to
disturb the Landau level spectrum, because
$$
\Phi_0/\Phi\propto\theta^2/B\,.
$$
If $\theta < \theta_c$, this line of reasoning becomes invalid
because the Dirac cones in the spectrum disappear
even at $B=0$).
Figure~\ref{FigButterfly}
shows the spectrum of tBLG for rational
$\Phi_0/\Phi<1$,
calculated at
$\theta=2^{\circ}$.
It is seen that, in general, the gaps between sub-bands increase when
$\Phi_0/\Phi$
decreases.

C.-K.~Lu and H.\,A.~Fertig in Ref.~\cite{FertigLLoptAbs2014} used a similar low-energy model,
Eq.~\eqref{HPNASbutterfly}, to analyze the effect of the bias voltage on the LL in tBLG.
Besides the electronic spectrum in a magnetic field, they calculated the optical absorption spectra. They~\cite{FertigLLoptAbs2014} claimed that the bias voltage can help to distinguish the transitions involving the states localized in one particular layer from the transitions involving the layer-delocalized states.

Tight-binding calculations of the tBLG spectrum in a magnetic field were performed in Refs.~\cite{MoonButterfly,WangButterfly,HasegawaButterfly}. The authors of Refs.~\cite{WangButterfly,HasegawaButterfly} used the exponential dependence of the hopping amplitudes $t(\mathbf{r};\mathbf{r}')$ on the distance between sites, while P.~Moon and M.~Koshino~\cite{MoonButterfly} applied the somewhat more complicated parametrization given by Eqs.~\eqref{SlaterKoster}--\eqref{Fcutoff}. In magnetic field, the hopping amplitude $t(\mathbf{r};\mathbf{r}')$ acquires an addition phase factor and becomes
\begin{equation}
t(\mathbf{r};\mathbf{r}')
\exp\left(\frac{ie}{\hbar c}\int_{\mathbf{r}}^{\mathbf{r}'}\!\!\!\!\!
\mathbf{A}d\mathbf{r}\right)\,.
\end{equation}
The gauge in which the vector potential
$\mathbf{A}$
is parallel to the superlattice vector
$\mathbf{R}_2$
is used in
Ref.~\cite{MoonButterfly}. When
$\Phi/\Phi_0=p/q$,
where $p$ and $q$ are co-prime
integers, the tight-binding tBLG
Hamiltonian~\eqref{HTB}
has a periodicity described by the supercell vectors
$q\mathbf{R}_1$
and
$\mathbf{R}_2$.
Consequently, the area of the magnetic Brillouin zone is $q$ times smaller
than the Brillouin zone area for
$\mathbf{B}=0$,
and the number of bands for
$(m_0,r)$
superstructure is
$qN(m_0,r)$.

\begin{figure}[t]
\centering
\includegraphics[height=0.58\textheight]{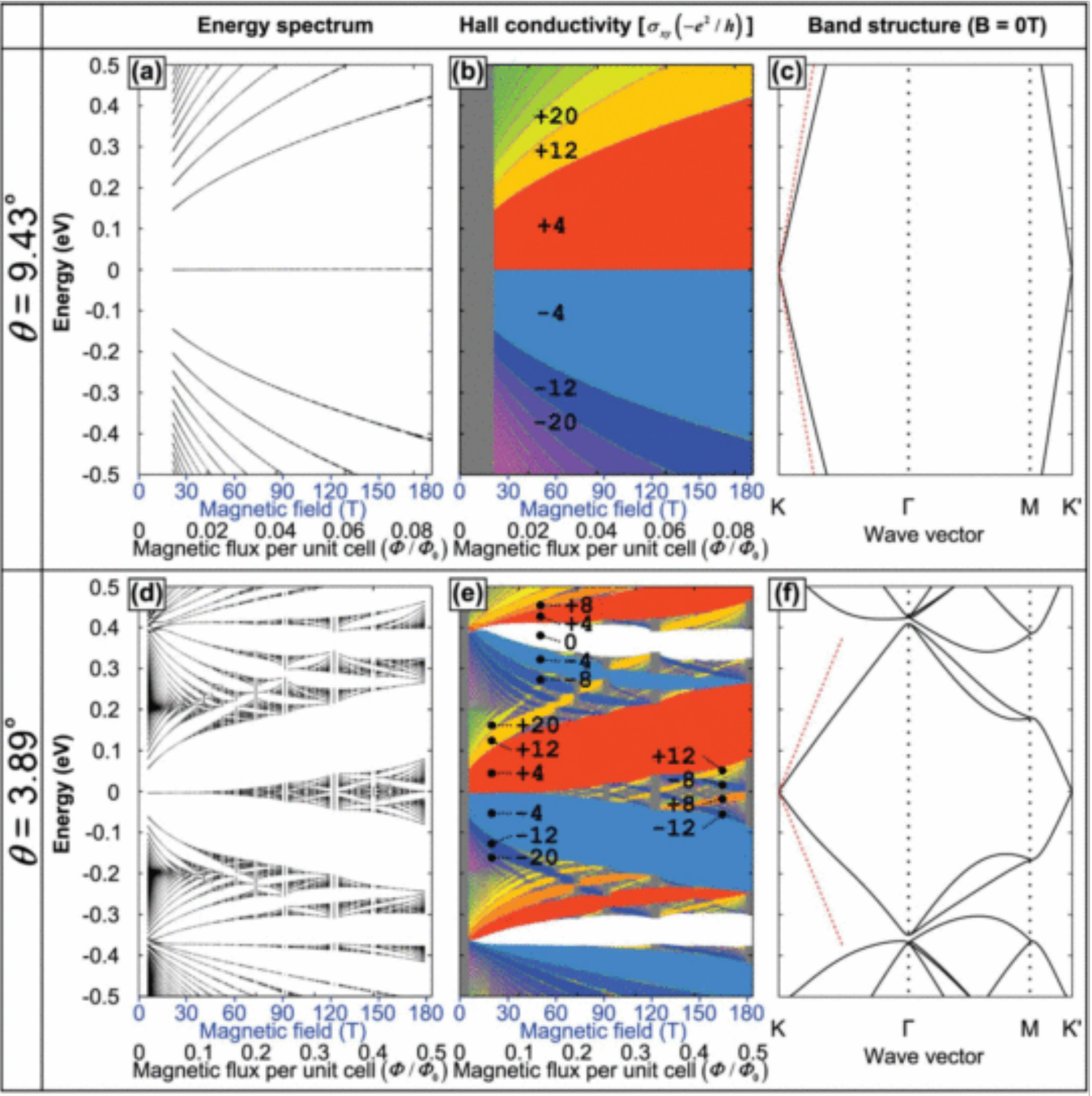}
\caption{(Color online)
Numerically calculated energy spectrum (a,d) and Hall conductivity (b,e)
versus magnetic field for tBLG samples with
$\theta\cong9.43^{\circ}$
(upper panels) and
$\theta\cong3.89^{\circ}$
(lower panels). The right panels (c,f) show the band structure at $B=0$.
The dashed (red) lines show the dispersion of monolayer graphene near the Dirac point. In panels (b) and (e) the values of the Hall conductivity are indicated by numbers and colors.
Reprinted figure with permission from P.~Moon and M.~Koshino, Phys. Rev. B,
{\bf 85}, 195458 (2012). Copyright 2012 by the American Physical Society.
\url{http://dx.doi.org/10.1103/PhysRevB.85.195458}.\label{FigButterflyTB}}
\end{figure}

The authors of Ref.~\cite{MoonButterfly} considered a spectrum in the energy window $-0.5$\,eV$<E<0.5$\,eV, for several superstructures with $r=1$ and twist angles in the range $1.5^{\circ}\lesssim\theta\lesssim10^{\circ}$, in magnetic fields up to $180$\,T. In addition to the spectrum, they calculated the Hall conductivity $\sigma_{xy}$.
When the Fermi level
$\mu$
lies in the band gap, the conductivity
$\sigma_{xy}$
can be found according
to~\cite{stone1992quantum}
\begin{equation}
\sigma_{xy}=-e\left(\frac{\partial n}{\partial B}\right)_{\!\!\mu},
\end{equation}
where $n$ is the number of electrons per unit area for a given $\mu$.
Some results of this study are presented in
Fig~\ref{FigButterflyTB}.
Panels (a,d) and (b,e) show the dependencies
of the spectrum and the Hall conductivity on the magnetic field. Right
panels (c,f) present the band structure in the case $\mathbf{B}=0$.
When $\Phi/\Phi_0\ll1$,
the spectrum consists of a sequence of nearly-degenerate Landau levels
[described by the square-root dependence
Eq.~\eqref{ELLmassless}],
and the Hall conductivity in units of
$e^2/2\pi\hbar$
follows the law
Eq.~\eqref{QHE_2slg},
corresponding to uncoupled graphene layers [see panels (a,b) in
Fig.~\ref{FigButterflyTB}].
For large twist angles, the regime of small
$\Phi/\Phi_0$
is maintained up to very high magnetic fields. The deviation from the
square-root dependence of the sequence of Landau levels occurs when
the absolute value of the LL energy, $E_n$,
exceeds the energy of the van Hove singularities (if $\theta\cong3.89^{\circ}$
this happens when $|E_n|\gtrsim0.2$\,eV).
Simultaneously, the Hall conductivity drops down to negative values and
then grows in discrete increments equal four conductivity quanta
[see panel (e) in Fig.~\ref{FigButterflyTB}]. The behavior of the tBLG spectrum in magnetic fields near the van Hove singularities has been also studied~\cite{LLnearVHSemiClassic2014} using a semiclassical approach.

Even for small $\Phi/\Phi_0$, the superlattice periodicity breaks the degeneracy of each Landau level:
the density of states spreads in the energy domain to form several
sub-bands separated by small gaps. These gaps increase when the magnetic
field grows. As a result, the spectrum acquires a complicated fractal
structure dissimilar to the original Landau level picture
[Fig.~\ref{FigButterflyTB}(d)].
Simultaneously, the Hall conductivity becomes a non-monotonic function of
the Fermi level. Characteristic magnetic fields of the crossover from the
Landau level to fractal descriptions can be estimated from
Eq.~\eqref{BB}. For example,
$B=50$,
$23$,
and
$7.2$,\,T for
$\theta=3.89^{\circ}$,
$2.65^{\circ}$,
and
$1.47^{\circ}$,
respectively~\cite{MoonButterfly}.

Similar results for the tBLG electronic spectrum where obtained by Y.~Hasegawa and M.~Kohmoto in
Ref.~\cite{HasegawaButterfly}.
The authors used the so-called periodic Landau gauge for the vector
potential
$\mathbf{A}$.
The spectrum they calculated consists of many bands separated by the gaps.
Each gap  can be characterized by two integers
$(s,\nu)$,
where
$\nu$
is related to the Hall conductivity as
$\sigma_{xy}=(e^2/2\pi\hbar)\nu$.
The authors~\cite{HasegawaButterfly} analyzed the evolution of the spectrum when changing of the magnetic field and the twist angle. For example, they showed that for considerably small twist angles, there are several narrow low-energy bands  separated by gaps having the same value of $\nu$; that is, the passing of the chemical potential trough these bands does not change the Hall conductivity.

Reference~\cite{WangButterfly} by Z.\,F.~Wang et~al. presents tight-binding calculations of the spectrum in a magnetic field for finite samples with sizes up to
$140\times140$\,nm$^2$. The Lanczos algorithm was used for calculating eigenstates
in the energy window $|E|<0.2$\,eV.
Because of the finite size of the samples, the authors were able to
consider both commensurate and incommensurate twist angles. The authors
distinguished three regions of twist angles where the spectra are qualitatively
different from each other. In the `quasi-Bernal' regime,
$0\leqslant\theta\lesssim0.3^{\circ}$,
the spectrum consists of nearly-degenerate Landau levels with energies corresponding to the AB bilayer. In the intermediate regime,
$0.3^{\circ}\lesssim\theta\lesssim3^{\circ}$,
the spectrum has a complex fractal-like structure (for
$B\sim10$\,T).
At larger angles, the spectrum consists of Landau levels corresponding to
effectively uncoupled layers; the energies of these levels are
described by
Eq.~\eqref{ELLmassless}
with a renormalized Fermi velocity, which monotonously decreases with the
twist angle both for commensurate and incommensurate structures. The
Supporting information to
Ref.~\cite{WangButterfly}
presents a movie showing the evolution of the spectrum when changing
the twist angle. This video can be downloaded from \url{http://pubs.acs.org/doi/suppl/10.1021/nl301794t}.

\subsection{Electronic transport in twisted bilayer graphene}
\label{SubSectTBLGTransport}

\subsubsection{Dynamical conductivity}

\begin{figure}[t]
\centering
\includegraphics[height=0.58\textheight]{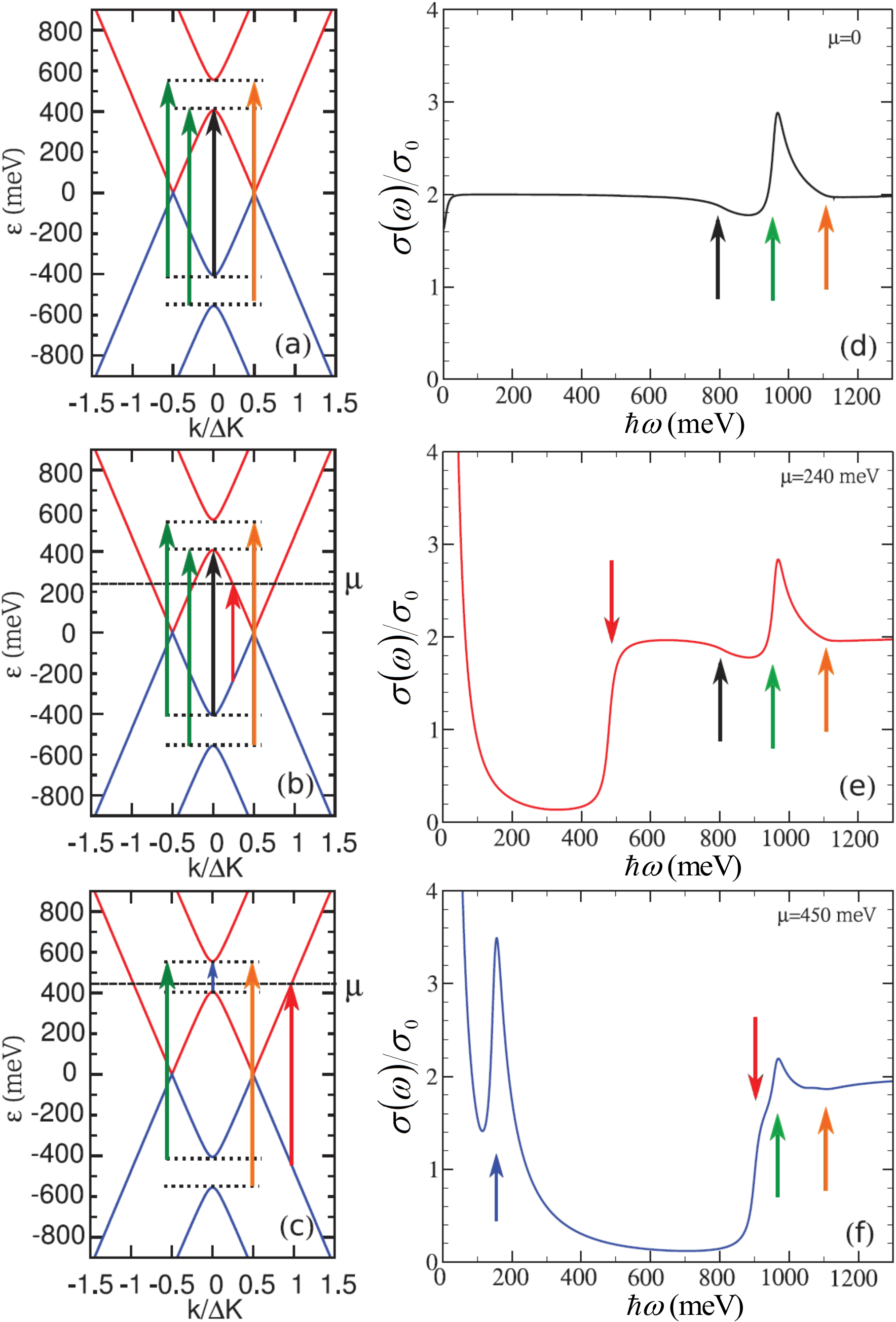}
\caption{(Color online) Dynamical conductivity of twisted bilayer
graphene. Panels (a)--(c): the low-energy single-electron tBLG spectrum
along the line connecting the
$\mathbf{K}$
and
$\mathbf{K}_{\theta}$
points calculated in
Ref.~\cite{SigmaNicol} for
$\theta=5^{\circ}$.
Different significant optical transitions are indicated by arrows. Panels
(d)--(f): the frequency dependence of the longitudinal conductivity
calculated for different values of the chemical potential $\mu$.  The
arrows indicate the frequency of the transitions shown in panels (a)--(c).
Reprinted figure with permission from C.\,J.~Tabert and E.\,J.~Nicol, Phys.
Rev. B, {\bf 87}, 121402 (2013). Copyright 2013 by the American Physical
Society.
\url{http://dx.doi.org/10.1103/PhysRevB.87.121402}.\label{FigTBLGSigma}}
\end{figure}

The response of the tBLG on the ac electromagnetic field has been studied
both experimentally~\cite{ExpSigmaNano,ExpSigmaPRL,ExpSigmaPerp}
and
theoretically~\cite{SigmaNicol,PNASsigma,SigmaMoon,SigmaPlasmons}.
In
Ref.~\cite{SigmaNicol} C.\,J.~Tabert and E.\,J.~Nicol calculated the optical conductivity in the framework of the
simplified low-energy theory proposed by R. de Gail
et~al.~\cite{deGail}.
They used the $4\times4$
Hamiltonian
$\hat{H}(\mathbf{k})$
given by
Eq.~\eqref{HdeGail0}
with the interlayer term
$\hat{H}_{\bot}$
in the form of
Eq.~\eqref{TdeGail0}.
The use of the simplified expression for the interlayer hopping allows them
to calculate the spectrum analytically. The results are shown in
Fig.~\ref{FigTBLGSigma}(a)--(c).
The real part of the dynamical conductivity at zero temperature is
calculated numerically using
Eq.~\eqref{condAA},
which for
$T=0$
has the form
\begin{equation}
\sigma_{xx}(\omega)=\frac{2e^2}{\omega}\!\!\!\int\limits_{\mu-\omega}^{\mu}\!\!\!\frac{d\omega'}{2\pi}\!\!\int\frac{d^2\mathbf{k}}{(2\pi)^2}
\Tr\left[\hat{v}_x\hat{A}(\mathbf{k},\omega'+\omega)\hat{v}_x\hat{A}(\mathbf{k},\omega')\right]\,,
\end{equation}
where
$\hat{v}_x=\partial\hat{H}/\partial k_x$,
and
$\hat{A}(\mathbf{k},\omega)$
is the spectral function (both are
$4\times4$
matrices). As in the case of the AA and AB bilayers considered in the
previous sections, the authors~\cite{SigmaNicol} use the Lorentzian broadening for the
$\delta$-function in
$\hat{A}(\mathbf{k},\omega)$,
$\pi\delta(x)\to\Gamma/(x^2+\Gamma^2)$,
where $\Gamma$ is associated with the transport scattering rate on the impurities.

Figures~\ref{FigTBLGSigma}(d)--(f) show the in-plane dynamical conductivity as a function of frequency calculated for the tBLG with
$\theta=5^{\circ}$
at different values of the chemical potential
$\mu$.
At zero doping, the conductivity is approximately constant at low frequencies and equal to twice the background conductivity of single-layer graphene
$e^2/(4\hbar)$.
At higher frequencies, the conductivity has a dip-and-peak structure
associated with the transitions from/to electron states near the low-energy
van Hove singularities [see the black, green, and orange arrows in
Fig.~\ref{FigTBLGSigma}(a)--(f)].
Such a peak in the conductivity is similar to that for the AB
bilayer~\cite{NicCarb1,Abergel2007}
(see Figs.~\ref{DynCond_ABfig}
and~\ref{DynCondExpfig}) but is located at higher energies, near the energy of the VHS splitting,
$\Delta E_{\text{vHs}}$
(for definition, see
subsection~\ref{VHS_VfRenorm}).
Its position is closer to zero energy for smaller twist angles. At large
twist angles, $\Delta E_{\text{vHs}}\sim1$\,eV,
and the corresponding frequencies lie in the visible range. For example, the
contrast spectroscopy measurements done
in Ref.~\cite{ExpSigmaNano}
revealed the peak in conductivity located at wavelength
$\lambda\approx595$\,nm
($\hbar\omega\cong2.08$\,eV)
for tBLG sample with
$\theta\cong13.7^{\circ}$.
At smaller twist angles, the peak can lie in the infrared and even terahertz
frequency range. Thus, the peak at
$\omega\cong2.64$\,THz
was observed experimentally using terahertz time-domain spectroscopy in
Ref.~\cite{ExpSigmaPRL}.
The twist angle of the sample studied in the latter work was estimated as
$\theta\cong1.16^{\circ}$.

At finite doping, the conductivity is modified in two aspects. First, it
acquires a Drude peak centered at zero frequency. Second, at frequencies
$\omega<2\mu$
the interband transitions are not allowed due to the Pauli exclusion
principle, and this leads to the suppression of the dynamical conductivity
in the range
$0\lesssim\omega<2\mu$
[see
Fig.~\ref{FigTBLGSigma}(b,e)].
At larger frequencies, the conductivity is the same as for
$\mu=0$.
For larger doping, when
$\mu$
lies between the van Hove singularity and the bottom of the upper band, the conductivity
acquires an addition low-energy peak associated with the transitions
between these two states [see
Fig.~\ref{FigTBLGSigma}(c,f), c.f. with
Fig.~\ref{DynCond_ABfig}(b)].

\begin{figure}[t]
\centering
\includegraphics[width=0.47\textwidth]{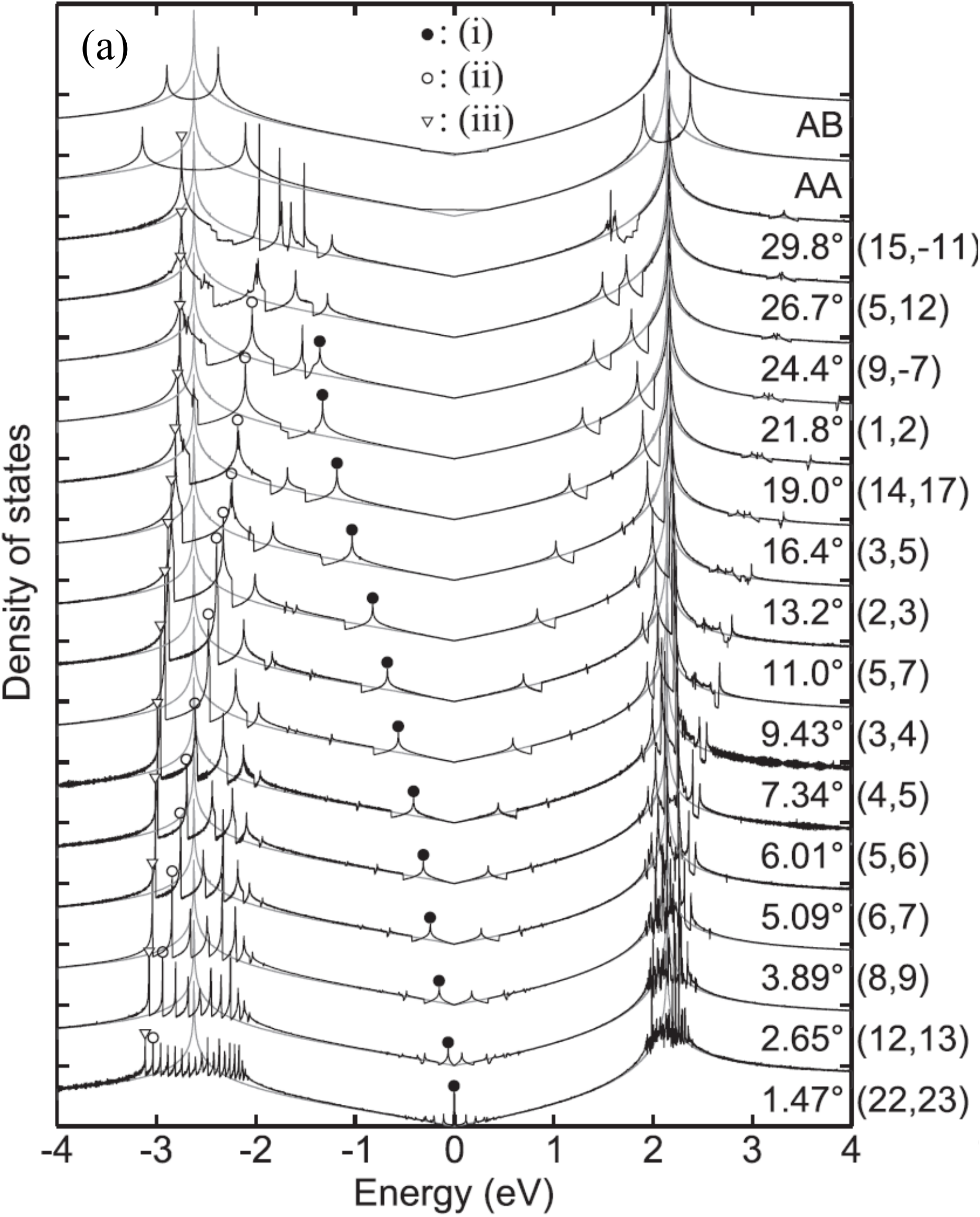}
\includegraphics[width=0.5\textwidth]{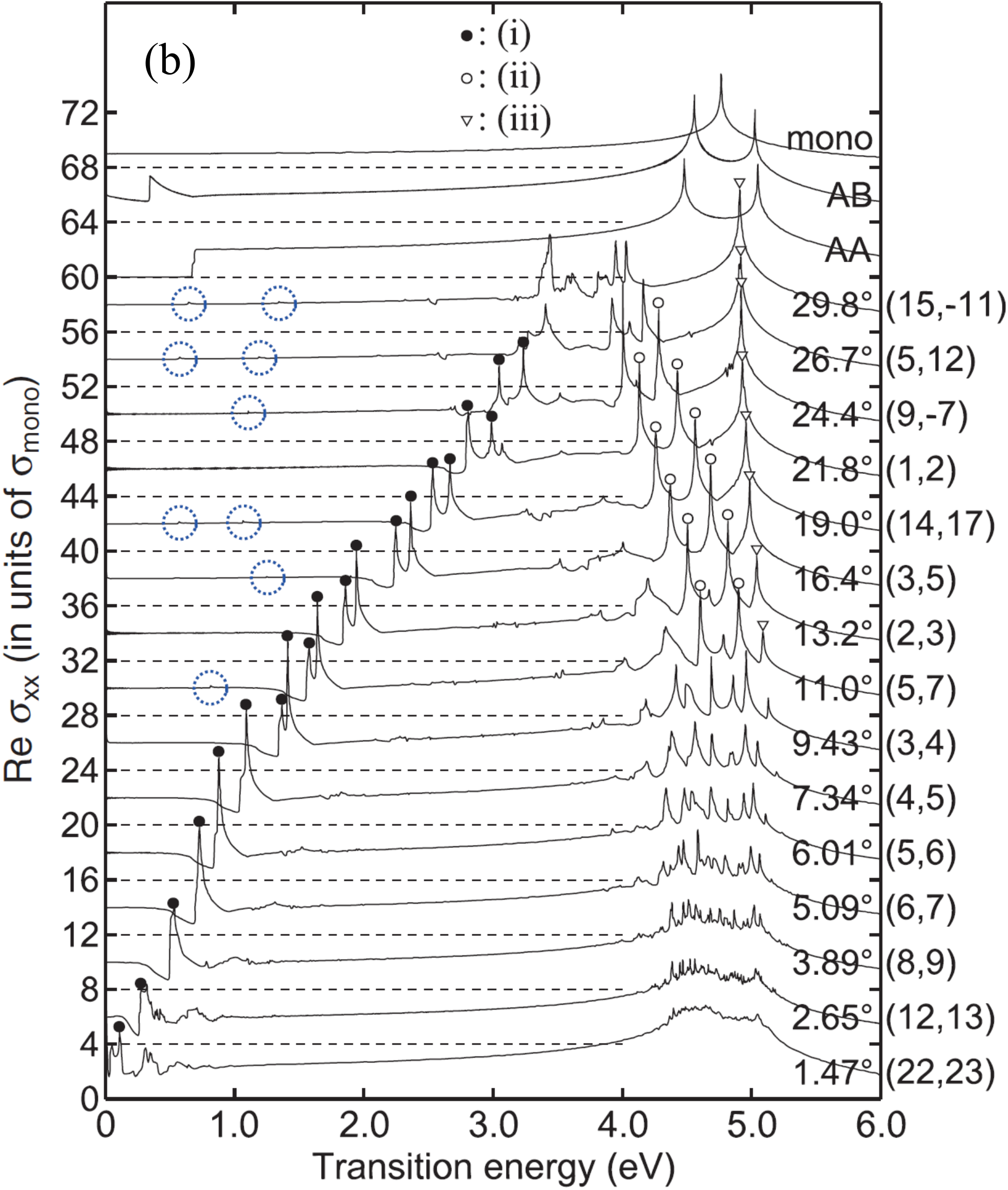}
\caption{(Color online) The density of states (a) and the dynamical
conductivity (b) calculated in tight-binding approximation for several
undoped tBLG superstructures, as well for AB- and AA-stacked bilayers in
Ref.~\cite{SigmaMoon}.
Peaks marked by ($\bullet$), ($\circ$), and ($\triangledown$) symbols
correspond to the three types of the VHS described in the text. The
superstructures
$(n,m)$
in the figure correspond to structures
$(n,m-n)$
in our notation. Small peaks encircled by dashed blue lines in panel (b)
exist only for
$r\neq1$
($m\ne n+1$)
superstructures.
Reprinted figures with permission from P.~Moon and M.~Koshino, Phys. Rev.
B, {\bf 87}, 205404 (2013). Copyright 2013 by the American Physical Society.
\url{http://dx.doi.org/10.1103/PhysRevB.87.205404}.\label{FigTBLGSigmaTB}}
\end{figure}

Tight-binding calculations of the dynamical conductivity were performed by P.~Moon and M.~Koshino in Ref.~\cite{SigmaMoon}.
The hopping amplitudes are calculated according to the exponential
parametrization
Eqs.~\eqref{SlaterKoster}--\eqref{Fcutoff}.
The Lorentzian broadening of the spectral functions with
$\Gamma=7$\,meV
was used for calculating the dynamical conductivity. The authors
considered undoped bilayers, and focused on the change of the
dynamical conductivity with the twist angle. The dependencies of $\sigma$
on $\omega$ calculated for several superstructures are presented in
Fig.~\ref{FigTBLGSigmaTB}.
Since the tight-binding approach involves much more energy bands into
consideration, in comparison to the low-energy theories, the obtained curves
have a more complicated structure than those presented in
Fig.~\ref{FigTBLGSigma}.
The curves
$\sigma(\omega)$
have several peaks associated with the electron transitions between
different VHS. Each singularity manifests itself as a peak in the density
of states, shown in
Fig.~\ref{FigTBLGSigmaTB}(a).
Authors~\cite{SigmaMoon} distinguish three major types of VHS. Singularities of the types
(i) and (ii) result from the intersection of the single-layer graphene
bands caused by the interlayer coupling. Specifically, singularities of
types (i) appear at momenta near the points
$\mathbf{K}_0=(\mathbf{K}+\mathbf{K}_{\theta})/2$
and
$\mathbf{K}'_0=(\mathbf{K}'+\mathbf{K}'_{\theta})/2$,
while the type (ii) VHS exist near the middle point of the line connecting
the points
$\mathbf{K}_0$
and
$\mathbf{K}'_0$.
Finally, the type (iii) singularities originate from the VHS of
single-layer graphene. Since VHS of different types appear at different
momenta, and the momentum of the absorbed photon is negligible, there can
exist only transitions between electron states near the VHS of the same
type, and the peaks in the conductivity can be classified in the same
manner. The peaks of type (i) in the conductivity curves in
Fig.~\ref{FigTBLGSigmaTB}(b)
approach zero energy for structures with smaller twist angles. These
peaks are similar to those given by the low-energy theories [see, Fig.~\ref{FigTBLGSigma}(d)].
On the other hand, the peaks of type (ii) move to higher energies with decreasing twist angle. The position of the type (iii) peaks is almost
independent of $\theta$ because these peaks are related to single-layer
van Hove singularities. The peaks of types (ii) and (iii) are beyond the
scope of the low-energy theories.

In addition, superstructures with
$r\neq1$
have extra small peaks in the conductivity. They are marked by the dashed
blue circles in
Fig.~\ref{FigTBLGSigmaTB}(b).
The authors~\cite{SigmaMoon} associated these minor peaks with the electron transitions at
the corners of the Brillouin zone of the superlattice. Indeed, the Brillouin
zone of the superstructure
$(m,r)$
with
$r\neq1$
is approximately $r$ times smaller than that for the superstructure
$(m_0,1)$
having similar
$\theta$,
where
$m_0=[m/r]$.
According to calculations done in
Ref.~\cite{ourTBLG}, the spectrum of the superstructure
$(m,r)$
is similar to the spectrum of the
$(m_0,1)$
superstructure calculated in the $r$-times-folded Brillouin zone. The folding leads to an additional band crossings, which for the
$(m,r)$
superstructure can give rise to additional van Hove singularities resulting in these conductivity peaks. These peaks, when observed, would be a fingerprint of the
$r\neq1$ superstructures. Similar calculations, but for biased bilayers, have been performed in Ref.~\cite{MoonTBgate2014}.

\begin{figure}[t]
\centering
\includegraphics[width=0.6\textwidth]{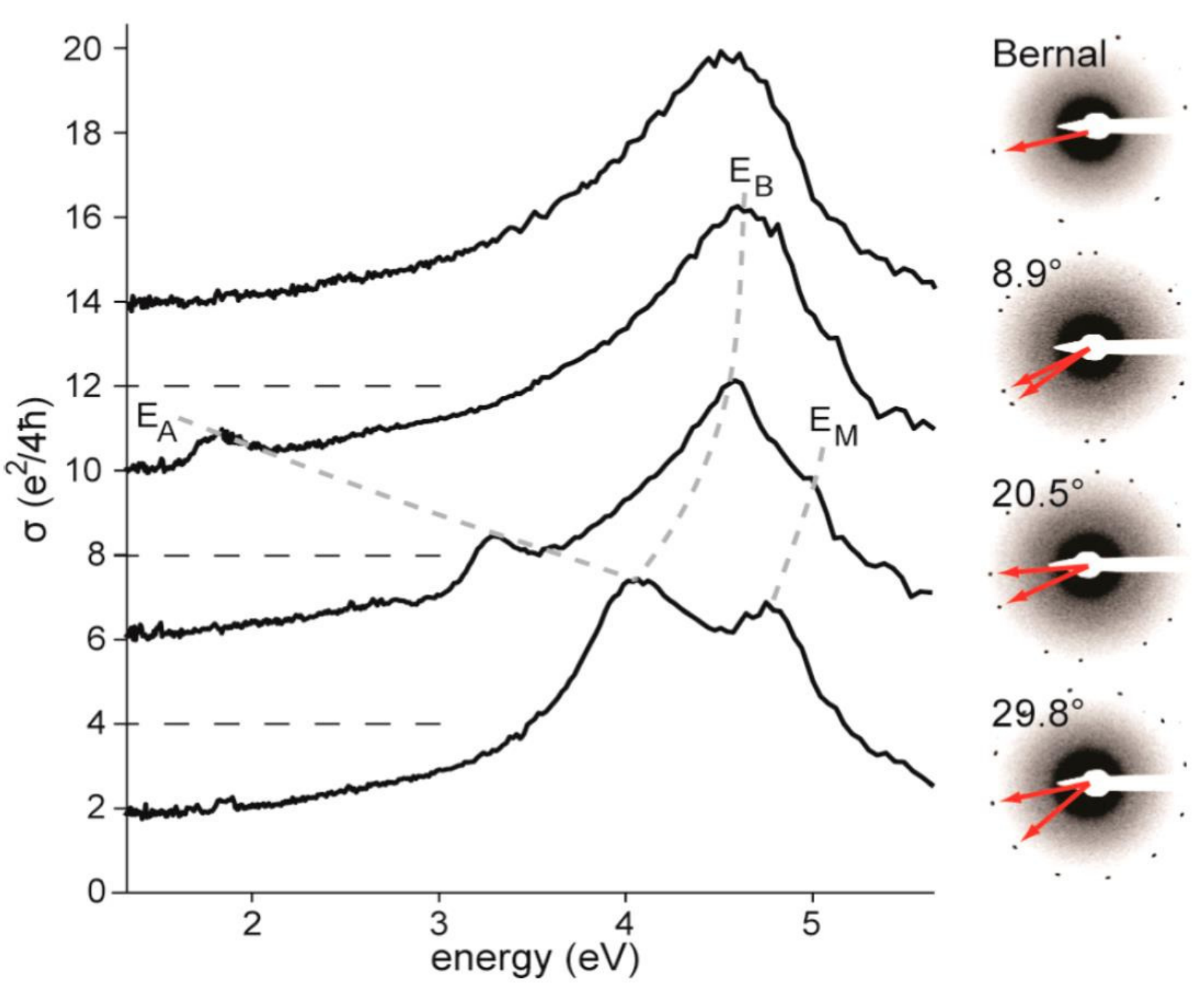}
\caption{(Color online) Optical conductivity spectra of the AB BLG (top)
and for three tBLG samples with different twist angles $\theta$ from
Ref.~\cite{OptCondVHS2014}. For clarity, neighboring curves are offset by
the value $4(e^2/4\hbar)$. The dashed lines connect the peaks with energies
$E_A$, $E_B$, and $E_M$ corresponding, respectively, to peaks of the types
(i), (ii), and (iii) in the notation of P.~Moon and
M.~Koshino~\cite{SigmaMoon} (see Fig.~\ref{FigTBLGSigmaTB}). Images on
the right show the diffraction patterns of the bilayers and the
corresponding values of the twist angle.
Reprinted with permission from R.\,W.~Havener et al.,
Nano Letters, {\bf 14}, 3353 (2014).
Copyright 2014 American Chemical Society.
\url{http://dx.doi.org/10.1021/nl500823k}.\label{FigTBLGSigmaExp}}
\end{figure}

Detailed experimental studies of the optical conductivity of tBLG have been performed by R.\,W.~Havener et~al. in Ref.~\cite{OptCondVHS2014}. The authors measured the optical conductivity in the energy range $1.2<\hbar\omega<5.6$\,eV for bilayers with different twist angles and for AB samples as well. Some results are presented in Fig.~\ref{FigTBLGSigmaExp}. The conductivity spectra of the twisted bilayers have three peaks which Ref.~\cite{OptCondVHS2014} associated with the peaks of the types (i), (ii), and (iii) mentioned above. The authors~\cite{OptCondVHS2014} compared their results with the tight-binding calculations by P.~Moon and M.~Koshino, Ref.~\cite{SigmaMoon}, and found a qualitative agreement between theory and experiment. However, according to the analysis done in Ref.~\cite{OptCondVHS2014}, for better fitting of the obtained data one has to take into account the effects of the electron-hole interactions and the existence of excitonic states.

The possibility of the formation of excitons in tBLG has been proposed theoretically in Ref.~\cite{Excitons2014}. The exciton is formed by the hole, located near the VHS above the Fermi level and the electron located close to the bottom of the upper electron band. These quasiparticles can be excited by optical transitions shown by the two left (green) arrows in Fig.~\ref{FigTBLGSigma}(a). According to calculations done in Ref.~\cite{Excitons2014}, the electron-hole interaction leads to the formation of the bound exciton with binding energy $E_b\sim0.5$\,eV. The existence of these excitonic states should result in increasing the lifetime of optical excitations. In particular, the decay time of the photocurrent in the tBLG should be much larger than that for the AB bilayer. This effect has been observed experimentally~\cite{OpticalExcitation2015} using transient absorption microscopy.

The real and imaginary parts of the dynamical conductivity were calculated by T.~Stauber et~al.~\cite{SigmaPlasmons} in the framework of the continuum (low-energy) approximation developed in~\cite{dSPRL,dSPRB,PNAS,NonAbelianGaugePot}. The direct calculation of
$\Imag[\sigma(\omega)]$
is impossible because it requires the integration over a wide energy range, including energies far beyond the low-energy approximation. The authors~\cite{SigmaPlasmons} obtained a regularized Kramers--Kronig relation allowing to calculate the imaginary part of the dynamical conductivity in this case. Their approach is based on the observation that at high frequencies, the real part of the conductivity is equal to
$2\sigma_0$,  where $\sigma_0=e^2/4\hbar$ is the background conductivity of single-layer graphene. As a result, they obtained the following equation
\begin{equation}\label{RegKK}
\Imag\sigma(\omega)=\frac{2}{\pi\omega}\fint\limits_{0}^{\infty}d\omega'\frac{\omega'^2[\Real\sigma_{\text{reg}}(\omega')-2\sigma_0]}{\omega^2-\omega'^2}\,,
\end{equation}
where the bar in the integral represents the principal part. In this equation,
$\sigma_{\text{reg}}$
is the regular part of the total conductivity not including the
$\delta$-peak at zero frequency. The authors~\cite{SigmaPlasmons} calculated
$\Real(\sigma_{\text{reg}})$
numerically according to the relation
\begin{equation}
\Real\left[\sigma_{\text{reg}}(\omega)\right]=\frac{4\pi e^2}{\hbar\omega}\sum_{nm}\int\frac{d^2\mathbf{k}}{(2\pi)^2}\left[n_F(\varepsilon_{n\mathbf{k}})-n_F(\varepsilon_{m\mathbf{k}})\right]%
\left|\langle m\mathbf{k}|\hat{j}_x|n\mathbf{k}\rangle\right|^2\delta[\omega-(\varepsilon_{m\mathbf{k}}-\varepsilon_{n\mathbf{k}})/\hbar]\,,
\end{equation}
where
$\varepsilon_{n\mathbf{k}}$
and
$|n\mathbf{k}\rangle$
are the eigenenergy and the eigenstate of the low-energy Hamiltonian, with
$n$
being the band index, and
$\hat{j}_x=-\partial\hat{H}/\partial k_x$.
The total conductivity, then becomes
\begin{equation}
\sigma(\omega)=\pi D\delta(\omega)+\sigma_{\text{reg}}(\omega)\,,
\end{equation}
where
$D$
is the the Drude weight, which is related to the imaginary part of
$\sigma$
as
\begin{equation}\label{DrudeTBLG}
D=\lim_{\omega\to0}\left[\omega\Imag\sigma(\omega)\right]\,.
\end{equation}

Using
Eqs.~\eqref{RegKK}--\eqref{DrudeTBLG}, the authors of
Ref.~\cite{SigmaPlasmons}
numerically calculated the Drude weight as a function of the chemical
potential $\mu$. When the twist angle exceeds a critical value (which was
estimated~\cite{SigmaPlasmons} as $\theta_c\cong1.05^{\circ}$),
for small $\mu$, the Drude weight is approximately twice of its value for
single-layer graphene,
\begin{equation}\label{Drude0TBLG}
D(\mu)\approx2D_0(\mu)\,,\;\;D_0(\mu)=\frac{e^2\mu}{\pi\hbar^2}\,.
\end{equation}
If the chemical potential $\mu$ grows, the linear increase eventually
changes to a sharp decrease. This happens when the chemical potential
becomes larger than the low-energy VHS energy. After this, $D$ starts
increasing again until $\mu$ reaches the next VHS, where
$D(\mu)$
also has a dip, and so on. The function
$D(\mu)$
never exceeds twice of its value for single-layer graphene for any $\mu$.

The situation changes drastically for angles
$\theta < \theta_c$.
In this case, the Drude weight at low $\mu$ increases initially much faster
than
$2D_0(\mu)$,
but then, at larger values of $\mu$ (about the width of the peak in the DOS
existing at
$\theta<\theta_c$),
it drops down to almost zero, indicating the existence of a flat band at
low energies. At some twist angles, the Drude weight vanishes completely
and the transport gap exists in the system at non-zero $\mu$.

Experimental data for {\it interlayer} conductivity as a function of
$\theta$ for a tBLG sample was reported recently in
Ref.~\cite{Koren2016}.
The measured room-temperature conductivity demonstrated pronounced peaks at
$\theta=21.8^\circ$
and
$\theta=38.2^\circ$.
It is easy to check that the former angle corresponds to the
$(1, 1)$
structure with the smallest possible supercell size, the latter angle is
for the
$(1, 3)$
structure, which is conjugate to 
$(1, 1)$.

\subsubsection{Plasmons}

Knowledge of the imaginary part of the dynamical conductivity allows, in
particular, to calculate the plasmon excitation spectrum in twisted
bilayer. This was done in
Ref.~\cite{SigmaPlasmons}
in the random-phase approximation. The plasmon spectrum is derived from the
solution of the equation
$\epsilon(\mathbf{k},\omega)=0$,
where
$\epsilon(\mathbf{k},\omega)$
is the dielectric function. The RPA result for the tBLG
is~\cite{SigmaPlasmons}
\begin{equation}\label{TBLGepsilon}
\epsilon(\mathbf{k},\omega)=1+\frac{2\pi i}{\epsilon_s}\frac{|\mathbf{k}|}{\omega}\sigma(\omega)\,,
\end{equation}
where $\epsilon_s$ is the dielectric constant of the substrate. When the chemical potential
lies below the low-energy VHS, the real part of the conductivity is
strongly suppressed in the frequency range
$0<\omega<2\mu$
[see Fig.~\ref{FigTBLGSigma}(e)],
$\Real\sigma_{\text{reg}}(\omega)\approx0$.
Thus, well defined (low damped) plasmons can exist at these frequencies. At small frequencies, one can use the expansion
$\Imag\sigma(\omega)\approx D/\omega$,
and obtain from
Eq.~\eqref{TBLGepsilon} the following dispersion relation for this type of plasmons
\begin{equation}
\omega_{p}(\mathbf{k})
\cong
\sqrt{\frac{2\pi D|\mathbf{k}|}{\epsilon_s}}\,,
\end{equation}
where $D$ is given by
Eq.~\eqref{Drude0TBLG}.
In general, the real part of the conductivity is non-zero, and
the plasmons have a non-zero damping. The dispersion relation in this case is
determined from the peak in the loss function
$$
S(\mathbf{k},\omega)=-\Imag[1/\epsilon(\mathbf{k},\omega)]\,.
$$
The authors~\cite{SigmaPlasmons} discussed different types of plasmonic
excitations in twisted bilayers, in particular, ``transverse plasmons''
mentioned in
Section~\ref{sec::plasmon}.

\subsubsection{Interlayer conductivity}

\begin{figure}[t]
\centering
\includegraphics[width=0.6\textwidth]{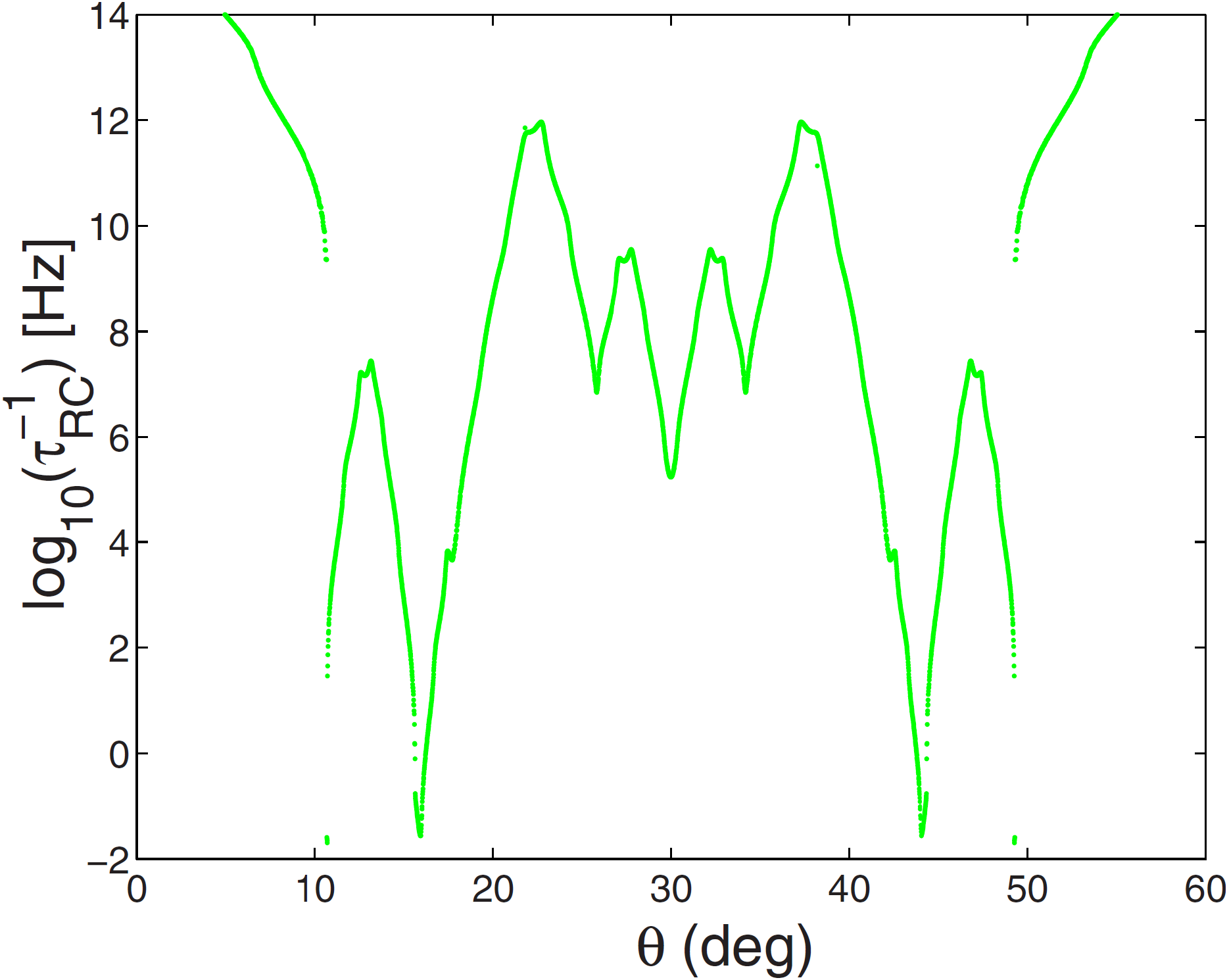}
\caption{(Color online) The interlayer RC-circuit equilibration rate,
$\tau_{RC}$,
versus twist angle, calculated for slightly doped tBLG samples with an excess
electron density
$n=5\times10^{12}$\,cm$^{-2}$
and
$\varepsilon_{F}/\hbar\tau=3$,
where
$\varepsilon_{F}$
is the Fermi energy, and
$\tau$
is the particle's lifetime due to scattering on impurities.
Reprinted figure with permission from R.~Bistritzer and A.\,H.~MacDonald,
Phys. Rev. B, {\bf 81}, 245412 (2010).
Copyright 2010 by the American Physical Society.
\url{http://dx.doi.org/10.1103/PhysRevB.81.245412}.
\label{FigTBLGtauRC}}
\end{figure}

Interlayer transport has been studied by R.~Bistritzer and A.\,H.~MacDonald~\cite{PNASsigma}
in the low-energy approximation. The authors considered both linear and non-linear
regimes. The current $I$, flowing between two layers with a non-zero
potential difference $V$, was found using second-order perturbation
theory in powers of the interlayer coupling. Both commensurate and
incommensurate structures were discussed. In the linear regime, one can
introduce the conductance
$G=[I(V)/V]_{V\to0}$,
and the time constant
$\tau_{RC}$
of the effective resistor-capacitor circuit ({\it RC}-circuit)
corresponding to the bilayer,
$$
\tau_{RC}=S/(4\pi Gc_0)\cong0.027S/G\,,
$$
where $S$ is the sample's area and
$c_0\cong3.3$\,{\AA}
is the interlayer distance. The rate
$1/\tau_{RC}\propto G$
as a function of the twist angle is shown in
Fig.~\ref{FigTBLGtauRC}.
The conductance depends substantially on $\theta$: it increases for
commensurate angles. For example, the curve in
Fig.~\ref{FigTBLGtauRC}
shows six maxima corresponding to the superstructures
$(2,1)$
($\theta\cong13.2^{\circ}$),
$(1,1)$
($\theta\cong21.8^{\circ}$),
$(2,3)$
($\theta\cong27.8^{\circ}$),
$(1,2)$
($\theta\cong32.2^{\circ}$),
$(1,3)$
($\theta\cong38.2^{\circ}$), and
$(1,6)$
($\theta\cong46.8^{\circ}$).
Reference~\cite{PNASsigma} also considered a non-linear regime, and calculated the $IV$
characteristics for different twist angles. The $IV$ curves are very
sensitive to the commensuration of the underlying structure. In particular,
at small bias voltage, the current for commensurate angles is several
orders of magnitude larger than that for close incommensurate angles.
For larger $V$, a region of negative differential conductance exists for
commensurate angles, whereas for incommensurate angles the value of
$dI/dV$ is always positive.

\subsubsection{Klein tunneling}
\label{subsect::twisted::klein}

\begin{figure}[t]
\centering
\includegraphics[height=0.18\textheight]{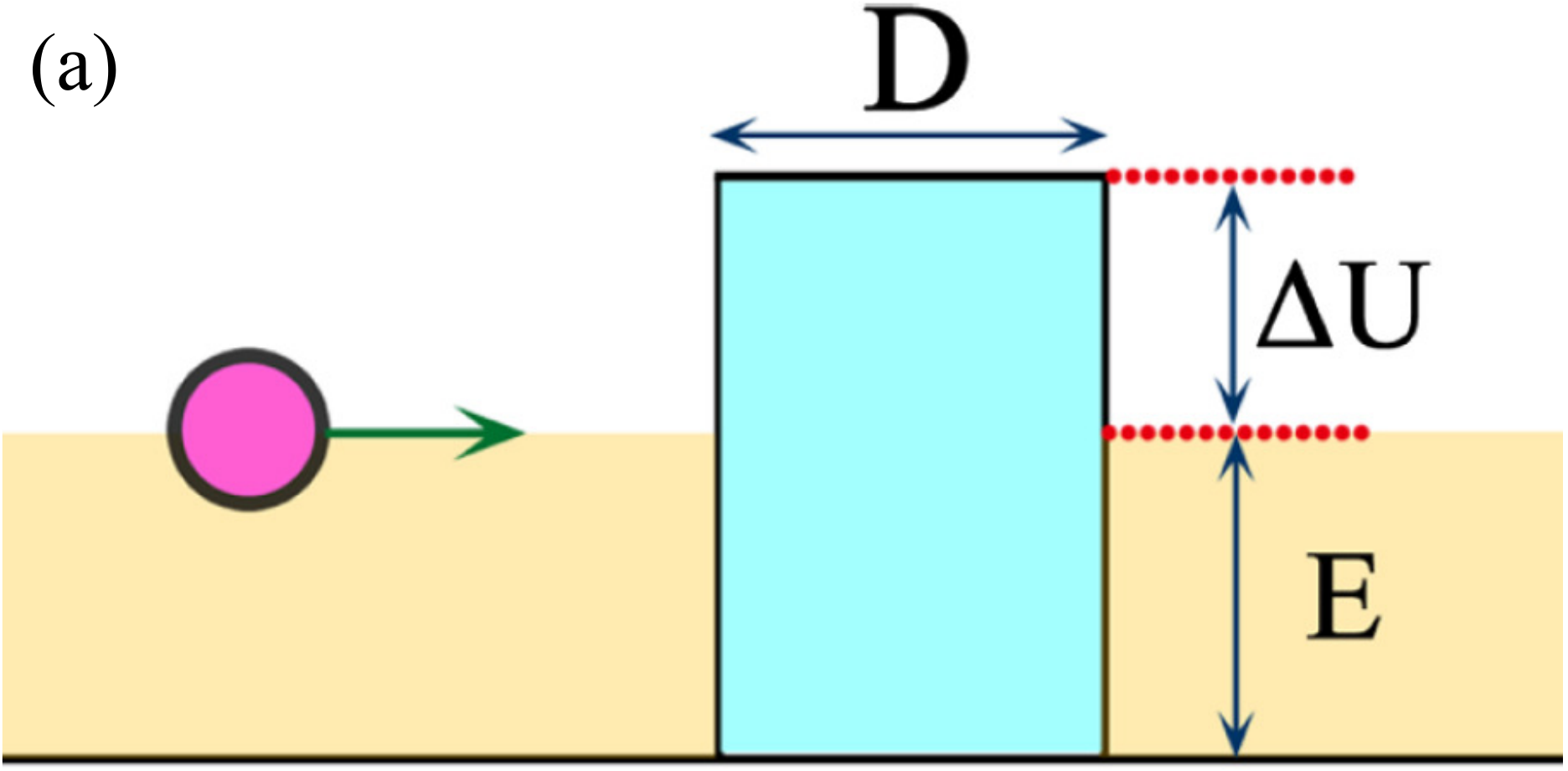}\\\vspace{2mm}
\includegraphics[height=0.3\textheight]{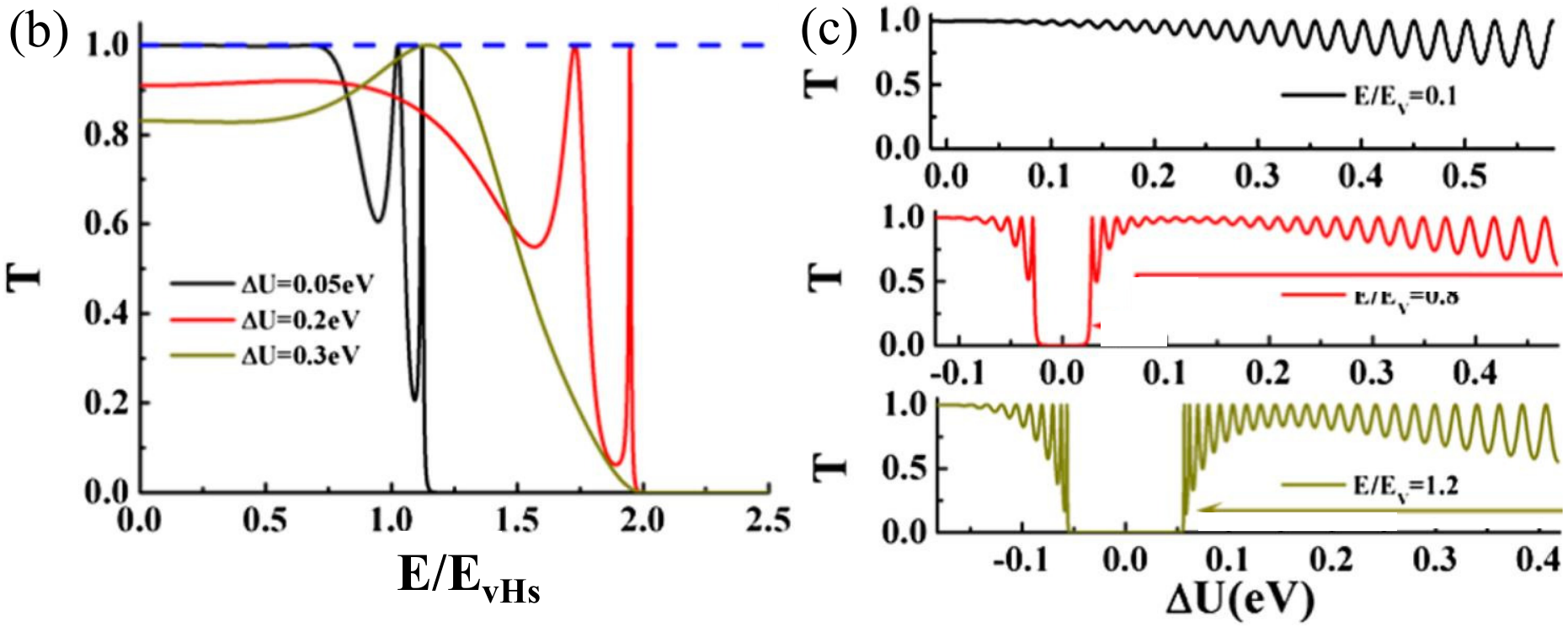}
\caption{(Color online) Klein tunneling in twisted bilayer graphene. (a)
Schematic diagram of the system. The electron with energy $E$
tunnels through the potential barrier of height
$U_0=E+\Delta U$
and width $D$ in the $x$ direction. The barrier is infinite along the $y$
direction. (b) Transmission probability of the normally-incident electrons
versus incident energy $E$, calculated for different values of  $\Delta U$.
(c) Transmission probability as a function of
$\Delta U$
calculated for three values of the energy of the incident electrons.
Reprinted figures with permission from W.-Y.~He et al., Phys. Rev. Lett.,
{\bf 111}, 066803 (2013). Copyright 2013 by the American Physical Society.
\url{http://dx.doi.org/10.1103/PhysRevLett.111.066803}.
\label{FigTBLGKlein}}
\end{figure}

The transmission probability of the tunneling through the potential barrier
in twisted bilayer graphene was calculated by W.-Y.~He et~al. in Ref.~\cite{KleinTBLG}.
The authors used a simple
$2\times2$
model Hamiltonian given by
Eq.~\eqref{HTBLGEff}.
The potential barrier has a rectangular shape. It is infinite
in the $y$ direction and has a width $D$ in the $x$
direction [see
Fig.~\ref{FigTBLGKlein}(a)].
Despite the simplicity of the model Hamiltonian, no analytical expression
was derived for the transmission probability $T$. They~\cite{KleinTBLG} calculated $T$ as a
function of the incident angle $\varphi$, energy of incident electrons $E$,
and the height of the barrier
$U_0$.
Analyzing the dependence
$T = T(\varphi)$,
the authors revealed the existence of angles, at which
$T=1$.
Such a behavior is similar to that for single-layer graphene, as well
as for AA- and AB-stacked bilayers, as we discussed in
subsection~\ref{Klein}.
The transmission probability of normally-incident electrons,
$\varphi=0$,
as a function of the electron energy is shown in
Fig.~\ref{FigTBLGKlein}(b).
At small energies, $T$ is nearly $1$ as in the case of single-layer
graphene or AA-stacked bilayer graphene.
When $E$ increases,
$T(E)$
exhibits several oscillations and then vanishes at higher energies.
It was noted~\cite{KleinTBLG} that the normal tunneling becomes zero for
$E\geqslant2E_{\text{vHs}}$,
where
$E_{\text{vHs}}$
is the energy of the VHS. Thus, at high energies the twisted bilayer
behaves as a Bernal-stacked bilayer.
They~\cite{KleinTBLG} also analyzed the behavior of the transmission
probability $T$ as a function of barrier's height.
Figure~\ref{FigTBLGKlein}(c)
presents the results for the normal transmission probability as a function
of the parameter
$\Delta U=U_0-E$.
The transmission probability oscillates with
$\Delta U$,
which can be explained by a geometrical resonance. The period of such
oscillations scales as
$1/D$.
It is interesting that at relatively high energies the tunneling is
suppressed in some region of
$\Delta U$
near
$\Delta U=0$.
The authors~\cite{KleinTBLG} attribute this fact to the absence of propagating wave
solutions to the Schr\"{o}dinger equation inside the barrier when $\Delta U$ is close to zero.

\section{Conclusions}

Bilayers of graphene are interesting objects for further investigations.
Their electronic properties are stacking-dependent, and, in many ways,
differ from those of the single-layer graphene. The bilayers have a number
electronic and electrodynamic features, which may be of importance for
fundamental research and possible applications.

The bilayers are studied intensively in recent years. In 2014 and in the
first half of 2015 only, more than 1100 papers on this subject were
published. However, despite significant efforts, several problems
concerning the electronic properties of the graphene bilayers are still
unsolved. For example, we discussed the issues related to identification of
ordered states in AB bilayer graphene and a gap in its electronic spectrum.
Different theoretical approaches predict different results. Moreover,
experimental observations of the gap opening are also controversial in many
cases. Whether these discrepancies are due to the sample quality or the
reasons lie in the intrinsic properties of the bilayer graphene is not
clear. 

Naturally, we are unable to overview here all questions concerning the
properties of graphene bilayers. Analysis of effects of dopants, adsorbed
atoms, special substrates and change in chemical composition of the layers
is the next step in the study of bilayer graphene-based systems. In this
review we have restricted ourselves to pure carbon bilayer graphene. We
also presented here only a brief discussion of some specific fields (for
example, photonics) referring the interested reader to recent reviews on
these subjects.

\section*{Acknowledgements}
We acknowledge partial support of the Russian Science Support Foundation,
RFBR Grant No. 15-02-02128.
FN is partially supported by
the RIKEN iTHES Project, MURI Center for Dynamic Magneto-Optics
via the AFOSR award number FA9550-14-1-0040, the Impact Program of JST,
and a Grant-in-Aid for Scientific Research (A),
and a grant from the John Templeton Foundation.
We are grateful to Ya.~Rodionov for his help with figures.
\newpage
\begin{appendix}

\end{appendix}
\newpage



\begin{thebibliography}{450}
\expandafter\ifx\csname natexlab\endcsname\relax\def\natexlab#1{#1}\fi
\expandafter\ifx\csname bibnamefont\endcsname\relax
  \def\bibnamefont#1{#1}\fi
\expandafter\ifx\csname bibfnamefont\endcsname\relax
  \def\bibfnamefont#1{#1}\fi
\expandafter\ifx\csname citenamefont\endcsname\relax
  \def\citenamefont#1{#1}\fi

\bibitem[{\citenamefont{Novoselov et~al.}(2004)\citenamefont{Novoselov, Geim,
  Morozov, Jiang, Zhang, Dubonos, Grigorieva, and Firsov}}]{NG1}
\bibinfo{author}{\bibfnamefont{K.}~\bibnamefont{Novoselov}},
  \bibinfo{author}{\bibfnamefont{A.}~\bibnamefont{Geim}},
  \bibinfo{author}{\bibfnamefont{S.}~\bibnamefont{Morozov}},
  \bibinfo{author}{\bibfnamefont{D.}~\bibnamefont{Jiang}},
  \bibinfo{author}{\bibfnamefont{Y.}~\bibnamefont{Zhang}},
  \bibinfo{author}{\bibfnamefont{S.}~\bibnamefont{Dubonos}},
  \bibinfo{author}{\bibfnamefont{I.}~\bibnamefont{Grigorieva}},
  \bibnamefont{and} \bibinfo{author}{\bibfnamefont{A.}~\bibnamefont{Firsov}},
  {``}\bibinfo{title}{Electric field effect in atomically thin carbon
  films},{''} \bibinfo{journal}{Science} \textbf{\bibinfo{volume}{306}},
  \bibinfo{pages}{666} (\bibinfo{year}{2004}).

\bibitem[{\citenamefont{Liu et~al.}(2009)\citenamefont{Liu, Suenaga, Harris,
  and Iijima}}]{liu_aa_exp2009}
\bibinfo{author}{\bibfnamefont{Z.}~\bibnamefont{Liu}},
  \bibinfo{author}{\bibfnamefont{K.}~\bibnamefont{Suenaga}},
  \bibinfo{author}{\bibfnamefont{P.~J.~F.} \bibnamefont{Harris}},
  \bibnamefont{and} \bibinfo{author}{\bibfnamefont{S.}~\bibnamefont{Iijima}},
  {``}\bibinfo{title}{Open and Closed Edges of Graphene Layers},{''}
  \bibinfo{journal}{Phys. Rev. Lett.} \textbf{\bibinfo{volume}{102}},
  \bibinfo{pages}{015501} (\bibinfo{year}{2009}).

\bibitem[{\citenamefont{Lopes~dos Santos et~al.}(2007)\citenamefont{Lopes~dos
  Santos, Peres, and Castro~Neto}}]{dSPRL}
\bibinfo{author}{\bibfnamefont{J.~M.~B.} \bibnamefont{Lopes~dos Santos}},
  \bibinfo{author}{\bibfnamefont{N.~M.~R.} \bibnamefont{Peres}},
  \bibnamefont{and} \bibinfo{author}{\bibfnamefont{A.~H.}
  \bibnamefont{Castro~Neto}}, {``}\bibinfo{title}{Graphene Bilayer with a
  Twist: Electronic Structure},{''} \bibinfo{journal}{Phys. Rev. Lett.}
  \textbf{\bibinfo{volume}{99}}, \bibinfo{pages}{256802}
  (\bibinfo{year}{2007}).

\bibitem[{\citenamefont{Castro~Neto et~al.}(2009)\citenamefont{Castro~Neto,
  Guinea, Peres, Novoselov, and Geim}}]{CastrNrev}
\bibinfo{author}{\bibfnamefont{A.~H.} \bibnamefont{Castro~Neto}},
  \bibinfo{author}{\bibfnamefont{F.}~\bibnamefont{Guinea}},
  \bibinfo{author}{\bibfnamefont{N.~M.~R.} \bibnamefont{Peres}},
  \bibinfo{author}{\bibfnamefont{K.~S.} \bibnamefont{Novoselov}},
  \bibnamefont{and} \bibinfo{author}{\bibfnamefont{A.~K.} \bibnamefont{Geim}},
  {``}\bibinfo{title}{The electronic properties of graphene},{''}
  \bibinfo{journal}{Rev. Mod. Phys.} \textbf{\bibinfo{volume}{81}},
  \bibinfo{pages}{109} (\bibinfo{year}{2009}).

\bibitem[{\citenamefont{McCann and Koshino}(2013)}]{mccann_kosh_rev2013}
\bibinfo{author}{\bibfnamefont{E.}~\bibnamefont{McCann}} \bibnamefont{and}
  \bibinfo{author}{\bibfnamefont{M.}~\bibnamefont{Koshino}},
  {``}\bibinfo{title}{The electronic properties of bilayer graphene},{''}
  \bibinfo{journal}{Rep. Prog. Phys.} \textbf{\bibinfo{volume}{76}},
  \bibinfo{pages}{056503} (\bibinfo{year}{2013}).

\bibitem[{\citenamefont{Brihuega et~al.}(2012)\citenamefont{Brihuega, Mallet,
  Gonz\'alez-Herrero, Trambly~de Laissardi\`ere, Ugeda, Magaud,
  G\'omez-Rodr\'{\i}guez, Yndur\'ain, and Veuillen}}]{STM1}
\bibinfo{author}{\bibfnamefont{I.}~\bibnamefont{Brihuega}},
  \bibinfo{author}{\bibfnamefont{P.}~\bibnamefont{Mallet}},
  \bibinfo{author}{\bibfnamefont{H.}~\bibnamefont{Gonz\'alez-Herrero}},
  \bibinfo{author}{\bibfnamefont{G.}~\bibnamefont{Trambly~de Laissardi\`ere}},
  \bibinfo{author}{\bibfnamefont{M.~M.} \bibnamefont{Ugeda}},
  \bibinfo{author}{\bibfnamefont{L.}~\bibnamefont{Magaud}},
  \bibinfo{author}{\bibfnamefont{J.~M.} \bibnamefont{G\'omez-Rodr\'{\i}guez}},
  \bibinfo{author}{\bibfnamefont{F.}~\bibnamefont{Yndur\'ain}},
  \bibnamefont{and} \bibinfo{author}{\bibfnamefont{J.-Y.}
  \bibnamefont{Veuillen}}, {``}\bibinfo{title}{Unraveling the Intrinsic and
  Robust Nature of van Hove Singularities in Twisted Bilayer Graphene by
  Scanning Tunneling Microscopy and Theoretical Analysis},{''}
  \bibinfo{journal}{Phys. Rev. Lett.} \textbf{\bibinfo{volume}{109}},
  \bibinfo{pages}{196802} (\bibinfo{year}{2012}).

\bibitem[{\citenamefont{de~Andres et~al.}(2008)\citenamefont{de~Andres,
  Ram\'{i}rez, and Verg\'es}}]{bi1}
\bibinfo{author}{\bibfnamefont{P.~L.} \bibnamefont{de~Andres}},
  \bibinfo{author}{\bibfnamefont{R.}~\bibnamefont{Ram\'{i}rez}},
  \bibnamefont{and} \bibinfo{author}{\bibfnamefont{J.~A.}
  \bibnamefont{Verg\'es}}, {``}\bibinfo{title}{Strong covalent bonding between
  two graphene layers},{''} \bibinfo{journal}{Phys. Rev. B}
  \textbf{\bibinfo{volume}{77}}, \bibinfo{pages}{045403}
  (\bibinfo{year}{2008}).

\bibitem[{\citenamefont{Shibuta and Elliott}(2011)}]{bi2}
\bibinfo{author}{\bibfnamefont{Y.}~\bibnamefont{Shibuta}} \bibnamefont{and}
  \bibinfo{author}{\bibfnamefont{J.~A.} \bibnamefont{Elliott}},
  {``}\bibinfo{title}{Interaction between two graphene sheets with a
  turbostratic orientational relationship},{''} \bibinfo{journal}{Chem. Phys.
  Lett.} \textbf{\bibinfo{volume}{512}}, \bibinfo{pages}{146 }
  (\bibinfo{year}{2011}).

\bibitem[{\citenamefont{Alam et~al.}(2011)\citenamefont{Alam, Lin, and
  Saito}}]{bi3}
\bibinfo{author}{\bibfnamefont{M.~S.} \bibnamefont{Alam}},
  \bibinfo{author}{\bibfnamefont{J.}~\bibnamefont{Lin}}, \bibnamefont{and}
  \bibinfo{author}{\bibfnamefont{M.}~\bibnamefont{Saito}},
  {``}\bibinfo{title}{First-Principles Calculation of the Interlayer Distance
  of the Two-Layer Graphene},{''} \bibinfo{journal}{Jpn. J. Appl. Phys.}
  \textbf{\bibinfo{volume}{50}}, \bibinfo{pages}{080213}
  (\bibinfo{year}{2011}).

\bibitem[{\citenamefont{Xu et~al.}(2013)\citenamefont{Xu, Li, Yakobson, and
  Ding}}]{bi4}
\bibinfo{author}{\bibfnamefont{Z.}~\bibnamefont{Xu}},
  \bibinfo{author}{\bibfnamefont{X.}~\bibnamefont{Li}},
  \bibinfo{author}{\bibfnamefont{B.~I.} \bibnamefont{Yakobson}},
  \bibnamefont{and} \bibinfo{author}{\bibfnamefont{F.}~\bibnamefont{Ding}},
  {``}\bibinfo{title}{Interaction between graphene layers and the mechanisms of
  graphite{'}s superlubricity and self-retraction},{''}
  \bibinfo{journal}{Nanoscale} \textbf{\bibinfo{volume}{5}},
  \bibinfo{pages}{6736} (\bibinfo{year}{2013}).

\bibitem[{\citenamefont{Berashevich and Chakraborty}(2011)}]{bi5}
\bibinfo{author}{\bibfnamefont{J.}~\bibnamefont{Berashevich}} \bibnamefont{and}
  \bibinfo{author}{\bibfnamefont{T.}~\bibnamefont{Chakraborty}},
  {``}\bibinfo{title}{On the Nature of Interlayer Interactions in a System of
  Two Graphene Fragments},{''} \bibinfo{journal}{J. Phys. Chem. C}
  \textbf{\bibinfo{volume}{115}}, \bibinfo{pages}{24666}
  (\bibinfo{year}{2011}).

\bibitem[{\citenamefont{Das~Sarma et~al.}(2011)\citenamefont{Das~Sarma, Adam,
  Hwang, and Rossi}}]{das_sarma_rev2011}
\bibinfo{author}{\bibfnamefont{S.}~\bibnamefont{Das~Sarma}},
  \bibinfo{author}{\bibfnamefont{S.}~\bibnamefont{Adam}},
  \bibinfo{author}{\bibfnamefont{E.~H.} \bibnamefont{Hwang}}, \bibnamefont{and}
  \bibinfo{author}{\bibfnamefont{E.}~\bibnamefont{Rossi}},
  {``}\bibinfo{title}{Electronic transport in two-dimensional graphene},{''}
  \bibinfo{journal}{Rev. Mod. Phys.} \textbf{\bibinfo{volume}{83}},
  \bibinfo{pages}{407} (\bibinfo{year}{2011}).

\bibitem[{\citenamefont{Yan}(2014)}]{yanRev}
\bibinfo{author}{\bibfnamefont{H.}~\bibnamefont{Yan}},
  {``}\bibinfo{title}{Bilayer graphene: physics and application outlook in
  photonics},{''} \bibinfo{journal}{Nanophotonics} \textbf{\bibinfo{volume}{4}}
  (\bibinfo{year}{2014}).

\bibitem[{\citenamefont{Stauber}(2014)}]{plasmonicsRev}
\bibinfo{author}{\bibfnamefont{T.}~\bibnamefont{Stauber}},
  {``}\bibinfo{title}{Plasmonics in Dirac systems: from graphene to topological
  insulators},{''} \bibinfo{journal}{J. Phys.: Condens. Matter}
  \textbf{\bibinfo{volume}{26}}, \bibinfo{pages}{123201}
  (\bibinfo{year}{2014}).

\bibitem[{\citenamefont{Glazov and Ganichev}(2014)}]{glazovRev}
\bibinfo{author}{\bibfnamefont{M.}~\bibnamefont{Glazov}} \bibnamefont{and}
  \bibinfo{author}{\bibfnamefont{S.}~\bibnamefont{Ganichev}},
  {``}\bibinfo{title}{High frequency electric field induced nonlinear effects
  in graphene},{''} \bibinfo{journal}{Phys. Rep.}
  \textbf{\bibinfo{volume}{535}}, \bibinfo{pages}{101} (\bibinfo{year}{2014}).

\bibitem[{\citenamefont{Mele}(2012)}]{MeleReview}
\bibinfo{author}{\bibfnamefont{E.~J.} \bibnamefont{Mele}},
  {``}\bibinfo{title}{Interlayer coupling in rotationally faulted multilayer
  graphenes},{''} \bibinfo{journal}{J. Phys. D: Appl. Phys.}
  \textbf{\bibinfo{volume}{45}}, \bibinfo{pages}{154004}
  (\bibinfo{year}{2012}).

\bibitem[{\citenamefont{Abergel et~al.}(2010)\citenamefont{Abergel, Apalkov,
  Berashevich, Ziegler, and Chakraborty}}]{Abergel}
\bibinfo{author}{\bibfnamefont{D.}~\bibnamefont{Abergel}},
  \bibinfo{author}{\bibfnamefont{V.}~\bibnamefont{Apalkov}},
  \bibinfo{author}{\bibfnamefont{J.}~\bibnamefont{Berashevich}},
  \bibinfo{author}{\bibfnamefont{K.}~\bibnamefont{Ziegler}}, \bibnamefont{and}
  \bibinfo{author}{\bibfnamefont{T.}~\bibnamefont{Chakraborty}},
  {``}\bibinfo{title}{Properties of graphene: a theoretical perspective},{''}
  \bibinfo{journal}{Adv. Phys.} \textbf{\bibinfo{volume}{59}},
  \bibinfo{pages}{261} (\bibinfo{year}{2010}).

\bibitem[{\citenamefont{Rozhkov et~al.}(2011)\citenamefont{Rozhkov, Giavaras,
  Bliokh, Freilikher, and Nori}}]{meso_review}
\bibinfo{author}{\bibfnamefont{A.}~\bibnamefont{Rozhkov}},
  \bibinfo{author}{\bibfnamefont{G.}~\bibnamefont{Giavaras}},
  \bibinfo{author}{\bibfnamefont{Y.~P.} \bibnamefont{Bliokh}},
  \bibinfo{author}{\bibfnamefont{V.}~\bibnamefont{Freilikher}},
  \bibnamefont{and} \bibinfo{author}{\bibfnamefont{F.}~\bibnamefont{Nori}},
  {``}\bibinfo{title}{Electronic properties of mesoscopic graphene structures:
  Charge confinement and control of spin and charge transport},{''}
  \bibinfo{journal}{Phys. Rep.} \textbf{\bibinfo{volume}{503}},
  \bibinfo{pages}{77 } (\bibinfo{year}{2011}).

\bibitem[{\citenamefont{Avouris et~al.}(2007)\citenamefont{Avouris, Chen, and
  Perebeinos}}]{Avouris2007}
\bibinfo{author}{\bibfnamefont{P.}~\bibnamefont{Avouris}},
  \bibinfo{author}{\bibfnamefont{Z.}~\bibnamefont{Chen}}, \bibnamefont{and}
  \bibinfo{author}{\bibfnamefont{V.}~\bibnamefont{Perebeinos}},
  {``}\bibinfo{title}{Carbon-based electronics},{''} \bibinfo{journal}{Nat.
  Nanotechnol.} \textbf{\bibinfo{volume}{2}}, \bibinfo{pages}{605}
  (\bibinfo{year}{2007}).

\bibitem[{\citenamefont{Lee et~al.}(2008)\citenamefont{Lee, Lee, Ahn, Kim,
  Wilson, and John}}]{lee_aa_exp2008}
\bibinfo{author}{\bibfnamefont{J.-K.} \bibnamefont{Lee}},
  \bibinfo{author}{\bibfnamefont{S.-C.} \bibnamefont{Lee}},
  \bibinfo{author}{\bibfnamefont{J.-P.} \bibnamefont{Ahn}},
  \bibinfo{author}{\bibfnamefont{S.-C.} \bibnamefont{Kim}},
  \bibinfo{author}{\bibfnamefont{J.~I.~B.} \bibnamefont{Wilson}},
  \bibnamefont{and} \bibinfo{author}{\bibfnamefont{P.}~\bibnamefont{John}},
  {``}\bibinfo{title}{The growth of AA graphite on (111) diamond},{''}
  \bibinfo{journal}{J. Chem. Phys.} \textbf{\bibinfo{volume}{129}},
  \bibinfo{eid}{234709} (\bibinfo{year}{2008}).

\bibitem[{\citenamefont{Lauffer et~al.}(2008)\citenamefont{Lauffer, Emtsev,
  Graupner, Seyller, Ley, Reshanov, and Weber}}]{lauffer_aa_exp2008}
\bibinfo{author}{\bibfnamefont{P.}~\bibnamefont{Lauffer}},
  \bibinfo{author}{\bibfnamefont{K.~V.} \bibnamefont{Emtsev}},
  \bibinfo{author}{\bibfnamefont{R.}~\bibnamefont{Graupner}},
  \bibinfo{author}{\bibfnamefont{T.}~\bibnamefont{Seyller}},
  \bibinfo{author}{\bibfnamefont{L.}~\bibnamefont{Ley}},
  \bibinfo{author}{\bibfnamefont{S.~A.} \bibnamefont{Reshanov}},
  \bibnamefont{and} \bibinfo{author}{\bibfnamefont{H.~B.} \bibnamefont{Weber}},
  {``}\bibinfo{title}{Atomic and electronic structure of few-layer graphene on
  SiC(0001) studied with scanning tunneling microscopy and spectroscopy},{''}
  \bibinfo{journal}{Phys. Rev. B} \textbf{\bibinfo{volume}{77}},
  \bibinfo{pages}{155426} (\bibinfo{year}{2008}).

\bibitem[{\citenamefont{Borysiuk et~al.}(2011)\citenamefont{Borysiuk, Soltys,
  and Piechota}}]{borysiuk2011_aa}
\bibinfo{author}{\bibfnamefont{J.}~\bibnamefont{Borysiuk}},
  \bibinfo{author}{\bibfnamefont{J.}~\bibnamefont{Soltys}}, \bibnamefont{and}
  \bibinfo{author}{\bibfnamefont{J.}~\bibnamefont{Piechota}},
  {``}\bibinfo{title}{Stacking sequence dependence of graphene layers on SiC
  (0001) - Experimental and theoretical investigation},{''}
  \bibinfo{journal}{J. Appl. Phys.} \textbf{\bibinfo{volume}{109}},
  \bibinfo{eid}{093523} (\bibinfo{year}{2011}).

\bibitem[{\citenamefont{Martin et~al.}(2010)\citenamefont{Martin, Feldman,
  Weitz, Allen, and Yacoby}}]{Martin2010}
\bibinfo{author}{\bibfnamefont{J.}~\bibnamefont{Martin}},
  \bibinfo{author}{\bibfnamefont{B.~E.} \bibnamefont{Feldman}},
  \bibinfo{author}{\bibfnamefont{R.~T.} \bibnamefont{Weitz}},
  \bibinfo{author}{\bibfnamefont{M.~T.} \bibnamefont{Allen}}, \bibnamefont{and}
  \bibinfo{author}{\bibfnamefont{A.}~\bibnamefont{Yacoby}},
  {``}\bibinfo{title}{Local Compressibility Measurements of Correlated States
  in Suspended Bilayer Graphene},{''} \bibinfo{journal}{Phys. Rev. Lett.}
  \textbf{\bibinfo{volume}{105}}, \bibinfo{pages}{256806}
  (\bibinfo{year}{2010}).

\bibitem[{\citenamefont{Novoselov et~al.}(2005)\citenamefont{Novoselov, Geim,
  Morozov, Jiang, Katsnelson, Grigorieva, Dubonos, and Firsov}}]{Novoselov2005}
\bibinfo{author}{\bibfnamefont{K.~S.} \bibnamefont{Novoselov}},
  \bibinfo{author}{\bibfnamefont{A.~K.} \bibnamefont{Geim}},
  \bibinfo{author}{\bibfnamefont{S.~V.} \bibnamefont{Morozov}},
  \bibinfo{author}{\bibfnamefont{D.}~\bibnamefont{Jiang}},
  \bibinfo{author}{\bibfnamefont{M.~I.} \bibnamefont{Katsnelson}},
  \bibinfo{author}{\bibfnamefont{I.~V.} \bibnamefont{Grigorieva}},
  \bibinfo{author}{\bibfnamefont{S.~V.} \bibnamefont{Dubonos}},
  \bibnamefont{and} \bibinfo{author}{\bibfnamefont{A.~A.}
  \bibnamefont{Firsov}}, {``}\bibinfo{title}{Two-dimensional gas of massless
  Dirac fermions in graphene},{''} \bibinfo{journal}{Nature}
  \textbf{\bibinfo{volume}{438}}, \bibinfo{pages}{197} (\bibinfo{year}{2005}).

\bibitem[{\citenamefont{Bolotin et~al.}(2008)\citenamefont{Bolotin, Sikes,
  Jiang, Klima, Fudenberg, Hone, Kim, and Stormer}}]{Bolotin2008}
\bibinfo{author}{\bibfnamefont{K.}~\bibnamefont{Bolotin}},
  \bibinfo{author}{\bibfnamefont{K.}~\bibnamefont{Sikes}},
  \bibinfo{author}{\bibfnamefont{Z.}~\bibnamefont{Jiang}},
  \bibinfo{author}{\bibfnamefont{M.}~\bibnamefont{Klima}},
  \bibinfo{author}{\bibfnamefont{G.}~\bibnamefont{Fudenberg}},
  \bibinfo{author}{\bibfnamefont{J.}~\bibnamefont{Hone}},
  \bibinfo{author}{\bibfnamefont{P.}~\bibnamefont{Kim}}, \bibnamefont{and}
  \bibinfo{author}{\bibfnamefont{H.}~\bibnamefont{Stormer}},
  {``}\bibinfo{title}{Ultrahigh electron mobility in suspended graphene},{''}
  \bibinfo{journal}{Solid State Commun.} \textbf{\bibinfo{volume}{146}},
  \bibinfo{pages}{351 } (\bibinfo{year}{2008}).

\bibitem[{\citenamefont{Mayorov et~al.}(2011)\citenamefont{Mayorov, Elias,
  Mucha-Kruczynski, Gorbachev, Tudorovskiy, Zhukov, Morozov, Katsnelson,
  Fal'ko, Geim et~al.}}]{Mayorov2011}
\bibinfo{author}{\bibfnamefont{A.~S.} \bibnamefont{Mayorov}},
  \bibinfo{author}{\bibfnamefont{D.~C.} \bibnamefont{Elias}},
  \bibinfo{author}{\bibfnamefont{M.}~\bibnamefont{Mucha-Kruczynski}},
  \bibinfo{author}{\bibfnamefont{R.~V.} \bibnamefont{Gorbachev}},
  \bibinfo{author}{\bibfnamefont{T.}~\bibnamefont{Tudorovskiy}},
  \bibinfo{author}{\bibfnamefont{A.}~\bibnamefont{Zhukov}},
  \bibinfo{author}{\bibfnamefont{S.~V.} \bibnamefont{Morozov}},
  \bibinfo{author}{\bibfnamefont{M.~I.} \bibnamefont{Katsnelson}},
  \bibinfo{author}{\bibfnamefont{V.~I.} \bibnamefont{Fal'ko}},
  \bibinfo{author}{\bibfnamefont{A.~K.} \bibnamefont{Geim}},
  \bibnamefont{et~al.}, {``}\bibinfo{title}{Interaction-Driven Spectrum
  Reconstruction in Bilayer Graphene},{''} \bibinfo{journal}{Science}
  \textbf{\bibinfo{volume}{333}}, \bibinfo{pages}{860} (\bibinfo{year}{2011}).

\bibitem[{\citenamefont{Feldman et~al.}(2009)\citenamefont{Feldman, Martin, and
  Yacoby}}]{Feldman2009}
\bibinfo{author}{\bibfnamefont{B.~E.} \bibnamefont{Feldman}},
  \bibinfo{author}{\bibfnamefont{J.}~\bibnamefont{Martin}}, \bibnamefont{and}
  \bibinfo{author}{\bibfnamefont{A.}~\bibnamefont{Yacoby}},
  {``}\bibinfo{title}{Broken-symmetry states and divergent resistance in
  suspended bilayer graphene},{''} \bibinfo{journal}{Nat. Phys.}
  \textbf{\bibinfo{volume}{5}}, \bibinfo{pages}{889} (\bibinfo{year}{2009}).

\bibitem[{\citenamefont{Bao et~al.}(2012)\citenamefont{Bao, Velasco, Zhang,
  Jing, Standley, Smirnov, Bockrath, MacDonald, and Lau}}]{Bao2012}
\bibinfo{author}{\bibfnamefont{W.}~\bibnamefont{Bao}},
  \bibinfo{author}{\bibfnamefont{J.}~\bibnamefont{Velasco}},
  \bibinfo{author}{\bibfnamefont{F.}~\bibnamefont{Zhang}},
  \bibinfo{author}{\bibfnamefont{L.}~\bibnamefont{Jing}},
  \bibinfo{author}{\bibfnamefont{B.}~\bibnamefont{Standley}},
  \bibinfo{author}{\bibfnamefont{D.}~\bibnamefont{Smirnov}},
  \bibinfo{author}{\bibfnamefont{M.}~\bibnamefont{Bockrath}},
  \bibinfo{author}{\bibfnamefont{A.~H.} \bibnamefont{MacDonald}},
  \bibnamefont{and} \bibinfo{author}{\bibfnamefont{C.~N.} \bibnamefont{Lau}},
  {``}\bibinfo{title}{Evidence for a spontaneous gapped state in ultraclean
  bilayer graphene},{''} \bibinfo{journal}{PNAS}
  \textbf{\bibinfo{volume}{109}}, \bibinfo{pages}{10802}
  (\bibinfo{year}{2012}).

\bibitem[{\citenamefont{Morozov et~al.}(2008)\citenamefont{Morozov, Novoselov,
  Katsnelson, Schedin, Elias, Jaszczak, and Geim}}]{Morozov2008}
\bibinfo{author}{\bibfnamefont{S.~V.} \bibnamefont{Morozov}},
  \bibinfo{author}{\bibfnamefont{K.~S.} \bibnamefont{Novoselov}},
  \bibinfo{author}{\bibfnamefont{M.~I.} \bibnamefont{Katsnelson}},
  \bibinfo{author}{\bibfnamefont{F.}~\bibnamefont{Schedin}},
  \bibinfo{author}{\bibfnamefont{D.~C.} \bibnamefont{Elias}},
  \bibinfo{author}{\bibfnamefont{J.~A.} \bibnamefont{Jaszczak}},
  \bibnamefont{and} \bibinfo{author}{\bibfnamefont{A.~K.} \bibnamefont{Geim}},
  {``}\bibinfo{title}{Giant intrinsic carrier mobilities in graphene and its
  bilayer},{''} \bibinfo{journal}{Phys. Rev. Lett.}
  \textbf{\bibinfo{volume}{100}}, \bibinfo{pages}{016602}
  (\bibinfo{year}{2008}).

\bibitem[{\citenamefont{Oostinga et~al.}(2008)\citenamefont{Oostinga, Heersche,
  Liu, Morpurgo, and Vandersypen}}]{Oostinga2008}
\bibinfo{author}{\bibfnamefont{J.~B.} \bibnamefont{Oostinga}},
  \bibinfo{author}{\bibfnamefont{H.~B.} \bibnamefont{Heersche}},
  \bibinfo{author}{\bibfnamefont{X.}~\bibnamefont{Liu}},
  \bibinfo{author}{\bibfnamefont{A.~F.} \bibnamefont{Morpurgo}},
  \bibnamefont{and} \bibinfo{author}{\bibfnamefont{L.~M.~K.}
  \bibnamefont{Vandersypen}}, {``}\bibinfo{title}{Gate-induced insulating state
  in bilayer graphene devices},{''} \bibinfo{journal}{Nat. Mater.}
  \textbf{\bibinfo{volume}{7}}, \bibinfo{pages}{151} (\bibinfo{year}{2008}).

\bibitem[{\citenamefont{Dean et~al.}(2010)\citenamefont{Dean, Young, Meric,
  Lee, Wang, Sorgenfrei, Watanabe, Taniguchi, Kim, Shepard
  et~al.}}]{dean2010boron}
\bibinfo{author}{\bibfnamefont{C.}~\bibnamefont{Dean}},
  \bibinfo{author}{\bibfnamefont{A.}~\bibnamefont{Young}},
  \bibinfo{author}{\bibfnamefont{I.}~\bibnamefont{Meric}},
  \bibinfo{author}{\bibfnamefont{C.}~\bibnamefont{Lee}},
  \bibinfo{author}{\bibfnamefont{L.}~\bibnamefont{Wang}},
  \bibinfo{author}{\bibfnamefont{S.}~\bibnamefont{Sorgenfrei}},
  \bibinfo{author}{\bibfnamefont{K.}~\bibnamefont{Watanabe}},
  \bibinfo{author}{\bibfnamefont{T.}~\bibnamefont{Taniguchi}},
  \bibinfo{author}{\bibfnamefont{P.}~\bibnamefont{Kim}},
  \bibinfo{author}{\bibfnamefont{K.}~\bibnamefont{Shepard}},
  \bibnamefont{et~al.}, {``}\bibinfo{title}{Boron nitride substrates for
  high-quality graphene electronics},{''} \bibinfo{journal}{Nat. Nanotechnol.}
  \textbf{\bibinfo{volume}{5}}, \bibinfo{pages}{722} (\bibinfo{year}{2010}).

\bibitem[{\citenamefont{Velasco~Jr. et~al.}(2012)\citenamefont{Velasco~Jr.,
  Jing, Bao, Lee, Kratz, Aji, Bockrath, Lau, Varma, Stillwell
  et~al.}}]{Velasco2012}
\bibinfo{author}{\bibfnamefont{J.}~\bibnamefont{Velasco~Jr.}},
  \bibinfo{author}{\bibfnamefont{L.}~\bibnamefont{Jing}},
  \bibinfo{author}{\bibfnamefont{W.}~\bibnamefont{Bao}},
  \bibinfo{author}{\bibfnamefont{Y.}~\bibnamefont{Lee}},
  \bibinfo{author}{\bibfnamefont{P.}~\bibnamefont{Kratz}},
  \bibinfo{author}{\bibfnamefont{V.}~\bibnamefont{Aji}},
  \bibinfo{author}{\bibfnamefont{M.}~\bibnamefont{Bockrath}},
  \bibinfo{author}{\bibfnamefont{C.}~\bibnamefont{Lau}},
  \bibinfo{author}{\bibfnamefont{C.}~\bibnamefont{Varma}},
  \bibinfo{author}{\bibfnamefont{R.}~\bibnamefont{Stillwell}},
  \bibnamefont{et~al.}, {``}\bibinfo{title}{Transport spectroscopy of
  symmetry-broken insulating states in bilayer graphene},{''}
  \bibinfo{journal}{Nat. Nanotechnol.} \textbf{\bibinfo{volume}{7}},
  \bibinfo{pages}{156} (\bibinfo{year}{2012}).

\bibitem[{\citenamefont{van Elferen et~al.}(2012)\citenamefont{van Elferen,
  Veligura, Kurganova, Zeitler, Maan, Tombros, Vera-Marun, and van
  Wees}}]{Elferen2012}
\bibinfo{author}{\bibfnamefont{H.~J.} \bibnamefont{van Elferen}},
  \bibinfo{author}{\bibfnamefont{A.}~\bibnamefont{Veligura}},
  \bibinfo{author}{\bibfnamefont{E.~V.} \bibnamefont{Kurganova}},
  \bibinfo{author}{\bibfnamefont{U.}~\bibnamefont{Zeitler}},
  \bibinfo{author}{\bibfnamefont{J.~C.} \bibnamefont{Maan}},
  \bibinfo{author}{\bibfnamefont{N.}~\bibnamefont{Tombros}},
  \bibinfo{author}{\bibfnamefont{I.~J.} \bibnamefont{Vera-Marun}},
  \bibnamefont{and} \bibinfo{author}{\bibfnamefont{B.~J.} \bibnamefont{van
  Wees}}, {``}\bibinfo{title}{Field-induced quantum Hall ferromagnetism in
  suspended bilayer graphene},{''} \bibinfo{journal}{Phys. Rev. B}
  \textbf{\bibinfo{volume}{85}}, \bibinfo{pages}{115408}
  (\bibinfo{year}{2012}).

\bibitem[{\citenamefont{Semenoff}(1984)}]{Semenoff}
\bibinfo{author}{\bibfnamefont{G.~W.} \bibnamefont{Semenoff}},
  {``}\bibinfo{title}{Condensed-Matter Simulation of a Three-Dimensional
  Anomaly},{''} \bibinfo{journal}{Phys. Rev. Lett.}
  \textbf{\bibinfo{volume}{53}}, \bibinfo{pages}{2449} (\bibinfo{year}{1984}).

\bibitem[{\citenamefont{Hobson and Nierenberg}(1953)}]{Hobson}
\bibinfo{author}{\bibfnamefont{J.~P.} \bibnamefont{Hobson}} \bibnamefont{and}
  \bibinfo{author}{\bibfnamefont{W.~A.} \bibnamefont{Nierenberg}},
  {``}\bibinfo{title}{The Statistics of a Two-Dimensional, Hexagonal Net},{''}
  \bibinfo{journal}{Phys. Rev.} \textbf{\bibinfo{volume}{89}},
  \bibinfo{pages}{662} (\bibinfo{year}{1953}).

\bibitem[{\citenamefont{Rakhmanov et~al.}(2012)\citenamefont{Rakhmanov,
  Rozhkov, Sboychakov, and Nori}}]{PrlOur}
\bibinfo{author}{\bibfnamefont{A.~L.} \bibnamefont{Rakhmanov}},
  \bibinfo{author}{\bibfnamefont{A.~V.} \bibnamefont{Rozhkov}},
  \bibinfo{author}{\bibfnamefont{A.~O.} \bibnamefont{Sboychakov}},
  \bibnamefont{and} \bibinfo{author}{\bibfnamefont{F.}~\bibnamefont{Nori}},
  {``}\bibinfo{title}{Instabilities of the $AA$-Stacked Graphene Bilayer},{''}
  \bibinfo{journal}{Phys. Rev. Lett.} \textbf{\bibinfo{volume}{109}},
  \bibinfo{pages}{206801} (\bibinfo{year}{2012}).

\bibitem[{\citenamefont{Charlier et~al.}(1992)\citenamefont{Charlier,
  Michenaud, and Gonze}}]{tunnel}
\bibinfo{author}{\bibfnamefont{J.-C.} \bibnamefont{Charlier}},
  \bibinfo{author}{\bibfnamefont{J.-P.} \bibnamefont{Michenaud}},
  \bibnamefont{and} \bibinfo{author}{\bibfnamefont{X.}~\bibnamefont{Gonze}},
  {``}\bibinfo{title}{First-principles study of the electronic properties of
  simple hexagonal graphite},{''} \bibinfo{journal}{Phys. Rev. B}
  \textbf{\bibinfo{volume}{46}}, \bibinfo{pages}{4531} (\bibinfo{year}{1992}).

\bibitem[{\citenamefont{Brandt et~al.}(1988)\citenamefont{Brandt, Chudinov, and
  Ponomarev}}]{tunnel1}
\bibinfo{author}{\bibfnamefont{N.~B.} \bibnamefont{Brandt}},
  \bibinfo{author}{\bibfnamefont{S.~M.} \bibnamefont{Chudinov}},
  \bibnamefont{and} \bibinfo{author}{\bibfnamefont{Y.~G.}
  \bibnamefont{Ponomarev}}, \emph{\bibinfo{title}{Modern Problems in Condensed
  Matter Sciences}}, vol. \bibinfo{volume}{20.1}
  (\bibinfo{publisher}{North-Holland, Amsterdam}, \bibinfo{year}{1988}).

\bibitem[{\citenamefont{Dresselhaus and Dresselhaus}(2002)}]{tunnel2}
\bibinfo{author}{\bibfnamefont{M.~S.} \bibnamefont{Dresselhaus}}
  \bibnamefont{and}
  \bibinfo{author}{\bibfnamefont{G.}~\bibnamefont{Dresselhaus}},
  {``}\bibinfo{title}{Intercalation compounds of graphite},{''}
  \bibinfo{journal}{Adv. Phys.} \textbf{\bibinfo{volume}{51}},
  \bibinfo{pages}{1} (\bibinfo{year}{2002}).

\bibitem[{\citenamefont{Reich et~al.}(2002)\citenamefont{Reich, Maultzsch,
  Thomsen, and Ordej\'on}}]{tunnel3}
\bibinfo{author}{\bibfnamefont{S.}~\bibnamefont{Reich}},
  \bibinfo{author}{\bibfnamefont{J.}~\bibnamefont{Maultzsch}},
  \bibinfo{author}{\bibfnamefont{C.}~\bibnamefont{Thomsen}}, \bibnamefont{and}
  \bibinfo{author}{\bibfnamefont{P.}~\bibnamefont{Ordej\'on}},
  {``}\bibinfo{title}{Tight-binding description of graphene},{''}
  \bibinfo{journal}{Phys. Rev. B} \textbf{\bibinfo{volume}{66}},
  \bibinfo{pages}{035412} (\bibinfo{year}{2002}).

\bibitem[{\citenamefont{Deacon et~al.}(2007)\citenamefont{Deacon, Chuang,
  Nicholas, Novoselov, and Geim}}]{tunnel4}
\bibinfo{author}{\bibfnamefont{R.~S.} \bibnamefont{Deacon}},
  \bibinfo{author}{\bibfnamefont{K.-C.} \bibnamefont{Chuang}},
  \bibinfo{author}{\bibfnamefont{R.~J.} \bibnamefont{Nicholas}},
  \bibinfo{author}{\bibfnamefont{K.~S.} \bibnamefont{Novoselov}},
  \bibnamefont{and} \bibinfo{author}{\bibfnamefont{A.~K.} \bibnamefont{Geim}},
  {``}\bibinfo{title}{Cyclotron resonance study of the electron and hole
  velocity in graphene monolayers},{''} \bibinfo{journal}{Phys. Rev. B}
  \textbf{\bibinfo{volume}{76}}, \bibinfo{pages}{081406(R)}
  (\bibinfo{year}{2007}).

\bibitem[{\citenamefont{Xiang et~al.}(2010)\citenamefont{Xiang, Kan, Wei, Gong,
  and Whangbo}}]{graphane1}
\bibinfo{author}{\bibfnamefont{H.~J.} \bibnamefont{Xiang}},
  \bibinfo{author}{\bibfnamefont{E.~J.} \bibnamefont{Kan}},
  \bibinfo{author}{\bibfnamefont{S.-H.} \bibnamefont{Wei}},
  \bibinfo{author}{\bibfnamefont{X.~G.} \bibnamefont{Gong}}, \bibnamefont{and}
  \bibinfo{author}{\bibfnamefont{M.-H.} \bibnamefont{Whangbo}},
  {``}\bibinfo{title}{Thermodynamically stable single-side hydrogenated
  graphene},{''} \bibinfo{journal}{Phys. Rev. B} \textbf{\bibinfo{volume}{82}},
  \bibinfo{pages}{165425} (\bibinfo{year}{2010}).

\bibitem[{\citenamefont{Pujari et~al.}(2011)\citenamefont{Pujari, Gusarov,
  Brett, and Kovalenko}}]{graphane2}
\bibinfo{author}{\bibfnamefont{B.~S.} \bibnamefont{Pujari}},
  \bibinfo{author}{\bibfnamefont{S.}~\bibnamefont{Gusarov}},
  \bibinfo{author}{\bibfnamefont{M.}~\bibnamefont{Brett}}, \bibnamefont{and}
  \bibinfo{author}{\bibfnamefont{A.}~\bibnamefont{Kovalenko}},
  {``}\bibinfo{title}{Single-side-hydrogenated graphene: Density functional
  theory predictions},{''} \bibinfo{journal}{Phys. Rev. B}
  \textbf{\bibinfo{volume}{84}}, \bibinfo{pages}{041402}
  (\bibinfo{year}{2011}).

\bibitem[{\citenamefont{Openov and Podlivaev}(2012)}]{graphane3}
\bibinfo{author}{\bibfnamefont{L.}~\bibnamefont{Openov}} \bibnamefont{and}
  \bibinfo{author}{\bibfnamefont{A.}~\bibnamefont{Podlivaev}},
  {``}\bibinfo{title}{Insulator band gap in single-side-hydrogenated graphene
  nanoribbons},{''} \bibinfo{journal}{Semiconductors}
  \textbf{\bibinfo{volume}{46}}, \bibinfo{pages}{199} (\bibinfo{year}{2012}).

\bibitem[{\citenamefont{Akzyanov et~al.}(2014)\citenamefont{Akzyanov,
  Sboychakov, Rozhkov, Rakhmanov, and Nori}}]{PrbVOur}
\bibinfo{author}{\bibfnamefont{R.~S.} \bibnamefont{Akzyanov}},
  \bibinfo{author}{\bibfnamefont{A.~O.} \bibnamefont{Sboychakov}},
  \bibinfo{author}{\bibfnamefont{A.~V.} \bibnamefont{Rozhkov}},
  \bibinfo{author}{\bibfnamefont{A.~L.} \bibnamefont{Rakhmanov}},
  \bibnamefont{and} \bibinfo{author}{\bibfnamefont{F.}~\bibnamefont{Nori}},
  {``}\bibinfo{title}{$AA$-stacked bilayer graphene in an applied electric
  field: Tunable antiferromagnetism and coexisting exciton order
  parameter},{''} \bibinfo{journal}{Phys. Rev. B}
  \textbf{\bibinfo{volume}{90}}, \bibinfo{pages}{155415}
  (\bibinfo{year}{2014}).

\bibitem[{\citenamefont{Henriksen and Eisenstein}(2010)}]{Henriksen2010}
\bibinfo{author}{\bibfnamefont{E.~A.} \bibnamefont{Henriksen}}
  \bibnamefont{and} \bibinfo{author}{\bibfnamefont{J.~P.}
  \bibnamefont{Eisenstein}}, {``}\bibinfo{title}{Measurement of the electronic
  compressibility of bilayer graphene},{''} \bibinfo{journal}{Phys. Rev. B}
  \textbf{\bibinfo{volume}{82}}, \bibinfo{pages}{041412}
  (\bibinfo{year}{2010}).

\bibitem[{\citenamefont{Malard et~al.}(2007)\citenamefont{Malard, Nilsson,
  Elias, Brant, Plentz, Alves, Castro~Neto, and Pimenta}}]{Malard2007}
\bibinfo{author}{\bibfnamefont{L.~M.} \bibnamefont{Malard}},
  \bibinfo{author}{\bibfnamefont{J.}~\bibnamefont{Nilsson}},
  \bibinfo{author}{\bibfnamefont{D.~C.} \bibnamefont{Elias}},
  \bibinfo{author}{\bibfnamefont{J.~C.} \bibnamefont{Brant}},
  \bibinfo{author}{\bibfnamefont{F.}~\bibnamefont{Plentz}},
  \bibinfo{author}{\bibfnamefont{E.~S.} \bibnamefont{Alves}},
  \bibinfo{author}{\bibfnamefont{A.~H.} \bibnamefont{Castro~Neto}},
  \bibnamefont{and} \bibinfo{author}{\bibfnamefont{M.~A.}
  \bibnamefont{Pimenta}}, {``}\bibinfo{title}{Probing the electronic structure
  of bilayer graphene by Raman scattering},{''} \bibinfo{journal}{Phys. Rev. B}
  \textbf{\bibinfo{volume}{76}}, \bibinfo{pages}{201401}
  (\bibinfo{year}{2007}).

\bibitem[{\citenamefont{Zhang et~al.}(2008)\citenamefont{Zhang, Li, Basov,
  Fogler, Hao, and Martin}}]{Zhang2008}
\bibinfo{author}{\bibfnamefont{L.~M.} \bibnamefont{Zhang}},
  \bibinfo{author}{\bibfnamefont{Z.~Q.} \bibnamefont{Li}},
  \bibinfo{author}{\bibfnamefont{D.~N.} \bibnamefont{Basov}},
  \bibinfo{author}{\bibfnamefont{M.~M.} \bibnamefont{Fogler}},
  \bibinfo{author}{\bibfnamefont{Z.}~\bibnamefont{Hao}}, \bibnamefont{and}
  \bibinfo{author}{\bibfnamefont{M.~C.} \bibnamefont{Martin}},
  {``}\bibinfo{title}{Determination of the electronic structure of bilayer
  graphene from infrared spectroscopy},{''} \bibinfo{journal}{Phys. Rev. B}
  \textbf{\bibinfo{volume}{78}}, \bibinfo{pages}{235408}
  (\bibinfo{year}{2008}).

\bibitem[{\citenamefont{Li et~al.}(2009)\citenamefont{Li, Henriksen, Jiang,
  Hao, Martin, Kim, Stormer, and Basov}}]{Li2009}
\bibinfo{author}{\bibfnamefont{Z.~Q.} \bibnamefont{Li}},
  \bibinfo{author}{\bibfnamefont{E.~A.} \bibnamefont{Henriksen}},
  \bibinfo{author}{\bibfnamefont{Z.}~\bibnamefont{Jiang}},
  \bibinfo{author}{\bibfnamefont{Z.}~\bibnamefont{Hao}},
  \bibinfo{author}{\bibfnamefont{M.~C.} \bibnamefont{Martin}},
  \bibinfo{author}{\bibfnamefont{P.}~\bibnamefont{Kim}},
  \bibinfo{author}{\bibfnamefont{H.~L.} \bibnamefont{Stormer}},
  \bibnamefont{and} \bibinfo{author}{\bibfnamefont{D.~N.} \bibnamefont{Basov}},
  {``}\bibinfo{title}{Band Structure Asymmetry of Bilayer Graphene Revealed by
  Infrared Spectroscopy},{''} \bibinfo{journal}{Phys. Rev. Lett.}
  \textbf{\bibinfo{volume}{102}}, \bibinfo{pages}{037403}
  (\bibinfo{year}{2009}).

\bibitem[{\citenamefont{McCann and Fal'ko}(2006)}]{mccann_falko2006}
\bibinfo{author}{\bibfnamefont{E.}~\bibnamefont{McCann}} \bibnamefont{and}
  \bibinfo{author}{\bibfnamefont{V.~I.} \bibnamefont{Fal'ko}},
  {``}\bibinfo{title}{Landau-Level Degeneracy and Quantum Hall Effect in a
  Graphite Bilayer},{''} \bibinfo{journal}{Phys. Rev. Lett.}
  \textbf{\bibinfo{volume}{96}}, \bibinfo{pages}{086805}
  (\bibinfo{year}{2006}).

\bibitem[{\citenamefont{Kuzmenko
  et~al.}(2009{\natexlab{a}})\citenamefont{Kuzmenko, Crassee, van~der Marel,
  Blake, and Novoselov}}]{Kuzmenko2009}
\bibinfo{author}{\bibfnamefont{A.~B.} \bibnamefont{Kuzmenko}},
  \bibinfo{author}{\bibfnamefont{I.}~\bibnamefont{Crassee}},
  \bibinfo{author}{\bibfnamefont{D.}~\bibnamefont{van~der Marel}},
  \bibinfo{author}{\bibfnamefont{P.}~\bibnamefont{Blake}}, \bibnamefont{and}
  \bibinfo{author}{\bibfnamefont{K.~S.} \bibnamefont{Novoselov}},
  {``}\bibinfo{title}{Determination of the gate-tunable band gap and
  tight-binding parameters in bilayer graphene using infrared
  spectroscopy},{''} \bibinfo{journal}{Phys. Rev. B}
  \textbf{\bibinfo{volume}{80}}, \bibinfo{pages}{165406}
  (\bibinfo{year}{2009}{\natexlab{a}}).

\bibitem[{\citenamefont{Charlier et~al.}(1991)\citenamefont{Charlier, Gonze,
  and Michenaud}}]{dft_graphIte1991}
\bibinfo{author}{\bibfnamefont{J.-C.} \bibnamefont{Charlier}},
  \bibinfo{author}{\bibfnamefont{X.}~\bibnamefont{Gonze}}, \bibnamefont{and}
  \bibinfo{author}{\bibfnamefont{J.-P.} \bibnamefont{Michenaud}},
  {``}\bibinfo{title}{First-principles study of the electronic properties of
  graphite},{''} \bibinfo{journal}{Phys. Rev. B} \textbf{\bibinfo{volume}{43}},
  \bibinfo{pages}{4579} (\bibinfo{year}{1991}).

\bibitem[{\citenamefont{Mafra et~al.}(2009)\citenamefont{Mafra, Malard, Doorn,
  Htoon, Nilsson, Castro~Neto, and Pimenta}}]{Mafra2009}
\bibinfo{author}{\bibfnamefont{D.~L.} \bibnamefont{Mafra}},
  \bibinfo{author}{\bibfnamefont{L.~M.} \bibnamefont{Malard}},
  \bibinfo{author}{\bibfnamefont{S.~K.} \bibnamefont{Doorn}},
  \bibinfo{author}{\bibfnamefont{H.}~\bibnamefont{Htoon}},
  \bibinfo{author}{\bibfnamefont{J.}~\bibnamefont{Nilsson}},
  \bibinfo{author}{\bibfnamefont{A.~H.} \bibnamefont{Castro~Neto}},
  \bibnamefont{and} \bibinfo{author}{\bibfnamefont{M.~A.}
  \bibnamefont{Pimenta}}, {``}\bibinfo{title}{Observation of the Kohn anomaly
  near the K point of bilayer graphene},{''} \bibinfo{journal}{Phys. Rev. B}
  \textbf{\bibinfo{volume}{80}}, \bibinfo{pages}{241414}
  (\bibinfo{year}{2009}).

\bibitem[{\citenamefont{Kuzmenko
  et~al.}(2009{\natexlab{b}})\citenamefont{Kuzmenko, van Heumen, van~der Marel,
  Lerch, Blake, Novoselov, and Geim}}]{Kuzmenko2009b}
\bibinfo{author}{\bibfnamefont{A.~B.} \bibnamefont{Kuzmenko}},
  \bibinfo{author}{\bibfnamefont{E.}~\bibnamefont{van Heumen}},
  \bibinfo{author}{\bibfnamefont{D.}~\bibnamefont{van~der Marel}},
  \bibinfo{author}{\bibfnamefont{P.}~\bibnamefont{Lerch}},
  \bibinfo{author}{\bibfnamefont{P.}~\bibnamefont{Blake}},
  \bibinfo{author}{\bibfnamefont{K.~S.} \bibnamefont{Novoselov}},
  \bibnamefont{and} \bibinfo{author}{\bibfnamefont{A.~K.} \bibnamefont{Geim}},
  {``}\bibinfo{title}{Infrared spectroscopy of electronic bands in bilayer
  graphene},{''} \bibinfo{journal}{Phys. Rev. B} \textbf{\bibinfo{volume}{79}},
  \bibinfo{pages}{115441} (\bibinfo{year}{2009}{\natexlab{b}}).

\bibitem[{\citenamefont{Zou et~al.}(2011)\citenamefont{Zou, Hong, and
  Zhu}}]{Zou2011}
\bibinfo{author}{\bibfnamefont{K.}~\bibnamefont{Zou}},
  \bibinfo{author}{\bibfnamefont{X.}~\bibnamefont{Hong}}, \bibnamefont{and}
  \bibinfo{author}{\bibfnamefont{J.}~\bibnamefont{Zhu}},
  {``}\bibinfo{title}{Effective mass of electrons and holes in bilayer
  graphene: Electron-hole asymmetry and electron-electron interaction},{''}
  \bibinfo{journal}{Phys. Rev. B} \textbf{\bibinfo{volume}{84}},
  \bibinfo{pages}{085408} (\bibinfo{year}{2011}).

\bibitem[{\citenamefont{McCann}(2006)}]{McCann2006}
\bibinfo{author}{\bibfnamefont{E.}~\bibnamefont{McCann}},
  {``}\bibinfo{title}{Asymmetry gap in the electronic band structure of bilayer
  graphene},{''} \bibinfo{journal}{Phys. Rev. B} \textbf{\bibinfo{volume}{74}},
  \bibinfo{pages}{161403} (\bibinfo{year}{2006}).

\bibitem[{\citenamefont{Ohta et~al.}(2006)\citenamefont{Ohta, Bostwick,
  Seyller, Horn, and Rotenberg}}]{Ohta2006}
\bibinfo{author}{\bibfnamefont{T.}~\bibnamefont{Ohta}},
  \bibinfo{author}{\bibfnamefont{A.}~\bibnamefont{Bostwick}},
  \bibinfo{author}{\bibfnamefont{T.}~\bibnamefont{Seyller}},
  \bibinfo{author}{\bibfnamefont{K.}~\bibnamefont{Horn}}, \bibnamefont{and}
  \bibinfo{author}{\bibfnamefont{E.}~\bibnamefont{Rotenberg}},
  {``}\bibinfo{title}{Controlling the electronic structure of bilayer
  graphene},{''} \bibinfo{journal}{Science} \textbf{\bibinfo{volume}{313}},
  \bibinfo{pages}{951} (\bibinfo{year}{2006}).

\bibitem[{\citenamefont{Castro et~al.}(2007{\natexlab{a}})\citenamefont{Castro,
  Novoselov, Morozov, Peres, Lopes~dos Santos, Nilsson, Guinea, Geim, and
  Castro~Neto}}]{Castro2007}
\bibinfo{author}{\bibfnamefont{E.~V.} \bibnamefont{Castro}},
  \bibinfo{author}{\bibfnamefont{K.~S.} \bibnamefont{Novoselov}},
  \bibinfo{author}{\bibfnamefont{S.~V.} \bibnamefont{Morozov}},
  \bibinfo{author}{\bibfnamefont{N.~M.~R.} \bibnamefont{Peres}},
  \bibinfo{author}{\bibfnamefont{J.~M.~B.} \bibnamefont{Lopes~dos Santos}},
  \bibinfo{author}{\bibfnamefont{J.}~\bibnamefont{Nilsson}},
  \bibinfo{author}{\bibfnamefont{F.}~\bibnamefont{Guinea}},
  \bibinfo{author}{\bibfnamefont{A.~K.} \bibnamefont{Geim}}, \bibnamefont{and}
  \bibinfo{author}{\bibfnamefont{A.~H.} \bibnamefont{Castro~Neto}},
  {``}\bibinfo{title}{Biased bilayer graphene: Semiconductor with a gap tunable
  by the electric field effect},{''} \bibinfo{journal}{Phys. Rev. Lett.}
  \textbf{\bibinfo{volume}{99}}, \bibinfo{pages}{216802}
  (\bibinfo{year}{2007}{\natexlab{a}}).

\bibitem[{\citenamefont{Jing et~al.}(2010)\citenamefont{Jing, Velasco, Kratz,
  Liu, Bao, Bockrath, and Lau}}]{Jing2010}
\bibinfo{author}{\bibfnamefont{L.}~\bibnamefont{Jing}},
  \bibinfo{author}{\bibfnamefont{J.}~\bibnamefont{Velasco~Jr.}},
  \bibinfo{author}{\bibfnamefont{Ph.}~\bibnamefont{Kratz}},
  \bibinfo{author}{\bibfnamefont{G.}~\bibnamefont{Liu}},
  \bibinfo{author}{\bibfnamefont{W.}~\bibnamefont{Bao}},
  \bibinfo{author}{\bibfnamefont{M.}~\bibnamefont{Bockrath}}, \bibnamefont{and}
  \bibinfo{author}{\bibfnamefont{Ch.~N.} \bibnamefont{Lau}},
  {``}\bibinfo{title}{Quantum Transport and Field-Induced Insulating States in
  Bilayer Graphene pnp Junctions},{''} \bibinfo{journal}{Nano Lett.}
  \textbf{\bibinfo{volume}{10}}, \bibinfo{pages}{4000} (\bibinfo{year}{2010}).

\bibitem[{\citenamefont{Hao et~al.}(2016)\citenamefont{Hao, Wang, Liu, Chen,
  Wang, Tan, Nie, Suk, Jiang, Liang et~al.}}]{Hao2016}
\bibinfo{author}{\bibfnamefont{Y.}~\bibnamefont{Hao}},
  \bibinfo{author}{\bibfnamefont{L.}~\bibnamefont{Wang}},
  \bibinfo{author}{\bibfnamefont{Y.}~\bibnamefont{Liu}},
  \bibinfo{author}{\bibfnamefont{H.}~\bibnamefont{Chen}},
  \bibinfo{author}{\bibfnamefont{X.}~\bibnamefont{Wang}},
  \bibinfo{author}{\bibfnamefont{C.}~\bibnamefont{Tan}},
  \bibinfo{author}{\bibfnamefont{S.}~\bibnamefont{Nie}},
  \bibinfo{author}{\bibfnamefont{J.~W.} \bibnamefont{Suk}},
  \bibinfo{author}{\bibfnamefont{T.}~\bibnamefont{Jiang}},
  \bibinfo{author}{\bibfnamefont{T.}~\bibnamefont{Liang}},
  \bibnamefont{et~al.}, {``}\bibinfo{title}{Oxygen-activated growth and bandgap
  tunability of large single-crystal bilayer graphene},{''}
  \bibinfo{journal}{Nat. Nano.} \textbf{\bibinfo{volume}{11}},
  \bibinfo{pages}{426} (\bibinfo{year}{2016}).

\bibitem[{\citenamefont{Nanda and Satpathy}(2009)}]{Nanda2009}
\bibinfo{author}{\bibfnamefont{B.~R.~K.} \bibnamefont{Nanda}} \bibnamefont{and}
  \bibinfo{author}{\bibfnamefont{S.}~\bibnamefont{Satpathy}},
  {``}\bibinfo{title}{Strain and electric field modulation of the electronic
  structure of bilayer graphene},{''} \bibinfo{journal}{Phys. Rev. B}
  \textbf{\bibinfo{volume}{80}}, \bibinfo{pages}{165430}
  (\bibinfo{year}{2009}).

\bibitem[{\citenamefont{Slawinska et~al.}(2010)\citenamefont{Slawinska, Zasada,
  and Klusek}}]{Slawinska2010}
\bibinfo{author}{\bibfnamefont{J.}~\bibnamefont{Slawinska}},
  \bibinfo{author}{\bibfnamefont{I.}~\bibnamefont{Zasada}}, \bibnamefont{and}
  \bibinfo{author}{\bibfnamefont{Z.}~\bibnamefont{Klusek}},
  {``}\bibinfo{title}{Energy gap tuning in graphene on hexagonal boron nitride
  bilayer system},{''} \bibinfo{journal}{Phys. Rev. B}
  \textbf{\bibinfo{volume}{81}}, \bibinfo{pages}{155433}
  (\bibinfo{year}{2010}).

\bibitem[{\citenamefont{Ramasubramaniam
  et~al.}(2011)\citenamefont{Ramasubramaniam, Naveh, and
  Towe}}]{Ramasubramaniam2011}
\bibinfo{author}{\bibfnamefont{A.}~\bibnamefont{Ramasubramaniam}},
  \bibinfo{author}{\bibfnamefont{D.}~\bibnamefont{Naveh}}, \bibnamefont{and}
  \bibinfo{author}{\bibfnamefont{E.}~\bibnamefont{Towe}},
  {``}\bibinfo{title}{Tunable Band Gaps in Bilayer Graphene-BN
  Heterostructures},{''} \bibinfo{journal}{Nano Lett.}
  \textbf{\bibinfo{volume}{11}}, \bibinfo{pages}{1070} (\bibinfo{year}{2011}).

\bibitem[{\citenamefont{Castro et~al.}(2009)\citenamefont{Castro,
  L{\'o}pez-Sancho, and Vozmediano}}]{dos2009}
\bibinfo{author}{\bibfnamefont{E.~V.} \bibnamefont{Castro}},
  \bibinfo{author}{\bibfnamefont{M.~P.} \bibnamefont{L{\'o}pez-Sancho}},
  \bibnamefont{and} \bibinfo{author}{\bibfnamefont{M.~A.~H.}
  \bibnamefont{Vozmediano}}, {``}\bibinfo{title}{Pinning and switching of
  magnetic moments in bilayer graphene},{''} \bibinfo{journal}{New J. Phys.}
  \textbf{\bibinfo{volume}{11}}, \bibinfo{pages}{095017}
  (\bibinfo{year}{2009}).

\bibitem[{\citenamefont{Castro et~al.}(2007{\natexlab{b}})\citenamefont{Castro,
  Peres, and Lopes~dos Santos}}]{Castro2007a}
\bibinfo{author}{\bibfnamefont{E.~V.} \bibnamefont{Castro}},
  \bibinfo{author}{\bibfnamefont{N.~M.~R.} \bibnamefont{Peres}},
  \bibnamefont{and} \bibinfo{author}{\bibfnamefont{J.~M.~B.}
  \bibnamefont{Lopes~dos Santos}}, {``}\bibinfo{title}{Gaped graphene bilayer:
  disorder and magnetic field effects},{''} \bibinfo{journal}{Phys. Status
  Solidi (b)} \textbf{\bibinfo{volume}{244}}, \bibinfo{pages}{2311}
  (\bibinfo{year}{2007}{\natexlab{b}}).

\bibitem[{\citenamefont{Castro et~al.}(2010)\citenamefont{Castro, Novoselov,
  Morozov, Peres, Lopes~dos Santos, Nilsson, Guinea, Geim, and
  Castro~Neto}}]{Castro2010}
\bibinfo{author}{\bibfnamefont{E.~V.} \bibnamefont{Castro}},
  \bibinfo{author}{\bibfnamefont{K.~S.} \bibnamefont{Novoselov}},
  \bibinfo{author}{\bibfnamefont{S.~V.} \bibnamefont{Morozov}},
  \bibinfo{author}{\bibfnamefont{N.~M.~R.} \bibnamefont{Peres}},
  \bibinfo{author}{\bibfnamefont{J.~M.~B.} \bibnamefont{Lopes~dos Santos}},
  \bibinfo{author}{\bibfnamefont{J.}~\bibnamefont{Nilsson}},
  \bibinfo{author}{\bibfnamefont{F.}~\bibnamefont{Guinea}},
  \bibinfo{author}{\bibfnamefont{A.~K.} \bibnamefont{Geim}}, \bibnamefont{and}
  \bibinfo{author}{\bibfnamefont{A.~H.} \bibnamefont{Castro~Neto}},
  {``}\bibinfo{title}{Electronic properties of a biased graphene bilayer},{''}
  \bibinfo{journal}{J. Phys.: Condens. Matter} \textbf{\bibinfo{volume}{22}},
  \bibinfo{pages}{175503} (\bibinfo{year}{2010}).

\bibitem[{\citenamefont{Nemec and Cuniberti}(2007)}]{Nemec2007}
\bibinfo{author}{\bibfnamefont{N.}~\bibnamefont{Nemec}} \bibnamefont{and}
  \bibinfo{author}{\bibfnamefont{G.}~\bibnamefont{Cuniberti}},
  {``}\bibinfo{title}{Hofstadter butterflies of bilayer graphene},{''}
  \bibinfo{journal}{Phys. Rev. B} \textbf{\bibinfo{volume}{75}},
  \bibinfo{pages}{201404} (\bibinfo{year}{2007}).

\bibitem[{\citenamefont{Choi et~al.}(2010)\citenamefont{Choi, Jhi, and
  Son}}]{Choi2010}
\bibinfo{author}{\bibfnamefont{S.-M.} \bibnamefont{Choi}},
  \bibinfo{author}{\bibfnamefont{S.-H.} \bibnamefont{Jhi}}, \bibnamefont{and}
  \bibinfo{author}{\bibfnamefont{Y.-W.} \bibnamefont{Son}},
  {``}\bibinfo{title}{Controlling Energy Gap of Bilayer Graphene by
  Strain},{''} \bibinfo{journal}{Nano Lett.} \textbf{\bibinfo{volume}{10}},
  \bibinfo{pages}{3486} (\bibinfo{year}{2010}).

\bibitem[{\citenamefont{Mucha-Kruczy{\'{n}}ski
  et~al.}(2011)\citenamefont{Mucha-Kruczy{\'{n}}ski, Aleiner, and
  Fal'ko}}]{Mucha-Kruczynski2011}
\bibinfo{author}{\bibfnamefont{M.}~\bibnamefont{Mucha-Kruczy{\'{n}}ski}},
  \bibinfo{author}{\bibfnamefont{I.~L.} \bibnamefont{Aleiner}},
  \bibnamefont{and} \bibinfo{author}{\bibfnamefont{V.~I.}
  \bibnamefont{Fal'ko}}, {``}\bibinfo{title}{Strained bilayer graphene: Band
  structure topology and Landau level spectrum},{''} \bibinfo{journal}{Phys.
  Rev. B} \textbf{\bibinfo{volume}{84}}, \bibinfo{pages}{041404}
  (\bibinfo{year}{2011}).

\bibitem[{\citenamefont{Rycerz et~al.}(2007)\citenamefont{Rycerz, Tworzyd{\l}o,
  and Beenakker}}]{Rycerz}
\bibinfo{author}{\bibfnamefont{A.}~\bibnamefont{Rycerz}},
  \bibinfo{author}{\bibfnamefont{J.}~\bibnamefont{Tworzyd{\l}o}},
  \bibnamefont{and}
  \bibinfo{author}{\bibfnamefont{C.}~\bibnamefont{Beenakker}},
  {``}\bibinfo{title}{Valley filter and valley valve in graphene},{''}
  \bibinfo{journal}{Nat. Phys.} \textbf{\bibinfo{volume}{3}},
  \bibinfo{pages}{172} (\bibinfo{year}{2007}).

\bibitem[{\citenamefont{Shimazaki et~al.}(2015)\citenamefont{Shimazaki,
  Yamamoto, Borzenets, Watanabe, Taniguchi, and Tarucha}}]{Shimazaki2015}
\bibinfo{author}{\bibfnamefont{Y.}~\bibnamefont{Shimazaki}},
  \bibinfo{author}{\bibfnamefont{M.}~\bibnamefont{Yamamoto}},
  \bibinfo{author}{\bibfnamefont{I.~V.} \bibnamefont{Borzenets}},
  \bibinfo{author}{\bibfnamefont{K.}~\bibnamefont{Watanabe}},
  \bibinfo{author}{\bibfnamefont{T.}~\bibnamefont{Taniguchi}},
  \bibnamefont{and} \bibinfo{author}{\bibfnamefont{S.}~\bibnamefont{Tarucha}},
  {``}\bibinfo{title}{Generation and detection of pure valley current by
  electrically induced Berry curvature in bilayer graphene},{''}
  \bibinfo{journal}{Nat. Phys.} \textbf{\bibinfo{volume}{11}},
  \bibinfo{pages}{1032} (\bibinfo{year}{2015}).

\bibitem[{\citenamefont{Sanderson et~al.}(2013)\citenamefont{Sanderson, Ang,
  and Zhang}}]{ChirAaKlein}
\bibinfo{author}{\bibfnamefont{M.}~\bibnamefont{Sanderson}},
  \bibinfo{author}{\bibfnamefont{Y.~S.} \bibnamefont{Ang}}, \bibnamefont{and}
  \bibinfo{author}{\bibfnamefont{C.}~\bibnamefont{Zhang}},
  {``}\bibinfo{title}{Klein tunneling and cone transport in AA-stacked bilayer
  graphene},{''} \bibinfo{journal}{Phys. Rev. B} \textbf{\bibinfo{volume}{88}},
  \bibinfo{pages}{245404} (\bibinfo{year}{2013}).

\bibitem[{\citenamefont{Sboychakov
  et~al.}(2013{\natexlab{a}})\citenamefont{Sboychakov, Rozhkov, Rakhmanov, and
  Nori}}]{PrbOur}
\bibinfo{author}{\bibfnamefont{A.~O.} \bibnamefont{Sboychakov}},
  \bibinfo{author}{\bibfnamefont{A.~V.} \bibnamefont{Rozhkov}},
  \bibinfo{author}{\bibfnamefont{A.~L.} \bibnamefont{Rakhmanov}},
  \bibnamefont{and} \bibinfo{author}{\bibfnamefont{F.}~\bibnamefont{Nori}},
  {``}\bibinfo{title}{Antiferromagnetic states and phase separation in doped
  AA-stacked graphene bilayers},{''} \bibinfo{journal}{Phys. Rev. B}
  \textbf{\bibinfo{volume}{88}}, \bibinfo{pages}{045409}
  (\bibinfo{year}{2013}{\natexlab{a}}).

\bibitem[{\citenamefont{Katsnelson et~al.}(2006)\citenamefont{Katsnelson,
  Novoselov, and Geim}}]{klein_tunn2006}
\bibinfo{author}{\bibfnamefont{M.~I.} \bibnamefont{Katsnelson}},
  \bibinfo{author}{\bibfnamefont{K.~S.} \bibnamefont{Novoselov}},
  \bibnamefont{and} \bibinfo{author}{\bibfnamefont{A.~K.} \bibnamefont{Geim}},
  {``}\bibinfo{title}{Chiral tunnelling and the Klein paradox in graphene},{''}
  \bibinfo{journal}{Nat. Phys.} \textbf{\bibinfo{volume}{2}},
  \bibinfo{pages}{620} (\bibinfo{year}{2006}).

\bibitem[{\citenamefont{Cheianov et~al.}(2007)\citenamefont{Cheianov, Fal'ko,
  and Altshuler}}]{graphene_veselago2007}
\bibinfo{author}{\bibfnamefont{V.~V.} \bibnamefont{Cheianov}},
  \bibinfo{author}{\bibfnamefont{V.}~\bibnamefont{Fal'ko}}, \bibnamefont{and}
  \bibinfo{author}{\bibfnamefont{B.~L.} \bibnamefont{Altshuler}},
  {``}\bibinfo{title}{The Focusing of Electron Flow and a Veselago Lens in
  Graphene p-n Junctions},{''} \bibinfo{journal}{Science}
  \textbf{\bibinfo{volume}{315}}, \bibinfo{pages}{1252} (\bibinfo{year}{2007}).

\bibitem[{\citenamefont{Bliokh et~al.}(2009)\citenamefont{Bliokh, Freilikher,
  Savel'ev, and Nori}}]{bliokhFrei}
\bibinfo{author}{\bibfnamefont{Y.~P.} \bibnamefont{Bliokh}},
  \bibinfo{author}{\bibfnamefont{V.}~\bibnamefont{Freilikher}},
  \bibinfo{author}{\bibfnamefont{S.}~\bibnamefont{Savel'ev}}, \bibnamefont{and}
  \bibinfo{author}{\bibfnamefont{F.}~\bibnamefont{Nori}},
  {``}\bibinfo{title}{Transport and localization in periodic and disordered
  graphene superlattices},{''} \bibinfo{journal}{Phys. Rev. B}
  \textbf{\bibinfo{volume}{79}}, \bibinfo{pages}{075123}
  (\bibinfo{year}{2009}).

\bibitem[{\citenamefont{Yampol'skii et~al.}(2011)\citenamefont{Yampol'skii,
  Apostolov, Maizelis, Levchenko, and Nori}}]{yam_klein1}
\bibinfo{author}{\bibfnamefont{V.~A.} \bibnamefont{Yampol'skii}},
  \bibinfo{author}{\bibfnamefont{S.~S.} \bibnamefont{Apostolov}},
  \bibinfo{author}{\bibfnamefont{Z.~A.} \bibnamefont{Maizelis}},
  \bibinfo{author}{\bibfnamefont{A.}~\bibnamefont{Levchenko}},
  \bibnamefont{and} \bibinfo{author}{\bibfnamefont{F.}~\bibnamefont{Nori}},
  {``}\bibinfo{title}{Voltage-driven quantum oscillations of conductance in
  graphene},{''} \bibinfo{journal}{EPL} \textbf{\bibinfo{volume}{96}},
  \bibinfo{pages}{67009} (\bibinfo{year}{2011}).

\bibitem[{\citenamefont{Yampol'skii et~al.}(2008)\citenamefont{Yampol'skii,
  Savel'ev, and Nori}}]{yam_klein2}
\bibinfo{author}{\bibfnamefont{V.~A.} \bibnamefont{Yampol'skii}},
  \bibinfo{author}{\bibfnamefont{S.}~\bibnamefont{Savel'ev}}, \bibnamefont{and}
  \bibinfo{author}{\bibfnamefont{F.}~\bibnamefont{Nori}},
  {``}\bibinfo{title}{Voltage-driven quantum oscillations in graphene},{''}
  \bibinfo{journal}{New J. Phys.} \textbf{\bibinfo{volume}{10}},
  \bibinfo{pages}{053024} (\bibinfo{year}{2008}).

\bibitem[{\citenamefont{Tudorovskiy et~al.}(2012)\citenamefont{Tudorovskiy,
  Reijnders, and Katsnelson}}]{chiral_tunn_tudorovskiy2012}
\bibinfo{author}{\bibfnamefont{T.}~\bibnamefont{Tudorovskiy}},
  \bibinfo{author}{\bibfnamefont{K.~J.~A.} \bibnamefont{Reijnders}},
  \bibnamefont{and} \bibinfo{author}{\bibfnamefont{M.~I.}
  \bibnamefont{Katsnelson}}, {``}\bibinfo{title}{Chiral tunneling in
  single-layer and bilayer graphene},{''} \bibinfo{journal}{Phys. Scripta}
  \textbf{\bibinfo{volume}{2012}}, \bibinfo{pages}{014010}
  (\bibinfo{year}{2012}).

\bibitem[{\citenamefont{Fogler et~al.}(2008)\citenamefont{Fogler, Novikov,
  Glazman, and Shklovskii}}]{fogler_pn_disorder}
\bibinfo{author}{\bibfnamefont{M.~M.} \bibnamefont{Fogler}},
  \bibinfo{author}{\bibfnamefont{D.~S.} \bibnamefont{Novikov}},
  \bibinfo{author}{\bibfnamefont{L.~I.} \bibnamefont{Glazman}},
  \bibnamefont{and} \bibinfo{author}{\bibfnamefont{B.~I.}
  \bibnamefont{Shklovskii}}, {``}\bibinfo{title}{Effect of disorder on a
  graphene p-n junction},{''} \bibinfo{journal}{Phys. Rev. B}
  \textbf{\bibinfo{volume}{77}}, \bibinfo{pages}{075420}
  (\bibinfo{year}{2008}).

\bibitem[{\citenamefont{Wang and Jin}(2013{\natexlab{a}})}]{DWang}
\bibinfo{author}{\bibfnamefont{D.}~\bibnamefont{Wang}} \bibnamefont{and}
  \bibinfo{author}{\bibfnamefont{G.}~\bibnamefont{Jin}},
  {``}\bibinfo{title}{Tunneling magnetoresistance tuned by a vertical electric
  field in an AA-stacked graphene bilayer with double magnetic barriers},{''}
  \bibinfo{journal}{J. Appl. Phys.} \textbf{\bibinfo{volume}{114}},
  \bibinfo{pages}{233701} (\bibinfo{year}{2013}{\natexlab{a}}).

\bibitem[{\citenamefont{Novoselov et~al.}(2007)\citenamefont{Novoselov, Jiang,
  Zhang, Morozov, Stormer, Zeitler, Maan, Boebinger, Kim, and Geim}}]{NovoQHE}
\bibinfo{author}{\bibfnamefont{K.~S.} \bibnamefont{Novoselov}},
  \bibinfo{author}{\bibfnamefont{Z.}~\bibnamefont{Jiang}},
  \bibinfo{author}{\bibfnamefont{Y.}~\bibnamefont{Zhang}},
  \bibinfo{author}{\bibfnamefont{S.~V.} \bibnamefont{Morozov}},
  \bibinfo{author}{\bibfnamefont{H.~L.} \bibnamefont{Stormer}},
  \bibinfo{author}{\bibfnamefont{U.}~\bibnamefont{Zeitler}},
  \bibinfo{author}{\bibfnamefont{J.~C.} \bibnamefont{Maan}},
  \bibinfo{author}{\bibfnamefont{G.~S.} \bibnamefont{Boebinger}},
  \bibinfo{author}{\bibfnamefont{P.}~\bibnamefont{Kim}}, \bibnamefont{and}
  \bibinfo{author}{\bibfnamefont{A.~K.} \bibnamefont{Geim}},
  {``}\bibinfo{title}{Room-Temperature Quantum Hall Effect in Graphene},{''}
  \bibinfo{journal}{Science} \textbf{\bibinfo{volume}{315}},
  \bibinfo{pages}{1379} (\bibinfo{year}{2007}).

\bibitem[{\citenamefont{MacDonald}(1989)}]{MacDonald}
\bibinfo{editor}{\bibfnamefont{A.~H.} \bibnamefont{MacDonald}}, ed.,
  \emph{\bibinfo{title}{Quantum Hall Effect: A Perspective}}
  (\bibinfo{publisher}{Kluwer, Boston MA}, \bibinfo{year}{1989}).

\bibitem[{\citenamefont{Zhang et~al.}(2005)\citenamefont{Zhang, Tan, Stormer,
  and Kim}}]{Dubonos2}
\bibinfo{author}{\bibfnamefont{Y.}~\bibnamefont{Zhang}},
  \bibinfo{author}{\bibfnamefont{Y.-W.} \bibnamefont{Tan}},
  \bibinfo{author}{\bibfnamefont{H.~L.} \bibnamefont{Stormer}},
  \bibnamefont{and} \bibinfo{author}{\bibfnamefont{P.}~\bibnamefont{Kim}},
  {``}\bibinfo{title}{Experimental observation of the quantum Hall effect and
  Berry's phase in graphene},{''} \bibinfo{journal}{Nature}
  \textbf{\bibinfo{volume}{438}}, \bibinfo{pages}{201} (\bibinfo{year}{2005}).

\bibitem[{\citenamefont{Hsu and Guo}(2010)}]{QE_AA}
\bibinfo{author}{\bibfnamefont{Y.-F.} \bibnamefont{Hsu}} \bibnamefont{and}
  \bibinfo{author}{\bibfnamefont{G.-Y.} \bibnamefont{Guo}},
  {``}\bibinfo{title}{Anomalous integer quantum Hall effect in
  {$\mathit{A}\mathit{A}$}-stacked bilayer graphene},{''}
  \bibinfo{journal}{Phys. Rev. B} \textbf{\bibinfo{volume}{82}},
  \bibinfo{pages}{165404} (\bibinfo{year}{2010}).

\bibitem[{\citenamefont{Ho et~al.}(2010{\natexlab{a}})\citenamefont{Ho, Wu,
  Chen, Chiu, and Lin}}]{Opt_LLAA}
\bibinfo{author}{\bibfnamefont{Y.-H.} \bibnamefont{Ho}},
  \bibinfo{author}{\bibfnamefont{J.-Y.} \bibnamefont{Wu}},
  \bibinfo{author}{\bibfnamefont{R.-B.} \bibnamefont{Chen}},
  \bibinfo{author}{\bibfnamefont{Y.-H.} \bibnamefont{Chiu}}, \bibnamefont{and}
  \bibinfo{author}{\bibfnamefont{M.-F.} \bibnamefont{Lin}},
  {``}\bibinfo{title}{Optical transitions between Landau levels: AA-stacked
  bilayer graphene},{''} \bibinfo{journal}{Appl. Phys. Lett.}
  \textbf{\bibinfo{volume}{97}}, \bibinfo{eid}{101905}
  (\bibinfo{year}{2010}{\natexlab{a}}).

\bibitem[{\citenamefont{Wang}(2011)}]{LL_QE_AA}
\bibinfo{author}{\bibfnamefont{D.}~\bibnamefont{Wang}},
  {``}\bibinfo{title}{Electric- and magnetic-field-tuned Landau levels and Hall
  conductivity in AA-stacked bilayer graphene},{''} \bibinfo{journal}{Phys.
  Lett. A} \textbf{\bibinfo{volume}{375}}, \bibinfo{pages}{4070 }
  (\bibinfo{year}{2011}).

\bibitem[{\citenamefont{Tsai et~al.}(2012)\citenamefont{Tsai, Chiu, Ho, and
  Lin}}]{LL_Gate_AA}
\bibinfo{author}{\bibfnamefont{S.-J.} \bibnamefont{Tsai}},
  \bibinfo{author}{\bibfnamefont{Y.-H.} \bibnamefont{Chiu}},
  \bibinfo{author}{\bibfnamefont{Y.-H.} \bibnamefont{Ho}}, \bibnamefont{and}
  \bibinfo{author}{\bibfnamefont{M.-F.} \bibnamefont{Lin}},
  {``}\bibinfo{title}{Gate-voltage-dependent Landau levels in AA-stacked
  bilayer graphene},{''} \bibinfo{journal}{Chem. Phys. Lett.}
  \textbf{\bibinfo{volume}{550}}, \bibinfo{pages}{104 } (\bibinfo{year}{2012}).

\bibitem[{\citenamefont{Wang and Jin}(2013{\natexlab{b}})}]{LL_NonUnF_AA}
\bibinfo{author}{\bibfnamefont{D.}~\bibnamefont{Wang}} \bibnamefont{and}
  \bibinfo{author}{\bibfnamefont{G.}~\bibnamefont{Jin}},
  {``}\bibinfo{title}{Effect of a nonuniform magnetic field on the Landau
  states in a biased AA-stacked graphene bilayer},{''} \bibinfo{journal}{Phys.
  Lett. A} \textbf{\bibinfo{volume}{377}}, \bibinfo{pages}{2901 }
  (\bibinfo{year}{2013}{\natexlab{b}}).

\bibitem[{\citenamefont{Mahan}(1990)}]{Mahan}
\bibinfo{author}{\bibfnamefont{G.~D.} \bibnamefont{Mahan}},
  \emph{\bibinfo{title}{Many Particle Physics}} (\bibinfo{publisher}{Plenum,
  New York}, \bibinfo{year}{1990}).

\bibitem[{\citenamefont{Pereira
  et~al.}(2007{\natexlab{a}})\citenamefont{Pereira, Peeters, and
  Vasilopoulos}}]{Pereira2007}
\bibinfo{author}{\bibfnamefont{J.}~\bibnamefont{Pereira},
  \bibfnamefont{J.~Milton}}, \bibinfo{author}{\bibfnamefont{F.~M.}
  \bibnamefont{Peeters}}, \bibnamefont{and}
  \bibinfo{author}{\bibfnamefont{P.}~\bibnamefont{Vasilopoulos}},
  {``}\bibinfo{title}{Landau levels and oscillator strength in a biased bilayer
  of graphene},{''} \bibinfo{journal}{Phys. Rev. B}
  \textbf{\bibinfo{volume}{76}}, \bibinfo{pages}{115419}
  (\bibinfo{year}{2007}{\natexlab{a}}).

\bibitem[{\citenamefont{Lai et~al.}(2008)\citenamefont{Lai, Ho, Chang, and
  Lin}}]{Lai}
\bibinfo{author}{\bibfnamefont{Y.~H.} \bibnamefont{Lai}},
  \bibinfo{author}{\bibfnamefont{J.~H.} \bibnamefont{Ho}},
  \bibinfo{author}{\bibfnamefont{C.~P.} \bibnamefont{Chang}}, \bibnamefont{and}
  \bibinfo{author}{\bibfnamefont{M.~F.} \bibnamefont{Lin}},
  {``}\bibinfo{title}{Magnetoelectronic properties of bilayer Bernal
  graphene},{''} \bibinfo{journal}{Phys. Rev. B} \textbf{\bibinfo{volume}{77}},
  \bibinfo{pages}{085426} (\bibinfo{year}{2008}).

\bibitem[{\citenamefont{Partoens and Peeters}(2006)}]{Partoens}
\bibinfo{author}{\bibfnamefont{B.}~\bibnamefont{Partoens}} \bibnamefont{and}
  \bibinfo{author}{\bibfnamefont{F.~M.} \bibnamefont{Peeters}},
  {``}\bibinfo{title}{From graphene to graphite: Electronic structure around
  the $K$ point},{''} \bibinfo{journal}{Phys. Rev. B}
  \textbf{\bibinfo{volume}{74}}, \bibinfo{pages}{075404}
  (\bibinfo{year}{2006}).

\bibitem[{\citenamefont{Gr{\"u}neis et~al.}(2008)\citenamefont{Gr{\"u}neis,
  Attaccalite, Wirtz, Shiozawa, Saito, Pichler, and Rubio}}]{Wirtz}
\bibinfo{author}{\bibfnamefont{A.}~\bibnamefont{Gr{\"u}neis}},
  \bibinfo{author}{\bibfnamefont{C.}~\bibnamefont{Attaccalite}},
  \bibinfo{author}{\bibfnamefont{L.}~\bibnamefont{Wirtz}},
  \bibinfo{author}{\bibfnamefont{H.}~\bibnamefont{Shiozawa}},
  \bibinfo{author}{\bibfnamefont{R.}~\bibnamefont{Saito}},
  \bibinfo{author}{\bibfnamefont{T.}~\bibnamefont{Pichler}}, \bibnamefont{and}
  \bibinfo{author}{\bibfnamefont{A.}~\bibnamefont{Rubio}},
  {``}\bibinfo{title}{Tight-binding description of the quasiparticle dispersion
  of graphite and few-layer graphene},{''} \bibinfo{journal}{Phys. Rev. B}
  \textbf{\bibinfo{volume}{78}}, \bibinfo{pages}{205425}
  (\bibinfo{year}{2008}).

\bibitem[{\citenamefont{Zhang et~al.}(2011)\citenamefont{Zhang, Fogler, and
  Arovas}}]{Arovas}
\bibinfo{author}{\bibfnamefont{L.~M.} \bibnamefont{Zhang}},
  \bibinfo{author}{\bibfnamefont{M.~M.} \bibnamefont{Fogler}},
  \bibnamefont{and} \bibinfo{author}{\bibfnamefont{D.~P.}
  \bibnamefont{Arovas}}, {``}\bibinfo{title}{Magnetoelectric coupling, Berry
  phase, and Landau level dispersion in a biased bilayer graphene},{''}
  \bibinfo{journal}{Phys. Rev. B} \textbf{\bibinfo{volume}{84}},
  \bibinfo{pages}{075451} (\bibinfo{year}{2011}).

\bibitem[{\citenamefont{Ozerin and Falkovsky}(2012)}]{Ozerin}
\bibinfo{author}{\bibfnamefont{A.~Y.} \bibnamefont{Ozerin}} \bibnamefont{and}
  \bibinfo{author}{\bibfnamefont{L.~A.} \bibnamefont{Falkovsky}},
  {``}\bibinfo{title}{Berry phase, semiclassical quantization, and Landau
  levels},{''} \bibinfo{journal}{Phys. Rev. B} \textbf{\bibinfo{volume}{85}},
  \bibinfo{pages}{205143} (\bibinfo{year}{2012}).

\bibitem[{\citenamefont{Falkovsky}(2013)}]{Falkovsky}
\bibinfo{author}{\bibfnamefont{L.~A.} \bibnamefont{Falkovsky}},
  {``}\bibinfo{title}{Magneto-optics of monolayer and bilayer graphene},{''}
  \bibinfo{journal}{JETP Lett.} \textbf{\bibinfo{volume}{97}},
  \bibinfo{pages}{429} (\bibinfo{year}{2013}).

\bibitem[{\citenamefont{Li and Andrei}(2007)}]{andrei}
\bibinfo{author}{\bibfnamefont{G.}~\bibnamefont{Li}} \bibnamefont{and}
  \bibinfo{author}{\bibfnamefont{E.~Y.} \bibnamefont{Andrei}},
  {``}\bibinfo{title}{Observation of Landau levels of Dirac fermions in
  graphite},{''} \bibinfo{journal}{Nat. Phys.} \textbf{\bibinfo{volume}{3}},
  \bibinfo{pages}{623} (\bibinfo{year}{2007}).

\bibitem[{\citenamefont{Chuang et~al.}(2009)\citenamefont{Chuang, Baker, and
  Nicholas}}]{Chuang}
\bibinfo{author}{\bibfnamefont{K.-C.} \bibnamefont{Chuang}},
  \bibinfo{author}{\bibfnamefont{A.~M.~R.} \bibnamefont{Baker}},
  \bibnamefont{and} \bibinfo{author}{\bibfnamefont{R.~J.}
  \bibnamefont{Nicholas}}, {``}\bibinfo{title}{Magnetoabsorption study of
  Landau levels in graphite},{''} \bibinfo{journal}{Phys. Rev. B}
  \textbf{\bibinfo{volume}{80}}, \bibinfo{pages}{161410}
  (\bibinfo{year}{2009}).

\bibitem[{\citenamefont{Morimoto et~al.}(2012)\citenamefont{Morimoto, Koshino,
  and Aoki}}]{Aoki}
\bibinfo{author}{\bibfnamefont{T.}~\bibnamefont{Morimoto}},
  \bibinfo{author}{\bibfnamefont{M.}~\bibnamefont{Koshino}}, \bibnamefont{and}
  \bibinfo{author}{\bibfnamefont{H.}~\bibnamefont{Aoki}},
  {``}\bibinfo{title}{Faraday rotation in bilayer and trilayer graphene in the
  quantum Hall regime},{''} \bibinfo{journal}{Phys. Rev. B}
  \textbf{\bibinfo{volume}{86}}, \bibinfo{pages}{155426}
  (\bibinfo{year}{2012}).

\bibitem[{\citenamefont{Falkovsky}(2011)}]{Falkovsky1}
\bibinfo{author}{\bibfnamefont{L.~A.} \bibnamefont{Falkovsky}},
  {``}\bibinfo{title}{Quantum magneto-optics of graphite with trigonal
  warping},{''} \bibinfo{journal}{Phys. Rev. B} \textbf{\bibinfo{volume}{84}},
  \bibinfo{pages}{115414} (\bibinfo{year}{2011}).

\bibitem[{\citenamefont{Berciaud et~al.}(2014)\citenamefont{Berciaud, Potemski,
  and Faugeras}}]{Berciaud}
\bibinfo{author}{\bibfnamefont{S.}~\bibnamefont{Berciaud}},
  \bibinfo{author}{\bibfnamefont{M.}~\bibnamefont{Potemski}}, \bibnamefont{and}
  \bibinfo{author}{\bibfnamefont{C.}~\bibnamefont{Faugeras}},
  {``}\bibinfo{title}{Probing Electronic Excitations in Mono- to Pentalayer
  Graphene by Micro Magneto-Raman Spectroscopy},{''} \bibinfo{journal}{Nano
  Lett.} \textbf{\bibinfo{volume}{14}}, \bibinfo{pages}{4548}
  (\bibinfo{year}{2014}).

\bibitem[{\citenamefont{Wang and Jin}(2010)}]{Wang_Jin}
\bibinfo{author}{\bibfnamefont{D.}~\bibnamefont{Wang}} \bibnamefont{and}
  \bibinfo{author}{\bibfnamefont{G.}~\bibnamefont{Jin}},
  {``}\bibinfo{title}{Combined effect of magnetic and electric fields on Landau
  level spectrum and magneto-optical absorption in bilayer graphene},{''}
  \bibinfo{journal}{Europhys. Lett.} \textbf{\bibinfo{volume}{92}},
  \bibinfo{pages}{57008} (\bibinfo{year}{2010}).

\bibitem[{\citenamefont{Henriksen et~al.}(2008)\citenamefont{Henriksen, Jiang,
  Tung, Schwartz, Takita, Wang, Kim, and Stormer}}]{Henriksen}
\bibinfo{author}{\bibfnamefont{E.~A.} \bibnamefont{Henriksen}},
  \bibinfo{author}{\bibfnamefont{Z.}~\bibnamefont{Jiang}},
  \bibinfo{author}{\bibfnamefont{L.-C.} \bibnamefont{Tung}},
  \bibinfo{author}{\bibfnamefont{M.~E.} \bibnamefont{Schwartz}},
  \bibinfo{author}{\bibfnamefont{M.}~\bibnamefont{Takita}},
  \bibinfo{author}{\bibfnamefont{Y.-J.} \bibnamefont{Wang}},
  \bibinfo{author}{\bibfnamefont{P.}~\bibnamefont{Kim}}, \bibnamefont{and}
  \bibinfo{author}{\bibfnamefont{H.~L.} \bibnamefont{Stormer}},
  {``}\bibinfo{title}{Cyclotron resonance in bilayer graphene},{''}
  \bibinfo{journal}{Phys. Rev. Lett.} \textbf{\bibinfo{volume}{100}},
  \bibinfo{pages}{087403} (\bibinfo{year}{2008}).

\bibitem[{\citenamefont{Lee et~al.}(2014)\citenamefont{Lee, Fallahazad, Xue,
  Dillen, Kim, Taniguchi, Watanabe, and Tutuc}}]{Dillen}
\bibinfo{author}{\bibfnamefont{K.}~\bibnamefont{Lee}},
  \bibinfo{author}{\bibfnamefont{B.}~\bibnamefont{Fallahazad}},
  \bibinfo{author}{\bibfnamefont{J.}~\bibnamefont{Xue}},
  \bibinfo{author}{\bibfnamefont{D.~C.} \bibnamefont{Dillen}},
  \bibinfo{author}{\bibfnamefont{K.}~\bibnamefont{Kim}},
  \bibinfo{author}{\bibfnamefont{T.}~\bibnamefont{Taniguchi}},
  \bibinfo{author}{\bibfnamefont{K.}~\bibnamefont{Watanabe}}, \bibnamefont{and}
  \bibinfo{author}{\bibfnamefont{E.}~\bibnamefont{Tutuc}},
  {``}\bibinfo{title}{Chemical potential and quantum Hall ferromagnetism in
  bilayer graphene},{''} \bibinfo{journal}{Science}
  \textbf{\bibinfo{volume}{345}}, \bibinfo{pages}{58} (\bibinfo{year}{2014}).

\bibitem[{\citenamefont{Velasco et~al.}(2014)\citenamefont{Velasco, Lee, Zhao,
  Jing, Kratz, Bockrath, and Lau}}]{Kratz}
\bibinfo{author}{\bibfnamefont{J.}~\bibnamefont{Velasco}},
  \bibinfo{author}{\bibfnamefont{Y.}~\bibnamefont{Lee}},
  \bibinfo{author}{\bibfnamefont{Z.}~\bibnamefont{Zhao}},
  \bibinfo{author}{\bibfnamefont{L.}~\bibnamefont{Jing}},
  \bibinfo{author}{\bibfnamefont{P.}~\bibnamefont{Kratz}},
  \bibinfo{author}{\bibfnamefont{M.}~\bibnamefont{Bockrath}}, \bibnamefont{and}
  \bibinfo{author}{\bibfnamefont{C.~N.} \bibnamefont{Lau}},
  {``}\bibinfo{title}{Transport Measurement of Landau Level Gaps in Bilayer
  Graphene with Layer Polarization Control},{''} \bibinfo{journal}{Nano Lett.}
  \textbf{\bibinfo{volume}{14}}, \bibinfo{pages}{1324} (\bibinfo{year}{2014}).

\bibitem[{\citenamefont{Hyun et~al.}(2012)\citenamefont{Hyun, Kim, Sochichiu,
  and Choi}}]{Hyun}
\bibinfo{author}{\bibfnamefont{Y.-H.} \bibnamefont{Hyun}},
  \bibinfo{author}{\bibfnamefont{Y.}~\bibnamefont{Kim}},
  \bibinfo{author}{\bibfnamefont{C.}~\bibnamefont{Sochichiu}},
  \bibnamefont{and} \bibinfo{author}{\bibfnamefont{M.-Y.} \bibnamefont{Choi}},
  {``}\bibinfo{title}{Landau level spectrum for bilayer graphene in a tilted
  magnetic field},{''} \bibinfo{journal}{J. Phys.: Condens. Matter}
  \textbf{\bibinfo{volume}{24}}, \bibinfo{pages}{045501}
  (\bibinfo{year}{2012}).

\bibitem[{\citenamefont{Kawarabayashi et~al.}(2012)\citenamefont{Kawarabayashi,
  Hatsugai, and Aoki}}]{Kawarabayashi}
\bibinfo{author}{\bibfnamefont{T.}~\bibnamefont{Kawarabayashi}},
  \bibinfo{author}{\bibfnamefont{Y.}~\bibnamefont{Hatsugai}}, \bibnamefont{and}
  \bibinfo{author}{\bibfnamefont{H.}~\bibnamefont{Aoki}},
  {``}\bibinfo{title}{Topologically protected Landau levels in bilayer graphene
  in finite electric fields},{''} \bibinfo{journal}{Phys. Rev. B}
  \textbf{\bibinfo{volume}{85}}, \bibinfo{pages}{165410}
  (\bibinfo{year}{2012}).

\bibitem[{\citenamefont{Koshino and Ando}(2008)}]{Koshino}
\bibinfo{author}{\bibfnamefont{M.}~\bibnamefont{Koshino}} \bibnamefont{and}
  \bibinfo{author}{\bibfnamefont{T.}~\bibnamefont{Ando}},
  {``}\bibinfo{title}{Magneto-optical properties of multilayer graphene},{''}
  \bibinfo{journal}{Phys. Rev. B} \textbf{\bibinfo{volume}{77}},
  \bibinfo{pages}{115313} (\bibinfo{year}{2008}).

\bibitem[{\citenamefont{Ho et~al.}(2010{\natexlab{b}})\citenamefont{Ho, Chiu,
  Lin, Chang, and Lin}}]{ChangLin}
\bibinfo{author}{\bibfnamefont{Y.-H.} \bibnamefont{Ho}},
  \bibinfo{author}{\bibfnamefont{Y.-H.} \bibnamefont{Chiu}},
  \bibinfo{author}{\bibfnamefont{D.-H.} \bibnamefont{Lin}},
  \bibinfo{author}{\bibfnamefont{C.-P.} \bibnamefont{Chang}}, \bibnamefont{and}
  \bibinfo{author}{\bibfnamefont{M.-F.} \bibnamefont{Lin}},
  {``}\bibinfo{title}{Magneto-optical Selection Rules in Bilayer Bernal
  Graphene},{''} \bibinfo{journal}{ACS Nano} \textbf{\bibinfo{volume}{4}},
  \bibinfo{pages}{1465} (\bibinfo{year}{2010}{\natexlab{b}}).

\bibitem[{\citenamefont{Novoselov et~al.}(2006)\citenamefont{Novoselov, McCann,
  Morozov, Fal'ko, Katsnelson, Zeitler, Jiang, Schedin, and
  Geim}}]{Novoselov2006}
\bibinfo{author}{\bibfnamefont{K.~S.} \bibnamefont{Novoselov}},
  \bibinfo{author}{\bibfnamefont{E.}~\bibnamefont{McCann}},
  \bibinfo{author}{\bibfnamefont{S.~V.} \bibnamefont{Morozov}},
  \bibinfo{author}{\bibfnamefont{V.~I.} \bibnamefont{Fal'ko}},
  \bibinfo{author}{\bibfnamefont{M.~I.} \bibnamefont{Katsnelson}},
  \bibinfo{author}{\bibfnamefont{U.}~\bibnamefont{Zeitler}},
  \bibinfo{author}{\bibfnamefont{D.}~\bibnamefont{Jiang}},
  \bibinfo{author}{\bibfnamefont{F.}~\bibnamefont{Schedin}}, \bibnamefont{and}
  \bibinfo{author}{\bibfnamefont{A.~K.} \bibnamefont{Geim}},
  {``}\bibinfo{title}{Unconventional quantum Hall effect and Berry's phase of
  2{$\pi$} in bilayer graphene},{''} \bibinfo{journal}{Nat. Phys.}
  \textbf{\bibinfo{volume}{2}}, \bibinfo{pages}{177} (\bibinfo{year}{2006}).

\bibitem[{\citenamefont{Nakamura et~al.}(2008)\citenamefont{Nakamura, Hirasawa,
  and Imura}}]{Nakamura2008}
\bibinfo{author}{\bibfnamefont{M.}~\bibnamefont{Nakamura}},
  \bibinfo{author}{\bibfnamefont{L.}~\bibnamefont{Hirasawa}}, \bibnamefont{and}
  \bibinfo{author}{\bibfnamefont{K.-I.} \bibnamefont{Imura}},
  {``}\bibinfo{title}{Quantum Hall effect in bilayer and multilayer graphene
  with finite Fermi energy},{''} \bibinfo{journal}{Phys. Rev. B}
  \textbf{\bibinfo{volume}{78}}, \bibinfo{pages}{033403}
  (\bibinfo{year}{2008}).

\bibitem[{\citenamefont{Du et~al.}(2009)\citenamefont{Du, Skachko, Duerr,
  Luican, and Andrei}}]{Du}
\bibinfo{author}{\bibfnamefont{X.}~\bibnamefont{Du}},
  \bibinfo{author}{\bibfnamefont{I.}~\bibnamefont{Skachko}},
  \bibinfo{author}{\bibfnamefont{F.}~\bibnamefont{Duerr}},
  \bibinfo{author}{\bibfnamefont{A.}~\bibnamefont{Luican}}, \bibnamefont{and}
  \bibinfo{author}{\bibfnamefont{E.~Y.} \bibnamefont{Andrei}},
  {``}\bibinfo{title}{Fractional quantum Hall effect and insulating phase of
  Dirac electrons in graphene},{''} \bibinfo{journal}{Nature}
  \textbf{\bibinfo{volume}{462}}, \bibinfo{pages}{192} (\bibinfo{year}{2009}).

\bibitem[{\citenamefont{Bolotin et~al.}(2009)\citenamefont{Bolotin, Ghahari,
  Shulman, Stormer, and Kim}}]{Bolotin}
\bibinfo{author}{\bibfnamefont{K.~I.} \bibnamefont{Bolotin}},
  \bibinfo{author}{\bibfnamefont{F.}~\bibnamefont{Ghahari}},
  \bibinfo{author}{\bibfnamefont{M.~D.} \bibnamefont{Shulman}},
  \bibinfo{author}{\bibfnamefont{H.~L.} \bibnamefont{Stormer}},
  \bibnamefont{and} \bibinfo{author}{\bibfnamefont{P.}~\bibnamefont{Kim}},
  {``}\bibinfo{title}{Observation of the fractional quantum Hall effect in
  graphene},{''} \bibinfo{journal}{Nature} \textbf{\bibinfo{volume}{462}},
  \bibinfo{pages}{196} (\bibinfo{year}{2009}).

\bibitem[{\citenamefont{Dean et~al.}(2011)\citenamefont{Dean, Young,
  Cadden-Zimansky, Wang, Ren, Watanabe, Taniguchi, Kim, Hone, and
  Shepard}}]{Dean}
\bibinfo{author}{\bibfnamefont{C.~R.} \bibnamefont{Dean}},
  \bibinfo{author}{\bibfnamefont{A.~F.} \bibnamefont{Young}},
  \bibinfo{author}{\bibfnamefont{P.}~\bibnamefont{Cadden-Zimansky}},
  \bibinfo{author}{\bibfnamefont{L.}~\bibnamefont{Wang}},
  \bibinfo{author}{\bibfnamefont{H.}~\bibnamefont{Ren}},
  \bibinfo{author}{\bibfnamefont{K.}~\bibnamefont{Watanabe}},
  \bibinfo{author}{\bibfnamefont{T.}~\bibnamefont{Taniguchi}},
  \bibinfo{author}{\bibfnamefont{P.}~\bibnamefont{Kim}},
  \bibinfo{author}{\bibfnamefont{J.}~\bibnamefont{Hone}}, \bibnamefont{and}
  \bibinfo{author}{\bibfnamefont{K.~L.} \bibnamefont{Shepard}},
  {``}\bibinfo{title}{Multicomponent fractional quantum Hall effect in
  graphene},{''} \bibinfo{journal}{Nat. Phys.} \textbf{\bibinfo{volume}{7}},
  \bibinfo{pages}{693} (\bibinfo{year}{2011}).

\bibitem[{\citenamefont{Bao et~al.}(2010)\citenamefont{Bao, Zhao, Zhang, Liu,
  Kratz, Jing, Velasco, Smirnov, and Lau}}]{Bao2010}
\bibinfo{author}{\bibfnamefont{W.}~\bibnamefont{Bao}},
  \bibinfo{author}{\bibfnamefont{Z.}~\bibnamefont{Zhao}},
  \bibinfo{author}{\bibfnamefont{H.}~\bibnamefont{Zhang}},
  \bibinfo{author}{\bibfnamefont{G.}~\bibnamefont{Liu}},
  \bibinfo{author}{\bibfnamefont{P.}~\bibnamefont{Kratz}},
  \bibinfo{author}{\bibfnamefont{L.}~\bibnamefont{Jing}},
  \bibinfo{author}{\bibfnamefont{J.}~\bibnamefont{Velasco}},
  \bibinfo{author}{\bibfnamefont{D.}~\bibnamefont{Smirnov}}, \bibnamefont{and}
  \bibinfo{author}{\bibfnamefont{C.~N.} \bibnamefont{Lau}},
  {``}\bibinfo{title}{Magnetoconductance Oscillations and Evidence for
  Fractional Quantum Hall States in Suspended Bilayer and Trilayer
  Graphene},{''} \bibinfo{journal}{Phys. Rev. Lett.}
  \textbf{\bibinfo{volume}{105}}, \bibinfo{pages}{246601}
  (\bibinfo{year}{2010}).

\bibitem[{\citenamefont{Kou et~al.}(2014)\citenamefont{Kou, Feldman, Levin,
  Halperin, Watanabe, Taniguchi, and Yacoby}}]{Feldman2}
\bibinfo{author}{\bibfnamefont{A.}~\bibnamefont{Kou}},
  \bibinfo{author}{\bibfnamefont{B.~E.} \bibnamefont{Feldman}},
  \bibinfo{author}{\bibfnamefont{A.~J.} \bibnamefont{Levin}},
  \bibinfo{author}{\bibfnamefont{B.~I.} \bibnamefont{Halperin}},
  \bibinfo{author}{\bibfnamefont{K.}~\bibnamefont{Watanabe}},
  \bibinfo{author}{\bibfnamefont{T.}~\bibnamefont{Taniguchi}},
  \bibnamefont{and} \bibinfo{author}{\bibfnamefont{A.}~\bibnamefont{Yacoby}},
  {``}\bibinfo{title}{Electron-hole asymmetric integer and fractional quantum
  Hall effect in bilayer graphene},{''} \bibinfo{journal}{Science}
  \textbf{\bibinfo{volume}{345}}, \bibinfo{pages}{55} (\bibinfo{year}{2014}).

\bibitem[{\citenamefont{Ki et~al.}(2014)\citenamefont{Ki, Fal'ko, Abanin, and
  Morpurgo}}]{Ki}
\bibinfo{author}{\bibfnamefont{D.-K.} \bibnamefont{Ki}},
  \bibinfo{author}{\bibfnamefont{V.~I.} \bibnamefont{Fal'ko}},
  \bibinfo{author}{\bibfnamefont{D.~A.} \bibnamefont{Abanin}},
  \bibnamefont{and} \bibinfo{author}{\bibfnamefont{A.~F.}
  \bibnamefont{Morpurgo}}, {``}\bibinfo{title}{Observation of Even Denominator
  Fractional Quantum Hall Effect in Suspended Bilayer Graphene},{''}
  \bibinfo{journal}{Nano Lett.} \textbf{\bibinfo{volume}{14}},
  \bibinfo{pages}{2135} (\bibinfo{year}{2014}).

\bibitem[{\citenamefont{Shibata and Nomura}(2009)}]{Shibata}
\bibinfo{author}{\bibfnamefont{N.}~\bibnamefont{Shibata}} \bibnamefont{and}
  \bibinfo{author}{\bibfnamefont{K.}~\bibnamefont{Nomura}},
  {``}\bibinfo{title}{Fractional Quantum Hall Effects in Graphene and Its
  Bilayer},{''} \bibinfo{journal}{J. Phys. Soc. Japan}
  \textbf{\bibinfo{volume}{78}}, \bibinfo{pages}{104708}
  (\bibinfo{year}{2009}).

\bibitem[{\citenamefont{Papi{\'{c}} and Abanin}(2014)}]{Abanin1}
\bibinfo{author}{\bibfnamefont{Z.}~\bibnamefont{Papi{\'{c}}}} \bibnamefont{and}
  \bibinfo{author}{\bibfnamefont{D.~A.} \bibnamefont{Abanin}},
  {``}\bibinfo{title}{Topological Phases in the Zeroth Landau Level of Bilayer
  Graphene},{''} \bibinfo{journal}{Phys. Rev. Lett.}
  \textbf{\bibinfo{volume}{112}}, \bibinfo{pages}{046602}
  (\bibinfo{year}{2014}).

\bibitem[{\citenamefont{Apalkov and Chakraborty}(2010)}]{Apalkov}
\bibinfo{author}{\bibfnamefont{V.~M.} \bibnamefont{Apalkov}} \bibnamefont{and}
  \bibinfo{author}{\bibfnamefont{T.}~\bibnamefont{Chakraborty}},
  {``}\bibinfo{title}{Controllable Driven Phase Transitions in Fractional
  Quantum Hall States in Bilayer Graphene},{''} \bibinfo{journal}{Phys. Rev.
  Lett.} \textbf{\bibinfo{volume}{105}}, \bibinfo{pages}{036801}
  (\bibinfo{year}{2010}).

\bibitem[{\citenamefont{Papi{\'{c}}
  et~al.}(2011{\natexlab{a}})\citenamefont{Papi{\'{c}}, Thomale, and
  Abanin}}]{Abanin2}
\bibinfo{author}{\bibfnamefont{Z.}~\bibnamefont{Papi{\'{c}}}},
  \bibinfo{author}{\bibfnamefont{R.}~\bibnamefont{Thomale}}, \bibnamefont{and}
  \bibinfo{author}{\bibfnamefont{D.~A.} \bibnamefont{Abanin}},
  {``}\bibinfo{title}{Tunable Electron Interactions and Fractional Quantum Hall
  States in Graphene},{''} \bibinfo{journal}{Phys. Rev. Lett.}
  \textbf{\bibinfo{volume}{107}}, \bibinfo{pages}{176602}
  (\bibinfo{year}{2011}{\natexlab{a}}).

\bibitem[{\citenamefont{Papi{\'{c}}
  et~al.}(2011{\natexlab{b}})\citenamefont{Papi{\'{c}}, Abanin, Barlas, and
  Bhatt}}]{Barlas}
\bibinfo{author}{\bibfnamefont{Z.}~\bibnamefont{Papi{\'{c}}}},
  \bibinfo{author}{\bibfnamefont{D.~A.} \bibnamefont{Abanin}},
  \bibinfo{author}{\bibfnamefont{Y.}~\bibnamefont{Barlas}}, \bibnamefont{and}
  \bibinfo{author}{\bibfnamefont{R.~N.} \bibnamefont{Bhatt}},
  {``}\bibinfo{title}{Tunable interactions and phase transitions in Dirac
  materials in a magnetic field},{''} \bibinfo{journal}{Phys. Rev. B}
  \textbf{\bibinfo{volume}{84}}, \bibinfo{pages}{241306}
  (\bibinfo{year}{2011}{\natexlab{b}}).

\bibitem[{\citenamefont{Apalkov and
  Chakraborty}(2011{\natexlab{a}})}]{Apalkov1}
\bibinfo{author}{\bibfnamefont{V.~M.} \bibnamefont{Apalkov}} \bibnamefont{and}
  \bibinfo{author}{\bibfnamefont{T.}~\bibnamefont{Chakraborty}},
  {``}\bibinfo{title}{Stable Pfaffian State in Bilayer Graphene},{''}
  \bibinfo{journal}{Phys. Rev. Lett.} \textbf{\bibinfo{volume}{107}},
  \bibinfo{pages}{186803} (\bibinfo{year}{2011}{\natexlab{a}}).

\bibitem[{\citenamefont{Snizhko et~al.}(2012)\citenamefont{Snizhko, Cheianov,
  and Simon}}]{Snizhko}
\bibinfo{author}{\bibfnamefont{K.}~\bibnamefont{Snizhko}},
  \bibinfo{author}{\bibfnamefont{V.}~\bibnamefont{Cheianov}}, \bibnamefont{and}
  \bibinfo{author}{\bibfnamefont{S.~H.} \bibnamefont{Simon}},
  {``}\bibinfo{title}{Importance of interband transitions for the fractional
  quantum Hall effect in bilayer graphene},{''} \bibinfo{journal}{Phys. Rev. B}
  \textbf{\bibinfo{volume}{85}}, \bibinfo{pages}{201415}
  (\bibinfo{year}{2012}).

\bibitem[{\citenamefont{Maher et~al.}(2014)\citenamefont{Maher, Wang, Gao,
  Forsythe, Taniguchi, Watanabe, Abanin, Papi{\'c}, Cadden-Zimansky, Hone
  et~al.}}]{maher2014tunable}
\bibinfo{author}{\bibfnamefont{P.}~\bibnamefont{Maher}},
  \bibinfo{author}{\bibfnamefont{L.}~\bibnamefont{Wang}},
  \bibinfo{author}{\bibfnamefont{Y.}~\bibnamefont{Gao}},
  \bibinfo{author}{\bibfnamefont{C.}~\bibnamefont{Forsythe}},
  \bibinfo{author}{\bibfnamefont{T.}~\bibnamefont{Taniguchi}},
  \bibinfo{author}{\bibfnamefont{K.}~\bibnamefont{Watanabe}},
  \bibinfo{author}{\bibfnamefont{D.}~\bibnamefont{Abanin}},
  \bibinfo{author}{\bibfnamefont{Z.}~\bibnamefont{Papi{\'c}}},
  \bibinfo{author}{\bibfnamefont{P.}~\bibnamefont{Cadden-Zimansky}},
  \bibinfo{author}{\bibfnamefont{J.}~\bibnamefont{Hone}}, \bibnamefont{et~al.},
  {``}\bibinfo{title}{Tunable fractional quantum Hall phases in bilayer
  graphene},{''} \bibinfo{journal}{Science} \textbf{\bibinfo{volume}{345}},
  \bibinfo{pages}{61} (\bibinfo{year}{2014}).

\bibitem[{\citenamefont{Geraedts et~al.}(2015)\citenamefont{Geraedts, Zaletel,
  Papi\ifmmode~\acute{c}\else \'{c}\fi{}, and Mong}}]{Geraedts}
\bibinfo{author}{\bibfnamefont{S.}~\bibnamefont{Geraedts}},
  \bibinfo{author}{\bibfnamefont{M.~P.} \bibnamefont{Zaletel}},
  \bibinfo{author}{\bibfnamefont{Z.}~\bibnamefont{Papi\ifmmode~\acute{c}\else
  \'{c}\fi{}}}, \bibnamefont{and} \bibinfo{author}{\bibfnamefont{R.~S.~K.}
  \bibnamefont{Mong}}, {``}\bibinfo{title}{Competing Abelian and non-Abelian
  topological orders in $\ensuremath{\nu}=1/3+1/3$ quantum Hall bilayers},{''}
  \bibinfo{journal}{Phys. Rev. B} \textbf{\bibinfo{volume}{91}},
  \bibinfo{pages}{205139} (\bibinfo{year}{2015}).

\bibitem[{\citenamefont{Abergel and Chakraborty}(2009)}]{Abergeland}
\bibinfo{author}{\bibfnamefont{D.~S.~L.} \bibnamefont{Abergel}}
  \bibnamefont{and}
  \bibinfo{author}{\bibfnamefont{T.}~\bibnamefont{Chakraborty}},
  {``}\bibinfo{title}{Long-Range Coulomb Interaction in Bilayer Graphene},{''}
  \bibinfo{journal}{Phys. Rev. Lett.} \textbf{\bibinfo{volume}{102}},
  \bibinfo{pages}{056807} (\bibinfo{year}{2009}).

\bibitem[{\citenamefont{Kharitonov}(2012{\natexlab{a}})}]{Kharitonov}
\bibinfo{author}{\bibfnamefont{M.}~\bibnamefont{Kharitonov}},
  {``}\bibinfo{title}{Canted Antiferromagnetic Phase of the $\nu${}=0 Quantum
  Hall State in Bilayer Graphene},{''} \bibinfo{journal}{Phys. Rev. Lett.}
  \textbf{\bibinfo{volume}{109}}, \bibinfo{pages}{046803}
  (\bibinfo{year}{2012}{\natexlab{a}}).

\bibitem[{\citenamefont{Kharitonov}(2012{\natexlab{b}})}]{haritonov_afm2012}
\bibinfo{author}{\bibfnamefont{M.}~\bibnamefont{Kharitonov}},
  {``}\bibinfo{title}{Antiferromagnetic state in bilayer graphene},{''}
  \bibinfo{journal}{Phys. Rev. B} \textbf{\bibinfo{volume}{86}},
  \bibinfo{pages}{195435} (\bibinfo{year}{2012}{\natexlab{b}}).

\bibitem[{\citenamefont{Gorbar et~al.}(2010)\citenamefont{Gorbar, Gusynin, and
  Miransky}}]{Gorbar}
\bibinfo{author}{\bibfnamefont{E.~V.} \bibnamefont{Gorbar}},
  \bibinfo{author}{\bibfnamefont{V.~P.} \bibnamefont{Gusynin}},
  \bibnamefont{and} \bibinfo{author}{\bibfnamefont{V.~A.}
  \bibnamefont{Miransky}}, {``}\bibinfo{title}{Dynamics and phase diagram of
  the $\nu=0$ quantum Hall state in bilayer graphene},{''} \bibinfo{journal}{Phys.
  Rev. B} \textbf{\bibinfo{volume}{81}}, \bibinfo{pages}{155451}
  (\bibinfo{year}{2010}).

\bibitem[{\citenamefont{Gorbar et~al.}(2011)\citenamefont{Gorbar, Gusynin, Jia,
  and Miransky}}]{Gorbar1}
\bibinfo{author}{\bibfnamefont{E.~V.} \bibnamefont{Gorbar}},
  \bibinfo{author}{\bibfnamefont{V.~P.} \bibnamefont{Gusynin}},
  \bibinfo{author}{\bibfnamefont{J.}~\bibnamefont{Jia}}, \bibnamefont{and}
  \bibinfo{author}{\bibfnamefont{V.~A.} \bibnamefont{Miransky}},
  {``}\bibinfo{title}{Broken-symmetry states and phase diagram of the lowest
  Landau level in bilayer graphene},{''} \bibinfo{journal}{Phys. Rev. B}
  \textbf{\bibinfo{volume}{84}}, \bibinfo{pages}{235449}
  (\bibinfo{year}{2011}).

\bibitem[{\citenamefont{Toke and Fal'ko}(2011)}]{Toke2011}
\bibinfo{author}{\bibfnamefont{C.}~\bibnamefont{Toke}} \bibnamefont{and}
  \bibinfo{author}{\bibfnamefont{V.~I.} \bibnamefont{Fal'ko}},
  {``}\bibinfo{title}{Intra-Landau-level magnetoexcitons and the transition
  between quantum Hall states in undoped bilayer graphene},{''}
  \bibinfo{journal}{Phys. Rev. B} \textbf{\bibinfo{volume}{83}},
  \bibinfo{pages}{115455} (\bibinfo{year}{2011}).

\bibitem[{\citenamefont{Haldane}(1988)}]{Haldane}
\bibinfo{author}{\bibfnamefont{F.~D.~M.} \bibnamefont{Haldane}},
  {``}\bibinfo{title}{Model for a Quantum Hall Effect without Landau Levels:
  Condensed-Matter Realization of the `Parity Anomaly'},{''}
  \bibinfo{journal}{Phys. Rev. Lett.} \textbf{\bibinfo{volume}{61}},
  \bibinfo{pages}{2015} (\bibinfo{year}{1988}).

\bibitem[{\citenamefont{Kane and Mele}(2005{\natexlab{a}})}]{KaneMele1}
\bibinfo{author}{\bibfnamefont{C.~L.} \bibnamefont{Kane}} \bibnamefont{and}
  \bibinfo{author}{\bibfnamefont{E.~J.} \bibnamefont{Mele}},
  {``}\bibinfo{title}{Quantum Spin Hall Effect in Graphene},{''}
  \bibinfo{journal}{Phys. Rev. Lett.} \textbf{\bibinfo{volume}{95}},
  \bibinfo{pages}{226801} (\bibinfo{year}{2005}{\natexlab{a}}).

\bibitem[{\citenamefont{Kane and Mele}(2005{\natexlab{b}})}]{KaneMele2}
\bibinfo{author}{\bibfnamefont{C.~L.} \bibnamefont{Kane}} \bibnamefont{and}
  \bibinfo{author}{\bibfnamefont{E.~J.} \bibnamefont{Mele}},
  {``}\bibinfo{title}{${Z}_{2}$ Topological Order and the Quantum Spin Hall
  Effect},{''} \bibinfo{journal}{Phys. Rev. Lett.}
  \textbf{\bibinfo{volume}{95}}, \bibinfo{pages}{146802}
  (\bibinfo{year}{2005}{\natexlab{b}}).

\bibitem[{\citenamefont{Weeks et~al.}(2011)\citenamefont{Weeks, Hu, Alicea,
  Franz, and Wu}}]{Weeks}
\bibinfo{author}{\bibfnamefont{C.}~\bibnamefont{Weeks}},
  \bibinfo{author}{\bibfnamefont{J.}~\bibnamefont{Hu}},
  \bibinfo{author}{\bibfnamefont{J.}~\bibnamefont{Alicea}},
  \bibinfo{author}{\bibfnamefont{M.}~\bibnamefont{Franz}}, \bibnamefont{and}
  \bibinfo{author}{\bibfnamefont{R.}~\bibnamefont{Wu}},
  {``}\bibinfo{title}{Engineering a Robust Quantum Spin Hall State in Graphene
  via Adatom Deposition},{''} \bibinfo{journal}{Phys. Rev. X}
  \textbf{\bibinfo{volume}{1}}, \bibinfo{pages}{021001} (\bibinfo{year}{2011}).

\bibitem[{\citenamefont{Zhang et~al.}(2012{\natexlab{a}})\citenamefont{Zhang,
  Lazo, Bl\"ugel, Heinze, and Mokrousov}}]{Lazo1}
\bibinfo{author}{\bibfnamefont{H.}~\bibnamefont{Zhang}},
  \bibinfo{author}{\bibfnamefont{C.}~\bibnamefont{Lazo}},
  \bibinfo{author}{\bibfnamefont{S.}~\bibnamefont{Bl\"ugel}},
  \bibinfo{author}{\bibfnamefont{S.}~\bibnamefont{Heinze}}, \bibnamefont{and}
  \bibinfo{author}{\bibfnamefont{Y.}~\bibnamefont{Mokrousov}},
  {``}\bibinfo{title}{Electrically Tunable Quantum Anomalous Hall Effect in
  Graphene Decorated by $5d$ Transition-Metal Adatoms},{''}
  \bibinfo{journal}{Phys. Rev. Lett.} \textbf{\bibinfo{volume}{108}},
  \bibinfo{pages}{056802} (\bibinfo{year}{2012}{\natexlab{a}}).

\bibitem[{\citenamefont{Qiao et~al.}(2010)\citenamefont{Qiao, Yang, Feng, Tse,
  Ding, Yao, Wang, and Niu}}]{Lazo2}
\bibinfo{author}{\bibfnamefont{Z.}~\bibnamefont{Qiao}},
  \bibinfo{author}{\bibfnamefont{S.~A.} \bibnamefont{Yang}},
  \bibinfo{author}{\bibfnamefont{W.}~\bibnamefont{Feng}},
  \bibinfo{author}{\bibfnamefont{W.-K.} \bibnamefont{Tse}},
  \bibinfo{author}{\bibfnamefont{J.}~\bibnamefont{Ding}},
  \bibinfo{author}{\bibfnamefont{Y.}~\bibnamefont{Yao}},
  \bibinfo{author}{\bibfnamefont{J.}~\bibnamefont{Wang}}, \bibnamefont{and}
  \bibinfo{author}{\bibfnamefont{Q.}~\bibnamefont{Niu}},
  {``}\bibinfo{title}{Quantum anomalous Hall effect in graphene from Rashba and
  exchange effects},{''} \bibinfo{journal}{Phys. Rev. B}
  \textbf{\bibinfo{volume}{82}}, \bibinfo{pages}{161414}
  (\bibinfo{year}{2010}).

\bibitem[{\citenamefont{Ding et~al.}(2011)\citenamefont{Ding, Qiao, Feng, Yao,
  and Niu}}]{Lazo3}
\bibinfo{author}{\bibfnamefont{J.}~\bibnamefont{Ding}},
  \bibinfo{author}{\bibfnamefont{Z.}~\bibnamefont{Qiao}},
  \bibinfo{author}{\bibfnamefont{W.}~\bibnamefont{Feng}},
  \bibinfo{author}{\bibfnamefont{Y.}~\bibnamefont{Yao}}, \bibnamefont{and}
  \bibinfo{author}{\bibfnamefont{Q.}~\bibnamefont{Niu}},
  {``}\bibinfo{title}{Engineering quantum anomalous/valley Hall states in
  graphene via metal-atom adsorption: An \textit{ab-initio} study},{''}
  \bibinfo{journal}{Phys. Rev. B} \textbf{\bibinfo{volume}{84}},
  \bibinfo{pages}{195444} (\bibinfo{year}{2011}).

\bibitem[{\citenamefont{Varykhalov et~al.}(2008)\citenamefont{Varykhalov,
  S\'anchez-Barriga, Shikin, Biswas, Vescovo, Rybkin, Marchenko, and
  Rader}}]{Varykhalov1}
\bibinfo{author}{\bibfnamefont{A.}~\bibnamefont{Varykhalov}},
  \bibinfo{author}{\bibfnamefont{J.}~\bibnamefont{S\'anchez-Barriga}},
  \bibinfo{author}{\bibfnamefont{A.~M.} \bibnamefont{Shikin}},
  \bibinfo{author}{\bibfnamefont{C.}~\bibnamefont{Biswas}},
  \bibinfo{author}{\bibfnamefont{E.}~\bibnamefont{Vescovo}},
  \bibinfo{author}{\bibfnamefont{A.}~\bibnamefont{Rybkin}},
  \bibinfo{author}{\bibfnamefont{D.}~\bibnamefont{Marchenko}},
  \bibnamefont{and} \bibinfo{author}{\bibfnamefont{O.}~\bibnamefont{Rader}},
  {``}\bibinfo{title}{Electronic and Magnetic Properties of Quasifreestanding
  Graphene on Ni},{''} \bibinfo{journal}{Phys. Rev. Lett.}
  \textbf{\bibinfo{volume}{101}}, \bibinfo{pages}{157601}
  (\bibinfo{year}{2008}).

\bibitem[{\citenamefont{Dedkov et~al.}(2008)\citenamefont{Dedkov, Fonin,
  R\"udiger, and Laubschat}}]{Varykhalov2}
\bibinfo{author}{\bibfnamefont{Y.~S.} \bibnamefont{Dedkov}},
  \bibinfo{author}{\bibfnamefont{M.}~\bibnamefont{Fonin}},
  \bibinfo{author}{\bibfnamefont{U.}~\bibnamefont{R\"udiger}},
  \bibnamefont{and}
  \bibinfo{author}{\bibfnamefont{C.}~\bibnamefont{Laubschat}},
  {``}\bibinfo{title}{Rashba Effect in the Graphene/Ni(111) System},{''}
  \bibinfo{journal}{Phys. Rev. Lett.} \textbf{\bibinfo{volume}{100}},
  \bibinfo{pages}{107602} (\bibinfo{year}{2008}).

\bibitem[{\citenamefont{Rader et~al.}(2009)\citenamefont{Rader, Varykhalov,
  S\'anchez-Barriga, Marchenko, Rybkin, and Shikin}}]{Varykhalov3}
\bibinfo{author}{\bibfnamefont{O.}~\bibnamefont{Rader}},
  \bibinfo{author}{\bibfnamefont{A.}~\bibnamefont{Varykhalov}},
  \bibinfo{author}{\bibfnamefont{J.}~\bibnamefont{S\'anchez-Barriga}},
  \bibinfo{author}{\bibfnamefont{D.}~\bibnamefont{Marchenko}},
  \bibinfo{author}{\bibfnamefont{A.}~\bibnamefont{Rybkin}}, \bibnamefont{and}
  \bibinfo{author}{\bibfnamefont{A.~M.} \bibnamefont{Shikin}},
  {``}\bibinfo{title}{Is There a Rashba Effect in Graphene on $3d$
  Ferromagnets?},{''} \bibinfo{journal}{Phys. Rev. Lett.}
  \textbf{\bibinfo{volume}{102}}, \bibinfo{pages}{057602}
  (\bibinfo{year}{2009}).

\bibitem[{\citenamefont{Varykhalov and Rader}(2009)}]{Varykhalov4}
\bibinfo{author}{\bibfnamefont{A.}~\bibnamefont{Varykhalov}} \bibnamefont{and}
  \bibinfo{author}{\bibfnamefont{O.}~\bibnamefont{Rader}},
  {``}\bibinfo{title}{Graphene grown on Co(0001) films and islands: Electronic
  structure and its precise magnetization dependence},{''}
  \bibinfo{journal}{Phys. Rev. B} \textbf{\bibinfo{volume}{80}},
  \bibinfo{pages}{035437} (\bibinfo{year}{2009}).

\bibitem[{\citenamefont{Guinea}(2010)}]{Guinea2010}
\bibinfo{author}{\bibfnamefont{F.}~\bibnamefont{Guinea}},
  {``}\bibinfo{title}{Spin-orbit coupling in a graphene bilayer and in
  graphite},{''} \bibinfo{journal}{New J. Phys.} \textbf{\bibinfo{volume}{12}},
  \bibinfo{pages}{083063} (\bibinfo{year}{2010}).

\bibitem[{\citenamefont{van Gelderen and Smith}(2010)}]{Gelderen}
\bibinfo{author}{\bibfnamefont{R.}~\bibnamefont{van Gelderen}}
  \bibnamefont{and} \bibinfo{author}{\bibfnamefont{C.~M.} \bibnamefont{Smith}},
  {``}\bibinfo{title}{Rashba and intrinsic spin-orbit interactions in biased
  bilayer graphene},{''} \bibinfo{journal}{Phys. Rev. B}
  \textbf{\bibinfo{volume}{81}}, \bibinfo{pages}{125435}
  (\bibinfo{year}{2010}).

\bibitem[{\citenamefont{Prada et~al.}(2011)\citenamefont{Prada, San-Jose, Brey,
  and Fertig}}]{Prada}
\bibinfo{author}{\bibfnamefont{E.}~\bibnamefont{Prada}},
  \bibinfo{author}{\bibfnamefont{P.}~\bibnamefont{San-Jose}},
  \bibinfo{author}{\bibfnamefont{L.}~\bibnamefont{Brey}}, \bibnamefont{and}
  \bibinfo{author}{\bibfnamefont{H.~A.} \bibnamefont{Fertig}},
  {``}\bibinfo{title}{Band topology and the quantum spin Hall effect in bilayer
  graphene},{''} \bibinfo{journal}{Solid State Commun.}
  \textbf{\bibinfo{volume}{151}}, \bibinfo{pages}{1075} (\bibinfo{year}{2011}).

\bibitem[{\citenamefont{Qiao et~al.}(2011{\natexlab{a}})\citenamefont{Qiao,
  Tse, Jiang, Yao, and Niu}}]{Qiao}
\bibinfo{author}{\bibfnamefont{Z.}~\bibnamefont{Qiao}},
  \bibinfo{author}{\bibfnamefont{W.-K.} \bibnamefont{Tse}},
  \bibinfo{author}{\bibfnamefont{H.}~\bibnamefont{Jiang}},
  \bibinfo{author}{\bibfnamefont{Y.}~\bibnamefont{Yao}}, \bibnamefont{and}
  \bibinfo{author}{\bibfnamefont{Q.}~\bibnamefont{Niu}},
  {``}\bibinfo{title}{Two-Dimensional Topological Insulator State and
  Topological Phase Transition in Bilayer Graphene},{''}
  \bibinfo{journal}{Phys. Rev. Lett.} \textbf{\bibinfo{volume}{107}},
  \bibinfo{pages}{256801} (\bibinfo{year}{2011}{\natexlab{a}}).

\bibitem[{\citenamefont{Mireles and Schliemann}(2012)}]{Mireles}
\bibinfo{author}{\bibfnamefont{F.}~\bibnamefont{Mireles}} \bibnamefont{and}
  \bibinfo{author}{\bibfnamefont{J.}~\bibnamefont{Schliemann}},
  {``}\bibinfo{title}{Energy spectrum and Landau levels in bilayer graphene
  with spin-orbit interaction},{''} \bibinfo{journal}{New J. Phys.}
  \textbf{\bibinfo{volume}{14}}, \bibinfo{pages}{093026}
  (\bibinfo{year}{2012}).

\bibitem[{\citenamefont{Konschuh et~al.}(2012)\citenamefont{Konschuh, Gmitra,
  Kochan, and Fabian}}]{Konschuh}
\bibinfo{author}{\bibfnamefont{S.}~\bibnamefont{Konschuh}},
  \bibinfo{author}{\bibfnamefont{M.}~\bibnamefont{Gmitra}},
  \bibinfo{author}{\bibfnamefont{D.}~\bibnamefont{Kochan}}, \bibnamefont{and}
  \bibinfo{author}{\bibfnamefont{J.}~\bibnamefont{Fabian}},
  {``}\bibinfo{title}{Theory of spin-orbit coupling in bilayer graphene},{''}
  \bibinfo{journal}{Phys. Rev. B} \textbf{\bibinfo{volume}{85}},
  \bibinfo{pages}{115423} (\bibinfo{year}{2012}).

\bibitem[{\citenamefont{Qiao et~al.}(2013)\citenamefont{Qiao, Li, Tse, Jiang,
  Yao, and Niu}}]{Qiao1}
\bibinfo{author}{\bibfnamefont{Z.}~\bibnamefont{Qiao}},
  \bibinfo{author}{\bibfnamefont{X.}~\bibnamefont{Li}},
  \bibinfo{author}{\bibfnamefont{W.-K.} \bibnamefont{Tse}},
  \bibinfo{author}{\bibfnamefont{H.}~\bibnamefont{Jiang}},
  \bibinfo{author}{\bibfnamefont{Y.}~\bibnamefont{Yao}}, \bibnamefont{and}
  \bibinfo{author}{\bibfnamefont{Q.}~\bibnamefont{Niu}},
  {``}\bibinfo{title}{Topological phases in gated bilayer graphene: Effects of
  Rashba spin-orbit coupling and exchange field},{''} \bibinfo{journal}{Phys.
  Rev. B} \textbf{\bibinfo{volume}{87}}, \bibinfo{pages}{125405}
  (\bibinfo{year}{2013}).

\bibitem[{\citenamefont{Zhai and Jin}(2014)}]{Zhai}
\bibinfo{author}{\bibfnamefont{X.}~\bibnamefont{Zhai}} \bibnamefont{and}
  \bibinfo{author}{\bibfnamefont{G.}~\bibnamefont{Jin}},
  {``}\bibinfo{title}{Reversing Berry phase and modulating Andreev reflection
  by Rashba spin-orbit coupling in graphene mono- and bilayers},{''}
  \bibinfo{journal}{Phys. Rev. B} \textbf{\bibinfo{volume}{89}},
  \bibinfo{pages}{085430} (\bibinfo{year}{2014}).

\bibitem[{\citenamefont{Dyrda{\l} and Barna{\'s}}(2014)}]{dyrdal2014spin}
\bibinfo{author}{\bibfnamefont{A.}~\bibnamefont{Dyrda{\l}}} \bibnamefont{and}
  \bibinfo{author}{\bibfnamefont{J.}~\bibnamefont{Barna{\'s}}},
  {``}\bibinfo{title}{Spin Hall effect in AA-stacked bilayer graphene},{''}
  \bibinfo{journal}{Solid State Commun.} \textbf{\bibinfo{volume}{188}},
  \bibinfo{pages}{27} (\bibinfo{year}{2014}).

\bibitem[{\citenamefont{Landgraf et~al.}(2013)\citenamefont{Landgraf,
  Shallcross, T\"urschmann, Weckbecker, and Pankratov}}]{PankratovFlakes}
\bibinfo{author}{\bibfnamefont{W.}~\bibnamefont{Landgraf}},
  \bibinfo{author}{\bibfnamefont{S.}~\bibnamefont{Shallcross}},
  \bibinfo{author}{\bibfnamefont{K.}~\bibnamefont{T\"urschmann}},
  \bibinfo{author}{\bibfnamefont{D.}~\bibnamefont{Weckbecker}},
  \bibnamefont{and}
  \bibinfo{author}{\bibfnamefont{O.}~\bibnamefont{Pankratov}},
  {``}\bibinfo{title}{Electronic structure of twisted graphene flakes},{''}
  \bibinfo{journal}{Phys. Rev. B} \textbf{\bibinfo{volume}{87}},
  \bibinfo{pages}{075433} (\bibinfo{year}{2013}).

\bibitem[{\citenamefont{Su\'arez~Morell
  et~al.}(2014)\citenamefont{Su\'arez~Morell, Vergara, Pacheco, Brey, and
  Chico}}]{MorellNanoribons2014}
\bibinfo{author}{\bibfnamefont{E.}~\bibnamefont{Su\'arez~Morell}},
  \bibinfo{author}{\bibfnamefont{R.}~\bibnamefont{Vergara}},
  \bibinfo{author}{\bibfnamefont{M.}~\bibnamefont{Pacheco}},
  \bibinfo{author}{\bibfnamefont{L.}~\bibnamefont{Brey}}, \bibnamefont{and}
  \bibinfo{author}{\bibfnamefont{L.}~\bibnamefont{Chico}},
  {``}\bibinfo{title}{Electronic properties of twisted bilayer
  nanoribbons},{''} \bibinfo{journal}{Phys. Rev. B}
  \textbf{\bibinfo{volume}{89}}, \bibinfo{pages}{205405}
  (\bibinfo{year}{2014}).

\bibitem[{\citenamefont{Qi et~al.}(2015)\citenamefont{Qi, Daniels, Hong, Park,
  Meunier, Drndi{\'{c}}, and Johnson}}]{nanoribbon_exper_joule}
\bibinfo{author}{\bibfnamefont{Z.~J.} \bibnamefont{Qi}},
  \bibinfo{author}{\bibfnamefont{C.}~\bibnamefont{Daniels}},
  \bibinfo{author}{\bibfnamefont{S.~J.} \bibnamefont{Hong}},
  \bibinfo{author}{\bibfnamefont{Y.~W.} \bibnamefont{Park}},
  \bibinfo{author}{\bibfnamefont{V.}~\bibnamefont{Meunier}},
  \bibinfo{author}{\bibfnamefont{M.}~\bibnamefont{Drndi{\'{c}}}},
  \bibnamefont{and} \bibinfo{author}{\bibfnamefont{A.~T.~C.}
  \bibnamefont{Johnson}}, {``}\bibinfo{title}{Electronic Transport of
  Recrystallized Freestanding Graphene Nanoribbons},{''} \bibinfo{journal}{ACS
  Nano} \textbf{\bibinfo{volume}{9}}, \bibinfo{pages}{3510}
  (\bibinfo{year}{2015}).

\bibitem[{\citenamefont{Sahu et~al.}(2010)\citenamefont{Sahu, Min, and
  Banerjee}}]{Sahu2010}
\bibinfo{author}{\bibfnamefont{B.}~\bibnamefont{Sahu}},
  \bibinfo{author}{\bibfnamefont{H.}~\bibnamefont{Min}}, \bibnamefont{and}
  \bibinfo{author}{\bibfnamefont{S.~K.} \bibnamefont{Banerjee}},
  {``}\bibinfo{title}{Effects of magnetism and electric field on the energy gap
  of bilayer graphene nanoflakes},{''} \bibinfo{journal}{Phys. Rev. B}
  \textbf{\bibinfo{volume}{81}}, \bibinfo{pages}{045414}
  (\bibinfo{year}{2010}).

\bibitem[{\citenamefont{Sahu et~al.}(2008)\citenamefont{Sahu, Min, MacDonald,
  and Banerjee}}]{Sahu2008}
\bibinfo{author}{\bibfnamefont{B.}~\bibnamefont{Sahu}},
  \bibinfo{author}{\bibfnamefont{H.}~\bibnamefont{Min}},
  \bibinfo{author}{\bibfnamefont{A.~H.} \bibnamefont{MacDonald}},
  \bibnamefont{and} \bibinfo{author}{\bibfnamefont{S.~K.}
  \bibnamefont{Banerjee}}, {``}\bibinfo{title}{Energy gaps, magnetism, and
  electric-field effects in bilayer graphene nanoribbons},{''}
  \bibinfo{journal}{Phys. Rev. B} \textbf{\bibinfo{volume}{78}},
  \bibinfo{pages}{045404} (\bibinfo{year}{2008}).

\bibitem[{\citenamefont{Matulis and Peeters}(2008)}]{Matulis2008}
\bibinfo{author}{\bibfnamefont{A.}~\bibnamefont{Matulis}} \bibnamefont{and}
  \bibinfo{author}{\bibfnamefont{F.~M.} \bibnamefont{Peeters}},
  {``}\bibinfo{title}{Quasibound states of quantum dots in single and bilayer
  graphene},{''} \bibinfo{journal}{Phys. Rev. B} \textbf{\bibinfo{volume}{77}},
  \bibinfo{pages}{115423} (\bibinfo{year}{2008}).

\bibitem[{\citenamefont{Pereira
  et~al.}(2007{\natexlab{b}})\citenamefont{Pereira, Vasilopoulos, and
  Peeters}}]{Pereira2007a}
\bibinfo{author}{\bibfnamefont{J.}~\bibnamefont{Pereira},
  \bibfnamefont{J.~Milton}},
  \bibinfo{author}{\bibfnamefont{P.}~\bibnamefont{Vasilopoulos}},
  \bibnamefont{and} \bibinfo{author}{\bibfnamefont{F.~M.}
  \bibnamefont{Peeters}}, {``}\bibinfo{title}{Tunable quantum dots in bilayer
  graphene},{''} \bibinfo{journal}{Nano Lett.} \textbf{\bibinfo{volume}{7}},
  \bibinfo{pages}{946} (\bibinfo{year}{2007}{\natexlab{b}}).

\bibitem[{\citenamefont{Giavaras and Nori}(2011)}]{giavaras_nori2011}
\bibinfo{author}{\bibfnamefont{G.}~\bibnamefont{Giavaras}} \bibnamefont{and}
  \bibinfo{author}{\bibfnamefont{F.}~\bibnamefont{Nori}},
  {``}\bibinfo{title}{Dirac gap-induced graphene quantum dot in an
  electrostatic potential},{''} \bibinfo{journal}{Phys. Rev. B}
  \textbf{\bibinfo{volume}{83}}, \bibinfo{pages}{165427}
  (\bibinfo{year}{2011}).

\bibitem[{\citenamefont{Allen et~al.}(2012)\citenamefont{Allen, Martin, and
  Yacoby}}]{Allen2012}
\bibinfo{author}{\bibfnamefont{M.~T.} \bibnamefont{Allen}},
  \bibinfo{author}{\bibfnamefont{J.}~\bibnamefont{Martin}}, \bibnamefont{and}
  \bibinfo{author}{\bibfnamefont{A.}~\bibnamefont{Yacoby}},
  {``}\bibinfo{title}{Gate-defined quantum confinement in suspended bilayer
  graphene},{''} \bibinfo{journal}{Nat. Commun.} \textbf{\bibinfo{volume}{3}},
  \bibinfo{pages}{934} (\bibinfo{year}{2012}).

\bibitem[{\citenamefont{Goossens et~al.}(2012)\citenamefont{Goossens, Driessen,
  Baart, Watanabe, Taniguchi, and
  Vandersypen}}]{bilayer_dot_exper_bor_nitr_gate2012}
\bibinfo{author}{\bibfnamefont{A.~S.~M.} \bibnamefont{Goossens}},
  \bibinfo{author}{\bibfnamefont{S.~C.~M.} \bibnamefont{Driessen}},
  \bibinfo{author}{\bibfnamefont{T.~A.} \bibnamefont{Baart}},
  \bibinfo{author}{\bibfnamefont{K.}~\bibnamefont{Watanabe}},
  \bibinfo{author}{\bibfnamefont{T.}~\bibnamefont{Taniguchi}},
  \bibnamefont{and} \bibinfo{author}{\bibfnamefont{L.~M.~K.}
  \bibnamefont{Vandersypen}}, {``}\bibinfo{title}{Gate-Defined Confinement in
  Bilayer Graphene-Hexagonal Boron Nitride Hybrid Devices},{''}
  \bibinfo{journal}{Nano Lett.} \textbf{\bibinfo{volume}{12}},
  \bibinfo{pages}{4656} (\bibinfo{year}{2012}).

\bibitem[{\citenamefont{M{\"{u}}ller et~al.}(2014)\citenamefont{M{\"{u}}ller,
  Kaestner, Hohls, Weimann, Pierz, and Schumacher}}]{gate_def_dot_exp}
\bibinfo{author}{\bibfnamefont{A.}~\bibnamefont{M{\"{u}}ller}},
  \bibinfo{author}{\bibfnamefont{B.}~\bibnamefont{Kaestner}},
  \bibinfo{author}{\bibfnamefont{F.}~\bibnamefont{Hohls}},
  \bibinfo{author}{\bibfnamefont{T.}~\bibnamefont{Weimann}},
  \bibinfo{author}{\bibfnamefont{K.}~\bibnamefont{Pierz}}, \bibnamefont{and}
  \bibinfo{author}{\bibfnamefont{H.~W.} \bibnamefont{Schumacher}},
  {``}\bibinfo{title}{Bilayer graphene quantum dot defined by topgates},{''}
  \bibinfo{journal}{J. Appl. Phys.} \textbf{\bibinfo{volume}{115}},
  \bibinfo{eid}{233710} (\bibinfo{year}{2014}).

\bibitem[{\citenamefont{De~Martino et~al.}(2007)\citenamefont{De~Martino,
  Dell'Anna, and Egger}}]{martino_qdot2007}
\bibinfo{author}{\bibfnamefont{A.}~\bibnamefont{De~Martino}},
  \bibinfo{author}{\bibfnamefont{L.}~\bibnamefont{Dell'Anna}},
  \bibnamefont{and} \bibinfo{author}{\bibfnamefont{R.}~\bibnamefont{Egger}},
  {``}\bibinfo{title}{Magnetic Confinement of Massless Dirac Fermions in
  Graphene},{''} \bibinfo{journal}{Phys. Rev. Lett.}
  \textbf{\bibinfo{volume}{98}}, \bibinfo{pages}{066802}
  (\bibinfo{year}{2007}).

\bibitem[{\citenamefont{Giavaras et~al.}(2009)\citenamefont{Giavaras, Maksym,
  and Roy}}]{giavaras2009}
\bibinfo{author}{\bibfnamefont{G.}~\bibnamefont{Giavaras}},
  \bibinfo{author}{\bibfnamefont{P.~A.} \bibnamefont{Maksym}},
  \bibnamefont{and} \bibinfo{author}{\bibfnamefont{M.}~\bibnamefont{Roy}},
  {``}\bibinfo{title}{Magnetic field induced confinement-deconfinement
  transition in graphene quantum dots},{''} \bibinfo{journal}{J. Phys.:
  Condens. Matter} \textbf{\bibinfo{volume}{21}}, \bibinfo{pages}{102201}
  (\bibinfo{year}{2009}).

\bibitem[{\citenamefont{Giavaras and Nori}(2012)}]{elect_magn_qdot_single2012}
\bibinfo{author}{\bibfnamefont{G.}~\bibnamefont{Giavaras}} \bibnamefont{and}
  \bibinfo{author}{\bibfnamefont{F.}~\bibnamefont{Nori}},
  {``}\bibinfo{title}{Tunable quantum dots in monolayer graphene},{''}
  \bibinfo{journal}{Phys. Rev. B} \textbf{\bibinfo{volume}{85}},
  \bibinfo{pages}{165446} (\bibinfo{year}{2012}).

\bibitem[{\citenamefont{Zarenia et~al.}(2009)\citenamefont{Zarenia, Pereira,
  Peeters, and Farias}}]{zarenia_blg_ring2009}
\bibinfo{author}{\bibfnamefont{M.}~\bibnamefont{Zarenia}},
  \bibinfo{author}{\bibfnamefont{J.~M.} \bibnamefont{Pereira}},
  \bibinfo{author}{\bibfnamefont{F.~M.} \bibnamefont{Peeters}},
  \bibnamefont{and} \bibinfo{author}{\bibfnamefont{G.~A.}
  \bibnamefont{Farias}}, {``}\bibinfo{title}{Electrostatically Confined Quantum
  Rings in Bilayer Graphene},{''} \bibinfo{journal}{Nano Lett.}
  \textbf{\bibinfo{volume}{9}}, \bibinfo{pages}{4088} (\bibinfo{year}{2009}).

\bibitem[{\citenamefont{Xavier et~al.}(2010)\citenamefont{Xavier, Pereira~Jr.,
  Chaves, Farias, and Peeters}}]{Xavier2010}
\bibinfo{author}{\bibfnamefont{L.~J.~P.} \bibnamefont{Xavier}},
  \bibinfo{author}{\bibfnamefont{J.~M.} \bibnamefont{Pereira~Jr.}},
  \bibinfo{author}{\bibfnamefont{A.}~\bibnamefont{Chaves}},
  \bibinfo{author}{\bibfnamefont{G.~A.} \bibnamefont{Farias}},
  \bibnamefont{and} \bibinfo{author}{\bibfnamefont{F.~M.}
  \bibnamefont{Peeters}}, {``}\bibinfo{title}{Topological confinement in
  graphene bilayer quantum rings},{''} \bibinfo{journal}{Appl. Phys. Lett.}
  \textbf{\bibinfo{volume}{96}}, \bibinfo{pages}{212108}
  (\bibinfo{year}{2010}).

\bibitem[{\citenamefont{Martin et~al.}(2008{\natexlab{a}})\citenamefont{Martin,
  Blanter, and Morpurgo}}]{Martin2008}
\bibinfo{author}{\bibfnamefont{I.}~\bibnamefont{Martin}},
  \bibinfo{author}{\bibfnamefont{Y.~M.} \bibnamefont{Blanter}},
  \bibnamefont{and} \bibinfo{author}{\bibfnamefont{A.~F.}
  \bibnamefont{Morpurgo}}, {``}\bibinfo{title}{Topological confinement in
  bilayer graphene},{''} \bibinfo{journal}{Phys. Rev. Lett.}
  \textbf{\bibinfo{volume}{100}}, \bibinfo{pages}{036804}
  (\bibinfo{year}{2008}{\natexlab{a}}).

\bibitem[{\citenamefont{Sun and Chang}(2008)}]{sun_transport_aa_ab2008}
\bibinfo{author}{\bibfnamefont{S.-J.} \bibnamefont{Sun}} \bibnamefont{and}
  \bibinfo{author}{\bibfnamefont{C.}~\bibnamefont{Chang}},
  {``}\bibinfo{title}{Ballistic transport in bilayer nano-graphite ribbons
  under gate and magnetic fields},{''} \bibinfo{journal}{Eur. Phys. J. B}
  \textbf{\bibinfo{volume}{64}}, \bibinfo{pages}{249} (\bibinfo{year}{2008}).

\bibitem[{\citenamefont{Xu et~al.}(2012)\citenamefont{Xu, Wang, Shi, and
  Zhang}}]{xu_transport_aa2012}
\bibinfo{author}{\bibfnamefont{N.}~\bibnamefont{Xu}},
  \bibinfo{author}{\bibfnamefont{B.}~\bibnamefont{Wang}},
  \bibinfo{author}{\bibfnamefont{D.}~\bibnamefont{Shi}}, \bibnamefont{and}
  \bibinfo{author}{\bibfnamefont{C.}~\bibnamefont{Zhang}},
  {``}\bibinfo{title}{Transport properties of AA-stacking bilayer graphene
  nanoribbons},{''} \bibinfo{journal}{Solid State Commun.}
  \textbf{\bibinfo{volume}{152}}, \bibinfo{pages}{994 } (\bibinfo{year}{2012}).

\bibitem[{\citenamefont{Gonzalez et~al.}(2010)\citenamefont{Gonzalez, Santos,
  Pacheco, Chico, and Brey}}]{Gonzalez2010}
\bibinfo{author}{\bibfnamefont{J.~W.} \bibnamefont{Gonzalez}},
  \bibinfo{author}{\bibfnamefont{H.}~\bibnamefont{Santos}},
  \bibinfo{author}{\bibfnamefont{M.}~\bibnamefont{Pacheco}},
  \bibinfo{author}{\bibfnamefont{L.}~\bibnamefont{Chico}}, \bibnamefont{and}
  \bibinfo{author}{\bibfnamefont{L.}~\bibnamefont{Brey}},
  {``}\bibinfo{title}{Electronic transport through bilayer graphene
  flakes},{''} \bibinfo{journal}{Phys. Rev. B} \textbf{\bibinfo{volume}{81}},
  \bibinfo{pages}{195406} (\bibinfo{year}{2010}).

\bibitem[{\citenamefont{Benameur et~al.}(2015)\citenamefont{Benameur, Gargiulo,
  Manzeli, Autes, Tosun, Yazyev, and Kis}}]{Benameur2015}
\bibinfo{author}{\bibfnamefont{M.~M.} \bibnamefont{Benameur}},
  \bibinfo{author}{\bibfnamefont{F.}~\bibnamefont{Gargiulo}},
  \bibinfo{author}{\bibfnamefont{S.}~\bibnamefont{Manzeli}},
  \bibinfo{author}{\bibfnamefont{G.}~\bibnamefont{Autes}},
  \bibinfo{author}{\bibfnamefont{M.}~\bibnamefont{Tosun}},
  \bibinfo{author}{\bibfnamefont{O.~V.} \bibnamefont{Yazyev}},
  \bibnamefont{and} \bibinfo{author}{\bibfnamefont{A.}~\bibnamefont{Kis}},
  {``}\bibinfo{title}{Electromechanical oscillations in bilayer graphene},{''}
  \bibinfo{journal}{Nat. Commun.} \textbf{\bibinfo{volume}{6}}
  (\bibinfo{year}{2015}).

\bibitem[{\citenamefont{Habib et~al.}(2011)\citenamefont{Habib, Zahid, and
  Lake}}]{Habib2011}
\bibinfo{author}{\bibfnamefont{K.~M.~M.} \bibnamefont{Habib}},
  \bibinfo{author}{\bibfnamefont{F.}~\bibnamefont{Zahid}}, \bibnamefont{and}
  \bibinfo{author}{\bibfnamefont{R.~K.} \bibnamefont{Lake}},
  {``}\bibinfo{title}{Negative differential resistance in bilayer graphene
  nanoribbons},{''} \bibinfo{journal}{Appl. Phys. Lett.}
  \textbf{\bibinfo{volume}{98}}, \bibinfo{pages}{192112}
  (\bibinfo{year}{2011}).

\bibitem[{\citenamefont{Maksimov et~al.}(2013)\citenamefont{Maksimov, Rozhkov,
  and Sboychakov}}]{Pavel_armchair}
\bibinfo{author}{\bibfnamefont{P.~A.} \bibnamefont{Maksimov}},
  \bibinfo{author}{\bibfnamefont{A.~V.} \bibnamefont{Rozhkov}},
  \bibnamefont{and} \bibinfo{author}{\bibfnamefont{A.~O.}
  \bibnamefont{Sboychakov}}, {``}\bibinfo{title}{Localized electron states near
  the armchair edge of graphene},{''} \bibinfo{journal}{Phys. Rev. B}
  \textbf{\bibinfo{volume}{88}}, \bibinfo{pages}{245421}
  (\bibinfo{year}{2013}).

\bibitem[{\citenamefont{Castro et~al.}(2008{\natexlab{a}})\citenamefont{Castro,
  Peres, Lopes~dos Santos, Castro~Neto, and Guinea}}]{bilayer_zigzag_loc2008}
\bibinfo{author}{\bibfnamefont{E.~V.} \bibnamefont{Castro}},
  \bibinfo{author}{\bibfnamefont{N.~M.~R.} \bibnamefont{Peres}},
  \bibinfo{author}{\bibfnamefont{J.~M.~B.} \bibnamefont{Lopes~dos Santos}},
  \bibinfo{author}{\bibfnamefont{A.~H.} \bibnamefont{Castro~Neto}},
  \bibnamefont{and} \bibinfo{author}{\bibfnamefont{F.}~\bibnamefont{Guinea}},
  {``}\bibinfo{title}{Localized States at Zigzag Edges of Bilayer
  Graphene},{''} \bibinfo{journal}{Phys. Rev. Lett.}
  \textbf{\bibinfo{volume}{100}}, \bibinfo{pages}{026802}
  (\bibinfo{year}{2008}{\natexlab{a}}).

\bibitem[{\citenamefont{Cortijo et~al.}(2010)\citenamefont{Cortijo,
  Oroszl\'any, and Schomerus}}]{bilayer_zigzag_loc_gen2010}
\bibinfo{author}{\bibfnamefont{A.}~\bibnamefont{Cortijo}},
  \bibinfo{author}{\bibfnamefont{L.}~\bibnamefont{Oroszl\'any}},
  \bibnamefont{and}
  \bibinfo{author}{\bibfnamefont{H.}~\bibnamefont{Schomerus}},
  {``}\bibinfo{title}{Fast and slow edges in bilayer graphene nanoribbons:
  Tuning the transition from band to Mott insulator},{''}
  \bibinfo{journal}{Phys. Rev. B} \textbf{\bibinfo{volume}{81}},
  \bibinfo{pages}{235422} (\bibinfo{year}{2010}).

\bibitem[{\citenamefont{Zhang et~al.}(2010{\natexlab{a}})\citenamefont{Zhang,
  Chen, Zeng, and Guo}}]{Zhang_first_principle_2010a}
\bibinfo{author}{\bibfnamefont{Z.}~\bibnamefont{Zhang}},
  \bibinfo{author}{\bibfnamefont{C.}~\bibnamefont{Chen}},
  \bibinfo{author}{\bibfnamefont{X.~C.} \bibnamefont{Zeng}}, \bibnamefont{and}
  \bibinfo{author}{\bibfnamefont{W.}~\bibnamefont{Guo}},
  {``}\bibinfo{title}{Tuning the magnetic and electronic properties of bilayer
  graphene nanoribbons on Si(001) by bias voltage},{''} \bibinfo{journal}{Phys.
  Rev. B} \textbf{\bibinfo{volume}{81}}, \bibinfo{pages}{155428}
  (\bibinfo{year}{2010}{\natexlab{a}}).

\bibitem[{\citenamefont{Lima et~al.}(2009)\citenamefont{Lima, Fazzio, and
  da~Silva}}]{Lima2009}
\bibinfo{author}{\bibfnamefont{M.~P.} \bibnamefont{Lima}},
  \bibinfo{author}{\bibfnamefont{A.}~\bibnamefont{Fazzio}}, \bibnamefont{and}
  \bibinfo{author}{\bibfnamefont{A.~J.~R.} \bibnamefont{da~Silva}},
  {``}\bibinfo{title}{Edge effects in bilayer graphene nanoribbons: Ab initio
  total-energy density functional theory calculations},{''}
  \bibinfo{journal}{Phys. Rev. B} \textbf{\bibinfo{volume}{79}},
  \bibinfo{pages}{153401} (\bibinfo{year}{2009}).

\bibitem[{\citenamefont{Wright et~al.}(2009)\citenamefont{Wright, Cao, and
  Zhang}}]{Wright2009}
\bibinfo{author}{\bibfnamefont{A.~R.} \bibnamefont{Wright}},
  \bibinfo{author}{\bibfnamefont{J.~C.} \bibnamefont{Cao}}, \bibnamefont{and}
  \bibinfo{author}{\bibfnamefont{C.}~\bibnamefont{Zhang}},
  {``}\bibinfo{title}{Enhanced Optical Conductivity of Bilayer Graphene
  Nanoribbons in the Terahertz Regime},{''} \bibinfo{journal}{Phys. Rev. Lett.}
  \textbf{\bibinfo{volume}{103}}, \bibinfo{pages}{207401}
  (\bibinfo{year}{2009}).

\bibitem[{\citenamefont{Li et~al.}(2008{\natexlab{a}})\citenamefont{Li, Wang,
  Zhang, Lee, and Dai}}]{nanoribbon_production2008}
\bibinfo{author}{\bibfnamefont{X.}~\bibnamefont{Li}},
  \bibinfo{author}{\bibfnamefont{X.}~\bibnamefont{Wang}},
  \bibinfo{author}{\bibfnamefont{L.}~\bibnamefont{Zhang}},
  \bibinfo{author}{\bibfnamefont{S.}~\bibnamefont{Lee}}, \bibnamefont{and}
  \bibinfo{author}{\bibfnamefont{H.}~\bibnamefont{Dai}},
  {``}\bibinfo{title}{Chemically Derived, Ultrasmooth Graphene Nanoribbon
  Semiconductors},{''} \bibinfo{journal}{Science}
  \textbf{\bibinfo{volume}{319}}, \bibinfo{pages}{1229}
  (\bibinfo{year}{2008}{\natexlab{a}}).

\bibitem[{\citenamefont{Xu et~al.}(2009)\citenamefont{Xu, Heinzel, and
  Zozoulenko}}]{landauer_ab_ideal_disord_zozoul2009}
\bibinfo{author}{\bibfnamefont{H.}~\bibnamefont{Xu}},
  \bibinfo{author}{\bibfnamefont{T.}~\bibnamefont{Heinzel}}, \bibnamefont{and}
  \bibinfo{author}{\bibfnamefont{I.~V.} \bibnamefont{Zozoulenko}},
  {``}\bibinfo{title}{Edge disorder and localization regimes in bilayer
  graphene nanoribbons},{''} \bibinfo{journal}{Phys. Rev. B}
  \textbf{\bibinfo{volume}{80}}, \bibinfo{pages}{045308}
  (\bibinfo{year}{2009}).

\bibitem[{\citenamefont{Fujita et~al.}(1997)\citenamefont{Fujita, Igami, and
  Nakada}}]{lattice_distortion}
\bibinfo{author}{\bibfnamefont{M.}~\bibnamefont{Fujita}},
  \bibinfo{author}{\bibfnamefont{M.}~\bibnamefont{Igami}}, \bibnamefont{and}
  \bibinfo{author}{\bibfnamefont{K.}~\bibnamefont{Nakada}},
  {``}\bibinfo{title}{Lattice distortion in nanographite ribbons},{''}
  \bibinfo{journal}{J. Phys. Soc. Japan} \textbf{\bibinfo{volume}{66}},
  \bibinfo{pages}{1864} (\bibinfo{year}{1997}).

\bibitem[{\citenamefont{Zarea and Sandler}(2007)}]{sandler_coulomb_gap}
\bibinfo{author}{\bibfnamefont{M.}~\bibnamefont{Zarea}} \bibnamefont{and}
  \bibinfo{author}{\bibfnamefont{N.}~\bibnamefont{Sandler}},
  {``}\bibinfo{title}{Electron-electron and spin-orbit interactions in armchair
  graphene ribbons},{''} \bibinfo{journal}{Phys. Rev. Lett.}
  \textbf{\bibinfo{volume}{99}}, \bibinfo{pages}{256804}
  (\bibinfo{year}{2007}).

\bibitem[{\citenamefont{Rozhkov et~al.}(2009)\citenamefont{Rozhkov, Savel'ev,
  and Nori}}]{our_nanoribbon_paper_2009}
\bibinfo{author}{\bibfnamefont{A.~V.} \bibnamefont{Rozhkov}},
  \bibinfo{author}{\bibfnamefont{S.}~\bibnamefont{Savel'ev}}, \bibnamefont{and}
  \bibinfo{author}{\bibfnamefont{F.}~\bibnamefont{Nori}},
  {``}\bibinfo{title}{Electronic properties of armchair graphene
  nanoribbons},{''} \bibinfo{journal}{Phys. Rev. B}
  \textbf{\bibinfo{volume}{79}}, \bibinfo{pages}{125420}
  (\bibinfo{year}{2009}).

\bibitem[{\citenamefont{Li et~al.}(2015)\citenamefont{Li, Hsieh, Yang, and
  Chang}}]{nribb_theor_magn_landauer2015}
\bibinfo{author}{\bibfnamefont{T.}~\bibnamefont{Li}},
  \bibinfo{author}{\bibfnamefont{C.}~\bibnamefont{Hsieh}},
  \bibinfo{author}{\bibfnamefont{S.}~\bibnamefont{Yang}}, \bibnamefont{and}
  \bibinfo{author}{\bibfnamefont{S.}~\bibnamefont{Chang}},
  {``}\bibinfo{title}{Ballistic transport of bilayer graphene nanoribbons in a
  spatially modulated magnetic field},{''} \bibinfo{journal}{Solid State
  Commun.} \textbf{\bibinfo{volume}{206}}, \bibinfo{pages}{6 }
  (\bibinfo{year}{2015}).

\bibitem[{\citenamefont{Zarenia et~al.}(2011)\citenamefont{Zarenia, Pereira,
  Farias, and Peeters}}]{Zarenia2011a}
\bibinfo{author}{\bibfnamefont{M.}~\bibnamefont{Zarenia}},
  \bibinfo{author}{\bibfnamefont{J.}~\bibnamefont{Pereira},
  \bibfnamefont{J.~M.}}, \bibinfo{author}{\bibfnamefont{G.~A.}
  \bibnamefont{Farias}}, \bibnamefont{and}
  \bibinfo{author}{\bibfnamefont{F.~M.} \bibnamefont{Peeters}},
  {``}\bibinfo{title}{Chiral states in bilayer graphene: Magnetic field
  dependence and gap opening},{''} \bibinfo{journal}{Phys. Rev. B}
  \textbf{\bibinfo{volume}{84}}, \bibinfo{pages}{125451}
  (\bibinfo{year}{2011}).

\bibitem[{\citenamefont{Schroer et~al.}(2015)\citenamefont{Schroer, Silvestrov,
  and Recher}}]{schroer_cooper_pair_split2015}
\bibinfo{author}{\bibfnamefont{A.}~\bibnamefont{Schroer}},
  \bibinfo{author}{\bibfnamefont{P.~G.} \bibnamefont{Silvestrov}},
  \bibnamefont{and} \bibinfo{author}{\bibfnamefont{P.}~\bibnamefont{Recher}},
  {``}\bibinfo{title}{Valley-based Cooper pair splitting via topologically
  confined channels in bilayer graphene},{''} \bibinfo{journal}{Phys. Rev. B}
  \textbf{\bibinfo{volume}{92}}, \bibinfo{pages}{241404}
  (\bibinfo{year}{2015}).

\bibitem[{\citenamefont{Qiao et~al.}(2011{\natexlab{b}})\citenamefont{Qiao,
  Jung, Niu, and MacDonald}}]{Qiao2011}
\bibinfo{author}{\bibfnamefont{Z.}~\bibnamefont{Qiao}},
  \bibinfo{author}{\bibfnamefont{J.}~\bibnamefont{Jung}},
  \bibinfo{author}{\bibfnamefont{Q.}~\bibnamefont{Niu}}, \bibnamefont{and}
  \bibinfo{author}{\bibfnamefont{A.~H.} \bibnamefont{MacDonald}},
  {``}\bibinfo{title}{Electronic Highways in Bilayer Graphene},{''}
  \bibinfo{journal}{Nano Lett.} \textbf{\bibinfo{volume}{11}},
  \bibinfo{pages}{3453} (\bibinfo{year}{2011}{\natexlab{b}}).

\bibitem[{\citenamefont{Killi et~al.}(2010)\citenamefont{Killi, Wei, Affleck,
  and Paramekanti}}]{Killi2010}
\bibinfo{author}{\bibfnamefont{M.}~\bibnamefont{Killi}},
  \bibinfo{author}{\bibfnamefont{T.-C.} \bibnamefont{Wei}},
  \bibinfo{author}{\bibfnamefont{I.}~\bibnamefont{Affleck}}, \bibnamefont{and}
  \bibinfo{author}{\bibfnamefont{A.}~\bibnamefont{Paramekanti}},
  {``}\bibinfo{title}{Tunable Luttinger Liquid Physics in Biased Bilayer
  Graphene},{''} \bibinfo{journal}{Phys. Rev. Lett.}
  \textbf{\bibinfo{volume}{104}}, \bibinfo{pages}{216406}
  (\bibinfo{year}{2010}).

\bibitem[{\citenamefont{Zhang et~al.}(2013)\citenamefont{Zhang, MacDonald, and
  Mele}}]{zhang_defect_topolog_channel2013}
\bibinfo{author}{\bibfnamefont{F.}~\bibnamefont{Zhang}},
  \bibinfo{author}{\bibfnamefont{A.~H.} \bibnamefont{MacDonald}},
  \bibnamefont{and} \bibinfo{author}{\bibfnamefont{E.~J.} \bibnamefont{Mele}},
  {``}\bibinfo{title}{Valley Chern numbers and boundary modes in gapped bilayer
  graphene},{''} \bibinfo{journal}{PNAS} \textbf{\bibinfo{volume}{110}},
  \bibinfo{pages}{10546} (\bibinfo{year}{2013}).

\bibitem[{\citenamefont{Vaezi et~al.}(2013)\citenamefont{Vaezi, Liang, Ngai,
  Yang, and Kim}}]{vaezi_defect_topolog_channel2013}
\bibinfo{author}{\bibfnamefont{A.}~\bibnamefont{Vaezi}},
  \bibinfo{author}{\bibfnamefont{Y.}~\bibnamefont{Liang}},
  \bibinfo{author}{\bibfnamefont{D.~H.} \bibnamefont{Ngai}},
  \bibinfo{author}{\bibfnamefont{L.}~\bibnamefont{Yang}}, \bibnamefont{and}
  \bibinfo{author}{\bibfnamefont{E.-A.} \bibnamefont{Kim}},
  {``}\bibinfo{title}{Topological Edge States at a Tilt Boundary in Gated
  Multilayer Graphene},{''} \bibinfo{journal}{Phys. Rev. X}
  \textbf{\bibinfo{volume}{3}}, \bibinfo{pages}{021018} (\bibinfo{year}{2013}).

\bibitem[{\citenamefont{Alden et~al.}(2013)\citenamefont{Alden, Tsen, Huang,
  Hovden, Brown, Park, Muller, and McEuen}}]{Alden2013}
\bibinfo{author}{\bibfnamefont{J.~S.} \bibnamefont{Alden}},
  \bibinfo{author}{\bibfnamefont{A.~W.} \bibnamefont{Tsen}},
  \bibinfo{author}{\bibfnamefont{P.~Y.} \bibnamefont{Huang}},
  \bibinfo{author}{\bibfnamefont{R.}~\bibnamefont{Hovden}},
  \bibinfo{author}{\bibfnamefont{L.}~\bibnamefont{Brown}},
  \bibinfo{author}{\bibfnamefont{J.}~\bibnamefont{Park}},
  \bibinfo{author}{\bibfnamefont{D.~A.} \bibnamefont{Muller}},
  \bibnamefont{and} \bibinfo{author}{\bibfnamefont{P.~L.}
  \bibnamefont{McEuen}}, {``}\bibinfo{title}{Strain solitons and topological
  defects in bilayer graphene},{''} \bibinfo{journal}{PNAS}
  \textbf{\bibinfo{volume}{110}}, \bibinfo{pages}{11256}
  (\bibinfo{year}{2013}).

\bibitem[{\citenamefont{Ju et~al.}(2015)\citenamefont{Ju, Shi, Nair, Lv, Jin,
  Velasco~Jr, Ojeda-Aristizabal, Bechtel, Martin, Zettl
  et~al.}}]{topological_experiment}
\bibinfo{author}{\bibfnamefont{L.}~\bibnamefont{Ju}},
  \bibinfo{author}{\bibfnamefont{Z.}~\bibnamefont{Shi}},
  \bibinfo{author}{\bibfnamefont{N.}~\bibnamefont{Nair}},
  \bibinfo{author}{\bibfnamefont{Y.}~\bibnamefont{Lv}},
  \bibinfo{author}{\bibfnamefont{C.}~\bibnamefont{Jin}},
  \bibinfo{author}{\bibfnamefont{J.}~\bibnamefont{Velasco~Jr}},
  \bibinfo{author}{\bibfnamefont{C.}~\bibnamefont{Ojeda-Aristizabal}},
  \bibinfo{author}{\bibfnamefont{H.~A.} \bibnamefont{Bechtel}},
  \bibinfo{author}{\bibfnamefont{M.~C.} \bibnamefont{Martin}},
  \bibinfo{author}{\bibfnamefont{A.}~\bibnamefont{Zettl}},
  \bibnamefont{et~al.}, {``}\bibinfo{title}{Topological valley transport at
  bilayer graphene domain walls},{''} \bibinfo{journal}{Nature}
  \textbf{\bibinfo{volume}{520}}, \bibinfo{pages}{650} (\bibinfo{year}{2015}).

\bibitem[{\citenamefont{Li et~al.}(2011)\citenamefont{Li, Martin,
  B{\"{u}}ttiker, and Morpurgo}}]{Li2011}
\bibinfo{author}{\bibfnamefont{J.}~\bibnamefont{Li}},
  \bibinfo{author}{\bibfnamefont{I.}~\bibnamefont{Martin}},
  \bibinfo{author}{\bibfnamefont{M.}~\bibnamefont{B{\"{u}}ttiker}},
  \bibnamefont{and} \bibinfo{author}{\bibfnamefont{A.~F.}
  \bibnamefont{Morpurgo}}, {``}\bibinfo{title}{Topological origin of subgap
  conductance in insulating bilayer graphene},{''} \bibinfo{journal}{Nat.
  Phys.} \textbf{\bibinfo{volume}{7}}, \bibinfo{pages}{38}
  (\bibinfo{year}{2011}).

\bibitem[{\citenamefont{Nandkishore and
  Levitov}(2011{\natexlab{a}})}]{Nandkishore2011}
\bibinfo{author}{\bibfnamefont{R.}~\bibnamefont{Nandkishore}} \bibnamefont{and}
  \bibinfo{author}{\bibfnamefont{L.}~\bibnamefont{Levitov}},
  {``}\bibinfo{title}{Common-path interference and oscillatory Zener tunneling
  in bilayer graphene p-n junctions},{''} \bibinfo{journal}{PNAS}
  \textbf{\bibinfo{volume}{108}}, \bibinfo{pages}{14021}
  (\bibinfo{year}{2011}{\natexlab{a}}).

\bibitem[{\citenamefont{Gradinar et~al.}(2012)\citenamefont{Gradinar,
  Schomerus, and Fal'ko}}]{gradinar_pn_bilayer2012}
\bibinfo{author}{\bibfnamefont{D.~A.} \bibnamefont{Gradinar}},
  \bibinfo{author}{\bibfnamefont{H.}~\bibnamefont{Schomerus}},
  \bibnamefont{and} \bibinfo{author}{\bibfnamefont{V.~I.}
  \bibnamefont{Fal'ko}}, {``}\bibinfo{title}{Conductance anomaly near the
  Lifshitz transition in strained bilayer graphene},{''}
  \bibinfo{journal}{Phys. Rev. B} \textbf{\bibinfo{volume}{85}},
  \bibinfo{pages}{165429} (\bibinfo{year}{2012}).

\bibitem[{\citenamefont{Van~Duppen and
  Peeters}(2013)}]{four_band_tunn_duppen2013}
\bibinfo{author}{\bibfnamefont{B.}~\bibnamefont{Van~Duppen}} \bibnamefont{and}
  \bibinfo{author}{\bibfnamefont{F.~M.} \bibnamefont{Peeters}},
  {``}\bibinfo{title}{Four-band tunneling in bilayer graphene},{''}
  \bibinfo{journal}{Phys. Rev. B} \textbf{\bibinfo{volume}{87}},
  \bibinfo{pages}{205427} (\bibinfo{year}{2013}).

\bibitem[{\citenamefont{Kleptsyn et~al.}(2015)\citenamefont{Kleptsyn, Okunev,
  Schurov, Zubov, and Katsnelson}}]{kleptsyn_no_magic2015}
\bibinfo{author}{\bibfnamefont{V.}~\bibnamefont{Kleptsyn}},
  \bibinfo{author}{\bibfnamefont{A.}~\bibnamefont{Okunev}},
  \bibinfo{author}{\bibfnamefont{I.}~\bibnamefont{Schurov}},
  \bibinfo{author}{\bibfnamefont{D.}~\bibnamefont{Zubov}}, \bibnamefont{and}
  \bibinfo{author}{\bibfnamefont{M.~I.} \bibnamefont{Katsnelson}},
  {``}\bibinfo{title}{Chiral tunneling through generic one-dimensional
  potential barriers in bilayer graphene},{''} \bibinfo{journal}{Phys. Rev. B}
  \textbf{\bibinfo{volume}{92}}, \bibinfo{pages}{165407}
  (\bibinfo{year}{2015}).

\bibitem[{\citenamefont{Park}(2015)}]{smooth_barrier2015}
\bibinfo{author}{\bibfnamefont{C.-S.} \bibnamefont{Park}},
  {``}\bibinfo{title}{Two-dimensional transmission through modified
  P\"oschl-Teller potential in bilayer graphene},{''} \bibinfo{journal}{Phys.
  Rev. B} \textbf{\bibinfo{volume}{92}}, \bibinfo{pages}{165422}
  (\bibinfo{year}{2015}).

\bibitem[{\citenamefont{Ryzhii and Ryzhii}(2009)}]{Ryzhii2009}
\bibinfo{author}{\bibfnamefont{V.}~\bibnamefont{Ryzhii}} \bibnamefont{and}
  \bibinfo{author}{\bibfnamefont{M.}~\bibnamefont{Ryzhii}},
  {``}\bibinfo{title}{Graphene bilayer field-effect phototransistor for
  terahertz and infrared detection},{''} \bibinfo{journal}{Phys. Rev. B}
  \textbf{\bibinfo{volume}{79}}, \bibinfo{pages}{245311}
  (\bibinfo{year}{2009}).

\bibitem[{\citenamefont{Spirito et~al.}(2014)\citenamefont{Spirito, Coquillat,
  De~Bonis, Lombardo, Bruna, Ferrari, Pellegrini, Tredicucci, Knap, and
  Vitiello}}]{terahertz_exp2014}
\bibinfo{author}{\bibfnamefont{D.}~\bibnamefont{Spirito}},
  \bibinfo{author}{\bibfnamefont{D.}~\bibnamefont{Coquillat}},
  \bibinfo{author}{\bibfnamefont{S.~L.} \bibnamefont{De~Bonis}},
  \bibinfo{author}{\bibfnamefont{A.}~\bibnamefont{Lombardo}},
  \bibinfo{author}{\bibfnamefont{M.}~\bibnamefont{Bruna}},
  \bibinfo{author}{\bibfnamefont{A.~C.} \bibnamefont{Ferrari}},
  \bibinfo{author}{\bibfnamefont{V.}~\bibnamefont{Pellegrini}},
  \bibinfo{author}{\bibfnamefont{A.}~\bibnamefont{Tredicucci}},
  \bibinfo{author}{\bibfnamefont{W.}~\bibnamefont{Knap}}, \bibnamefont{and}
  \bibinfo{author}{\bibfnamefont{M.~S.} \bibnamefont{Vitiello}},
  {``}\bibinfo{title}{High performance bilayer-graphene terahertz
  detectors},{''} \bibinfo{journal}{Appl. Phys. Lett.}
  \textbf{\bibinfo{volume}{104}}, \bibinfo{eid}{061111} (\bibinfo{year}{2014}).

\bibitem[{\citenamefont{Chakraborty et~al.}(2009)\citenamefont{Chakraborty,
  Das, and Sood}}]{Chakraborty2009}
\bibinfo{author}{\bibfnamefont{B.}~\bibnamefont{Chakraborty}},
  \bibinfo{author}{\bibfnamefont{A.}~\bibnamefont{Das}}, \bibnamefont{and}
  \bibinfo{author}{\bibfnamefont{A.~K.} \bibnamefont{Sood}},
  {``}\bibinfo{title}{The formation of a p-n junction in a polymer electrolyte
  top-gated bilayer graphene transistor},{''} \bibinfo{journal}{Nanotechnology}
  \textbf{\bibinfo{volume}{20}}, \bibinfo{pages}{365203}
  (\bibinfo{year}{2009}).

\bibitem[{\citenamefont{Szafranek et~al.}(2010)\citenamefont{Szafranek, Schall,
  Otto, Neumaier, and Kurz}}]{Szafranek2010}
\bibinfo{author}{\bibfnamefont{B.~N.} \bibnamefont{Szafranek}},
  \bibinfo{author}{\bibfnamefont{D.}~\bibnamefont{Schall}},
  \bibinfo{author}{\bibfnamefont{M.}~\bibnamefont{Otto}},
  \bibinfo{author}{\bibfnamefont{D.}~\bibnamefont{Neumaier}}, \bibnamefont{and}
  \bibinfo{author}{\bibfnamefont{H.}~\bibnamefont{Kurz}},
  {``}\bibinfo{title}{Electrical observation of a tunable band gap in bilayer
  graphene nanoribbons at room temperature},{''} \bibinfo{journal}{Appl. Phys.
  Lett.} \textbf{\bibinfo{volume}{96}}, \bibinfo{pages}{112103}
  (\bibinfo{year}{2010}).

\bibitem[{\citenamefont{Szafranek et~al.}(2011)\citenamefont{Szafranek, Schall,
  Otto, Neumaier, and Kurz}}]{Szafranek2011}
\bibinfo{author}{\bibfnamefont{B.~N.} \bibnamefont{Szafranek}},
  \bibinfo{author}{\bibfnamefont{D.}~\bibnamefont{Schall}},
  \bibinfo{author}{\bibfnamefont{M.}~\bibnamefont{Otto}},
  \bibinfo{author}{\bibfnamefont{D.}~\bibnamefont{Neumaier}}, \bibnamefont{and}
  \bibinfo{author}{\bibfnamefont{H.}~\bibnamefont{Kurz}},
  {``}\bibinfo{title}{High On/Off Ratios in Bilayer Graphene Field Effect
  Transistors Realized by Surface Dopants},{''} \bibinfo{journal}{Nano Lett.}
  \textbf{\bibinfo{volume}{11}}, \bibinfo{pages}{2640} (\bibinfo{year}{2011}).

\bibitem[{\citenamefont{Szafranek et~al.}(2012)\citenamefont{Szafranek, Fiori,
  Schall, Neumaier, and Kurz}}]{Szafranek2012}
\bibinfo{author}{\bibfnamefont{B.~N.} \bibnamefont{Szafranek}},
  \bibinfo{author}{\bibfnamefont{G.}~\bibnamefont{Fiori}},
  \bibinfo{author}{\bibfnamefont{D.}~\bibnamefont{Schall}},
  \bibinfo{author}{\bibfnamefont{D.}~\bibnamefont{Neumaier}}, \bibnamefont{and}
  \bibinfo{author}{\bibfnamefont{H.}~\bibnamefont{Kurz}},
  {``}\bibinfo{title}{Current Saturation and Voltage Gain in Bilayer Graphene
  Field Effect Transistors},{''} \bibinfo{journal}{Nano Lett.}
  \textbf{\bibinfo{volume}{12}}, \bibinfo{pages}{1324} (\bibinfo{year}{2012}).

\bibitem[{\citenamefont{Alymov et~al.}(2016)\citenamefont{Alymov, Vyurkov,
  Ryzhii, and Svintsov}}]{Alymov2016}
\bibinfo{author}{\bibfnamefont{G.}~\bibnamefont{Alymov}},
  \bibinfo{author}{\bibfnamefont{V.}~\bibnamefont{Vyurkov}},
  \bibinfo{author}{\bibfnamefont{V.}~\bibnamefont{Ryzhii}}, \bibnamefont{and}
  \bibinfo{author}{\bibfnamefont{D.}~\bibnamefont{Svintsov}},
  {``}\bibinfo{title}{Abrupt current switching in graphene bilayer tunnel
  transistors enabled by van Hove singularities},{''} \bibinfo{journal}{Sci.
  Rep.} \textbf{\bibinfo{volume}{6}}, \bibinfo{pages}{24654}
  (\bibinfo{year}{2016}).

\bibitem[{\citenamefont{Miyazaki et~al.}(2012)\citenamefont{Miyazaki, Li,
  Nakaharai, and Tsukagoshi}}]{pn_miyazaki2012}
\bibinfo{author}{\bibfnamefont{H.}~\bibnamefont{Miyazaki}},
  \bibinfo{author}{\bibfnamefont{S.-L.} \bibnamefont{Li}},
  \bibinfo{author}{\bibfnamefont{S.}~\bibnamefont{Nakaharai}},
  \bibnamefont{and}
  \bibinfo{author}{\bibfnamefont{K.}~\bibnamefont{Tsukagoshi}},
  {``}\bibinfo{title}{Unipolar transport in bilayer graphene controlled by
  multiple p-n interfaces},{''} \bibinfo{journal}{Appl. Phys. Lett.}
  \textbf{\bibinfo{volume}{100}}, \bibinfo{eid}{163115} (\bibinfo{year}{2012}).

\bibitem[{\citenamefont{Kanayama and Nagashio}(2015)}]{bilayer_gap_exp2015}
\bibinfo{author}{\bibfnamefont{K.}~\bibnamefont{Kanayama}} \bibnamefont{and}
  \bibinfo{author}{\bibfnamefont{K.}~\bibnamefont{Nagashio}},
  {``}\bibinfo{title}{Gap state analysis in electric-field-induced band gap for
  bilayer graphene},{''} \bibinfo{journal}{Sci. Rep.}
  \textbf{\bibinfo{volume}{5}}, \bibinfo{pages}{15789} (\bibinfo{year}{2015}).

\bibitem[{\citenamefont{Koshino and Ando}(2006)}]{Koshino2006}
\bibinfo{author}{\bibfnamefont{M.}~\bibnamefont{Koshino}} \bibnamefont{and}
  \bibinfo{author}{\bibfnamefont{T.}~\bibnamefont{Ando}},
  {``}\bibinfo{title}{Transport in bilayer graphene: Calculations within a
  self-consistent Born approximation},{''} \bibinfo{journal}{Phys. Rev. B}
  \textbf{\bibinfo{volume}{73}}, \bibinfo{pages}{245403}
  (\bibinfo{year}{2006}).

\bibitem[{\citenamefont{Koshino}(2009)}]{Koshino2009}
\bibinfo{author}{\bibfnamefont{M.}~\bibnamefont{Koshino}},
  {``}\bibinfo{title}{Electronic transport in bilayer graphene},{''}
  \bibinfo{journal}{New J. Phys.} \textbf{\bibinfo{volume}{11}},
  \bibinfo{pages}{095010} (\bibinfo{year}{2009}).

\bibitem[{\citenamefont{Ando}(2011)}]{ando_transport2011}
\bibinfo{author}{\bibfnamefont{T.}~\bibnamefont{Ando}},
  {``}\bibinfo{title}{Bilayer Graphene with Long-range Scatterers Studied in a
  Self-Consistent Born Approximation},{''} \bibinfo{journal}{J. Phys. Soc.
  Japan} \textbf{\bibinfo{volume}{80}}, \bibinfo{pages}{014707}
  (\bibinfo{year}{2011}).

\bibitem[{\citenamefont{Trushin et~al.}(2010)\citenamefont{Trushin,
  Kailasvuori, Schliemann, and MacDonald}}]{Trushin2010}
\bibinfo{author}{\bibfnamefont{M.}~\bibnamefont{Trushin}},
  \bibinfo{author}{\bibfnamefont{J.}~\bibnamefont{Kailasvuori}},
  \bibinfo{author}{\bibfnamefont{J.}~\bibnamefont{Schliemann}},
  \bibnamefont{and} \bibinfo{author}{\bibfnamefont{A.~H.}
  \bibnamefont{MacDonald}}, {``}\bibinfo{title}{Finite conductivity minimum in
  bilayer graphene without charge inhomogeneities},{''} \bibinfo{journal}{Phys.
  Rev. B} \textbf{\bibinfo{volume}{82}}, \bibinfo{pages}{155308}
  (\bibinfo{year}{2010}).

\bibitem[{\citenamefont{Trushin and
  Schliemann}(2007)}]{trushin_single_layer2007}
\bibinfo{author}{\bibfnamefont{M.}~\bibnamefont{Trushin}} \bibnamefont{and}
  \bibinfo{author}{\bibfnamefont{J.}~\bibnamefont{Schliemann}},
  {``}\bibinfo{title}{Minimum Electrical and Thermal Conductivity of Graphene:
  A Quasiclassical Approach},{''} \bibinfo{journal}{Phys. Rev. Lett.}
  \textbf{\bibinfo{volume}{99}}, \bibinfo{pages}{216602}
  (\bibinfo{year}{2007}).

\bibitem[{\citenamefont{Auslender and
  Katsnelson}(2007)}]{auslander_single_layer2007}
\bibinfo{author}{\bibfnamefont{M.}~\bibnamefont{Auslender}} \bibnamefont{and}
  \bibinfo{author}{\bibfnamefont{M.~I.} \bibnamefont{Katsnelson}},
  {``}\bibinfo{title}{Generalized kinetic equations for charge carriers in
  graphene},{''} \bibinfo{journal}{Phys. Rev. B} \textbf{\bibinfo{volume}{76}},
  \bibinfo{pages}{235425} (\bibinfo{year}{2007}).

\bibitem[{\citenamefont{Nilsson
  et~al.}(2006{\natexlab{a}})\citenamefont{Nilsson, Castro~Neto, Guinea, and
  Peres}}]{Nilsson2006_transp}
\bibinfo{author}{\bibfnamefont{J.}~\bibnamefont{Nilsson}},
  \bibinfo{author}{\bibfnamefont{A.~H.} \bibnamefont{Castro~Neto}},
  \bibinfo{author}{\bibfnamefont{F.}~\bibnamefont{Guinea}}, \bibnamefont{and}
  \bibinfo{author}{\bibfnamefont{N.~M.~R.} \bibnamefont{Peres}},
  {``}\bibinfo{title}{Electronic Properties of Graphene Multilayers},{''}
  \bibinfo{journal}{Phys. Rev. Lett.} \textbf{\bibinfo{volume}{97}},
  \bibinfo{pages}{266801} (\bibinfo{year}{2006}{\natexlab{a}}).

\bibitem[{\citenamefont{Adam and Das~Sarma}(2008)}]{Adam2008}
\bibinfo{author}{\bibfnamefont{S.}~\bibnamefont{Adam}} \bibnamefont{and}
  \bibinfo{author}{\bibfnamefont{S.}~\bibnamefont{Das~Sarma}},
  {``}\bibinfo{title}{Boltzmann transport and residual conductivity in bilayer
  graphene},{''} \bibinfo{journal}{Phys. Rev. B} \textbf{\bibinfo{volume}{77}},
  \bibinfo{pages}{115436} (\bibinfo{year}{2008}).

\bibitem[{\citenamefont{Das~Sarma et~al.}(2010)\citenamefont{Das~Sarma, Hwang,
  and Rossi}}]{DasSarma2010}
\bibinfo{author}{\bibfnamefont{S.}~\bibnamefont{Das~Sarma}},
  \bibinfo{author}{\bibfnamefont{E.~H.} \bibnamefont{Hwang}}, \bibnamefont{and}
  \bibinfo{author}{\bibfnamefont{E.}~\bibnamefont{Rossi}},
  {``}\bibinfo{title}{Theory of carrier transport in bilayer graphene},{''}
  \bibinfo{journal}{Phys. Rev. B} \textbf{\bibinfo{volume}{81}},
  \bibinfo{pages}{161407} (\bibinfo{year}{2010}).

\bibitem[{\citenamefont{Xiao et~al.}(2010)\citenamefont{Xiao, Chen, Adam,
  Williams, and Fuhrer}}]{Xiao2010}
\bibinfo{author}{\bibfnamefont{S.}~\bibnamefont{Xiao}},
  \bibinfo{author}{\bibfnamefont{J.-H.} \bibnamefont{Chen}},
  \bibinfo{author}{\bibfnamefont{S.}~\bibnamefont{Adam}},
  \bibinfo{author}{\bibfnamefont{E.~D.} \bibnamefont{Williams}},
  \bibnamefont{and} \bibinfo{author}{\bibfnamefont{M.~S.}
  \bibnamefont{Fuhrer}}, {``}\bibinfo{title}{Charged impurity scattering in
  bilayer graphene},{''} \bibinfo{journal}{Phys. Rev. B}
  \textbf{\bibinfo{volume}{82}}, \bibinfo{pages}{041406}
  (\bibinfo{year}{2010}).

\bibitem[{\citenamefont{Lv and Wan}(2010)}]{Lv2010}
\bibinfo{author}{\bibfnamefont{M.}~\bibnamefont{Lv}} \bibnamefont{and}
  \bibinfo{author}{\bibfnamefont{S.}~\bibnamefont{Wan}},
  {``}\bibinfo{title}{Screening-induced transport at finite temperature in
  bilayer graphene},{''} \bibinfo{journal}{Phys. Rev. B}
  \textbf{\bibinfo{volume}{81}}, \bibinfo{pages}{195409}
  (\bibinfo{year}{2010}).

\bibitem[{\citenamefont{Ferreira et~al.}(2011)\citenamefont{Ferreira,
  Viana-Gomes, Nilsson, Mucciolo, Peres, and Castro~Neto}}]{Ferreira2011}
\bibinfo{author}{\bibfnamefont{A.}~\bibnamefont{Ferreira}},
  \bibinfo{author}{\bibfnamefont{J.}~\bibnamefont{Viana-Gomes}},
  \bibinfo{author}{\bibfnamefont{J.}~\bibnamefont{Nilsson}},
  \bibinfo{author}{\bibfnamefont{E.~R.} \bibnamefont{Mucciolo}},
  \bibinfo{author}{\bibfnamefont{N.~M.~R.} \bibnamefont{Peres}},
  \bibnamefont{and} \bibinfo{author}{\bibfnamefont{A.~H.}
  \bibnamefont{Castro~Neto}}, {``}\bibinfo{title}{Unified description of the dc
  conductivity of monolayer and bilayer graphene at finite densities based on
  resonant scatterers},{''} \bibinfo{journal}{Phys. Rev. B}
  \textbf{\bibinfo{volume}{83}}, \bibinfo{pages}{165402}
  (\bibinfo{year}{2011}).

\bibitem[{\citenamefont{Ni et~al.}(2010)\citenamefont{Ni, Ponomarenko, Nair,
  Yang, Anissimova, Grigorieva, Schedin, Blake, Shen, Hill et~al.}}]{Ni2010}
\bibinfo{author}{\bibfnamefont{Z.~H.} \bibnamefont{Ni}},
  \bibinfo{author}{\bibfnamefont{L.~A.} \bibnamefont{Ponomarenko}},
  \bibinfo{author}{\bibfnamefont{R.~R.} \bibnamefont{Nair}},
  \bibinfo{author}{\bibfnamefont{R.}~\bibnamefont{Yang}},
  \bibinfo{author}{\bibfnamefont{S.}~\bibnamefont{Anissimova}},
  \bibinfo{author}{\bibfnamefont{I.~V.} \bibnamefont{Grigorieva}},
  \bibinfo{author}{\bibfnamefont{F.}~\bibnamefont{Schedin}},
  \bibinfo{author}{\bibfnamefont{P.}~\bibnamefont{Blake}},
  \bibinfo{author}{\bibfnamefont{Z.~X.} \bibnamefont{Shen}},
  \bibinfo{author}{\bibfnamefont{E.~H.} \bibnamefont{Hill}},
  \bibnamefont{et~al.}, {``}\bibinfo{title}{On Resonant Scatterers As a Factor
  Limiting Carrier Mobility in Graphene},{''} \bibinfo{journal}{Nano Lett.}
  \textbf{\bibinfo{volume}{10}}, \bibinfo{pages}{3868} (\bibinfo{year}{2010}).

\bibitem[{\citenamefont{Katoch et~al.}(2010)\citenamefont{Katoch, Chen,
  Tsuchikawa, Smith, Mucciolo, and Ishigami}}]{Katoch2010}
\bibinfo{author}{\bibfnamefont{J.}~\bibnamefont{Katoch}},
  \bibinfo{author}{\bibfnamefont{J.-H.} \bibnamefont{Chen}},
  \bibinfo{author}{\bibfnamefont{R.}~\bibnamefont{Tsuchikawa}},
  \bibinfo{author}{\bibfnamefont{C.}~\bibnamefont{Smith}},
  \bibinfo{author}{\bibfnamefont{E.}~\bibnamefont{Mucciolo}}, \bibnamefont{and}
  \bibinfo{author}{\bibfnamefont{M.}~\bibnamefont{Ishigami}},
  {``}\bibinfo{title}{Uncovering the dominant scatterer in graphene sheets on
  ${\text{SiO}}_{2}$},{''} \bibinfo{journal}{Phys. Rev. B}
  \textbf{\bibinfo{volume}{82}}, \bibinfo{pages}{081417}
  (\bibinfo{year}{2010}).

\bibitem[{\citenamefont{Monteverde et~al.}(2010)\citenamefont{Monteverde,
  Ojeda-Aristizabal, Weil, Bennaceur, Ferrier, Gueron, Glattli, Bouchiat,
  Fuchs, and Maslov}}]{Monteverde2010}
\bibinfo{author}{\bibfnamefont{M.}~\bibnamefont{Monteverde}},
  \bibinfo{author}{\bibfnamefont{C.}~\bibnamefont{Ojeda-Aristizabal}},
  \bibinfo{author}{\bibfnamefont{R.}~\bibnamefont{Weil}},
  \bibinfo{author}{\bibfnamefont{K.}~\bibnamefont{Bennaceur}},
  \bibinfo{author}{\bibfnamefont{M.}~\bibnamefont{Ferrier}},
  \bibinfo{author}{\bibfnamefont{S.}~\bibnamefont{Gueron}},
  \bibinfo{author}{\bibfnamefont{C.}~\bibnamefont{Glattli}},
  \bibinfo{author}{\bibfnamefont{H.}~\bibnamefont{Bouchiat}},
  \bibinfo{author}{\bibfnamefont{J.~N.} \bibnamefont{Fuchs}}, \bibnamefont{and}
  \bibinfo{author}{\bibfnamefont{D.~L.} \bibnamefont{Maslov}},
  {``}\bibinfo{title}{Transport and Elastic Scattering Times as Probes of the
  Nature of Impurity Scattering in Single-Layer and Bilayer Graphene},{''}
  \bibinfo{journal}{Phys. Rev. Lett.} \textbf{\bibinfo{volume}{104}},
  \bibinfo{pages}{126801} (\bibinfo{year}{2010}).

\bibitem[{\citenamefont{Yuan et~al.}(2010{\natexlab{a}})\citenamefont{Yuan,
  De~Raedt, and Katsnelson}}]{Yuan2010}
\bibinfo{author}{\bibfnamefont{S.}~\bibnamefont{Yuan}},
  \bibinfo{author}{\bibfnamefont{H.}~\bibnamefont{De~Raedt}}, \bibnamefont{and}
  \bibinfo{author}{\bibfnamefont{M.~I.} \bibnamefont{Katsnelson}},
  {``}\bibinfo{title}{Electronic transport in disordered bilayer and trilayer
  graphene},{''} \bibinfo{journal}{Phys. Rev. B} \textbf{\bibinfo{volume}{82}},
  \bibinfo{pages}{235409} (\bibinfo{year}{2010}{\natexlab{a}}).

\bibitem[{\citenamefont{Yuan et~al.}(2010{\natexlab{b}})\citenamefont{Yuan,
  De~Raedt, and Katsnelson}}]{time-evol}
\bibinfo{author}{\bibfnamefont{S.}~\bibnamefont{Yuan}},
  \bibinfo{author}{\bibfnamefont{H.}~\bibnamefont{De~Raedt}}, \bibnamefont{and}
  \bibinfo{author}{\bibfnamefont{M.}~\bibnamefont{Katsnelson}},
  {``}\bibinfo{title}{Modeling electronic structure and transport properties of
  graphene with resonant scattering centers},{''} \bibinfo{journal}{Phys. Rev.
  B} \textbf{\bibinfo{volume}{82}}, \bibinfo{pages}{115448}
  (\bibinfo{year}{2010}{\natexlab{b}}).

\bibitem[{\citenamefont{Zhu et~al.}(2009)\citenamefont{Zhu, Perebeinos,
  Freitag, and Avouris}}]{Zhu2009}
\bibinfo{author}{\bibfnamefont{W.}~\bibnamefont{Zhu}},
  \bibinfo{author}{\bibfnamefont{V.}~\bibnamefont{Perebeinos}},
  \bibinfo{author}{\bibfnamefont{M.}~\bibnamefont{Freitag}}, \bibnamefont{and}
  \bibinfo{author}{\bibfnamefont{P.}~\bibnamefont{Avouris}},
  {``}\bibinfo{title}{Carrier scattering, mobilities, and electrostatic
  potential in monolayer, bilayer, and trilayer graphene},{''}
  \bibinfo{journal}{Phys. Rev. B} \textbf{\bibinfo{volume}{80}},
  \bibinfo{pages}{235402} (\bibinfo{year}{2009}).

\bibitem[{\citenamefont{Kisslinger et~al.}(2015)\citenamefont{Kisslinger, Ott,
  Heide, Kampert, Butz, Spiecker, Shallcross, and Weber}}]{Kissl}
\bibinfo{author}{\bibfnamefont{F.}~\bibnamefont{Kisslinger}},
  \bibinfo{author}{\bibfnamefont{C.}~\bibnamefont{Ott}},
  \bibinfo{author}{\bibfnamefont{C.}~\bibnamefont{Heide}},
  \bibinfo{author}{\bibfnamefont{E.}~\bibnamefont{Kampert}},
  \bibinfo{author}{\bibfnamefont{B.}~\bibnamefont{Butz}},
  \bibinfo{author}{\bibfnamefont{E.}~\bibnamefont{Spiecker}},
  \bibinfo{author}{\bibfnamefont{S.}~\bibnamefont{Shallcross}},
  \bibnamefont{and} \bibinfo{author}{\bibfnamefont{H.~B.} \bibnamefont{Weber}},
  {``}\bibinfo{title}{Linear magnetoresistance in mosaic-like bilayer
  graphene},{''} \bibinfo{journal}{Nat. Phys.} \textbf{\bibinfo{volume}{11}},
  \bibinfo{pages}{650} (\bibinfo{year}{2015}).

\bibitem[{\citenamefont{Hwang et~al.}(2007)\citenamefont{Hwang, Adam, and
  Das~Sarma}}]{Hwang_puddles2007}
\bibinfo{author}{\bibfnamefont{E.}~\bibnamefont{Hwang}},
  \bibinfo{author}{\bibfnamefont{S.}~\bibnamefont{Adam}}, \bibnamefont{and}
  \bibinfo{author}{\bibfnamefont{S.}~\bibnamefont{Das~Sarma}},
  {``}\bibinfo{title}{Carrier Transport in Two-Dimensional Graphene
  Layers},{''} \bibinfo{journal}{Phys. Rev. Lett.}
  \textbf{\bibinfo{volume}{98}}, \bibinfo{pages}{186806}
  (\bibinfo{year}{2007}).

\bibitem[{\citenamefont{Martin et~al.}(2008{\natexlab{b}})\citenamefont{Martin,
  Akerman, Ulbricht, Lohmann, Smet, von Klitzing, and
  Yacoby}}]{single_layer_puddles2008}
\bibinfo{author}{\bibfnamefont{J.}~\bibnamefont{Martin}},
  \bibinfo{author}{\bibfnamefont{N.}~\bibnamefont{Akerman}},
  \bibinfo{author}{\bibfnamefont{G.}~\bibnamefont{Ulbricht}},
  \bibinfo{author}{\bibfnamefont{T.}~\bibnamefont{Lohmann}},
  \bibinfo{author}{\bibfnamefont{J.~H.} \bibnamefont{Smet}},
  \bibinfo{author}{\bibfnamefont{K.}~\bibnamefont{von Klitzing}},
  \bibnamefont{and} \bibinfo{author}{\bibfnamefont{A.}~\bibnamefont{Yacoby}},
  {``}\bibinfo{title}{Observation of electron-hole puddles in graphene using a
  scanning single-electron transistor},{''} \bibinfo{journal}{Nat. Phys.}
  \textbf{\bibinfo{volume}{4}}, \bibinfo{pages}{144}
  (\bibinfo{year}{2008}{\natexlab{b}}).

\bibitem[{\citenamefont{Adam and Stiles}(2010)}]{Adam2010}
\bibinfo{author}{\bibfnamefont{S.}~\bibnamefont{Adam}} \bibnamefont{and}
  \bibinfo{author}{\bibfnamefont{M.~D.} \bibnamefont{Stiles}},
  {``}\bibinfo{title}{Temperature dependence of the diffusive conductivity of
  bilayer graphene},{''} \bibinfo{journal}{Phys. Rev. B}
  \textbf{\bibinfo{volume}{82}}, \bibinfo{pages}{075423}
  (\bibinfo{year}{2010}).

\bibitem[{\citenamefont{Hwang and Das~Sarma}(2010)}]{Hwang2010}
\bibinfo{author}{\bibfnamefont{E.~H.} \bibnamefont{Hwang}} \bibnamefont{and}
  \bibinfo{author}{\bibfnamefont{S.}~\bibnamefont{Das~Sarma}},
  {``}\bibinfo{title}{Insulating behavior in metallic bilayer graphene:
  Interplay between density inhomogeneity and temperature},{''}
  \bibinfo{journal}{Phys. Rev. B} \textbf{\bibinfo{volume}{82}},
  \bibinfo{pages}{081409} (\bibinfo{year}{2010}).

\bibitem[{\citenamefont{Katsnelson}(2006)}]{Katsnelson2006}
\bibinfo{author}{\bibfnamefont{M.~I.} \bibnamefont{Katsnelson}},
  {``}\bibinfo{title}{Minimal conductivity in bilayer graphene},{''}
  \bibinfo{journal}{Eur. Phys. J. B} \textbf{\bibinfo{volume}{52}},
  \bibinfo{pages}{151} (\bibinfo{year}{2006}).

\bibitem[{\citenamefont{Cserti}(2007)}]{Cserti2007}
\bibinfo{author}{\bibfnamefont{J.}~\bibnamefont{Cserti}},
  {``}\bibinfo{title}{Minimal longitudinal dc conductivity of perfect bilayer
  graphene},{''} \bibinfo{journal}{Phys. Rev. B} \textbf{\bibinfo{volume}{75}},
  \bibinfo{pages}{033405} (\bibinfo{year}{2007}).

\bibitem[{\citenamefont{Snyman and Beenakker}(2007)}]{Snyman2007}
\bibinfo{author}{\bibfnamefont{I.}~\bibnamefont{Snyman}} \bibnamefont{and}
  \bibinfo{author}{\bibfnamefont{C.~W.~J.} \bibnamefont{Beenakker}},
  {``}\bibinfo{title}{Ballistic transmission through a graphene bilayer},{''}
  \bibinfo{journal}{Phys. Rev. B} \textbf{\bibinfo{volume}{75}},
  \bibinfo{pages}{045322} (\bibinfo{year}{2007}).

\bibitem[{\citenamefont{Cserti et~al.}(2007)\citenamefont{Cserti, Csordas, and
  David}}]{Cserti_trig2007}
\bibinfo{author}{\bibfnamefont{J.}~\bibnamefont{Cserti}},
  \bibinfo{author}{\bibfnamefont{A.}~\bibnamefont{Csordas}}, \bibnamefont{and}
  \bibinfo{author}{\bibfnamefont{G.}~\bibnamefont{David}},
  {``}\bibinfo{title}{Role of the trigonal warping on the minimal conductivity
  of bilayer graphene},{''} \bibinfo{journal}{Phys. Rev. Lett.}
  \textbf{\bibinfo{volume}{99}}, \bibinfo{pages}{066802}
  (\bibinfo{year}{2007}).

\bibitem[{\citenamefont{San-Jose et~al.}(2014)\citenamefont{San-Jose,
  Gorbachev, Geim, Novoselov, and Guinea}}]{san2014stacking}
\bibinfo{author}{\bibfnamefont{P.}~\bibnamefont{San-Jose}},
  \bibinfo{author}{\bibfnamefont{R.}~\bibnamefont{Gorbachev}},
  \bibinfo{author}{\bibfnamefont{A.}~\bibnamefont{Geim}},
  \bibinfo{author}{\bibfnamefont{K.}~\bibnamefont{Novoselov}},
  \bibnamefont{and} \bibinfo{author}{\bibfnamefont{F.}~\bibnamefont{Guinea}},
  {``}\bibinfo{title}{Stacking boundaries and transport in bilayer
  graphene},{''} \bibinfo{journal}{Nano Lett.} \textbf{\bibinfo{volume}{14}},
  \bibinfo{pages}{2052} (\bibinfo{year}{2014}).

\bibitem[{\citenamefont{Nilsson and Castro~Neto}(2007)}]{Nilsson2007}
\bibinfo{author}{\bibfnamefont{J.}~\bibnamefont{Nilsson}} \bibnamefont{and}
  \bibinfo{author}{\bibfnamefont{A.~H.} \bibnamefont{Castro~Neto}},
  {``}\bibinfo{title}{Impurities in a biased graphene bilayer},{''}
  \bibinfo{journal}{Phys. Rev. Lett.} \textbf{\bibinfo{volume}{98}},
  \bibinfo{pages}{126801} (\bibinfo{year}{2007}).

\bibitem[{\citenamefont{Zou and Zhu}(2010)}]{Zou2010}
\bibinfo{author}{\bibfnamefont{K.}~\bibnamefont{Zou}} \bibnamefont{and}
  \bibinfo{author}{\bibfnamefont{J.}~\bibnamefont{Zhu}},
  {``}\bibinfo{title}{Transport in gapped bilayer graphene: The role of
  potential fluctuations},{''} \bibinfo{journal}{Phys. Rev. B}
  \textbf{\bibinfo{volume}{82}}, \bibinfo{pages}{081407}
  (\bibinfo{year}{2010}).

\bibitem[{\citenamefont{Taychatanapat and
  Jarillo-Herrero}(2010)}]{Taychatanapat2010}
\bibinfo{author}{\bibfnamefont{T.}~\bibnamefont{Taychatanapat}}
  \bibnamefont{and}
  \bibinfo{author}{\bibfnamefont{P.}~\bibnamefont{Jarillo-Herrero}},
  {``}\bibinfo{title}{Electronic Transport in Dual-Gated Bilayer Graphene at
  Large Displacement Fields},{''} \bibinfo{journal}{Phys. Rev. Lett.}
  \textbf{\bibinfo{volume}{105}}, \bibinfo{pages}{166601}
  (\bibinfo{year}{2010}).

\bibitem[{\citenamefont{Rossi and Das~Sarma}(2011)}]{rossi2011}
\bibinfo{author}{\bibfnamefont{E.}~\bibnamefont{Rossi}} \bibnamefont{and}
  \bibinfo{author}{\bibfnamefont{S.}~\bibnamefont{Das~Sarma}},
  {``}\bibinfo{title}{Inhomogenous Electronic Structure, Transport Gap, and
  Percolation Threshold in Disordered Bilayer Graphene},{''}
  \bibinfo{journal}{Phys. Rev. Lett.} \textbf{\bibinfo{volume}{107}},
  \bibinfo{pages}{155502} (\bibinfo{year}{2011}).

\bibitem[{\citenamefont{Trushin}(2012)}]{trushin2012}
\bibinfo{author}{\bibfnamefont{M.}~\bibnamefont{Trushin}},
  {``}\bibinfo{title}{Thermally activated conductivity in gapped bilayer
  graphene},{''} \bibinfo{journal}{EPL} \textbf{\bibinfo{volume}{98}},
  \bibinfo{pages}{47007} (\bibinfo{year}{2012}).

\bibitem[{\citenamefont{Park et~al.}(2015{\natexlab{a}})\citenamefont{Park,
  Ryou, Hong, Sumpter, Kim, and Yoon}}]{Park}
\bibinfo{author}{\bibfnamefont{C.}~\bibnamefont{Park}},
  \bibinfo{author}{\bibfnamefont{J.}~\bibnamefont{Ryou}},
  \bibinfo{author}{\bibfnamefont{S.}~\bibnamefont{Hong}},
  \bibinfo{author}{\bibfnamefont{B.~G.} \bibnamefont{Sumpter}},
  \bibinfo{author}{\bibfnamefont{G.}~\bibnamefont{Kim}}, \bibnamefont{and}
  \bibinfo{author}{\bibfnamefont{M.}~\bibnamefont{Yoon}},
  {``}\bibinfo{title}{Electronic Properties of Bilayer Graphene Strongly
  Coupled to Interlayer Stacking and an External Electric Field},{''}
  \bibinfo{journal}{Phys. Rev. Lett.} \textbf{\bibinfo{volume}{115}},
  \bibinfo{pages}{015502} (\bibinfo{year}{2015}{\natexlab{a}}).

\bibitem[{\citenamefont{Zhou et~al.}(2014)\citenamefont{Zhou, Zhang, Yu, and
  Liao}}]{YZouIn}
\bibinfo{author}{\bibfnamefont{Y.-B.} \bibnamefont{Zhou}},
  \bibinfo{author}{\bibfnamefont{L.}~\bibnamefont{Zhang}},
  \bibinfo{author}{\bibfnamefont{D.-P.} \bibnamefont{Yu}}, \bibnamefont{and}
  \bibinfo{author}{\bibfnamefont{Z.-M.} \bibnamefont{Liao}},
  {``}\bibinfo{title}{Magnetic field induced insulating state in bilayer
  graphene at charge neutral point},{''} \bibinfo{journal}{Appl. Phys. Lett.}
  \textbf{\bibinfo{volume}{104}}, \bibinfo{pages}{153103}
  (\bibinfo{year}{2014}).

\bibitem[{\citenamefont{Ando et~al.}(2002)\citenamefont{Ando, Zheng, and
  Suzuura}}]{DyCondSLGT1}
\bibinfo{author}{\bibfnamefont{T.}~\bibnamefont{Ando}},
  \bibinfo{author}{\bibfnamefont{Y.}~\bibnamefont{Zheng}}, \bibnamefont{and}
  \bibinfo{author}{\bibfnamefont{H.}~\bibnamefont{Suzuura}},
  {``}\bibinfo{title}{Dynamical Conductivity and Zero-Mode Anomaly in Honeycomb
  Lattices},{''} \bibinfo{journal}{J. Phys. Soc. Japan}
  \textbf{\bibinfo{volume}{71}}, \bibinfo{pages}{1318} (\bibinfo{year}{2002}).

\bibitem[{\citenamefont{Gusynin and Sharapov}(2006)}]{DyCondSLGT2}
\bibinfo{author}{\bibfnamefont{V.~P.} \bibnamefont{Gusynin}} \bibnamefont{and}
  \bibinfo{author}{\bibfnamefont{S.~G.} \bibnamefont{Sharapov}},
  {``}\bibinfo{title}{Transport of Dirac quasiparticles in graphene: Hall and
  optical conductivities},{''} \bibinfo{journal}{Phys. Rev. B}
  \textbf{\bibinfo{volume}{73}}, \bibinfo{pages}{245411}
  (\bibinfo{year}{2006}).

\bibitem[{\citenamefont{Gusynin et~al.}(2006)\citenamefont{Gusynin, Sharapov,
  and Carbotte}}]{DyCondSLGT3}
\bibinfo{author}{\bibfnamefont{V.~P.} \bibnamefont{Gusynin}},
  \bibinfo{author}{\bibfnamefont{S.~G.} \bibnamefont{Sharapov}},
  \bibnamefont{and} \bibinfo{author}{\bibfnamefont{J.~P.}
  \bibnamefont{Carbotte}}, {``}\bibinfo{title}{Unusual Microwave Response of
  Dirac Quasiparticles in Graphene},{''} \bibinfo{journal}{Phys. Rev. Lett.}
  \textbf{\bibinfo{volume}{96}}, \bibinfo{pages}{256802}
  (\bibinfo{year}{2006}).

\bibitem[{\citenamefont{Falkovsky and Varlamov}(2007)}]{DyCondSLGT4}
\bibinfo{author}{\bibfnamefont{L.~A.} \bibnamefont{Falkovsky}}
  \bibnamefont{and} \bibinfo{author}{\bibfnamefont{A.~A.}
  \bibnamefont{Varlamov}}, {``}\bibinfo{title}{Space-time dispersion of
  graphene conductivity},{''} \bibinfo{journal}{Eur. Phys. J. B}
  \textbf{\bibinfo{volume}{56}}, \bibinfo{pages}{281} (\bibinfo{year}{2007}).

\bibitem[{\citenamefont{Peres et~al.}(2006)\citenamefont{Peres, Guinea, and
  Castro~Neto}}]{DyCondSLGT5}
\bibinfo{author}{\bibfnamefont{N.~M.~R.} \bibnamefont{Peres}},
  \bibinfo{author}{\bibfnamefont{F.}~\bibnamefont{Guinea}}, \bibnamefont{and}
  \bibinfo{author}{\bibfnamefont{A.~H.} \bibnamefont{Castro~Neto}},
  {``}\bibinfo{title}{Electronic properties of disordered two-dimensional
  carbon},{''} \bibinfo{journal}{Phys. Rev. B} \textbf{\bibinfo{volume}{73}},
  \bibinfo{pages}{125411} (\bibinfo{year}{2006}).

\bibitem[{\citenamefont{Li et~al.}(2008{\natexlab{b}})\citenamefont{Li,
  Henriksen, Jiang, Hao, Martin, Kim, Stormer, and Basov}}]{DyCondSLGExp1}
\bibinfo{author}{\bibfnamefont{Z.}~\bibnamefont{Li}},
  \bibinfo{author}{\bibfnamefont{E.~A.} \bibnamefont{Henriksen}},
  \bibinfo{author}{\bibfnamefont{Z.}~\bibnamefont{Jiang}},
  \bibinfo{author}{\bibfnamefont{Z.}~\bibnamefont{Hao}},
  \bibinfo{author}{\bibfnamefont{M.~C.} \bibnamefont{Martin}},
  \bibinfo{author}{\bibfnamefont{P.}~\bibnamefont{Kim}},
  \bibinfo{author}{\bibfnamefont{H.}~\bibnamefont{Stormer}}, \bibnamefont{and}
  \bibinfo{author}{\bibfnamefont{D.~N.} \bibnamefont{Basov}},
  {``}\bibinfo{title}{Dirac charge dynamics in graphene by infrared
  spectroscopy},{''} \bibinfo{journal}{Nat. Phys.}
  \textbf{\bibinfo{volume}{4}}, \bibinfo{pages}{532}
  (\bibinfo{year}{2008}{\natexlab{b}}).

\bibitem[{\citenamefont{Kuzmenko et~al.}(2008)\citenamefont{Kuzmenko, van
  Heumen, Carbone, and van~der Marel}}]{DyCondSLGExp2}
\bibinfo{author}{\bibfnamefont{A.~B.} \bibnamefont{Kuzmenko}},
  \bibinfo{author}{\bibfnamefont{E.}~\bibnamefont{van Heumen}},
  \bibinfo{author}{\bibfnamefont{F.}~\bibnamefont{Carbone}}, \bibnamefont{and}
  \bibinfo{author}{\bibfnamefont{D.}~\bibnamefont{van~der Marel}},
  {``}\bibinfo{title}{Universal Optical Conductance of Graphite},{''}
  \bibinfo{journal}{Phys. Rev. Lett.} \textbf{\bibinfo{volume}{100}},
  \bibinfo{pages}{117401} (\bibinfo{year}{2008}).

\bibitem[{\citenamefont{Nair et~al.}(2008)\citenamefont{Nair, Blake,
  Grigorenko, Novoselov, Booth, Stauber, Peres, and Geim}}]{DyCondSLGExp3}
\bibinfo{author}{\bibfnamefont{R.~R.} \bibnamefont{Nair}},
  \bibinfo{author}{\bibfnamefont{P.}~\bibnamefont{Blake}},
  \bibinfo{author}{\bibfnamefont{A.~N.} \bibnamefont{Grigorenko}},
  \bibinfo{author}{\bibfnamefont{K.~S.} \bibnamefont{Novoselov}},
  \bibinfo{author}{\bibfnamefont{T.~J.} \bibnamefont{Booth}},
  \bibinfo{author}{\bibfnamefont{T.}~\bibnamefont{Stauber}},
  \bibinfo{author}{\bibfnamefont{N.~M.~R.} \bibnamefont{Peres}},
  \bibnamefont{and} \bibinfo{author}{\bibfnamefont{A.~K.} \bibnamefont{Geim}},
  {``}\bibinfo{title}{Fine Structure Constant Defines Visual Transparency of
  Graphene},{''} \bibinfo{journal}{Science} \textbf{\bibinfo{volume}{320}},
  \bibinfo{pages}{1308} (\bibinfo{year}{2008}).

\bibitem[{\citenamefont{Mak et~al.}(2008)\citenamefont{Mak, Sfeir, Wu, Lui,
  Misewich, and Heinz}}]{DyCondSLGExp4}
\bibinfo{author}{\bibfnamefont{K.~F.} \bibnamefont{Mak}},
  \bibinfo{author}{\bibfnamefont{M.~Y.} \bibnamefont{Sfeir}},
  \bibinfo{author}{\bibfnamefont{Y.}~\bibnamefont{Wu}},
  \bibinfo{author}{\bibfnamefont{C.~H.} \bibnamefont{Lui}},
  \bibinfo{author}{\bibfnamefont{J.~A.} \bibnamefont{Misewich}},
  \bibnamefont{and} \bibinfo{author}{\bibfnamefont{T.~F.} \bibnamefont{Heinz}},
  {``}\bibinfo{title}{Measurement of the Optical Conductivity of Graphene},{''}
  \bibinfo{journal}{Phys. Rev. Lett.} \textbf{\bibinfo{volume}{101}},
  \bibinfo{pages}{196405} (\bibinfo{year}{2008}).

\bibitem[{\citenamefont{Tabert and Nicol}(2012)}]{NicCarb2}
\bibinfo{author}{\bibfnamefont{C.~J.} \bibnamefont{Tabert}} \bibnamefont{and}
  \bibinfo{author}{\bibfnamefont{E.~J.} \bibnamefont{Nicol}},
  {``}\bibinfo{title}{Dynamical conductivity of AA-stacked bilayer
  graphene},{''} \bibinfo{journal}{Phys. Rev. B} \textbf{\bibinfo{volume}{86}},
  \bibinfo{pages}{075439} (\bibinfo{year}{2012}).

\bibitem[{\citenamefont{Nicol and Carbotte}(2008)}]{NicCarb1}
\bibinfo{author}{\bibfnamefont{E.~J.} \bibnamefont{Nicol}} \bibnamefont{and}
  \bibinfo{author}{\bibfnamefont{J.~P.} \bibnamefont{Carbotte}},
  {``}\bibinfo{title}{Optical conductivity of bilayer graphene with and without
  an asymmetry gap},{''} \bibinfo{journal}{Phys. Rev. B}
  \textbf{\bibinfo{volume}{77}}, \bibinfo{pages}{155409}
  (\bibinfo{year}{2008}).

\bibitem[{\citenamefont{McCann et~al.}(2007)\citenamefont{McCann, Abergel, and
  Fal'ko}}]{McCann2007}
\bibinfo{author}{\bibfnamefont{E.}~\bibnamefont{McCann}},
  \bibinfo{author}{\bibfnamefont{D.~S.~L.} \bibnamefont{Abergel}},
  \bibnamefont{and} \bibinfo{author}{\bibfnamefont{V.~I.}
  \bibnamefont{Fal'ko}}, {``}\bibinfo{title}{Electrons in bilayer
  graphene},{''} \bibinfo{journal}{Solid State Commun.}
  \textbf{\bibinfo{volume}{143}}, \bibinfo{pages}{110} (\bibinfo{year}{2007}).

\bibitem[{\citenamefont{Nilsson et~al.}(2008)\citenamefont{Nilsson,
  Castro~Neto, Guinea, and Peres}}]{Nilsson2008}
\bibinfo{author}{\bibfnamefont{J.}~\bibnamefont{Nilsson}},
  \bibinfo{author}{\bibfnamefont{A.~H.} \bibnamefont{Castro~Neto}},
  \bibinfo{author}{\bibfnamefont{F.}~\bibnamefont{Guinea}}, \bibnamefont{and}
  \bibinfo{author}{\bibfnamefont{N.~M.~R.} \bibnamefont{Peres}},
  {``}\bibinfo{title}{Electronic properties of bilayer and multilayer
  graphene},{''} \bibinfo{journal}{Phys. Rev. B} \textbf{\bibinfo{volume}{78}},
  \bibinfo{pages}{045405} (\bibinfo{year}{2008}).

\bibitem[{\citenamefont{Abergel and Fal'ko}(2007)}]{Abergel2007}
\bibinfo{author}{\bibfnamefont{D.~S.~L.} \bibnamefont{Abergel}}
  \bibnamefont{and} \bibinfo{author}{\bibfnamefont{V.~I.}
  \bibnamefont{Fal'ko}}, {``}\bibinfo{title}{Optical and magneto-optical
  far-infrared properties of bilayer graphene},{''} \bibinfo{journal}{Phys.
  Rev. B} \textbf{\bibinfo{volume}{75}}, \bibinfo{pages}{155430}
  (\bibinfo{year}{2007}).

\bibitem[{\citenamefont{Benfatto et~al.}(2008)\citenamefont{Benfatto, Sharapov,
  and Carbotte}}]{Benfatto}
\bibinfo{author}{\bibfnamefont{L.}~\bibnamefont{Benfatto}},
  \bibinfo{author}{\bibfnamefont{S.~G.} \bibnamefont{Sharapov}},
  \bibnamefont{and} \bibinfo{author}{\bibfnamefont{J.~P.}
  \bibnamefont{Carbotte}}, {``}\bibinfo{title}{Robustness of the optical
  conductivity sum rule in bilayer graphene},{''} \bibinfo{journal}{Phys. Rev.
  B} \textbf{\bibinfo{volume}{77}}, \bibinfo{pages}{125422}
  (\bibinfo{year}{2008}).

\bibitem[{\citenamefont{Abergel and Mucha-Kruczy{\'{n}}ski}(2015)}]{optics2015}
\bibinfo{author}{\bibfnamefont{D.~S.~L.} \bibnamefont{Abergel}}
  \bibnamefont{and}
  \bibinfo{author}{\bibfnamefont{M.}~\bibnamefont{Mucha-Kruczy{\'{n}}ski}},
  {``}\bibinfo{title}{Infrared absorption of closely aligned heterostructures
  of monolayer and bilayer graphene with hexagonal boron nitride},{''}
  \bibinfo{journal}{Phys. Rev. B} \textbf{\bibinfo{volume}{92}},
  \bibinfo{pages}{115430} (\bibinfo{year}{2015}).

\bibitem[{\citenamefont{Drut and L\"ahde}(2009)}]{single_layer_afm2009}
\bibinfo{author}{\bibfnamefont{J.~E.} \bibnamefont{Drut}} \bibnamefont{and}
  \bibinfo{author}{\bibfnamefont{T.~A.} \bibnamefont{L\"ahde}},
  {``}\bibinfo{title}{Is Graphene in Vacuum an Insulator?},{''}
  \bibinfo{journal}{Phys. Rev. Lett.} \textbf{\bibinfo{volume}{102}},
  \bibinfo{pages}{026802} (\bibinfo{year}{2009}).

\bibitem[{\citenamefont{Ulybyshev et~al.}(2013)\citenamefont{Ulybyshev,
  Buividovich, Katsnelson, and Polikarpov}}]{polikarpov2013}
\bibinfo{author}{\bibfnamefont{M.~V.} \bibnamefont{Ulybyshev}},
  \bibinfo{author}{\bibfnamefont{P.~V.} \bibnamefont{Buividovich}},
  \bibinfo{author}{\bibfnamefont{M.~I.} \bibnamefont{Katsnelson}},
  \bibnamefont{and} \bibinfo{author}{\bibfnamefont{M.~I.}
  \bibnamefont{Polikarpov}}, {``}\bibinfo{title}{Monte Carlo Study of the
  Semimetal-Insulator Phase Transition in Monolayer Graphene with a Realistic
  Interelectron Interaction Potential},{''} \bibinfo{journal}{Phys. Rev. Lett.}
  \textbf{\bibinfo{volume}{111}}, \bibinfo{pages}{056801}
  (\bibinfo{year}{2013}).

\bibitem[{\citenamefont{Wehling et~al.}(2011)\citenamefont{Wehling,
  {\c{S}}a{\c{s}}{\i}o{\u{g}}lu, Friedrich, Lichtenstein, Katsnelson, and
  Bl{\"u}gel}}]{Wehling}
\bibinfo{author}{\bibfnamefont{T.~O.} \bibnamefont{Wehling}},
  \bibinfo{author}{\bibfnamefont{E.}~\bibnamefont{{\c{S}}a{\c{s}}{\i}o{\u{g}}lu}},
  \bibinfo{author}{\bibfnamefont{C.}~\bibnamefont{Friedrich}},
  \bibinfo{author}{\bibfnamefont{A.~I.} \bibnamefont{Lichtenstein}},
  \bibinfo{author}{\bibfnamefont{M.~I.} \bibnamefont{Katsnelson}},
  \bibnamefont{and}
  \bibinfo{author}{\bibfnamefont{S.}~\bibnamefont{Bl{\"u}gel}},
  {``}\bibinfo{title}{Strength of Effective Coulomb Interactions in Graphene
  and Graphite},{''} \bibinfo{journal}{Phys. Rev. Lett.}
  \textbf{\bibinfo{volume}{106}}, \bibinfo{pages}{236805}
  (\bibinfo{year}{2011}).

\bibitem[{\citenamefont{Sboychakov
  et~al.}(2013{\natexlab{b}})\citenamefont{Sboychakov, Rakhmanov, Rozhkov, and
  Nori}}]{PrbROur}
\bibinfo{author}{\bibfnamefont{A.~O.} \bibnamefont{Sboychakov}},
  \bibinfo{author}{\bibfnamefont{A.~L.} \bibnamefont{Rakhmanov}},
  \bibinfo{author}{\bibfnamefont{A.~V.} \bibnamefont{Rozhkov}},
  \bibnamefont{and} \bibinfo{author}{\bibfnamefont{F.}~\bibnamefont{Nori}},
  {``}\bibinfo{title}{Metal-insulator transition and phase separation in doped
  AA-stacked graphene bilayer},{''} \bibinfo{journal}{Phys. Rev. B}
  \textbf{\bibinfo{volume}{87}}, \bibinfo{pages}{121401}
  (\bibinfo{year}{2013}{\natexlab{b}}).

\bibitem[{\citenamefont{Sch{\"u}ler et~al.}(2013)\citenamefont{Sch{\"u}ler,
  R{\"o}sner, Wehling, Lichtenstein, and Katsnelson}}]{EffCoulomb}
\bibinfo{author}{\bibfnamefont{M.}~\bibnamefont{Sch{\"u}ler}},
  \bibinfo{author}{\bibfnamefont{M.}~\bibnamefont{R{\"o}sner}},
  \bibinfo{author}{\bibfnamefont{T.~O.} \bibnamefont{Wehling}},
  \bibinfo{author}{\bibfnamefont{A.~I.} \bibnamefont{Lichtenstein}},
  \bibnamefont{and} \bibinfo{author}{\bibfnamefont{M.~I.}
  \bibnamefont{Katsnelson}}, {``}\bibinfo{title}{Optimal Hubbard Models for
  Materials with Nonlocal Coulomb Interactions: Graphene, Silicene, and
  Benzene},{''} \bibinfo{journal}{Phys. Rev. Lett.}
  \textbf{\bibinfo{volume}{111}}, \bibinfo{pages}{036601}
  (\bibinfo{year}{2013}).

\bibitem[{\citenamefont{Rice}(1970)}]{Rice}
\bibinfo{author}{\bibfnamefont{T.~M.} \bibnamefont{Rice}},
  {``}\bibinfo{title}{Band-Structure Effects in Itinerant
  Antiferromagnetism},{''} \bibinfo{journal}{Phys. Rev. B}
  \textbf{\bibinfo{volume}{2}}, \bibinfo{pages}{3619} (\bibinfo{year}{1970}).

\bibitem[{\citenamefont{Fulde and Ferrell}(1964)}]{FFLO1}
\bibinfo{author}{\bibfnamefont{P.}~\bibnamefont{Fulde}} \bibnamefont{and}
  \bibinfo{author}{\bibfnamefont{R.~A.} \bibnamefont{Ferrell}},
  {``}\bibinfo{title}{Superconductivity in a Strong Spin-Exchange Field},{''}
  \bibinfo{journal}{Phys. Rev. } \textbf{\bibinfo{volume}{135}},
  \bibinfo{pages}{A550} (\bibinfo{year}{1964}).

\bibitem[{\citenamefont{Larkin and Ovchinnikov}(1965)}]{FFLO2}
\bibinfo{author}{\bibfnamefont{A.~I.} \bibnamefont{Larkin}} \bibnamefont{and}
  \bibinfo{author}{\bibfnamefont{Y.~N.} \bibnamefont{Ovchinnikov}},
  {``}\bibinfo{title}{Nonuniform state of superconductors},{''}
  \bibinfo{journal}{Sov. Phys. JETP} \textbf{\bibinfo{volume}{20}},
  \bibinfo{pages}{762} (\bibinfo{year}{1965}).

\bibitem[{\citenamefont{Aslamazov}(1969)}]{FFLO3}
\bibinfo{author}{\bibfnamefont{L.}~\bibnamefont{Aslamazov}},
  {``}\bibinfo{title}{Influence of impurities on the existence of an
  inhomogeneous state in a ferromagnetic superconduuctor},{''}
  \bibinfo{journal}{Sov. Phys. JETP} \textbf{\bibinfo{volume}{28}},
  \bibinfo{pages}{773} (\bibinfo{year}{1969}).

\bibitem[{\citenamefont{Sheehy and Radzihovsky}(2007)}]{FFLO4}
\bibinfo{author}{\bibfnamefont{D.~E.} \bibnamefont{Sheehy}} \bibnamefont{and}
  \bibinfo{author}{\bibfnamefont{L.}~\bibnamefont{Radzihovsky}},
  {``}\bibinfo{title}{BEC-BCS crossover, phase transitions and phase separation
  in polarized resonantly-paired superfluids},{''} \bibinfo{journal}{Ann.
  Phys.} \textbf{\bibinfo{volume}{322}}, \bibinfo{pages}{1790 }
  (\bibinfo{year}{2007}).

\bibitem[{\citenamefont{Takada}(1970)}]{FFLO5}
\bibinfo{author}{\bibfnamefont{S.}~\bibnamefont{Takada}},
  {``}\bibinfo{title}{Superconductivity in a Molecular Field. II: Stability of
  Fulde-Ferrel Phase},{''} \bibinfo{journal}{Progr. Theoret. Phys. Suppl.}
  \textbf{\bibinfo{volume}{43}}, \bibinfo{pages}{27} (\bibinfo{year}{1970}).

\bibitem[{\citenamefont{Le~Bellac et~al.}(2004)\citenamefont{Le~Bellac,
  Mortessagne, and Batrouni}}]{Thermo}
\bibinfo{author}{\bibfnamefont{M.}~\bibnamefont{Le~Bellac}},
  \bibinfo{author}{\bibfnamefont{F.}~\bibnamefont{Mortessagne}},
  \bibnamefont{and} \bibinfo{author}{\bibfnamefont{G.~G.}
  \bibnamefont{Batrouni}}, \emph{\bibinfo{title}{Equilibrium and
  Non-Equilibrium Statistical Thermodynamics}} (\bibinfo{publisher}{Cambridge
  University Press, Cambridge}, \bibinfo{year}{2004}).

\bibitem[{\citenamefont{Lorenzana
  et~al.}(2001{\natexlab{a}})\citenamefont{Lorenzana, Castellani, and
  Castro}}]{PSsructure1}
\bibinfo{author}{\bibfnamefont{J.}~\bibnamefont{Lorenzana}},
  \bibinfo{author}{\bibfnamefont{C.}~\bibnamefont{Castellani}},
  \bibnamefont{and} \bibinfo{author}{\bibfnamefont{C.~Di}
  \bibnamefont{Castro}}, {``}\bibinfo{title}{Phase separation frustrated by the
  long-range Coulomb interaction. I. Theory},{''} \bibinfo{journal}{Phys. Rev.
  B} \textbf{\bibinfo{volume}{64}}, \bibinfo{pages}{235127}
  (\bibinfo{year}{2001}{\natexlab{a}}).

\bibitem[{\citenamefont{Lorenzana
  et~al.}(2001{\natexlab{b}})\citenamefont{Lorenzana, Castellani, and
  Di~Castro}}]{PSsructure2}
\bibinfo{author}{\bibfnamefont{J.}~\bibnamefont{Lorenzana}},
  \bibinfo{author}{\bibfnamefont{C.}~\bibnamefont{Castellani}},
  \bibnamefont{and}
  \bibinfo{author}{\bibfnamefont{C.}~\bibnamefont{Di~Castro}},
  {``}\bibinfo{title}{Phase separation frustrated by the long-range Coulomb
  interaction. II. Applications},{''} \bibinfo{journal}{Phys. Rev. B}
  \textbf{\bibinfo{volume}{64}}, \bibinfo{pages}{235128}
  (\bibinfo{year}{2001}{\natexlab{b}}).

\bibitem[{\citenamefont{Jamei et~al.}(2005)\citenamefont{Jamei, Kivelson, and
  Spivak}}]{PSsructure3}
\bibinfo{author}{\bibfnamefont{R.}~\bibnamefont{Jamei}},
  \bibinfo{author}{\bibfnamefont{S.}~\bibnamefont{Kivelson}}, \bibnamefont{and}
  \bibinfo{author}{\bibfnamefont{B.}~\bibnamefont{Spivak}},
  {``}\bibinfo{title}{Universal Aspects of Coulomb-Frustrated Phase
  Separation},{''} \bibinfo{journal}{Phys. Rev. Lett.}
  \textbf{\bibinfo{volume}{94}}, \bibinfo{pages}{056805}
  (\bibinfo{year}{2005}).

\bibitem[{\citenamefont{Kugel et~al.}(2005)\citenamefont{Kugel, Rakhmanov, and
  Sboychakov}}]{PSsructure4}
\bibinfo{author}{\bibfnamefont{K.~I.} \bibnamefont{Kugel}},
  \bibinfo{author}{\bibfnamefont{A.~L.} \bibnamefont{Rakhmanov}},
  \bibnamefont{and} \bibinfo{author}{\bibfnamefont{A.~O.}
  \bibnamefont{Sboychakov}}, {``}\bibinfo{title}{Phase Separation in
  Jahn-Teller Systems with Localized and Itinerant Electrons},{''}
  \bibinfo{journal}{Phys. Rev. Lett.} \textbf{\bibinfo{volume}{95}},
  \bibinfo{pages}{267210} (\bibinfo{year}{2005}).

\bibitem[{\citenamefont{Su and MacDonald}(2008)}]{ExcDrag1}
\bibinfo{author}{\bibfnamefont{J.-J.} \bibnamefont{Su}} \bibnamefont{and}
  \bibinfo{author}{\bibfnamefont{A.~H.} \bibnamefont{MacDonald}},
  {``}\bibinfo{title}{How to make a bilayer exciton condensate flow},{''}
  \bibinfo{journal}{Nat. Phys.} \textbf{\bibinfo{volume}{4}},
  \bibinfo{pages}{799} (\bibinfo{year}{2008}).

\bibitem[{\citenamefont{Hwang et~al.}(2011)\citenamefont{Hwang, Sensarma, and
  Das~Sarma}}]{ExcDrag2}
\bibinfo{author}{\bibfnamefont{E.~H.} \bibnamefont{Hwang}},
  \bibinfo{author}{\bibfnamefont{R.}~\bibnamefont{Sensarma}}, \bibnamefont{and}
  \bibinfo{author}{\bibfnamefont{S.}~\bibnamefont{Das~Sarma}},
  {``}\bibinfo{title}{Coulomb drag in monolayer and bilayer graphene},{''}
  \bibinfo{journal}{Phys. Rev. B} \textbf{\bibinfo{volume}{84}},
  \bibinfo{pages}{245441} (\bibinfo{year}{2011}).

\bibitem[{\citenamefont{Nandi et~al.}(2012)\citenamefont{Nandi, Finck,
  Eisenstein, Pfeiffer, and West}}]{ExcDrag3}
\bibinfo{author}{\bibfnamefont{D.}~\bibnamefont{Nandi}},
  \bibinfo{author}{\bibfnamefont{A.~D.~K.} \bibnamefont{Finck}},
  \bibinfo{author}{\bibfnamefont{J.~P.} \bibnamefont{Eisenstein}},
  \bibinfo{author}{\bibfnamefont{L.~N.} \bibnamefont{Pfeiffer}},
  \bibnamefont{and} \bibinfo{author}{\bibfnamefont{K.~W.} \bibnamefont{West}},
  {``}\bibinfo{title}{Exciton condensation and perfect Coulomb drag},{''}
  \bibinfo{journal}{Nature} \textbf{\bibinfo{volume}{488}},
  \bibinfo{pages}{481} (\bibinfo{year}{2012}).

\bibitem[{\citenamefont{Gorbachev et~al.}(2012)\citenamefont{Gorbachev, Geim,
  Katsnelson, Novoselov, Tudorovskiy, Grigorieva, MacDonald, Morozov, Watanabe,
  Taniguchi et~al.}}]{DragGr}
\bibinfo{author}{\bibfnamefont{R.~V.} \bibnamefont{Gorbachev}},
  \bibinfo{author}{\bibfnamefont{A.~K.} \bibnamefont{Geim}},
  \bibinfo{author}{\bibfnamefont{M.~I.} \bibnamefont{Katsnelson}},
  \bibinfo{author}{\bibfnamefont{R.~S.} \bibnamefont{Novoselov}},
  \bibinfo{author}{\bibfnamefont{T.}~\bibnamefont{Tudorovskiy}},
  \bibinfo{author}{\bibfnamefont{I.~V.} \bibnamefont{Grigorieva}},
  \bibinfo{author}{\bibfnamefont{A.~H.} \bibnamefont{MacDonald}},
  \bibinfo{author}{\bibfnamefont{S.}~\bibnamefont{Morozov}},
  \bibinfo{author}{\bibfnamefont{K.}~\bibnamefont{Watanabe}},
  \bibinfo{author}{\bibfnamefont{T.}~\bibnamefont{Taniguchi}},
  \bibnamefont{et~al.}, {``}\bibinfo{title}{Strong Coulomb drag and broken
  symmetry in double-layer graphene},{''} \bibinfo{journal}{Nat. Phys.}
  \textbf{\bibinfo{volume}{8}}, \bibinfo{pages}{896} (\bibinfo{year}{2012}).

\bibitem[{\citenamefont{de~la Pe{\~n}a et~al.}(2014)\citenamefont{de~la
  Pe{\~n}a, Scherer, and Honerkamp}}]{Honerkamp}
\bibinfo{author}{\bibfnamefont{D.~S.} \bibnamefont{de~la Pe{\~n}a}},
  \bibinfo{author}{\bibfnamefont{M.~M.} \bibnamefont{Scherer}},
  \bibnamefont{and}
  \bibinfo{author}{\bibfnamefont{C.}~\bibnamefont{Honerkamp}},
  {``}\bibinfo{title}{Electronic instabilities of the AA-honeycomb
  bilayer},{''} \bibinfo{journal}{Ann. Phys. (Leipzig)}
  \textbf{\bibinfo{volume}{526}}, \bibinfo{pages}{366} (\bibinfo{year}{2014}).

\bibitem[{\citenamefont{Brey and Fertig}(2013)}]{BreyFertig}
\bibinfo{author}{\bibfnamefont{L.}~\bibnamefont{Brey}} \bibnamefont{and}
  \bibinfo{author}{\bibfnamefont{H.~A.} \bibnamefont{Fertig}},
  {``}\bibinfo{title}{Gapped phase in {$AA$}-stacked bilayer graphene},{''}
  \bibinfo{journal}{Phys. Rev. B} \textbf{\bibinfo{volume}{87}},
  \bibinfo{pages}{115411} (\bibinfo{year}{2013}).

\bibitem[{\citenamefont{Nikolaev and
  Ulybyshev}(2014{\natexlab{a}})}]{ulybyshev}
\bibinfo{author}{\bibfnamefont{A.}~\bibnamefont{Nikolaev}} \bibnamefont{and}
  \bibinfo{author}{\bibfnamefont{M.}~\bibnamefont{Ulybyshev}},
  {``}\bibinfo{title}{Monte-Carlo study of the phase transition in the
  AA-stacked bilayer graphene},{''} \bibinfo{journal}{arXiv preprint
  arXiv:1412.1359}  (\bibinfo{year}{2014}{\natexlab{a}}).

\bibitem[{\citenamefont{Nikolaev and
  Ulybyshev}(2014{\natexlab{b}})}]{ulybyshev_conf}
\bibinfo{author}{\bibfnamefont{A.}~\bibnamefont{Nikolaev}} \bibnamefont{and}
  \bibinfo{author}{\bibfnamefont{M.}~\bibnamefont{Ulybyshev}}, in
  \emph{\bibinfo{booktitle}{PoS LAT2014}} (\bibinfo{year}{2014}{\natexlab{b}}),
  vol. \bibinfo{volume}{054}.

\bibitem[{\citenamefont{Weitz et~al.}(2010)\citenamefont{Weitz, Allen, Feldman,
  Martin, and Yacoby}}]{Weitz2010}
\bibinfo{author}{\bibfnamefont{R.~T.} \bibnamefont{Weitz}},
  \bibinfo{author}{\bibfnamefont{M.~T.} \bibnamefont{Allen}},
  \bibinfo{author}{\bibfnamefont{B.~E.} \bibnamefont{Feldman}},
  \bibinfo{author}{\bibfnamefont{J.}~\bibnamefont{Martin}}, \bibnamefont{and}
  \bibinfo{author}{\bibfnamefont{A.}~\bibnamefont{Yacoby}},
  {``}\bibinfo{title}{Broken-Symmetry States in Doubly Gated Suspended Bilayer
  Graphene},{''} \bibinfo{journal}{Science} \textbf{\bibinfo{volume}{330}},
  \bibinfo{pages}{812} (\bibinfo{year}{2010}).

\bibitem[{\citenamefont{Freitag
  et~al.}(2012{\natexlab{a}})\citenamefont{Freitag, Trbovic, Weiss, and
  Sch\"onenberger}}]{Freitag2012}
\bibinfo{author}{\bibfnamefont{F.}~\bibnamefont{Freitag}},
  \bibinfo{author}{\bibfnamefont{J.}~\bibnamefont{Trbovic}},
  \bibinfo{author}{\bibfnamefont{M.}~\bibnamefont{Weiss}}, \bibnamefont{and}
  \bibinfo{author}{\bibfnamefont{C.}~\bibnamefont{Sch\"onenberger}},
  {``}\bibinfo{title}{Spontaneously Gapped Ground State in Suspended Bilayer
  Graphene},{''} \bibinfo{journal}{Phys. Rev. Lett.}
  \textbf{\bibinfo{volume}{108}}, \bibinfo{pages}{076602}
  (\bibinfo{year}{2012}{\natexlab{a}}).

\bibitem[{\citenamefont{Freitag
  et~al.}(2012{\natexlab{b}})\citenamefont{Freitag, Weiss, Maurand, Trbovic,
  and SchÃÂ¶nenberger}}]{Freitag20122053}
\bibinfo{author}{\bibfnamefont{F.}~\bibnamefont{Freitag}},
  \bibinfo{author}{\bibfnamefont{M.}~\bibnamefont{Weiss}},
  \bibinfo{author}{\bibfnamefont{R.}~\bibnamefont{Maurand}},
  \bibinfo{author}{\bibfnamefont{J.}~\bibnamefont{Trbovic}}, \bibnamefont{and}
  \bibinfo{author}{\bibfnamefont{C.}~\bibnamefont{SchÃÂ¶nenberger}},
  {``}\bibinfo{title}{Homogeneity of bilayer graphene},{''}
  \bibinfo{journal}{Solid State Communications} \textbf{\bibinfo{volume}{152}},
  \bibinfo{pages}{2053 } (\bibinfo{year}{2012}{\natexlab{b}}).

\bibitem[{\citenamefont{Veligura et~al.}(2012)\citenamefont{Veligura, van
  Elferen, Tombros, Maan, Zeitler, and van Wees}}]{veligura2012}
\bibinfo{author}{\bibfnamefont{A.}~\bibnamefont{Veligura}},
  \bibinfo{author}{\bibfnamefont{H.~J.} \bibnamefont{van Elferen}},
  \bibinfo{author}{\bibfnamefont{N.}~\bibnamefont{Tombros}},
  \bibinfo{author}{\bibfnamefont{J.~C.} \bibnamefont{Maan}},
  \bibinfo{author}{\bibfnamefont{U.}~\bibnamefont{Zeitler}}, \bibnamefont{and}
  \bibinfo{author}{\bibfnamefont{B.~J.} \bibnamefont{van Wees}},
  {``}\bibinfo{title}{Transport gap in suspended bilayer graphene at zero
  magnetic field},{''} \bibinfo{journal}{Phys. Rev. B}
  \textbf{\bibinfo{volume}{85}}, \bibinfo{pages}{155412}
  (\bibinfo{year}{2012}).

\bibitem[{\citenamefont{Freitag et~al.}(2013)\citenamefont{Freitag, Weiss,
  Maurand, Trbovic, and Sch\"onenberger}}]{freitag2013}
\bibinfo{author}{\bibfnamefont{F.}~\bibnamefont{Freitag}},
  \bibinfo{author}{\bibfnamefont{M.}~\bibnamefont{Weiss}},
  \bibinfo{author}{\bibfnamefont{R.}~\bibnamefont{Maurand}},
  \bibinfo{author}{\bibfnamefont{J.}~\bibnamefont{Trbovic}}, \bibnamefont{and}
  \bibinfo{author}{\bibfnamefont{C.}~\bibnamefont{Sch\"onenberger}},
  {``}\bibinfo{title}{Spin symmetry of the bilayer graphene ground state},{''}
  \bibinfo{journal}{Phys. Rev. B} \textbf{\bibinfo{volume}{87}},
  \bibinfo{pages}{161402} (\bibinfo{year}{2013}).

\bibitem[{\citenamefont{Nilsson
  et~al.}(2006{\natexlab{b}})\citenamefont{Nilsson, Castro~Neto, Peres, and
  Guinea}}]{Nilsson2006}
\bibinfo{author}{\bibfnamefont{J.}~\bibnamefont{Nilsson}},
  \bibinfo{author}{\bibfnamefont{A.~H.} \bibnamefont{Castro~Neto}},
  \bibinfo{author}{\bibfnamefont{N.~M.~R.} \bibnamefont{Peres}},
  \bibnamefont{and} \bibinfo{author}{\bibfnamefont{F.}~\bibnamefont{Guinea}},
  {``}\bibinfo{title}{Electron-electron interactions and the phase diagram of a
  graphene bilayer},{''} \bibinfo{journal}{Phys. Rev. B}
  \textbf{\bibinfo{volume}{73}}, \bibinfo{pages}{214418}
  (\bibinfo{year}{2006}{\natexlab{b}}).

\bibitem[{\citenamefont{Bloch}(1929)}]{ferro_bloch}
\bibinfo{author}{\bibfnamefont{F.}~\bibnamefont{Bloch}},
  {``}\bibinfo{title}{Bemerkung zur Elektronentheorie des Ferromagnetismus und
  der elektrischen Leitf{\"a}higkeit},{''} \bibinfo{journal}{Z. Physik}
  \textbf{\bibinfo{volume}{57}}, \bibinfo{pages}{545} (\bibinfo{year}{1929}).

\bibitem[{\citenamefont{Castro et~al.}(2008{\natexlab{b}})\citenamefont{Castro,
  Peres, Stauber, and Silva}}]{Castro2008}
\bibinfo{author}{\bibfnamefont{E.~V.} \bibnamefont{Castro}},
  \bibinfo{author}{\bibfnamefont{N.~M.~R.} \bibnamefont{Peres}},
  \bibinfo{author}{\bibfnamefont{T.}~\bibnamefont{Stauber}}, \bibnamefont{and}
  \bibinfo{author}{\bibfnamefont{N.~A.~P.} \bibnamefont{Silva}},
  {``}\bibinfo{title}{Low-density ferromagnetism in biased bilayer
  graphene},{''} \bibinfo{journal}{Phys. Rev. Lett.}
  \textbf{\bibinfo{volume}{100}}, \bibinfo{pages}{186803}
  (\bibinfo{year}{2008}{\natexlab{b}}).

\bibitem[{\citenamefont{Stauber et~al.}(2007)\citenamefont{Stauber, Peres,
  Guinea, and Castro~Neto}}]{bilayer_ferro2007}
\bibinfo{author}{\bibfnamefont{T.}~\bibnamefont{Stauber}},
  \bibinfo{author}{\bibfnamefont{N.~M.~R.} \bibnamefont{Peres}},
  \bibinfo{author}{\bibfnamefont{F.}~\bibnamefont{Guinea}}, \bibnamefont{and}
  \bibinfo{author}{\bibfnamefont{A.~H.} \bibnamefont{Castro~Neto}},
  {``}\bibinfo{title}{Fermi liquid theory of a Fermi ring},{''}
  \bibinfo{journal}{Phys. Rev. B} \textbf{\bibinfo{volume}{75}},
  \bibinfo{pages}{115425} (\bibinfo{year}{2007}).

\bibitem[{\citenamefont{Min et~al.}(2008)\citenamefont{Min, Borghi, Polini, and
  MacDonald}}]{min_pseudo_fm2008}
\bibinfo{author}{\bibfnamefont{H.}~\bibnamefont{Min}},
  \bibinfo{author}{\bibfnamefont{G.}~\bibnamefont{Borghi}},
  \bibinfo{author}{\bibfnamefont{M.}~\bibnamefont{Polini}}, \bibnamefont{and}
  \bibinfo{author}{\bibfnamefont{A.~H.} \bibnamefont{MacDonald}},
  {``}\bibinfo{title}{Pseudospin magnetism in graphene},{''}
  \bibinfo{journal}{Phys. Rev. B} \textbf{\bibinfo{volume}{77}},
  \bibinfo{pages}{041407} (\bibinfo{year}{2008}).

\bibitem[{\citenamefont{Zhang et~al.}(2010{\natexlab{b}})\citenamefont{Zhang,
  Min, Polini, and MacDonald}}]{zhang_rg2010}
\bibinfo{author}{\bibfnamefont{F.}~\bibnamefont{Zhang}},
  \bibinfo{author}{\bibfnamefont{H.}~\bibnamefont{Min}},
  \bibinfo{author}{\bibfnamefont{M.}~\bibnamefont{Polini}}, \bibnamefont{and}
  \bibinfo{author}{\bibfnamefont{A.~H.} \bibnamefont{MacDonald}},
  {``}\bibinfo{title}{Spontaneous inversion symmetry breaking in graphene
  bilayers},{''} \bibinfo{journal}{Phys. Rev. B} \textbf{\bibinfo{volume}{81}},
  \bibinfo{pages}{041402} (\bibinfo{year}{2010}{\natexlab{b}}).

\bibitem[{\citenamefont{Zhang et~al.}(2012{\natexlab{b}})\citenamefont{Zhang,
  Min, and MacDonald}}]{zhang_rg_order2012}
\bibinfo{author}{\bibfnamefont{F.}~\bibnamefont{Zhang}},
  \bibinfo{author}{\bibfnamefont{H.}~\bibnamefont{Min}}, \bibnamefont{and}
  \bibinfo{author}{\bibfnamefont{A.~H.} \bibnamefont{MacDonald}},
  {``}\bibinfo{title}{Competing ordered states in bilayer graphene},{''}
  \bibinfo{journal}{Phys. Rev. B} \textbf{\bibinfo{volume}{86}},
  \bibinfo{pages}{155128} (\bibinfo{year}{2012}{\natexlab{b}}).

\bibitem[{\citenamefont{Nandkishore and
  Levitov}(2010{\natexlab{a}})}]{Nandkishore2010}
\bibinfo{author}{\bibfnamefont{R.}~\bibnamefont{Nandkishore}} \bibnamefont{and}
  \bibinfo{author}{\bibfnamefont{L.}~\bibnamefont{Levitov}},
  {``}\bibinfo{title}{Dynamical Screening and Excitonic Instability in Bilayer
  Graphene},{''} \bibinfo{journal}{Phys. Rev. Lett.}
  \textbf{\bibinfo{volume}{104}}, \bibinfo{pages}{156803}
  (\bibinfo{year}{2010}{\natexlab{a}}).

\bibitem[{\citenamefont{Nandkishore and
  Levitov}(2010{\natexlab{b}})}]{Nandkishore2010b}
\bibinfo{author}{\bibfnamefont{R.}~\bibnamefont{Nandkishore}} \bibnamefont{and}
  \bibinfo{author}{\bibfnamefont{L.}~\bibnamefont{Levitov}},
  {``}\bibinfo{title}{Quantum anomalous Hall state in bilayer graphene},{''}
  \bibinfo{journal}{Phys. Rev. B} \textbf{\bibinfo{volume}{82}},
  \bibinfo{pages}{115124} (\bibinfo{year}{2010}{\natexlab{b}}).

\bibitem[{\citenamefont{Jung et~al.}(2011)\citenamefont{Jung, Zhang, and
  MacDonald}}]{Jung2011}
\bibinfo{author}{\bibfnamefont{J.}~\bibnamefont{Jung}},
  \bibinfo{author}{\bibfnamefont{F.}~\bibnamefont{Zhang}}, \bibnamefont{and}
  \bibinfo{author}{\bibfnamefont{A.~H.} \bibnamefont{MacDonald}},
  {``}\bibinfo{title}{Lattice theory of pseudospin ferromagnetism in bilayer
  graphene: Competing interaction-induced quantum Hall states},{''}
  \bibinfo{journal}{Phys. Rev. B} \textbf{\bibinfo{volume}{83}},
  \bibinfo{pages}{115408} (\bibinfo{year}{2011}).

\bibitem[{\citenamefont{Nandkishore and
  Levitov}(2011{\natexlab{b}})}]{Nandkishore_kerr2011a}
\bibinfo{author}{\bibfnamefont{R.}~\bibnamefont{Nandkishore}} \bibnamefont{and}
  \bibinfo{author}{\bibfnamefont{L.}~\bibnamefont{Levitov}},
  {``}\bibinfo{title}{Polar Kerr Effect and Time Reversal Symmetry Breaking in
  Bilayer Graphene},{''} \bibinfo{journal}{Phys. Rev. Lett.}
  \textbf{\bibinfo{volume}{107}}, \bibinfo{pages}{097402}
  (\bibinfo{year}{2011}{\natexlab{b}}).

\bibitem[{\citenamefont{Zhang and MacDonald}(2012)}]{zhang_quant_hall2012}
\bibinfo{author}{\bibfnamefont{F.}~\bibnamefont{Zhang}} \bibnamefont{and}
  \bibinfo{author}{\bibfnamefont{A.~H.} \bibnamefont{MacDonald}},
  {``}\bibinfo{title}{Distinguishing Spontaneous Quantum Hall States in Bilayer
  Graphene},{''} \bibinfo{journal}{Phys. Rev. Lett.}
  \textbf{\bibinfo{volume}{108}}, \bibinfo{pages}{186804}
  (\bibinfo{year}{2012}).

\bibitem[{\citenamefont{Wang et~al.}(2013)\citenamefont{Wang, Wang, Gao, and
  Zhang}}]{af_1princ2013}
\bibinfo{author}{\bibfnamefont{Y.}~\bibnamefont{Wang}},
  \bibinfo{author}{\bibfnamefont{H.}~\bibnamefont{Wang}},
  \bibinfo{author}{\bibfnamefont{J.-H.} \bibnamefont{Gao}}, \bibnamefont{and}
  \bibinfo{author}{\bibfnamefont{F.-C.} \bibnamefont{Zhang}},
  {``}\bibinfo{title}{Layer antiferromagnetic state in bilayer graphene: A
  first-principles investigation},{''} \bibinfo{journal}{Phys. Rev. B}
  \textbf{\bibinfo{volume}{87}}, \bibinfo{pages}{195413}
  (\bibinfo{year}{2013}).

\bibitem[{\citenamefont{Vafek}(2010)}]{vafek_rg2010}
\bibinfo{author}{\bibfnamefont{O.}~\bibnamefont{Vafek}},
  {``}\bibinfo{title}{Interacting fermions on the honeycomb bilayer: From weak
  to strong coupling},{''} \bibinfo{journal}{Phys. Rev. B}
  \textbf{\bibinfo{volume}{82}}, \bibinfo{pages}{205106}
  (\bibinfo{year}{2010}).

\bibitem[{\citenamefont{Lang et~al.}(2012)\citenamefont{Lang, Meng, Scherer,
  Uebelacker, Assaad, Muramatsu, Honerkamp, and Wessel}}]{lang_af_hubb2012}
\bibinfo{author}{\bibfnamefont{T.~C.} \bibnamefont{Lang}},
  \bibinfo{author}{\bibfnamefont{Z.~Y.} \bibnamefont{Meng}},
  \bibinfo{author}{\bibfnamefont{M.~M.} \bibnamefont{Scherer}},
  \bibinfo{author}{\bibfnamefont{S.}~\bibnamefont{Uebelacker}},
  \bibinfo{author}{\bibfnamefont{F.~F.} \bibnamefont{Assaad}},
  \bibinfo{author}{\bibfnamefont{A.}~\bibnamefont{Muramatsu}},
  \bibinfo{author}{\bibfnamefont{C.}~\bibnamefont{Honerkamp}},
  \bibnamefont{and} \bibinfo{author}{\bibfnamefont{S.}~\bibnamefont{Wessel}},
  {``}\bibinfo{title}{Antiferromagnetism in the Hubbard Model on the
  Bernal-Stacked Honeycomb Bilayer},{''} \bibinfo{journal}{Phys. Rev. Lett.}
  \textbf{\bibinfo{volume}{109}}, \bibinfo{pages}{126402}
  (\bibinfo{year}{2012}).

\bibitem[{\citenamefont{Yuan et~al.}(2013)\citenamefont{Yuan, Xu, Wang, Zhou,
  Gao, and Zhang}}]{af_ab_hubb2013}
\bibinfo{author}{\bibfnamefont{J.}~\bibnamefont{Yuan}},
  \bibinfo{author}{\bibfnamefont{D.-H.} \bibnamefont{Xu}},
  \bibinfo{author}{\bibfnamefont{H.}~\bibnamefont{Wang}},
  \bibinfo{author}{\bibfnamefont{Y.}~\bibnamefont{Zhou}},
  \bibinfo{author}{\bibfnamefont{J.-H.} \bibnamefont{Gao}}, \bibnamefont{and}
  \bibinfo{author}{\bibfnamefont{F.-C.} \bibnamefont{Zhang}},
  {``}\bibinfo{title}{Possible half-metallic phase in bilayer graphene:
  Calculations based on mean-field theory applied to a two-layer Hubbard
  model},{''} \bibinfo{journal}{Phys. Rev. B} \textbf{\bibinfo{volume}{88}},
  \bibinfo{pages}{201109} (\bibinfo{year}{2013}).

\bibitem[{\citenamefont{Sun et~al.}(2014)\citenamefont{Sun, Xu, Zhou, and
  Zhang}}]{sun_hub_mc_afm2014}
\bibinfo{author}{\bibfnamefont{J.}~\bibnamefont{Sun}},
  \bibinfo{author}{\bibfnamefont{D.-H.} \bibnamefont{Xu}},
  \bibinfo{author}{\bibfnamefont{Y.}~\bibnamefont{Zhou}}, \bibnamefont{and}
  \bibinfo{author}{\bibfnamefont{F.-C.} \bibnamefont{Zhang}},
  {``}\bibinfo{title}{Electrically controllable magnetic order in the bilayer
  Hubbard model on honeycomb lattice: A determinant quantum Monte Carlo
  study},{''} \bibinfo{journal}{Phys. Rev. B} \textbf{\bibinfo{volume}{90}},
  \bibinfo{pages}{125429} (\bibinfo{year}{2014}).

\bibitem[{\citenamefont{Scherer et~al.}(2012)\citenamefont{Scherer, Uebelacker,
  and Honerkamp}}]{scherer_hub_frg2012}
\bibinfo{author}{\bibfnamefont{M.~M.} \bibnamefont{Scherer}},
  \bibinfo{author}{\bibfnamefont{S.}~\bibnamefont{Uebelacker}},
  \bibnamefont{and}
  \bibinfo{author}{\bibfnamefont{C.}~\bibnamefont{Honerkamp}},
  {``}\bibinfo{title}{Instabilities of interacting electrons on the honeycomb
  bilayer},{''} \bibinfo{journal}{Phys. Rev. B} \textbf{\bibinfo{volume}{85}},
  \bibinfo{pages}{235408} (\bibinfo{year}{2012}).

\bibitem[{\citenamefont{Vafek and Yang}(2010)}]{vafek_nemat_rg2010}
\bibinfo{author}{\bibfnamefont{O.}~\bibnamefont{Vafek}} \bibnamefont{and}
  \bibinfo{author}{\bibfnamefont{K.}~\bibnamefont{Yang}},
  {``}\bibinfo{title}{Many-body instability of Coulomb interacting bilayer
  graphene: Renormalization group approach},{''} \bibinfo{journal}{Phys. Rev.
  B} \textbf{\bibinfo{volume}{81}}, \bibinfo{pages}{041401}
  (\bibinfo{year}{2010}).

\bibitem[{\citenamefont{Lemonik et~al.}(2012)\citenamefont{Lemonik, Aleiner,
  and Fal'ko}}]{lemonic_rg_nemat_long2012}
\bibinfo{author}{\bibfnamefont{Y.}~\bibnamefont{Lemonik}},
  \bibinfo{author}{\bibfnamefont{I.}~\bibnamefont{Aleiner}}, \bibnamefont{and}
  \bibinfo{author}{\bibfnamefont{V.~I.} \bibnamefont{Fal'ko}},
  {``}\bibinfo{title}{Competing nematic, antiferromagnetic, and spin-flux
  orders in the ground state of bilayer graphene},{''} \bibinfo{journal}{Phys.
  Rev. B} \textbf{\bibinfo{volume}{85}}, \bibinfo{pages}{245451}
  (\bibinfo{year}{2012}).

\bibitem[{\citenamefont{Throckmorton and Vafek}(2012)}]{vafek_rg2012}
\bibinfo{author}{\bibfnamefont{R.~E.} \bibnamefont{Throckmorton}}
  \bibnamefont{and} \bibinfo{author}{\bibfnamefont{O.}~\bibnamefont{Vafek}},
  {``}\bibinfo{title}{Fermions on bilayer graphene: Symmetry breaking for B=0
  and $\nu${}=0},{''} \bibinfo{journal}{Phys. Rev. B}
  \textbf{\bibinfo{volume}{86}}, \bibinfo{pages}{115447}
  (\bibinfo{year}{2012}).

\bibitem[{\citenamefont{Cvetkovic et~al.}(2012)\citenamefont{Cvetkovic,
  Throckmorton, and Vafek}}]{cvetkovic_multi2012}
\bibinfo{author}{\bibfnamefont{V.}~\bibnamefont{Cvetkovic}},
  \bibinfo{author}{\bibfnamefont{R.~E.} \bibnamefont{Throckmorton}},
  \bibnamefont{and} \bibinfo{author}{\bibfnamefont{O.}~\bibnamefont{Vafek}},
  {``}\bibinfo{title}{Electronic multicriticality in bilayer graphene},{''}
  \bibinfo{journal}{Phys. Rev. B} \textbf{\bibinfo{volume}{86}},
  \bibinfo{pages}{075467} (\bibinfo{year}{2012}).

\bibitem[{\citenamefont{Lemonik et~al.}(2010)\citenamefont{Lemonik, Aleiner,
  Toke, and Fal'ko}}]{Lemonik2010}
\bibinfo{author}{\bibfnamefont{Y.}~\bibnamefont{Lemonik}},
  \bibinfo{author}{\bibfnamefont{I.~L.} \bibnamefont{Aleiner}},
  \bibinfo{author}{\bibfnamefont{C.}~\bibnamefont{Toke}}, \bibnamefont{and}
  \bibinfo{author}{\bibfnamefont{V.~I.} \bibnamefont{Fal'ko}},
  {``}\bibinfo{title}{Spontaneous symmetry breaking and Lifshitz transition in
  bilayer graphene},{''} \bibinfo{journal}{Phys. Rev. B}
  \textbf{\bibinfo{volume}{82}}, \bibinfo{pages}{201408}
  (\bibinfo{year}{2010}).

\bibitem[{\citenamefont{Gorbar et~al.}(2012)\citenamefont{Gorbar, Gusynin,
  Miransky, and Shovkovy}}]{gorbar_compet_nemta2012}
\bibinfo{author}{\bibfnamefont{E.~V.} \bibnamefont{Gorbar}},
  \bibinfo{author}{\bibfnamefont{V.~P.} \bibnamefont{Gusynin}},
  \bibinfo{author}{\bibfnamefont{V.~A.} \bibnamefont{Miransky}},
  \bibnamefont{and} \bibinfo{author}{\bibfnamefont{I.~A.}
  \bibnamefont{Shovkovy}}, {``}\bibinfo{title}{Coexistence and competition of
  nematic and gapped states in bilayer graphene},{''} \bibinfo{journal}{Phys.
  Rev. B} \textbf{\bibinfo{volume}{86}}, \bibinfo{pages}{125439}
  (\bibinfo{year}{2012}).

\bibitem[{\citenamefont{Ulstrup et~al.}(2014)\citenamefont{Ulstrup, Johannsen,
  Cilento, Miwa, Crepaldi, Zacchigna, Cacho, Chapman, Springate, Mammadov
  et~al.}}]{uultrafast}
\bibinfo{author}{\bibfnamefont{S.}~\bibnamefont{Ulstrup}},
  \bibinfo{author}{\bibfnamefont{J.~C.} \bibnamefont{Johannsen}},
  \bibinfo{author}{\bibfnamefont{F.}~\bibnamefont{Cilento}},
  \bibinfo{author}{\bibfnamefont{J.~A.} \bibnamefont{Miwa}},
  \bibinfo{author}{\bibfnamefont{A.}~\bibnamefont{Crepaldi}},
  \bibinfo{author}{\bibfnamefont{M.}~\bibnamefont{Zacchigna}},
  \bibinfo{author}{\bibfnamefont{C.}~\bibnamefont{Cacho}},
  \bibinfo{author}{\bibfnamefont{R.}~\bibnamefont{Chapman}},
  \bibinfo{author}{\bibfnamefont{E.}~\bibnamefont{Springate}},
  \bibinfo{author}{\bibfnamefont{S.}~\bibnamefont{Mammadov}},
  \bibnamefont{et~al.}, {``}\bibinfo{title}{Ultrafast dynamics of massive Dirac
  fermions in bilayer graphene},{''} \bibinfo{journal}{Phys. Rev. Lett.}
  \textbf{\bibinfo{volume}{112}}, \bibinfo{pages}{257401}
  (\bibinfo{year}{2014}).

\bibitem[{\citenamefont{Zhang et~al.}(2015)\citenamefont{Zhang, Nandkishore,
  and Rossi}}]{zhangDisorder}
\bibinfo{author}{\bibfnamefont{J.}~\bibnamefont{Zhang}},
  \bibinfo{author}{\bibfnamefont{R.}~\bibnamefont{Nandkishore}},
  \bibnamefont{and} \bibinfo{author}{\bibfnamefont{E.}~\bibnamefont{Rossi}},
  {``}\bibinfo{title}{Disorder-tuned selection of order in bilayer
  graphene},{''} \bibinfo{journal}{Phys. Rev. B} \textbf{\bibinfo{volume}{91}},
  \bibinfo{pages}{205425} (\bibinfo{year}{2015}).

\bibitem[{\citenamefont{Dillenschneider and Han}(2008)}]{Dillenschneider2008}
\bibinfo{author}{\bibfnamefont{R.}~\bibnamefont{Dillenschneider}}
  \bibnamefont{and} \bibinfo{author}{\bibfnamefont{J.~H.} \bibnamefont{Han}},
  {``}\bibinfo{title}{Exciton formation in graphene bilayer},{''}
  \bibinfo{journal}{Phys. Rev. B} \textbf{\bibinfo{volume}{78}},
  \bibinfo{pages}{045401} (\bibinfo{year}{2008}).

\bibitem[{\citenamefont{Dahal et~al.}(2010)\citenamefont{Dahal, Wehling,
  Bedell, Zhu, and Balatsky}}]{Dahal_cdw2010}
\bibinfo{author}{\bibfnamefont{H.~P.} \bibnamefont{Dahal}},
  \bibinfo{author}{\bibfnamefont{T.~O.} \bibnamefont{Wehling}},
  \bibinfo{author}{\bibfnamefont{K.~S.} \bibnamefont{Bedell}},
  \bibinfo{author}{\bibfnamefont{J.-X.} \bibnamefont{Zhu}}, \bibnamefont{and}
  \bibinfo{author}{\bibfnamefont{A.}~\bibnamefont{Balatsky}},
  {``}\bibinfo{title}{Charge inhomogeneity in a single and bilayer
  graphene},{''} \bibinfo{journal}{Physica B} \textbf{\bibinfo{volume}{405}},
  \bibinfo{pages}{2241 } (\bibinfo{year}{2010}).

\bibitem[{\citenamefont{Zhu et~al.}(2013)\citenamefont{Zhu, Aji, and
  Varma}}]{zhu_loop_order2013}
\bibinfo{author}{\bibfnamefont{L.}~\bibnamefont{Zhu}},
  \bibinfo{author}{\bibfnamefont{V.}~\bibnamefont{Aji}}, \bibnamefont{and}
  \bibinfo{author}{\bibfnamefont{C.~M.} \bibnamefont{Varma}},
  {``}\bibinfo{title}{Ordered loop current states in bilayer graphene},{''}
  \bibinfo{journal}{Phys. Rev. B} \textbf{\bibinfo{volume}{87}},
  \bibinfo{pages}{035427} (\bibinfo{year}{2013}).

\bibitem[{\citenamefont{Barlas and Yang}(2009)}]{Barlas2009}
\bibinfo{author}{\bibfnamefont{Y.}~\bibnamefont{Barlas}} \bibnamefont{and}
  \bibinfo{author}{\bibfnamefont{K.}~\bibnamefont{Yang}},
  {``}\bibinfo{title}{Non-Fermi-liquid behavior in neutral bilayer
  graphene},{''} \bibinfo{journal}{Phys. Rev. B} \textbf{\bibinfo{volume}{80}},
  \bibinfo{pages}{161408} (\bibinfo{year}{2009}).

\bibitem[{\citenamefont{Nandkishore and
  Levitov}(2010{\natexlab{c}})}]{Nandkishore2010a}
\bibinfo{author}{\bibfnamefont{R.}~\bibnamefont{Nandkishore}} \bibnamefont{and}
  \bibinfo{author}{\bibfnamefont{L.}~\bibnamefont{Levitov}},
  {``}\bibinfo{title}{Electron interactions in bilayer graphene: Marginal Fermi
  liquid and zero-bias anomaly},{''} \bibinfo{journal}{Phys. Rev. B}
  \textbf{\bibinfo{volume}{82}}, \bibinfo{pages}{115431}
  (\bibinfo{year}{2010}{\natexlab{c}}).

\bibitem[{\citenamefont{Kusminskiy et~al.}(2008)\citenamefont{Kusminskiy,
  Nilsson, Campbell, and Castro~Neto}}]{Kusminskiy2008}
\bibinfo{author}{\bibfnamefont{S.~V.} \bibnamefont{Kusminskiy}},
  \bibinfo{author}{\bibfnamefont{J.}~\bibnamefont{Nilsson}},
  \bibinfo{author}{\bibfnamefont{D.~K.} \bibnamefont{Campbell}},
  \bibnamefont{and} \bibinfo{author}{\bibfnamefont{A.~H.}
  \bibnamefont{Castro~Neto}}, {``}\bibinfo{title}{Electronic compressibility of
  a graphene bilayer},{''} \bibinfo{journal}{Phys. Rev. Lett.}
  \textbf{\bibinfo{volume}{100}}, \bibinfo{pages}{106805}
  (\bibinfo{year}{2008}).

\bibitem[{\citenamefont{Borghi et~al.}(2009)\citenamefont{Borghi, Polini,
  Asgari, and MacDonald}}]{Borghi2009a}
\bibinfo{author}{\bibfnamefont{G.}~\bibnamefont{Borghi}},
  \bibinfo{author}{\bibfnamefont{M.}~\bibnamefont{Polini}},
  \bibinfo{author}{\bibfnamefont{R.}~\bibnamefont{Asgari}}, \bibnamefont{and}
  \bibinfo{author}{\bibfnamefont{A.~H.} \bibnamefont{MacDonald}},
  {``}\bibinfo{title}{Fermi velocity enhancement in monolayer and bilayer
  graphene},{''} \bibinfo{journal}{Solid State Commun.}
  \textbf{\bibinfo{volume}{149}}, \bibinfo{pages}{1117} (\bibinfo{year}{2009}).

\bibitem[{\citenamefont{Sensarma et~al.}(2011)\citenamefont{Sensarma, Hwang,
  and Das~Sarma}}]{Sensarma2011}
\bibinfo{author}{\bibfnamefont{R.}~\bibnamefont{Sensarma}},
  \bibinfo{author}{\bibfnamefont{E.~H.} \bibnamefont{Hwang}}, \bibnamefont{and}
  \bibinfo{author}{\bibfnamefont{S.}~\bibnamefont{Das~Sarma}},
  {``}\bibinfo{title}{Quasiparticles, plasmarons, and quantum spectral function
  in bilayer graphene},{''} \bibinfo{journal}{Phys. Rev. B}
  \textbf{\bibinfo{volume}{84}}, \bibinfo{pages}{041408}
  (\bibinfo{year}{2011}).

\bibitem[{\citenamefont{Shizuya}(2010)}]{Shizuya2010}
\bibinfo{author}{\bibfnamefont{K.}~\bibnamefont{Shizuya}},
  {``}\bibinfo{title}{Many-body corrections to cyclotron resonance in monolayer
  and bilayer graphene},{''} \bibinfo{journal}{Phys. Rev. B}
  \textbf{\bibinfo{volume}{81}}, \bibinfo{pages}{075407}
  (\bibinfo{year}{2010}).

\bibitem[{\citenamefont{Cheng et~al.}(2015)\citenamefont{Cheng, Xie, Pachoud,
  Moser, Chen, Wee, Castro~Neto, Tsuei, and {\"O}zyilmaz}}]{cheng2015}
\bibinfo{author}{\bibfnamefont{C.-M.} \bibnamefont{Cheng}},
  \bibinfo{author}{\bibfnamefont{L.~F.} \bibnamefont{Xie}},
  \bibinfo{author}{\bibfnamefont{A.}~\bibnamefont{Pachoud}},
  \bibinfo{author}{\bibfnamefont{H.~O.} \bibnamefont{Moser}},
  \bibinfo{author}{\bibfnamefont{W.}~\bibnamefont{Chen}},
  \bibinfo{author}{\bibfnamefont{A.~T.~S.} \bibnamefont{Wee}},
  \bibinfo{author}{\bibfnamefont{A.~H.} \bibnamefont{Castro~Neto}},
  \bibinfo{author}{\bibfnamefont{K.-D.} \bibnamefont{Tsuei}}, \bibnamefont{and}
  \bibinfo{author}{\bibfnamefont{B.}~\bibnamefont{{\"O}zyilmaz}},
  {``}\bibinfo{title}{Anomalous Spectral Features of a Neutral Bilayer
  Graphene},{''} \bibinfo{journal}{Scientific reports}
  \textbf{\bibinfo{volume}{5}} (\bibinfo{year}{2015}).

\bibitem[{\citenamefont{Hwang and Das~Sarma}(2008)}]{Hwang2008}
\bibinfo{author}{\bibfnamefont{E.~H.} \bibnamefont{Hwang}} \bibnamefont{and}
  \bibinfo{author}{\bibfnamefont{S.}~\bibnamefont{Das~Sarma}},
  {``}\bibinfo{title}{Screening, Kohn Anomaly, Friedel Oscillation, and RKKY
  Interaction in Bilayer Graphene},{''} \bibinfo{journal}{Phys. Rev. Lett.}
  \textbf{\bibinfo{volume}{101}}, \bibinfo{pages}{156802}
  (\bibinfo{year}{2008}).

\bibitem[{\citenamefont{Sensarma et~al.}(2010)\citenamefont{Sensarma, Hwang,
  and Das~Sarma}}]{Sensarma2010}
\bibinfo{author}{\bibfnamefont{R.}~\bibnamefont{Sensarma}},
  \bibinfo{author}{\bibfnamefont{E.~H.} \bibnamefont{Hwang}}, \bibnamefont{and}
  \bibinfo{author}{\bibfnamefont{S.}~\bibnamefont{Das~Sarma}},
  {``}\bibinfo{title}{Dynamic screening and low-energy collective modes in
  bilayer graphene},{''} \bibinfo{journal}{Phys. Rev. B}
  \textbf{\bibinfo{volume}{82}}, \bibinfo{pages}{195428}
  (\bibinfo{year}{2010}).

\bibitem[{\citenamefont{Gamayun}(2011)}]{Gamayun2011}
\bibinfo{author}{\bibfnamefont{O.~V.} \bibnamefont{Gamayun}},
  {``}\bibinfo{title}{Dynamical screening in bilayer graphene},{''}
  \bibinfo{journal}{Phys. Rev. B} \textbf{\bibinfo{volume}{84}},
  \bibinfo{pages}{085112} (\bibinfo{year}{2011}).

\bibitem[{\citenamefont{Triola and Rossi}(2012)}]{Tirola2012}
\bibinfo{author}{\bibfnamefont{C.}~\bibnamefont{Triola}} \bibnamefont{and}
  \bibinfo{author}{\bibfnamefont{E.}~\bibnamefont{Rossi}},
  {``}\bibinfo{title}{Screening and collective modes in gapped bilayer
  graphene},{''} \bibinfo{journal}{Phys. Rev. B} \textbf{\bibinfo{volume}{86}},
  \bibinfo{pages}{161408} (\bibinfo{year}{2012}).

\bibitem[{\citenamefont{Wang and Chakraborty}(2007)}]{Wang2007}
\bibinfo{author}{\bibfnamefont{X.-F.} \bibnamefont{Wang}} \bibnamefont{and}
  \bibinfo{author}{\bibfnamefont{T.}~\bibnamefont{Chakraborty}},
  {``}\bibinfo{title}{Coulomb screening and collective excitations in a
  graphene bilayer},{''} \bibinfo{journal}{Phys. Rev. B}
  \textbf{\bibinfo{volume}{75}}, \bibinfo{pages}{041404}
  (\bibinfo{year}{2007}).

\bibitem[{\citenamefont{Wang and Chakraborty}(2010)}]{Wang2010a}
\bibinfo{author}{\bibfnamefont{X.-F.} \bibnamefont{Wang}} \bibnamefont{and}
  \bibinfo{author}{\bibfnamefont{T.}~\bibnamefont{Chakraborty}},
  {``}\bibinfo{title}{Coulomb screening and collective excitations in biased
  bilayer graphene},{''} \bibinfo{journal}{Phys. Rev. B}
  \textbf{\bibinfo{volume}{81}}, \bibinfo{pages}{081402}
  (\bibinfo{year}{2010}).

\bibitem[{\citenamefont{Pisarra et~al.}(2016)\citenamefont{Pisarra, Sindona,
  Gravina, Silkin, and Pitarke}}]{pisarra_screening_plasmon2016}
\bibinfo{author}{\bibfnamefont{M.}~\bibnamefont{Pisarra}},
  \bibinfo{author}{\bibfnamefont{A.}~\bibnamefont{Sindona}},
  \bibinfo{author}{\bibfnamefont{M.}~\bibnamefont{Gravina}},
  \bibinfo{author}{\bibfnamefont{V.~M.} \bibnamefont{Silkin}},
  \bibnamefont{and} \bibinfo{author}{\bibfnamefont{J.~M.}
  \bibnamefont{Pitarke}}, {``}\bibinfo{title}{Dielectric screening and plasmon
  resonances in bilayer graphene},{''} \bibinfo{journal}{Phys. Rev. B}
  \textbf{\bibinfo{volume}{93}}, \bibinfo{pages}{035440}
  (\bibinfo{year}{2016}).

\bibitem[{\citenamefont{Klier et~al.}(2015)\citenamefont{Klier, Shallcross,
  Sharma, and Pankratov}}]{klier_rkky2015}
\bibinfo{author}{\bibfnamefont{N.}~\bibnamefont{Klier}},
  \bibinfo{author}{\bibfnamefont{S.}~\bibnamefont{Shallcross}},
  \bibinfo{author}{\bibfnamefont{S.}~\bibnamefont{Sharma}}, \bibnamefont{and}
  \bibinfo{author}{\bibfnamefont{O.}~\bibnamefont{Pankratov}},
  {``}\bibinfo{title}{Ruderman-Kittel-Kasuya-Yosida interaction at finite
  temperature: Graphene and bilayer graphene},{''} \bibinfo{journal}{Phys. Rev.
  B} \textbf{\bibinfo{volume}{92}}, \bibinfo{pages}{205414}
  (\bibinfo{year}{2015}).

\bibitem[{\citenamefont{Mohammadi}(2015)}]{mohammadi}
\bibinfo{author}{\bibfnamefont{Y.}~\bibnamefont{Mohammadi}},
  {``}\bibinfo{title}{Charge screening and carrier transport in AA-stacked
  bilayer graphene: tuning via a perpendicular electric field},{''}
  \bibinfo{journal}{Solid State Commun.} \textbf{\bibinfo{volume}{202}},
  \bibinfo{pages}{14} (\bibinfo{year}{2015}).

\bibitem[{\citenamefont{Ho et~al.}(2006)\citenamefont{Ho, Lu, Hwang, Chang, and
  Lin}}]{plasmon_theor_aa_ab}
\bibinfo{author}{\bibfnamefont{J.~H.} \bibnamefont{Ho}},
  \bibinfo{author}{\bibfnamefont{C.~L.} \bibnamefont{Lu}},
  \bibinfo{author}{\bibfnamefont{C.~C.} \bibnamefont{Hwang}},
  \bibinfo{author}{\bibfnamefont{C.~P.} \bibnamefont{Chang}}, \bibnamefont{and}
  \bibinfo{author}{\bibfnamefont{M.~F.} \bibnamefont{Lin}},
  {``}\bibinfo{title}{Coulomb excitations in AA- and AB-stacked bilayer
  graphites},{''} \bibinfo{journal}{Phys. Rev. B}
  \textbf{\bibinfo{volume}{74}}, \bibinfo{pages}{085406}
  (\bibinfo{year}{2006}).

\bibitem[{\citenamefont{Lin et~al.}(2012)\citenamefont{Lin, Chuang, and
  Wu}}]{plasmon_theor_aa_mult2012}
\bibinfo{author}{\bibfnamefont{M.-F.} \bibnamefont{Lin}},
  \bibinfo{author}{\bibfnamefont{Y.-C.} \bibnamefont{Chuang}},
  \bibnamefont{and} \bibinfo{author}{\bibfnamefont{J.-Y.} \bibnamefont{Wu}},
  {``}\bibinfo{title}{Electrically tunable plasma excitations in AA-stacked
  multilayer graphene},{''} \bibinfo{journal}{Phys. Rev. B}
  \textbf{\bibinfo{volume}{86}}, \bibinfo{pages}{125434}
  (\bibinfo{year}{2012}).

\bibitem[{\citenamefont{Rold\'an and Brey}(2013)}]{Roldan2013}
\bibinfo{author}{\bibfnamefont{R.}~\bibnamefont{Rold\'an}} \bibnamefont{and}
  \bibinfo{author}{\bibfnamefont{L.}~\bibnamefont{Brey}},
  {``}\bibinfo{title}{Dielectric screening and plasmons in AA-stacked bilayer
  graphene},{''} \bibinfo{journal}{Phys. Rev. B} \textbf{\bibinfo{volume}{88}},
  \bibinfo{pages}{115420} (\bibinfo{year}{2013}).

\bibitem[{\citenamefont{Wang et~al.}(2016)\citenamefont{Wang, Xiao, and
  Mortensen}}]{wang_plasmon_aa_ab2016}
\bibinfo{author}{\bibfnamefont{W.}~\bibnamefont{Wang}},
  \bibinfo{author}{\bibfnamefont{S.}~\bibnamefont{Xiao}}, \bibnamefont{and}
  \bibinfo{author}{\bibfnamefont{N.~A.} \bibnamefont{Mortensen}},
  {``}\bibinfo{title}{Localized plasmons in bilayer graphene nanodisks},{''}
  \bibinfo{journal}{Phys. Rev. B} \textbf{\bibinfo{volume}{93}},
  \bibinfo{pages}{165407} (\bibinfo{year}{2016}).

\bibitem[{\citenamefont{Tahir and Sabeeh}(2008)}]{Tahir2008}
\bibinfo{author}{\bibfnamefont{M.}~\bibnamefont{Tahir}} \bibnamefont{and}
  \bibinfo{author}{\bibfnamefont{K.}~\bibnamefont{Sabeeh}},
  {``}\bibinfo{title}{Inter-band magnetoplasmons in mono- and bilayer
  graphene},{''} \bibinfo{journal}{J. Phys.: Condens. Matter}
  \textbf{\bibinfo{volume}{20}}, \bibinfo{pages}{425202}
  (\bibinfo{year}{2008}).

\bibitem[{\citenamefont{Mikhailov and
  Ziegler}(2007)}]{transvers_plasmon_slg2007}
\bibinfo{author}{\bibfnamefont{S.~A.} \bibnamefont{Mikhailov}}
  \bibnamefont{and} \bibinfo{author}{\bibfnamefont{K.}~\bibnamefont{Ziegler}},
  {``}\bibinfo{title}{New Electromagnetic Mode in Graphene},{''}
  \bibinfo{journal}{Phys. Rev. Lett.} \textbf{\bibinfo{volume}{99}},
  \bibinfo{pages}{016803} (\bibinfo{year}{2007}).

\bibitem[{\citenamefont{Fal'ko and
  Khmel'nitskii}(1989)}]{cond_vs_speed_of_light}
\bibinfo{author}{\bibfnamefont{V.~I.} \bibnamefont{Fal'ko}} \bibnamefont{and}
  \bibinfo{author}{\bibfnamefont{D.~E.} \bibnamefont{Khmel'nitskii}},
  {``}\bibinfo{title}{What if a film conductivity exceeds the speed of
  light?},{''} \bibinfo{journal}{Sov. Phys. JETP}
  \textbf{\bibinfo{volume}{68}}, \bibinfo{pages}{1150} (\bibinfo{year}{1989}).

\bibitem[{\citenamefont{Jablan et~al.}(2011)\citenamefont{Jablan, Buljan, and
  Soljacic}}]{Jablan2011}
\bibinfo{author}{\bibfnamefont{M.}~\bibnamefont{Jablan}},
  \bibinfo{author}{\bibfnamefont{H.}~\bibnamefont{Buljan}}, \bibnamefont{and}
  \bibinfo{author}{\bibfnamefont{M.}~\bibnamefont{Soljacic}},
  {``}\bibinfo{title}{Transverse electric plasmons in bilayer graphene},{''}
  \bibinfo{journal}{Opt. Express} \textbf{\bibinfo{volume}{19}},
  \bibinfo{pages}{11236} (\bibinfo{year}{2011}).

\bibitem[{\citenamefont{Stauber et~al.}(2013)\citenamefont{Stauber, San-Jose,
  and Brey}}]{SigmaPlasmons}
\bibinfo{author}{\bibfnamefont{T.}~\bibnamefont{Stauber}},
  \bibinfo{author}{\bibfnamefont{P.}~\bibnamefont{San-Jose}}, \bibnamefont{and}
  \bibinfo{author}{\bibfnamefont{L.}~\bibnamefont{Brey}},
  {``}\bibinfo{title}{Optical conductivity, Drude weight and plasmons in
  twisted graphene bilayers},{''} \bibinfo{journal}{New J. Phys.}
  \textbf{\bibinfo{volume}{15}}, \bibinfo{pages}{113050}
  (\bibinfo{year}{2013}).

\bibitem[{\citenamefont{Low et~al.}(2014)\citenamefont{Low, Guinea, Yan, Xia,
  and Avouris}}]{LowP}
\bibinfo{author}{\bibfnamefont{T.}~\bibnamefont{Low}},
  \bibinfo{author}{\bibfnamefont{F.}~\bibnamefont{Guinea}},
  \bibinfo{author}{\bibfnamefont{H.}~\bibnamefont{Yan}},
  \bibinfo{author}{\bibfnamefont{F.}~\bibnamefont{Xia}}, \bibnamefont{and}
  \bibinfo{author}{\bibfnamefont{P.}~\bibnamefont{Avouris}},
  {``}\bibinfo{title}{Novel midinfrared plasmonic properties of bilayer
  graphene},{''} \bibinfo{journal}{Phys. Rev. Lett.}
  \textbf{\bibinfo{volume}{112}}, \bibinfo{pages}{116801}
  (\bibinfo{year}{2014}).

\bibitem[{\citenamefont{Fei et~al.}(2015)\citenamefont{Fei, Iwinski, Ni, Zhang,
  Bao, Rodin, Lee, Wagner, Liu, Dai et~al.}}]{FeiBasov}
\bibinfo{author}{\bibfnamefont{Z.}~\bibnamefont{Fei}},
  \bibinfo{author}{\bibfnamefont{E.}~\bibnamefont{Iwinski}},
  \bibinfo{author}{\bibfnamefont{G.}~\bibnamefont{Ni}},
  \bibinfo{author}{\bibfnamefont{L.}~\bibnamefont{Zhang}},
  \bibinfo{author}{\bibfnamefont{W.}~\bibnamefont{Bao}},
  \bibinfo{author}{\bibfnamefont{A.}~\bibnamefont{Rodin}},
  \bibinfo{author}{\bibfnamefont{Y.}~\bibnamefont{Lee}},
  \bibinfo{author}{\bibfnamefont{M.}~\bibnamefont{Wagner}},
  \bibinfo{author}{\bibfnamefont{M.}~\bibnamefont{Liu}},
  \bibinfo{author}{\bibfnamefont{S.}~\bibnamefont{Dai}}, \bibnamefont{et~al.},
  {``}\bibinfo{title}{Tunneling plasmonics in bilayer graphene},{''}
  \bibinfo{journal}{Nano Lett.} \textbf{\bibinfo{volume}{15}},
  \bibinfo{pages}{4973} (\bibinfo{year}{2015}).

\bibitem[{\citenamefont{Yan et~al.}(2014{\natexlab{a}})\citenamefont{Yan, Low,
  Guinea, Xia, and Avouris}}]{YanTr}
\bibinfo{author}{\bibfnamefont{H.}~\bibnamefont{Yan}},
  \bibinfo{author}{\bibfnamefont{T.}~\bibnamefont{Low}},
  \bibinfo{author}{\bibfnamefont{F.}~\bibnamefont{Guinea}},
  \bibinfo{author}{\bibfnamefont{F.}~\bibnamefont{Xia}}, \bibnamefont{and}
  \bibinfo{author}{\bibfnamefont{P.}~\bibnamefont{Avouris}},
  {``}\bibinfo{title}{Tunable phonon-induced transparency in bilayer graphene
  nanoribbons},{''} \bibinfo{journal}{Nano Lett.}
  \textbf{\bibinfo{volume}{14}}, \bibinfo{pages}{4581}
  (\bibinfo{year}{2014}{\natexlab{a}}).

\bibitem[{\citenamefont{Hass et~al.}(2008)\citenamefont{Hass, Varchon,
  Mill\'an-Otoya, Sprinkle, Sharma, de~Heer, Berger, First, Magaud, and
  Conrad}}]{STM_DFT}
\bibinfo{author}{\bibfnamefont{J.}~\bibnamefont{Hass}},
  \bibinfo{author}{\bibfnamefont{F.}~\bibnamefont{Varchon}},
  \bibinfo{author}{\bibfnamefont{J.~E.} \bibnamefont{Mill\'an-Otoya}},
  \bibinfo{author}{\bibfnamefont{M.}~\bibnamefont{Sprinkle}},
  \bibinfo{author}{\bibfnamefont{N.}~\bibnamefont{Sharma}},
  \bibinfo{author}{\bibfnamefont{W.~A.} \bibnamefont{de~Heer}},
  \bibinfo{author}{\bibfnamefont{C.}~\bibnamefont{Berger}},
  \bibinfo{author}{\bibfnamefont{P.~N.} \bibnamefont{First}},
  \bibinfo{author}{\bibfnamefont{L.}~\bibnamefont{Magaud}}, \bibnamefont{and}
  \bibinfo{author}{\bibfnamefont{E.~H.} \bibnamefont{Conrad}},
  {``}\bibinfo{title}{Why Multilayer Graphene on
  $4H\mathrm{\text{-}}\mathrm{SiC}(000\overline{1})$ Behaves Like a Single
  Sheet of Graphene},{''} \bibinfo{journal}{Phys. Rev. Lett.}
  \textbf{\bibinfo{volume}{100}}, \bibinfo{pages}{125504}
  (\bibinfo{year}{2008}).

\bibitem[{\citenamefont{Meng et~al.}(2012)\citenamefont{Meng, Zhang, Yan, Feng,
  He, Dou, and Nie}}]{MengSTM}
\bibinfo{author}{\bibfnamefont{L.}~\bibnamefont{Meng}},
  \bibinfo{author}{\bibfnamefont{Y.}~\bibnamefont{Zhang}},
  \bibinfo{author}{\bibfnamefont{W.}~\bibnamefont{Yan}},
  \bibinfo{author}{\bibfnamefont{L.}~\bibnamefont{Feng}},
  \bibinfo{author}{\bibfnamefont{L.}~\bibnamefont{He}},
  \bibinfo{author}{\bibfnamefont{R.-F.} \bibnamefont{Dou}}, \bibnamefont{and}
  \bibinfo{author}{\bibfnamefont{J.-C.} \bibnamefont{Nie}},
  {``}\bibinfo{title}{Single-layer behavior and slow carrier density dynamic of
  twisted graphene bilayer},{''} \bibinfo{journal}{Appl. Phys. Lett.}
  \textbf{\bibinfo{volume}{100}}, \bibinfo{eid}{091601} (\bibinfo{year}{2012}).

\bibitem[{\citenamefont{Hicks et~al.}(2011)\citenamefont{Hicks, Sprinkle,
  Shepperd, Wang, Tejeda, Taleb-Ibrahimi, Bertran, Le~F\`evre, de~Heer, Berger
  et~al.}}]{HicksARPES}
\bibinfo{author}{\bibfnamefont{J.}~\bibnamefont{Hicks}},
  \bibinfo{author}{\bibfnamefont{M.}~\bibnamefont{Sprinkle}},
  \bibinfo{author}{\bibfnamefont{K.}~\bibnamefont{Shepperd}},
  \bibinfo{author}{\bibfnamefont{F.}~\bibnamefont{Wang}},
  \bibinfo{author}{\bibfnamefont{A.}~\bibnamefont{Tejeda}},
  \bibinfo{author}{\bibfnamefont{A.}~\bibnamefont{Taleb-Ibrahimi}},
  \bibinfo{author}{\bibfnamefont{F.}~\bibnamefont{Bertran}},
  \bibinfo{author}{\bibfnamefont{P.}~\bibnamefont{Le~F\`evre}},
  \bibinfo{author}{\bibfnamefont{W.~A.} \bibnamefont{de~Heer}},
  \bibinfo{author}{\bibfnamefont{C.}~\bibnamefont{Berger}},
  \bibnamefont{et~al.}, {``}\bibinfo{title}{Symmetry breaking in commensurate
  graphene rotational stacking: Comparison of theory and experiment},{''}
  \bibinfo{journal}{Phys. Rev. B} \textbf{\bibinfo{volume}{83}},
  \bibinfo{pages}{205403} (\bibinfo{year}{2011}).

\bibitem[{\citenamefont{Luican et~al.}(2011)\citenamefont{Luican, Li, Reina,
  Kong, Nair, Novoselov, Geim, and Andrei}}]{STM2}
\bibinfo{author}{\bibfnamefont{A.}~\bibnamefont{Luican}},
  \bibinfo{author}{\bibfnamefont{G.}~\bibnamefont{Li}},
  \bibinfo{author}{\bibfnamefont{A.}~\bibnamefont{Reina}},
  \bibinfo{author}{\bibfnamefont{J.}~\bibnamefont{Kong}},
  \bibinfo{author}{\bibfnamefont{R.~R.} \bibnamefont{Nair}},
  \bibinfo{author}{\bibfnamefont{K.~S.} \bibnamefont{Novoselov}},
  \bibinfo{author}{\bibfnamefont{A.~K.} \bibnamefont{Geim}}, \bibnamefont{and}
  \bibinfo{author}{\bibfnamefont{E.~Y.} \bibnamefont{Andrei}},
  {``}\bibinfo{title}{Single-Layer Behavior and Its Breakdown in Twisted
  Graphene Layers},{''} \bibinfo{journal}{Phys. Rev. Lett.}
  \textbf{\bibinfo{volume}{106}}, \bibinfo{pages}{126802}
  (\bibinfo{year}{2011}).

\bibitem[{\citenamefont{Yan et~al.}(2012)\citenamefont{Yan, Liu, Dou, Meng,
  Feng, Chu, Zhang, Liu, Nie, and He}}]{STM_VHS2}
\bibinfo{author}{\bibfnamefont{W.}~\bibnamefont{Yan}},
  \bibinfo{author}{\bibfnamefont{M.}~\bibnamefont{Liu}},
  \bibinfo{author}{\bibfnamefont{R.-F.} \bibnamefont{Dou}},
  \bibinfo{author}{\bibfnamefont{L.}~\bibnamefont{Meng}},
  \bibinfo{author}{\bibfnamefont{L.}~\bibnamefont{Feng}},
  \bibinfo{author}{\bibfnamefont{Z.-D.} \bibnamefont{Chu}},
  \bibinfo{author}{\bibfnamefont{Y.}~\bibnamefont{Zhang}},
  \bibinfo{author}{\bibfnamefont{Z.}~\bibnamefont{Liu}},
  \bibinfo{author}{\bibfnamefont{J.-C.} \bibnamefont{Nie}}, \bibnamefont{and}
  \bibinfo{author}{\bibfnamefont{L.}~\bibnamefont{He}},
  {``}\bibinfo{title}{Angle-Dependent van Hove Singularities in a Slightly
  Twisted Graphene Bilayer},{''} \bibinfo{journal}{Phys. Rev. Lett.}
  \textbf{\bibinfo{volume}{109}}, \bibinfo{pages}{126801}
  (\bibinfo{year}{2012}).

\bibitem[{\citenamefont{Brown et~al.}(2012)\citenamefont{Brown, Hovden, Huang,
  Wojcik, Muller, and Park}}]{DF_TEM1}
\bibinfo{author}{\bibfnamefont{L.}~\bibnamefont{Brown}},
  \bibinfo{author}{\bibfnamefont{R.}~\bibnamefont{Hovden}},
  \bibinfo{author}{\bibfnamefont{P.}~\bibnamefont{Huang}},
  \bibinfo{author}{\bibfnamefont{M.}~\bibnamefont{Wojcik}},
  \bibinfo{author}{\bibfnamefont{D.~A.} \bibnamefont{Muller}},
  \bibnamefont{and} \bibinfo{author}{\bibfnamefont{J.}~\bibnamefont{Park}},
  {``}\bibinfo{title}{Twinning and Twisting of Tri- and Bilayer Graphene},{''}
  \bibinfo{journal}{Nano Lett.} \textbf{\bibinfo{volume}{12}},
  \bibinfo{pages}{1609} (\bibinfo{year}{2012}).

\bibitem[{\citenamefont{Havener et~al.}(2012)\citenamefont{Havener, Zhuang,
  Brown, Hennig, and Park}}]{DF_TEM2}
\bibinfo{author}{\bibfnamefont{R.~W.} \bibnamefont{Havener}},
  \bibinfo{author}{\bibfnamefont{H.}~\bibnamefont{Zhuang}},
  \bibinfo{author}{\bibfnamefont{L.}~\bibnamefont{Brown}},
  \bibinfo{author}{\bibfnamefont{R.~G.} \bibnamefont{Hennig}},
  \bibnamefont{and} \bibinfo{author}{\bibfnamefont{J.}~\bibnamefont{Park}},
  {``}\bibinfo{title}{Angle-Resolved Raman Imaging of Interlayer Rotations and
  Interactions in Twisted Bilayer Graphene},{''} \bibinfo{journal}{Nano Lett.}
  \textbf{\bibinfo{volume}{12}}, \bibinfo{pages}{3162} (\bibinfo{year}{2012}).

\bibitem[{\citenamefont{Righi et~al.}(2011)\citenamefont{Righi, Costa, Chacham,
  Fantini, Venezuela, Magnuson, Colombo, Bacsa, Ruoff, and
  Pimenta}}]{RighiRaman1}
\bibinfo{author}{\bibfnamefont{A.}~\bibnamefont{Righi}},
  \bibinfo{author}{\bibfnamefont{S.~D.} \bibnamefont{Costa}},
  \bibinfo{author}{\bibfnamefont{H.}~\bibnamefont{Chacham}},
  \bibinfo{author}{\bibfnamefont{C.}~\bibnamefont{Fantini}},
  \bibinfo{author}{\bibfnamefont{P.}~\bibnamefont{Venezuela}},
  \bibinfo{author}{\bibfnamefont{C.}~\bibnamefont{Magnuson}},
  \bibinfo{author}{\bibfnamefont{L.}~\bibnamefont{Colombo}},
  \bibinfo{author}{\bibfnamefont{W.~S.} \bibnamefont{Bacsa}},
  \bibinfo{author}{\bibfnamefont{R.~S.} \bibnamefont{Ruoff}}, \bibnamefont{and}
  \bibinfo{author}{\bibfnamefont{M.~A.} \bibnamefont{Pimenta}},
  {``}\bibinfo{title}{Graphene Moir\'e patterns observed by umklapp
  double-resonance Raman scattering},{''} \bibinfo{journal}{Phys. Rev. B}
  \textbf{\bibinfo{volume}{84}}, \bibinfo{pages}{241409}
  (\bibinfo{year}{2011}).

\bibitem[{\citenamefont{Righi et~al.}(2013)\citenamefont{Righi, Venezuela,
  Chacham, Costa, Fantini, Ruoff, Colombo, Bacsa, and Pimenta}}]{RighiRaman2}
\bibinfo{author}{\bibfnamefont{A.}~\bibnamefont{Righi}},
  \bibinfo{author}{\bibfnamefont{P.}~\bibnamefont{Venezuela}},
  \bibinfo{author}{\bibfnamefont{H.}~\bibnamefont{Chacham}},
  \bibinfo{author}{\bibfnamefont{S.}~\bibnamefont{Costa}},
  \bibinfo{author}{\bibfnamefont{C.}~\bibnamefont{Fantini}},
  \bibinfo{author}{\bibfnamefont{R.}~\bibnamefont{Ruoff}},
  \bibinfo{author}{\bibfnamefont{L.}~\bibnamefont{Colombo}},
  \bibinfo{author}{\bibfnamefont{W.}~\bibnamefont{Bacsa}}, \bibnamefont{and}
  \bibinfo{author}{\bibfnamefont{M.}~\bibnamefont{Pimenta}},
  {``}\bibinfo{title}{Resonance Raman spectroscopy in twisted bilayer
  graphene},{''} \bibinfo{journal}{Solid State Commun.}
  \textbf{\bibinfo{volume}{175â176}}, \bibinfo{pages}{13 }
  (\bibinfo{year}{2013}), \bibinfo{note}{special Issue: Graphene V: Recent
  Advances in Studies of Graphene and Graphene analogues}.

\bibitem[{\citenamefont{Robinson et~al.}(2013)\citenamefont{Robinson,
  Schmucker, Diaconescu, Long, Culbertson, Ohta, Friedman, and
  Beechem}}]{RobinsonRaman}
\bibinfo{author}{\bibfnamefont{J.~T.} \bibnamefont{Robinson}},
  \bibinfo{author}{\bibfnamefont{S.~W.} \bibnamefont{Schmucker}},
  \bibinfo{author}{\bibfnamefont{C.~B.} \bibnamefont{Diaconescu}},
  \bibinfo{author}{\bibfnamefont{J.~P.} \bibnamefont{Long}},
  \bibinfo{author}{\bibfnamefont{J.~C.} \bibnamefont{Culbertson}},
  \bibinfo{author}{\bibfnamefont{T.}~\bibnamefont{Ohta}},
  \bibinfo{author}{\bibfnamefont{A.~L.} \bibnamefont{Friedman}},
  \bibnamefont{and} \bibinfo{author}{\bibfnamefont{T.~E.}
  \bibnamefont{Beechem}}, {``}\bibinfo{title}{Electronic Hybridization of
  Large-Area Stacked Graphene Films},{''} \bibinfo{journal}{ACS Nano}
  \textbf{\bibinfo{volume}{7}}, \bibinfo{pages}{637} (\bibinfo{year}{2013}).

\bibitem[{\citenamefont{Liu et~al.}(2015)\citenamefont{Liu, Li, Chen, Wang, Qi,
  He, Zheng, Zhou, Zhang, Gu et~al.}}]{Liu2015}
\bibinfo{author}{\bibfnamefont{J.-B.} \bibnamefont{Liu}},
  \bibinfo{author}{\bibfnamefont{P.-J.} \bibnamefont{Li}},
  \bibinfo{author}{\bibfnamefont{Y.-F.} \bibnamefont{Chen}},
  \bibinfo{author}{\bibfnamefont{Z.-G.} \bibnamefont{Wang}},
  \bibinfo{author}{\bibfnamefont{F.}~\bibnamefont{Qi}},
  \bibinfo{author}{\bibfnamefont{J.-R.} \bibnamefont{He}},
  \bibinfo{author}{\bibfnamefont{B.-J.} \bibnamefont{Zheng}},
  \bibinfo{author}{\bibfnamefont{J.-H.} \bibnamefont{Zhou}},
  \bibinfo{author}{\bibfnamefont{W.-L.} \bibnamefont{Zhang}},
  \bibinfo{author}{\bibfnamefont{L.}~\bibnamefont{Gu}}, \bibnamefont{et~al.},
  {``}\bibinfo{title}{Observation of tunable electrical bandgap in large-area
  twisted bilayer graphene synthesized by chemical vapor deposition},{''}
  \bibinfo{journal}{Sci. Rep.} \textbf{\bibinfo{volume}{5}},
  \bibinfo{pages}{15285} (\bibinfo{year}{2015}).

\bibitem[{\citenamefont{Poncharal et~al.}(2008)\citenamefont{Poncharal, Ayari,
  Michel, and Sauvajol}}]{PoncharalRaman}
\bibinfo{author}{\bibfnamefont{P.}~\bibnamefont{Poncharal}},
  \bibinfo{author}{\bibfnamefont{A.}~\bibnamefont{Ayari}},
  \bibinfo{author}{\bibfnamefont{T.}~\bibnamefont{Michel}}, \bibnamefont{and}
  \bibinfo{author}{\bibfnamefont{J.-L.} \bibnamefont{Sauvajol}},
  {``}\bibinfo{title}{Raman spectra of misoriented bilayer graphene},{''}
  \bibinfo{journal}{Phys. Rev. B} \textbf{\bibinfo{volume}{78}},
  \bibinfo{pages}{113407} (\bibinfo{year}{2008}).

\bibitem[{\citenamefont{Ni et~al.}(2008)\citenamefont{Ni, Wang, Yu, You, and
  Shen}}]{NiRamanVrenorm}
\bibinfo{author}{\bibfnamefont{Z.}~\bibnamefont{Ni}},
  \bibinfo{author}{\bibfnamefont{Y.}~\bibnamefont{Wang}},
  \bibinfo{author}{\bibfnamefont{T.}~\bibnamefont{Yu}},
  \bibinfo{author}{\bibfnamefont{Y.}~\bibnamefont{You}}, \bibnamefont{and}
  \bibinfo{author}{\bibfnamefont{Z.}~\bibnamefont{Shen}},
  {``}\bibinfo{title}{Reduction of Fermi velocity in folded graphene observed
  by resonance Raman spectroscopy},{''} \bibinfo{journal}{Phys. Rev. B}
  \textbf{\bibinfo{volume}{77}}, \bibinfo{pages}{235403}
  (\bibinfo{year}{2008}).

\bibitem[{\citenamefont{Carozo et~al.}(2013)\citenamefont{Carozo, Almeida,
  Fragneaud, Bed\^e, Moutinho, Ribeiro-Soares, Andrade, Souza~Filho, Matos,
  Wang et~al.}}]{CarozoRaman}
\bibinfo{author}{\bibfnamefont{V.}~\bibnamefont{Carozo}},
  \bibinfo{author}{\bibfnamefont{C.~M.} \bibnamefont{Almeida}},
  \bibinfo{author}{\bibfnamefont{B.}~\bibnamefont{Fragneaud}},
  \bibinfo{author}{\bibfnamefont{P.~M.} \bibnamefont{Bed\^e}},
  \bibinfo{author}{\bibfnamefont{M.~V.~O.} \bibnamefont{Moutinho}},
  \bibinfo{author}{\bibfnamefont{J.}~\bibnamefont{Ribeiro-Soares}},
  \bibinfo{author}{\bibfnamefont{N.~F.} \bibnamefont{Andrade}},
  \bibinfo{author}{\bibfnamefont{A.~G.} \bibnamefont{Souza~Filho}},
  \bibinfo{author}{\bibfnamefont{M.~J.~S.} \bibnamefont{Matos}},
  \bibinfo{author}{\bibfnamefont{B.}~\bibnamefont{Wang}}, \bibnamefont{et~al.},
  {``}\bibinfo{title}{Resonance effects on the Raman spectra of graphene
  superlattices},{''} \bibinfo{journal}{Phys. Rev. B}
  \textbf{\bibinfo{volume}{88}}, \bibinfo{pages}{085401}
  (\bibinfo{year}{2013}).

\bibitem[{\citenamefont{Ohta et~al.}(2012{\natexlab{a}})\citenamefont{Ohta,
  Beechem, Robinson, and Kellogg}}]{SynthesysMoire}
\bibinfo{author}{\bibfnamefont{T.}~\bibnamefont{Ohta}},
  \bibinfo{author}{\bibfnamefont{T.~E.} \bibnamefont{Beechem}},
  \bibinfo{author}{\bibfnamefont{J.~T.} \bibnamefont{Robinson}},
  \bibnamefont{and} \bibinfo{author}{\bibfnamefont{G.~L.}
  \bibnamefont{Kellogg}}, {``}\bibinfo{title}{Long-range atomic ordering and
  variable interlayer interactions in two overlapping graphene lattices with
  stacking misorientations},{''} \bibinfo{journal}{Phys. Rev. B}
  \textbf{\bibinfo{volume}{85}}, \bibinfo{pages}{075415}
  (\bibinfo{year}{2012}{\natexlab{a}}).

\bibitem[{\citenamefont{Othmen et~al.}(2015)\citenamefont{Othmen, Arezki,
  Ajlani, Cavanna, Boutchich, Oueslati, and Madouri}}]{Othmen2015}
\bibinfo{author}{\bibfnamefont{R.}~\bibnamefont{Othmen}},
  \bibinfo{author}{\bibfnamefont{H.}~\bibnamefont{Arezki}},
  \bibinfo{author}{\bibfnamefont{H.}~\bibnamefont{Ajlani}},
  \bibinfo{author}{\bibfnamefont{A.}~\bibnamefont{Cavanna}},
  \bibinfo{author}{\bibfnamefont{M.}~\bibnamefont{Boutchich}},
  \bibinfo{author}{\bibfnamefont{M.}~\bibnamefont{Oueslati}}, \bibnamefont{and}
  \bibinfo{author}{\bibfnamefont{A.}~\bibnamefont{Madouri}},
  {``}\bibinfo{title}{Direct transfer and Raman characterization of twisted
  graphene bilayer},{''} \bibinfo{journal}{Appl. Phys. Lett.}
  \textbf{\bibinfo{volume}{106}}, \bibinfo{eid}{103107} (\bibinfo{year}{2015}).

\bibitem[{\citenamefont{Rong and Kuiper}(1993)}]{MoirePattern}
\bibinfo{author}{\bibfnamefont{Z.~Y.} \bibnamefont{Rong}} \bibnamefont{and}
  \bibinfo{author}{\bibfnamefont{P.}~\bibnamefont{Kuiper}},
  {``}\bibinfo{title}{Electronic effects in scanning tunneling microscopy:
  Moir\'e pattern on a graphite surface},{''} \bibinfo{journal}{Phys. Rev. B}
  \textbf{\bibinfo{volume}{48}}, \bibinfo{pages}{17427} (\bibinfo{year}{1993}).

\bibitem[{\citenamefont{Li et~al.}(2010)\citenamefont{Li, Luican, Lopes~dos
  Santos, Castro~Neto, Reina, Kong, and Andrei}}]{STM_VHS1}
\bibinfo{author}{\bibfnamefont{G.}~\bibnamefont{Li}},
  \bibinfo{author}{\bibfnamefont{A.}~\bibnamefont{Luican}},
  \bibinfo{author}{\bibfnamefont{J.~M.~B.} \bibnamefont{Lopes~dos Santos}},
  \bibinfo{author}{\bibfnamefont{A.~H.} \bibnamefont{Castro~Neto}},
  \bibinfo{author}{\bibfnamefont{A.}~\bibnamefont{Reina}},
  \bibinfo{author}{\bibfnamefont{J.}~\bibnamefont{Kong}}, \bibnamefont{and}
  \bibinfo{author}{\bibfnamefont{E.~Y.} \bibnamefont{Andrei}},
  {``}\bibinfo{title}{Observation of Van Hove singularities in twisted graphene
  layers},{''} \bibinfo{journal}{Nat. Phys.} \textbf{\bibinfo{volume}{6}},
  \bibinfo{pages}{109} (\bibinfo{year}{2010}).

\bibitem[{\citenamefont{Cisternas and Correa}(2012)}]{CisternasSTM}
\bibinfo{author}{\bibfnamefont{E.}~\bibnamefont{Cisternas}} \bibnamefont{and}
  \bibinfo{author}{\bibfnamefont{J.}~\bibnamefont{Correa}},
  {``}\bibinfo{title}{Theoretical reproduction of superstructures revealed by
  STM on bilayer graphene},{''} \bibinfo{journal}{Chem. Phys.}
  \textbf{\bibinfo{volume}{409}}, \bibinfo{pages}{74 } (\bibinfo{year}{2012}).

\bibitem[{\citenamefont{Lopes~dos Santos et~al.}(2012)\citenamefont{Lopes~dos
  Santos, Peres, and Castro~Neto}}]{dSPRB}
\bibinfo{author}{\bibfnamefont{J.~M.~B.} \bibnamefont{Lopes~dos Santos}},
  \bibinfo{author}{\bibfnamefont{N.~M.~R.} \bibnamefont{Peres}},
  \bibnamefont{and} \bibinfo{author}{\bibfnamefont{A.~H.}
  \bibnamefont{Castro~Neto}}, {``}\bibinfo{title}{Continuum model of the
  twisted graphene bilayer},{''} \bibinfo{journal}{Phys. Rev. B}
  \textbf{\bibinfo{volume}{86}}, \bibinfo{pages}{155449}
  (\bibinfo{year}{2012}).

\bibitem[{\citenamefont{Shallcross et~al.}(2010)\citenamefont{Shallcross,
  Sharma, Kandelaki, and Pankratov}}]{Pankratov1}
\bibinfo{author}{\bibfnamefont{S.}~\bibnamefont{Shallcross}},
  \bibinfo{author}{\bibfnamefont{S.}~\bibnamefont{Sharma}},
  \bibinfo{author}{\bibfnamefont{E.}~\bibnamefont{Kandelaki}},
  \bibnamefont{and} \bibinfo{author}{\bibfnamefont{O.~A.}
  \bibnamefont{Pankratov}}, {``}\bibinfo{title}{Electronic structure of
  turbostratic graphene},{''} \bibinfo{journal}{Phys. Rev. B}
  \textbf{\bibinfo{volume}{81}}, \bibinfo{pages}{165105}
  (\bibinfo{year}{2010}).

\bibitem[{\citenamefont{Shallcross
  et~al.}(2008{\natexlab{a}})\citenamefont{Shallcross, Sharma, and
  Pankratov}}]{PankratovPRL}
\bibinfo{author}{\bibfnamefont{S.}~\bibnamefont{Shallcross}},
  \bibinfo{author}{\bibfnamefont{S.}~\bibnamefont{Sharma}}, \bibnamefont{and}
  \bibinfo{author}{\bibfnamefont{O.~A.} \bibnamefont{Pankratov}},
  {``}\bibinfo{title}{Quantum Interference at the Twist Boundary in
  Graphene},{''} \bibinfo{journal}{Phys. Rev. Lett.}
  \textbf{\bibinfo{volume}{101}}, \bibinfo{pages}{056803}
  (\bibinfo{year}{2008}{\natexlab{a}}).

\bibitem[{\citenamefont{Mele}(2010)}]{MelePRB1}
\bibinfo{author}{\bibfnamefont{E.~J.} \bibnamefont{Mele}},
  {``}\bibinfo{title}{Commensuration and interlayer coherence in twisted
  bilayer graphene},{''} \bibinfo{journal}{Phys. Rev. B}
  \textbf{\bibinfo{volume}{81}}, \bibinfo{pages}{161405}
  (\bibinfo{year}{2010}).

\bibitem[{\citenamefont{Mele}(2011)}]{MelePRB2}
\bibinfo{author}{\bibfnamefont{E.~J.} \bibnamefont{Mele}},
  {``}\bibinfo{title}{Band symmetries and singularities in twisted multilayer
  graphene},{''} \bibinfo{journal}{Phys. Rev. B} \textbf{\bibinfo{volume}{84}},
  \bibinfo{pages}{235439} (\bibinfo{year}{2011}).

\bibitem[{\citenamefont{Bistritzer and MacDonald}(2011{\natexlab{a}})}]{PNAS}
\bibinfo{author}{\bibfnamefont{R.}~\bibnamefont{Bistritzer}} \bibnamefont{and}
  \bibinfo{author}{\bibfnamefont{A.~H.} \bibnamefont{MacDonald}},
  {``}\bibinfo{title}{Moir\'{e} bands in twisted double-layer graphene},{''}
  \bibinfo{journal}{PNAS} \textbf{\bibinfo{volume}{108}},
  \bibinfo{pages}{12233} (\bibinfo{year}{2011}{\natexlab{a}}).

\bibitem[{\citenamefont{Shallcross
  et~al.}(2008{\natexlab{b}})\citenamefont{Shallcross, Sharma, and
  Pankratov}}]{PankratovDFT}
\bibinfo{author}{\bibfnamefont{S.}~\bibnamefont{Shallcross}},
  \bibinfo{author}{\bibfnamefont{S.}~\bibnamefont{Sharma}}, \bibnamefont{and}
  \bibinfo{author}{\bibfnamefont{O.~A.} \bibnamefont{Pankratov}},
  {``}\bibinfo{title}{Twist boundary in graphene: energetics and electric field
  effect},{''} \bibinfo{journal}{J. Phys.: Condens. Matter}
  \textbf{\bibinfo{volume}{20}}, \bibinfo{pages}{454224}
  (\bibinfo{year}{2008}{\natexlab{b}}).

\bibitem[{\citenamefont{Trambly~de Laissardi\`{e}re
  et~al.}(2010)\citenamefont{Trambly~de Laissardi\`{e}re, Mayou, and
  Magaud}}]{NanoLettTB}
\bibinfo{author}{\bibfnamefont{G.}~\bibnamefont{Trambly~de Laissardi\`{e}re}},
  \bibinfo{author}{\bibfnamefont{D.}~\bibnamefont{Mayou}}, \bibnamefont{and}
  \bibinfo{author}{\bibfnamefont{L.}~\bibnamefont{Magaud}},
  {``}\bibinfo{title}{Localization of Dirac Electrons in Rotated Graphene
  Bilayers},{''} \bibinfo{journal}{Nano Lett.} \textbf{\bibinfo{volume}{10}},
  \bibinfo{pages}{804} (\bibinfo{year}{2010}).

\bibitem[{\citenamefont{Trambly~de Laissardi\`ere
  et~al.}(2012)\citenamefont{Trambly~de Laissardi\`ere, Mayou, and
  Magaud}}]{TramblyTB_Loc}
\bibinfo{author}{\bibfnamefont{G.}~\bibnamefont{Trambly~de Laissardi\`ere}},
  \bibinfo{author}{\bibfnamefont{D.}~\bibnamefont{Mayou}}, \bibnamefont{and}
  \bibinfo{author}{\bibfnamefont{L.}~\bibnamefont{Magaud}},
  {``}\bibinfo{title}{Numerical studies of confined states in rotated bilayers
  of graphene},{''} \bibinfo{journal}{Phys. Rev. B}
  \textbf{\bibinfo{volume}{86}}, \bibinfo{pages}{125413}
  (\bibinfo{year}{2012}).

\bibitem[{\citenamefont{Su\'arez~Morell
  et~al.}(2010)\citenamefont{Su\'arez~Morell, Correa, Vargas, Pacheco, and
  Barticevic}}]{Morell1}
\bibinfo{author}{\bibfnamefont{E.}~\bibnamefont{Su\'arez~Morell}},
  \bibinfo{author}{\bibfnamefont{J.~D.} \bibnamefont{Correa}},
  \bibinfo{author}{\bibfnamefont{P.}~\bibnamefont{Vargas}},
  \bibinfo{author}{\bibfnamefont{M.}~\bibnamefont{Pacheco}}, \bibnamefont{and}
  \bibinfo{author}{\bibfnamefont{Z.}~\bibnamefont{Barticevic}},
  {``}\bibinfo{title}{Flat bands in slightly twisted bilayer graphene:
  Tight-binding calculations},{''} \bibinfo{journal}{Phys. Rev. B}
  \textbf{\bibinfo{volume}{82}}, \bibinfo{pages}{121407}
  (\bibinfo{year}{2010}).

\bibitem[{\citenamefont{Su\'arez~Morell
  et~al.}(2011)\citenamefont{Su\'arez~Morell, Vargas, Chico, and
  Brey}}]{Morell2}
\bibinfo{author}{\bibfnamefont{E.}~\bibnamefont{Su\'arez~Morell}},
  \bibinfo{author}{\bibfnamefont{P.}~\bibnamefont{Vargas}},
  \bibinfo{author}{\bibfnamefont{L.}~\bibnamefont{Chico}}, \bibnamefont{and}
  \bibinfo{author}{\bibfnamefont{L.}~\bibnamefont{Brey}},
  {``}\bibinfo{title}{Charge redistribution and interlayer coupling in twisted
  bilayer graphene under electric fields},{''} \bibinfo{journal}{Phys. Rev. B}
  \textbf{\bibinfo{volume}{84}}, \bibinfo{pages}{195421}
  (\bibinfo{year}{2011}).

\bibitem[{\citenamefont{Shallcross et~al.}(2013)\citenamefont{Shallcross,
  Sharma, and Pankratov}}]{PankratovPRB2013}
\bibinfo{author}{\bibfnamefont{S.}~\bibnamefont{Shallcross}},
  \bibinfo{author}{\bibfnamefont{S.}~\bibnamefont{Sharma}}, \bibnamefont{and}
  \bibinfo{author}{\bibfnamefont{O.}~\bibnamefont{Pankratov}},
  {``}\bibinfo{title}{Emergent momentum scale, localization, and van Hove
  singularities in the graphene twist bilayer},{''} \bibinfo{journal}{Phys.
  Rev. B} \textbf{\bibinfo{volume}{87}}, \bibinfo{pages}{245403}
  (\bibinfo{year}{2013}).

\bibitem[{\citenamefont{Latil et~al.}(2007)\citenamefont{Latil, Meunier, and
  Henrard}}]{LatilDFT}
\bibinfo{author}{\bibfnamefont{S.}~\bibnamefont{Latil}},
  \bibinfo{author}{\bibfnamefont{V.}~\bibnamefont{Meunier}}, \bibnamefont{and}
  \bibinfo{author}{\bibfnamefont{L.}~\bibnamefont{Henrard}},
  {``}\bibinfo{title}{Massless fermions in multilayer graphitic systems with
  misoriented layers: \textit{Ab initio} calculations and experimental
  fingerprints},{''} \bibinfo{journal}{Phys. Rev. B}
  \textbf{\bibinfo{volume}{76}}, \bibinfo{pages}{201402}
  (\bibinfo{year}{2007}).

\bibitem[{\citenamefont{Xian et~al.}(2011)\citenamefont{Xian, Barraza-Lopez,
  and Chou}}]{DeltaUeffDFT}
\bibinfo{author}{\bibfnamefont{L.}~\bibnamefont{Xian}},
  \bibinfo{author}{\bibfnamefont{S.}~\bibnamefont{Barraza-Lopez}},
  \bibnamefont{and} \bibinfo{author}{\bibfnamefont{M.~Y.} \bibnamefont{Chou}},
  {``}\bibinfo{title}{Effects of electrostatic fields and charge doping on the
  linear bands in twisted graphene bilayers},{''} \bibinfo{journal}{Phys. Rev.
  B} \textbf{\bibinfo{volume}{84}}, \bibinfo{pages}{075425}
  (\bibinfo{year}{2011}).

\bibitem[{\citenamefont{Sato et~al.}(2012)\citenamefont{Sato, Saito, Cong, Yu,
  and Dresselhaus}}]{RamanTheory}
\bibinfo{author}{\bibfnamefont{K.}~\bibnamefont{Sato}},
  \bibinfo{author}{\bibfnamefont{R.}~\bibnamefont{Saito}},
  \bibinfo{author}{\bibfnamefont{C.}~\bibnamefont{Cong}},
  \bibinfo{author}{\bibfnamefont{T.}~\bibnamefont{Yu}}, \bibnamefont{and}
  \bibinfo{author}{\bibfnamefont{M.~S.} \bibnamefont{Dresselhaus}},
  {``}\bibinfo{title}{Zone folding effect in Raman $G$-band intensity of
  twisted bilayer graphene},{''} \bibinfo{journal}{Phys. Rev. B}
  \textbf{\bibinfo{volume}{86}}, \bibinfo{pages}{125414}
  (\bibinfo{year}{2012}).

\bibitem[{\citenamefont{Uchida et~al.}(2014)\citenamefont{Uchida, Furuya,
  Iwata, and Oshiyama}}]{LargeScaleDFT_PRB2014}
\bibinfo{author}{\bibfnamefont{K.}~\bibnamefont{Uchida}},
  \bibinfo{author}{\bibfnamefont{S.}~\bibnamefont{Furuya}},
  \bibinfo{author}{\bibfnamefont{J.-I.} \bibnamefont{Iwata}}, \bibnamefont{and}
  \bibinfo{author}{\bibfnamefont{A.}~\bibnamefont{Oshiyama}},
  {``}\bibinfo{title}{Atomic corrugation and electron localization due to
  Moir\'e patterns in twisted bilayer graphenes},{''} \bibinfo{journal}{Phys.
  Rev. B} \textbf{\bibinfo{volume}{90}}, \bibinfo{pages}{155451}
  (\bibinfo{year}{2014}).

\bibitem[{\citenamefont{Oshiyama et~al.}(2015)\citenamefont{Oshiyama, Iwata,
  Uchida, and Matsushita}}]{LargeScaleDFT2015}
\bibinfo{author}{\bibfnamefont{A.}~\bibnamefont{Oshiyama}},
  \bibinfo{author}{\bibfnamefont{J.-I.} \bibnamefont{Iwata}},
  \bibinfo{author}{\bibfnamefont{K.}~\bibnamefont{Uchida}}, \bibnamefont{and}
  \bibinfo{author}{\bibfnamefont{Y.-I.} \bibnamefont{Matsushita}},
  {``}\bibinfo{title}{Large-scale real-space density-functional calculations:
  Moir\'e-induced electron localization in graphene},{''} \bibinfo{journal}{J.
  Appl. Phys.} \textbf{\bibinfo{volume}{117}}, \bibinfo{eid}{112811}
  (\bibinfo{year}{2015}).

\bibitem[{\citenamefont{Kindermann and First}(2012)}]{alaMele}
\bibinfo{author}{\bibfnamefont{M.}~\bibnamefont{Kindermann}} \bibnamefont{and}
  \bibinfo{author}{\bibfnamefont{P.~N.} \bibnamefont{First}},
  {``}\bibinfo{title}{Effective theory of rotationally faulted multilayer
  graphene - the local limit},{''} \bibinfo{journal}{J. Phys. D: Appl. Phys.}
  \textbf{\bibinfo{volume}{45}}, \bibinfo{pages}{154005}
  (\bibinfo{year}{2012}).

\bibitem[{\citenamefont{de~Gail et~al.}(2011)\citenamefont{de~Gail, Goerbig,
  Guinea, Montambaux, and Castro~Neto}}]{deGail}
\bibinfo{author}{\bibfnamefont{R.}~\bibnamefont{de~Gail}},
  \bibinfo{author}{\bibfnamefont{M.~O.} \bibnamefont{Goerbig}},
  \bibinfo{author}{\bibfnamefont{F.}~\bibnamefont{Guinea}},
  \bibinfo{author}{\bibfnamefont{G.}~\bibnamefont{Montambaux}},
  \bibnamefont{and} \bibinfo{author}{\bibfnamefont{A.~H.}
  \bibnamefont{Castro~Neto}}, {``}\bibinfo{title}{Topologically protected zero
  modes in twisted bilayer graphene},{''} \bibinfo{journal}{Phys. Rev. B}
  \textbf{\bibinfo{volume}{84}}, \bibinfo{pages}{045436}
  (\bibinfo{year}{2011}).

\bibitem[{\citenamefont{Choi et~al.}(2011)\citenamefont{Choi, Hyun, and
  Kim}}]{LL_lowEnergy}
\bibinfo{author}{\bibfnamefont{M.-Y.} \bibnamefont{Choi}},
  \bibinfo{author}{\bibfnamefont{Y.-H.} \bibnamefont{Hyun}}, \bibnamefont{and}
  \bibinfo{author}{\bibfnamefont{Y.}~\bibnamefont{Kim}},
  {``}\bibinfo{title}{Angle dependence of the Landau level spectrum in twisted
  bilayer graphene},{''} \bibinfo{journal}{Phys. Rev. B}
  \textbf{\bibinfo{volume}{84}}, \bibinfo{pages}{195437}
  (\bibinfo{year}{2011}).

\bibitem[{\citenamefont{San-Jose et~al.}(2012)\citenamefont{San-Jose,
  Gonz\'alez, and Guinea}}]{NonAbelianGaugePot}
\bibinfo{author}{\bibfnamefont{P.}~\bibnamefont{San-Jose}},
  \bibinfo{author}{\bibfnamefont{J.}~\bibnamefont{Gonz\'alez}},
  \bibnamefont{and} \bibinfo{author}{\bibfnamefont{F.}~\bibnamefont{Guinea}},
  {``}\bibinfo{title}{Non-Abelian Gauge Potentials in Graphene Bilayers},{''}
  \bibinfo{journal}{Phys. Rev. Lett.} \textbf{\bibinfo{volume}{108}},
  \bibinfo{pages}{216802} (\bibinfo{year}{2012}).

\bibitem[{\citenamefont{Chu et~al.}(2013)\citenamefont{Chu, He, and
  He}}]{ChuEffTheor}
\bibinfo{author}{\bibfnamefont{Z.-D.} \bibnamefont{Chu}},
  \bibinfo{author}{\bibfnamefont{W.-Y.} \bibnamefont{He}}, \bibnamefont{and}
  \bibinfo{author}{\bibfnamefont{L.}~\bibnamefont{He}},
  {``}\bibinfo{title}{Coexistence of van Hove singularities and superlattice
  Dirac points in a slightly twisted graphene bilayer},{''}
  \bibinfo{journal}{Phys. Rev. B} \textbf{\bibinfo{volume}{87}},
  \bibinfo{pages}{155419} (\bibinfo{year}{2013}).

\bibitem[{\citenamefont{Xian et~al.}(2013)\citenamefont{Xian, Wang, and
  Chou}}]{TBLGneutrino}
\bibinfo{author}{\bibfnamefont{L.}~\bibnamefont{Xian}},
  \bibinfo{author}{\bibfnamefont{Z.~F.} \bibnamefont{Wang}}, \bibnamefont{and}
  \bibinfo{author}{\bibfnamefont{M.~Y.} \bibnamefont{Chou}},
  {``}\bibinfo{title}{Coupled Dirac Fermions and Neutrino-like Oscillations in
  Twisted Bilayer Graphene},{''} \bibinfo{journal}{Nano Lett.}
  \textbf{\bibinfo{volume}{13}}, \bibinfo{pages}{5159} (\bibinfo{year}{2013}).

\bibitem[{\citenamefont{Weckbecker et~al.}(2016)\citenamefont{Weckbecker,
  Shallcross, Fleischmann, Ray, Sharma, and Pankratov}}]{low_en_pankratov2016}
\bibinfo{author}{\bibfnamefont{D.}~\bibnamefont{Weckbecker}},
  \bibinfo{author}{\bibfnamefont{S.}~\bibnamefont{Shallcross}},
  \bibinfo{author}{\bibfnamefont{M.}~\bibnamefont{Fleischmann}},
  \bibinfo{author}{\bibfnamefont{N.}~\bibnamefont{Ray}},
  \bibinfo{author}{\bibfnamefont{S.}~\bibnamefont{Sharma}}, \bibnamefont{and}
  \bibinfo{author}{\bibfnamefont{O.}~\bibnamefont{Pankratov}},
  {``}\bibinfo{title}{Low-energy theory for the graphene twist bilayer},{''}
  \bibinfo{journal}{Phys. Rev. B} \textbf{\bibinfo{volume}{93}},
  \bibinfo{pages}{035452} (\bibinfo{year}{2016}).

\bibitem[{\citenamefont{Tang et~al.}(1996)\citenamefont{Tang, Wang, Chan, and
  Ho}}]{Tang}
\bibinfo{author}{\bibfnamefont{M.~S.} \bibnamefont{Tang}},
  \bibinfo{author}{\bibfnamefont{C.~Z.} \bibnamefont{Wang}},
  \bibinfo{author}{\bibfnamefont{C.~T.} \bibnamefont{Chan}}, \bibnamefont{and}
  \bibinfo{author}{\bibfnamefont{K.~M.} \bibnamefont{Ho}},
  {``}\bibinfo{title}{Environment-dependent tight-binding potential model},{''}
  \bibinfo{journal}{Phys. Rev. B} \textbf{\bibinfo{volume}{53}},
  \bibinfo{pages}{979} (\bibinfo{year}{1996}).

\bibitem[{\citenamefont{He et~al.}(2014)\citenamefont{He, Su, Yang, and
  He}}]{PseudoMagneticField2014}
\bibinfo{author}{\bibfnamefont{W.-Y.} \bibnamefont{He}},
  \bibinfo{author}{\bibfnamefont{Y.}~\bibnamefont{Su}},
  \bibinfo{author}{\bibfnamefont{M.}~\bibnamefont{Yang}}, \bibnamefont{and}
  \bibinfo{author}{\bibfnamefont{L.}~\bibnamefont{He}},
  {``}\bibinfo{title}{Creating in-plane pseudomagnetic fields in excess of 1000
  T by misoriented stacking in a graphene bilayer},{''} \bibinfo{journal}{Phys.
  Rev. B} \textbf{\bibinfo{volume}{89}}, \bibinfo{pages}{125418}
  (\bibinfo{year}{2014}).

\bibitem[{\citenamefont{Mendez et~al.}(1980)\citenamefont{Mendez, Misu, and
  Dresselhaus}}]{AB::Mendez}
\bibinfo{author}{\bibfnamefont{E.}~\bibnamefont{Mendez}},
  \bibinfo{author}{\bibfnamefont{A.}~\bibnamefont{Misu}}, \bibnamefont{and}
  \bibinfo{author}{\bibfnamefont{M.~S.} \bibnamefont{Dresselhaus}},
  {``}\bibinfo{title}{Magnetoreflection study of graphite under pressure},{''}
  \bibinfo{journal}{Phys. Rev. B} \textbf{\bibinfo{volume}{21}},
  \bibinfo{pages}{827} (\bibinfo{year}{1980}).

\bibitem[{\citenamefont{Mucha-Kruczy{\ifmmode \acute{n}\else \'{n}\fi{}}ski
  et~al.}(2008)\citenamefont{Mucha-Kruczy{\ifmmode \acute{n}\else
  \'{n}\fi{}}ski, Tsyplyatyev, Grishin, McCann, Fal'ko, Bostwick, and
  Rotenberg}}]{AB::Mucha}
\bibinfo{author}{\bibfnamefont{M.}~\bibnamefont{Mucha-Kruczy{\ifmmode
  \acute{n}\else \'{n}\fi{}}ski}},
  \bibinfo{author}{\bibfnamefont{O.}~\bibnamefont{Tsyplyatyev}},
  \bibinfo{author}{\bibfnamefont{A.}~\bibnamefont{Grishin}},
  \bibinfo{author}{\bibfnamefont{E.}~\bibnamefont{McCann}},
  \bibinfo{author}{\bibfnamefont{V.~I.} \bibnamefont{Fal'ko}},
  \bibinfo{author}{\bibfnamefont{A.}~\bibnamefont{Bostwick}}, \bibnamefont{and}
  \bibinfo{author}{\bibfnamefont{E.}~\bibnamefont{Rotenberg}},
  {``}\bibinfo{title}{Characterization of graphene through anisotropy of
  constant-energy maps in angle-resolved photoemission},{''}
  \bibinfo{journal}{Phys. Rev. B} \textbf{\bibinfo{volume}{77}},
  \bibinfo{pages}{195403} (\bibinfo{year}{2008}).

\bibitem[{\citenamefont{Miller et~al.}(2009)\citenamefont{Miller, Kubista,
  Rutter, Ruan, de~Heer, First, and Stroscio}}]{MillerLLexp}
\bibinfo{author}{\bibfnamefont{D.~L.} \bibnamefont{Miller}},
  \bibinfo{author}{\bibfnamefont{K.~D.} \bibnamefont{Kubista}},
  \bibinfo{author}{\bibfnamefont{G.~M.} \bibnamefont{Rutter}},
  \bibinfo{author}{\bibfnamefont{M.}~\bibnamefont{Ruan}},
  \bibinfo{author}{\bibfnamefont{W.~A.} \bibnamefont{de~Heer}},
  \bibinfo{author}{\bibfnamefont{P.~N.} \bibnamefont{First}}, \bibnamefont{and}
  \bibinfo{author}{\bibfnamefont{J.~A.} \bibnamefont{Stroscio}},
  {``}\bibinfo{title}{Observing the Quantization of Zero Mass Carriers in
  Graphene},{''} \bibinfo{journal}{Science} \textbf{\bibinfo{volume}{324}},
  \bibinfo{pages}{924} (\bibinfo{year}{2009}).

\bibitem[{\citenamefont{Sprinkle et~al.}(2009)\citenamefont{Sprinkle, Siegel,
  Hu, Hicks, Tejeda, Taleb-Ibrahimi, Le~F\`evre, Bertran, Vizzini, Enriquez
  et~al.}}]{SprinkleARPES}
\bibinfo{author}{\bibfnamefont{M.}~\bibnamefont{Sprinkle}},
  \bibinfo{author}{\bibfnamefont{D.}~\bibnamefont{Siegel}},
  \bibinfo{author}{\bibfnamefont{Y.}~\bibnamefont{Hu}},
  \bibinfo{author}{\bibfnamefont{J.}~\bibnamefont{Hicks}},
  \bibinfo{author}{\bibfnamefont{A.}~\bibnamefont{Tejeda}},
  \bibinfo{author}{\bibfnamefont{A.}~\bibnamefont{Taleb-Ibrahimi}},
  \bibinfo{author}{\bibfnamefont{P.}~\bibnamefont{Le~F\`evre}},
  \bibinfo{author}{\bibfnamefont{F.}~\bibnamefont{Bertran}},
  \bibinfo{author}{\bibfnamefont{S.}~\bibnamefont{Vizzini}},
  \bibinfo{author}{\bibfnamefont{H.}~\bibnamefont{Enriquez}},
  \bibnamefont{et~al.}, {``}\bibinfo{title}{First Direct Observation of a
  Nearly Ideal Graphene Band Structure},{''} \bibinfo{journal}{Phys. Rev.
  Lett.} \textbf{\bibinfo{volume}{103}}, \bibinfo{pages}{226803}
  (\bibinfo{year}{2009}).

\bibitem[{\citenamefont{Razado-Colambo
  et~al.}(2016)\citenamefont{Razado-Colambo, Avila, Nys, Chen, Wallart,
  Asensio, and Vignaud}}]{razado2016nanoarpes_tblg}
\bibinfo{author}{\bibfnamefont{I.}~\bibnamefont{Razado-Colambo}},
  \bibinfo{author}{\bibfnamefont{J.}~\bibnamefont{Avila}},
  \bibinfo{author}{\bibfnamefont{J.-P.} \bibnamefont{Nys}},
  \bibinfo{author}{\bibfnamefont{C.}~\bibnamefont{Chen}},
  \bibinfo{author}{\bibfnamefont{X.}~\bibnamefont{Wallart}},
  \bibinfo{author}{\bibfnamefont{M.-C.} \bibnamefont{Asensio}},
  \bibnamefont{and} \bibinfo{author}{\bibfnamefont{D.}~\bibnamefont{Vignaud}},
  {``}\bibinfo{title}{NanoARPES of twisted bilayer graphene on SiC: absence of
  velocity renormalization for small angles},{''} \bibinfo{journal}{Sci. Rep.}
  \textbf{\bibinfo{volume}{6}}, \bibinfo{pages}{27261} (\bibinfo{year}{2016}).

\bibitem[{\citenamefont{Song et~al.}(2010)\citenamefont{Song, Otte, Kuk, Hu,
  Torrance, First, de~Heer, Min, Adam, Stiles et~al.}}]{NatureLL}
\bibinfo{author}{\bibfnamefont{Y.~J.} \bibnamefont{Song}},
  \bibinfo{author}{\bibfnamefont{A.~F.} \bibnamefont{Otte}},
  \bibinfo{author}{\bibfnamefont{Y.}~\bibnamefont{Kuk}},
  \bibinfo{author}{\bibfnamefont{Y.}~\bibnamefont{Hu}},
  \bibinfo{author}{\bibfnamefont{D.~B.} \bibnamefont{Torrance}},
  \bibinfo{author}{\bibfnamefont{P.~N.} \bibnamefont{First}},
  \bibinfo{author}{\bibfnamefont{W.~A.} \bibnamefont{de~Heer}},
  \bibinfo{author}{\bibfnamefont{H.}~\bibnamefont{Min}},
  \bibinfo{author}{\bibfnamefont{S.}~\bibnamefont{Adam}},
  \bibinfo{author}{\bibfnamefont{M.~D.} \bibnamefont{Stiles}},
  \bibnamefont{et~al.}, {``}\bibinfo{title}{High-resolution tunnelling
  spectroscopy of a graphene quartet},{''} \bibinfo{journal}{Nature}
  \textbf{\bibinfo{volume}{467}}, \bibinfo{pages}{185} (\bibinfo{year}{2010}).

\bibitem[{\citenamefont{Sanchez-Yamagishi
  et~al.}(2012)\citenamefont{Sanchez-Yamagishi, Taychatanapat, Watanabe,
  Taniguchi, Yacoby, and Jarillo-Herrero}}]{tBLG_QHE}
\bibinfo{author}{\bibfnamefont{J.~D.} \bibnamefont{Sanchez-Yamagishi}},
  \bibinfo{author}{\bibfnamefont{T.}~\bibnamefont{Taychatanapat}},
  \bibinfo{author}{\bibfnamefont{K.}~\bibnamefont{Watanabe}},
  \bibinfo{author}{\bibfnamefont{T.}~\bibnamefont{Taniguchi}},
  \bibinfo{author}{\bibfnamefont{A.}~\bibnamefont{Yacoby}}, \bibnamefont{and}
  \bibinfo{author}{\bibfnamefont{P.}~\bibnamefont{Jarillo-Herrero}},
  {``}\bibinfo{title}{Quantum Hall Effect, Screening, and Layer-Polarized
  Insulating States in Twisted Bilayer Graphene},{''} \bibinfo{journal}{Phys.
  Rev. Lett.} \textbf{\bibinfo{volume}{108}}, \bibinfo{pages}{076601}
  (\bibinfo{year}{2012}).

\bibitem[{\citenamefont{Yan et~al.}(2014{\natexlab{b}})\citenamefont{Yan, Meng,
  Liu, Qiao, Chu, Dou, Liu, Nie, Naugle, and He}}]{STM_VHS2014}
\bibinfo{author}{\bibfnamefont{W.}~\bibnamefont{Yan}},
  \bibinfo{author}{\bibfnamefont{L.}~\bibnamefont{Meng}},
  \bibinfo{author}{\bibfnamefont{M.}~\bibnamefont{Liu}},
  \bibinfo{author}{\bibfnamefont{J.-B.} \bibnamefont{Qiao}},
  \bibinfo{author}{\bibfnamefont{Z.-D.} \bibnamefont{Chu}},
  \bibinfo{author}{\bibfnamefont{R.-F.} \bibnamefont{Dou}},
  \bibinfo{author}{\bibfnamefont{Z.}~\bibnamefont{Liu}},
  \bibinfo{author}{\bibfnamefont{J.-C.} \bibnamefont{Nie}},
  \bibinfo{author}{\bibfnamefont{D.~G.} \bibnamefont{Naugle}},
  \bibnamefont{and} \bibinfo{author}{\bibfnamefont{L.}~\bibnamefont{He}},
  {``}\bibinfo{title}{Angle-dependent van Hove singularities and their
  breakdown in twisted graphene bilayers},{''} \bibinfo{journal}{Phys. Rev. B}
  \textbf{\bibinfo{volume}{90}}, \bibinfo{pages}{115402}
  (\bibinfo{year}{2014}{\natexlab{b}}).

\bibitem[{\citenamefont{Cherkez et~al.}(2015)\citenamefont{Cherkez,
  de~Laissardi\`ere, Mallet, and Veuillen}}]{STM_VHSdoping2015}
\bibinfo{author}{\bibfnamefont{V.}~\bibnamefont{Cherkez}},
  \bibinfo{author}{\bibfnamefont{G.} \bibnamefont{Trambly~de~Laissardi\`ere}},
  \bibinfo{author}{\bibfnamefont{P.}~\bibnamefont{Mallet}}, \bibnamefont{and}
  \bibinfo{author}{\bibfnamefont{J.-Y.} \bibnamefont{Veuillen}},
  {``}\bibinfo{title}{Van Hove singularities in doped twisted graphene bilayers
  studied by scanning tunneling spectroscopy},{''} \bibinfo{journal}{Phys. Rev.
  B} \textbf{\bibinfo{volume}{91}}, \bibinfo{pages}{155428}
  (\bibinfo{year}{2015}).

\bibitem[{\citenamefont{Yin et~al.}(2015{\natexlab{a}})\citenamefont{Yin, Qiao,
  Wang, Zuo, Yan, Xu, Dou, Nie, and He}}]{HePRB2015}
\bibinfo{author}{\bibfnamefont{L.-J.} \bibnamefont{Yin}},
  \bibinfo{author}{\bibfnamefont{J.-B.} \bibnamefont{Qiao}},
  \bibinfo{author}{\bibfnamefont{W.-X.} \bibnamefont{Wang}},
  \bibinfo{author}{\bibfnamefont{W.-J.} \bibnamefont{Zuo}},
  \bibinfo{author}{\bibfnamefont{W.}~\bibnamefont{Yan}},
  \bibinfo{author}{\bibfnamefont{R.}~\bibnamefont{Xu}},
  \bibinfo{author}{\bibfnamefont{R.-F.} \bibnamefont{Dou}},
  \bibinfo{author}{\bibfnamefont{J.-C.} \bibnamefont{Nie}}, \bibnamefont{and}
  \bibinfo{author}{\bibfnamefont{L.}~\bibnamefont{He}},
  {``}\bibinfo{title}{Landau quantization and Fermi velocity renormalization in
  twisted graphene bilayers},{''} \bibinfo{journal}{Phys. Rev. B}
  \textbf{\bibinfo{volume}{92}}, \bibinfo{pages}{201408}
  (\bibinfo{year}{2015}{\natexlab{a}}).

\bibitem[{\citenamefont{Ohta et~al.}(2012{\natexlab{b}})\citenamefont{Ohta,
  Robinson, Feibelman, Bostwick, Rotenberg, and Beechem}}]{OhtaARPES}
\bibinfo{author}{\bibfnamefont{T.}~\bibnamefont{Ohta}},
  \bibinfo{author}{\bibfnamefont{J.~T.} \bibnamefont{Robinson}},
  \bibinfo{author}{\bibfnamefont{P.~J.} \bibnamefont{Feibelman}},
  \bibinfo{author}{\bibfnamefont{A.}~\bibnamefont{Bostwick}},
  \bibinfo{author}{\bibfnamefont{E.}~\bibnamefont{Rotenberg}},
  \bibnamefont{and} \bibinfo{author}{\bibfnamefont{T.~E.}
  \bibnamefont{Beechem}}, {``}\bibinfo{title}{Evidence for Interlayer Coupling
  and Moir\'e Periodic Potentials in Twisted Bilayer Graphene},{''}
  \bibinfo{journal}{Phys. Rev. Lett.} \textbf{\bibinfo{volume}{109}},
  \bibinfo{pages}{186807} (\bibinfo{year}{2012}{\natexlab{b}}).

\bibitem[{\citenamefont{Ni et~al.}(2009)\citenamefont{Ni, Liu, Wang, Zheng, Li,
  Yu, and Shen}}]{NiRaman}
\bibinfo{author}{\bibfnamefont{Z.}~\bibnamefont{Ni}},
  \bibinfo{author}{\bibfnamefont{L.}~\bibnamefont{Liu}},
  \bibinfo{author}{\bibfnamefont{Y.}~\bibnamefont{Wang}},
  \bibinfo{author}{\bibfnamefont{Z.}~\bibnamefont{Zheng}},
  \bibinfo{author}{\bibfnamefont{L.-J.} \bibnamefont{Li}},
  \bibinfo{author}{\bibfnamefont{T.}~\bibnamefont{Yu}}, \bibnamefont{and}
  \bibinfo{author}{\bibfnamefont{Z.}~\bibnamefont{Shen}},
  {``}\bibinfo{title}{G-band Raman double resonance in twisted bilayer
  graphene: Evidence of band splitting and folding},{''}
  \bibinfo{journal}{Phys. Rev. B} \textbf{\bibinfo{volume}{80}},
  \bibinfo{pages}{125404} (\bibinfo{year}{2009}).

\bibitem[{\citenamefont{He et~al.}(2013{\natexlab{a}})\citenamefont{He, Chung,
  Delaney, Keiser, Jauregui, Shand, Chancey, Wang, Bao, and Chen}}]{HeRaman}
\bibinfo{author}{\bibfnamefont{R.}~\bibnamefont{He}},
  \bibinfo{author}{\bibfnamefont{T.-F.} \bibnamefont{Chung}},
  \bibinfo{author}{\bibfnamefont{C.}~\bibnamefont{Delaney}},
  \bibinfo{author}{\bibfnamefont{C.}~\bibnamefont{Keiser}},
  \bibinfo{author}{\bibfnamefont{L.~A.} \bibnamefont{Jauregui}},
  \bibinfo{author}{\bibfnamefont{P.~M.} \bibnamefont{Shand}},
  \bibinfo{author}{\bibfnamefont{C.~C.} \bibnamefont{Chancey}},
  \bibinfo{author}{\bibfnamefont{Y.}~\bibnamefont{Wang}},
  \bibinfo{author}{\bibfnamefont{J.}~\bibnamefont{Bao}}, \bibnamefont{and}
  \bibinfo{author}{\bibfnamefont{Y.~P.} \bibnamefont{Chen}},
  {``}\bibinfo{title}{Observation of Low Energy Raman Modes in Twisted Bilayer
  Graphene},{''} \bibinfo{journal}{Nano Lett.} \textbf{\bibinfo{volume}{13}},
  \bibinfo{pages}{3594} (\bibinfo{year}{2013}{\natexlab{a}}).

\bibitem[{\citenamefont{Jorio and Can\c{c}ado}(2013)}]{JorioRaman}
\bibinfo{author}{\bibfnamefont{A.}~\bibnamefont{Jorio}} \bibnamefont{and}
  \bibinfo{author}{\bibfnamefont{L.~G.} \bibnamefont{Can\c{c}ado}},
  {``}\bibinfo{title}{Raman spectroscopy of twisted bilayer graphene},{''}
  \bibinfo{journal}{Solid State Commun.}
  \textbf{\bibinfo{volume}{175â176}}, \bibinfo{pages}{3 }
  (\bibinfo{year}{2013}), \bibinfo{note}{special Issue: Graphene V: Recent
  Advances in Studies of Graphene and Graphene analogues}.

\bibitem[{\citenamefont{Beechem et~al.}(2014)\citenamefont{Beechem, Ohta,
  Diaconescu, and Robinson}}]{RotDisorder2014}
\bibinfo{author}{\bibfnamefont{T.~E.} \bibnamefont{Beechem}},
  \bibinfo{author}{\bibfnamefont{T.}~\bibnamefont{Ohta}},
  \bibinfo{author}{\bibfnamefont{B.}~\bibnamefont{Diaconescu}},
  \bibnamefont{and} \bibinfo{author}{\bibfnamefont{J.~T.}
  \bibnamefont{Robinson}}, {``}\bibinfo{title}{Rotational Disorder in Twisted
  Bilayer Graphene},{''} \bibinfo{journal}{ACS Nano}
  \textbf{\bibinfo{volume}{8}}, \bibinfo{pages}{1655} (\bibinfo{year}{2014}).

\bibitem[{\citenamefont{Wu et~al.}(2014)\citenamefont{Wu, Zhang, Ij{\"a}s, Han,
  Qiao, Li, Jiang, Ferrari, and Tan}}]{WuRaman2014}
\bibinfo{author}{\bibfnamefont{J.-B.} \bibnamefont{Wu}},
  \bibinfo{author}{\bibfnamefont{X.}~\bibnamefont{Zhang}},
  \bibinfo{author}{\bibfnamefont{M.}~\bibnamefont{Ij{\"a}s}},
  \bibinfo{author}{\bibfnamefont{W.-P.} \bibnamefont{Han}},
  \bibinfo{author}{\bibfnamefont{X.-F.} \bibnamefont{Qiao}},
  \bibinfo{author}{\bibfnamefont{X.-L.} \bibnamefont{Li}},
  \bibinfo{author}{\bibfnamefont{D.-S.} \bibnamefont{Jiang}},
  \bibinfo{author}{\bibfnamefont{A.~C.} \bibnamefont{Ferrari}},
  \bibnamefont{and} \bibinfo{author}{\bibfnamefont{P.-H.} \bibnamefont{Tan}},
  {``}\bibinfo{title}{Resonant Raman spectroscopy of twisted multilayer
  graphene},{''} \bibinfo{journal}{Nat. Commun.} \textbf{\bibinfo{volume}{5}}
  (\bibinfo{year}{2014}).

\bibitem[{\citenamefont{Wong et~al.}(2015)\citenamefont{Wong, Wang, Jung,
  Pezzini, DaSilva, Tsai, Jung, Khajeh, Kim, Lee et~al.}}]{wong_tblg2015}
\bibinfo{author}{\bibfnamefont{D.}~\bibnamefont{Wong}},
  \bibinfo{author}{\bibfnamefont{Y.}~\bibnamefont{Wang}},
  \bibinfo{author}{\bibfnamefont{J.}~\bibnamefont{Jung}},
  \bibinfo{author}{\bibfnamefont{S.}~\bibnamefont{Pezzini}},
  \bibinfo{author}{\bibfnamefont{A.~M.} \bibnamefont{DaSilva}},
  \bibinfo{author}{\bibfnamefont{H.-Z.} \bibnamefont{Tsai}},
  \bibinfo{author}{\bibfnamefont{H.~S.} \bibnamefont{Jung}},
  \bibinfo{author}{\bibfnamefont{R.}~\bibnamefont{Khajeh}},
  \bibinfo{author}{\bibfnamefont{Y.}~\bibnamefont{Kim}},
  \bibinfo{author}{\bibfnamefont{J.}~\bibnamefont{Lee}}, \bibnamefont{et~al.},
  {``}\bibinfo{title}{Local spectroscopy of Moir{\'{e}}-induced electronic
  structure in gate-tunable twisted bilayer graphene},{''}
  \bibinfo{journal}{Phys. Rev. B} \textbf{\bibinfo{volume}{92}},
  \bibinfo{pages}{155409} (\bibinfo{year}{2015}).

\bibitem[{\citenamefont{Sboychakov et~al.}(2015)\citenamefont{Sboychakov,
  Rakhmanov, Rozhkov, and Nori}}]{ourTBLG}
\bibinfo{author}{\bibfnamefont{A.~O.} \bibnamefont{Sboychakov}},
  \bibinfo{author}{\bibfnamefont{A.~L.} \bibnamefont{Rakhmanov}},
  \bibinfo{author}{\bibfnamefont{A.~V.} \bibnamefont{Rozhkov}},
  \bibnamefont{and} \bibinfo{author}{\bibfnamefont{F.}~\bibnamefont{Nori}},
  {``}\bibinfo{title}{Electronic spectrum of twisted bilayer graphene},{''}
  \bibinfo{journal}{Phys. Rev. B} \textbf{\bibinfo{volume}{92}},
  \bibinfo{pages}{075402} (\bibinfo{year}{2015}).

\bibitem[{\citenamefont{Yin et~al.}(2015{\natexlab{b}})\citenamefont{Yin, Qiao,
  Zuo, Li, and He}}]{SmallAngleSTM2015}
\bibinfo{author}{\bibfnamefont{L.-J.} \bibnamefont{Yin}},
  \bibinfo{author}{\bibfnamefont{J.-B.} \bibnamefont{Qiao}},
  \bibinfo{author}{\bibfnamefont{W.-J.} \bibnamefont{Zuo}},
  \bibinfo{author}{\bibfnamefont{W.-T.} \bibnamefont{Li}}, \bibnamefont{and}
  \bibinfo{author}{\bibfnamefont{L.}~\bibnamefont{He}},
  {``}\bibinfo{title}{Experimental evidence for non-Abelian gauge potentials in
  twisted graphene bilayers},{''} \bibinfo{journal}{Phys. Rev. B}
  \textbf{\bibinfo{volume}{92}}, \bibinfo{pages}{081406}
  (\bibinfo{year}{2015}{\natexlab{b}}).

\bibitem[{\citenamefont{Park et~al.}(2015{\natexlab{b}})\citenamefont{Park,
  Mitchel, Elhamri, Grazulis, Hoelscher, Mahalingam, Hwang, Mo, and
  Lee}}]{TransportGap2015}
\bibinfo{author}{\bibfnamefont{J.}~\bibnamefont{Park}},
  \bibinfo{author}{\bibfnamefont{W.~C.} \bibnamefont{Mitchel}},
  \bibinfo{author}{\bibfnamefont{S.}~\bibnamefont{Elhamri}},
  \bibinfo{author}{\bibfnamefont{L.}~\bibnamefont{Grazulis}},
  \bibinfo{author}{\bibfnamefont{J.}~\bibnamefont{Hoelscher}},
  \bibinfo{author}{\bibfnamefont{K.}~\bibnamefont{Mahalingam}},
  \bibinfo{author}{\bibfnamefont{C.}~\bibnamefont{Hwang}},
  \bibinfo{author}{\bibfnamefont{S.-K.} \bibnamefont{Mo}}, \bibnamefont{and}
  \bibinfo{author}{\bibfnamefont{J.}~\bibnamefont{Lee}},
  {``}\bibinfo{title}{Observation of the intrinsic bandgap behaviour in
  as-grown epitaxial twisted graphene},{''} \bibinfo{journal}{Nat. Commun.}
  \textbf{\bibinfo{volume}{6}}, \bibinfo{pages}{5677}
  (\bibinfo{year}{2015}{\natexlab{b}}).

\bibitem[{\citenamefont{Slater and Koster}(1954)}]{SlaterKoster}
\bibinfo{author}{\bibfnamefont{J.~C.} \bibnamefont{Slater}} \bibnamefont{and}
  \bibinfo{author}{\bibfnamefont{G.~F.} \bibnamefont{Koster}},
  {``}\bibinfo{title}{Simplified LCAO Method for the Periodic Potential
  Problem},{''} \bibinfo{journal}{Phys. Rev. }
  \textbf{\bibinfo{volume}{94}}, \bibinfo{pages}{1498} (\bibinfo{year}{1954}).

\bibitem[{\citenamefont{San-Jose and Prada}(2013)}]{DeltaUeff}
\bibinfo{author}{\bibfnamefont{P.}~\bibnamefont{San-Jose}} \bibnamefont{and}
  \bibinfo{author}{\bibfnamefont{E.}~\bibnamefont{Prada}},
  {``}\bibinfo{title}{Helical networks in twisted bilayer graphene under
  interlayer bias},{''} \bibinfo{journal}{Phys. Rev. B}
  \textbf{\bibinfo{volume}{88}}, \bibinfo{pages}{121408}
  (\bibinfo{year}{2013}).

\bibitem[{\citenamefont{Moon et~al.}(2014)\citenamefont{Moon, Son, and
  Koshino}}]{MoonTBgate2014}
\bibinfo{author}{\bibfnamefont{P.}~\bibnamefont{Moon}},
  \bibinfo{author}{\bibfnamefont{Y.-W.} \bibnamefont{Son}}, \bibnamefont{and}
  \bibinfo{author}{\bibfnamefont{M.}~\bibnamefont{Koshino}},
  {``}\bibinfo{title}{Optical absorption of twisted bilayer graphene with
  interlayer potential asymmetry},{''} \bibinfo{journal}{Phys. Rev. B}
  \textbf{\bibinfo{volume}{90}}, \bibinfo{pages}{155427}
  (\bibinfo{year}{2014}).

\bibitem[{\citenamefont{Muniz and Maroudas}(2012{\natexlab{a}})}]{Muniz1}
\bibinfo{author}{\bibfnamefont{A.~R.} \bibnamefont{Muniz}} \bibnamefont{and}
  \bibinfo{author}{\bibfnamefont{D.}~\bibnamefont{Maroudas}},
  {``}\bibinfo{title}{Opening and tuning of band gap by the formation of
  diamond superlattices in twisted bilayer graphene},{''}
  \bibinfo{journal}{Phys. Rev. B} \textbf{\bibinfo{volume}{86}},
  \bibinfo{pages}{075404} (\bibinfo{year}{2012}{\natexlab{a}}).

\bibitem[{\citenamefont{Muniz and Maroudas}(2012{\natexlab{b}})}]{Muniz2}
\bibinfo{author}{\bibfnamefont{A.~R.} \bibnamefont{Muniz}} \bibnamefont{and}
  \bibinfo{author}{\bibfnamefont{D.}~\bibnamefont{Maroudas}},
  {``}\bibinfo{title}{Formation of fullerene superlattices by interlayer
  bonding in twisted bilayer graphene},{''} \bibinfo{journal}{J. Appl. Phys.}
  \textbf{\bibinfo{volume}{111}}, \bibinfo{eid}{043513}
  (\bibinfo{year}{2012}{\natexlab{b}}).

\bibitem[{\citenamefont{Symalla et~al.}(2015)\citenamefont{Symalla, Shallcross,
  Beljakov, Fink, Wenzel, and Meded}}]{GapFabrication2015}
\bibinfo{author}{\bibfnamefont{F.}~\bibnamefont{Symalla}},
  \bibinfo{author}{\bibfnamefont{S.}~\bibnamefont{Shallcross}},
  \bibinfo{author}{\bibfnamefont{I.}~\bibnamefont{Beljakov}},
  \bibinfo{author}{\bibfnamefont{K.}~\bibnamefont{Fink}},
  \bibinfo{author}{\bibfnamefont{W.}~\bibnamefont{Wenzel}}, \bibnamefont{and}
  \bibinfo{author}{\bibfnamefont{V.}~\bibnamefont{Meded}},
  {``}\bibinfo{title}{Band-gap engineering with a twist: Formation of
  intercalant superlattices in twisted graphene bilayers},{''}
  \bibinfo{journal}{Phys. Rev. B} \textbf{\bibinfo{volume}{91}},
  \bibinfo{pages}{205412} (\bibinfo{year}{2015}).

\bibitem[{\citenamefont{Ulman and Narasimhan}(2014)}]{PointDefectsDFT2014}
\bibinfo{author}{\bibfnamefont{K.}~\bibnamefont{Ulman}} \bibnamefont{and}
  \bibinfo{author}{\bibfnamefont{S.}~\bibnamefont{Narasimhan}},
  {``}\bibinfo{title}{Point defects in twisted bilayer graphene: A density
  functional theory study},{''} \bibinfo{journal}{Phys. Rev. B}
  \textbf{\bibinfo{volume}{89}}, \bibinfo{pages}{245429}
  (\bibinfo{year}{2014}).

\bibitem[{\citenamefont{Landau and Lifshitz}(1981)}]{LandauQM}
\bibinfo{author}{\bibfnamefont{L.~D.} \bibnamefont{Landau}} \bibnamefont{and}
  \bibinfo{author}{\bibfnamefont{L.~M.} \bibnamefont{Lifshitz}},
  \emph{\bibinfo{title}{Quantum Mechanics Non-Relativistic Theory, Third
  Edition: Volume 3}} (\bibinfo{publisher}{Butterworth-Heinemann},
  \bibinfo{year}{1981}), \bibinfo{edition}{3rd} ed.

\bibitem[{\citenamefont{Lee et~al.}(2011)\citenamefont{Lee, Riedl, Beringer,
  Castro~Neto, von Klitzing, Starke, and Smet}}]{LeeQHE}
\bibinfo{author}{\bibfnamefont{D.~S.} \bibnamefont{Lee}},
  \bibinfo{author}{\bibfnamefont{C.}~\bibnamefont{Riedl}},
  \bibinfo{author}{\bibfnamefont{T.}~\bibnamefont{Beringer}},
  \bibinfo{author}{\bibfnamefont{A.~H.} \bibnamefont{Castro~Neto}},
  \bibinfo{author}{\bibfnamefont{K.}~\bibnamefont{von Klitzing}},
  \bibinfo{author}{\bibfnamefont{U.}~\bibnamefont{Starke}}, \bibnamefont{and}
  \bibinfo{author}{\bibfnamefont{J.~H.} \bibnamefont{Smet}},
  {``}\bibinfo{title}{Quantum Hall Effect in Twisted Bilayer Graphene},{''}
  \bibinfo{journal}{Phys. Rev. Lett.} \textbf{\bibinfo{volume}{107}},
  \bibinfo{pages}{216602} (\bibinfo{year}{2011}).

\bibitem[{\citenamefont{Fallahazad et~al.}(2012)\citenamefont{Fallahazad, Hao,
  Lee, Kim, Ruoff, and Tutuc}}]{FallahazadQHE}
\bibinfo{author}{\bibfnamefont{B.}~\bibnamefont{Fallahazad}},
  \bibinfo{author}{\bibfnamefont{Y.}~\bibnamefont{Hao}},
  \bibinfo{author}{\bibfnamefont{K.}~\bibnamefont{Lee}},
  \bibinfo{author}{\bibfnamefont{S.}~\bibnamefont{Kim}},
  \bibinfo{author}{\bibfnamefont{R.~S.} \bibnamefont{Ruoff}}, \bibnamefont{and}
  \bibinfo{author}{\bibfnamefont{E.}~\bibnamefont{Tutuc}},
  {``}\bibinfo{title}{Quantum Hall effect in Bernal stacked and twisted bilayer
  graphene grown on Cu by chemical vapor deposition},{''}
  \bibinfo{journal}{Phys. Rev. B} \textbf{\bibinfo{volume}{85}},
  \bibinfo{pages}{201408} (\bibinfo{year}{2012}).

\bibitem[{\citenamefont{Miller et~al.}(2010)\citenamefont{Miller, Kubista,
  Rutter, Ruan, de~Heer, Kindermann, First, and Stroscio}}]{MillerLL0split}
\bibinfo{author}{\bibfnamefont{D.~L.} \bibnamefont{Miller}},
  \bibinfo{author}{\bibfnamefont{K.~D.} \bibnamefont{Kubista}},
  \bibinfo{author}{\bibfnamefont{G.~M.} \bibnamefont{Rutter}},
  \bibinfo{author}{\bibfnamefont{M.}~\bibnamefont{Ruan}},
  \bibinfo{author}{\bibfnamefont{W.~A.} \bibnamefont{de~Heer}},
  \bibinfo{author}{\bibfnamefont{M.}~\bibnamefont{Kindermann}},
  \bibinfo{author}{\bibfnamefont{P.~N.} \bibnamefont{First}}, \bibnamefont{and}
  \bibinfo{author}{\bibfnamefont{J.~A.} \bibnamefont{Stroscio}},
  {``}\bibinfo{title}{Real-space mapping of magnetically quantized graphene
  states},{''} \bibinfo{journal}{Nat. Phys.} \textbf{\bibinfo{volume}{6}},
  \bibinfo{pages}{811} (\bibinfo{year}{2010}).

\bibitem[{\citenamefont{Kindermann and Mele}(2011)}]{KindermanLL}
\bibinfo{author}{\bibfnamefont{M.}~\bibnamefont{Kindermann}} \bibnamefont{and}
  \bibinfo{author}{\bibfnamefont{E.~J.} \bibnamefont{Mele}},
  {``}\bibinfo{title}{Landau quantization in twisted bilayer graphene: The
  Dirac comb},{''} \bibinfo{journal}{Phys. Rev. B}
  \textbf{\bibinfo{volume}{84}}, \bibinfo{pages}{161406}
  (\bibinfo{year}{2011}).

\bibitem[{\citenamefont{Pal et~al.}(2014)\citenamefont{Pal, Mele, and
  Kindermann}}]{PalLL}
\bibinfo{author}{\bibfnamefont{H.~K.} \bibnamefont{Pal}},
  \bibinfo{author}{\bibfnamefont{E.~J.} \bibnamefont{Mele}}, \bibnamefont{and}
  \bibinfo{author}{\bibfnamefont{M.}~\bibnamefont{Kindermann}},
  {``}\bibinfo{title}{Landau level splitting in rotationally faulted multilayer
  graphene},{''} \bibinfo{journal}{Phys. Rev. B} \textbf{\bibinfo{volume}{89}},
  \bibinfo{pages}{081403} (\bibinfo{year}{2014}).

\bibitem[{\citenamefont{Apalkov and
  Chakraborty}(2011{\natexlab{b}})}]{ApalkovPRB2011}
\bibinfo{author}{\bibfnamefont{V.~M.} \bibnamefont{Apalkov}} \bibnamefont{and}
  \bibinfo{author}{\bibfnamefont{T.}~\bibnamefont{Chakraborty}},
  {``}\bibinfo{title}{Optical transitions at commensurate angles in a
  misoriented bilayer graphene in an external magnetic field},{''}
  \bibinfo{journal}{Phys. Rev. B} \textbf{\bibinfo{volume}{84}},
  \bibinfo{pages}{033408} (\bibinfo{year}{2011}{\natexlab{b}}).

\bibitem[{\citenamefont{Kindermann and First}(2011)}]{KindermanLL2}
\bibinfo{author}{\bibfnamefont{M.}~\bibnamefont{Kindermann}} \bibnamefont{and}
  \bibinfo{author}{\bibfnamefont{P.~N.} \bibnamefont{First}},
  {``}\bibinfo{title}{Local sublattice-symmetry breaking in rotationally
  faulted multilayer graphene},{''} \bibinfo{journal}{Phys. Rev. B}
  \textbf{\bibinfo{volume}{83}}, \bibinfo{pages}{045425}
  (\bibinfo{year}{2011}).

\bibitem[{\citenamefont{Hofstadter}(1976)}]{Hofstadter}
\bibinfo{author}{\bibfnamefont{D.~R.} \bibnamefont{Hofstadter}},
  {``}\bibinfo{title}{Energy levels and wave functions of Bloch electrons in
  rational and irrational magnetic fields},{''} \bibinfo{journal}{Phys. Rev. B}
  \textbf{\bibinfo{volume}{14}}, \bibinfo{pages}{2239} (\bibinfo{year}{1976}).

\bibitem[{\citenamefont{Moon and Koshino}(2012)}]{MoonButterfly}
\bibinfo{author}{\bibfnamefont{P.}~\bibnamefont{Moon}} \bibnamefont{and}
  \bibinfo{author}{\bibfnamefont{M.}~\bibnamefont{Koshino}},
  {``}\bibinfo{title}{Energy spectrum and quantum Hall effect in twisted
  bilayer graphene},{''} \bibinfo{journal}{Phys. Rev. B}
  \textbf{\bibinfo{volume}{85}}, \bibinfo{pages}{195458}
  (\bibinfo{year}{2012}).

\bibitem[{\citenamefont{Hunt et~al.}(2013)\citenamefont{Hunt,
  Sanchez-Yamagishi, Young, Yankowitz, LeRoy, Watanabe, Taniguchi, Moon,
  Koshino, Jarillo-Herrero et~al.}}]{Hunt2013}
\bibinfo{author}{\bibfnamefont{B.}~\bibnamefont{Hunt}},
  \bibinfo{author}{\bibfnamefont{J.~D.} \bibnamefont{Sanchez-Yamagishi}},
  \bibinfo{author}{\bibfnamefont{A.~F.} \bibnamefont{Young}},
  \bibinfo{author}{\bibfnamefont{M.}~\bibnamefont{Yankowitz}},
  \bibinfo{author}{\bibfnamefont{B.~J.} \bibnamefont{LeRoy}},
  \bibinfo{author}{\bibfnamefont{K.}~\bibnamefont{Watanabe}},
  \bibinfo{author}{\bibfnamefont{T.}~\bibnamefont{Taniguchi}},
  \bibinfo{author}{\bibfnamefont{P.}~\bibnamefont{Moon}},
  \bibinfo{author}{\bibfnamefont{M.}~\bibnamefont{Koshino}},
  \bibinfo{author}{\bibfnamefont{P.}~\bibnamefont{Jarillo-Herrero}},
  \bibnamefont{et~al.}, {``}\bibinfo{title}{Massive Dirac Fermions and
  Hofstadter Butterfly in a van der Waals Heterostructure},{''}
  \bibinfo{journal}{Science} \textbf{\bibinfo{volume}{340}},
  \bibinfo{pages}{1427} (\bibinfo{year}{2013}).

\bibitem[{\citenamefont{Ponomarenko et~al.}(2013)\citenamefont{Ponomarenko,
  Gorbachev, Yu, Elias, Jalil, Patel, Mishchenko, Mayorov, Woods, Wallbank
  et~al.}}]{Ponomarenko2013}
\bibinfo{author}{\bibfnamefont{L.~A.} \bibnamefont{Ponomarenko}},
  \bibinfo{author}{\bibfnamefont{R.~V.} \bibnamefont{Gorbachev}},
  \bibinfo{author}{\bibfnamefont{G.~L.} \bibnamefont{Yu}},
  \bibinfo{author}{\bibfnamefont{D.~C.} \bibnamefont{Elias}},
  \bibinfo{author}{\bibfnamefont{R.}~\bibnamefont{Jalil}},
  \bibinfo{author}{\bibfnamefont{A.~A.} \bibnamefont{Patel}},
  \bibinfo{author}{\bibfnamefont{A.}~\bibnamefont{Mishchenko}},
  \bibinfo{author}{\bibfnamefont{A.~S.} \bibnamefont{Mayorov}},
  \bibinfo{author}{\bibfnamefont{C.~R.} \bibnamefont{Woods}},
  \bibinfo{author}{\bibfnamefont{J.~R.} \bibnamefont{Wallbank}},
  \bibnamefont{et~al.}, {``}\bibinfo{title}{Cloning of Dirac fermions in
  graphene superlattices},{''} \bibinfo{journal}{Nature}
  \textbf{\bibinfo{volume}{497}}, \bibinfo{pages}{594} (\bibinfo{year}{2013}).

\bibitem[{\citenamefont{Dean et~al.}(2013)\citenamefont{Dean, Wang, Maher,
  Forsythe, Ghahari, Gao, Katoch, Ishigami, Moon, Koshino et~al.}}]{Dean2013}
\bibinfo{author}{\bibfnamefont{C.~R.} \bibnamefont{Dean}},
  \bibinfo{author}{\bibfnamefont{L.}~\bibnamefont{Wang}},
  \bibinfo{author}{\bibfnamefont{P.}~\bibnamefont{Maher}},
  \bibinfo{author}{\bibfnamefont{C.}~\bibnamefont{Forsythe}},
  \bibinfo{author}{\bibfnamefont{F.}~\bibnamefont{Ghahari}},
  \bibinfo{author}{\bibfnamefont{Y.}~\bibnamefont{Gao}},
  \bibinfo{author}{\bibfnamefont{J.}~\bibnamefont{Katoch}},
  \bibinfo{author}{\bibfnamefont{M.}~\bibnamefont{Ishigami}},
  \bibinfo{author}{\bibfnamefont{P.}~\bibnamefont{Moon}},
  \bibinfo{author}{\bibfnamefont{M.}~\bibnamefont{Koshino}},
  \bibnamefont{et~al.}, {``}\bibinfo{title}{Hofstadter's butterfly and the
  fractal quantum Hall effect in moire superlattices},{''}
  \bibinfo{journal}{Nature} \textbf{\bibinfo{volume}{497}},
  \bibinfo{pages}{598} (\bibinfo{year}{2013}).

\bibitem[{\citenamefont{Schmidt et~al.}(2014)\citenamefont{Schmidt, Rode,
  Smirnov, and Haug}}]{alaHofstadter2014}
\bibinfo{author}{\bibfnamefont{H.}~\bibnamefont{Schmidt}},
  \bibinfo{author}{\bibfnamefont{J.~C.} \bibnamefont{Rode}},
  \bibinfo{author}{\bibfnamefont{D.}~\bibnamefont{Smirnov}}, \bibnamefont{and}
  \bibinfo{author}{\bibfnamefont{R.~J.} \bibnamefont{Haug}},
  {``}\bibinfo{title}{Superlattice structures in twisted bilayers of folded
  graphene},{''} \bibinfo{journal}{Nat. Commun.} \textbf{\bibinfo{volume}{5}},
  \bibinfo{pages}{5742} (\bibinfo{year}{2014}).

\bibitem[{\citenamefont{Bistritzer and
  MacDonald}(2011{\natexlab{b}})}]{PNASbutterfly}
\bibinfo{author}{\bibfnamefont{R.}~\bibnamefont{Bistritzer}} \bibnamefont{and}
  \bibinfo{author}{\bibfnamefont{A.~H.} \bibnamefont{MacDonald}},
  {``}\bibinfo{title}{Moir\'e butterflies in twisted bilayer graphene},{''}
  \bibinfo{journal}{Phys. Rev. B} \textbf{\bibinfo{volume}{84}},
  \bibinfo{pages}{035440} (\bibinfo{year}{2011}{\natexlab{b}}).

\bibitem[{\citenamefont{Wang et~al.}(2012)\citenamefont{Wang, Liu, and
  Chou}}]{WangButterfly}
\bibinfo{author}{\bibfnamefont{Z.~F.} \bibnamefont{Wang}},
  \bibinfo{author}{\bibfnamefont{F.}~\bibnamefont{Liu}}, \bibnamefont{and}
  \bibinfo{author}{\bibfnamefont{M.~Y.} \bibnamefont{Chou}},
  {``}\bibinfo{title}{Fractal Landau-Level Spectra in Twisted Bilayer
  Graphene},{''} \bibinfo{journal}{Nano Lett.} \textbf{\bibinfo{volume}{12}},
  \bibinfo{pages}{3833} (\bibinfo{year}{2012}).

\bibitem[{\citenamefont{Hasegawa and Kohmoto}(2013)}]{HasegawaButterfly}
\bibinfo{author}{\bibfnamefont{Y.}~\bibnamefont{Hasegawa}} \bibnamefont{and}
  \bibinfo{author}{\bibfnamefont{M.}~\bibnamefont{Kohmoto}},
  {``}\bibinfo{title}{Periodic Landau gauge and quantum Hall effect in twisted
  bilayer graphene},{''} \bibinfo{journal}{Phys. Rev. B}
  \textbf{\bibinfo{volume}{88}}, \bibinfo{pages}{125426}
  (\bibinfo{year}{2013}).

\bibitem[{\citenamefont{Lu and
  Fertig}(2014{\natexlab{a}})}]{FertigLLoptAbs2014}
\bibinfo{author}{\bibfnamefont{C.-K.} \bibnamefont{Lu}} \bibnamefont{and}
  \bibinfo{author}{\bibfnamefont{H.~A.} \bibnamefont{Fertig}},
  {``}\bibinfo{title}{Probing layer localization in twisted graphene bilayers
  via cyclotron resonance},{''} \bibinfo{journal}{Phys. Rev. B}
  \textbf{\bibinfo{volume}{90}}, \bibinfo{pages}{115436}
  (\bibinfo{year}{2014}{\natexlab{a}}).

\bibitem[{\citenamefont{Stone}(1992)}]{stone1992quantum}
\bibinfo{author}{\bibfnamefont{M.}~\bibnamefont{Stone}},
  \emph{\bibinfo{title}{Quantum Hall Effect}} (\bibinfo{publisher}{World
  Scientific}, \bibinfo{year}{1992}).

\bibitem[{\citenamefont{Lu and
  Fertig}(2014{\natexlab{b}})}]{LLnearVHSemiClassic2014}
\bibinfo{author}{\bibfnamefont{C.-K.} \bibnamefont{Lu}} \bibnamefont{and}
  \bibinfo{author}{\bibfnamefont{H.~A.} \bibnamefont{Fertig}},
  {``}\bibinfo{title}{Magnetic breakdown in twisted bilayer graphene},{''}
  \bibinfo{journal}{Phys. Rev. B} \textbf{\bibinfo{volume}{89}},
  \bibinfo{pages}{085408} (\bibinfo{year}{2014}{\natexlab{b}}).

\bibitem[{\citenamefont{Tabert and Nicol}(2013)}]{SigmaNicol}
\bibinfo{author}{\bibfnamefont{C.~J.} \bibnamefont{Tabert}} \bibnamefont{and}
  \bibinfo{author}{\bibfnamefont{E.~J.} \bibnamefont{Nicol}},
  {``}\bibinfo{title}{Optical conductivity of twisted bilayer graphene},{''}
  \bibinfo{journal}{Phys. Rev. B} \textbf{\bibinfo{volume}{87}},
  \bibinfo{pages}{121402} (\bibinfo{year}{2013}).

\bibitem[{\citenamefont{Wang et~al.}(2010)\citenamefont{Wang, Ni, Liu, Liu,
  Cong, Yu, Wang, Shen, and Shen}}]{ExpSigmaNano}
\bibinfo{author}{\bibfnamefont{Y.}~\bibnamefont{Wang}},
  \bibinfo{author}{\bibfnamefont{Z.}~\bibnamefont{Ni}},
  \bibinfo{author}{\bibfnamefont{L.}~\bibnamefont{Liu}},
  \bibinfo{author}{\bibfnamefont{Y.}~\bibnamefont{Liu}},
  \bibinfo{author}{\bibfnamefont{C.}~\bibnamefont{Cong}},
  \bibinfo{author}{\bibfnamefont{T.}~\bibnamefont{Yu}},
  \bibinfo{author}{\bibfnamefont{X.}~\bibnamefont{Wang}},
  \bibinfo{author}{\bibfnamefont{D.}~\bibnamefont{Shen}}, \bibnamefont{and}
  \bibinfo{author}{\bibfnamefont{Z.}~\bibnamefont{Shen}},
  {``}\bibinfo{title}{Stacking-Dependent Optical Conductivity of Bilayer
  Graphene},{''} \bibinfo{journal}{ACS Nano} \textbf{\bibinfo{volume}{4}},
  \bibinfo{pages}{4074} (\bibinfo{year}{2010}).

\bibitem[{\citenamefont{Zou et~al.}(2013)\citenamefont{Zou, Shang, Leaw, Luo,
  Luo, La{-}o{-}vorakiat, Cheng, Cheong, Su, Zhu et~al.}}]{ExpSigmaPRL}
\bibinfo{author}{\bibfnamefont{X.}~\bibnamefont{Zou}},
  \bibinfo{author}{\bibfnamefont{J.}~\bibnamefont{Shang}},
  \bibinfo{author}{\bibfnamefont{J.}~\bibnamefont{Leaw}},
  \bibinfo{author}{\bibfnamefont{Z.}~\bibnamefont{Luo}},
  \bibinfo{author}{\bibfnamefont{L.}~\bibnamefont{Luo}},
  \bibinfo{author}{\bibfnamefont{C.}~\bibnamefont{La{-}o{-}vorakiat}},
  \bibinfo{author}{\bibfnamefont{L.}~\bibnamefont{Cheng}},
  \bibinfo{author}{\bibfnamefont{S.~A.} \bibnamefont{Cheong}},
  \bibinfo{author}{\bibfnamefont{H.}~\bibnamefont{Su}},
  \bibinfo{author}{\bibfnamefont{J.-X.} \bibnamefont{Zhu}},
  \bibnamefont{et~al.}, {``}\bibinfo{title}{Terahertz Conductivity of Twisted
  Bilayer Graphene},{''} \bibinfo{journal}{Phys. Rev. Lett.}
  \textbf{\bibinfo{volume}{110}}, \bibinfo{pages}{067401}
  (\bibinfo{year}{2013}).

\bibitem[{\citenamefont{Kim et~al.}(2013)\citenamefont{Kim, Yun, Nam, Son, Lee,
  Kim, Seo, Choi, Lee, Lee et~al.}}]{ExpSigmaPerp}
\bibinfo{author}{\bibfnamefont{Y.}~\bibnamefont{Kim}},
  \bibinfo{author}{\bibfnamefont{H.}~\bibnamefont{Yun}},
  \bibinfo{author}{\bibfnamefont{S.-G.} \bibnamefont{Nam}},
  \bibinfo{author}{\bibfnamefont{M.}~\bibnamefont{Son}},
  \bibinfo{author}{\bibfnamefont{D.~S.} \bibnamefont{Lee}},
  \bibinfo{author}{\bibfnamefont{D.~C.} \bibnamefont{Kim}},
  \bibinfo{author}{\bibfnamefont{S.}~\bibnamefont{Seo}},
  \bibinfo{author}{\bibfnamefont{H.~C.} \bibnamefont{Choi}},
  \bibinfo{author}{\bibfnamefont{H.-J.} \bibnamefont{Lee}},
  \bibinfo{author}{\bibfnamefont{S.~W.} \bibnamefont{Lee}},
  \bibnamefont{et~al.}, {``}\bibinfo{title}{Breakdown of the Interlayer
  Coherence in Twisted Bilayer Graphene},{''} \bibinfo{journal}{Phys. Rev.
  Lett.} \textbf{\bibinfo{volume}{110}}, \bibinfo{pages}{096602}
  (\bibinfo{year}{2013}).

\bibitem[{\citenamefont{Bistritzer and MacDonald}(2010)}]{PNASsigma}
\bibinfo{author}{\bibfnamefont{R.}~\bibnamefont{Bistritzer}} \bibnamefont{and}
  \bibinfo{author}{\bibfnamefont{A.~H.} \bibnamefont{MacDonald}},
  {``}\bibinfo{title}{Transport between twisted graphene layers},{''}
  \bibinfo{journal}{Phys. Rev. B} \textbf{\bibinfo{volume}{81}},
  \bibinfo{pages}{245412} (\bibinfo{year}{2010}).

\bibitem[{\citenamefont{Moon and Koshino}(2013)}]{SigmaMoon}
\bibinfo{author}{\bibfnamefont{P.}~\bibnamefont{Moon}} \bibnamefont{and}
  \bibinfo{author}{\bibfnamefont{M.}~\bibnamefont{Koshino}},
  {``}\bibinfo{title}{Optical absorption in twisted bilayer graphene},{''}
  \bibinfo{journal}{Phys. Rev. B} \textbf{\bibinfo{volume}{87}},
  \bibinfo{pages}{205404} (\bibinfo{year}{2013}).

\bibitem[{\citenamefont{Havener et~al.}(2014)\citenamefont{Havener, Liang,
  Brown, Yang, and Park}}]{OptCondVHS2014}
\bibinfo{author}{\bibfnamefont{R.~W.} \bibnamefont{Havener}},
  \bibinfo{author}{\bibfnamefont{Y.}~\bibnamefont{Liang}},
  \bibinfo{author}{\bibfnamefont{L.}~\bibnamefont{Brown}},
  \bibinfo{author}{\bibfnamefont{L.}~\bibnamefont{Yang}}, \bibnamefont{and}
  \bibinfo{author}{\bibfnamefont{J.}~\bibnamefont{Park}},
  {``}\bibinfo{title}{Van Hove Singularities and Excitonic Effects in the
  Optical Conductivity of Twisted Bilayer Graphene},{''} \bibinfo{journal}{Nano
  Lett.} \textbf{\bibinfo{volume}{14}}, \bibinfo{pages}{3353}
  (\bibinfo{year}{2014}).

\bibitem[{\citenamefont{Liang et~al.}(2014)\citenamefont{Liang, Soklaski,
  Huang, Graham, Havener, Park, and Yang}}]{Excitons2014}
\bibinfo{author}{\bibfnamefont{Y.}~\bibnamefont{Liang}},
  \bibinfo{author}{\bibfnamefont{R.}~\bibnamefont{Soklaski}},
  \bibinfo{author}{\bibfnamefont{S.}~\bibnamefont{Huang}},
  \bibinfo{author}{\bibfnamefont{M.~W.} \bibnamefont{Graham}},
  \bibinfo{author}{\bibfnamefont{R.}~\bibnamefont{Havener}},
  \bibinfo{author}{\bibfnamefont{J.}~\bibnamefont{Park}}, \bibnamefont{and}
  \bibinfo{author}{\bibfnamefont{L.}~\bibnamefont{Yang}},
  {``}\bibinfo{title}{Strongly bound excitons in gapless two-dimensional
  structures},{''} \bibinfo{journal}{Phys. Rev. B}
  \textbf{\bibinfo{volume}{90}}, \bibinfo{pages}{115418}
  (\bibinfo{year}{2014}).

\bibitem[{\citenamefont{Patel et~al.}(2015)\citenamefont{Patel, Havener, Brown,
  Liang, Yang, Park, and Graham}}]{OpticalExcitation2015}
\bibinfo{author}{\bibfnamefont{H.}~\bibnamefont{Patel}},
  \bibinfo{author}{\bibfnamefont{R.~W.} \bibnamefont{Havener}},
  \bibinfo{author}{\bibfnamefont{L.}~\bibnamefont{Brown}},
  \bibinfo{author}{\bibfnamefont{Y.}~\bibnamefont{Liang}},
  \bibinfo{author}{\bibfnamefont{L.}~\bibnamefont{Yang}},
  \bibinfo{author}{\bibfnamefont{J.}~\bibnamefont{Park}}, \bibnamefont{and}
  \bibinfo{author}{\bibfnamefont{M.~W.} \bibnamefont{Graham}},
  {``}\bibinfo{title}{Tunable Optical Excitations in Twisted Bilayer Graphene
  Form Strongly Bound Excitons},{''} \bibinfo{journal}{Nano Lett.}
  \textbf{\bibinfo{volume}{15}}, \bibinfo{pages}{5932} (\bibinfo{year}{2015}).

\bibitem[{\citenamefont{Koren et~al.}(2016)\citenamefont{Koren, Leven,
  L{\"{o}}rtscher, Knoll, Hod, and Duerig}}]{Koren2016}
\bibinfo{author}{\bibfnamefont{E.}~\bibnamefont{Koren}},
  \bibinfo{author}{\bibfnamefont{I.}~\bibnamefont{Leven}},
  \bibinfo{author}{\bibfnamefont{E.}~\bibnamefont{L{\"{o}}rtscher}},
  \bibinfo{author}{\bibfnamefont{A.}~\bibnamefont{Knoll}},
  \bibinfo{author}{\bibfnamefont{O.}~\bibnamefont{Hod}}, \bibnamefont{and}
  \bibinfo{author}{\bibfnamefont{U.}~\bibnamefont{Duerig}},
  {``}\bibinfo{title}{Coherent commensurate electronic states at the interface
  between misoriented graphene layers},{''} \bibinfo{journal}{Nat. Nano.}
  \textbf{\bibinfo{volume}{advance online publication}},
  (\bibinfo{year}{2016}).

\bibitem[{\citenamefont{He et~al.}(2013{\natexlab{b}})\citenamefont{He, Chu,
  and He}}]{KleinTBLG}
\bibinfo{author}{\bibfnamefont{W.-Y.} \bibnamefont{He}},
  \bibinfo{author}{\bibfnamefont{Z.-D.} \bibnamefont{Chu}}, \bibnamefont{and}
  \bibinfo{author}{\bibfnamefont{L.}~\bibnamefont{He}},
  {``}\bibinfo{title}{Chiral Tunneling in a Twisted Graphene Bilayer},{''}
  \bibinfo{journal}{Phys. Rev. Lett.} \textbf{\bibinfo{volume}{111}},
  \bibinfo{pages}{066803} (\bibinfo{year}{2013}{\natexlab{b}}).

\end{thebibliography}

\end{document}